\documentclass[review]{elsarticle}

\usepackage{hyperref}
\journal{Journal of \LaTeX\ Templates}

\usepackage{color}
\usepackage{braket}
\usepackage{amsfonts}
\usepackage{amssymb}
\usepackage{amsmath}
\usepackage{mathtools}
\usepackage{floatrow}
\usepackage{tikz}
\usepackage[margin=2.0cm]{geometry}

\newcommand{\op}[1]{\ensuremath{\hat{#1}}}
\newcommand{\ladderdown}{\ensuremath{\op{a}^{\vphantom{\dagger}}}}
\newcommand{\ladderup}{\ensuremath{\op{a}^\dagger}}
\newcommand{\annihilop}{\ladderdown}
\newcommand{\creationop}{\ladderup}

\begin{document}

\begin{frontmatter}

\title{The Uniform Electron Gas at Warm Dense Matter Conditions}

\author{Tobias Dornheim$^\dagger$}
\ead{dornheim@theo-physik.uni-kiel.de}
\author{Simon Groth$^\dagger$}
\ead{groth@theo-physik.uni-kiel.de}
\author{Michael Bonitz}
\ead{bonitz@physik.uni-kiel.de}
\address{$^\dagger$These authors contributed equally to this work.\\
Institut f\"ur Theoretische Physik und Astrophysik, Christian-Albrechts-Universit\"at zu Kiel, Leibnizstr.~15, 24098 Kiel, Germany}




\begin{abstract}
Motivated by the current high interest in the field of warm dense matter research, in this article we review the uniform electron gas (UEG) at finite temperature and over a broad density range relevant for warm dense matter applications. We provide an exhaustive overview of different simulation techniques, focusing on recent developments in the dielectric formalism (linear response theory) and quantum Monte Carlo (QMC) methods. 
Our primary focus is on two novel QMC methods that have recently allowed us to achieve breakthroughs in the thermodynamics of the warm dense electron gas: Permutation blocking path integral MC (PB-PIMC) and configuration path integral MC (CPIMC). In fact, a combination of PB-PIMC and CPIMC has allowed for a highly accurate description of the warm dense UEG over a broad density-temperature range. We are able to effectively avoid the notorious fermion sign problem, without invoking uncontrolled approximations such as the fixed node approximation.
Furthermore, a new finite-size correction scheme is presented that makes it possible to treat the UEG in the thermodynamic limit without loss of accuracy. In addition, we in detail discuss the construction of a parametrization of the exchange-correlation free energy, on the basis of these data -- the central thermodynamic quantity that provides a complete description of the UEG and is of crucial importance as input for the simulation of real warm dense matter applications, e.g., via thermal density functional theory.

A second major aspect of this review is the use of our  \textit{ab inito} simulation results to test 
previous theories, including restricted PIMC, finite-temperature Green functions, the classical mapping by Perrot and Dharma-wardana, and various dielectric methods such as the random phase approximation, or the Singwi-Tosi-Land-Sj\"olander (both in the static and quantum versions), Vashishta-Singwi and the recent Tanaka scheme for the local field correction. Thus, for the first time, thorough benchmarks of the accuracy of important approximation schemes regarding various quantities such as different energies, in particular the exchange-correlation free energy, and the static structure factor, are possible.
In the final part of this paper, we outline a way how to rigorously extend our QMC studies to the inhomogeneous electron gas. We present first \textit{ab initio} data for the static density response and for the static local field correction.
\end{abstract}

\begin{keyword}
\texttt{elsarticle.cls}\sep \LaTeX\sep Elsevier \sep template
\MSC[2010] 00-01\sep  99-00
\end{keyword}

\end{frontmatter}


\tableofcontents

\section{Introduction}


\subsection{The uniform electron gas at zero temperature}
The uniform electron gas (UEG), often referred to as ``jellium'', is one of the most important model systems in physics and quantum chemistry, and consists of Coulomb interacting electrons in a positive neutralizing background~\cite{giuliani2005quantum}. Therefore, it constitutes the quantum mechanical analogue of the classical one-component plasma (OCP)~\cite{ott_2018} and qualitatively reproduces many physical phenomena~\cite{loos_uniform_2016} such as Wigner crystallization, spin-polarization transitions, and screening. Often, it is used as a simple model system for conducting electrons in alkali metals~\cite{giuliani2005quantum,mahan1990many}.
The investigation of the UEG at zero temperature has lead to several key insights, like the BCS theory of superconductivity~\cite{bardeen_theory_1957}, Fermi liquid theory~\cite{baym2008landau,giuliani2005quantum}, and the quasi-particle picture of collective excitations~\cite{pines_collective_1952,bohm_collective_1953}.
Further, as a continuous correlated electronic quantum system, it has served as a workbench for the development of countless computational many-body methods, most prominently dielectric approximations, e.g., Refs.~\cite{bohm_collective_1953,singwi_electron_1968,vashishta_electron_1972,kugler_theory_1975,kugler_collective_????,ichimaru_strongly_1982,nozieres_theory_1999} and quantum Monte Carlo (QMC) methods~\cite{ceperley_ground_1978,ceperley_ground_1980,foulkes_quantum_2001,shepherd_convergence_2012,shepherd_full_2012,shepherd_investigation_2012,lopez_rios_inhomogeneous_2006,holzmann_backflow_2003}.
Even though the UEG itself does not represent a real physical system, its accurate description has been of paramount importance for the unrivaled success of density functional theory (DFT)~\cite{kohn_self-consistent_1965,hohenberg_inhomogeneous_1964}, the working horse of modern many-body simulations of realistic materials in solid state physics, quantum chemistry, and beyond~\cite{jones_density_2015,burke_perspective_2012,jones_density_1989}. Within the DFT framework, the complicated interacting many-electron system is mapped onto an effective one-particle (non-interacting) system via the introduction of an effective potential containing all exchange and correlation effects. While exact knowledge of the latter would require a complete solution of the many-body problem so that nothing was gained, it can often be accurately approximated by the exchange-correlation energy of the UEG, using a parametrization in dependence of density~\cite{vosko_accurate_1980,perdew_self-interaction_1981,chachiyo_communication:_2016}.

The first accurate data of the ferromagnetic and paramagnetic UEG were obtained in 1980 by Ceperley and Alder~\cite{ceperley_ground_1980}, who carried out ground state QMC simulations (see Ref.~\cite{foulkes_quantum_2001} for a review) covering a wide range of densities. Subsequently, these data were used as input for parametrizations, most notably by Vosko \textit{et al.}~\cite{vosko_accurate_1980} and Perdew and Zunger~\cite{perdew_self-interaction_1981}. Since then, these seminal works have been used thousands of times for DFT calculations in the local (spin-)density approximation (L(S)DA) and as the basis for more sophisticated gradient approximations, e.g., Refs.~\cite{perdew_generalizedd_1996,perdew_generalized_1996}.
Note that, in the mean time, there have been carried out more sophisticated QMC simulations~\cite{ortiz_correlation_1993,ortiz_correlation_1994,ortiz_erratum:_1997,ortiz_zero_1999,drummond_diffusion_2004,spink_quantum_2013}, with Spink \textit{et al.}~\cite{spink_quantum_2013} providing the most accurate energies available.

In addition to the exchange-correlation energy, there exist many parametrizations of other quantities on the basis of QMC simulations such as pair distribution functions and static structure factors~\cite{overhauser_1995,perdew_pair-distribution_1992,gori-giorgi_analytic_2000,gori-giorgi_pair_2002} and the momentum distribution~\cite{ortiz_correlation_1994,ortiz_erratum:_1997,holzmann_momentum_2011,kimball_short-range_1975,yasuhara_note_1976,starostin_quantum_2002,takada_momentum_1991,takada_kita_1991,ziesche_momentum_2012,ziesche_high-density_2010,ziesche_three-dimensional_2005,gori-giorgi_momentum_2002,maebashi_analysis_2011,takada_emergence_2016}. Finally, we mention the QMC investigation of the inhomogeneous electron gas~\cite{moroni_static_1992,moroni_static_1995,sugiyama_static_1992,bowen_static_1994,corradini_analytical_1998}, which gives important insights into the density response formalism, see Sec.~\ref{sec:response} for more details.

\subsection{Warm dense matter}

\begin{figure}\vspace*{-1.5cm}\hspace*{-0.95cm}
\includegraphics[width=1\textwidth]{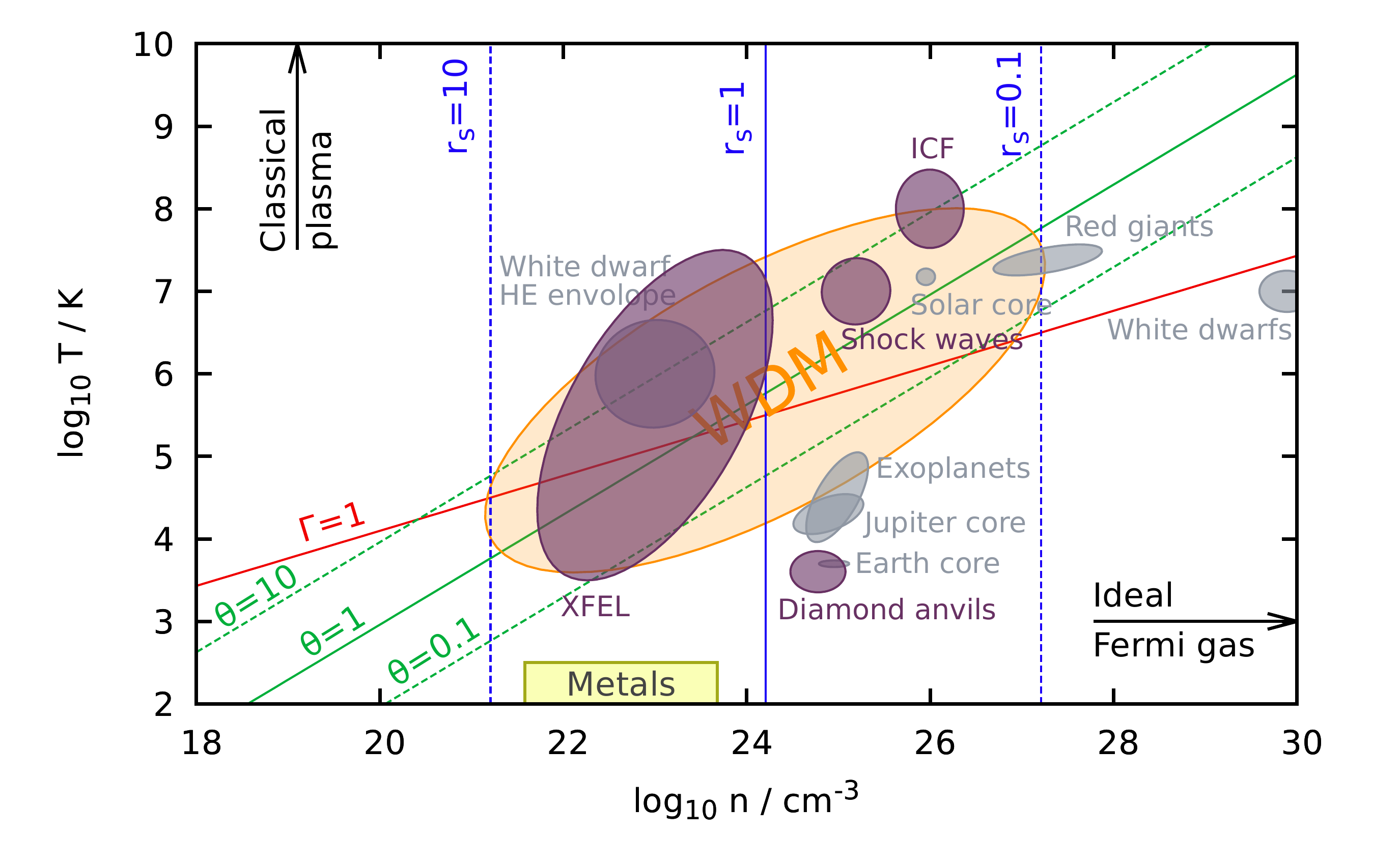}\hspace*{-0.95cm}

\caption{\label{fig:overview_wdm}Temperature-density plane around the warm dense matter (WDM, orange) regime -- Shown are lines of constant density parameter $r_s$ (blue) and reduced temperature $\theta$ (green). Purple and grey bubbles schematically sketch experimental and astrophysical applications, respectively. The various parameter ranges have been taken from Refs.~\cite{fortov_extreme_2009,gov}.
}
\end{figure}

Over the last decades, there has emerged a growing interest in the properties of matter under extreme conditions, i.e., at high temperature and densities exceeding those in solids by several orders of magnitude. This exotic state is usually referred to as \emph{warm dense matter} (WDM) and is characterized by two parameters being of the order of unity: (i) the density parameter (Wigner-Seitz radius) $r_s$, and (ii) the reduced temperature $\theta$
\begin{eqnarray}
r_s\ a_\text{B} = \left( \frac{3}{4\pi n} \right)^{1/3} \quad , \quad \theta = \frac{k_\text{B}T}{E_\text{F}} \quad ,
\end{eqnarray}
with $E_\text{F}$ being the Fermi energy defined in Eq.~(\ref{eq:fermi_energy_definition}). Here $r_s$ plays the role of a quantum coupling parameter: at high density ($r_s\to0$), the electrons behave as an ideal Fermi gas and towards low density, the Coulomb repulsion predominates, eventually leading to a Wigner crystal~\cite{wigner_interaction_1934,filinov_wigner_2001, filinov_pss_2000, drummond_diffusion_2004}. Further, $\theta$ can be understood as the quantum degeneracy parameter, where $\theta\gg 1 $ indicates a classical system (typically characterized by the classical coupling parameter $\Gamma = 1/(r_s a_\text{B} k_\text{B}T)$, cf.~the red line in Fig.~\ref{fig:overview_wdm}); for an overview on Coulomb correlation effects in classical systems, see ref.~\cite{bonitz_ropp_10}. On the other hand, 
the case $\theta\lesssim1$ characterizes a strongly degenerate quantum system. 
Thus, in the WDM regime, Coulomb coupling correlations, thermal excitations, and fermionic exchange effects are equally important at the same time. Naturally, this makes an accurate theoretical description of such systems most challenging~\cite{graziani2014frontiers}.

In nature, WDM occurs in astrophysical objects such as giant planet interiors \cite{pustow_h/he_2016,nettelmann_saturn_2013,nettelmann_uranus_2016,militzer_massive_2008,wilson_sequestration_2010,soubiran_properties_2017,vorberger_hydrogen-helium_2007,vorberger_properties_2007,nettelmann_ab_2008,french_equation_2009,knudson_probing_2012}, brown and white dwarfs~\cite{saumon_the_role_1992,hubbard_liquid_1997,collins_measurements_1998,glenzer_matter_2016,chabrier_cooling_2000}
and neutron star crusts~\cite{daligault_electronion_2009}, see Refs.~\cite{fortov_extreme_2009,shukla_colloquium_2011} for a recent review. Further areas of interest contain the physics of meteor impacts~\cite{glenzer_matter_2016} and nuclear stewardship~\cite{brumfiel_nuclear_2010}.
Another highly important aspect of warm dense matter research is the concept of inertial confinement fusion~\cite{hu_first-principles_2011,kritcher_-flight_2011,gomez_experimental_2014,schmit_understanding_2014,nora_gigabar_2015,hurricane_inertially_2016}, which could become a potentially nearly infinite source of clean energy in the future.

WDM conditions are now routinely realized at large research facilities such as the national ignition facility (NIF) at Lawrence Livermore National Lab, California~\cite{moses_national_2009,hammel_high-mode_2010}, the Z-machine at the Sandia National Labs in New Mexico~\cite{root_shock_2010,knudson_use_2003,knudson_direct_2015,matzen_pulsed-power-driven_2005,magyar_equations_2012}, the Linac Coherent Light Source (LCLS) in Stanford, California~\cite{ding_measurements_2009,fletcher_ultrabright_2015,sperling_free-electron_2015},  FLASH and the European X-FEL (free electron laser) in Hamburg, Germany~\cite{zastrau_resolving_2014,tschentscher_photon_2017} and other laser and free electron laser laboratories.
Moreover, we mention shock-compression experiments, e.g.~\cite{fortov_phase_2007,fortov_shock_2010,knudson_use_2003}. Of particular importance is X-ray Thomson scattering (XRTS), e.g.~Refs.~\cite{glenzer_observations_2007,fortmann_theory_2012,clerouin_evidence_2015,kritcher_ultrafast_2008,kraus_nanosecond_2016,zastrau_resolving_2014,davis_x-ray_2016}, which provides a widespread diagnostics for warm dense matter experiments, see Ref.~\cite{glenzer_x-ray_2009} for a review. More specifically, it allows for the direct measurement of the dynamic structure factor, which can subsequently be used to obtain, for example, the temperature~\cite{glenzer_x-ray_2009}.
Finally, we stress that WDM experiments allow for the investigation of many other quantities, such as the dielectric function~\cite{ng_outstanding_2012,ping_broadband_2006}, electrical and thermal conductivities~\cite{ng_dc_2016,ping_differential_2015,chen_evolution_2013,chen_single-shot_2016}, the electron-ion temperature equilibration~\cite{hartley_electron-ion_2015} and even the formation of transient nonequilibrium states \cite{ernstorfer_formation_2009, chen_evolution_2013}.
As a schematic overview, in Fig.~\ref{fig:overview_wdm} various important applications are depicted in the density-temperature plane around the warm dense matter regime. For a recent text book overview we refer to \cite{ebeling_fortov_filinov_17}.

Despite the remarkable experimental progress, a thorough theoretical description of warm dense matter is still lacking (even in the case of thermodynamic equilibrium), and it is well-known that simple analytic models do not sufficiently reproduce experimental measurements~\cite{desilva_electrical_1998,mostovych_reflective_1997}. Naturally, an exact quantum mechanical treatment that incorporates all correlation and excitation effects is not feasible. Unfortunately, quantum Monte Carlo methods which often allow for accurate results in the ground state are not straightforwardly extended to the simulation of fermionic matter at finite temperature. More specifically, exact fermionic path integral Monte Carlo (PIMC) simulations (see Sec.~\ref{sec:PIMC}) are severely hampered by the so-called fermion sign problem; nevertheless, there has been made some progress in direct fermionic QMC simulations by Filinov and co-workers~\cite{filinov_construction_1986,filinov_thermodynamics_2001,filinov_phase_2001,filinov_thermodynamic_2004,filinov_correlation_2007,filinov_proton_2012,filinov_fermionic_2015,filinov_thermodynamics_2015,filinov_total_2015}. To avoid the fermion sign problem, usually the fixed node approximation is utilized~\cite{ceperley_fermion_1991,ceperley_path-integral_1992,militzer_path_2000} (also ``Restricted PIMC'', RPIMC, see Sec.~\ref{sec:RPIMC}) breaks down at low temperature and high density. Therefore, RPIMC is not available over substantial parts of the warm dense regime, and the accuracy is, in general, unknown.

The probably most widespread simulation technique for warm dense matter is the combination of molecular dynamics (for the heavy ions) with a thermal density functional theory description of the electrons~\cite{mermin_thermal_1965,gupta_density_1982,Pribram-Jones2014}, usually denoted as DFT-MD~\cite{balbuena1999molecular,desjarlais_density-functional_2003,holst_thermophysical_2008,holst_electronic_2011,witte_warm_2017}. Naturally, the decoupling of the ionic and electronic systems according to the Born-Oppenheimer approximation might not be appropriate in all situations. In addition, similar to the ground state, the accuracy of the DFT calculation itself strongly relies on the specific choice of the exchange-correlation functional~\cite{clay_benchmarking_2014,clay_benchmarking_2016}. 
An additional obstacle for thermal DFT calculations is the explicit dependence of the exchange-correlation functional on temperature~\cite{karasiev_importance_2016,dharma-wardana_current_2016}, a topic which has only recently attracted serious attention, but might be crucial to achieve real predictive capability~\cite{graziani2014frontiers,pribram-jones_dft:_2015}. 
Even worse, at moderate to high temperature, the usual thermal Kohn-Sham (KS) treatment of DFT becomes unfeasible, due to the increasing number of orbitals necessary to reach convergence. 
For this reason, Militzer and co-workers have proposed to combine RPIMC at high temperature with DFT elsewhere, and successfully applied this idea to the simulations of many different materials at warm dense matter conditions~\cite{driver_all-electron_2012,militzer_development_2015,driver_first-principles_2016,zhang_first-principles_2017,driver_comparison_2017,driver_first-principles_2017}.
A possible extension of KS-DFT towards stronger excitations is given by the so-called orbital free (OF) DFT~\cite{lambert_structural_2006,lambert_properties_2007,karasiev_generalized-gradient-approximation_2012,sjostrom_fast_2014,karasiev_finite-temperature_2014,gao_validity_2016}, where the total electronic density is not represented by Kohn-Sham orbitals. While being computationally cheap and, in principle, still exact, in practice orbital free DFT relies on an approximation for the ideal part of the (free) energy~\cite{dufty_scaling_2011}, whereas the latter is treated exactly within KS-DFT. Since the ideal part usually constitutes the largest contribution, it is widely agreed that OF-DFT does not provide sufficient accuracy, and, therefore, cannot give a suitable description of warm dense matter~\cite{gao_validity_2016}.
A recent, more promising, strategy to extend KS-DFT towards higher temperature has been introduced by Zhang and co-workers, see Refs.~\cite{zhang_link_2016,gao_validity_2016,zhang_extended_2016} for details.

On the other hand, even at relatively low temperature, when the electrons are in the ground state, a DFT description for the electronic component is often not sufficient~\cite{clay_benchmarking_2014,clay_benchmarking_2016}. For this reason, Ceperley, Pierleoni and co-workers proposed to combine a classical Monte Carlo (instead of MD) for the heavy ions, with highly accurate ground-state QMC calculations for the electrons. This so-called coupled electron-ion QMC (CEIMC) method~\cite{pierleoni_computational_2005,ceperley_coupled_2002,pierleoni_coupled_2004,tubman_molecular-atomic_2015} has subsequently been applied, e.g., to the (controversially discussed, see also the recent experiments in Ref.~\cite{dias_observation_2017}) liquid-liquid phase transition in hydrogen~\cite{morales_equation_2010,pierleoni_liquidliquid_2016}. Note that, within the CEIMC approach, quantum effects of the ions can easily be included, e.g., Refs.~\cite{pierleoni_liquidliquid_2016}. In a similar spirit, Sorella and co-workers~\cite{luo_ab_2015,attaccalite_stable_2008,mazzola_distinct_2015,zen_ab_2015,mazzola_unexpectedly_2014,mazzola_finite-temperature_2012} introduced a combination of electronic ground state QMC calculations with classical MD for the ions, although, to our knowledge, no consensus with CEIMC (and, for that matter, with DFT-MD) simulations has been reached so far regarding liquid hydrogen.

In addition, there has been remarkable recent progress in the development of real time-dependent DFT calculations~\cite{ullrich2012time,Ullrich2014,baczewski_x-ray_2016,magyar_stopping_2016}, which would also give direct access to the dynamic properties of the electrons, although this topic remains in its infancy due to the high computational cost of accurate exchange correlation functionals.



Finally, we mention the possiblity of so-called quantum-classical mappings employed by Dharma-wardana \textit{et al.}~~\cite{dharma-wardana_spin-_2004,dharma-wardana_static_2006,dharma-wardana_pair-distribution_2008,dharma-wardana_classical-map_2012}, where the complicated quantum mechanical system of interest is mapped onto a classical model system with an effective ``quantum temperature'', see Sec.~\ref{sec:map} for more details.

\subsection{The warm dense electron gas}

Of particular interest for the theoretical description of WDM are the properties of the warm dense uniform electron gas. As mentioned above, an accurate parametrization of the exchange-correlation free energy with respect to temperature $\theta$, density $r_s$, and spin-polarization $\xi$ is of paramount importance for thermal DFT simulation both in the local (spin) density approximation or as a basis for more sophisticated gradient approximations~\cite{karasiev_nonempirical_2016,perdew_generalized_1996}. 
Further, direct applications of such a functional include astrophysical models~\cite{saumon_fluid_1991,saumon_fluid_1992,chabrier_quantum_1993,chabrier_equation_1998,potekhin_thermodynamic_2010,potekhin_equation_2013}, quantum hydrodynamics~\cite{crouseilles_quantum_2008,michta_quantum_2015,diaw_viscous_2017}, and the benchmark for approximations, such as finite-temperature Green function methods~\cite{kremp2006quantum,vorberger_equation_2004}, for a recent study see Ref.~\cite{kas_finite_2017}.

However, even the description of this simple model system, without an explicit treatment of the ionic component, has turned out to be surprisingly difficult. Throughout the eighties of the last century, Ebeling and co-workers~\cite{ebeling_thermodynamic_1982,richert_thermodynamic_1984,ebeling_plasma_1985,ebeling_nonideal_1989,ebeling_free_1990} proposed various interpolations between different known limits (i.e., high temperature, weak coupling, and the ground state). A more sophisticated approach is given by the dielectric formalism, which, at finite temperature, has been extensively developed and applied to the UEG by Ichimaru, Tanaka, and co-workers, see Refs.~\cite{tanaka_parametrized_1985,tanaka_thermodynamics_1986,tanaka_spin-dependent_1989,ichimaru_statistical_1987,ichimaru2004statistical_1,ichimaru2004statistical_2}. For a more comprehensive discussion of recent improvements in this field, see Sec.~\ref{sec:LRT}. In addition, we mention the classical-mapping based scheme by Perrot and Dharma-wardana~\cite{dharma-wardana_simple_2000,perrot_spin-polarized_2000}, the application of which is discussed in Sec.~\ref{sec:PDW}.
Unfortunately, all aforementioned results contain uncontrolled approximations and systematic errors of varying degrees, so that their respective accuracy has remained unclear.

While, in principle, thermodynamic QMC methods allow for a potentially exact description, their application to the warm dense UEG has long been prevented by the so-called fermion sign problem, see Sec.~\ref{sec:QMC}. For this reason, the first QMC results for this system were obtained by Brown \textit{et al.}~\cite{brown_path-integral_2013} in 2013 by employing the fixed node approximation (i.e., RPIMC). While this strategy allows for QMC simulations without a sign problem, this comes at the cost of the exact ab-initio character and it has been shown that results for different thermodynamic quantities are not consistent~\cite{karasiev_accurate_2014}. Nevertheless, these data have subsequently been used as the basis for various parametrization~\cite{sjostrom_uniform_2013,karasiev_accurate_2014,brown_exchange-correlation_2013}.

This overall unsatisfactory situation has sparked remarkable recent progress in the field of fermionic QMC simulations of the UEG at finite temperature. 
The first new development in this direction has been the configuration path integral Monte Carlo method (CPIMC, see Sec.~\ref{sec:CPIMC}), which, in contrast to standard PIMC, is formulated in second quantization with respect to plane waves, and has been developed by Schoof, Groth and co-workers~\cite{schoof_configuration_2011,schoof_towards_2015,groth_abinitio_2016}. In principle, CPIMC can be viewed as performing a Monte Carlo simulation on the exact, infinite perturbation expansion around the ideal system. Therefore, it excels at high density and strong degeneracy, but breaks down around $r_s\sim1$ and, thus, exhibits a complementary nature with respect to standard PIMC in coordinate space. Surprisingly, the comparison of exact CPIMC data~\cite{schoof_textitab_2015} for $N=33$ spin-polarized electrons with the RPIMC data by Brown \textit{et al.}~\cite{brown_path-integral_2013} revealed systematic deviations exceeding $10\%$ towards low temperature and high density, thereby highlighting the need for further improved simulations. 
Therefore, Dornheim and co-workers~\cite{dornheim_permutation_2015,dornheim_permutation_2015-1} introduced the so-called permutation blocking PIMC (PB-PIMC, see Sec.~\ref{sec:PB-PIMC}) paradigm, which significantly extends standard PIMC both towards lower temperature and higher density. 
In combination, CPIMC and PB-PIMC allow for an accurate description of the UEG over the entire density range down to half the Fermi temperature~\cite{groth_abinitio_2016,dornheim_abinitio_2016}.
Soon thereafter, these results were fully confirmed by a third independent method. This density matrix QMC (DMQMC, see Sec.~\ref{sec:DMQMC})~\cite{malone_interaction_2015,malone_accurate_2016,blunt_density-matrix_2014} is akin to CPIMC by being formulated in Fock space. Hence, there has emerged a consensus regarding the description of the electron gas with a finite number of particles~\cite{dornheim_abinitio_2017}.
The next logical step is the extrapolation to the thermodynamic limit, i.e., to the infinite system at a constant density, see Sec.~\ref{sec:FSC}. As it turned out, the extrapolation scheme utilized by Brown \textit{et al.}~\cite{brown_path-integral_2013} is not appropriate over substantial parts of the warm dense regime. Therefore, Dornheim, Groth and co-workers~\cite{dornheim_abinitio_2016,dornheim_abinitio_2016-1} have developed an improved formalism that allows to approach the thermodynamic limit without the loss of accuracy over the entire density-temperature plane.

Finally, these first \textit{ab initio} results have very recently been used by the same authors to construct a highly accurate parametrization of the exchange-correlation free energy of the UEG covering the entire WDM regime~\cite{groth_ab_2017}, see Sec.~\ref{sec:fxc}. 
Thereby, a complete thermodynamic description of the uniform electron gas at warm dense matter conditions has been achieved.

\subsection{Outline of this article}

\begin{itemize}

\item In Sec.~\ref{sec:definitions}, we start by providing some important definitions and physical quantities that are of high relevance for the warm dense UEG. Further, we discuss the jellium Hamiltonian for a finite number of electrons in a box with periodic boundary conditions, and the corresponding Ewald summation. 

\item In Sec.~\ref{sec:LRT}, we give an exhaustive introduction to the dielectric formalism within the density-density linear response theory and its application to the uniform electron gas, both in the ground state and at finite temperature. 
Particular emphasis is put on the STLS approach, which is extensively used throughout this work. Most importantly, it is a crucial ingredient for the accurate extrapolation of QMC data to the thermodynamic limit, see Sec.~\ref{sec:FSC}.
In addition, we summarize all relevant equations that are required for the implementation and numerical evaluation of various dielectric approximations.

\item In Sec.~\ref{sec:other}, we briefly discuss other approximate methods that have been applied to the warm dense UEG.
This includes the finite-temperature Green function approach, as well as two different classical mapping formalisms.

\item In Sec.~\ref{sec:QMC}, we provide an all-encompassing discussion of the application of quantum Monte Carlo methods to the uniform electron gas at warm dense matter conditions. We start with a brief problem statement regarding the calculation of thermodynamic expectation values in statistical physics. The solution is given by the famous Metropolis algorithm, which constitutes the backbone of most finite-temperature quantum Monte Carlo methods (Sec.~\ref{sec:Metropolis}). 
Undoubtedly, the most successful among these is the path integral Monte Carlo method (Sec.~\ref{sec:PIMC}), which, unfortunately, breaks down for electrons in the warm dense matter regime due to the notorious fermion sign problem (Sec.~\ref{sec:FSP}).
Two possible workarounds are given by our novel permutation blocking PIMC (Sec.~\ref{sec:PB-PIMC}) and configuration PIMC (Sec.~\ref{sec:CPIMC}) methods, which we both introduce in detail. Further mentioned are the approximate restricted PIMC method (Sec.~\ref{sec:RPIMC}) and the recent independent density matrix QMC approach (Sec.~\ref{sec:DMQMC}). The section is concluded with a thorough comparison between results for different quantities  by all of these methods for a finite number of electrons (Sec.~\ref{sec:comparison_finite_N}).

\item In Sec.~\ref{sec:FSC}, we discuss the extrapolation of QMC data that has been obtained for a finite number of electrons to the thermodynamic limit. A brief introduction and problem statement (Sec.~\ref{sec:quellbrunn}) is followed by an exhaustive discussion of the theory of finite-size effects (Sec.~\ref{sec:theory_of_finite_size_effects}). Due to the demonstrated failure of pre-existing extrapolation schemes, in Sec.~\ref{sec:improved_fsc} we present our improved finite-size correction and subsequently illustrate its utility over the entire warm dense matter regime (Sec.~\ref{sec:examples_fsc}). 

\item In Sec.~\ref{sec:comparison}, we use our new data for the thermodynamic limit to gauge the accuracy of the most important existing approaches, both for the interaction energy and the static structure factor.

\item In Sec.~\ref{sec:fxc}, we give a concise introduction (Sec.~\ref{sec:fsc_intro}) of the state of the art of parametrizations of the exchange-correlation energy of the warm dense uniform electron gas, and of their respective construction (Sec.~\ref{sec:questionable}).  Particular emphasis is put on the parametrization of the spin-dependence, Sec.~\ref{sec:spinnn}.
Finally, we provide exhaustive comparisons (Sec.~\ref{sec:param_results}) of $f_\text{xc}$ itself, and of derived quantities, which allows us to gauge the accuracy of the most widely used functionals.

\item In Sec.~\ref{sec:response}, we extend our QMC simulations to the inhomogenous electron gas. This allows us to obtain highly accurate results for the static density response function and the corresponding local field correction (Sec.~\ref{sec:response_intro}). As a demonstration, we give two practical examples at strong coupling using PB-PIMC (Sec.~\ref{sec:response_pbpimc}) and at intermediate coupling using CPIMC (Sec.~\ref{sec:cpimc_response}). 
Further, we employ our parametrization of $f_\text{xc}$ to compute the long-range asymptotic behavior of the local field correction via the compressibility sum-rule and find excellent agreement to our QMC results.

\item In Sec.~\ref{sec:outlook}, we provide a summary and give an outlook about future tasks and open questions regarding the warm dense electron gas.

\end{itemize}

\section{Important quantities and definitions\label{sec:definitions}}
\subsection{Basic parameters of the warm dense UEG}
In the following, we introduce the most important parameters and quantities regarding the warm dense electron gas. Observe, that Hartree atomic units are assumed throughout this work, unless explicitly stated otherwise.
Of high importance is the above mentioned density parameter (often denoted as Wigner-Seitz radius, or Brueckner parameter)
\begin{eqnarray}
r_s = \left( \frac{3}{4\pi n} \right)^{1/3} \quad ,
\end{eqnarray}
which is independent of temperature and spin-polarization and solely depends on the combined density of both spin-up and -down electrons, $n=n^\uparrow + n^\downarrow$. 
The spin-polarization parameter $\xi$ is defined as
\begin{eqnarray}
\xi = \frac{ n^\uparrow - n^\downarrow }{n} \in [0,1] \quad ,
\end{eqnarray}
where it is implicitly assumed that $n^\uparrow \geq n^\downarrow$. Thus,  $\xi=0$ corresponds to the unpolarized (paramagnetic) case, whereas $\xi=1$ is being referred to as the spin-polarized (ferromagnetic) case. 
For completeness, we mention that $r_s$ and $\xi$ are sufficient to fully determine the thermodynamics of the UEG in the ground state.
At warm dense matter conditions, we also require information about the temperature, usually characterized by the quantum degeneracy parameter
\begin{eqnarray}\label{eq:theta_definition}
\theta = \frac{ k_\textnormal{B} T }{ E_\textnormal{F}} \quad ,
\end{eqnarray}
with
\begin{eqnarray}\label{eq:fermi_energy_definition}
E_\text{F} = \frac{(k^\uparrow_\text{F})^2}{2}
\end{eqnarray}
denoting the Fermi energy. Observe that we always define $E_\text{F}$ with respect to the Fermi wave vector of the spin-up electrons,
\begin{eqnarray}\label{eq:fermi_wave_vector}
k^\uparrow_\text{F} = (6\pi^2 n^\uparrow)^{1/3} \quad .
\end{eqnarray}
Hence, for an ideal electron gas at zero temperature $E_\text{F}$ defines the maximum energy of the occupied one-particle orbitals. Note that in the relevant literature, there exists another possible definition of $E_\text{F}$, where the Fermi wave vector is computed with respect to the total electron density, i.e., using $k_\text{F}=(3\pi^2 n)^{1/3} $ in Eq.~(\ref{eq:theta_definition}).

The warm dense matter regime, to which the present work is devoted, is roughly characterized by $0.1 \leq r_s \leq 10$ and $0\leq \theta \leq 10$.

\subsection{The Jellium Hamiltonian:
Coordinate representation}
The description of an infinite system based on a quantum Monte Carlo simulation of a finite number of electrons $N$ in a finite simulation box with volume $V=L^3$ is usually realized by making use of periodic boundary conditions. In addition to the Coulomb interaction of the electrons in the simulation cell, one also includes the interaction with all electrons in the  infinitely many images (the same applies to the positive homogeneous background). Unfortunately, such an infinite sum with diverging positive and negative terms is only conditionally convergent, i.e., the result depends on the ordering of the terms and is not well defined~\cite{leeuw_simulation_1980}. In practice, one usually employs the Ewald summation technique (see Ref.~\cite{ballenegger_communication:_2014} for a recent accessible discussion), which corresponds to the solution of Poisson's equation in periodic boundary conditions~\cite{toukmaji_ewald_1996,fraser_finite-size_1996}. The full Hamiltonian is then given by 
\begin{eqnarray}\label{eq:UEG_Ham}
\hat H = -\frac{1}{2} \sum_{i=1}^N \nabla^2_i + \sum_{i=1}^N \sum_{k>i}^N W_\text{E}(\mathbf{r}_i, \mathbf{r}_k) + \frac{N}{2} \xi_\text{M} \quad ,
\end{eqnarray}
with the periodic Ewald pair potential being defined as~\cite{fraser_finite-size_1996}
\begin{eqnarray}\label{eq:Ewald_potential}
W_\text{E}(\mathbf{r},\mathbf{s}) = \frac{1}{V\pi} \sum_{\mathbf{G}\neq 0} \left( G^{-2}  e^{-\frac{\pi^2G^2}{\kappa^2} +2\pi \text{i} \mathbf{G}\cdot ( \mathbf{r}-\mathbf{s} ) } \right) - \frac{\pi}{\kappa^2V} + \sum_\mathbf{R}\frac{ \text{erfc}(\kappa|\mathbf{r}-\mathbf{s}+\mathbf{R}|)}{|\mathbf{r}-\mathbf{s}+\mathbf{R}|}\quad ,
\end{eqnarray}
where $\mathbf{G}=\mathbf{n} L$ and $\mathbf{R}=\mathbf{m} L^{-1}$ denote reciprocal and real lattice vectors, respectively ($\mathbf{n},\mathbf{m}\in\mathbb{Z}^3$). Furthermore, $\xi_\text{M}$ is the so-called Madelung constant, which takes into account the interaction of a charge with its own background and array of images,
\begin{eqnarray}\label{eq:Madelung}
\xi_\text{M} &=& \lim_{\mathbf{r}\to\mathbf{s}} \left( W_\text{E}(\mathbf{r},\mathbf{s}) - \frac{1}{|\mathbf{r}-\mathbf{s}|} \right)\\
&=& \frac{1}{V\pi} \sum_{\mathbf{G}\neq0} G^{-2} e^{-\frac{\pi^2G^2}{\kappa^2}} - \frac{\pi}{\kappa^2 V} + \sum_{\mathbf{R}\neq0} \frac{ \text{erfc}(\kappa R) }{R} - 2\kappa \pi^{-\frac{1}{2}}\quad .
\end{eqnarray}
Observe that both Eqs.~(\ref{eq:Ewald_potential}) and (\ref{eq:Madelung}) are independent of the specific choice for the Ewald parameter $\kappa$, which can be exploited for optimization.
Further, we note that in Eq.~(\ref{eq:UEG_Ham}) there appear no additional terms describing the uniform positive background as the average value of $W_\text{E}(\mathbf{r},\mathbf{s})$ within the simulation box vanishes~\cite{fraser_finite-size_1996}.

Let us conclude this section with some practical remarks. Obviously, a direct evaluation of the infinite sums in reciprocal and real space in Eq.~(\ref{eq:Ewald_potential}) is not possible. Fortunately, the optimal choice of the free parameter $\kappa$ leads to a rapid convergence of both sums. Furthermore, there exist numerous schemes to accelerate the computation of the Ewald potential that are advisable in different situations, such as multipole expansions~\cite{duan_ewald_2000} or using a basis of Hermite interpolants~\cite{natoli_optimized_1995}, see, e.g., Refs.~\cite{toukmaji_ewald_1996,fennell_is_2006} for an overview.
Finally, we mention the possibility for \textit{pre-averaged} pair potentials, e.g., Refs.~\cite{yakub_efficient_2003,yakub_new_2005,yakub_effective_2006,vernizzi_coulomb_2011}, which can potentially get rid of ``artificial crystal effects'' due to the infinite periodic array of images, and are computationally cheap. Recently, this idea has been applied to quantum Monte Carlo simulations of an electron gas by Filinov and co-workers~\cite{filinov_fermionic_2015}.

\subsection{The Jellium Hamiltonian: Second quantization}\label{sec:second_quant}
%
Second quantization is an efficient way to incorporate the symmetry or anti-symmetry of quantum particles in a many-particle description. Due to the indistinguishability of quantum particles the relevant observables are the occupation numbers of individual single-particle orbitals $|i\rangle$ which are solutions of the one-particle problem. Here we will concentrate on the UEG where the natural choice of orbitals are plane waves spin states. For a general introduction to the theory of second quantization we refer the reader to standard text books, e.g. \cite{kremp2006quantum, bonitz2015quantum}.

In case of the UEG, the quantization is naturally performed with respect to plane wave spin orbitals, 
$|i\rangle \to |\mathbf{k}_i\sigma_i\rangle$, with the momentum and spin eigenvalues $\mathbf{k}_i$ and $\sigma_i$, respectively.
In coordinate representation they are written as $\langle \mathbf{r} \sigma \;|\mathbf{k}_i\sigma_i\rangle = \frac{1}{L^{3/2}} e^{i\mathbf{k}_i \cdot \mathbf{r}}\delta_{\sigma,\sigma_i}$ with $\mathbf{k}=\frac{2\pi}{L}\mathbf{m}$, $\mathbf{m}\in \mathbb{Z}^3$ and $\sigma_i\in\{\uparrow,\downarrow\}$ so that the UEG Hamiltonian, Eq.~(\ref{eq:UEG_Ham}), becomes 
\begin{eqnarray}\label{eq:UEG_Ham_second}
\op{H}=
\frac{1}{2}\sum_{i}\mathbf{k}_i^2 \creationop_{i}\annihilop_{i} + \smashoperator{\sum_{\substack{i<j,k<l \\ i\neq k,j\neq l}}} 
w^-_{ijkl}\creationop_{i}\creationop_{j} \annihilop_{l} \annihilop_{k} + N\frac{\xi_\text{M}}{2}\ .
\end{eqnarray}
Here, the creation (annihilation) operator $\creationop_i$ ($\annihilop_i$) creates (annihilates) an electron in the $i$-th spin orbital, and for electrons (fermions) the operators obey the standard anti-commutation relations. Also,  $w^-_{ijkl} =w_{ijkl}-w_{ijlk}$ denotes the antisymmetrized two-electron integral with
\begin{eqnarray}\label{eq:two_ints}
w_{ijkl}=\frac{4\pi e^2}{L^3 (\mathbf{k}_{i} - \mathbf{k}_{k})^2}\delta_{\mathbf{k}_i+\mathbf{k}_j, \mathbf{k}_k + \mathbf{k}_l}\delta_{\sigma_i,\sigma_k}\delta_{\sigma_j,\sigma_l}\ ,
\end{eqnarray}
and we used the Fourier representation of the Coulomb potential.
Further, the $N-$particle states are given by Slater determinants
\begin{eqnarray}\label{eq:fock_state}
|\{n\}\rangle=|n_1, n_2, \dots\rangle\ ,
\end{eqnarray}
with the fermionic occupation number $n_i\in\{0,1\}$ of the $i$-th plane wave spin-orbital. Obviously, the second quantization representation of the UEG Hamiltonian has two practical advantageous compared to its coordinate representation: 1) the Ewald interaction only enters in a trivial way via the Madelung constant, $\xi_\text{M}$, thus not requiring any elaborate evaluation of the interaction part, and 2), the correct Fermi statistics are automatically incorporated via the usual fermionic anti-commutator relations of the creation and annihilation operators.


\section{Dielectric Approximations and Linear Response Theory\label{sec:LRT}}

\subsection{Introduction}
Before the advent of the first exact but computationally highly demanding quantum Monte Carlo simulations of the UEG in the late 1970s, the approximate approaches based on the dielectric formulation~\cite{kugler_theory_1975,kugler_collective_????,ichimaru_strongly_1982,nozieres_theory_1999,giuliani2005quantum} have arguably constituted the most vital tool for gaining crucial insights into correlated quantum many-body systems. 
In the ground state, a seminal work in this direction have been provided by Bohm and Pines with the formulation of the random phase approximation (RPA)~\cite{bohm_collective_1953,nozieres_theory_1999,klimontowich_1952},
which becomes exact in both the long wavelength and high density limit and thus sufficiently describes long-range phenomena. Later, an alternative derivation of the RPA has been performed by Gell-Mann and Brueckner~\cite{gell-mann_correlation_1957} through a summation of Feynman diagrams leading to the first exact expansion of the correlation energy of the UEG in the high density regime. However, at metallic densities, $r_s\approx1.5,\dots,7$, the RPA dramatically overestimates short-range correlations between the electrons resulting in significantly too low correlation energies and an unphysical negative value of the pair-correlation function at zero distance. To overcome these shortcomings, Singwi, Tosi, Land and Sj\"olander (STLS)~\cite{singwi_electron_1968} proposed a self-consistent scheme that allows for an approximate but greatly improved treatment of the short-range exchange and correlation effects. Most notably, the STLS scheme predicted the exchange-correlation energies that have later been accurately computed by Ceperley and Alder~\cite{ceperley_ground_1978,ceperley_ground_1980} with an impressive accuracy of $\sim 1\%$ even up to densities $r_s\sim 20$ (see e.g. Ref.~\cite{tanaka_correlational_2016}). Nevertheless, the obtained pair-correlation functions still become slightly negative at densities $r_s\geq 4$, but, compared to the RPA, the magnitude of this error is strongly reduced. A further issue regarding the STLS scheme is the violation of the exact compressibilty sum rule, Eq.~(\ref{eq:CSR}). Vashishta and Singwi (VS) could modify the self-consistent scheme by also taking the density derivative of the pair-correlation function into account so that the compressibility sum rule is almost exactly verified~\cite{vashishta_electron_1972, hayashi_electron_1980}, though this lead to a reduced quality of the pair-correlation function and exchange-correlation energy. 

All of the mentioned schemes beyond RPA rely on a static (frequency-independent) approximation of the so-called local field correction, the central quantity in the dielectric formulation. There have been many attempts to further increase the overall accuracy of the static dielectric methods (for an overview see e.g.~Ref.~\cite{ichimaru_strongly_1982}), and even the extension to a more consistent formulation based on a dynamical local field correction has been achieved~\cite{hasegawa_electron_1975,holas_dynamic_1987}. However, regarding the interaction energy, the static STLS scheme turned out to give the most accurate results. 

Due to a former lack of experimental motivation, the extension of some of the dielectric approaches to finite temperature and their application to the UEG were carried out much later. The first calculations in the RPA have been carried out by Gupta and Rajagopal~\cite{gupta_inhomogeneous_1980,gupta_exchange-correlation_1980,gupta_density_1982}, which have later been revised and parametrized by Perrot and Dharma-wardana~\cite{perrot_exchange_1984}. After that, countless important contributions to this field have been made by Tanaka and Ichimaru~\cite{tanaka_parametrized_1985,tanaka_thermodynamics_1986,tanaka_spin-dependent_1989,ichimaru_statistical_1987,ichimaru2004statistical_1,ichimaru2004statistical_2}, who applied many of the static dielectric methods, i.e. with some static ansatz for the LFC, to the quantum and classical UEG at finite temperature. Among these works is the finite temperature STLS scheme~\cite{tanaka_thermodynamics_1986}, which, likewise to the ground state, predicted the exact exchange-correlation energy~\cite{groth_ab_2017,tanaka_correlational_2016,tanaka_improved_2017} with a similar impressive accuracy of $\sim 1\%$, cf.~Sec.~\ref{sec:comparison}. However, a consistent extension of the static finite temperature VS scheme~\cite{stolzmann_static_2001} could only be achieved much later~\cite{sjostrom_uniform_2013}, since the fulfillment of the compressibility sum rule turned out to be more elaborate here. Furthermore, Schweng and B\"ohm developed the finite temperature version of the dynamical STLS scheme~\cite{schweng_finite-temperature_1993} and successfully used it for a detailed investigation of the static LFC of the UEG, while a generalization to arbitrary spin-polarization of this formalism has been provided only very recently~\cite{arora_spin-resolved_2017}. 

We mention that, regarding the benefits and merits of the specific variants of the dielectric methods, the qualitative statements for the ground state given above also apply to their finite temperature extensions. Moreover, in addition to its predictive capabilities prior to the advent of the more accurate QMC simulations, in particular the RPA and STLS approach played an important role in the extrapolation of the results obtained from a finite simulation system (finite particle number $N$ and simulation box with volume $V$) to the thermodynamic limit, i.e.~$N,V\xrightarrow[]{n=\textnormal{const}}\infty$ (see Sec.~\ref{sec:FSC}). In addition, very recently, the temperature dependence of the STLS interaction energy has been successfully used to bridge the gap between the ground state and finite temperature QMC data which are available only above half the Fermi temperature~(see Sec.~\ref{sec:fxc}).

\subsection{Density response, dielectric function, local field correction, and structure factor}
The dielectric formulation is derived within the framework of the linear density-density response theory, where we are interested in the change of the electron density when a periodic (both in space and time) external potential with wavenumber $\mathbf{q}$, frequency $\omega$, and  amplitude $\Phi(\mathbf{q},\omega)$ is applied to the system, i.e., 
\begin{eqnarray}\label{eq:externalPot}
\Phi_\text{ext}(\mathbf{r},t)=\frac{1}{V}\Phi(\mathbf{q},\omega)\,e^{i[\mathbf{q}\mathbf{r}-(\omega-i\eta)t]} + \text{c.c.}\ 
\end{eqnarray}
The infinitesemal positive constant $\eta=0^+$ ensures that the perturbation vanishes at $t\to -\infty$ so that we can assume that the system has been in thermal equilibrium in the past and the external field has been switched on  adiabatically. Provided that the amplitude is sufficiently small and the unperturbed system is homogeneous, one can show that the resulting change in the electron density is given by~\cite{giuliani2005quantum,ichimaru2004statistical_1}
\begin{eqnarray}\label{eq:densityChange}
\delta n(\mathbf{r},t) =n(\mathbf{r},t) - n(\mathbf{r})_0 =  \frac{1}{V}\Phi(\mathbf{q},\omega)\chi(\mathbf{q},\omega)e^{i(\mathbf{q}\mathbf{r}-\omega t)} + \text{c.c.}\ ,
\end{eqnarray}
where we have introduced the Fourier transform of the density-density response function
\begin{eqnarray}\label{eq:chi_ft}
\chi(\omega,\mathbf{q}) = \lim_{\eta\to0}\int_{-\infty}^\infty \text{d}\tau\ 
e^{(i\omega-\eta)\tau}\tilde\chi(\mathbf{q},\tau)\ .
\end{eqnarray}
with its standard definition\footnote{Note that we restrict ourselves to the unpolarized case throughout the present section. Therefore, the response function $\chi$ is equal to the total response function of both spin-up and -down electrons.}
\begin{eqnarray}
\tilde\chi(\mathbf{q},\tau) = -i \braket{ \left[ \op{n}(\mathbf{q},\tau),\op{n}(-\mathbf{q},0) \right] }_0 \Theta(\tau)\ .
\end{eqnarray}
Here, $\braket{\cdot}_0$ denotes the ensemble average of the unperturbed system, and the time dependence of the Fourier transform of the density operator $\op{n}(\mathbf{q})=\sum_ie^{-i\mathbf{q}\op{\mathbf{r}}_i}$ is determined by the Heisenberg picture with respect to the unperturbed Hamiltonian, i.e., $\op{n}(\mathbf{q},t)=e^{i\op{H}_0t}\op{n}(\mathbf{q})e^{-i\op{H}_0t}$. From Eq.~(\ref{eq:densityChange}) we immediately see that the amplitude of the induced density fluctuations is simply
\begin{eqnarray}
n(\mathbf{q},\omega) =\frac{1}{V} \Phi(\mathbf{q},\omega)\chi(\mathbf{q},\omega)\ . 
\end{eqnarray}
Hence, all information of the system's response to the external perturbation, Eq.~(\ref{eq:externalPot}), is contained in the density-density response function $\chi(\mathbf{q},\omega)$.  Via the polarization potential approach~\cite{ichimaru2004statistical_1} it can be shown that the exact density response function can always be expressed in terms of the ideal (Lindhard) response function, $\chi_0$, and the so-called local field correction (LFC), $G$, as
\begin{eqnarray}\label{eq:chi_lfc}
\chi(\mathbf{q},\omega) = \frac{\chi_0(\mathbf{q},\omega)}{1-\frac{4\pi}{q^2}[1-G(\mathbf{q},\omega)]\chi_0(\mathbf{q},\omega)}\ ,
\end{eqnarray}
where the RPA response function is recovered when setting $G\equiv 0$, i.e.,
\begin{eqnarray}\label{eq:chi_rpa}
\chi_\text{RPA}(\mathbf{q},\omega) = \frac{\chi_0(\mathbf{q},\omega)}{1-\frac{4\pi}{q^2}\chi_0(\mathbf{q},\omega)} \ .
\end{eqnarray}
Thus, the LFC covers all correlation effects in the response of the system to a weak external potential. The imaginary part of the response function is linked to the dynamic structure factor $S(\mathbf{q},\omega)$ via the fluctuation dissipation theorem~\cite{giuliani2005quantum}
\begin{eqnarray}\label{eq:flucDiss}
\text{Im}\chi(\mathbf{q},\omega)= -\frac{\pi}{V} \left(1-e^{-\beta\omega}\right)S(\mathbf{q},\omega)\ ,
\end{eqnarray}
which can in turn be utilized to express the static structure factor
\begin{eqnarray}
S(\mathbf{q}) =\frac{1}{N}\int \mathrm{d}\omega \,S(\mathbf{q},\omega) = \frac{1}{N}\braket{\op{n}(\mathbf{q})\op{n}(\mathbf{-q})}_0
\end{eqnarray}
in terms of the response function
\begin{eqnarray}\label{eq:flucDissStatic}
S(\mathbf{q}) = - \frac{1}{2\pi n}\mathcal{P}\int_{-\infty}^\infty\mathrm{d}\omega
\coth{\left(\frac{\omega}{2T}\right)} \text{Im}\chi(\mathbf{q},\omega)\ ,
\end{eqnarray}
where $\mathcal{P}$ denotes the principal value, which is necessary due to the poles of the integrand on the real axis. Thereby we have obtained a direct connection between the dynamic properties of the system, i.e., within the linear response regime, and its thermodynamic properties. Note that the response function obeys the Kramers-Kronig relations
\begin{align}\label{eq:kramers}
\text{Re} \chi(\mathbf{q},\omega) &= \frac{2}{\pi} \mathcal{P}\int_0^\infty\mathrm{d}\nu \, \frac{\nu \,\text{Im}\chi(\mathbf{q},\omega)}{\nu^2-\omega^2}\ ,\\
\text{Im} \chi(\mathbf{q},\omega) &=- \frac{2\omega}{\pi} \mathcal{P}\int_0^\infty\mathrm{d}\nu \, \frac{ \text{Re}\chi(\mathbf{q},\omega)}{\nu^2-\omega^2}\ ,\nonumber
\end{align}
and hence, the real part of the response function can always be computed from its imaginary part and vice versa. The central idea of all dielectric approaches consists in deriving an approximate expression for the LFC so that it is expressed as a functional of the static structure factor, i.e.~$G=G[S]$. Then, together with Eqs.~(\ref{eq:chi_lfc}) and (\ref{eq:kramers}), one has a closed set of equations, which, in principal can be solved iteratively starting from the RPA ($G=0$), where the real and imaginary part of the ideal (Lindhard) response function, $\chi_0$, are readily evaluated numerically~\cite{giuliani2005quantum}. However, from a numerical point of view this approach is highly inconvenient due to the infinitely many poles of the integrands in the Eqs.~(\ref{eq:flucDissStatic}) and (\ref{eq:kramers}). A solution to this problem has been provided by Tanaka and Ichimaru~\cite{tanaka_thermodynamics_1986}, who reformulated the aforementioned set of equations for the complex valued density-density response function defined by 
\begin{eqnarray}
\tilde{\chi}(\mathbf{q},z):=\int_{-\infty}^\infty\frac{\mathrm{d}\nu}{\pi} \frac{\text{Im}\chi(\mathbf{q},\nu)}{\nu-z}\ ,
\end{eqnarray}
which, under the frequency integral, fulfills $2i\text{Im}\chi(\mathbf{q},\omega)=\lim_{\eta\to 0^+}\tilde{\chi}(\mathbf{q},\omega+i\eta)-\tilde{\chi}(\mathbf{q},\omega-i\eta)$, so that Eq.~(\ref{eq:flucDissStatic}) becomes 
\begin{align}
S(\mathbf{q}) &= - \frac{1}{4\pi i n}\mathcal{P}\lim_{\eta\to 0}\int_{-\infty}^\infty\mathrm{d}\omega
\coth{\left(\frac{\omega}{2T}\right)}\left[ \tilde{\chi}(\mathbf{q},\omega+i\eta)-\tilde{\chi}(\mathbf{q},\omega-i\eta)\right]\ .
\end{align}
Now the integral can be interpreted as a closed contour integral
\begin{align}\label{eq:contour}
 S(\mathbf{q})= -\frac{1}{4\pi i n} \lim_{\epsilon\to 0^+}\lim_{\eta\to 0^+}\lim_{R\to \infty}
\oint_{\text{C}}\mathrm{d}z \coth{\left(\frac{ z}{2T}\right)} \text{Im}\tilde{\chi}(\mathbf{q},z)\ ,
\end{align}
\begin{figure}
\center{\includegraphics[width=0.5\textwidth]{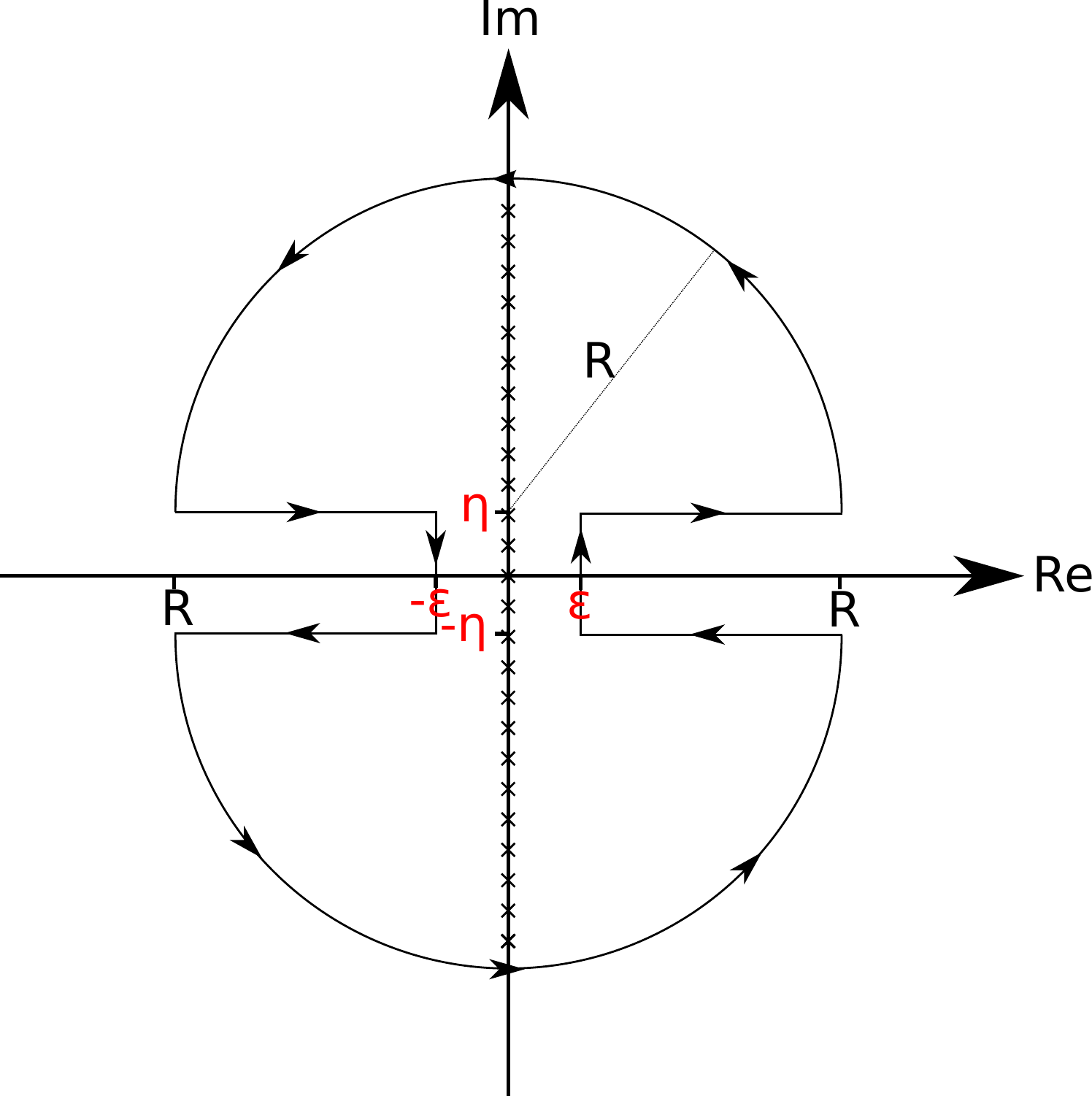}}
\caption{Illustration of the integration contour $\textnormal{C}$ in Eq.~(\ref{eq:contour}). Crosses indicate the poles of the hyperbolic cotangent.}
\label{fig:contour}
\end{figure}
with the explicit form of the contour being depicted in Fig.~\ref{fig:contour}, where the limit $R\to \infty$ is taken prior to the integration whereas $\epsilon,\eta\to 0^+$ is taken afterwards. Since the integrand is analytic on $C$, the contour integral can be solved by applying the residue theorem yielding
\begin{eqnarray}\label{eq:flucDiss_Matsubara_sum}
S(\mathbf{q}) = -\frac{T}{n}\sum_{l=-\infty}^{\infty}\tilde{\chi}(\mathbf{q}, z_l)\ ,
\end{eqnarray}
with the Matsubara frequencies $z_l$ representing the poles of the cotangent hyperbolic function on the imaginary axis,
\begin{eqnarray}
z_l= 2\pi i l T \ .
\end{eqnarray}
Hence, the frequency integral in Eq.~(\ref{eq:flucDissStatic}) can be replaced by a sum over the Matsubara frequencies, which is much more convenient for numerical evaluation.

Similar to the real frequency dependent response function, cf.~Eq.~(\ref{eq:chi_lfc}), the exact complex valued response function can be rewritten in terms of the complex valued ideal response function and LFC~\cite{tanaka_parametrized_1985,schweng_finite-temperature_1993},
\begin{eqnarray}\label{eq:chiComplex_lfc}
\tilde{\chi}(\mathbf{q},z) = \frac{\tilde{\chi}_0(\mathbf{q},z)}{1-\frac{4\pi}{q^2}[1-\tilde{G}(\mathbf{q},z)]\tilde{\chi}_0(\mathbf{q},z)}\ .
\end{eqnarray}
In the thermodynamic limit\footnote{As usual, replacing $\frac{1}{V}\sum_\mathbf{q}$ by $\int\frac{\mathrm{d}\mathbf{q}}{(2\pi)^3}$ transforms the expressions for the finite system (with periodic boundary conditions) to the thermodynamic limit.}, the finite temperature complex valued ideal response function is given by
\begin{eqnarray}
\tilde{\chi}_0(\mathbf{q},z)=-2\int\frac{\mathrm{d}\mathbf{k}}{(2\pi)^3}\frac{f(\mathbf{k}+\mathbf{q})-f(\mathbf{k})}{z-\epsilon_{\mathbf{k}+\mathbf{q}}+\epsilon_{\mathbf{k}}}\ ,
\end{eqnarray}
with $\epsilon_\mathbf{k}=k^2/2$ and $f$ being the Fermi distribution
\begin{eqnarray}
f(\mathbf{k})=\frac{1}{e^{k^2/(2T)-\alpha}+1}\ ,
\end{eqnarray}
where the reduced chemical potential $\alpha=\mu/T$ is determined by the normalization condition
\begin{eqnarray}\label{eq:normailzation_condition}
\int\frac{\mathrm{d}\mathbf{k}}{(2\pi)^3} f(\mathbf{k}) = \frac{n}{2}\ .
\end{eqnarray}
For numerical evaluation of the ideal response function at the different Matsubara frequencies  the following form is most suitable~\cite{tanaka_parametrized_1985, tanaka_thermodynamics_1986}:
\begin{eqnarray}\label{eq:complex_ideal_chi} 
\tilde{\chi}_0(\mathbf{q},z_l) = - \frac{2}{q} \int_0^\infty \frac{\mathrm{d}k}{(2\pi)^2} \frac{k}{e^{k^2/(T2)-\alpha}+1}
\ln{\left[\frac{(4\pi l T)^2 + (q^2+ 2qk)^2}{(4\pi l T)^2 + (q^2- 2qk)^2}\right]}\ .
\end{eqnarray}
\subsection{Approximations for the local field correction}
In the static dielectric approaches one approximates the dynamic LFC by its static value, i.e., replacing $\tilde{G}(\mathbf{q},z)$ by $\tilde{G}(\mathbf{q},0)$ in Eq.~(\ref{eq:chiComplex_lfc}), which turns out to be highly accurate in many cases. The most successful and widely used approximation for the static LFC is given by the one utilized in the STLS scheme~\cite{singwi_electron_1968}
\begin{align}\label{eq:stls_LFC}
G_{\textnormal{STLS}}(\mathbf{q},0) &= -\frac{1}{n}\int \frac{\mathrm{d}\mathbf{k}}{(2\pi)^3}\; \frac{\mathbf{q}\cdot\mathbf{k}}{\mathbf{k}^{ 2}} \left[S(\mathbf{q}-\mathbf{k}) - 1\right] \\
&= - \frac{1}{n}\int_0^\infty \frac{\mathrm{d}k}{(2\pi)^2}\; k^2 [S(k)-1] \left[\frac{q^2-k^2}{4kq}\ln\left(\frac{(q+k)^2}{(q-k)^2}\right) +1\right]\nonumber\ .
\end{align}
This expression is derived from the classical equation of motion of the one-particle distribution function, $f(\mathbf{r}_1,\mathbf{p}_1,t)$, by making the following product ansatz for the two-particle distribution function\footnote{Note that this ansatz can be further improved by considering an explicitly time-dependent pair distribution function, see Refs.~\cite{kahlert_dynamics_2014,kahlert_linear_2015}.}:
\begin{eqnarray}\label{eq:stls_derivation_ansatz}
f(\mathbf{r}_1,\mathbf{p}_1, \mathbf{r}_2,\mathbf{p}_2,t)\approx f(\mathbf{r}_1,\mathbf{p}_1,t)f(\mathbf{r}_2,\mathbf{p}_2,t)g_\text{eq}(\mathbf{r}_1-\mathbf{r}_2)\ ,
\end{eqnarray}
where $g_\text{eq}(\mathbf{r})$ denotes the exact equilibrium pair-distribution function. Since the two-particle distribution function couples to the three-particle distribution function and so on, Eq.~(\ref{eq:stls_derivation_ansatz}) serves as a closure relation of the hierarchy. 

The equations~(\ref{eq:flucDiss_Matsubara_sum}), (\ref{eq:chiComplex_lfc}), and (\ref{eq:stls_LFC}) now form a closed set of equations, which are self-consistently solved as follows:
\begin{enumerate}
    \item Compute the reduced chemical potential $\alpha$ by solving Eq.~(\ref{eq:normailzation_condition}).
    \item Compute and store the values of the ideal response function, $\chi_0(\mathbf{q},z_l)$, for sufficiently large values of $l$ ensuring that Eq.~(\ref{eq:flucDiss_Matsubara_sum}) always converges throughout the iteration.
    \item Compute the response function from Eq.~(\ref{eq:chiComplex_lfc}), initially by setting $G=0$.
    \item Compute the static structure factor $S(\mathbf{q})$ from Eq.~(\ref{eq:flucDiss_Matsubara_sum}).
    \item Compute the new LFC $G_\text{STLS}(\mathbf{q},0)$ from Eq.~(\ref{eq:stls_LFC}).
    \item Repeat steps 3 to 5 until convergence is reached. 
\end{enumerate}
For completeness we mention that, in particular at low temperature, the sum in Eq.~(\ref{eq:flucDiss_Matsubara_sum}) may only converge for extremely large values of $l$, but this obstacle can be overcome by separating those contributions for which the summation can be performed analytically beforehand, see Ref.~\cite{tanaka_parametrized_1985} for details.

Naturally, from the converged static structure factor we directly obtain the interaction energy (per particle) for the corresponding temperature and density parameter,
\begin{eqnarray}\label{eq:interaction_en_from_s}
v(\theta,r_s) = \frac{1}{\pi}\int_0^\infty\mathrm{d}k\;[S(k;r_s,\theta)-1]\ ,
\end{eqnarray}
which can in turn be used to compute the exchange-correlation free energy via the standard coupling constant integration 
\begin{eqnarray}\label{eq:fxc_from_v}
f_\textnormal{xc}(r_s,\theta) = \frac{1}{r_s^2}\int_0^{r_s}\mathrm{d}\bar{r}_s \;v(\theta,\bar{r}_s)\ .
\end{eqnarray}
As mentioned before, both in the ground state and at finite temperature, the STLS scheme provides highly accurate interaction energies, which is partly the result of a favourable error cancellation in Eq.~(\ref{eq:interaction_en_from_s}) as the STLS static structure factor tends to be slightly too large for small k-vector and vice versa, see Fig.~\ref{fig:SSF_comparison_dielectric} in Sec.~\ref{sec:FSC}. It is important to note that, compared to the RPA, the STLS structure factor and related thermodynamic properties are of substantially higher accuracy. In particular, the negative values of the pair-distribution function at zero distance, $g(0)$, are significantly reduced, although it still becomes slightly negative at lower densities~\cite{tanaka_thermodynamics_1986}. However, there is a well-known drawback regarding the consistency of the STLS results: the compressibility sum rule (CSR) is violated. The CSR is an exact property of the UEG linking the long-wavelength limit of the static LFC $G(q,0)$ to the second derivative of the exchange-correlation free energy:
\begin{eqnarray}\label{eq:CSR}
\lim_{q\to0} G(q,0) = - \frac{q^2}{4\pi} \frac{\partial^2}{\partial n^2} \left( n f_\text{xc} \right)\ .
\end{eqnarray}
Substituting $G^\text{STLS}(q,0)$ and $f^{\textnormal{STLS}}_\text{xc}$ on the left- and right-hand side of Eq.~(\ref{eq:CSR}) gives different results, which demonstrates that the STLS scheme does not provide a consistent physical description of the UEG. Moreover, the long range limit of the LFC differs significantly from the exact QMC result, which is shown in Fig.~\ref{fig:LFC_compressibility} in Sec.~\ref{sec:response}. In the ground state, Vashishta and Singwi~\cite{vashishta_electron_1972} proposed to modify the STLS expression, Eq.~(\ref{eq:stls_LFC}), for the LFC such that
\begin{eqnarray}
G^\textnormal{VS}(\mathbf{q},0) = \left(1+an\frac{\partial}{\partial n}\right)G^\textnormal{STLS}(\mathbf{q},0)\ ,
\end{eqnarray}
where the right choice of the additional free parameter $a$, in principle, allows for the exact fulfillment of Eq.~(\ref{eq:CSR}). In fact, they empirically found that setting $a=2/3$ reasonably satisfies the CSR for all densities in the ground state. Only recently, Sjostrom and Dufty~\cite{sjostrom_uniform_2013} successfully extended this approach to the finite temperature UEG. They even refined the approach by making the free parameter dependent on density and temperature, i.e., $a=a(r_s,\theta)$, and actually included the CSR, Eq.~(\ref{eq:CSR}), into the self-consistent scheme, which requires to simultaneously perform calculations for different values of $r_s$. Thereby, the obtained results are physically more consistent in that they do exactly fulfill the CSR. However, the overall quality of the thermodynamic quantities is decreased compared to the STLS scheme; for example, $g(0)$ becomes more negative~\cite{sjostrom_uniform_2013}. 

Since the accuracy of the STLS scheme decreases when the density parameter becomes too large, $r_s\gtrsim 20$, there have been many attempts to derive more refined expressions for the static LFC that perform better in the strong coupling regime (see e.g.~\cite{ichimaru_statistical_1987,ichimaru_strongly_1982}). Among them are the so-called (modified) convolution [(M)CA] and hypernetted chain approximations [(M)HNC] for the LFC. Both are known to be highly accurate for the description of the classical one-component plasma over the entire fluid regime~\cite{ichimaru_statistical_1987}. While the MCA scheme has been used earlier for the construction of a temperature, density, and spin-dependent parametrization of the exchange-correlation free energy of the UEG~\cite{tanaka_spin-dependent_1989}, the HNC scheme has only recently been applied to the UEG at warm dense matter conditions~\cite{tanaka_correlational_2016} and, compared to the STLS scheme, showed overall improved results for the thermodynamic properties but not for the interaction energy. The LFC in the HNC approximation is derived from the hypernetted chain equation for classical liquids~\cite{springer_integral_1973,ng_hypernetted_1974}, which yields~\cite{tanaka_correlational_2016} 
\begin{eqnarray}\label{eq:HNC_LFC}
G^{\text{HNC}}(\mathbf{q},0) = G^\text{STLS}(\mathbf{q},0) + \frac{1}{n}\int \frac{\mathrm{d}\mathbf{k}}{(2\pi)^3} \frac{\mathbf{q}\cdot \mathbf{k}}{k^2} [S(\mathbf{q}-\mathbf{k})-1][G(\mathbf{k},0)-1][S(\mathbf{k})-1]\ ,
\end{eqnarray}
where the CA expression is recovered by setting $G(\mathbf{k},0)\equiv 0$ on the left-hand side of Eq.~(\ref{eq:HNC_LFC}). Further, the corresponding modified versions, MCA and MHNC, are obtained by replacing $S(\mathbf{q}-\mathbf{k})$ by a screening function
\begin{eqnarray}
\bar{S}(q)=\frac{q^2}{q^2+q^2_s}\ .
\end{eqnarray}
The screening parameter $q_s$ is determined consistently from the condition
\begin{eqnarray}
\frac{1}{2}\int\frac{\mathrm{d}\mathbf{q}}{(2\pi)^3} \frac{4\pi}{q^2}[\bar{S}(q)-1] = 
\frac{1}{2}\int\frac{\mathrm{d}\mathbf{q}}{(2\pi)^3} \frac{4\pi}{q^2}[S(q)-1]\ ,
\end{eqnarray}
so that $S$ and $\bar{S}$ must correspond to the same interaction energy. Using the modified versions with the screening function has the practical advantage that, like the STLS contribution to the total LFC, cf.~Eq.~(\ref{eq:stls_LFC}), also the second term in Eq.~(\ref{eq:HNC_LFC}) can be recast into a one-dimensional integral~\cite{tanaka_correlational_2016}, i.e.,
\begin{eqnarray}
G^{\text{MHNC}}(\mathbf{q},0) 
= G^\text{STLS}(\mathbf{q},0)+\frac{q_s^2}{n}\int_0^\infty \frac{\mathrm{d}k}{(2\pi)^2}\left[
1 + \frac{k^2+q^2+q_s^2}{4qk}\ln{\left(\frac{(k-q)^2+q_s^2}{(k+q)^2+q_s^2}\right)}
\right]
[G(k)-1][S(k)-1]\ ,
\end{eqnarray}
which significantly speeds up the convergence process. Unfortunately, this is not possible for the full HNC LFC, Eq.~(\ref{eq:HNC_LFC}), and thus, one actually must carry out the three dimensional integration.

At this point it is important to note that all the above static dielectric schemes are somewhat classical in spirit since the utilized approximate expressions for the static LFC are all derived within purely classical theories. In other words, the discussed methods may be interpreted as being quantum mechanically only on the level of the RPA, while correlation effects are treated classically. In accordance to Eqs.~(\ref{eq:chi_lfc}) and~(\ref{eq:complex_ideal_chi}) the exact LFC must also depend on the frequency. First, Hasegawa and Shimizu~\cite{hasegawa_electron_1975} performed the formal derivation of the dynamic STLS LFC by closing the hierarchy for the equation of motion of the Wigner distribution with the same product ansatz, Eq.~(\ref{eq:stls_derivation_ansatz}), that has been used in the static STLS formalism for the classical distribution function. Due to its consistent quantum mechanical derivation, this approach has been termed quantum STLS (qSTLS). In the ground state, the first numerical calculations and detailed investigations have been carried out by Holas and Rahman~\cite{holas_dynamic_1987}. Later, the qSTLS scheme has also been applied to the finite temperature UEG, and more recently, it has been generalized to allow for the calculation of spin-resolved quantities~\cite{schweng_finite-temperature_1993,arora_spin-resolved_2017}. The dynamical LFC in the qSTLS scheme is given by
\begin{eqnarray}\label{eq:lfc_qstls}
G^\textnormal{qSTLS}(\mathbf{q},z_l)= -\frac{1}{n}\int \frac{\mathrm{d}\mathbf{k}}{(2\pi)^3}\frac{\bar{\chi}_0(\mathbf{q},\mathbf{k},z_l)}{\bar{\chi}_0(\mathbf{q},z_l)}\frac{k^2}{q^2}[S(\mathbf{k-\mathbf{q}})-1]\ ,
\end{eqnarray}
with the generalized response function (two arguments) being defined as
\begin{align}
\chi_0(\mathbf{q},\mathbf{k},z_l)&=-2\int\frac{\mathrm{d}\mathbf{p}}{(2\pi)^3}\frac{f(\mathbf{p}+\mathbf{k}/2)-f(\mathbf{p}-\mathbf{k}/2)}{z-\epsilon_{\mathbf{p}+\mathbf{q}/2}+\epsilon_{\mathbf{p}-\mathbf{q}/2}}\\
&= -\frac{2}{q}\int_0^\infty \frac{\mathrm{d}p}{(2\pi)^2}\,p\,f(p)\ln{\left[
\frac{(4\pi l T)^2 + (2pq+\mathbf{q}\mathbf{k})}{(4\pi l T)^2 + (2pq-\mathbf{q}\mathbf{k})}\nonumber
\right]}\quad .
\end{align}
For practical purposes, the qSTLS LFC, Eq.~(\ref{eq:lfc_qstls}), can be reduced to a three-dimensional integral~\cite{schweng_finite-temperature_1993}. Overall, compared to the static STLS approach, the qSTLS approach significantly improves the short-range behavior of the pair-correlation function. Most notably, the obtained results for the static LFC are physically more reasonable as they can exhibit important physical features. For example, they can actually have a maximum larger than one, a necessary condition for the occurrence of charge density waves~\cite{schweng_finite-temperature_1993}. This is in stark contrast to the static dielectric approaches where the static LFC usually converges monotonically to unity with increasing $k-$vector. Yet, the improvement of the interaction energy due to the qSTLS scheme is rather small.

An exhaustive overview of comparison between the dielectric approximation and recent, highly accurate quantum Monte Carlo data can be found in Sec.~\ref{sec:comparison} for the static structure factor and the interaction energy and, in Sec.~\ref{sec:response}, for the static density response function and local field correction.


\section{Other approximate approaches\label{sec:other}}

\subsection{Finite-temperature (Matsubara) Green functions}
An alternative derivation of the dielectric function encountered in the previous section can be achieved within the framework of quantum kinetic theory~\cite{bonitz2015quantum}. In this formalism, correlation effects are usually incorporated by approximating the collision integrals, which take the role of the local field correction in the dielectric formulation. 
For instance, completely neglecting collisions gives the random phase approximation, whereas invoking the relaxation time approximation~\cite{mermin_lindhard_1970,selchow_extended_2002} leads to the well-known Mermin dielectric function.

A closely related strategy is used in Green functions theory where  a suitable approximation of the so-called self-energy is used to truncate the Martin-Schwinger hierarchy~\cite{stefanucci2013nonequilibrium}. In the following, we briefly outline the approximation introduced by Montroll and Ward~\cite{montroll_quantum_1958} and also the additional $e^4$-contribution (see Ref.~\cite{schoof_textitab_2015} for a recent application to the warm dense UEG). For simplicity, we restrict ourselves to the spin-polarized case and write the total energy as a perturbation expansion with respect to coupling strength (dropping terms beyond second order) as~\cite{kremp2006quantum,vorberger_equation_2004}
\begin{eqnarray}
E &=& E_0^{\rm id}(T,\alpha_e)+E^{\rm HF}
+E^{\rm MW}+
E^{\rm e^4}\,.\label{uecorr}
\end{eqnarray}
Here $E_0$ denotes the ideal energy
\begin{eqnarray}
E_0 = \frac{3}{2} \frac{T}{\lambda_\text{DB}^3} I_{3/2}(\alpha) \quad ,
\end{eqnarray}
with $\lambda_\text{DB}=\sqrt{2\pi\hbar^2\beta/m}$ being the thermal wavelength, and $E^\text{HF}$ corresponds to the well-known Hartree-Fock energy
\begin{eqnarray}
E^\text{HF} = \lambda_\text{DB}^{-4} \int_{-\infty}^\alpha \text{d}\alpha'\ I^2_{-1/2}(\alpha') - \frac{3}{2\lambda_\text{DB}^4} I_{-1/2}(\alpha) I_{1/2}(\alpha) \quad ,
\end{eqnarray} 
where $I_k$ is the Fermi integral of order $k$, see, e.g., Ref.~\cite{kremp2006quantum}, and $\alpha=\beta\mu$. As usual, the chemical potential $\mu$ is defined by the normalization of the Fermi function to the total density, see Eq.~(\ref{eq:normailzation_condition}).
To compute the Montroll-Ward (MW) and $e^4$-contribution, it is convenient to utilize the pressure $p$, which is connected to the different parts of the total energy by
\begin{equation}
E^\text{j} =-p^{j}+T\frac{\partial}{\partial T}p^{j}, \qquad j={\rm MW}, \; e^4.
\label{eq:u-p}
\end{equation}
The MW-component of the pressure is then given by
\begin{eqnarray}\label{eq:p_e4}
p^{\rm MW}=\frac{-1}{4\pi^3}\int\limits_0^{\infty} \! dp\,p^2\,
{\cal P}\!\!
\int\limits_{0}^{\infty}d\omega\,\coth\left(\frac{\beta\omega}{2}\right)
\!\!\!\left[\arctan\frac{\mbox{Im}\,\varepsilon_\text{RPA}(p,\omega)}{\mbox{Re}\,\varepsilon_\text{RPA}(p,\omega)}
-\mbox{Im}\,\varepsilon_\text{RPA}(p,\omega)\right] \quad ,
\end{eqnarray}
with $\varepsilon_\text{RPA}(p,\omega)$ denoting the dielectric function in the random phase approximation, see Sec.~\ref{sec:LRT}.
Therefore, neglecting the $e^4$-term in Eq.~(\ref{uecorr}) gives the total energy in the RPA.
To include second order contributions, we compute
\begin{eqnarray}
p^{e^4} = \!
\int\frac{d{\bf p}d{\bf q}_1d{\bf q}_2}{64\pi^7}
\frac{1}{p^2 ({\bf p}+{\bf q}_1+{\bf q}_2)^2}
\frac{f_{q_1}f_{q_2}\bar{f}_{{\bf q}_1+{\bf p}}\bar{f}_{{\bf q}_2+{\bf
p}}-f_{{\bf q}_1+{\bf p}}f_{{\bf q}_2+{\bf
p}}\bar{f}_{q_1}\bar{f}_{q_2}}
{
q_1^2+q_2^2-({\bf p}+{\bf q}_1)^2-({\bf p}+{\bf q}_2)^2}\,
\end{eqnarray}
with $f_p \!=\! [\exp(\beta p^2/2 \!-\! \beta\mu) \!+\! 1]^{-1}$ being the Fermi function, and $\bar{f}_p \!=\! [1 \!-\! f_p]$
denotes the Pauli blocking factor.  Detailed benchmarks of the energy computed from Eq.~(\ref{uecorr}) will be presented in Sec.~\ref{sec:comparison}.

For completeness, we also mention the recent finite-temperature extension of the retarded cumulant Green function approach~\cite{kas_finite_2017} that is predicted to allow, both, for the computation of thermodynamic properties of the UEG (see Sec.~\ref{sec:param_results} for a comparison to QMC data) and, in addition, spectral properties.

\subsection{Classical mapping approaches\label{sec:map}}
In addition to the dielectric formalism (Sec.~\ref{sec:LRT}) and the quantum Monte Carlo methods introduced in Sec.~\ref{sec:QMC}, quantum-classical mappings constitute a third independent class of approaches to a thermodynamic description of the electron gas.
In this section, we give a concise overview of two different formulations, namely the works by F.~Perrot and M.W.C~Dharma-wardana~\cite{perrot_spin-polarized_2000,dharma-wardana_simple_2000} and the more recent and rigorous works by S.~Dutta and J.W.~Dufty~\cite{dutta_uniform_2013,dutta_classical_2013,dufty_classical_2013}.

    \subsubsection{Classical mapping approach by Perrot and Dharma-wardana\label{sec:PDW}}
    The basic idea of the formalism by Perrot and Dharma-wardana~\cite{perrot_spin-polarized_2000,dharma-wardana_simple_2000} (hereafter denoted as \textit{PDW}) is to define a classical system of charged particles at an effective \textit{quantum temperature} $T_q$, such that an input value for the ground state exchange-correlation energy $E_\text{xc}$ obtained from outside the theory is reproduced. While, in principle, data from any theory could be used, PDW chose the then most accurate data based on quantum Monte Carlo calculations by Ortiz and Ballone~\cite{ortiz_correlation_1994}. The properties of the effective classical system are approximately computed by solving the corresponding hyppernetted chain (HNC) equations~\cite{springer_integral_1973,ng_hypernetted_1974}. A potentially more accurate albeit computationally considerably more demanding treatment using the classical Monte Carlo or Molecular Dynamics methods, e.g.~Ref.~\cite{binder1995monte}, was deemed unnecessary as the error due to the HNC approximation was expected to be negligible for the densities of interest. For completeness, we mention that this assumption was somewhat contradicted by the recent works of Liu and Wu~\cite{liu_bridge-functional-based_2014}, who found that a more accurate inclusion of short-range correlations is important to describe the first peak in the pair distribution function at low density.
    Once the classical system is solved (thereby recovering the input value for $E_\text{xc}$), it is straightforward to obtain other observables such as the pair distribution function (or, equivalently, the static structure factor, cf.~Sec.~\ref{sec:FSC}) or the static density response function $\chi(\mathbf{k})$, cf.~Sec.~\ref{sec:LRT}. A particular advantage of the classical mapping approach is that the resulting PDF is always positive. This is in stark contrast to the dielectric approximations from Sec.~\ref{sec:LRT}, where the PDF tends to become negative at small distances for intermediate to strong coupling. Further, a comparison of the classical mapping with the ground state QMC results revealed quantitative agreement.

    To extend this formalism to finite temperature $T$, for which back in the early 2000s no accurate data for $E_\text{xc}(r_s,T)$ existed, PDW introduced a modified classical temperature
    \begin{eqnarray}
    T_\text{cf} = \left( T^2 + T_q^2 \right)^{1/2} \quad , \label{eq:PDW_T}
    \end{eqnarray}
    which is motivated by the fact that the leading dependence of the energy on $T$ is quadratic. 
    Note that the expression for $T_q$ in Eq.~(\ref{eq:PDW_T}) depends only on the density parameter $r_s$, 
    \begin{eqnarray}
    T_q = \frac{1}{a+b\sqrt{r_s}+cr_s}\quad ,
    \end{eqnarray}
    where the free parameters $a$, $b$, and $c$ were obtained to reproduce the ground state data for $E_\text{xc}$ as explained above.
    It is easy to see that Eq.~(\ref{eq:PDW_T}) becomes exact for high and low temperature, but constitutes an uncontrolled approximation for intermediate temperatures, most notably in the warm dense matter regime.
   
    In their seminal paper from 2000, PDW~\cite{perrot_spin-polarized_2000} provided extensive results for the uniform electron gas
    at finite temperature, including a parametrization of the exchange-correlation free energy $f_\text{xc}$ with respect to temperature, density, and spin-polarization. A concise introduction of the latter is presented in Sec.~\ref{sec:PDW}, where it is compared to the recent, highly accurate parametrization by Groth, Dornheim and co-workers~\cite{groth_ab_2017}.

    Further, the PDW formalism for the classical-mapping has subsequently been employed in numerous calculations of more realistic (and, thus, more complicated) systems, e.g., Refs.~\cite{dharma-wardana_spin-_2004,dharma-wardana_static_2006,dharma-wardana_pair-distribution_2008}, and an excellent review can be found in Ref.~\cite{dharma-wardana_classical-map_2012}.
     Finally, we mention that the shortcoming of the PDW classical-mapping at intermediate temperature was recently somewhat remedied by Liu and Wu~\cite{liu_improved_2014}, who replaced the simple interpolation for $T_\text{cf}$ from Eq.~(\ref{eq:PDW_T}) by the explicitly temperature-dependent expression
    \begin{eqnarray}
    T_\text{cf} = \frac{1}{ a(T) + b(T)\sqrt{r_s} + c(T)r_s }\quad ,
    \end{eqnarray}
    where the functions $a(T)$, $b(T)$, and $c(T)$ where chosen to reproduce the RPIMC data by Brown \textit{et al.}~\cite{brown_path-integral_2013} for $E_\text{xc}$, see Ref.~\cite{liu_improved_2014} for more details. 
    It was found that this gives better data for the pair correlation function, in particular for the description of long-range correlations.

    \subsubsection{Classical mapping approach by Dutta and Dufty}

Recently, Dufty and Dutta~\cite{dufty_classical_2012,dufty_classical_2013} presented a more rigorous classical-mapping formalism operating in the grand canonical ensemble (volume $V$, chemical potential $\mu$, and inverse temperature $\beta$ are fixed). While the volume $V$ is equal both for the \textit{true} quantum system and the \textit{effective} classical one, a modified inverse temperature $\beta_c$, chemical potential $\mu_c$, and pair potential $\phi_c(r)$ are introduced. To determine these two parameters and one function, we enforce the equivalence of pressure $p$, electron number density $n$ and of the pair distribution function $g(r)$ for the true and effective systems,
\begin{eqnarray}\label{eq:sandipan}
p_c(\beta_c, V, \mu_c | \phi_c(r)) &\equiv& p(\beta, V, \mu | \phi(r)) \\ \nonumber
n_c(\beta_c, V, \mu_c | \phi_c(r)) &\equiv& n(\beta, V, \mu | \phi(r)) \\ \nonumber
g_c(r,\beta_c,V,\mu_c | \phi_c(r)) &\equiv& g(r,\beta,V,\mu | \phi(r)) \quad ,
\end{eqnarray}
where, for the uniform electron gas, $\phi(r)$ is simply given by the Coulomb potential.
Observe that the vertical bars in Eq.~(\ref{eq:sandipan}) indicate that all three quantities are in fact functionals of the classical or quantum pair potentials, in addition to the functional dependence on the three thermodynamic variables. In practice, one has to provide expressions for $p$, $n$ and $g(r)$ of the quantum system, starting from which the relations in Eq.~(\ref{eq:sandipan}) can be inverted for $\mu_c$, $\beta_c$, and $\phi_c(r)$. 

Since providing two thermodynamic and one structural property of the system of interest as input for an approximate many-body formalism might admittedly seem like circular reasoning, we must ask ourselves what kind of information has been gained at which point. The answer is as follows: in practice, we provide the quantum input computed from the random phase approximation (see Sec.~\ref{sec:LRT}), and subsequently compute the classical parameters $\beta_c$, $\mu_c$, and $\phi_c(r)$ by solving Eq.~(\ref{eq:sandipan}) in the classical weak-coupling approximation. The main assumption is that the quantum effects are either local (such as diffraction) or weakly nonideal (such as antisymmetry under particle-exchange). In this case, the bulk of the more pronounced nonideality effects would be captured by subsequently feeding the obtained results for $\beta_c$, $\mu_c$, and $\phi_c(r)$ into a more accurate classical many-body method, such as the classical Monte-Carlo method, molecular dynamics, or, like in the PDW approach, the hypernetted chain approximation.

Overall, the application of the Dufty-Dutta formalism to the UEG at warm dense matter conditions~\cite{dutta_classical_2013,dutta_uniform_2013} has given results of similar accuracy as the PDW formalism, although not nearly as extensive data have been presented. For completeness, we mention that this approach is not limited to the UEG or, in general, to homogeneous systems. For example, first results for charges in a harmonic confinement have been reported in Refs.~\cite{dutta_classical_2013,wrighton_finite-temperature_2016}. The application to a realistic electron-ion plasma remains an important task for the future.

\section{Quantum Monte Carlo Methods\label{sec:QMC}}

In the following section, we will discuss in detail various quantum Monte Carlo methods and discuss the fermion sign problem, which emerges for the simulations of electrons. In particular, we introduce the Metropolis algorithm~\cite{metropolis_equation_1953}, which constitutes the backbone of all subsequent path inegral Monte Carlo methods except the density matrix QMC paradigm. Not mentioned are the multilevel blocking idea by Mak, Egger and co-workers~\cite{mak_multilevel_1998,egger_path-integral_2000,egger_multilevel_2001,dikovsky_analysis_2001,muhlbacher_crossover_2003} and the expanded-ensemble approach by Vorontsov-Velyaminov \textit{et al.}~\cite{vorontsov-velyaminov_path_2006,voznesenskiy_path-integralchar21expanded-ensemble_2009}.

\subsection{The Metropolis algorithm\label{sec:Metropolis}}
Due to its fundamental importance for the understanding of the quantum Monte Carlo methods introduced below, in this section we give a comprehensive introduction of the widely used Metropolis algorithm~\cite{metropolis_equation_1953}.
\subsubsection{Problem statement}
In statistical many-body physics, we often encounter probabilities of the form
\begin{eqnarray}\label{eq:simple_prob}
P(\mathbf{X}) = \frac{W(\mathbf{X})}{Z}\ .
\end{eqnarray}
For example, the multi-dimensional variable $\mathbf{X}$ might correspond to a configuration of classical particles, or spin-alignments in an Ising model, and $W=\textnormal{exp}(-E(\mathbf{X})\beta)$ to the corresponding "Boltzmann distribution" describing the probability of $\mathbf{X}$ to occur (with $E(\mathbf{X})$ being the energy of said configuration). The aim of a Monte Carlo simulation is then to generate a set of random configurations $\{\mathbf{X_i}\}$ that are distributed according to Eq.~(\ref{eq:simple_prob}), which can subsequently be used to compute averages such as the internal energy.

Usually, the problem with such a statistical description of a system is that the normalization of Eq.~(\ref{eq:simple_prob}),
\begin{eqnarray}
Z = \int \textnormal{d}\mathbf{X}\ W(\mathbf{X})\ ,
\end{eqnarray}
is not readily known. For the canonical ensemble (volume $V$, particle number $N$ and inverse temperature $\beta$ are fixed), to which we will restrict ourselves throughout this work, $Z$ corresponds to the canonical partition function. In this case, the exact knowledge of $Z$ allows to directly compute all observables (e.g., energies, pressure, etc.) via thermodynamic relations, thereby eliminating the need for a Monte Carlo simulation in the first place. The paramount achievement by Metropolis \textit{et al.}~\cite{metropolis_equation_1953} was to introduce an algorithm that allows to generate a set of random variables $\{\mathbf{X}_i\}$ with an unknown normalization $Z$. The significance of this accomplishment can hardly be overstated and the Metropolis algorithm has emerged as one of the most successful algorithms in computational physics and beyond.

\subsubsection{The detailed balance condition}

The starting point is the imposition of the so-called \textit{detailed balance condition},
\begin{eqnarray}\label{eq:detailed_balance}
T(\mathbf{X}\to\tilde{\mathbf{X}}) = T(\tilde{\mathbf{X}}\to\mathbf{X})\ ,
\end{eqnarray}
which states that the transition probability $T$ to go from a state $\mathbf{X}$ to another state $\mathbf{\tilde{X}}$ is equal to the same probability the other way around. While Eq.~(\ref{eq:detailed_balance}) constitutes an unnecessary rigorous restriction, it allows for a simple straightforward solution. Prior to that, we split the transition probability into a product of three separate parts,
\begin{eqnarray}\label{eq:generalized}
T(\mathbf{X}\to\mathbf{\tilde{X}}) = P(\mathbf{X})\ S(\mathbf{X}\to\mathbf{\tilde{X}})\ A(\mathbf{X}\to \mathbf{\tilde{X}})\ ,
\end{eqnarray}
specifically the probabilities to occupy the initial state $\mathbf{X}$, $P(\mathbf{X})$, to propose the target state $\tilde{\mathbf{X}}$ starting from $\mathbf{X}$, $S(\mathbf{X}\to\tilde{\mathbf{X}})$, and finally to accept the proposed transition, $A(\mathbf{X}\to\mathbf{\tilde{X}})$.
Inserting Eq.~(\ref{eq:generalized}) into (\ref{eq:detailed_balance}) leads to the generalized form of the detailed balance equation,
\begin{eqnarray}\label{eq:central}
P(\mathbf{X})\ S(\mathbf{X}\to\mathbf{\tilde{X}})\ A(\mathbf{X}\to\mathbf{\tilde{X}}) = P(\mathbf{\tilde{X}})\ 
S(\mathbf{\tilde{X}}\to\mathbf{X})\ A(\mathbf{\tilde{X}}\to\mathbf{X})\ ,
\end{eqnarray}
which is of central importance for the development and design of state of the art quantum Monte Carlo algorithms.
The solution of Eq.~(\ref{eq:central}) for the acceptance probability by Metropolis \textit{et al.}~\cite{metropolis_equation_1953}
is given by
\begin{eqnarray}\label{eq:acceptance}
A(\mathbf{X}\to\tilde{\mathbf{X}}) &=& \textnormal{min}\left( 1, \frac{P(\mathbf{\tilde{X}})}{P(\mathbf{X})}\ \frac{S(\mathbf{\tilde{X}}\to\mathbf{X})}{S(\mathbf{X}\to\mathbf{\tilde{X}})} \right)\ , \\ \nonumber
&=& \textnormal{min}\left( 1, \frac{W(\mathbf{\tilde{X}})}{W(\mathbf{X})}\ \frac{S(\mathbf{\tilde{X}}\to\mathbf{X})}{S(\mathbf{X}\to\mathbf{\tilde{X}})} \right)\ ,
\end{eqnarray}
which can be easily verified by considering Eq.~(\ref{eq:generalized}) for the cases $P(\mathbf{\tilde{X}})S(\mathbf{\tilde{X}}\to\mathbf{X}) > P(\mathbf{X}) S(\mathbf{X}\to\mathbf{\tilde{X}})$ and vice versa.
Observe that the unknown normalization $Z$ cancels in Eq.~(\ref{eq:acceptance}), which means that the acceptance probability can be readily evaluated.

We conclude this section with a sketch of a practical implementation of the Metropolis algorithm:
\begin{enumerate}
    \item Start with an (in principle arbitrary) initial configuration $\mathbf{X}_0$.
    \item\label{StepTwo} Propose a new configuration $\tilde{\mathbf{X}}$ according to some pre-defined sampling probability $S(\mathbf{X}_i\to\tilde{\mathbf{X}})$.
    \item\label{StepThree} Evaluate the corresponding acceptance probability $A(\mathbf{X}_i\to\tilde{\mathbf{X}})$, see Eq.~(\ref{eq:acceptance}), and subsequently draw a uniform random number $y\in[0,1)$. If we have $y \leq A(\mathbf{X}_i\to\tilde{\mathbf{X}})$, the update is accepted and the configuration is updated to $\mathbf{X}_{i+1}=\tilde{\mathbf{X}}$. Otherwise, we reject the update and the "new" configuration is equal to the old one, $\mathbf{X}_{i+1}=\mathbf{X}_i$.
    \item Repeat steps \ref{StepTwo} and \ref{StepThree} until we have generated sufficiently many configurations.
\end{enumerate}
Assuming an ergodic set of Monte Carlo updates (random ways to change between different configurations), the outlined algorithm can be used to generate a Markov chain of configurations $\{\mathbf{X}_i\}$ that are distributed according to $P(\mathbf{X})$, as asked in the problem statement. The concept of ergodicity is of central importance for the design of QMC algorithms and updates and means that (i) all possible configurations must be reachable in a finite (though, in principle, arbitrarily large) number of updates and (ii) the probability to go from one configuration $\mathbf{X}$ to another configuration $\tilde{\mathbf{X}}$ must only depend on $\mathbf{X}$ itself (no memory effects).
A possible segment of such a Markov chain as generated by the Metropolis algorithm is given by
\begin{eqnarray}
\nonumber \mathbf{X}_0 = \mathbf{a} \to \mathbf{X}_1 = \mathbf{a} \to \mathbf{X}_2 = \mathbf{b} \to  \mathbf{X}_3 = \dots \quad .
\end{eqnarray}
Starting at an initial configuration $\mathbf{X}_0$, a new configuration is proposed, but the update is rejected. Therefore the second element of the Markov chain is equal to the first one, $\mathbf{X}_0=\mathbf{X}_1=\mathbf{a}$. The second update is accepted, meaning that the third element is changed to the new configuration, $\mathbf{X}_2=\mathbf{b}$. 
It is important to understand that, even if a proposed update from $\mathbf{X}$ to $\mathbf{\tilde{X}}$ is rejected, the old configuration must still be counted as a new element in the Markov chain. Appending the Markov chain only after an update has been accepted is plainly wrong.

\subsection{Path Integral Monte Carlo\label{sec:PIMC}}

The path integral Monte Carlo approach~\cite{herman_path_1982} (see Ref.~\cite{ceperley_path_1995} for a review) is one of the most successful methods in quantum many body physics at finite temperature. The underlying basic idea is to map the complicated quantum system onto a classical system of interacting ring polymers~\cite{chandler_exploiting_1981}. The high dimensionality of the resulting partition function (each particle is now represented by an entire ring polymer consisting of potentially hundreds of parts) requires a stochastic treatment, i.e., the application of the Metropolis Monte Carlo method~\cite{metropolis_equation_1953}.
In particular, PIMC allows for quasi-exact simulations of up to $N\sim10^4$ bosons (and distinguishable, spinless particles, often referred to as boltzmannons, e.g., Ref.~\cite{pollock_simulation_1984}) and has played a crucial role for the theoretical understanding of such important phenomena as superfluidity~\cite{pollock_path-integral_1987,sindzingre_path-integral_1989,kwon_local_2006,dornheim_superfluidity_2015}, Bose-Einstein condensation~\cite{gruter_critical_1997,pilati_dilute_2010,noauthor_path-integral_2016} or the theory of collective excitations~\cite{filinov_collective_2012,filinov_correlation_2016}.
Unfortunately, as we will see, PIMC simulations of electrons (and fermions, in general) are severely limited by the so-called fermion sign problem~\cite{loh_sign_1990,troyer_computational_2005}.

\subsubsection{Distinguishable particles }
Let us start the discussion of the PIMC method by considering the partition function of $N$ distinguishable particles (so-called boltzmannons), in the canonical ensemble (i.e., volume $V$ and and inverse temperature $\beta=1/k_\textnormal{B}T$ are fixed)
\begin{eqnarray}
Z = \textnormal{Tr}\ \hat\rho\ . \label{eq:Z}
\end{eqnarray}
Here $\hat\rho=e^{-\beta\hat H}$ denotes the canonical density operator and the Hamiltonian is given by the sum of a kinetic and potential part,
\begin{eqnarray}
\hat H = \hat K + \hat V\ .
\end{eqnarray}
In coordinate space, Eq.~(\ref{eq:Z}) reads
\begin{eqnarray}
Z = \int \textnormal{d}\mathbf{R}\ \bra{\mathbf{R}} e^{-\beta\hat H}\ket{\mathbf{R}}\ ,
\end{eqnarray}
with $\mathbf{R} = \{ \mathbf{r}_1,\dots,\mathbf{r}_N\}$ containing all $3N$ particle coordinates. The problem is that the matrix elements are not known, as $\hat K$ and $\hat V$ do not commute
\begin{eqnarray}
e^{-\beta(\hat K + \hat V)} = e^{-\beta\hat V}e^{-\beta\hat K}e^{-\beta^2\hat C}\quad ,
\end{eqnarray}
where the error term is obtained from the Baker-Campbell-Hausdorff formula as~\cite{kleinert_path_2009}
\begin{eqnarray}\label{eq:bch}
\hat C = \frac{1}{2}[\hat V, \hat K] - \beta\left( \frac{1}{6}[\hat V, [\hat V, \hat K]] - \frac{1}{3}[[\hat V,\hat K],\hat K] \right) + \dots\ .
\end{eqnarray}
To overcome this obstacle, we exploit the group property of the exponential function
\begin{eqnarray}
e^{-\beta\hat H} = \prod_{\alpha=0}^{P-1} e^{-\epsilon\hat H}\ , \label{eq:group}
\end{eqnarray}
where $\epsilon = \beta / P$. 
By using Eq.~(\ref{eq:group}) and simultaneously inserting $P-1$ unity operators of the form
\begin{eqnarray}
\hat 1 = \int \textnormal{d}\mathbf{R}_\alpha\ \ket{\mathbf{R}_\alpha} \bra{\mathbf{R}_\alpha}\ ,
\end{eqnarray}
we obtain
\begin{eqnarray}\label{eq:Z2}
Z = \int\textnormal{d}\mathbf{X}\ \bra{\mathbf{R}_0} e^{-\epsilon\hat H}\ket{\mathbf{R}_1}\bra{\mathbf{R}_1} \dots 
\ket{\mathbf{R}_{P-1}}\bra{\mathbf{R}_{P-1}} e^{-\epsilon\hat H} \ket{\mathbf{R}_0}\ .
\end{eqnarray}
Observe that Eq.~(\ref{eq:Z2}) is still exact and the integration is carried out over $P$ sets of particle coordinates, $\textnormal{d}\mathbf{X} = \textnormal{d}\mathbf{R}_0 \dots \textnormal{d}\mathbf{R}_{P-1}$. Despite the increased dimensionality of the integral, this re-casting proves to be advantageous since each of the matrix elements must now be evaluated at a $P$ times higher temperature, and for sufficiently many factors we can introduce a high temperature approximation, e.g., the primitive factorization 
\begin{eqnarray}\label{eq:primitive}
e^{-\epsilon\hat H} \approx e^{-\epsilon\hat K}e^{-\epsilon \hat V}\ ,
\end{eqnarray}
which, according to the Trotter formula~\cite{trotter_product_1959,de_raedt_applications_1983}, becomes exact in the limit of $P\to\infty$
\begin{eqnarray}
e^{-\beta(\hat K + \hat V)} = \lim_{P\to\infty}\left( e^{-\epsilon\hat K} e^{-\epsilon\hat V} \right)^P\ .
\end{eqnarray}
A more vivid interpretation of Eq.~(\ref{eq:group}) is given in terms of imaginary time path integrals. In particular, we note that the density operator is equivalent to a propagation in imaginary time by $\tau = - \textnormal{i}\beta$ (henceforth, we shall adopt the more conventional definition $\tau\to\tau/(-\textnormal{i})\in[0,\beta]$). Therefore, Eq.~(\ref{eq:group}) corresponds to the introduction of $P$ imaginary "time slices" of length $\epsilon$ and a factorization like Eq.~(\ref{eq:primitive}) to an imaginary time propagator.
Inserting Eq.~(\ref{eq:primitive}) into (\ref{eq:Z2}) finally gives
\begin{eqnarray}\label{eq:zfinal}
Z = \int \textnormal{d}\mathbf{X}\ \prod_{\alpha=0}^{P-1} \left( e^{-\epsilon V(\mathbf{R}_\alpha)}\rho_0(\mathbf{R}_\alpha,\mathbf{R}_{\alpha+1},\epsilon)\right)\ ,
\end{eqnarray}
where $V(\mathbf{R}_\alpha)$ denotes all potential energy terms on time slice $\alpha$,
\begin{eqnarray}
V(\mathbf{R}_\alpha) = \sum_{i=1}^N V_\textnormal{ext}(\mathbf{r}_{\alpha,i}) + \sum_{k>i}^N W(|\mathbf{r}_{\alpha,i}-\mathbf{r}_{\alpha,k}|)\ ,
\end{eqnarray}
and $W(r)$ is an arbitrary pair interaction, e.g., the Coulomb repulsion, $W_\textnormal{C}(r)=1 / r$, and $V_\textnormal{ext}(\mathbf{r})$ denotes an external potential.
The ideal part of the density matrix is given by
\begin{eqnarray}\label{eq:rho_ideal}
\rho_0(\mathbf{R}_\alpha,\mathbf{R}_{\alpha+1},\epsilon) = \frac{1}{\lambda_\epsilon^{3N}} \prod_{i=1}^N \left[ \sum_{\mathbf{n}}
\textnormal{exp}\left( - \frac{\pi}{\lambda^2_\epsilon}(\mathbf{r}_{\alpha,k}-\mathbf{r}_{\alpha+1,k}+\mathbf{n}L)^2
\right)\right]\ ,
\end{eqnarray}
with $\lambda_\epsilon = \sqrt{2\pi\epsilon}$ being the thermal wavelength corresponding to the $P$-fold increased temperature.
The sum over $\mathbf{n}=(n_x,n_y,n_z)^T$, $n_i\in  \mathbb{Z}$, is due to the periodic boundary conditions. For completeness, we note that, technically, Eq.~(\ref{eq:rho_ideal}) constitutes an approximation as the correct ideal density matrix in a periodic box is given by an elliptic theta function~\cite{ceperley_path_1995}. However, this difference is of no practical consequence and, for $P\to\infty$, Eq.~(\ref{eq:rho_ideal}) becomes exact.

\begin{figure}\vspace*{-1.5cm}\hspace*{-0.95cm}
\includegraphics[width=0.5\textwidth]{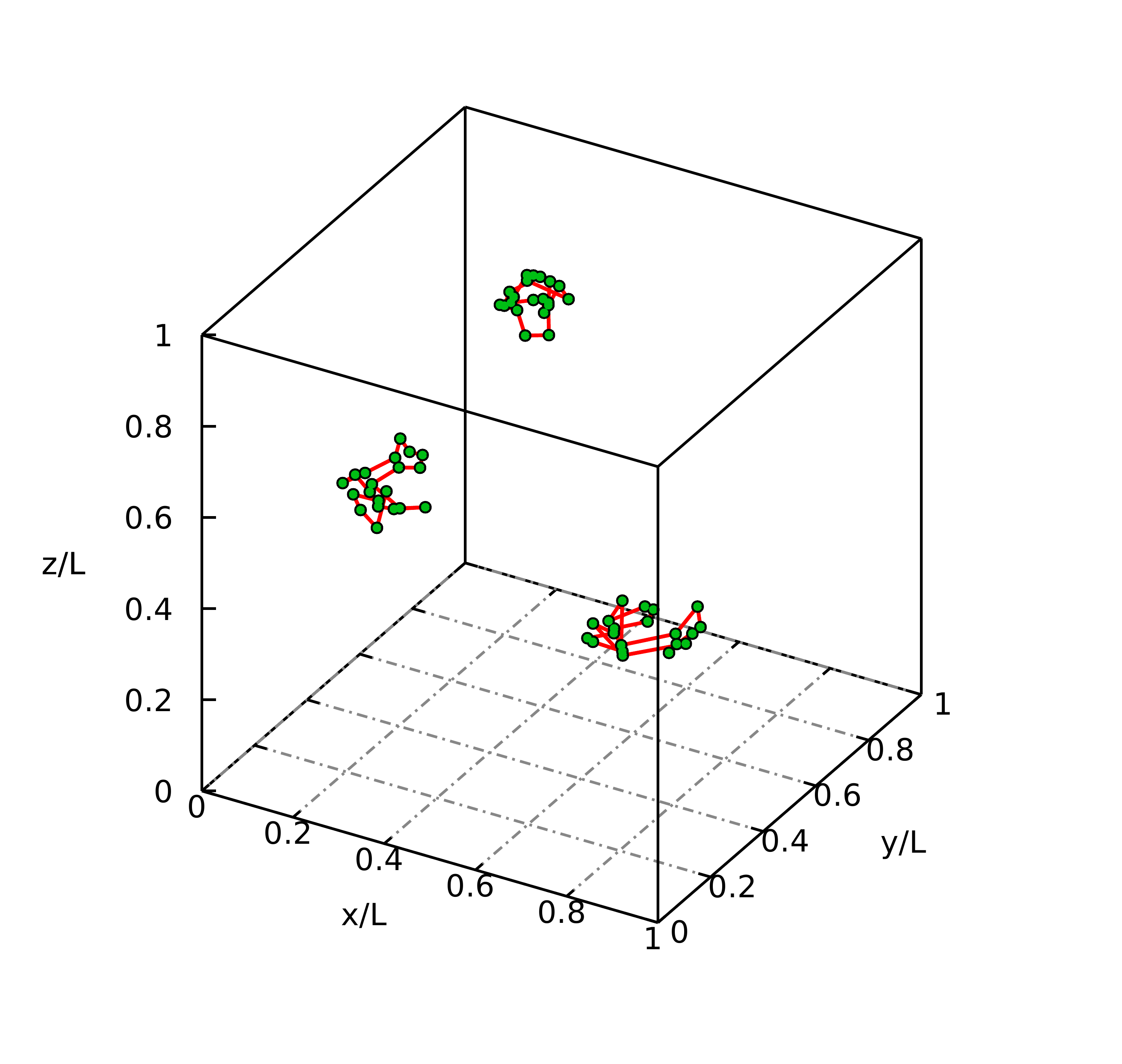}\hspace*{-0.95cm}
\includegraphics[width=0.5\textwidth]{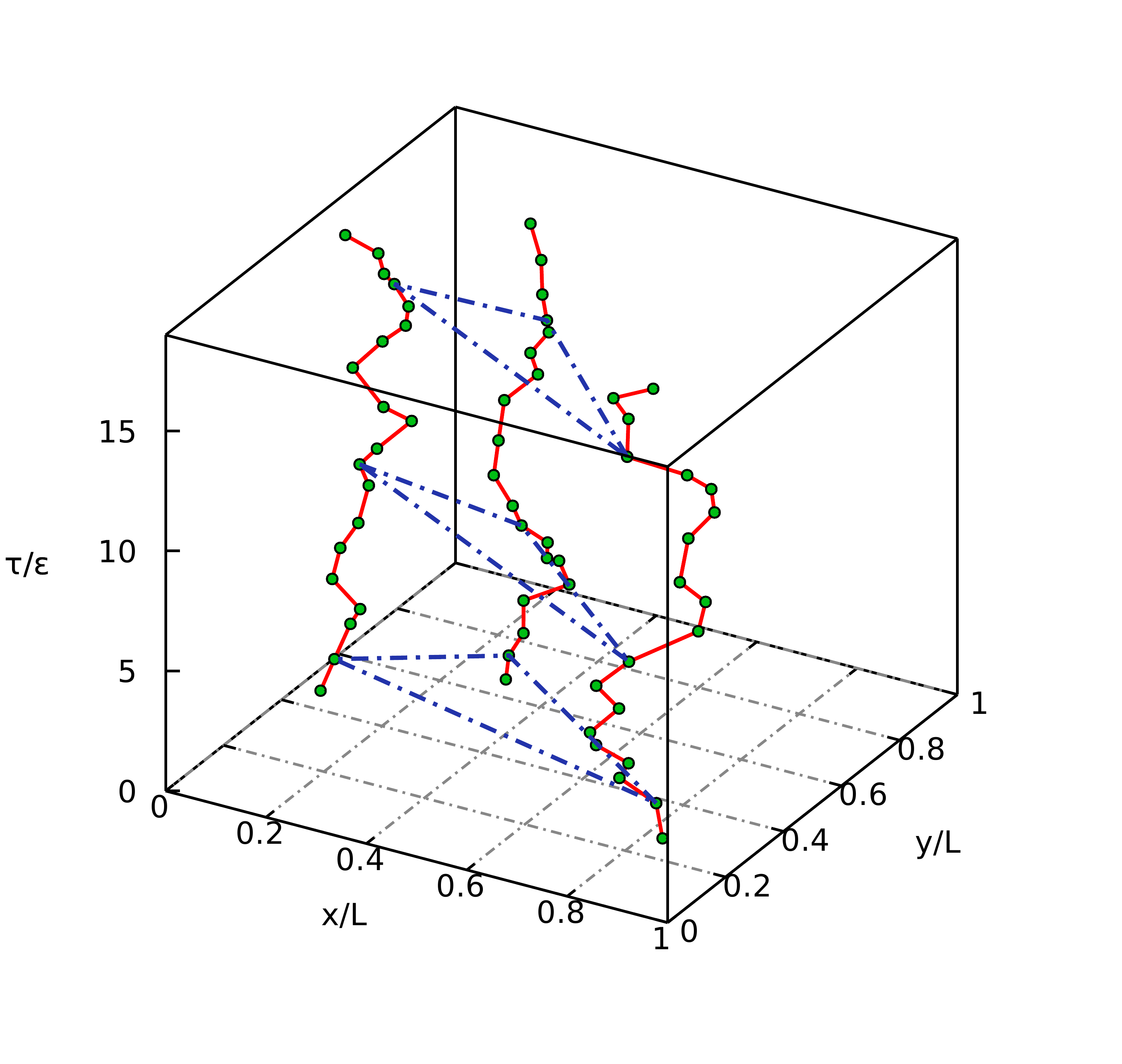}\vspace*{-1.cm}

\caption{\label{fig:PIMC_illustration}Schematic illustration of path integral Monte Carlo: The left panel shows a random configuration of three particles in a $3D$ simulation box. The right panel shows the same configuration, but the $z$-axis has been replaced by the imaginary time $\tau$. Beads on adjacent slices are harmonically linked by the free particle density matrix (see Eq.~(\ref{eq:rho_ideal}), red lines) and beads from different particles on the same time slice are subject to the pair interactions (dashed blue lines).
}
\end{figure}

Following Chandler and Wolynes~\cite{chandler_exploiting_1981}, Eq.~(\ref{eq:zfinal}) can be visualized as interacting ring polymers via the so-called ``classical isomorphism'', which is illustrated in Fig~\ref{fig:PIMC_illustration}. The complicated quantum many-body system has been mapped onto a classical system of interacting ring polymers. Each particle is represented by a closed path of $P$ ``beads'' (i.e., the polymer), see the left panel. Beads on adjacent time slices are effectively linked by a harmonic interaction, see Eq.~(\ref{eq:rho_ideal}). This is further illustrated in the right panel of Fig.~\ref{fig:PIMC_illustration}, where the $z$-axis has been replaced by the imaginary time $\tau$. In addition, we note that beads from different particles on the same time slice interact via the given pair interaction $W(r)$, cf.~the dashed blue lines. The extension of the paths of each particle roughly corresponds to the thermal wavelength $\lambda_\beta$. At high temperature, the paths resemble point particles and quantum effects are negligible. With increasing $\beta$, however, the ring polymers become more extended and the quantum nature of the system of interest starts to dominate.
In practice, Eq.~(\ref{eq:zfinal}) requires a high dimensional integration, which is most effectively achieved using Monte Carlo methods. In particular,
we employ the Metropolis algorithm to generate all possible configurations $\mathbf{X}$ according to the corresponding configuration weight $W$,
\begin{eqnarray}
Z = \int \textnormal{d}\mathbf{X}\ W(\mathbf{X})\ ,
\end{eqnarray}
where $W(\mathbf{X})$ is defined by Eq.~(\ref{eq:zfinal}).

Furthermore, we stress that we are not interested in the partition function itself, but instead in thermodynamic expectation values of an (in principle arbitrary) observable $\hat A$,
\begin{eqnarray}
\braket{\hat  A} = \frac{1}{Z} \int \textnormal{d}\mathbf{R}\ \bra{\mathbf{R}} \hat A \hat \rho \ket{\mathbf{R}}\ .
\end{eqnarray}
In practice, we have to derive a Monte Carlo estimator $A(\mathbf{X})$ so that we can estimate $\braket{\hat A}$ from the set of $N_\textnormal{MC}$ randomly generated configurations $\{\mathbf{X}\}_\textnormal{MC}$
\begin{eqnarray}\label{eq:estimation}
\braket{\hat A} &\approx& A_\textnormal{MC}\qquad \textnormal{and} \\
A_\textnormal{MC} &=& \frac{1}{N_\textnormal{MC}}\sum_{\mathbf{X}} A(\mathbf{X})\ .
\end{eqnarray}
Eq.~(\ref{eq:estimation}) seems to imply that the path integral Monte Carlo approach does not allow to obtain the exact thermodynamic expectation value of interest, but, instead, constitutes an approximation. More precisely, the MC estimate from a PIMC calculation is afflicted with a statistical uncertainty
\begin{equation}\label{eq:MCerror}
\Delta A = \sqrt{\frac{ \braket{\hat A^2} - \braket{\hat A}^2}{ N_\textnormal{MC} }}\ .
\end{equation}
The statistical interpretation of Eq.~(\ref{eq:MCerror}) is that $A_\textnormal{MC}$ is with a probability of $66\%$ within $\pm\Delta A$ of the exact result. Furthermore, this uncertainty interval decreases with an increasing number of MC samples $N_\textnormal{MC}$ so that, in principle, an arbitrary accuracy is possible. Therefore, PIMC is often described as ``quasi-exact''.

\subsubsection{PIMC simulations of fermions}
Let us now extend our considerations to the PIMC simulation of $N=N^\uparrow+N^\downarrow$ electrons, with $N^\uparrow$ and $N^\downarrow$ denoting the number of spin-up and spin-down electrons, respectively. To take into account the antisymmetric nature due to the indistinguishability of fermions, we must extend the PIMC partition function from Eq.~(\ref{eq:zfinal}) by the sum over all permutations of electrons from the same species ($S_{N^\uparrow}$ and $S_{N^\downarrow}$)
\begin{eqnarray}\label{eq:zfermion}
Z = \frac{1}{(N^\uparrow!\ N^\downarrow!)^P}  \int \textnormal{d}\mathbf{X}\ \prod_{\alpha=0}^{P-1} \left( \sum_{\sigma_\alpha^\uparrow\in S_{N^\uparrow}}
\sum_{\sigma_\alpha^\downarrow\in S_{N^\downarrow}}
\textnormal{sgn}(\sigma_\alpha^\uparrow) \textnormal{sgn}\left(\sigma_\alpha^\downarrow\right) e^{-\epsilon V(\mathbf{R}_\alpha)}\rho_0(\mathbf{R}_\alpha, \hat \pi_{\sigma_\alpha^\uparrow}  \hat \pi_{\sigma_\alpha^\downarrow}\mathbf{R}_{\alpha+1},\epsilon)\right)\ .
\end{eqnarray}
Here $\hat \pi_{\sigma_\alpha^{\uparrow,\downarrow}}$ denotes the exchange operator corresponding to a particular permutation $\sigma_\alpha^{\uparrow,\downarrow}$ and $\textnormal{sgn}(\sigma_\alpha^{\uparrow,\downarrow})$ denotes the corresponding signum. Note that, due to the idempotency of the antisymmetry operator, the sum over all permutations can be carried out on each time slice without changing the result. In practice, the sum over all possible configurations $\mathbf{X}$ in the PIMC simulation must now be extended to include paths incorporating more than a single particle.
\begin{figure}
\includegraphics[width=0.45\textwidth]{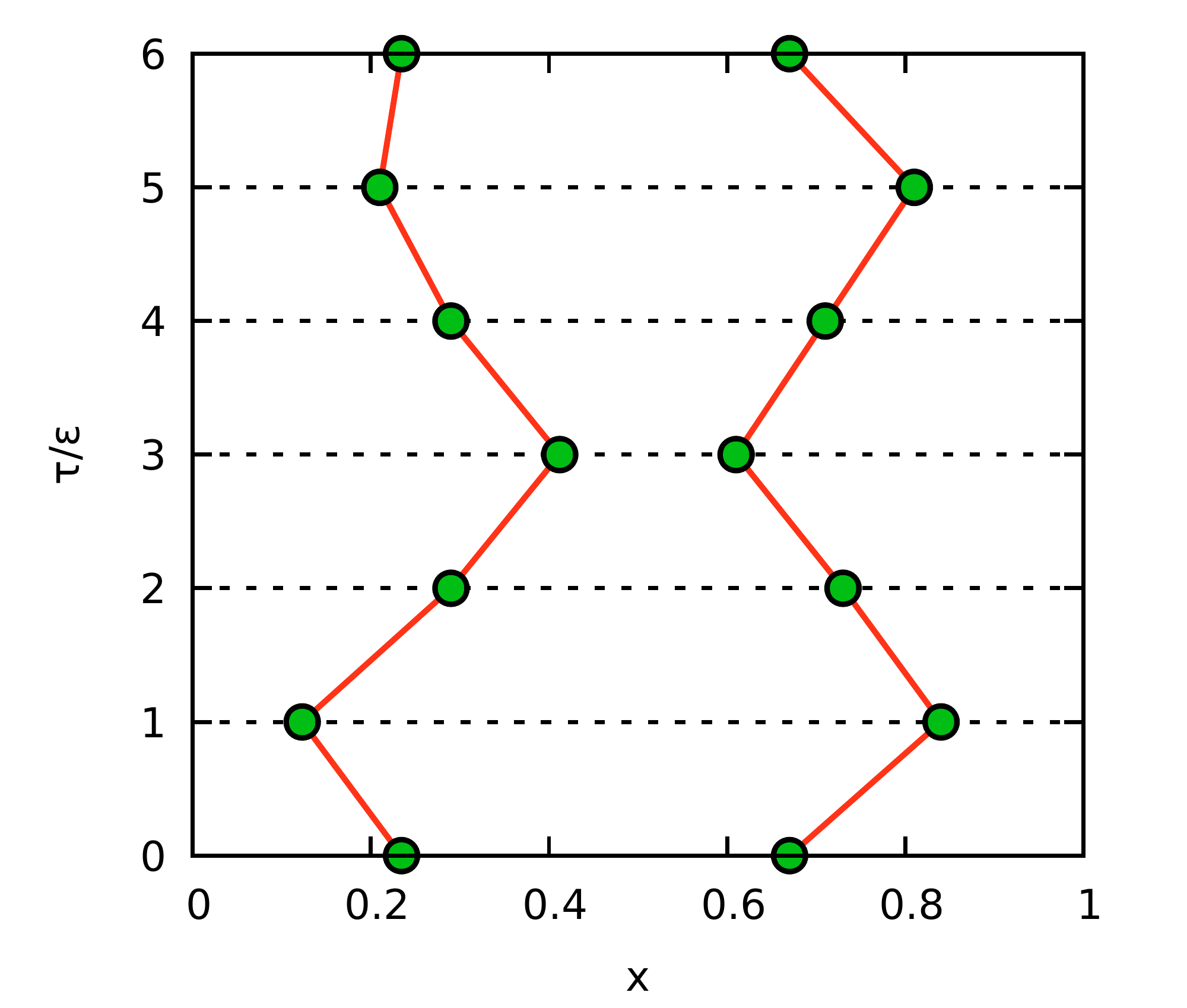}
\includegraphics[width=0.45\textwidth]{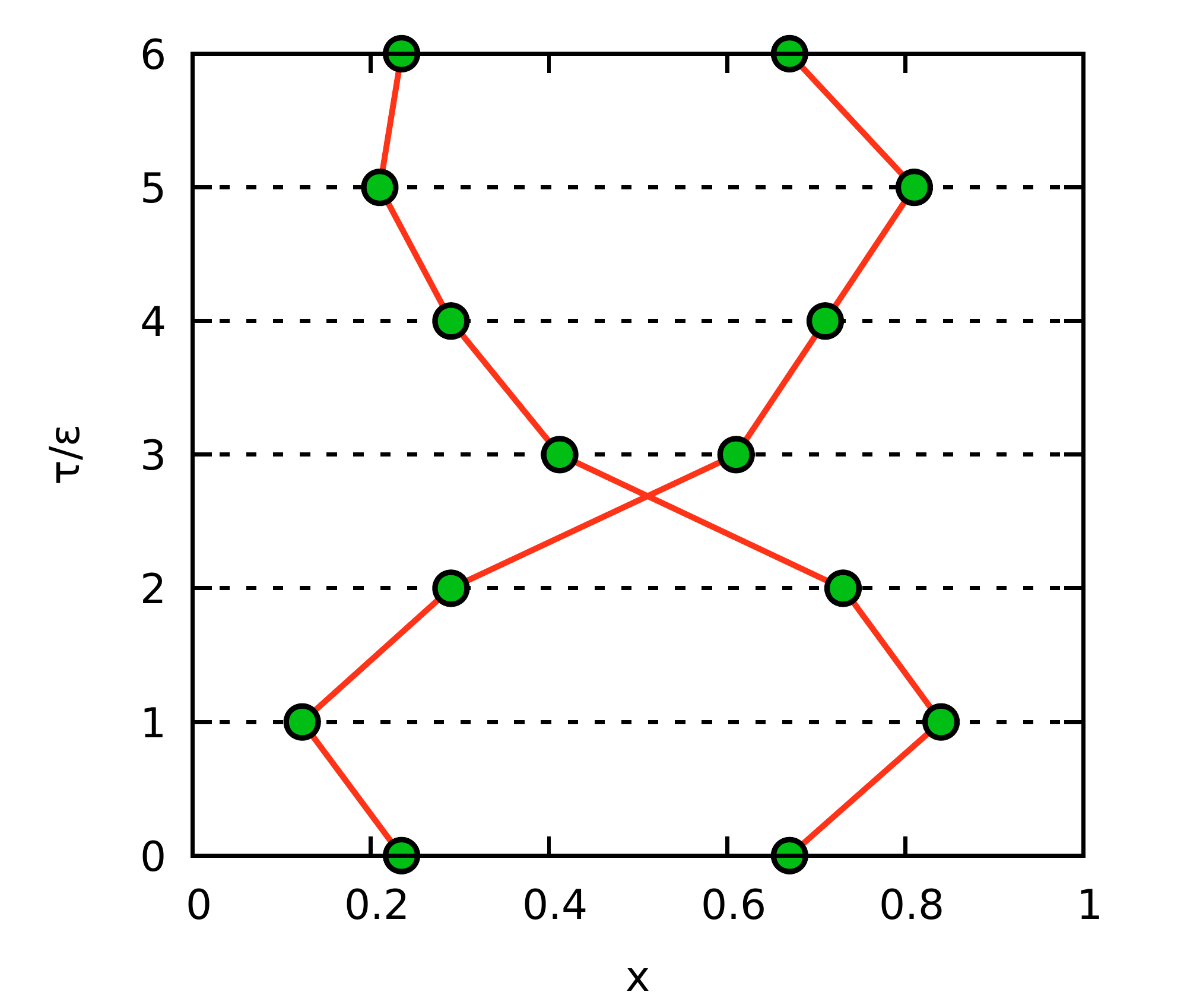}

\caption{\label{fig:PIMC_exchange}Schematic illustration of path integral Monte Carlo: Shown are two PIMC configurations in the $\tau$-$x$-plane with no pair exchange (left) and a single pair exchange (right). The corresponding configuration weights $W(\mathbf{X})$ are positive and negative.
}
\end{figure}
This is illustrated in Fig.~\ref{fig:PIMC_exchange} where two PIMC configurations with $N=N^\uparrow=2$ spin-polarized electrons are shown. In the left panel, there are two distinct paths. Hence, there is no pair exchange and the sign $\textnormal{sgn}(W(\mathbf{X}))$ is positive. In contrast, in the right panel, the paths cross and a single path incorporates both particles. Due to this single pair exchange, the sign of the configuration weight is negative.

\subsubsection{The fermion sign problem\label{sec:FSP}}
\begin{figure}
\includegraphics[width=0.45\textwidth]{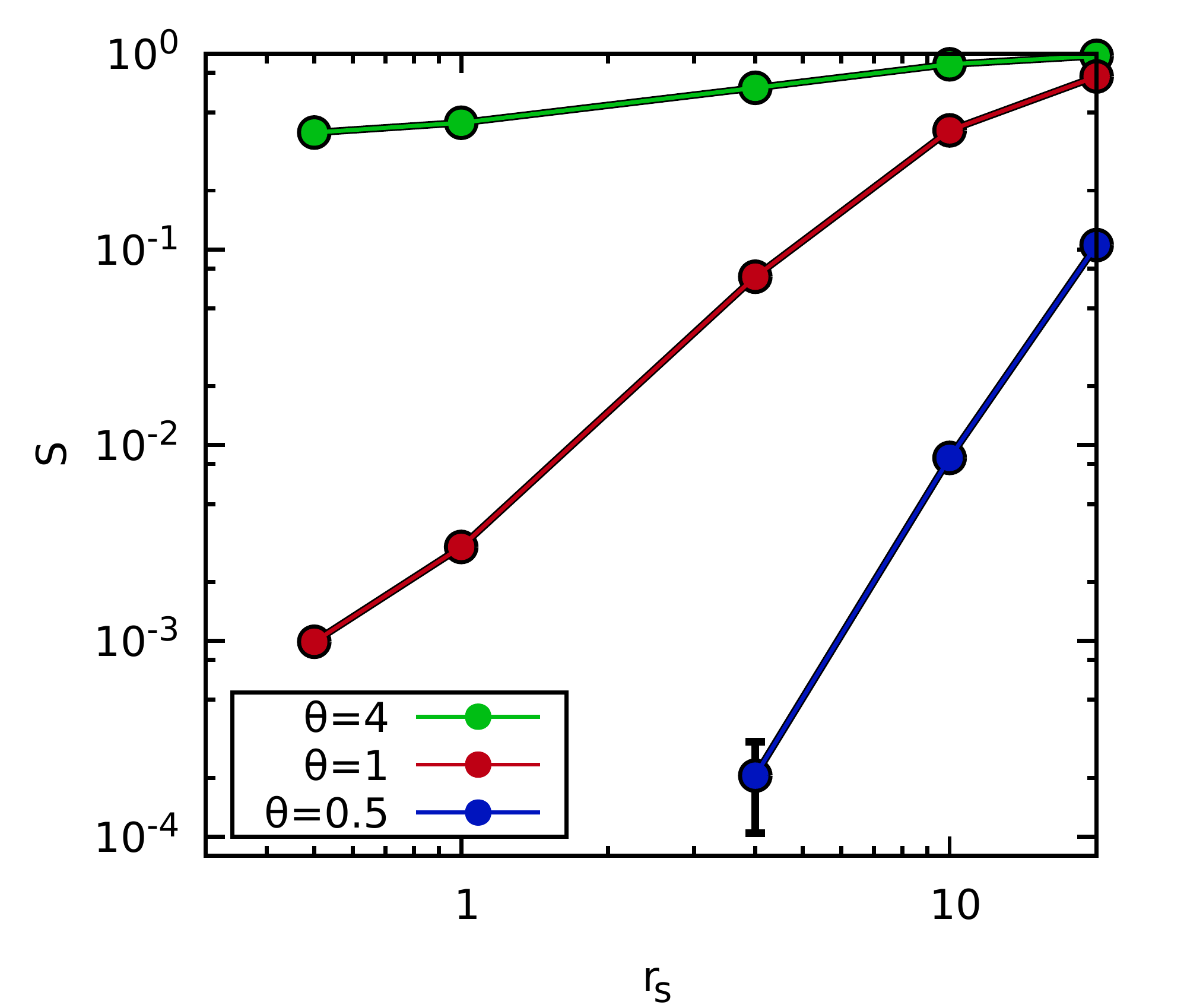}
\includegraphics[width=0.45\textwidth]{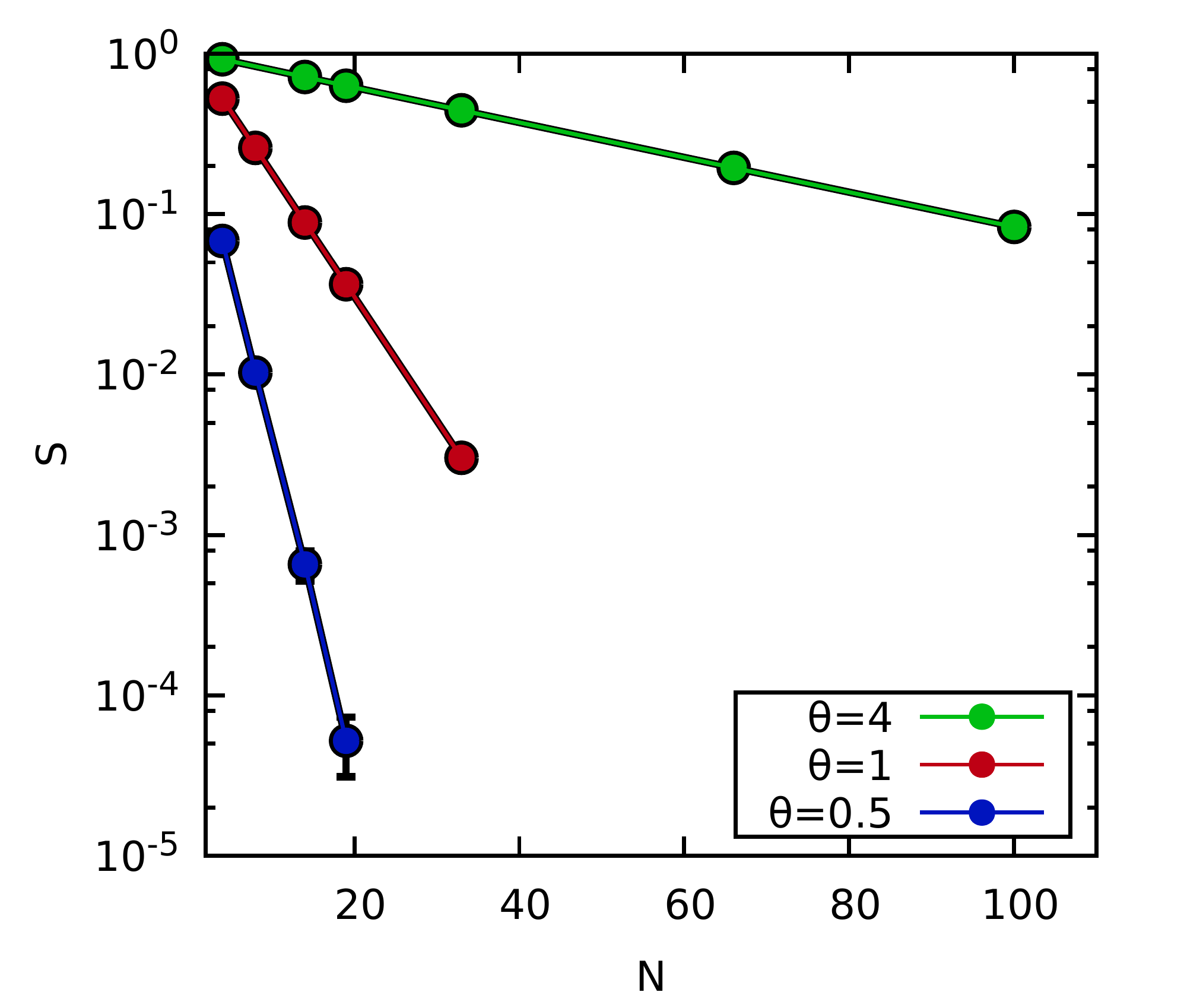}

\caption{\label{fig:PIMC_sign}Average sign of PIMC simulations of the spin-polarized UEG: The left panel shows $S$ in dependence of the density parameter $r_s$ for $N=33$ electrons for $\theta=4$ (green), $\theta=1$ (red), and $\theta=0.5$.
The right panel shows the dependence on system size for a fixed density parameter $r_s=1$. All points have been obtained with $P=50$ imaginary time propagators.
}
\end{figure}

At this point, we must ask ourselves how to generate the configurations $\mathbf{X}$ when the corresponding weights can be both positive and negative. Obviously, this cannot be done using the Metropolis algorithm in a straightforward way, since probabilities must be strictly positive.
To circumvent this issue, we switch to a modified configuration space, where we generate paths according to the absolute value of their weights, and define the modified partition function
\begin{eqnarray}
Z' &=& \int \textnormal{d}\mathbf{X}\ W'(\mathbf{X}) \label{eq:Z_prime} 
= \int \textnormal{d}\mathbf{X}\ |W(\mathbf{X})| \ .
\end{eqnarray}
The correct fermionic observables are then calculated as
\begin{eqnarray}\label{eq:ratio}
\braket{\hat A} = \frac{ \braket{ \hat S \hat A }' }{ \braket{\hat S}' }\ , 
\end{eqnarray}
where $\braket{\dots}'$ denotes the expectation value corresponding to the modulus weights
\begin{eqnarray}
\braket{\hat A}' = \frac{1}{Z'} \int \textnormal{d}\mathbf{X}\ |W(\mathbf{X})| A(\mathbf{X})\ ,
\end{eqnarray}
and $\hat S$ measures the sign of a configuration,
\begin{eqnarray}
\braket{\hat S}' = \frac{1}{Z'} \int \textnormal{d}\mathbf{X}\ S(\mathbf{X}) |W(\mathbf{X})| = \frac{Z}{Z'} \ ,
\end{eqnarray}
with $S(\mathbf{X}) = W(\mathbf{X}) / |W(\mathbf{X})|$.
The problem with Eq.~(\ref{eq:ratio}) is that for a decreasing average sign $S=\braket{\hat S}'$, both the enumerator and the denominator vanish simultaneously. This, in turn, leads to an exponentially increasing statistical uncertainty~\cite{dornheim_abinitio_2017,sign_cite}
\begin{eqnarray}\label{eq:FSP}
\frac{\Delta A}{A} \sim \frac{1}{\sqrt{N_\textnormal{MC}}\braket{S}'  } \sim \frac{ e^{\beta N (f-f')} }{ \sqrt{N_\textnormal{MC} } } \ ,
\end{eqnarray}
where $f$ and $f'$ denote the free energies per particle of the original and modified systems, respectively.
In particular, Eq.~(\ref{eq:FSP}) implies that the statistical uncertainty exponentially increases with the particle number $N$.
However, even for a fixed system size the simulations can become infeasible towards low temperature and weak coupling.
Note that Troyer and Wiese~\cite{troyer_computational_2005} have shown that the FSP is $NP$-hard for a certain class of Hamiltonians. Therefore, a general solution to this problem is unlikely.
\begin{figure}\vspace*{-2.5cm}\centering
\includegraphics[width=0.74\textwidth]{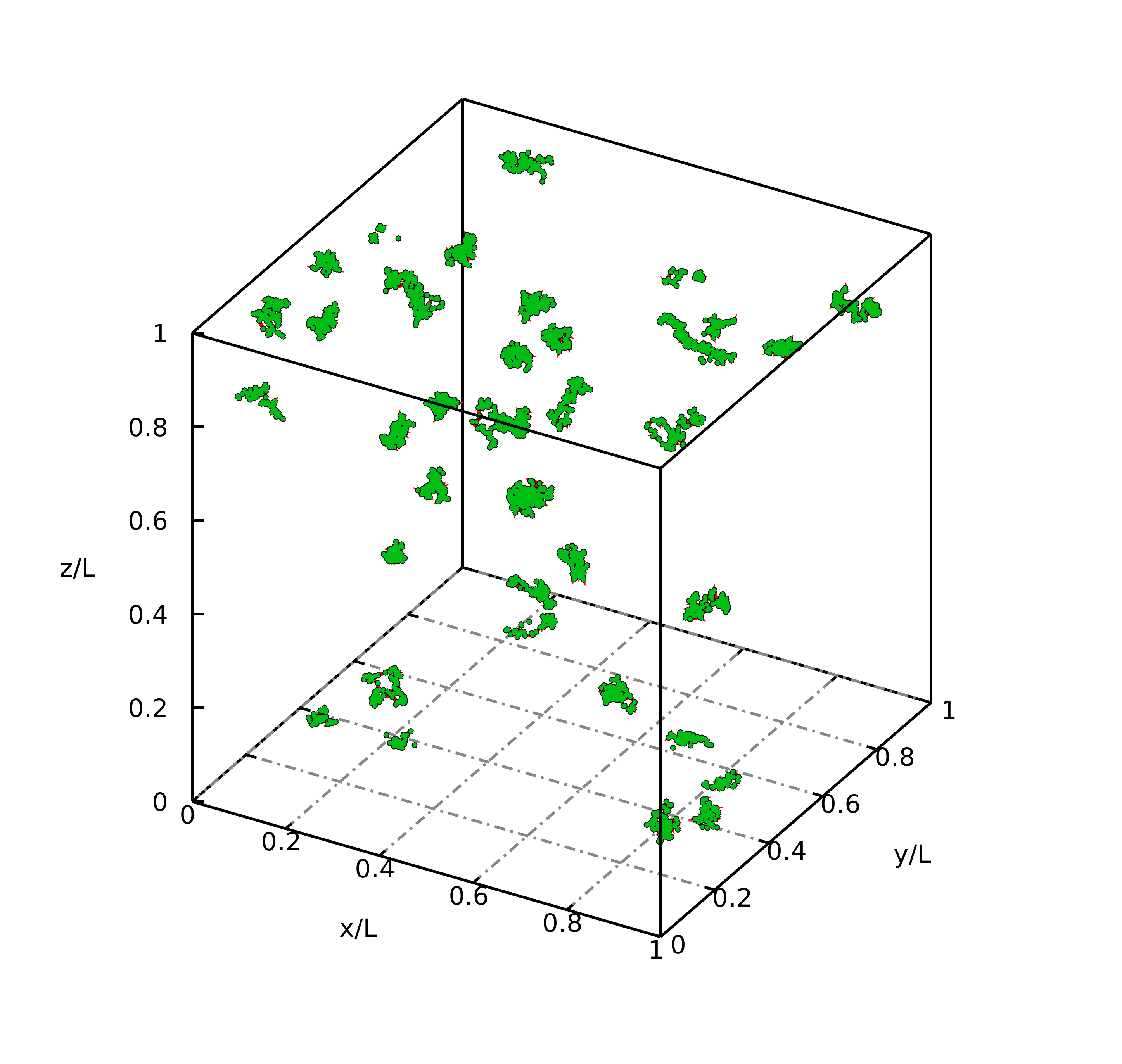}\vspace*{-1.cm} \\ 
\includegraphics[width=0.74\textwidth]{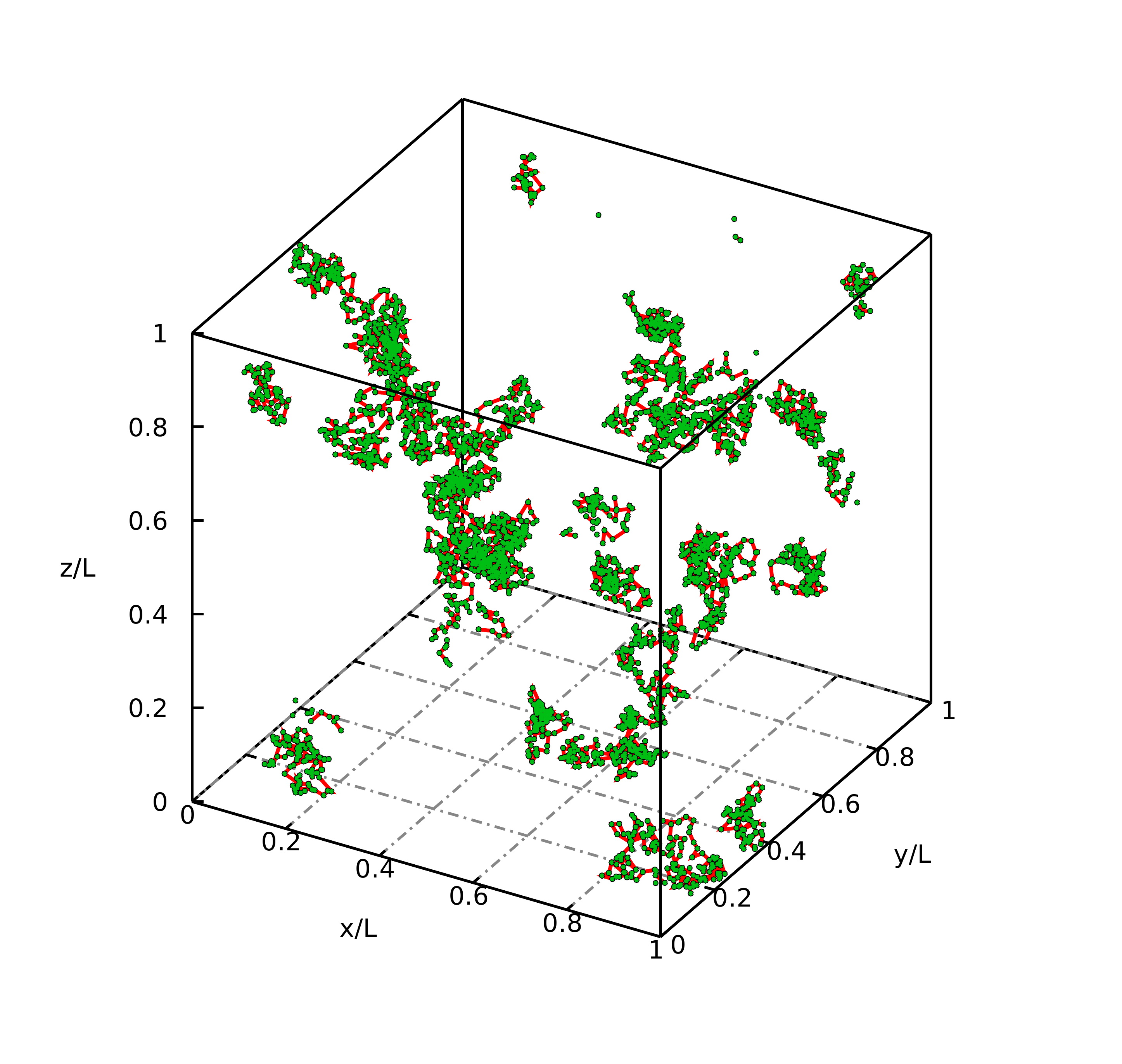}\vspace*{-1.cm}

\caption{\label{fig:PIMC_33}Screenshots from PIMC simulations of the warm dense electron gas with $N=33$ spin-polarized electrons, $P=100$, and $r_s=1$ with $\theta=4$ (top) and $\theta=1$ (bottom).
}
\end{figure}
The FSP within fermionic path integral Monte Carlo simulations is illustrated in Fig.~\ref{fig:PIMC_33}, where we show two random configurations from a PIMC simulation of the uniform electron gas with $N=33$ spin-polarized electrons, $P=100$ imaginary time slices and a density parameter $r_s=1$ (for completeness, we mention that we use a sampling scheme based on the worm algorithm~\cite{boninsegni_worm_2006,boninsegni_worm_2006-1}).
In the top panel, we chose $\theta=4$, i.e., a relatively high temperature. Therefore, the particle paths are only slightly extended and the thermal wavelength is significantly smaller than the average inter-particle distance. This, in turn, means that pair exchange only seldom occurs within the simulation and the average sign is large, rendering such conditions perfectly suitable for PIMC simulations.
In the bottom panel, the temperature is decreased to $\theta=1$. At such conditions, $\lambda_\beta$ is comparable to the particle distance and fermionic exchange plays an important role. This is manifest in the many exchange cycles, i.e., the paths that contain more than a single particle. Since each pair exchange leads to a sign change in the weight function, positive and negative weights occur with a nearly equal frequency, resulting an average sign of $S\sim10^{-3}$, cf.~Fig.~\ref{fig:PIMC_sign}. For this reason, standard PIMC simulations are confined to relatively high temperature or strong coupling where the exchange effects are suppressed by the Coulomb repulsion of the electrons.

This is investigated more quantitatively in Fig.~\ref{fig:PIMC_sign}.
In the left panel, we show the $r_s$-dependence of the average sign of PIMC simulations of the UEG of $N=33$ spin-polarized electrons, which corresponds to a closed momentum shell and, therefore, is often used in QMC studies~\cite{brown_path-integral_2013,schoof_textitab_2015,dornheim_permutation_2015-1,groth_abinitio_2016,malone_accurate_2016}. The number of imaginary time propagators was chosen as $P=50$ and the green, red, and blue points correspond to $\theta=4$, $\theta=1$, and $\theta=0.5$, respectively. All three curves exhibit the same qualitative behavior, that is, a decreasing sign towards smaller $r_s$ (i.e, towards high density). This can be understood by recalling that the density parameter $r_s$ plays the role of the coupling parameter for the UEG~\cite{kraeft_quantum_1986}: For strong coupling, the paths of different particles in the PIMC simulation are spatially separated and, hence, exchange cycles are not very probable. With decreasing $r_s$, the system becomes more ideal and the occurring pair exchanges lead to smaller values of $S$. Furthermore, we observe that this effect is significantly increased for lower temperatures, see the discussion of Fig.~\ref{fig:PIMC_33} above. For $\theta=4$, the sign does not drop below $S=0.3$ and standard PIMC simulations are efficient over the entire density range. For $\theta=1$, simulations for $r_s=4$ are barely feasible with reasonable computational effort, whereas for $\theta=0.5$, even $r_s=10$, which corresponds to relatively strong coupling, is difficult.

In the right panel, we show the dependence of the average sign on system size for a constant density parameter $r_s=1$. For all three depicted temperatures, $S$ exhibits an exponential decay with $N$ as predicted by Eq.~(\ref{eq:FSP}), which becomes significantly more steep for low $\theta$. For $\theta=4$, simulations of $N\sim100$ spin-polarized electrons are feasible. Yet, we stress that even at such high temperatures, fermionic exchange leads to an exponential increase of computation time with respect to $N$.
For $\theta=1$, the situation is considerably worse and the decay of $S$ restricts PIMC simulations to $N<20$. Finally, for $\theta=0.5$, even simulations of $N=10$ electrons are not feasible.

We thus conclude that standard PIMC cannot be used to obtain an accurate description of the UEG at warm dense matter conditions since the FSP renders simulations unfeasible towards high density and low temperature.


\subsection{Restricted Path Integral Monte Carlo\label{sec:RPIMC}}
A relatively common strategy to avoid the fermion sign problem is the so-called \textit{fixed node approximation}, which is also known as the restricted PIMC (RPIMC) method~\cite{ceperley_fermion_1991}. On the one hand, RPIMC gets completely rid of the FSP and, therefore, simulations are feasible at low temperature and strong degeneracy. On the other hand, as we shall see, this comes at the cost of an uncontrollable systematic error so that the exact \textit{ab initio} character of the quantum Monte Carlo paradigm is lost.

In statistical mechanics, the fermionic density matrix elements in coordinate space $\rho(\mathbf{R},\mathbf{R}',\beta)$ are often introduced as the solution to the Bloch equation
\begin{eqnarray}\label{eq:bloch_rpimc}
-\frac{\textnormal{d}}{\textnormal{d}\beta}\ \rho(\mathbf{R},\mathbf{R}',\beta)  = \hat H \rho(\mathbf{R},\mathbf{R}',\beta)\ ,
\end{eqnarray}
with the initial condition
\begin{eqnarray}\label{eq:initial_cond}
\rho(\mathbf{R},\mathbf{R}',0) = \hat A\delta(\mathbf{R}-\mathbf{R}')\ ,
\end{eqnarray}
where $\hat A$ denotes the antisymmetrization operator. For the restricted path integral Monte Carlo approach developed by Ceperley~\cite{ceperley_fermion_1991,ceperley_path-integral_1992}, the initial condition from Eq.~(\ref{eq:initial_cond}) is replaced with a zero boundary condition. 
Following Ref.~\cite{ceperley_fermion_1991}, we denote the second argument of the density matrix as the reference slice $\mathbf{R}_0$. Assuming that Eq.~(\ref{eq:initial_cond}) holds, we can define a nodal surface
\begin{eqnarray}
\gamma(\mathbf{R}_0,\tau) = \{ \mathbf{R}\ |\ \rho(\mathbf{R},\mathbf{R}_0,\tau)=0 \}\ , \label{eq:nodal_surface}
\end{eqnarray}
for all imaginary times $0\leq\tau\leq\beta$. Obviously, Eq.~(\ref{eq:nodal_surface}) divides the total configuration space into sub-regions of a fixed sign, described by the so-called \textit{reach}
\begin{eqnarray}
\Gamma(\mathbf{R}_0,\tau) = \{ \mathbf{R}_\tau\ |\ \rho(\mathbf{R},\mathbf{R}_0,\tau)\neq0\}\ . \label{eq:reach}
\end{eqnarray}
Equation~(\ref{eq:reach}) can be interpreted as the set of all paths $\mathbf{R}_\tau\to\mathbf{R}_0$ avoiding the nodes, which are the only paths contributing to the thermal density matrix.
Odd permutations cross the nodal surface an odd number of times and, therefore, do not satisfy Eq.~(\ref{eq:reach}). They do not contribute to $\rho(\mathbf{R},\mathbf{R}_0,\tau)$ as they cancel with the node-crossing paths of even permutation, which is sometimes denoted as the \textit{tiling property} proved in Ref.~\cite{ceperley_fermion_1991}.
This, in turn, means that all contributions to the thermal density matrix of a fixed reference slice $\mathbf{R}_0$ are strictly positive and, thus, perfectly suited for a Metropolis Monte Carlo simulation similar to Sec.~\ref{sec:PIMC} without the sign problem.
The fermionic expectation value of an arbitrary observable can then be computed by averaging over $\mathbf{R}_0$ itself.
In principle, this re-casting of the fermionic path integral Monte Carlo scheme in terms of different nodal regions is exact, given complete knowledge of the nodes. However, this information can only be obtained from a solution of the full fermionic many-body problem in the first place and, thus, little seems to be gained. In practice, we introduce an approximate \textit{trial ansatz} for the density matrix, most commonly from the ideal system (i.e., a Slater determinant or, for multiple particle species, a product thereof).
Naturally, one would assume that the ideal nodes work best for weak coupling, i.e., at high temperature and density. In particular, RPIMC simulations of the UEG should become exact for $r_s\to0$.

In practice, within a RPIMC simulation we propose a new path and subsequently enforce the nodal constraint, Eq.~(\ref{eq:reach}), by computing the sign of the new configuration weight and by rejecting the move if the sign is negative. This becomes particularly problematic when the reference slice $\mathbf{R}_0$ is changed (remember that RPIMC simulations require us to average over $\mathbf{R}_0$) since the constraint then has to be checked on all time slices. The problem is that for low temperature (i.e., for long paths, see Sec.~\ref{sec:PIMC}) the nodal surface for large distances in imaginary time $\tau$ to the reference slice can significantly change for small changes of the latter. This means that even small updates of $\mathbf{R}_0$ can be rejected most of the time and the reference point \textit{freezes}. This purely practical ergodicity problem potentially introduces a second source of systematic bias to RPIMC simulations.
A comprehensive comparison of RPIMC data to other QMC methods can be found in Sec.~\ref{sec:comparison_finite_N}.

As a final note, we mention that, in contrast to the ground state, the fixed node approximation as outlined above constitutes an uncontrolled approximation since the total energy is not variational. A possible strategy to overcome this issue is to perform an additional coupling constant integration (see Sec.~\ref{sec:fxc}) to compute the free energy $f$. The next step would then be to introduce a parametrization of the nodes with respect to a set of free parameters, which can be used to minimize $f$. However, this is substantially more complicated than at $T=0$ and, to the best of our knowledge, has not yet been pursued in practice.
Furthermore, we mention that RPIMC has nevertheless been applied to various realistic systems (such as deuterium, neon, or carbon plasmas) at warm dense matter conditions, e.g., Refs.~\cite{militzer_path_2000,driver_all-electron_2012,militzer_development_2015,driver_first-principles_2016}.

\subsection{Permutation Blocking Path Integral Monte Carlo\label{sec:PB-PIMC}}

The permutation blocking PIMC (PB-PIMC) approach~\cite{dornheim_permutation_2015,dornheim_permutation_2015-1,groth_abinitio_2016,dornheim_abinitio_2016,dornheim_analyzing_2016} can be viewed as a further development of the standard PIMC method from Sec.~\ref{sec:PIMC} and allows to go both towards lower temperature and increased density, i.e., towards the WDM regime where fermionic exchange is crucial. Here 'blocking' refers to the combination of multiple configurations with different signs into a single weight, which means that some part of the cancellation due to the fermion sign problem is carried out analytically. To further explore this point, let us consider an illustrative example. Let us split the partition function into the two parts
\begin{eqnarray}
Z = \int_{\mathbf{X}^-} \textnormal{d}\mathbf{X}\ W(\mathbf{X} ) + \int_{\mathbf{X}^+} \textnormal{d}\mathbf{X}\ W(\mathbf{X} )\ ,
\end{eqnarray}
where $\mathbf{X}^-$ ($\mathbf{X}^+$) denotes those configurations with a negative (positive) weight $W$. Now suppose that you could pair each negative weight $\mathbf{X}^-_i$ with a positive weight $\mathbf{X}^+_i$ with a larger (or equal) modulus weight and, in this way, obtain a  new 'meta-configuration' $\tilde{\mathbf{X}}_i$ with a meta-configuration weight
\begin{eqnarray}
\tilde{W}(\tilde{\mathbf{X}}_i) = W(\mathbf{X}^-_i) + W(\mathbf{X}^+_i) \geq 0 \ .
\end{eqnarray}
In this way, we have recasted the partition function as the integral over terms that are strictly positive,
\begin{eqnarray}\label{eq:blockingZ}
Z = \int \textnormal{d}\tilde{\mathbf{X}}\ \tilde{W} (\tilde{\mathbf{X}}) ,
\end{eqnarray}
and the fermion sign problem would be solved. Unfortunately, in practice, such a perfect implementation of the blocking idea is not possible. Instead, we combine positive and negative permutations from the fermionic partition function, Eq.~(\ref{eq:zfermion}), within determinants both for the spin-up and down electrons. The benefits due to such intrinsically antisymmetric imaginary time propagators have long been known, see e.g. Refs.~\cite{filinov_book_1977,takahashi_monte_1984,lyubartsev_simulation_2005,chin_high-order_2015}. In particular, they have been successfully exploited within the PIMC simulations by Filinov and co-workers~\cite{filinov_construction_1986,filinov_thermodynamics_2001,filinov_phase_2001,filinov_thermodynamic_2004,filinov_correlation_2007,filinov_proton_2012,filinov_fermionic_2015,filinov_thermodynamics_2015,filinov_total_2015}. As we will see, the problem with this approach is  that with an increasing number of time slices $P$ [which are needed to decrease the commutator errors due to the primitive factorization, cf.~Eq.~(\ref{eq:bch})], the effect of the blocking due to the determinant vanishes and the original sign problem is recovered. For this reason, the second key ingredient of the PB-PIMC approach is the introduction of a more sophisticated fourth-order factorization scheme that allows for sufficient accuracy with fewer time slices~\cite{takahashi_monte_1984-1,chin_gradient_2002,brualla_higher_2004,sakkos_high_2009,zillich_extrapolated_2010}.
The simulation scheme is completed by an efficient update scheme that allows for ergodic sampling in the new configuration space~\cite{dornheim_permutation_2015}.

Let us begin the derivation of the PB-PIMC partition function with an introduction of the fourth-order factorization of the density matrix~\cite{chin_gradient_2002}
\begin{eqnarray}\label{eq:fourth_order}
e^{-\epsilon\hat H} \approx e^{-v_1\epsilon \hat W_{a_1}} e^{-t_1\epsilon\hat K} e^{-v_2\epsilon\hat W_{1-2a_1}} e^{-t_1\epsilon\hat K} e^{-v_1\epsilon\hat W_{a_1}} e^{-2t_0\epsilon\hat K}\ ,
\end{eqnarray}
which has been studied extensively by Sakkos~\textit{et al.}~\cite{sakkos_high_2009}. 
\begin{figure}
\includegraphics[width=0.45\textwidth]{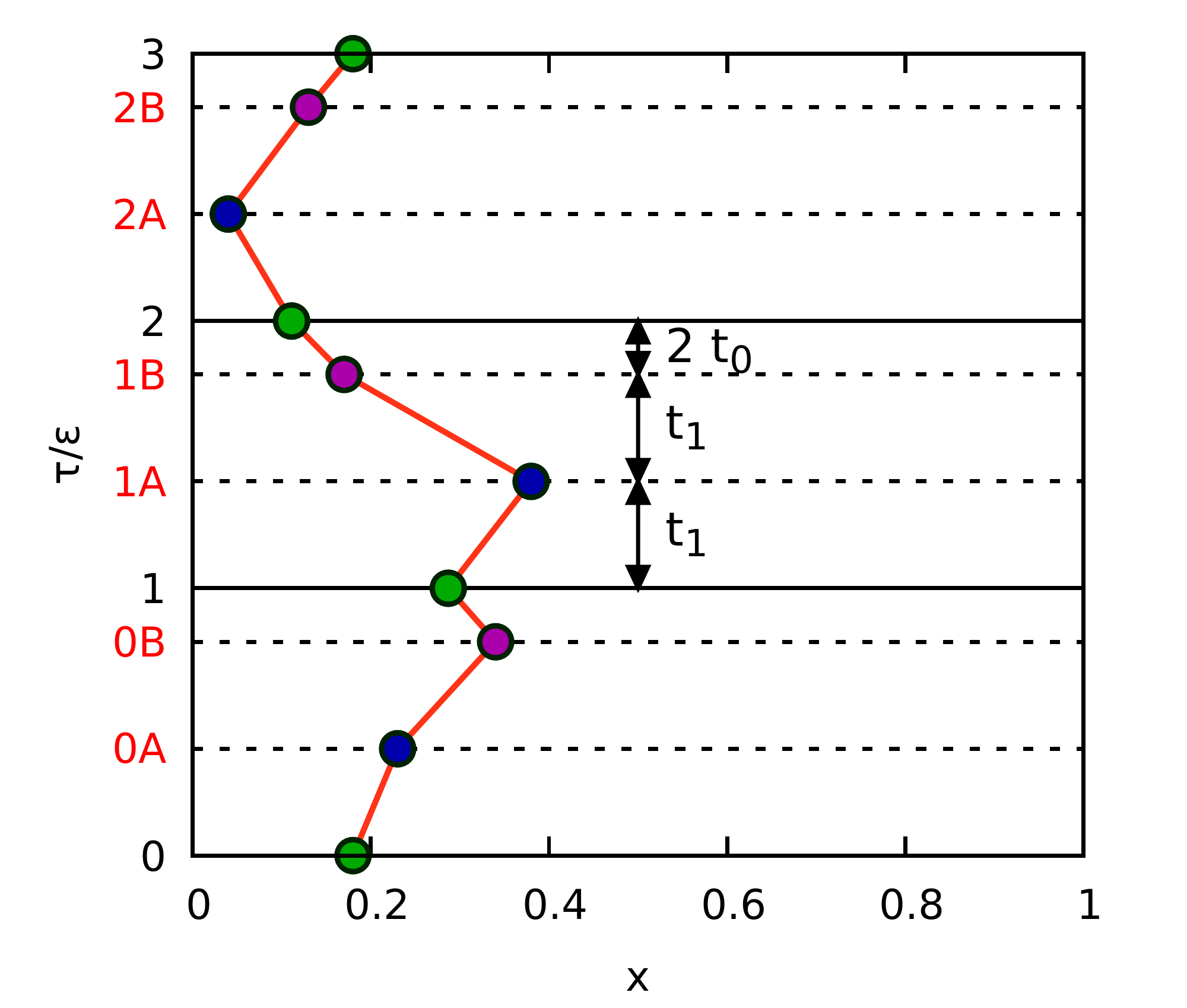}
\includegraphics[width=0.45\textwidth]{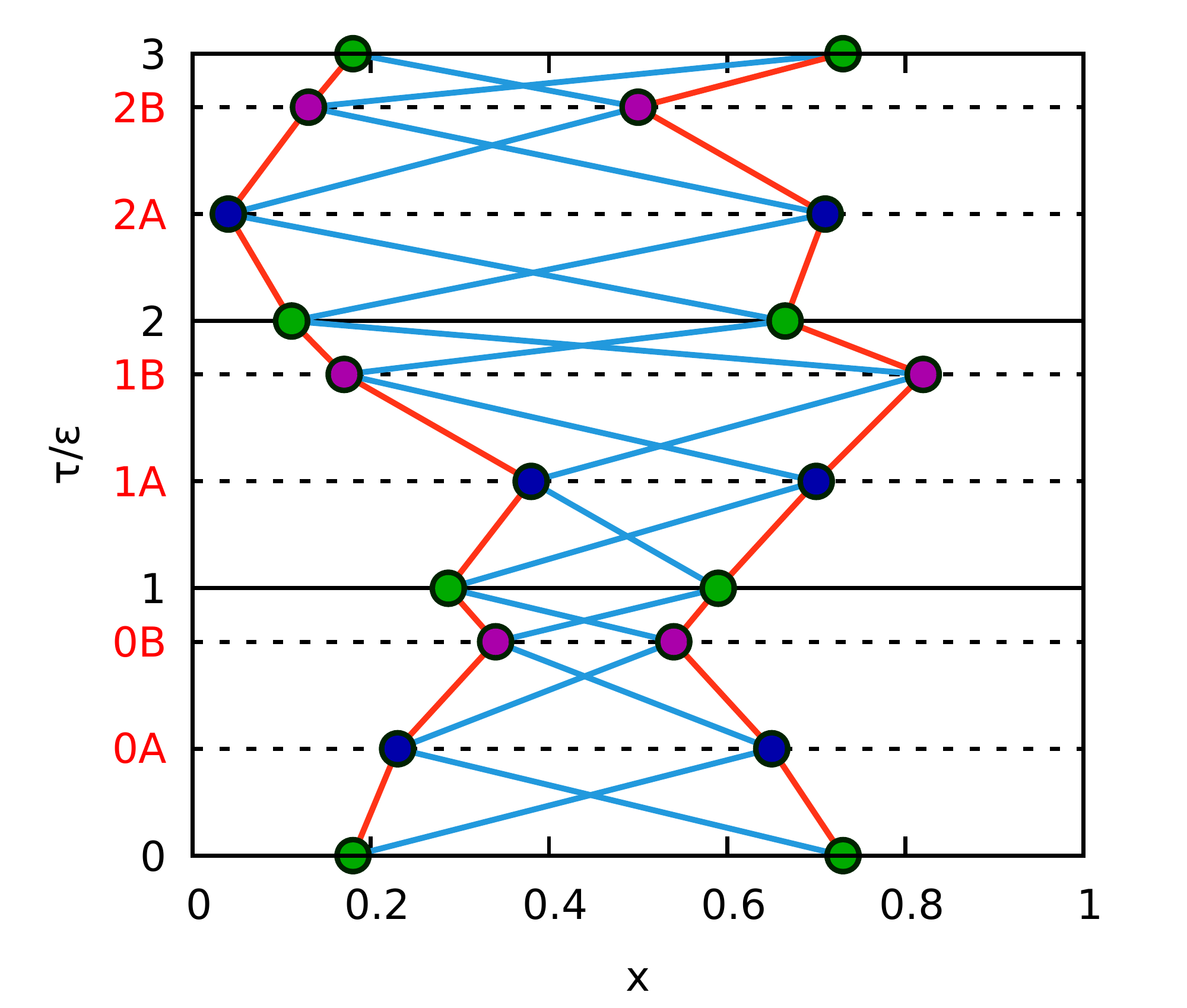}
\caption{\label{fig:PBPIMC_illustration}Schematic illustration of the PB-PIMC approach -- Left panel: Illustration of the fourth-order factorization from Eq.~(\ref{eq:fourth_order}) in the $\tau$-$x$-plane. Beads of different colors correspond to the main (green), ancilla A (blue), and ancilla B (purple) slices, which occur for each of the $P=3$ imaginary time propagators. The ratio $t_0/t_1$ is not fixed and can be used for optimization.
Right panel: Combination of $3PN!$ configurations from standard PIMC into a new 'meta-configuration' due to the determinants on all time slices.
}
\end{figure}
First and foremost, we note that there occur three factors involving the kinetic energy operator $\hat K$. Therefore, for each imaginary time propagator there are three time slices. This is illustrated in the left panel of Fig.~\ref{fig:PBPIMC_illustration}, where the path of a single particle is shown in the $\tau$-$x$-plane with $P=3$ fourth-order factors. For each propagator of length $\epsilon$, there are two equidistant slices of length $t_1\epsilon$, which we denote as the main slice (green beads) and ancilla slice A (blue beads). Furthermore, there is a third slice of length $2t_0\epsilon = \epsilon(1-2t_1)$, i.e., ancilla slice B (purple beads). Note that the ratio of $t_0/t_1$ is not fixed and $t_0$ can be chosen freely within $0\leq t_0 \leq (1-1/\sqrt{3})$, which can be exploited to further accelerate the convergence with $P$~\cite{sakkos_high_2009}.
In order to fully cancel the first error terms from the factorization error, Eq.~(\ref{eq:bch}), the $\hat W$-operators in Eq.~(\ref{eq:fourth_order}) combine the potential energy $\hat V$ with double commutator terms
\begin{eqnarray}
[[\hat V,\hat K],\hat V] = \sum_{i=1}^N |\mathbf{F}_i|^2 \ ,
\end{eqnarray}
with $\mathbf{F}_i = -\nabla_i V(\mathbf{R})$ denoting the entire force on particle $i$. In particular, it holds
\begin{eqnarray}\label{eq:hatW}
\hat W_{a_1} &=& \hat V + \frac{u_0}{v_1} a_1 \epsilon^2 \sum_{i=1}^N |\mathbf{F}_i|^2\ , \\
\hat W_{1-2a_1} &=& \hat V + \frac{u_0}{v_2}(1-2a_1)\epsilon^2 \sum_{i=1}^N |\mathbf{F}_i|^2 \ , \nonumber
\end{eqnarray}
and the coefficients $u_0$, $v_1$, and $v_2$ are fully determined by the choice for $t_0$ and $0\leq a_1\leq 1$,
\begin{eqnarray}
u_0 &=& \frac{1}{12} \left( 1 - \frac{1}{1-2t_0} + \frac{1}{6(1-2t_0)^3} \right)\ , \\
v_1 &=& \frac{1}{6(1-2t_0)^2} \ , \\
v_2 &=& 1 - 2v_1 \ .
\end{eqnarray}
Eq.~(\ref{eq:hatW}) implies that, in addition to the potential energy, we have to evaluate all forces (both due to an external potential and pair interactions) on all slices for each propagator, albeit the weight of the individual contributions from the different kind of slices can be adjusted. For example, by choosing $a_1=0$, the forces are only relevant on ancilla slice A, whereas for $a_1=1/3$ all three slices contribute equally. Again, we stress that this second free parameter (in addition to $t_0$) can be used for optimization.

Incorporating the fourth-order partition function into the expression for $Z$ from Eq.~(\ref{eq:zfermion}) leads to the final result for the PB-PIMC partition function~\cite{dornheim_abinitio_2016}
\begin{eqnarray}\label{eq:PBPIMC_Z}
Z = \frac{1}{(N^\uparrow!N^\downarrow!)^{3P}} \int \textnormal{d}\tilde{\mathbf{X}}\ \prod_{\alpha=0}^{P-1} \left( e^{-\epsilon\tilde {V}_\alpha} e^{-\epsilon^3 u_0 \tilde{F}_\alpha} D^\uparrow_\alpha D^\downarrow_\alpha \right) \ ,
\end{eqnarray}
where $\tilde{V}_\alpha$ and $\tilde{F}_\alpha$ contain all contributions due to the potential energy and the forces for a specific propagator $\alpha$,
\begin{eqnarray}
\tilde{V}_\alpha &=& v_1 V(\mathbf{R}_\alpha) + v_2 V(\mathbf{R}_{\alpha A}) + v_1 V(\mathbf{R}_{\alpha B}) \ , \\
\tilde{F}_\alpha &=& \sum_{i=1}^N( a_1|\mathbf{F}_{\alpha,i}|^2 + (1-2a_1)|\mathbf{F}_{\alpha A,i}|^2 + a_1|\mathbf{F}_{\alpha B,i}|^2)\ .
\end{eqnarray}
Further, we stress that the integration has to be carried out over all possible coordinates on the ancilla slices as well, i.e.,
\begin{eqnarray}
\textnormal{d}\tilde{\mathbf{X}}\ = \prod_{\alpha=0}^{P-1}\textnormal{d}\mathbf{R}_\alpha\textnormal{d}\mathbf{R}_{\alpha A}\textnormal{d}\mathbf{R}_{\alpha B} \ .
\end{eqnarray}
All fermionic exchange is contained within the exchange-diffusion functions
\begin{eqnarray}
D_\alpha^\uparrow &=& \textnormal{det}(\rho_{\alpha}^\uparrow)\textnormal{det}(\rho_{\alpha A}^\uparrow) \textnormal{det}(\rho_{\alpha B}^\uparrow)\ , \\
D_\alpha^\downarrow &=& \textnormal{det}(\rho_{\alpha}^\downarrow)\textnormal{det}(\rho_{\alpha A}^\downarrow) \textnormal{det}(\rho_{\alpha B}^\downarrow) \ , 
\end{eqnarray}
which constitute a product of the determinants of the free particle (diffusion) matrices between particles $i$ and $j$ on two adjacent time slices (not propagators)
\begin{eqnarray}\label{eq:free_diffusion}
\rho_{\alpha}^\uparrow (i,j) &=& \frac{1}{\lambda_{t_1\epsilon}^3 } \sum_\mathbf{n} \textnormal{exp}\left( -\frac{ \pi}{\lambda^2_{t_1\epsilon}} (\mathbf{r}^\uparrow_{\alpha, j} - \mathbf{r}^\uparrow_{\alpha A, i} + \mathbf{n}L)^2   \right) \ , \\
\rho_{\alpha A}^\uparrow (i,j) &=& \frac{1}{\lambda_{t_1\epsilon}^3 } \sum_\mathbf{n}\textnormal{exp}\left( -\frac{ \pi}{\lambda^2_{t_1\epsilon}} (\mathbf{r}^\uparrow_{\alpha A, j} - \mathbf{r}^\uparrow_{\alpha B, i} + \mathbf{n}L)^2   \right) \ , \\
\rho_{\alpha B}^\uparrow (i,j) &=& \frac{1}{\lambda_{2t_0\epsilon}^3 } \sum_\mathbf{n}\textnormal{exp}\left( -\frac{ \pi}{\lambda^2_{2t_0\epsilon}} (\mathbf{r}^\uparrow_{\alpha B, j} - \mathbf{r}^\uparrow_{\alpha+1, i} + \mathbf{n}L)^2   \right) \ ,
\end{eqnarray}
with an analogous definition for the spin-down electrons.
\begin{figure}
\includegraphics[width=0.8\textwidth]{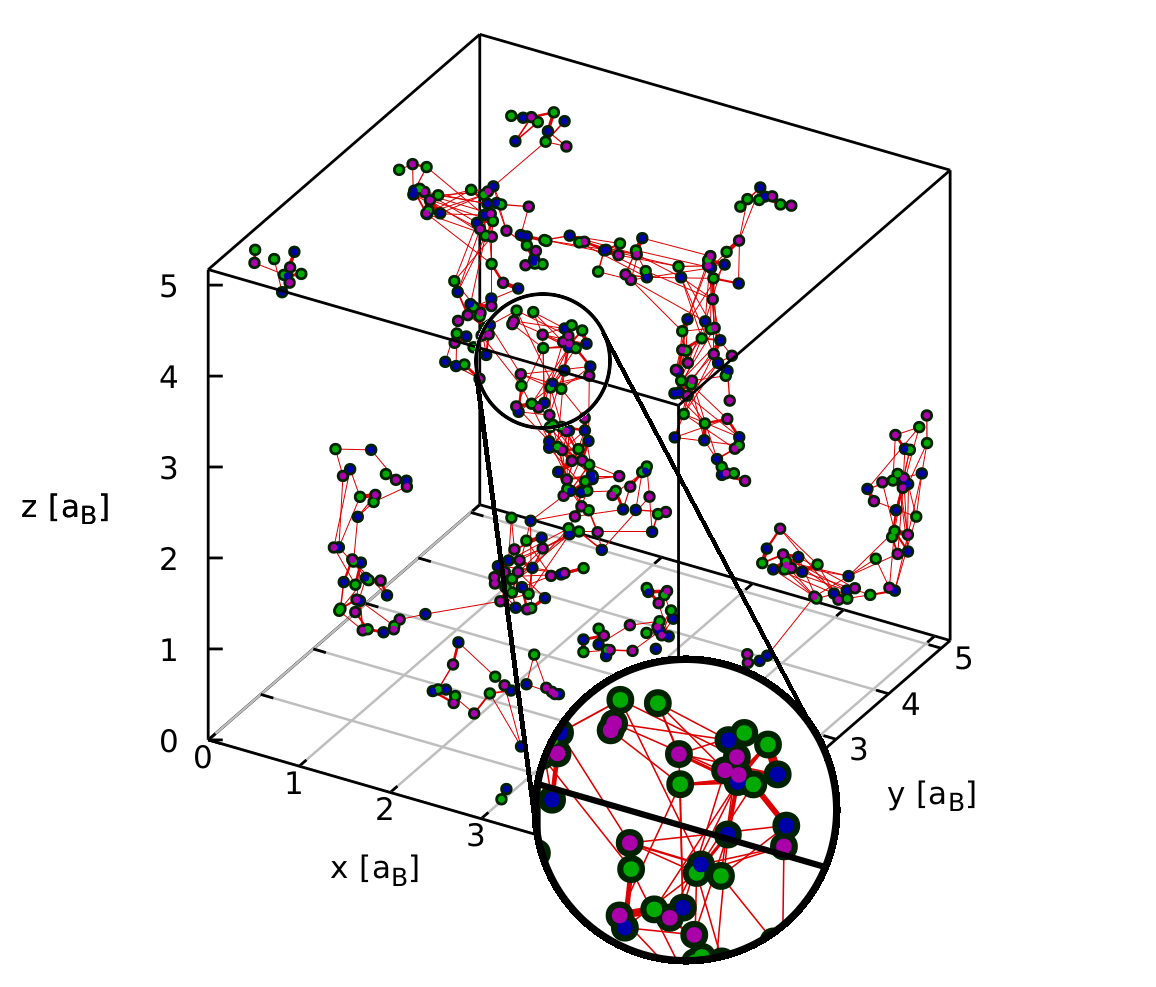}
\caption{\label{fig:PBPIMC_33}Screenshot of a PB-PIMC simulation of the spin-polarized UEG with $N=33$, $P=4$, $r_s=1$, and $\theta=1$. The green, blue, and purple beads correspond to main, ancilla A, and ancilla B slices, respectively. The different width of the red connections symbolizes the magnitude of the diffusion matrix elements, cf.~Eq.~(\ref{eq:free_diffusion}). Beads with more than two visible links significantly contribute to the permutation blocking.
}
\end{figure}
Note that we have again exploited the idempotency property of the antisymmetrization operator to introduce determinants on all the ancilla time slices as well. The reason for this choice becomes obvious by closely examining the new configuration space, which is illustrated in the right panel of Fig.~\ref{fig:PBPIMC_illustration}.
Shown is a configuration of two particles in the $\tau$-$x$-plane and beads on different types of time slices are distinguished by the different colors. For standard PIMC, a typical configuration would be given by the two red paths, which would correspond to two separate paths without a pair exchange. In addition, one would also have to consider all configurations with the same positions of the individual beads, but different connections between beads on adjacent slices, which would lead to contributions with different signs.
By introducing the determinants within the PB-PIMC scheme, we combine all $N!$ possible connections between beads on adjacent slices into a single configuration weight. As explained in the beginning of this section, this analytic blocking of configurations with different signs results in a drastically less severe sign problem and, therefore, to perform simulations in substantial parts of the WDM regime.

This is further illustrated in Fig.~\ref{fig:PBPIMC_33}, where we show a random screenshot from a PB-PIMC simulation with $P=4$ fourth-order propagators and $N=33$ spin-polarized electrons at $r_s=1$ and $\theta=1$. Again, the beads of different color correspond to different kind of time slices. The different line width of the red connections between some beads on adjacent slices symbolize the magnitude of the diffusion matrix elements, Eq.~(\ref{eq:free_diffusion}). Without the determinants, each bead would have exactly two connections. Hence, beads with more than two visible links in Fig.~\ref{fig:PBPIMC_33} significantly contribute to the permutation blocking, which, in stark contrast to standard PIMC, makes simulations feasible under such conditions.

As explained in Sec.~\ref{sec:Metropolis}, we use the Metropolis Monte Carlo algorithm~\cite{metropolis_equation_1953} to generate all possible paths $\tilde{\mathbf{X}}$ according to the appropriate configuration weight defined by Eq.~(\ref{eq:PBPIMC_Z}).
Let us now discuss how we can compute physical expectation values from this Markov chain of configurations.
For example, the total energy of the system can be computed from the partition function via the well-known relation
\begin{eqnarray}\label{eq:E_from_Z}
E = - \frac{1}{Z} \frac{\partial Z}{\partial\beta}\quad ,
\end{eqnarray}
and plugging in the PB-PIMC expression for $Z$, Eq.~(\ref{eq:PBPIMC_Z}), into (\ref{eq:E_from_Z}) gives the desired Monte Carlo estimator (for $N$ spin-polarized electrons, the generalization to an arbitrary degree of spin polarization is obvious),
\begin{eqnarray}\label{eq:PBPIMC_energy_estimator}
E = \frac{1}{P} \sum_{\alpha=0}^{P-1} \left( \tilde{V}_\alpha + 3\epsilon^2 u_0 \tilde{F}_k \right)
+ \frac{3DN}{2\epsilon} - \frac{\pi}{\beta}
\sum_{\alpha=0}^{P-1}\sum_{i=1}^N\sum_{k=1}^N \left( \eta^\alpha_{k,i} \lambda_{t_1\epsilon}^{-2}
+\eta^{\alpha A}_{k,i} \lambda_{t_1\epsilon}^{-2}
+\eta^{\alpha B}_{k,i} \lambda_{2t_0\epsilon}^{-2}  \right) \quad ,
\end{eqnarray}
with the definitions
\begin{eqnarray}
\eta^{\alpha}_{k,i} &=& \frac{ (\rho_{\alpha}^{-1})_{k,i} }{ \lambda_{t_1\epsilon}^3 } \sum_\mathbf{n} \left(
e^{-\frac{\pi}{\lambda^2_{t_1\epsilon}}(\mathbf{r}_{\alpha,k} - \mathbf{r}_{\alpha A,i} + L\mathbf{n})^2 }
(\mathbf{r}_{\alpha,k} - \mathbf{r}_{\alpha A,i} + L\mathbf{n})^2 \right) \\
\eta^{\alpha A}_{k,i} &=& \frac{ (\rho_{\alpha A}^{-1})_{k,i} }{ \lambda_{t_1\epsilon}^3 } \sum_\mathbf{n} \left(
e^{-\frac{\pi}{\lambda^2_{t_1\epsilon}}(\mathbf{r}_{\alpha A,k} - \mathbf{r}_{\alpha B,i} + L\mathbf{n})^2 }
(\mathbf{r}_{\alpha A,k} - \mathbf{r}_{\alpha B,i} + L\mathbf{n})^2 \right) \\
\eta^{\alpha B}_{k,i} &=& \frac{ (\rho_{\alpha B}^{-1})_{k,i} }{ \lambda_{2t_0\epsilon}^3 } \sum_\mathbf{n} \left(
e^{-\frac{\pi}{\lambda^2_{2t_0\epsilon}}(\mathbf{r}_{\alpha B,k} - \mathbf{r}_{\alpha+1,i} + L\mathbf{n})^2 }
(\mathbf{r}_{\alpha B,k} - \mathbf{r}_{\alpha +1,i} + L\mathbf{n})^2 \right) \quad .
\end{eqnarray}
Here the notation $(\rho_{\alpha}^{-1})_{k,i}$ indicates the $(k,i)$-element of the inverse diffusion matrix. 
Interestingly, the contribution of the force-terms to $E$ in Eq.~(\ref{eq:PBPIMC_energy_estimator}) splits to both the kinetic and potential energy, see Refs.~\cite{sakkos_high_2009,dornheim_permutation_2015} for more details.

\begin{figure}
\includegraphics[width=0.45\textwidth]{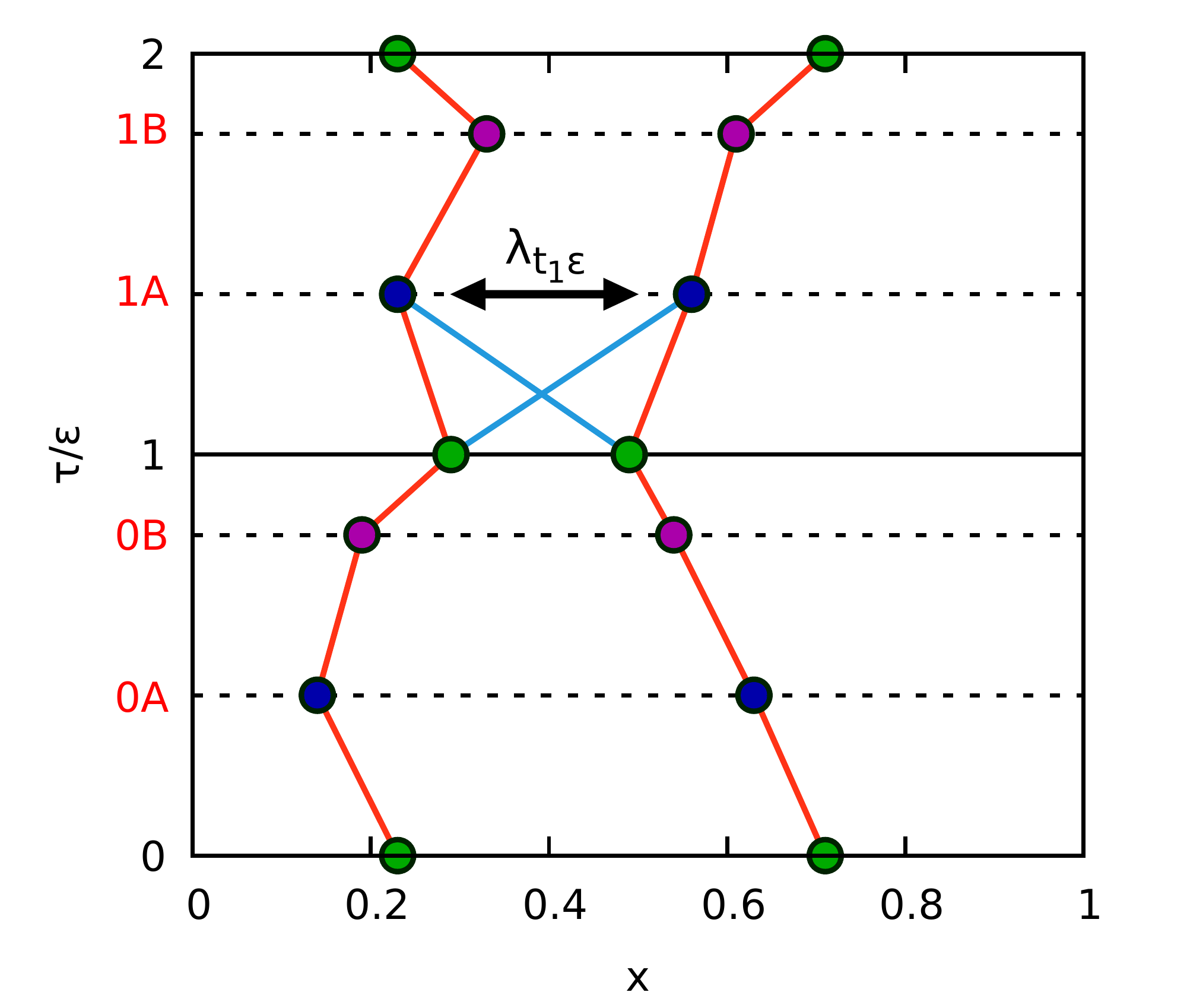}
\includegraphics[width=0.45\textwidth]{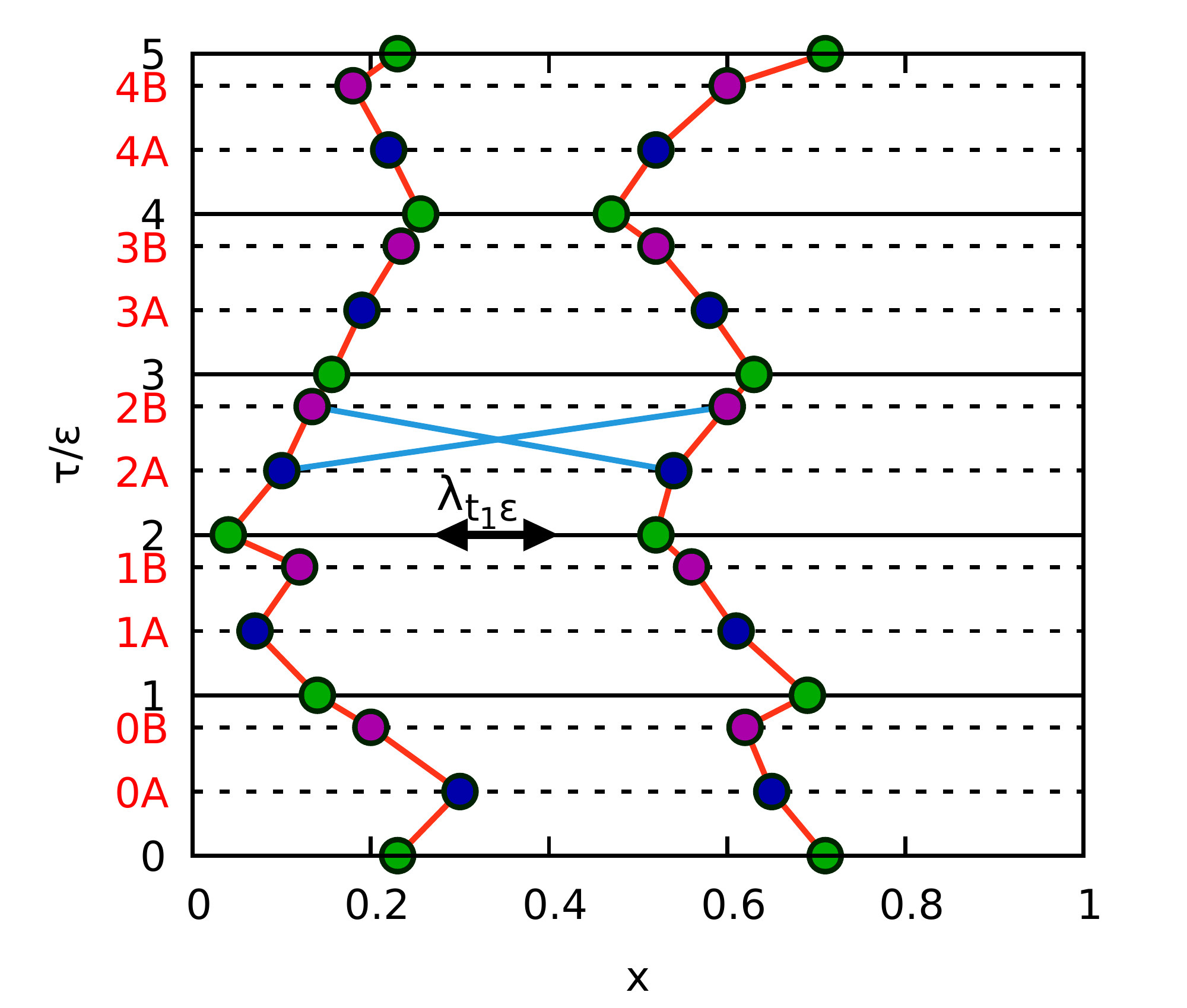}
\caption{\label{fig:PBPIMC_blocking}Effect of an increasing number of imaginary time slices on the permutation blocking -- Shown are configurations with two spin-polarized electrons in the $\tau$-$x$-plane with $P=2$ (left) and $P=5$ (right) fourth-order propagators. For $P=2$, the thermal wavelength of a single time slice, $\lambda_{t_1\epsilon}=\sqrt{2\pi t_1\epsilon}$ , is comparable to the average particle distance. Therefore, the off-diagonal (blue) diffusion matrix elements [cf.~Eq.~(\ref{eq:free_diffusion})] are comparable in magnitude to the diagonal (red) elements, and the permutation blocking within the determinants is efficient. In contrast, for $P=5$ there are either large diagonal (as in the depicted configuration) or large off-diagonal elements, but not both simultaneously, and the permutation blocking will have almost no effect.
}
\end{figure}

Finally, let us consider the effect on the permutation blocking of an increasing number of imaginary time propagators $P$, which is illustrated in Fig.~\ref{fig:PBPIMC_blocking}. In the left panel, we show a configuration of two spin-polarized electrons in the $\tau$-$x$-plane with $P=2$ fourth-order propagators. In this case, the thermal wavelength of a single time slice, $\lambda_{t_1\epsilon}=\sqrt{2\pi t_1\epsilon}$, is comparable to the average particle distance. Hence, the off-diagonal diffusion matrix elements (blue connections) are similar in magnitude to the diagonal elements (red connections) and the permutation blocking within the determinants is effective. However, this situation is drastically changed for increasing $P$, cf.~the right panel where a similar configuration is depicted for $P=5$. Evidently, in this case $\lambda_{t_1\epsilon}$ is much smaller than the particle distance and there are either large diagonal [which is the case in the depicted configuration] or off-diagonal diffusion matrix elements, but not both simultaneously. Therefore, the permutation blocking will be ineffective and for $P\to\infty$ the original sign problem from standard PIMC will be recovered.
In a nutshell, the introduction of antisymmetric imaginary time propagators allows to significantly alleviate the FSP and therefore to extend standard PIMC towards more degenerate systems. However, since this effect vanishes with increasing $P$, it is vital to combine the permutation blocking with a sophisticated factorization of the density matrix that allows for sufficient accuracy with only few propagators.

\begin{figure}
\includegraphics[width=0.45\textwidth]{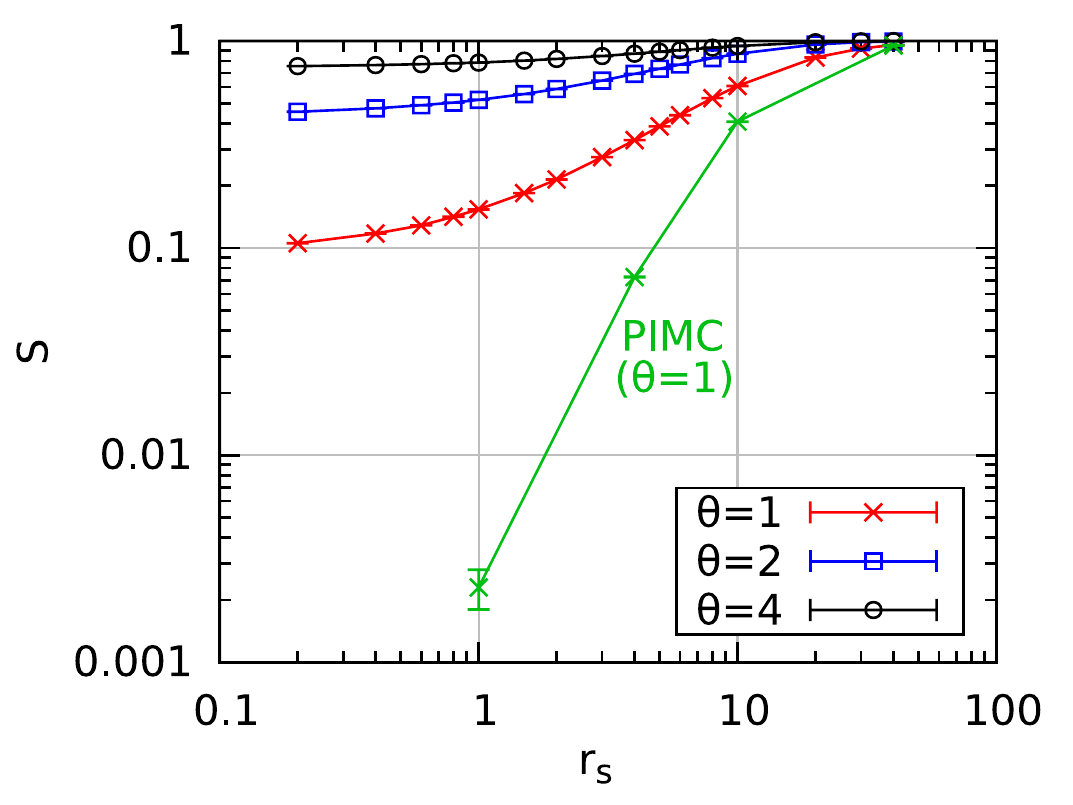}
\includegraphics[width=0.45\textwidth]{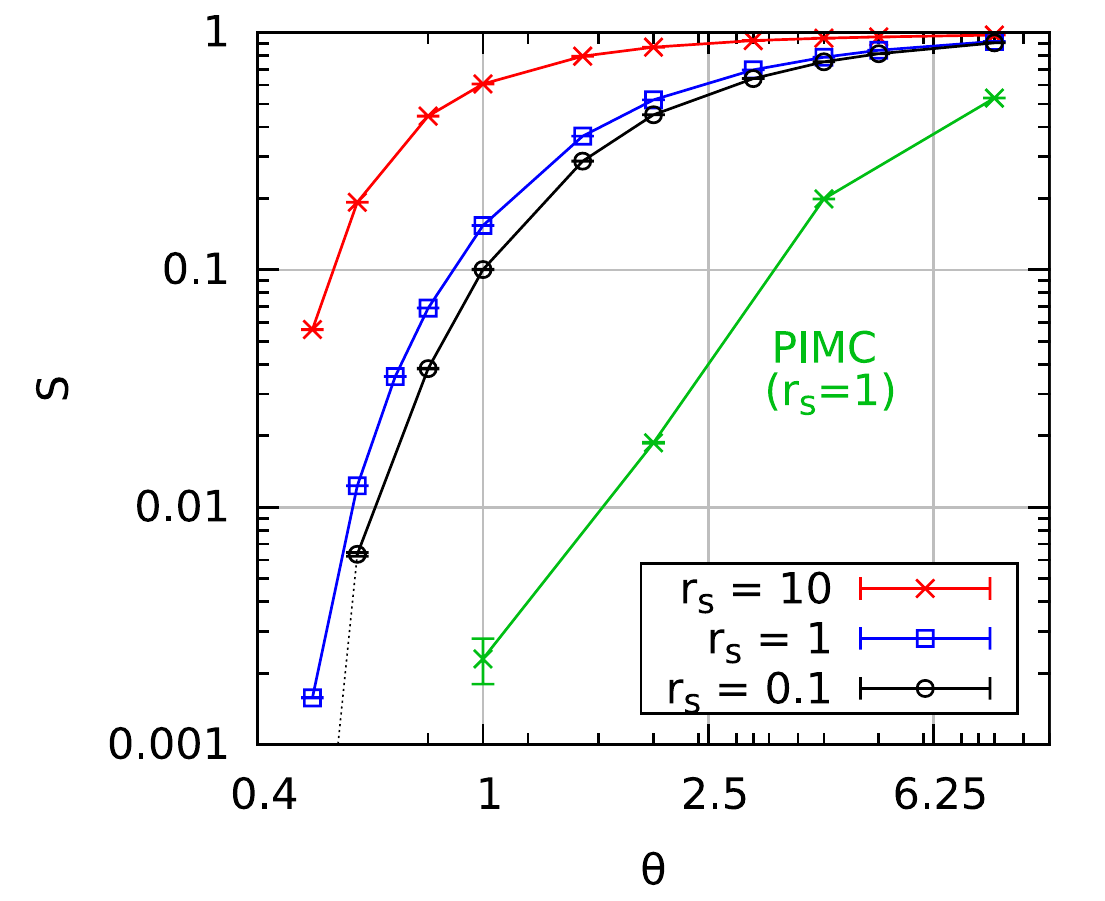}
\caption{\label{fig:PBPIMC_sign}Average sign in PB-PIMC simulations of $N=33$ spin-polarized electrons at warm dense matter conditions -- Left panel: Density-dependence of $S$ for $P=2$ propagators for $\theta=1$ (red), $\theta=2$ (blue), and $\theta=4$ (black). Right panel: Temperature-dependence of $S$ for $P=2$ for $r_s=10$ (red), $r_s=1$ (blue), and $r_s=0.1$ (black). All standard PIMC results for $S$ (green curves) have been taken from the Supplemental Material of Ref.~\cite{brown_path-integral_2013}. Both panels are reproduced with the permissions of the authors of Ref.~\cite{dornheim_permutation_2015-1}.
}
\end{figure}

Let us conclude this section with a more quantitative discussion of the fermion sign problem within PB-PIMC simulations of the spin-polarized UEG at warm dense matter conditions. In the left panel of Fig.~\ref{fig:PBPIMC_sign}, we show the dependence of the average sign on the density parameter $r_s$ for PB-PIMC simulations of $N=33$ spin-polarized electrons with $P=2$ imaginary time propagators at $\theta=1$ (red), $\theta=2$ (blue), and $\theta=4$ (black). All three curves exhibit a qualitatively similar behavior, i.e., a decreasing sign towards higher density, see also the discussion of Fig.~\ref{fig:PIMC_sign} above. However, in stark contrast to standard PIMC (green curve for $\theta=1$), the sign stays finite for all $r_s$. Thus, it has been demonstrated that, for the present conditions, PB-PIMC simulations are feasible over the entire density range.
In the right panel, the dependence of the average sign on $\theta$ is shown for the same system for $r_s=10$ (red), $r_s=1$ (blue), and $r_s=0.1$ (black). For large temperatures, the sign is nearly equal to unity and the computational effort is small. With decreasing $\theta$, both the diagonal and off-diagonal diffusion matrix elements become larger and both positive and negative determinants appear within the PB-PIMC simulations, eventually leading to a steep drop of $S$, which is more pronounced at weak coupling. Still, we stress that it is precisely at such conditions that the permutation blocking is most effective as well. Therefore, the sign problem is much less severe compared to standard PIMC (green curve). Overall, it can be seen that for warm dense matter conditions, i.e., for $r_s=1,\dots,6$, PB-PIMC simulations are feasible down to $\theta=0.5$.

\subsection{Configuration Path Integral Monte Carlo\label{sec:CPIMC}}
Another PIMC variant that has been proven to be highly valuable for the simulation of the UEG is the Configuration PIMC (CPIMC) method~\cite{schoof_configuration_2011,schoof_towards_2015,schoof_textitab_2015,groth_ab_2017}. It belongs to the class of continuous time world line Monte Carlo algorithms (CTWL-MC), which avoid the imaginary time discretization error by switching to the interaction picture with respect to a suitable part of the Hamiltonian. The basic idea of CTWL-MC stems from the works of Prokof’ev \emph{et al.}~\cite{prokofev_exact_1996} and Beard and Wiese~\cite{beard_simulations_1996}. Subsequently, many system specific CTWL-MC algorithms had been developed and highly optimized for fermionic as well as bosonic lattice models, most importantly for different variants of Hubbard and impurity models. A comprehensive review of the existing CTWL-MC algorithms and their applications can be found in Ref.~\cite{gull_continuous-time_2011}. However, until the development of CPIMC, continuous fermionic systems with long range Coulomb interactions have not been tackled with the CTWL-MC formailism mainly for two reasons: 1) the long range Coulomb interaction causes a severe sign problem and 2) it introduces new complex classes of diagrams which require a significantly more elaborate Monte-Carlo algorithm. 

Essentially, CPIMC can be viewed as performing Metropolis Monte Carlo with the complete (infinite) perturbation expansion of the partition function with respect to the coupling strength of the system. As such, this method is most efficient at weak coupling and becomes infeasible at strong coupling where it suffers from a severe sign problem; yet, the critical coupling parameter lies well beyond the failure of analytical approaches. Moreover, CPIMC is practically applicable over the entire temperature range, even down to the ground state. Thus, regarding the range of applicability with respect to density and temperature, CPIMC is highly complementary to the PB-PIMC approach discussed in Sec.~\ref{sec:PB-PIMC}. 

\subsubsection{CPIMC representation of the partition function}
For the derivation of both the standard PIMC and the PB-PIMC expansion of the partition function we started with utilizing $N-$particle states in coordinate representation to perform the trace over the density operator in Eq.~(\ref{eq:Z}). The correct Fermi statistics are then taken into account via a subsequent anti-symmetrization of the density operator, which causes the weight function to alter the sign with each pair exchange and, hence, can be regarded as the source of the FSP. To avoid this particular source, in CPIMC, we switch gears by making use of the second quantization representation of quantum mechanics for the UEG, which has been introduced in Sec.~\ref{sec:second_quant}. Here, the N-particle states, Eq.~(\ref{eq:fock_state}), are given by Slater determinants, which form a complete basis set of the $N-$particle states in Fock space. Thus, we can compute the partition function, Eq.~(\ref{eq:Z}), by carrying out the trace over the density operator with these states, yielding
\begin{eqnarray}
Z=\sum_{\{n\}}\braket{\{n\}|e^{-\beta\op{H}}|\{n\}}\ .
\end{eqnarray}
Unfortunately, the evaluation of the matrix elements of the density operator is not straightforward since the Slater determinants of plane waves are no eigenstates of the interacting UEG Hamiltonian, Eq.~(\ref{eq:UEG_Ham_second}), but only of the ideal UEG. One solution to this problem is to use the series expansion of the exponential function 
\begin{align}
Z&=\sum_{K=0}^\infty \sum_{\{n\}}\braket{\{n\}|\frac{(-\beta)^K}{K!}\op{H}^K|\{n\}} \nonumber\\ 
&= \sum_{K=0}^\infty \sum_{\{n\}^{(0)}} \sum_{\{n\}^{(1)}} \dots \sum_{\{n\}^{((K-1))}} \frac{(-\beta)^K}{K!}
\braket{\{n\}^{(0)}|\op{H}|\{n\}^{(1)}} \braket{\{n\}^{(1)}|\op{H}|\{n\}^{(2)}} \cdot\dots\cdot 
\braket{\{n\}^{(K-1)}|\op{H}|\{n\}^{(K)}}\ ,\label{eq:Z_SSE}
\end{align}
where we have inserted $K-1$ unities of the form $\op{1}=\sum_{\{n\}^{(i)}}\ket{\{n\}^{(i)}} \bra{\{n\}^{(i)}}$ so that $\{n\}^{(0)}=\{n\}^{(K)}$ holds implicitly. Applying the Slater-Condon rules to the UEG Hamiltonian we readily compute its matrix elements according to
\begin{align}\label{eq:matrix_elements}
\braket{\{n\}|\op{H}|\{\bar{n}\}} &=\begin{cases}
 \displaystyle D_{\{n\}}=\frac{1}{2}\sum_l \mathbf{k}_l^2 n_{l} + \frac{1}{2}\sum_{l<k}w^-_{lklk}n_{l}n_{k}, &\{n\}=\{\bar{n}\}\ ,\\[0.2cm]
 Y_{\{n\},\{\bar{n}\}} =w_{pqrs}^- (-1)^{\alpha_{\{n\},pq}+\alpha_{\{\bar{n}\},rs}}, & \{n\}=\{\bar{n}\}_{r<s}^{p<q}\ ,
\end{cases}
\end{align}  
with the phase factor
\begin{eqnarray}\label{eq:phase_factor}
\alpha_{\{n\},pq} &= \displaystyle\sum_{l=\min(p,q)+1}^{\max(p,q)-1}n_l\ ,
\end{eqnarray}
and the two-particle integrals being defined in Eq.~(\ref{eq:two_ints}). In this notation, $\ket{\{\bar{n}\}_{r<s}^{p<q}}$ refers to the Slater determinant that is obtained by exciting two electrons from the orbitals $r$ and $s$ to $p$ and $q$ in  $\ket{\{\bar{n}\}}$. Performing Metropolis Monte Carlo with the derived expression for the partition function, Eq.~(\ref{eq:Z_SSE}), has been termed the Stochastic Series Expansion (SSE) method. In particular, this approach has been successfully used for the simulation of the Heisenberg model~\cite{sandvik_quantum_1991,sandvik_finite_1997,sandvik_stochastic_1999,sandvik_mulichain_1999,shevchenko_double_2000}, for which Eq.~(\ref{eq:Z_SSE}) can be recast into a form that has solely positive addends, thereby completely avoiding the sign problem. However, this is not possible for the UEG and, in addition to the factor $(-\beta)^K$, we observe that the matrix elements can also attain both positive and negative values, which causes a serious sign problem. In CPIMC, we therefore follow a different strategy and separate the diagonal part $\op{D}$ of the Hamiltonian by exploiting the following identity of the density operator
\begin{align}\label{eq:do_identity}
 e^{-\beta\op{H}}=e^{-\beta\op{D}}\op{T}_\tau e^{-\int_0^\beta\op{Y}(\tau)\mathrm{d}\tau}\ 
 =e^{-\beta\op{D}}\sum_{K=0}^\infty \int\limits_{0}^{\beta} d\tau_1 \int\limits_{\tau_1}^{\beta} d\tau_2 \ldots \int\limits_{\tau_{K-1}}^\beta d\tau_K
(-1)^K\op{Y}(\tau_K)\op{Y}(\tau_{K-1})\cdot\ldots\cdot\op{Y}(\tau_1)\ ,
\end{align}
where $\op{T}_\tau$ denotes the time-ordering operator and the time-dependence of the off-diagonal operator $\op{Y}$ refers to the interaction picture in imaginary time with respect to the diagonal operator $\op{D}$,
\begin{eqnarray}
\op{Y}(\tau)&=e^{\tau\op{D}}\op{Y}e^{-\tau\op{D}}\ .
\end{eqnarray}
Note that, independent of the underlying one-particle basis of the quantization, according to the Slater-Condon rules the Hamiltonian can always be split into a diagonal and off-diagonal contribution such that $\op{H}=\op{D}+\op{Y}$. 
After inserting Eq.~(\ref{eq:do_identity}) into Eq.~(\ref{eq:Z_SSE}) and re-ordering some terms, the partition function becomes 
\begin{eqnarray}\label{eq:Z_almost}
Z =
\sum_{\substack{K=0 \\ K \neq 1}}^{\infty} \sum_{\{n\}^{(0)}} \sum_{\{n\}^{(1)}} \dots \sum_{\{n\}^{(K-1)}}\,
\int\limits_{0}^{\beta} d\tau_1 \int\limits_{\tau_1}^{\beta} d\tau_2 \ldots \int\limits_{\tau_{K-1}}^\beta d\tau_K 
(-1)^K  e^{-\sum\limits_{i=0}^{K} D_{\{n^{(i)}\}} \left(\tau_{i+1}-\tau_i\right) } 
\prod_{i=1}^{K} Y_{\{n^{(i)}\},\{n^{(i-1)}\} }\ .
\end{eqnarray}
Taking into account that the off-diagonal matrix elements do not vanish only if the occupation numbers of the left and right state, i.e. $\{n\}^{(i)}$ and $\{n\}^{(i-1)}$, differ in exactly four orbitals p,q,r,s, cf.~Eq.~(\ref{eq:matrix_elements}), we may introduce a multi-index $s_i=(pqrs)$ defining these four orbitals and re-write the summation as follows
\begin{eqnarray}\label{eq:Z_final}
Z =
\sum_{\substack{K=0 \\ K \neq 1}}^{\infty} \sum_{\{n\}}
\sum_{s_1\ldots s_{K-1}}\,
\int\limits_{0}^{\beta} d\tau_1 \int\limits_{\tau_1}^{\beta} d\tau_2 \ldots \int\limits_{\tau_{K-1}}^\beta d\tau_K 
(-1)^K  e^{-\sum\limits_{i=0}^{K} D_{\{n^{(i)}\}} \left(\tau_{i+1}-\tau_i\right) } 
\prod_{i=1}^{K} Y_{\{n^{(i)}\},\{n^{(i-1)}\} }(s_i)\ ,
\end{eqnarray}
where $\{n\}=\{n\}^{(0)}=\{n\}^{(K)}$ always holds. This is the exact CPIMC expansion of the partition function. Regarding the application of the Metropolis algorithm, the benefit of Eq.~(\ref{eq:Z_final}) over the SSE, Eq.~(\ref{eq:Z_SSE}), is obvious: by switching to the interaction picture we got rid of all sign changes that are caused by the diagonal matrix elements since in Eq.~(\ref{eq:Z_final}) these solely enter in the exponential function, which is always positive. Nevertheless, the sign changes due to the off-diagonal matrix elements are still present and are the source of the sign problem in the CPIMC method. 

\begin{figure}
\center{\includegraphics[width=85mm]{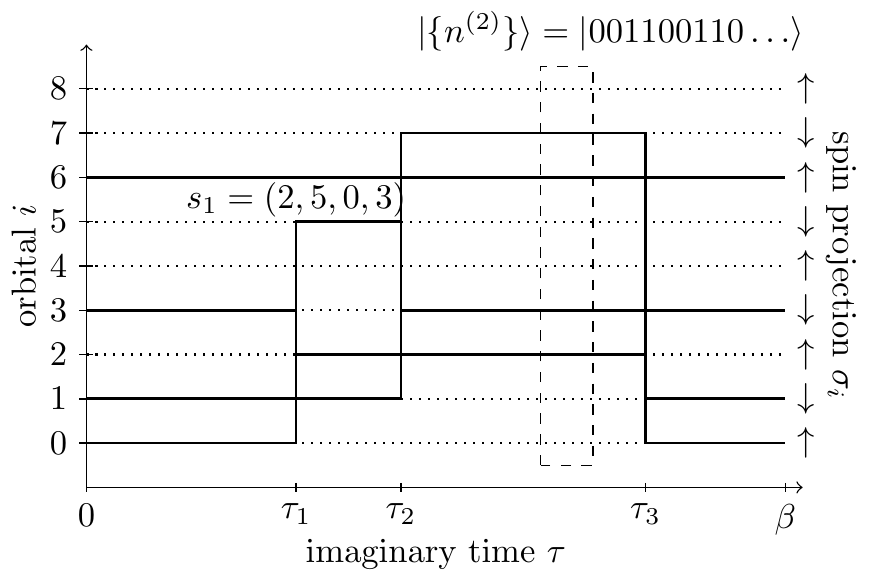}}
 \caption{Sketch of a typical CPIMC path of $N=4$ unpolarized electrons in Slater determinant (Fock) space in imaginary time. The starting determinant $\{n\}^{(0}$ at $\tau=0$ undergoes three two-particle excitations at times $\tau_1,\tau_2,$ and $\tau_3$, where the last excitation defined by the involved orbitals $s_3=(0,1,2,7)$ must always ensure that the last state $\{n\}^{(3)}$ is equivalent to $\{n\}^{(0)}$. Reproduced from Ref.~\cite{dornheim_abinitio_2016} with permission of the authors. 
}
 \label{fig:CPIMC_sketch}
\end{figure}

\begin{figure}
\center{
\includegraphics[width=0.4\textwidth]{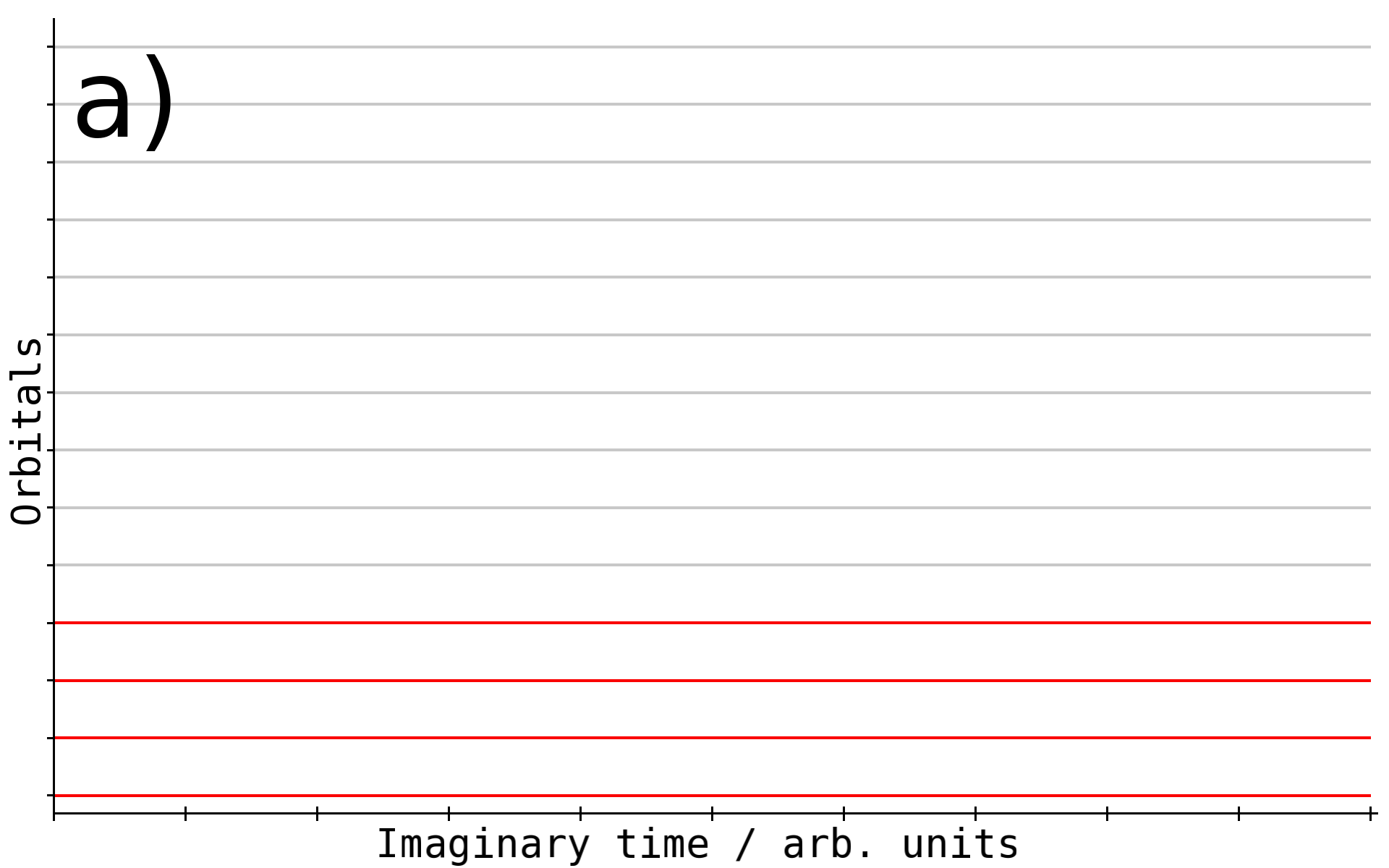}
\hspace{0.5cm}
\includegraphics[width=0.4\textwidth]{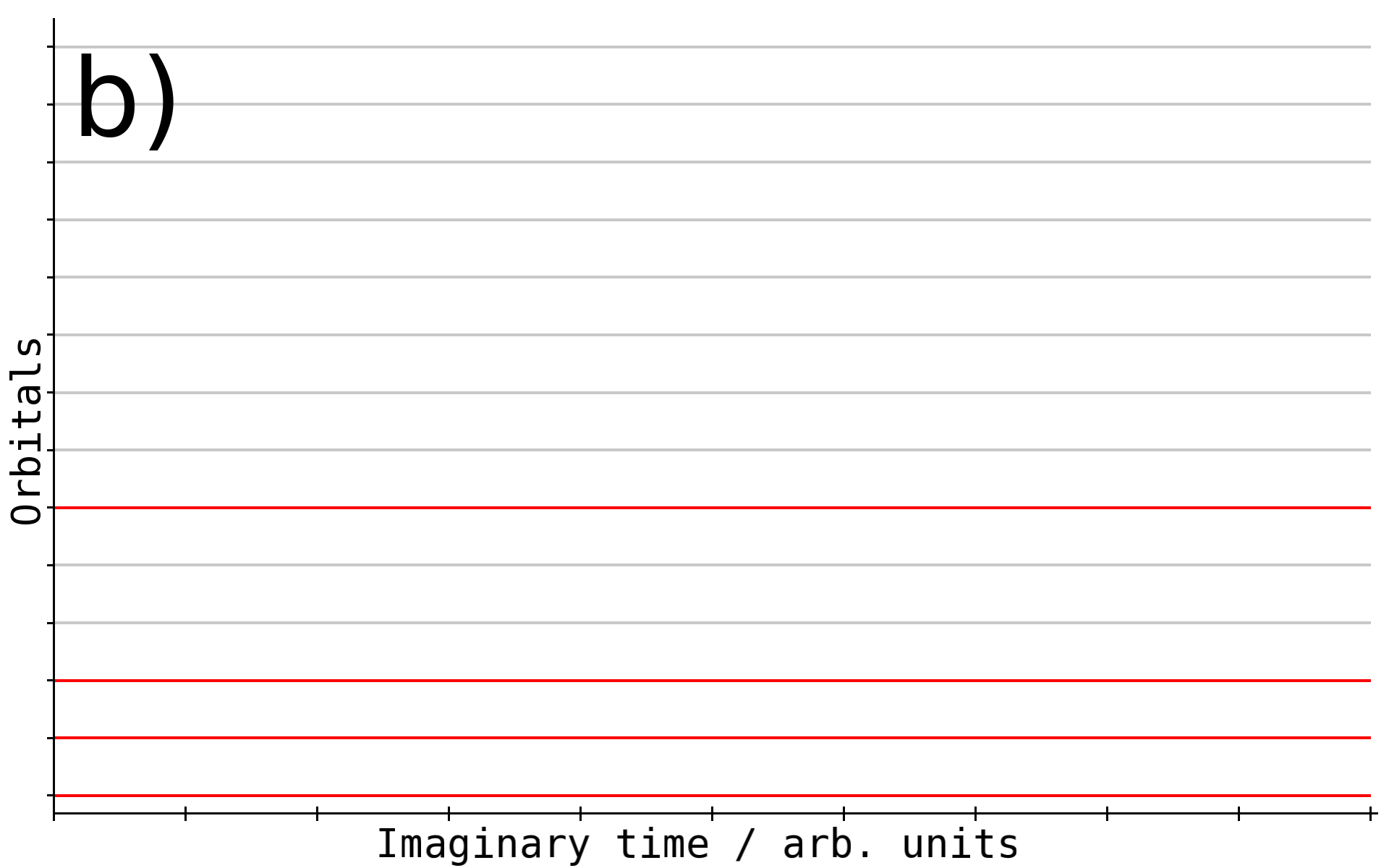}
}
\center{
\includegraphics[width=0.4\textwidth]{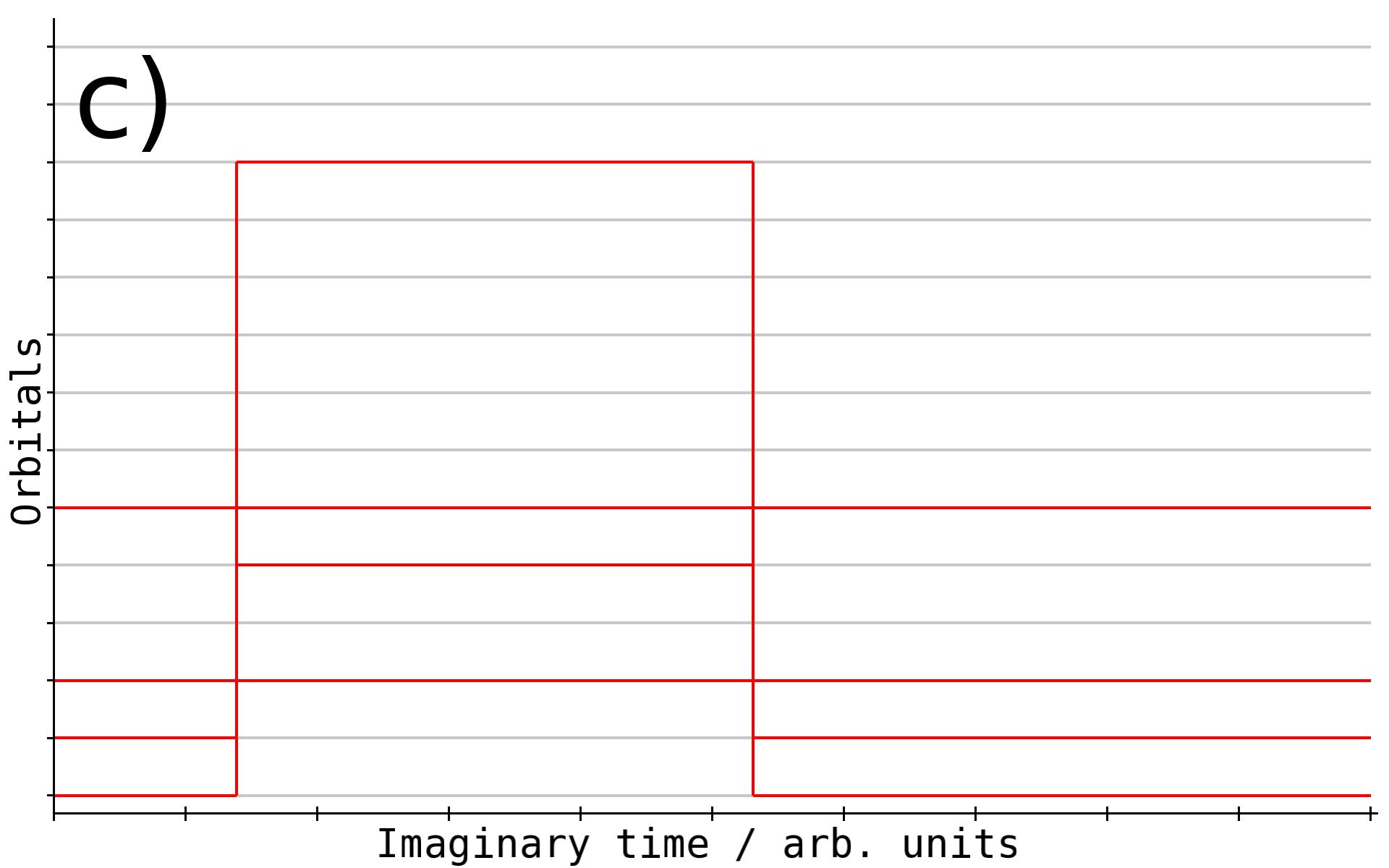}
\hspace{0.5cm}
\includegraphics[width=0.4\textwidth]{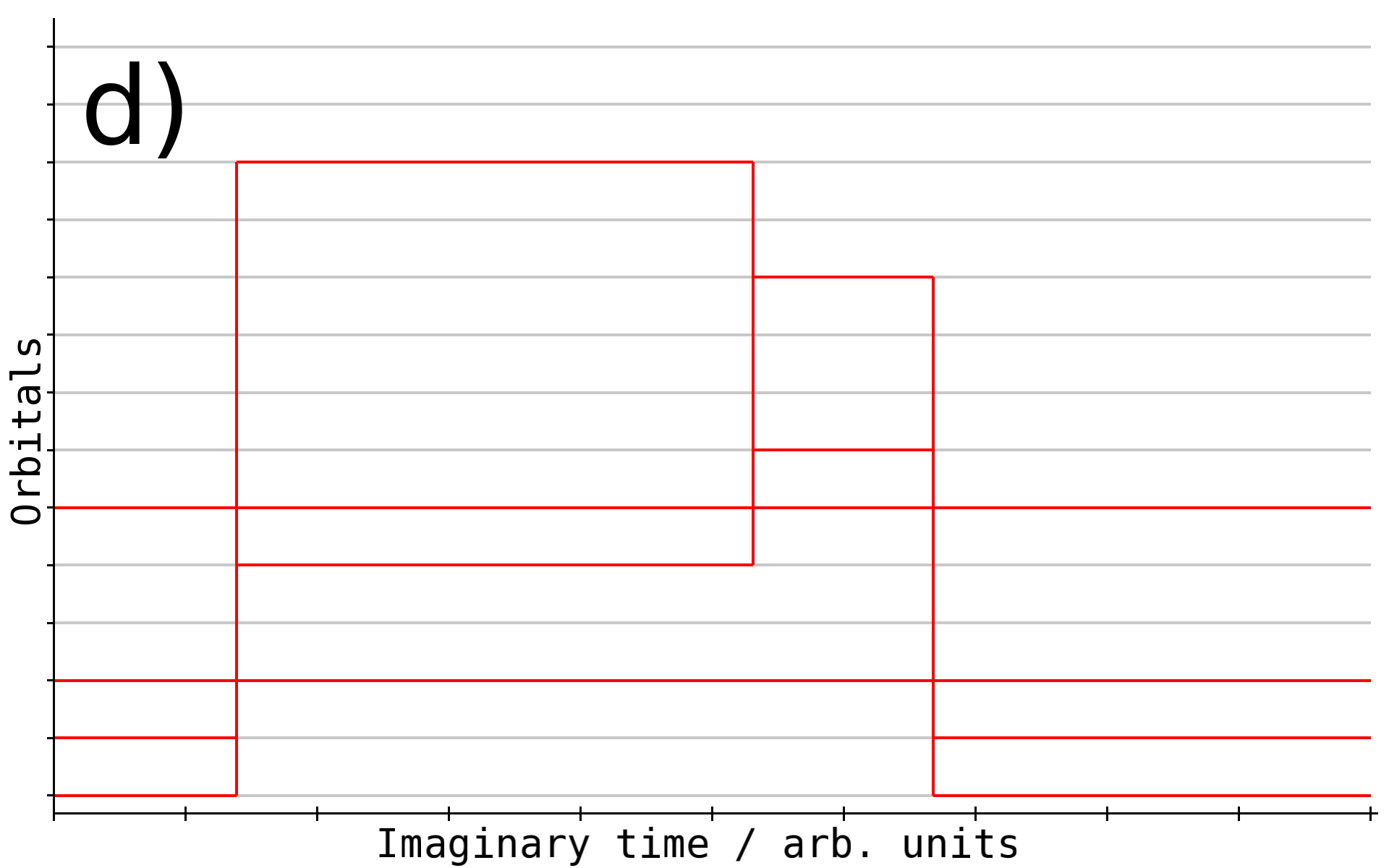}
}

\center{
\includegraphics[width=0.4\textwidth]{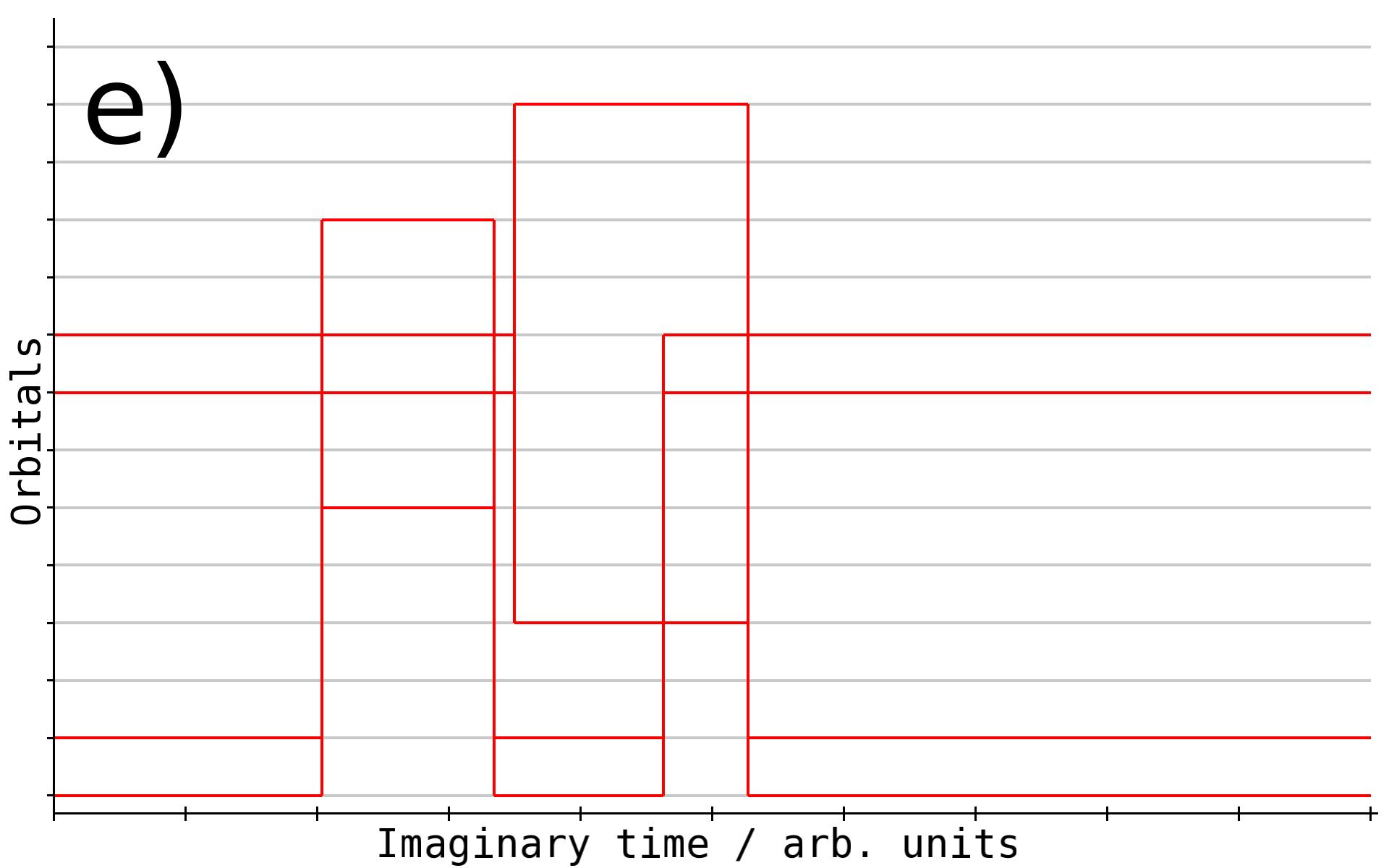}
\hspace{0.5cm}
\includegraphics[width=0.4\textwidth]{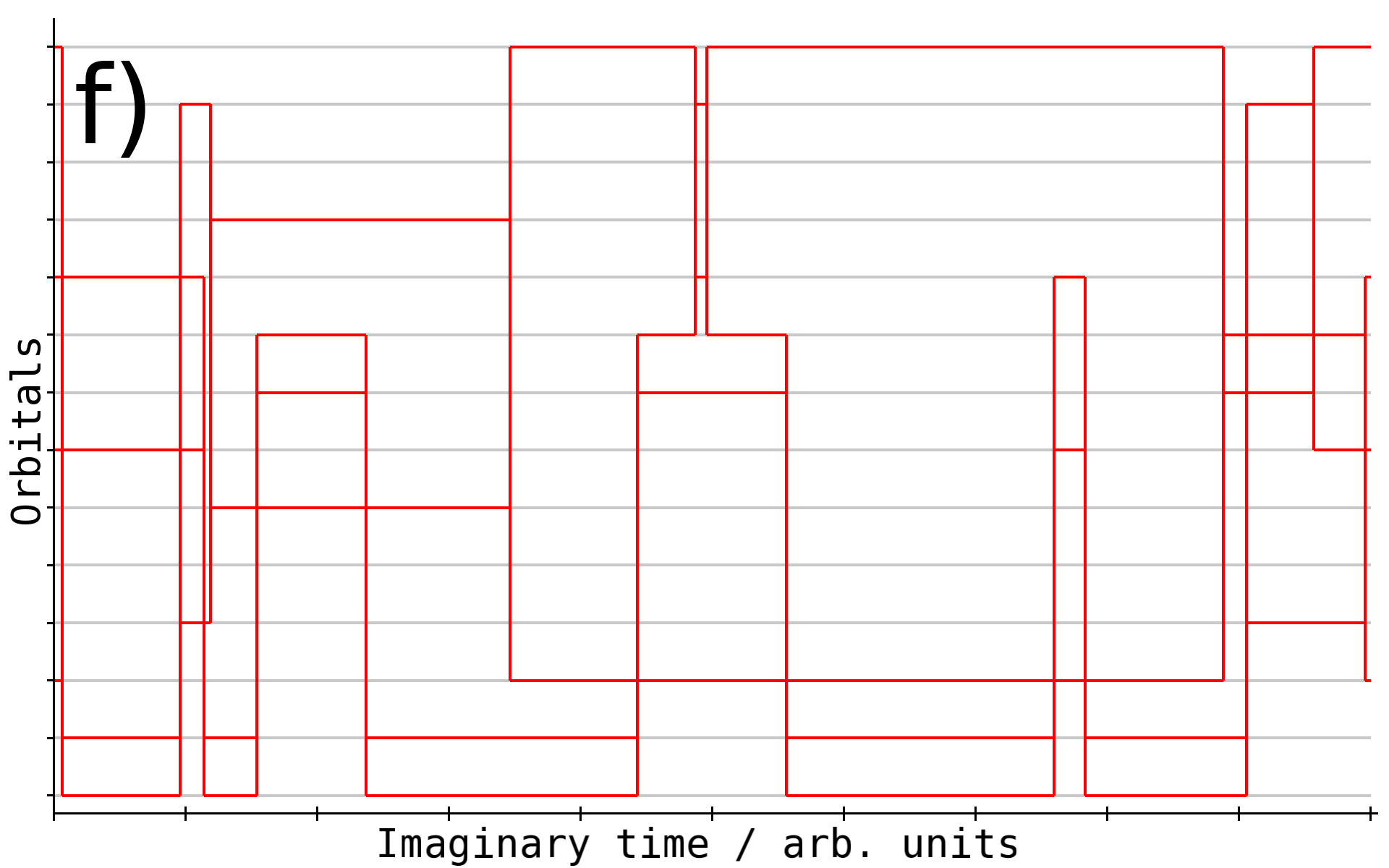}
}
 \caption{Snap shots of CPIMC paths from the simulation of $N=4$ unpolarized electrons at $r_s=1$ ant $\theta=1$ in $N_B=14$ plane wave spin-orbitals (indicated by the grey lines). The orbitals are ordered according to their corresponding kinetic energy $\mathbf{k}_i^2/2$. Depicted are the occupied orbitals (red lines) in dependence of the imaginary time, which sum up to 4 at any specific time $\tau\in[0,\beta]$.
 Panel \textbf{a)} shows the initial path that is used as the starting configuration in the Markov chain: no kinks with the lowest orbitals being occupied. In panel \textbf{b)} an entire orbital is excited, after which a pair of kinks is added in panel \textbf{c)}. Only then is it possible add single kinks by changing another kink in the path, which is depicted in panel \textbf{d)}. This way, depending on the density and temperature, the CPIMC algorithm eventually generates paths with more complicated structures as shown in panels \textbf{e)} and \textbf{d)}.
}
 \label{fig:CPIMC_simulation_algorithm}
\end{figure}

Similar to the standard PIMC and PB-PIMC approach, each contribution to the CPIMC expansion of the partition function, Eq.~(\ref{eq:Z_final}), can be interpreted as a path in imaginary time, $\mathbf{X}$, that is entirely defined by the starting set of occupation numbers $\{n\}$ and all subsequent excitations $\{s_1,s_2\dots,s_K\}$ with their corresponding times $\{\tau_1,\tau_2\dots,\tau_K\}$, i.e., 
\begin{eqnarray}
\mathbf{X} =\left(K,\{n\}, s_1,\ldots,s_{K-1},\tau_1,\ldots,\tau_K\right)\ .
\end{eqnarray}
In contrast to the standard PIMC formulation, these paths now evolve in the discrete Fock space instead of the continuous coordinate space. Moreover, there is no time discretization in the CPIMC formulation as the excitations occur at continuous times $\tau_i$. Hence, unlike PIMC in coordinate space, there is no time discretization error. A sketch of a typical path occurring in the simulation of $N=4$ unpolarized electrons is depicted in Fig.~\ref{fig:CPIMC_sketch}, where we chose the ordering of the spin orbitals such that even (odd) numbers correspond to up (down) spin projections. In correspondence to their visual appearance in these paths we refer to the excitations as ``kinks''. According to Eq.~(\ref{eq:Z_final}), the corresponding weight of each paths is given by
\begin{eqnarray}\label{eq:CPIMC_weight}
W(\mathbf{X})=(-1)^K  e^{-\sum\limits_{i=0}^{K} D_{\{n^{(i)}\}} \left(\tau_{i+1}-\tau_i\right) } 
\prod_{i=1}^{K} Y_{\{n^{(i)}\},\{n^{(i-1)}\} }(s_i)\ .
\end{eqnarray}
Note that, as discussed in detail in Sec.~\ref{sec:FSP}, the Metropolos algorithm can only be applied when using the modulus of the weight function. As usual, the Monte Carlo estimator of an observable, cf.~Eq.~(\ref{eq:estimation}), is derived from its thermodynamic relation to the partition function. For example, for the energy we have
\begin{eqnarray}
\langle\op{H}\rangle= - \frac{\partial}{\partial \beta} \ln Z 
=
\sum_{\substack{K=0 \\ K \neq 1}}^{\infty} \sum_{\{n\}}
\sum_{s_1\ldots s_{K-1}}\,
\int\limits_{0}^{\beta} d\tau_1 \int\limits_{\tau_1}^{\beta} d\tau_2 \ldots \int\limits_{\tau_{K-1}}^\beta d\tau_K 
\biggl(\frac{1}{\beta} \sum_{i=0}^K D_{\{n^{(i)}\}}(\tau_{i+1}-\tau_i) -\frac{K}{\beta}\biggr) W(\mathbf{X})\ . \label{eq:CPIMC_energy_estimator}
\end{eqnarray}
In practice, in CPIMC simulations, we start the generation of the Markov chain from an initial path without kinks and with the lowest $N$ plane wave spin-orbitals being occupied, where we choose the ordering of the orbitals in accordance to their kinetic energy $\mathbf{k}^2_i/2$. Fig.~\ref{fig:CPIMC_simulation_algorithm}~a) shows a snap shot of such a starting path from a CPIMC simulation of $N=4$ unpolarized electrons in $N_B=14$ spin orbitals. Due to the fact that there are no $\beta-$periodic (closed) paths containing only a single kink, only two possible changes can be proposed to proceed: either an entire occupied orbital can be excited to an unoccupied orbital, see Fig.~\ref{fig:CPIMC_simulation_algorithm} b), or a symmetric pair of kinks can be added at once, see Fig.~\ref{fig:CPIMC_simulation_algorithm} c). These proposed changes are accepted or rejected with the corresponding Metropolis acceptance probability, cf.~Eq.~(\ref{eq:acceptance}), which is computed using the modulus of the weight function $|W(\mathbf{X})|$. Only after a symmetric pair of kinks has been successfully added is it possible to add single kinks by changing another as demonstrated in Fig.~\ref{fig:CPIMC_simulation_algorithm} d). Depending on the temperature and density parameter in the simulation, the CPIMC algorithm eventually generates paths containing more kinks and more complex structures, see Figs.~\ref{fig:CPIMC_simulation_algorithm} e) and f).

\subsubsection{The sign problem in the CPIMC approach}
\begin{figure}
\center{
\includegraphics[width=0.55\textwidth]{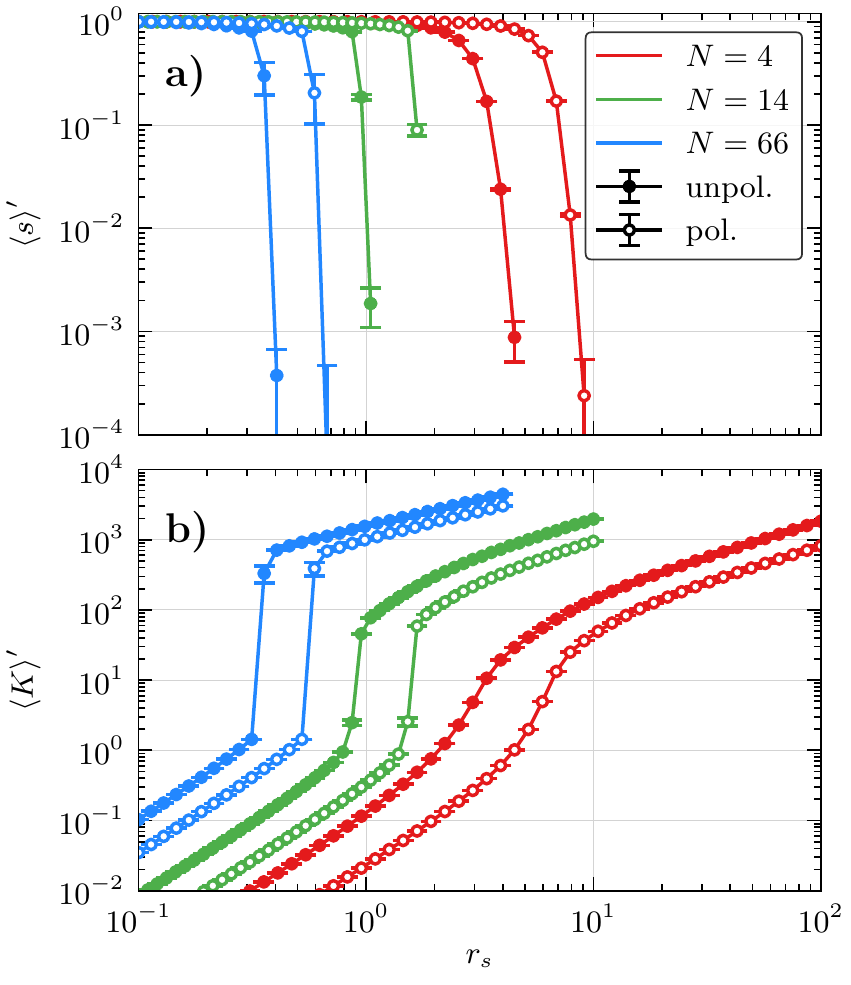}
}
 \caption{Average sign \textbf{a)} and average number of kinks \textbf{b)} in CPIMC simulation in dependence on the density parameter for $N=4,14,66$ at $\theta=1$. Shown are the results from the simulation of the spin-polarized (circles) and unpolarized (dots) UEG. Reproduced from Ref.~\cite{dornheim_abinitio_2016} with permission of the authors. 
}
 \label{fig:CPIMC_fsp}
\end{figure}

\begin{figure}
\center{
\includegraphics[width=0.8\textwidth]{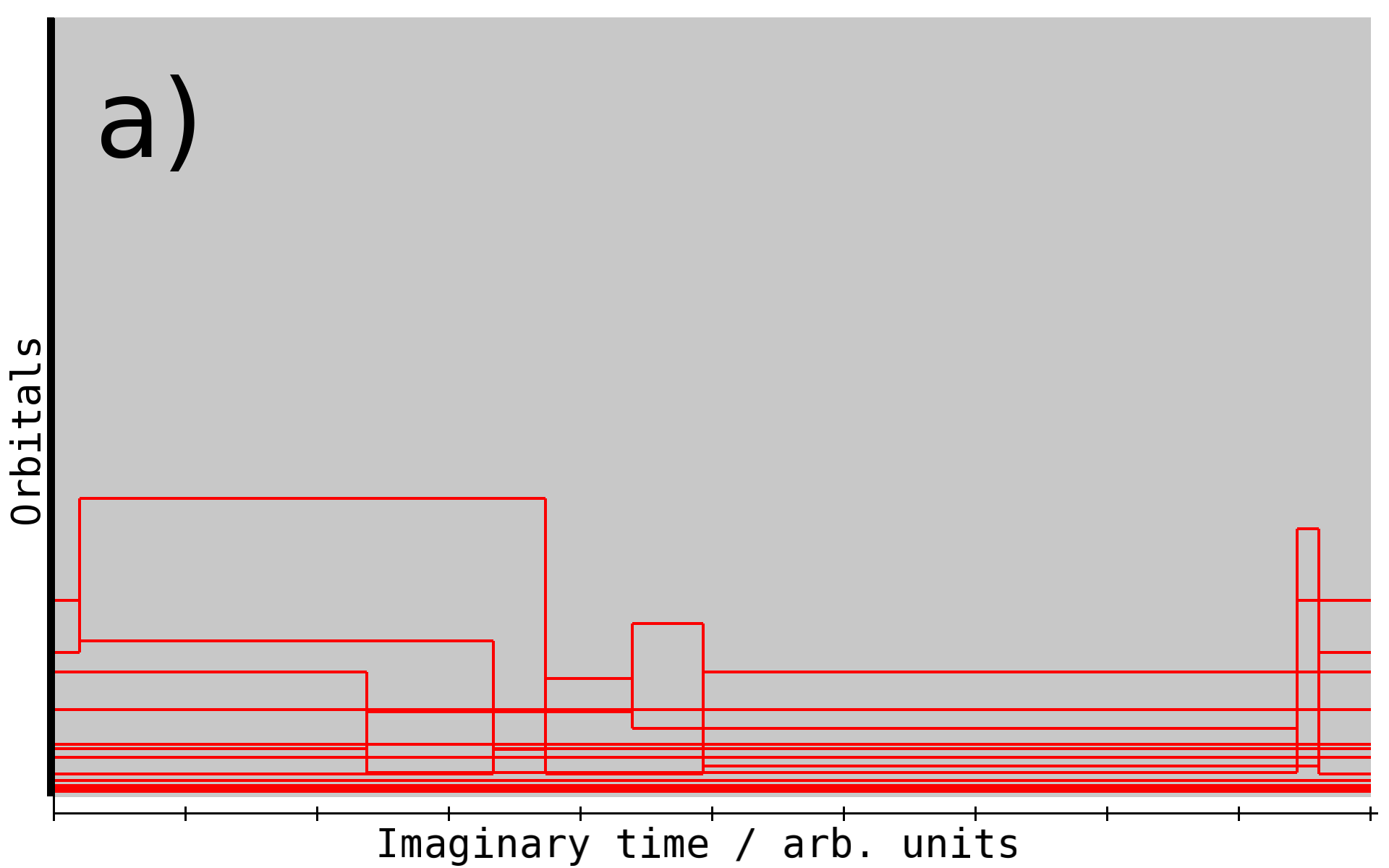}
}
\vspace{0.5cm}
\center{
\includegraphics[width=0.8\textwidth]{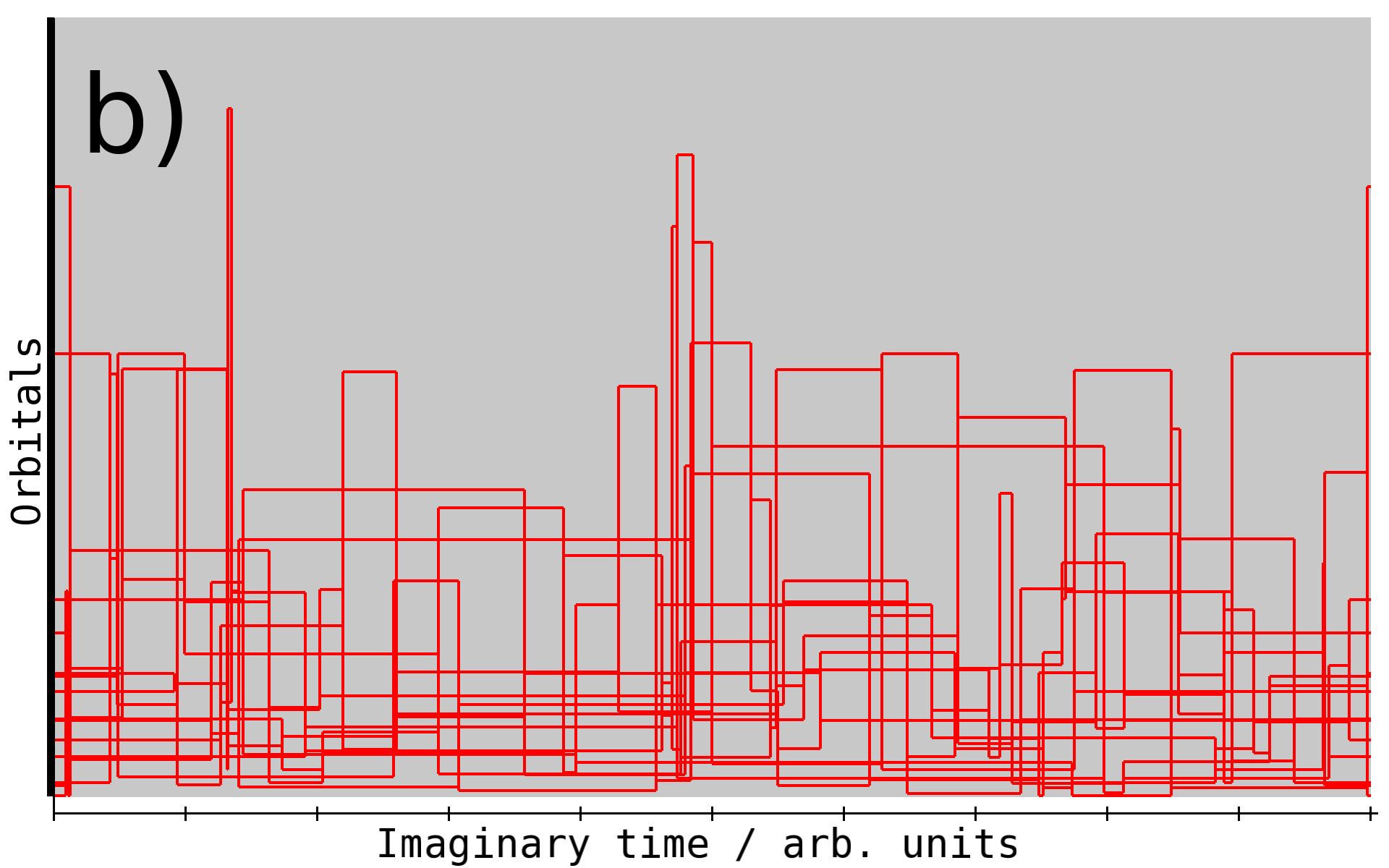}
}
 \caption{Snapshot of a typical path occurring in a CPIMC simulation of $N=14$ unpolarized electrons
 at $r_s=0.7$ (panel \textbf{a)}) and $r_s=1$ (panel \textbf{b)}), both at $\theta=1$ in $N_B=778$ plane wave spin-orbitals, which are ordered according to their corresponding kinetic energy $\mathbf{k}_i^2/2$. Plotted is the occupation of each orbital (red and grey indicate occupied and unoccupied orbitals, respectively) in dependence on the imaginary time. Note that the density of the $778$ orbitals (grey lines) appears to be continuous on this scale but when further zooming into the path it is of course discrete like in Fig.~\ref{fig:CPIMC_simulation_algorithm} where $N_B=14$.  
}
 \label{fig:CPIMC_simulation_highT}
\end{figure}

\begin{figure}
\center{
\includegraphics[width=0.8\textwidth]{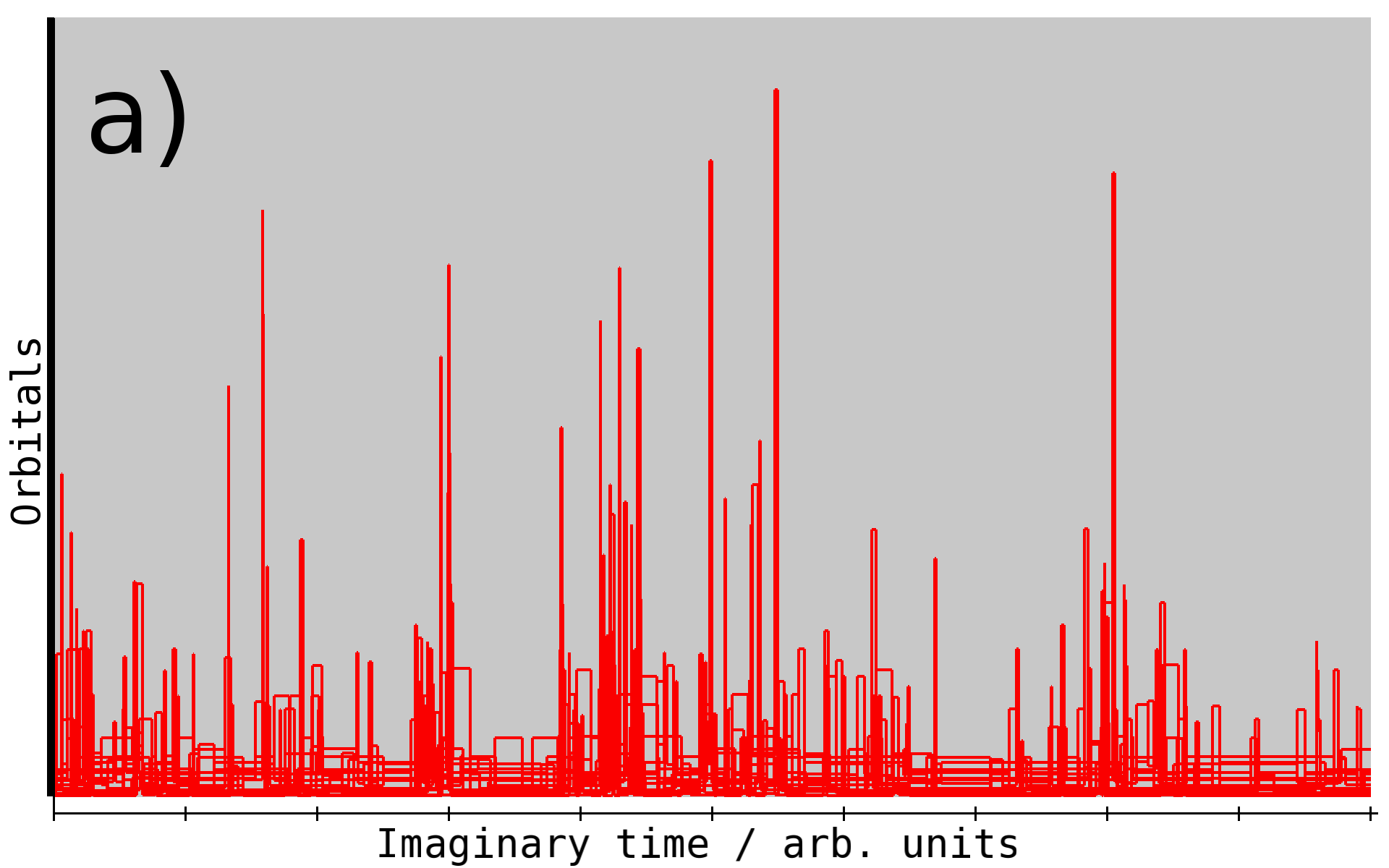}
}
\vspace{0.5cm}
\center{
\includegraphics[width=0.8\textwidth]{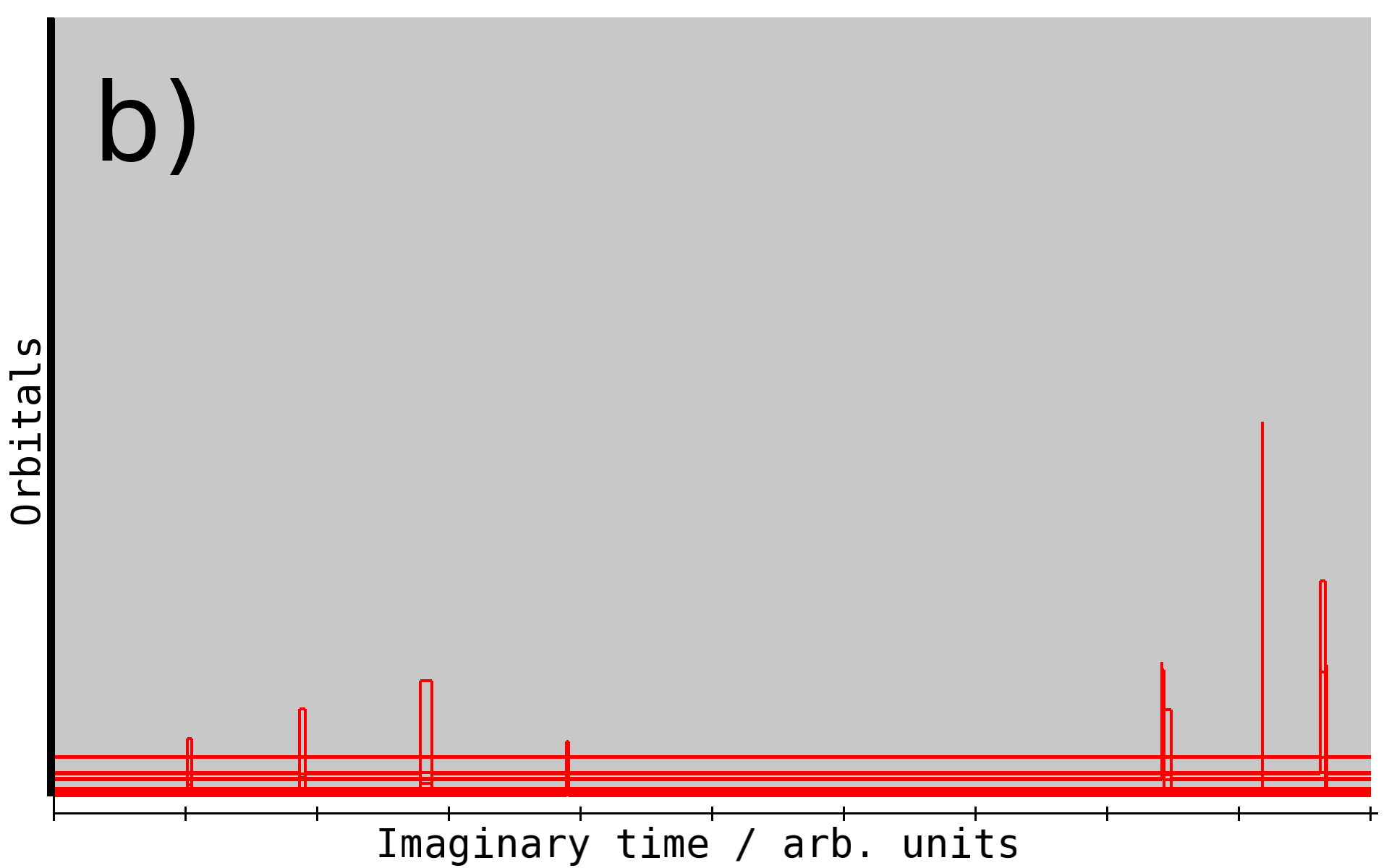}
}
 \caption{Snapshot of a typical path occurring in a CPIMC simulation of $N=14$ unpolarized electrons
 at $r_s=0.7$ (panel \textbf{a)}) and $r_s=0.4$ (panel \textbf{b)}) both at $\theta=0.01$ in $N_B=778$ plane wave spin-orbitals, which are ordered according to their corresponding kinetic energy $\mathbf{k}_i^2/2$. Plotted is the occupation of each orbital (red and grey indicate occupied and unoccupied orbitals, respectively) in dependence on the imaginary time.
}
 \label{fig:CPIMC_simulation_lowT}
\end{figure}
As discussed in Sec.~\ref{sec:FSP}, we can only apply the Metropolis algorithm to a partition function that has a weight function with alternating signs by simulating a modified system defined by the modulus of the weight function, cf.~Eq.~(\ref{eq:Z_prime}). Yet, this procedure comes at the cost of introducing the FSP. It is important to note that each kink enters the CPMC weight function, Eq.~(\ref{eq:Z_final}), with three possible sign changes: 1) the factor $(-1)^K$, 2) the sign of the corresponding two-particle integral, Eq.~(\ref{eq:two_ints}), and 3) the phase factor, Eq.~(\ref{eq:phase_factor}), that depends on the set of occupation numbers at the time of the kink. To investigate the FSP in the CPIMC approach, Fig.~\ref{fig:CPIMC_fsp} shows the average sign, a), and the average number of kinks, b), of all sampled paths in the generated Markov chain for simulations of $N=4$ (red), $N=14$ (green), and $N=66$ (blue) electrons at $\theta=1$ in dependence of the density parameter $r_s$, both for the polarized (circles) and unpolarized (dots) UEG. Since simulations with an average sign below $\sim 10^{-3}$ are not feasible, these quantities determine the applicable regime of the basic CPIMC method in the density-temperature plane. Independent from the number of electrons, the average sign is always unity in the ideal limit $r_s\to 0$, since here the UEG Hamiltonian is diagonal in the utilized plane wave basis. Hence, there cannot be any kinks in the paths and their weight is always positive. 

However, with decreasing density, i.e., increasing $r_s$, we observe that the average sign drops drastically at some critical density that strongly depends on the number of electrons, temperature, as well as the spin-polarization. This drop is caused by an enormous increment of the average number of kinks at this critical density (note the logarithmic scale). For example, in case of $N=14$ unpolarized electrons (green), at this temperature, the critical density is at $r_s\sim 0.8$. In Fig.~\ref{fig:CPIMC_simulation_highT}, we further explore this case by showing snap shots of typical CPIMC paths occurring in the simulation of $N=14$ electrons in $N_B=778$ basis functions at $r_s=0.7$, a), and $r_s=1$, b), both at $\theta=1$. While at $r_s=0.7$ the paths contain only very few kinks, at $r_s=1$, many paths contain $\sim 100$ kinks which are highly entangled and thereby induce many sign changes. When lowering the temperature while keeping the other system parameters constant this critical value of $r_s$ becomes even smaller, wich is illustrated by the two simulation snap shots in Fig.~\ref{fig:CPIMC_simulation_lowT} for $r_s=0.7$, a), and $r_s=0.4$, b), now at $\theta=0.01$. At these low temperatures close to the ground state, even a density parameter of $r_s=0.7$ is clearly not feasible with the basic CPIMC method as the paths typically contain about $500$ kinks, while, at $r_s=0.4$, the average number of kinks is reduced by two orders of magnitude so that simulations pose no problem here. Further, we point out that the structure of the generated CPIMC paths changes significantly with the temperature: at high temperature, see Fig.~\ref{fig:CPIMC_simulation_highT}, the average occupation of higher orbitals is much larger due to the increased kinetic energy of the electrons, while at low temperatures, see Fig.~\ref{fig:CPIMC_simulation_lowT}, most of the kinks tend to occur in symmetric pairs with only very short imaginary time in between, so that theses structures appear as needles in the paths. Interestingly, the overall sign change of these symmetric pairs always exactly compensates to one and thus they do not worsen the FSP. 

Finally, we stress that the linear dependence of the average number of kinks in Fig.~\ref{fig:CPIMC_fsp} b) before and after the critical density is not an artefact due to the inevitable practical restriction to a finite number of basis functions in the simulation. In particular, this demonstrates that the modified CPIMC partition function with the modulus weight function is actually a convergent sum for any finite system parameters of the UEG. Mathematically this must not necessarily be the case, since if a sum with alternating signs of its summands converges, of course, the same sum with the modulus of the summands can be divergent. Nevertheless, the fact that the FSP in the basic CPIMC approach has a "hard-wall-like" character is rather unsatisfactory: there is either none when there are on average less than $\sim2$ kinks in the paths or it is so strong that simulations are not feasible due to hundreds or even thousands of kinks. A problem which we will strongly mitigate in the next section.

\subsubsection{Reduction of the FSP with an auxiliary kink potential}
\begin{figure}
\center{
\includegraphics[width=0.45\textwidth]{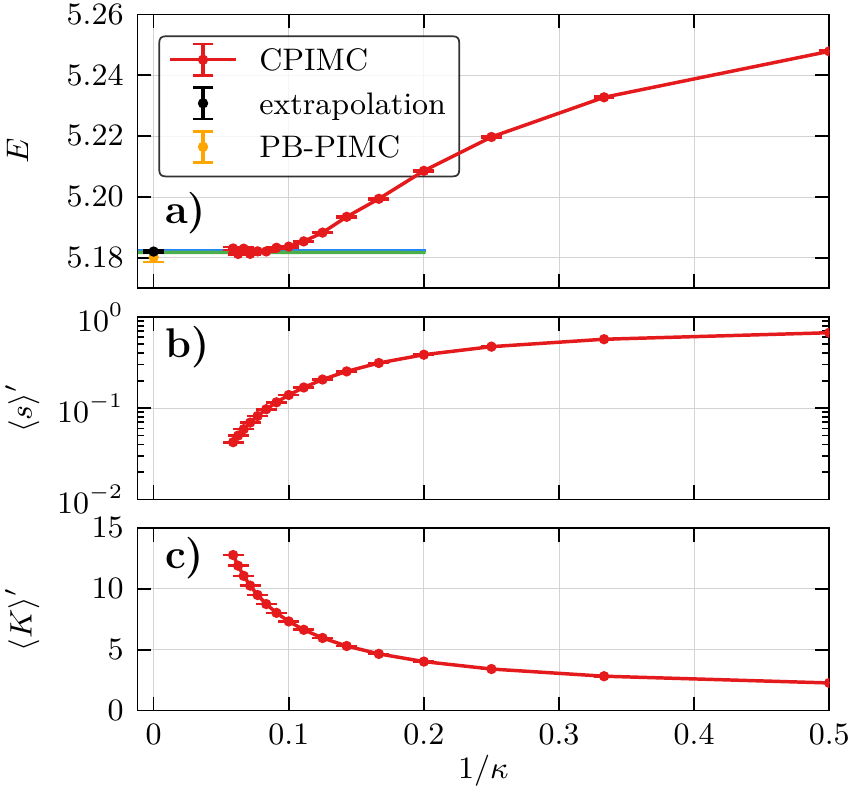}
\hspace{0.5cm}
\includegraphics[width=0.45\textwidth]{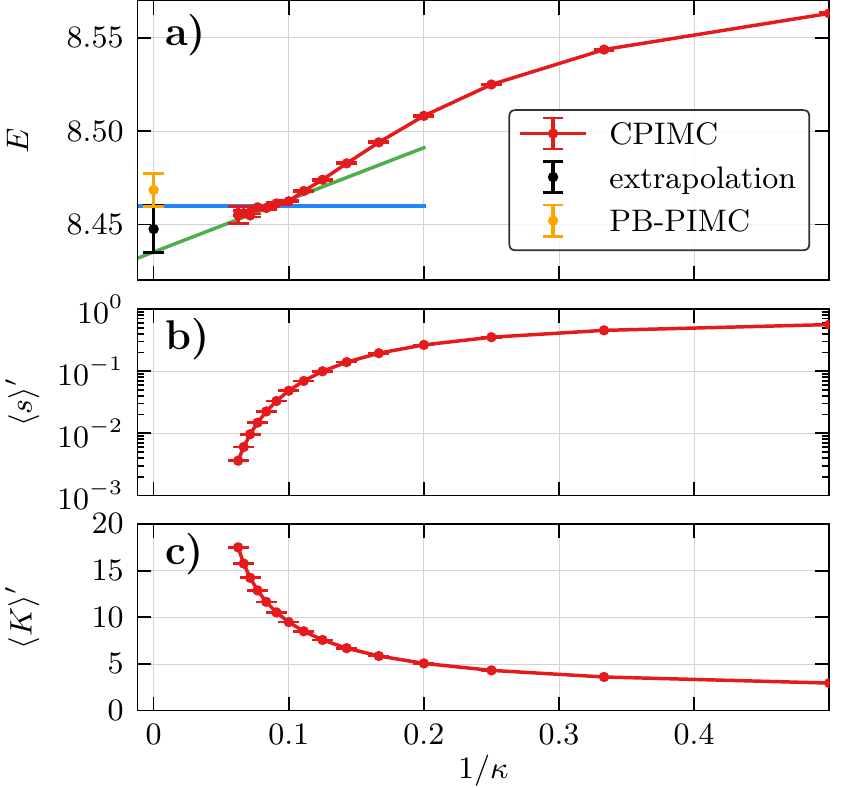}
}
 \caption{Convergence of \textbf{a)} the internal energy, \textbf{b)} the average sign and \textbf{c)} the average number of kinks with the kink potential parameter $\kappa$. Each point results from a complete CPIMC simulation of $N=66$ unpolarized electrons at $r_s=2$ and $\theta=4$ (left) and $r_s=0.8$ and $\theta=1$ (right). The blue (green) line shows a horizontal (linear) fit to the last points. The asymptotic value (black point) in the limit $1/\kappa \to 0$ is enclosed between the blue and green lines and, within error bars, coincides with the PB-PIMC result (orange points). Left (right) graphic reproduced (modified) from Ref.~\cite{dornheim_abinitio_2016} with permission of the authors.}
 \label{fig:CPIMC_kinkpot}
\end{figure}

The restriction of the CPIMC approach to the nearly ideal regime, i.e. very large densities, due to a severe FSP at some critical value of $r_s$ can be significantly alleviated by the use of a Fermi-like auxiliary kink potential 
\begin{eqnarray}
V_{\kappa}(K)=\frac{1}{e^{-(\kappa-K+0.5)}+1}\ ,
\label{eq:fermiPot}
\end{eqnarray}
by replacing the modulus of the weight function $|W(\mathbf{X})|$, cf.~Eq.~(\ref{eq:CPIMC_weight}), by the modified weight 
\begin{eqnarray}
|W_\kappa(\mathbf{X})| = |W(\mathbf{X})\cdot V_{\kappa}(K)|\ .
\end{eqnarray}
When performing simulations for fixed values of $\kappa$, this potential acts as a smoothly increasing penalty of paths with a large number of kinks $K$, thereby effectively suppressing the occurrence of these paths in the simulation. Since it is $\lim_{\kappa\to\infty} V_{\kappa}(K)=1$, we can extrapolate the results from CPIMC simulation with different values of $\kappa$ to the exact limit $1/\kappa\to 0$, which is illustrated in in the left panel of Fig.~\ref{fig:CPIMC_kinkpot} for $N=66$ electrons at $r_s=2$ and $\theta=4$. Indeed we observe that the total energy, a), is well converged at $\kappa\sim 10$ while the average sign, b), and the average number of kinks, c), are clearly not. In fact, for these parameters, the basic CPIMC simulation without the kink potential equilibrates at an average number of several hundreds of kinks. This fortunate behavior can be explained by a complete cancellation of all contributions to the energy of all paths that contain a larger number of kinks than about $10$. In other words, the simulated modified partition function with the modulus of the weight function converges at much larger values of $K$ than the physical partition function due to a complete cancellation of the weights. In this sense one may also call this circumstance a "sign blessing" rather than a "sign problem". 

Since the convergence with the potential parameter $1/\kappa$ is monotonic, we can obtain a highly accurate upper and lower bound of the exact result even in those cases where convergence is not entirely reached, which is shown in the right panel of Fig.~\ref{fig:CPIMC_kinkpot} for the example of $N=66$ electrons at $r_s=0.8$ and $\theta=1$. For these parameters the bare CPIMC method generates paths that contain about a thousand kinks [see solid blue points in Fig.~\ref{fig:CPIMC_fsp} b)]. Nevertheless,  within the given error bars, the resulting value (black) agrees well with that from the PB-PIMC simulation (orange). Overall, at a fixed number of electrons $N$ and temperature $\theta$, the usage of the kink potential, Eq.~(\ref{eq:fermiPot}), increases the feasible $r_s$ parameter in CPIMC simulations by at least a factor of two. Thus, the  applicability of the method is pushed into density regimes where common analytical perturbation theories break down.

\subsection{Density Matrix Quantum Monte Carlo\label{sec:DMQMC}}

The density matrix quantum Monte Carlo (DMQMC) approach developed by Foulkes, Malone, and co-workers~\cite{blunt_density-matrix_2014,malone_interaction_2015,malone_accurate_2016} is similar to the CPIMC method from the previous section in so far as both are formulated in antisymmetrized Fock space. As we shall see, this leads to a similar range of applicability (see Sec.~\ref{sec:comparison_finite_N}). However, in contrast to the path integral Monte Carlo paradigm, in DMQMC we directly sample the unnormalized thermal density matrix (expanded in a basis of Slater determinants). Therefore, it constitutes a direct extension of the full configuration interaction quantum Monte Carlo (FCIQMC) method~\cite{booth_fermion_2009,shepherd_full_2012,shepherd_convergence_2012,shepherd_investigation_2012}, which has proven to be highly successful in the ground state~\cite{booth_towards_2013}, to finite temperature. Furthermore, it can be viewed as the diffusion Monte Carlo analogue of CPIMC.

Following Ref.~\cite{malone_interaction_2015}, we write the Bloch equation [cf.~Eq.~(\ref{eq:bloch_rpimc})] in a symmetrized form,
\begin{eqnarray}
\frac{\textnormal{d}\hat\rho}{\textnormal{d}\beta} = -\frac{1}{2}(\hat H\hat\rho+\hat\rho\hat H)\ . \label{eq:block_symmetrized}
\end{eqnarray}
Thus, propagating the density matrix in imaginary time by an amount of $\Delta \beta$ using a simple (explicit) Euler scheme gives
\begin{eqnarray}\label{eq:euler}
\hat\rho(\beta + \Delta\beta) = \hat\rho(\beta) - \frac{\Delta\beta}{2}(\hat H \hat\rho(\beta) + \hat\rho(\beta)\hat H) + \mathcal{O}(\Delta\beta^2)\ .
\end{eqnarray}
The basic idea of the density matrix QMC method is to stochastically solve Eq.~(\ref{eq:euler}) by evolving a \textit{population} of positive and negative walkers (sometimes denoted as "particles", "psi-particles", or "psips") in the operator space spanned by tensor products of Slater determinants. Writing down Eq.~(\ref{eq:euler}) in terms of matrix elements $\rho_{ij}=\bra{i}\hat\rho\ket{j}$ (with $\ket{i}$ being a Slater determinant of plane waves) leads to  
\begin{eqnarray}\label{eq:dmqmc_tmp}
\rho_{ij}(\beta+\Delta\beta) = \rho_{ij}(\beta) - \frac{\Delta\beta}{2}\sum_k[(H_{ik}-S\delta_{ik})\rho_{kj} - \rho_{ik}(H_{kj}-S\delta_{kj})]\ ,
\end{eqnarray}
with $S$ being an, in principle, arbitrary shift that can be used to control the population of walkers~\cite{umrigar_diffusion_1993,booth_fermion_2009,malone_interaction_2015}.
Furthermore, it is convenient to introduce the \textit{update matrix}
\begin{eqnarray}
T_{ij} = - (H_{ij}-S\delta_{ij})\ ,
\end{eqnarray}
which allows us to write Eq.~(\ref{eq:dmqmc_tmp}) as
\begin{eqnarray}
\rho_{ij}(\beta+\Delta\beta) = \rho_{ij}(\beta) + \frac{\Delta\beta}{2}\sum_k(T_{ik}\rho_{kj}+\rho_{ik}T_{kj})\ .
\end{eqnarray}

The update scheme governing the stochastic evolution of the walkers can be summarized in three straightforward rules:
\begin{enumerate}
    \item \textit{Spawning} -- A walker can spawn from matrix element $\rho_{ik}$ to $\rho_{ij}$ with the probability $p_\textnormal{spawn}(ik\to ij) = \Delta\beta|T_{kj}|/2$ (the spawning process from $\rho_{kj}$ to $\rho_{ij}$ is similar).
    \item \textit{Clone/Die} -- Walkers on $\rho_{ij}$ can clone or die, leading to an increase or decrease of the population with the probability
    $p_\textnormal{d}(ij)=\Delta\beta|T_{ii}+T_{jj}|/2$. In particular, the population is increased if $\textnormal{sign}(T_{ii}+T_{jj})\times \textnormal{sign}(\rho_{ij}) > 0$ and decreased otherwise.
    \item \textit{Annihilation} -- Walkers on the same matrix elements, but with an opposite sign, are annihilated. This drastically improves the efficiency of the algorithm.
\end{enumerate}
Starting at $\beta=0$ (where $\rho_{ij}=\delta_{ij}$, realized by populating the diagonal density matrix elements with uniform probability), the above algorithm is used to propagate $\rho$ to the desired (inverse) temperature of interest. The full DMQMC simulation, i.e., the computation of thermodynamic expectation values, is then given by averaging over many independent of such "$\beta$-loops".

Regarding simulations of the electron gas using this basic version of DMQMC there appear two practical problems: (i) the distribution within the thermal density matrix changes rapidly with $\beta$ and (ii) important determinants are often not present in the initial configuration. To overcome these obstacles, Malone and co-workers~\cite{malone_interaction_2015} proposed to solve a different differential equation, describing the evolution of a mean-field density matrix to the exact, fully correlated density matrix, both at inverse temperature $\beta$. This so-called \textit{interaction picture} DMQMC method has turned out to be dramatically more efficient and was used to obtain all DMQMC data shown in Sec.~\ref{sec:comparison_finite_N}.

As a final note, we mention that the fermion sign problem in DMQMC manifests as an exponential growth of the number of walkers needed to resolve the exact thermal density matrix, eventually rendering even a stochastical approach unfeasible. 
To delay this "exponential wall",
the exact DMQMC simulation scheme can be used as a starting point for approximations. In particular, one can exploit the extreme degree of sparsity of the thermal density matrix to reduce the computational demands~\cite{malone_accurate_2016}. This, in turn, allows to significantly increase the range of applicability in terms of coupling strength, similar to the controlled kink extrapolation in the CPIMC method, see Sec.~\ref{sec:CPIMC}.
The basic idea of this \textit{initiator approximation}~\cite{malone_accurate_2016} is to prevent walkers on density matrix elements with a comparatively small weight from spawning off-spring on other small elements. Spawning events to unpopulated matrix elements are only possible from the set of so-called \textit{initiator determinants}, which are occupied by a number of walkers above a certain threshold $n_\textnormal{init}$, or if they result from multiple sign-coherent spawning events from other determinants. It is important to note that the bias due to the initiator approximation can be reduced by increasing the total number of walkers within the simulation, $N_\textnormal{walker}$, and vanishes completely in the limit $N_\textnormal{walker}\to\infty$. Therefore, this "i-DMQMC" algorithm can be viewed as a controlled approximation, although a non-monotonic convergence towards the exact result with $N_\textnormal{walker}$ is possible. Furthermore, the accuracy for any finite number $N_\textnormal{walker}$ is significantly reduced for quantities that do not commute with the Hamiltonian.

\subsection{Comparison of QMC methods\label{sec:comparison_finite_N}}

In this section, we present comparisons between data from different QMC methods in a chronological order, starting with the investigation by Schoof \textit{et al.}~\cite{schoof_textitab_2015} and finishing with the most recent comparison in Ref.~\cite{dornheim_abinitio_2017}, where all four methods had been included into the same plot. It is important to note that all results in this section have been obtained for a finite model system of $N=33$ (spin-polarized) or $N=66$ (unpolarized) electrons. An exhaustive introduction, explanation and discussion of finite-size errors, i.e., the extrapolation to the thermodynamic limit, can be found in section~\ref{sec:FSC}.

\subsubsection{The limits of the fixed node approximation}

\begin{figure}\centering
\includegraphics[width=0.5\textwidth]{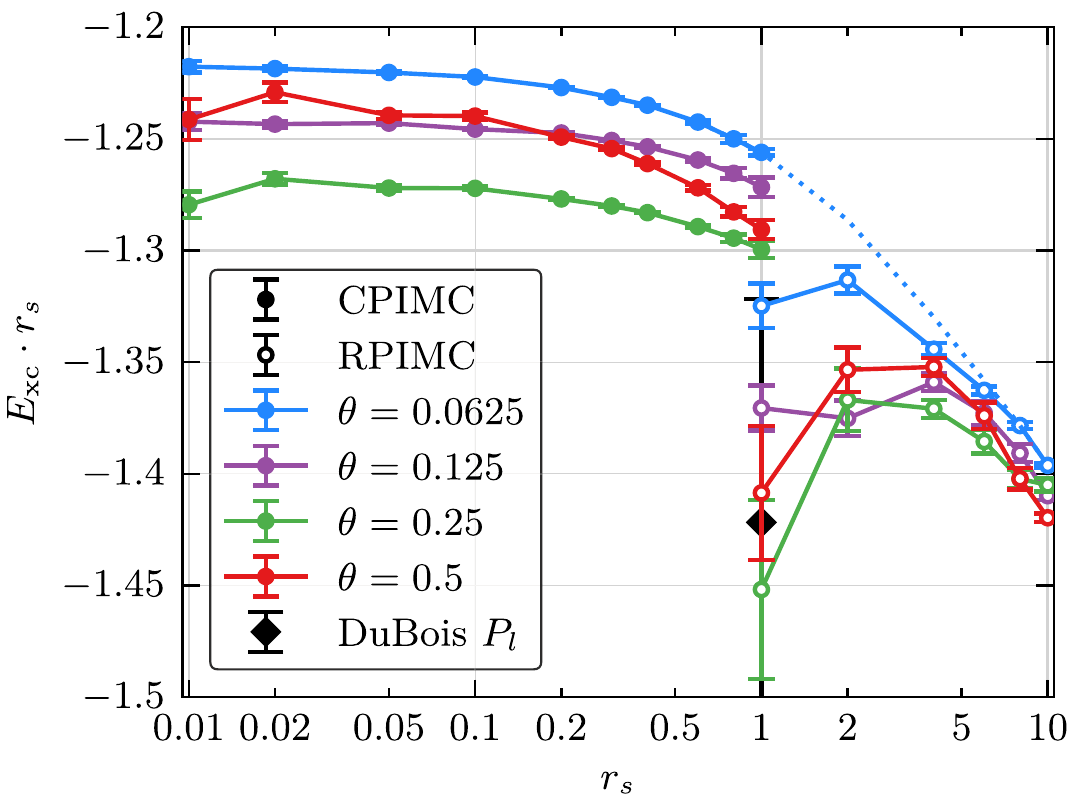}
\caption{\label{fig:tim_prl}Low-temperature results for the exchange-correlation energy of the spin-polarized UEG with $N=33$ electrons. The filled circles correspond to the configuration PIMC data by Schoof \textit{et al.}~\cite{schoof_textitab_2015} and the empty circles have been obtained by subtracting the finite-size correction from the restricted PIMC data in the Supplemental Material of Ref.~\cite{brown_path-integral_2013}. The black diamond corresponds to $\theta=0.0625$ and has been obtained via an approximation based on the extrapolation of permutation cycles introduced by DuBois \textit{et al.}~\cite{dubois_overcoming_2014}.
Reproduced from Ref.~\cite{schoof_textitab_2015} with the permission of the authors.
}
\end{figure}

In 2013, Brown and co-workers~\cite{brown_path-integral_2013} published the first QMC data for the UEG using the restricted PIMC method both for $\xi=0$ and $\xi=1$ covering substantial parts of the warm dense matter regime ($\theta=0.0625,0.125,0.25,\dots ,8$ and $1\leq r_s \leq 40$). It is well known that employing a nodal constraint (using the free particle nodes) constitutes an uncontrolled approximation so that the accuracy of the RPIMC data was not clear. However, the remarkably high accuracy of the fixed node approximation in ground state calculations~\cite{ceperley_ground_1980,spink_quantum_2013,foulkes_quantum_2001} lead to a high confidence in their results, which were subsequently used as input for various applications, e.g., Refs.~\cite{sjostrom_uniform_2013,karasiev_accurate_2014,brown_exchange-correlation_2013,sjostrom_gradient_2014,karasiev_nonempirical_2016}. In their seminal 2015 paper, Schoof \textit{et al.}~\cite{schoof_textitab_2015} were able to obtain \emph{exact} CPIMC data for the spin-polarized electron gas up to $r_s=1,\dots,4$ (depending on temperature), thus enabling them to gauge the bias in the RPIMC data.
The results are shown in Fig.~\ref{fig:tim_prl}, where the exchange-correlation energy $E_\textnormal{xc}=E-U_0$ (with $U_0$ being the energy of the ideal system) is plotted versus $r_s$ for $N=33$ electrons and four different temperatures in the low temperature regime, $\theta=0.0625,0.125,0.25,0.5$. The filled and empty circles correspond to the CPIMC and RPIMC data, respectively.
For completeness, we mention that the black diamond corresponds to a single data point for $\theta=0.0625$ from Ref.~\cite{dubois_overcoming_2014}, which was obtained by performing an approximate extrapolation over the permutation cycles in the PIMC simulation; yet, it is not relevant in the present context. 
Although the sign problem is practically absent in the CPIMC simulations at $r_s<0.1$, the statistical uncertainty (error bars) increases towards even higher density. The explanation for this behaviour is simple: with decreasing $r_s$ the system becomes more similar to the ideal case, thereby making $E_\textnormal{xc}$ the difference between two large numbers, which naturally leads to an increased relative error. On the other hand, the relative CPIMC errors also increase in magnitude for $r_s\geq 0.6$ due to the fermion sign problem, which eventually leads to an exponential wall at some critical value of $r_s$, at which CPIMC simulations are no longer feasible. However, at $r_s=1$ the error bars in the CPIMC data is clearly an order of magnitude smaller than those of the RPIMC data.

The most interesting feature of Fig.~\ref{fig:tim_prl} is the striking disagreement between the exact CPIMC and RPIMC points where the data overlap. In particular, the fixed node approximation leads to an unphysical drop towards high density and the bias in $E_\textnormal{xc}$ exceeds $10\%$. This is in stark contrast to ground state results, where already the data by Ceperley and Alder from 1980~\cite{ceperley_ground_1980} had an accuracy of the order of $0.1\%$. Furthermore, the decreasing quality of the RPIMC data towards high density and weaker coupling contradicts the usual assumption that the systematic error due to the free particle nodes should be most pronounced at stronger nonideality, but vanish for $r_s=0$ (ideal case). While we do not have a definitive explanation of this finding, a possible answer might be a lack of ergodicity within the RPIMC simulation due to the reference point freezing, see Sec.~\ref{sec:RPIMC}, an explanation that would be in good agreement with the observed increment of the RPIMC error bars towards higher density. Finally, we mention that Filinov~\cite{filinov_cluster_2001,filinov_analytical_2014} called into question the validity of the fixed node approximation even for the ideal case.

\subsubsection{Combining CPIMC and PB-PIMC}

The important findings by Schoof \textit{et al.}~\cite{schoof_textitab_2015} from the previous section seriously called into question the utility of the RPIMC data as a basis for density functional theory or other applications at warm dense matter conditions (even more so when considering the additional need for a sufficiently accurate finite-size correction, see Sec.~\ref{sec:FSC}). The problem is that the exact CPIMC method (see Sec.~\ref{sec:CPIMC}), due to its formulation as an infinite perturbation expansion around the ideal system, is limited to moderate coupling (around $r_s=1$, depending on temperature) and, therefore, cannot be used over substantial parts of the relevant WDM regime.
To overcome this issue, Dornheim \textit{et al.}~\cite{dornheim_permutation_2015} introduced the permutation blocking PIMC idea (see Sec.~\ref{sec:PB-PIMC} for a detailed introduction) and subsequently demonstrated its utility for simulation of the electron gas~\cite{dornheim_permutation_2015-1}. In particular, it was suggested that the combination of CPIMC and PB-PIMC at complementary parameters could be used to obtain highly accurate results over the entire density range~\cite{groth_abinitio_2016,dornheim_abinitio_2016}.

\begin{figure}\centering
\includegraphics[width=0.45\textwidth]{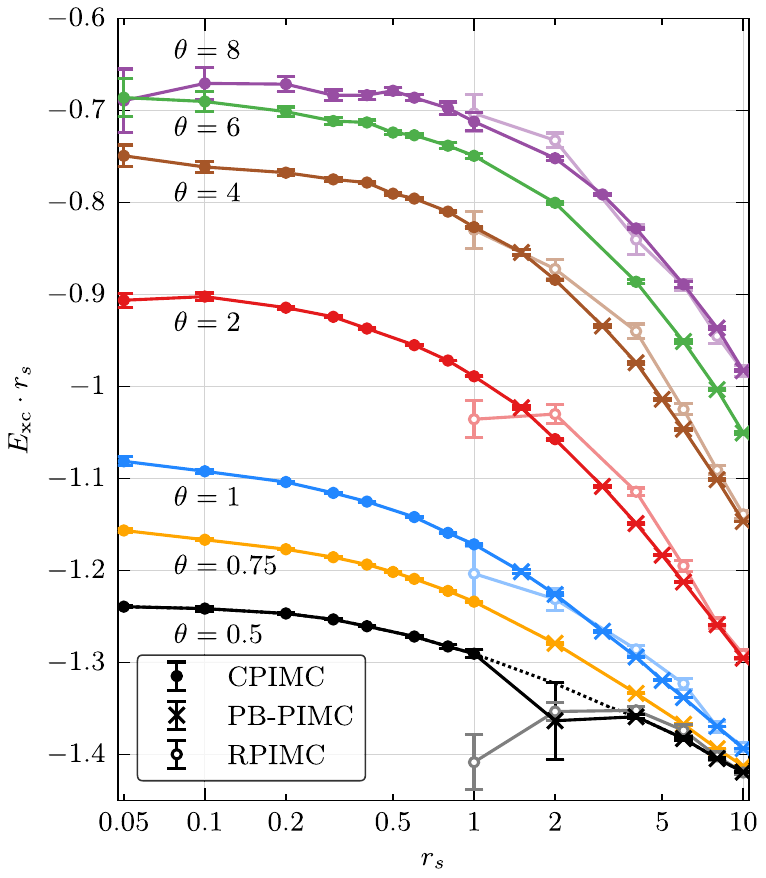}
\includegraphics[width=0.45\textwidth]{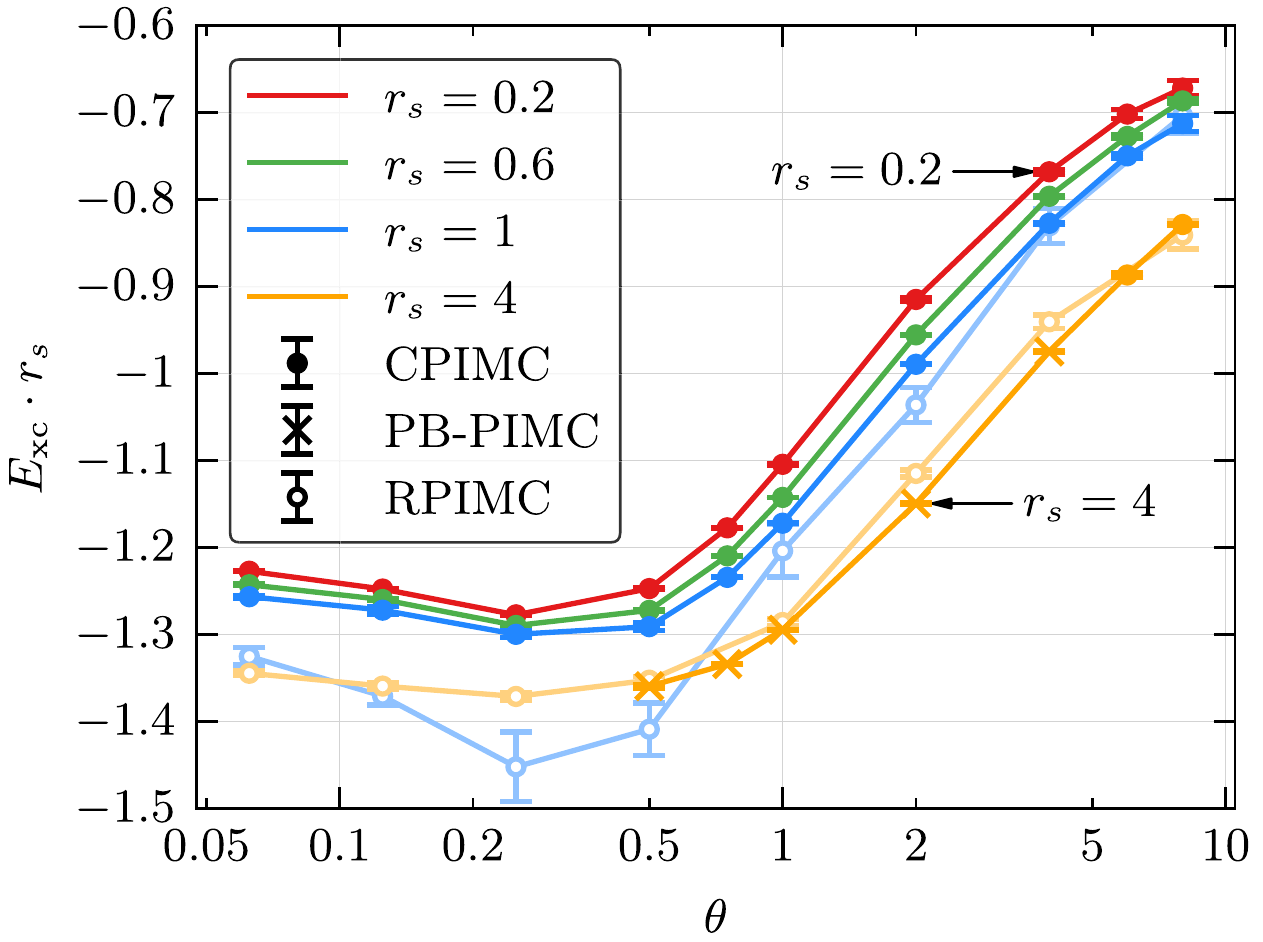}
\caption{\label{fig:prb_polarized}Combination of the configuration PIMC and permutation blocking PIMC methods. Shown is the exchange-correlation energy of $N=33$ spin-polarized electrons in dependence of the density parameter $r_s$ (left) and the reduced temperature $\theta$ (right). The colored filled circles and crosses correspond to the CPIMC and PB-PIMC data, respectively, and the faded empty circles to the RPIMC data by Brown \textit{et al.}~\cite{brown_path-integral_2013}. Reproduced from Ref.~\cite{groth_abinitio_2016} with the permission of the authors.
} 
\end{figure}

This is demonstrated in the left panel of Fig.~\ref{fig:prb_polarized}, where the exchange-correlation energy is shown in dependence of the density parameter $r_s$~\cite{groth_abinitio_2016}. The faded empty circles correspond to the RPIMC data by Brown \textit{et al.}~\cite{brown_path-integral_2013}, the filled circles to CPIMC and the crosses to PB-PIMC data. Note that we show either a CPIMC or a PB-PIMC point, depending on which method provides the smaller statistical uncertainty at a given $r_s$-$\theta$-combination. Again, we mention that the comparatively large error bars in $E_\textnormal{xc}$ at small $r_s$ and high temperature are due to its nature as the difference between two large numbers, the total and ideal energies $E$ and $U_0$, respectively. 
Evidently, the PB-PIMC data is in excellent agreement with and smoothly connects to the CPIMC results for all depicted temperatures. This means that the combination allows for a highly accurate description down to $\theta=0.5$. While CPIMC is also available for lower temperature, cf.~Fig.~\ref{fig:tim_prl}, the permutation blocking PIMC approach eventually becomes infeasible due to the FSP, which is the reason for the relatively large error bar at $r_s=2$ and $\theta=0.5$. For completeness, we mention that the interaction energy $V$, which is sufficient to construct a parametrization of the exchange-correlation free energy $f_\textnormal{xc}$ (see Sec.~\ref{sec:fxc}), can be obtained with a significantly higher accuracy at $\theta=0.5$, see Refs.~\cite{dornheim_permutation_2015-1,dornheim_abinitio_2016,dornheim_abinitio_2016-1}.

The RPIMC data, on the other hand, exhibit an unphysical behavior even at moderate to high temperature. In particular, both for $\theta=0.5$ and $\theta=1$ there occurs a drop in $E_\textnormal{xc}$, while for $\theta=2$ and $\theta=4$ there are pronounced bumps in the region $1\leq r_s \leq 6$.

In the right panel of Fig.~\ref{fig:prb_polarized}, we show the temperature dependence of $E_\textnormal{xc}$ for four different values of the density parameter, $r_s=0.2,0.6,1,4$. The RPIMC data are available for the two largest $r_s$-values, but again there appears a substantial disagreement to the combined CPIMC and PB-PIMC data. While all methods find a minimum in $E_\textnormal{xc}$ around $\theta=0.3$ for all depicted densities, the fixed node approximation leads to a drastically deeper minimum for $r_s=1$ (see also Fig.~\ref{fig:tim_prl} above). Groth and co-workers~\cite{groth_abinitio_2016} gave a possible explanation of this non-monotonic behavior as the competition of two effects: on the one hand, thermal broadening of the particle density leads to a reduction of the interaction energy with temperature, while, on the other hand, Coulomb interactions might be partly increased as the thermal deBroglie wavelength (see Sec.~\ref{sec:PIMC}) decreases with increasing $\theta$. Note that a similar trend has been predicted in the vicinity of Wigner crystallization in $2D$, see Ref.~\cite{clark_hexatic_2009}.

\begin{figure}\centering
\includegraphics[width=0.96\textwidth]{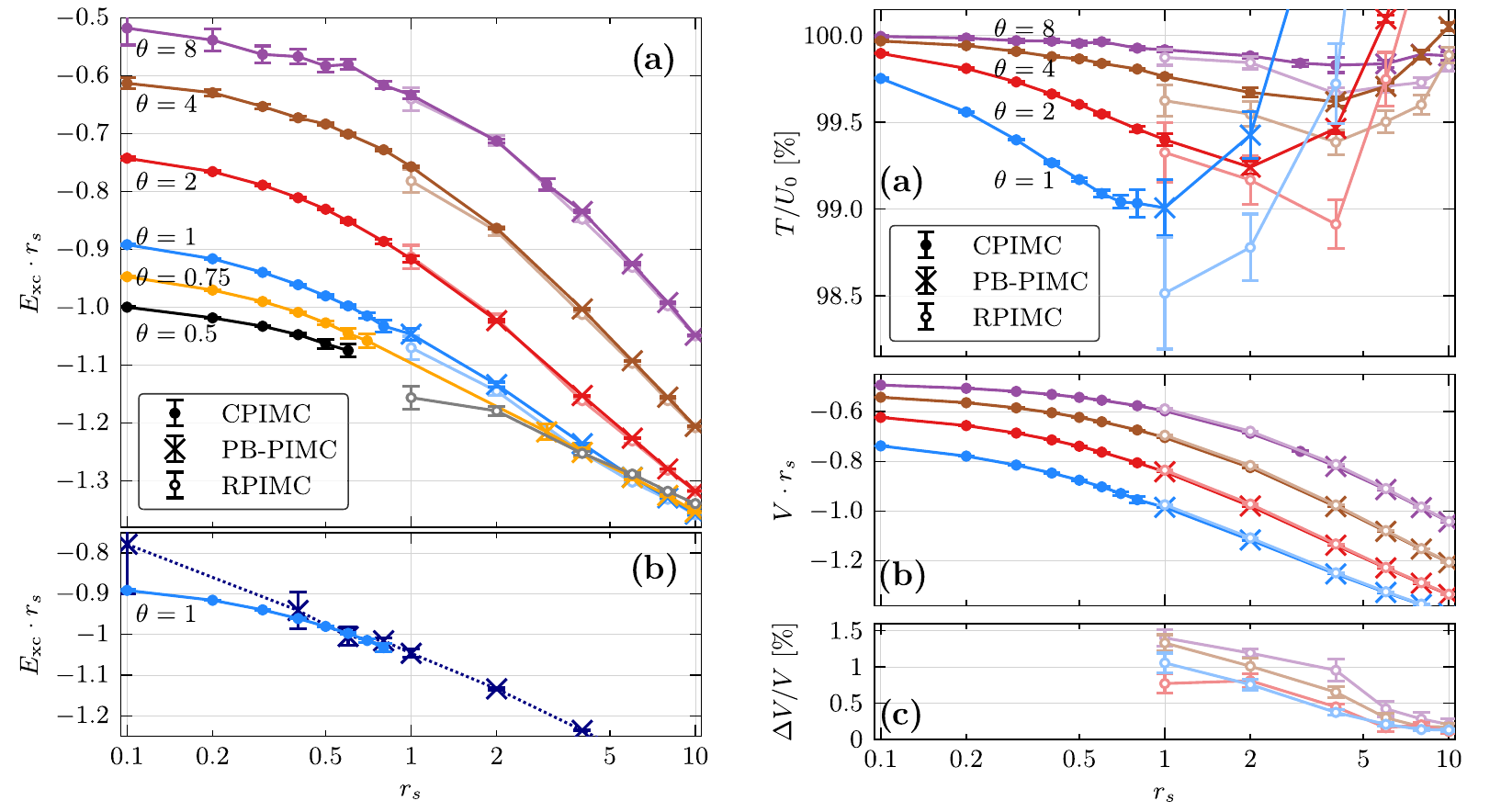}
\caption{\label{fig:prb_unpolarized}Combination of the configuration PIMC and permutation blocking PIMC methods for the unpolarized electron gas with $N=66$ electrons. In the top left panel, we show results for the density-dependence of the exchange-correlation energy from configuration PIMC (filled circles), permutation blocking PIMC (crosses), and restricted PIMC (empty circles, taken from Ref.~\cite{brown_path-integral_2013}). The bottom left panel shows all $E_{xc}$ data for $\theta=1$ both from PB-PIMC and CPIMC, where they are available.
In the top right and center right panel, we show the kinetic energy (in units of the ideal result, $U_0$) and interaction energy from all three methods. Finally, the bottom right panel shows the relative deviation between RPIMC and our data for $V$.
Reproduced from Ref.~\cite{dornheim_abinitio_2016} with the permission of the authors.
} 
\end{figure}

Up to this point, all depicted results had been obtained for the spin-polarized case, i.e., $\xi=1$. However, as real systems are found predominantly in an unpolarized state, the $\xi=0$ case is arguably even more important for real applications. For this reason, in the left panel of Fig.~\ref{fig:prb_unpolarized}, we show the $r_s$-dependence of $E_\textnormal{xc}$ for $N=66$ unpolarized electrons. Again, we show either a CPIMC or PB-PIMC data point, depending on the statistical uncertainty.
Due to the two-fold increase in system size (it is conventional to use a closed momentum shell, i.e., $N_\uparrow=N_\downarrow=33$ spin-up and -down electrons), PB-PIMC results for the exchange-correlation energy are only available above half the Fermi temperature. Regarding the CPIMC approach, there is an additional issue which further reduces the feasible $r_s$ parameter: electrons with opposite spin do not exchange which leads to an increased weight of kinks between those electrons (compared to the same corresponding to two electrons of equal spin)~\cite{dornheim_abinitio_2016}.
The bottom left panel of Fig.~\ref{fig:prb_unpolarized} shows data for $\theta=1$ only, but both from PB-PIMC and CPIMC where they are available. Again, we stress the excellent agreement between the two independent methods as all data agree within error bars and no systematic deviations can be resolved.
The comparion to the RPIMC data by Brown and co-workers~\cite{brown_path-integral_2013} reveals that, for the unpolarized case and for moderate temperatures, there is no systematic bias of the same order as for the spin-polarized case. Only for the lowest depicted temperature, $\theta=0.5$, there seems to appear a systematic drop of the RPIMC data towards high density.

In the right part of Fig.~\ref{fig:prb_unpolarized}, we consider separately both the kinetic and the potential (interaction) contribution to the total energy. Specifically, in the top right panel, we plot the $r_s$-dependence of the kinetic energy (here labelled $T$ and given in units of the ideal energy $U_0$) for $\theta=1,2,4,8$. Surprisingly, we find significantly larger disagreement than in $E_\textnormal{xc}$ for all depicted temperatures as the RPIMC data are systematically too small. Furthermore, these deviations do not vanish entirely even for large $r_s$.

The center right panel of the same figure shows the same information for the Ewald interaction energy $V$, although, on the given scale, no deviations are visible with the naked eye.
For this reason, in the bottom right panel, we show the relative deviation between our data and RPIMC in $V$. Unsurprisingly, we find deviations of a similar magnitude than in the kinetic part, but of an opposite sign, i.e., here the RPIMC data are always too large.

In a nutshell, our analysis of the unpolarized electron gas has revealed that (i) the fixed node approximation gives significantly more accurate results for the exchange-correlation energy than for the spin-polarized case, but (ii) the separate kinetic and potential contributions are systematically biased for all temperatures, even for large $r_s$.
Finding (ii) is a common property of approximations in quantum Monte Carlo methods for quantities that do not commute with the Hamiltonian.
Similar behaviors have been reported in ground state diffusion Monte Carlo calculations using the fixed node approximation\footnote{In DMC, the bias can be removed by the \textit{Hellmann-Feynman} operator sampling~\cite{gaudoin_hellman-feynman_2007}.}, e.g., Refs.~\cite{gaudoin_hellman-feynman_2007,gurtubay_benchmark_2010}, or in finite-temperature DMQMC calculations employing the initiator approximation~\cite{malone_accurate_2016}.

\subsubsection{Emerging consensus of QMC methods}

\begin{figure}\centering
\includegraphics[width=0.5\textwidth]{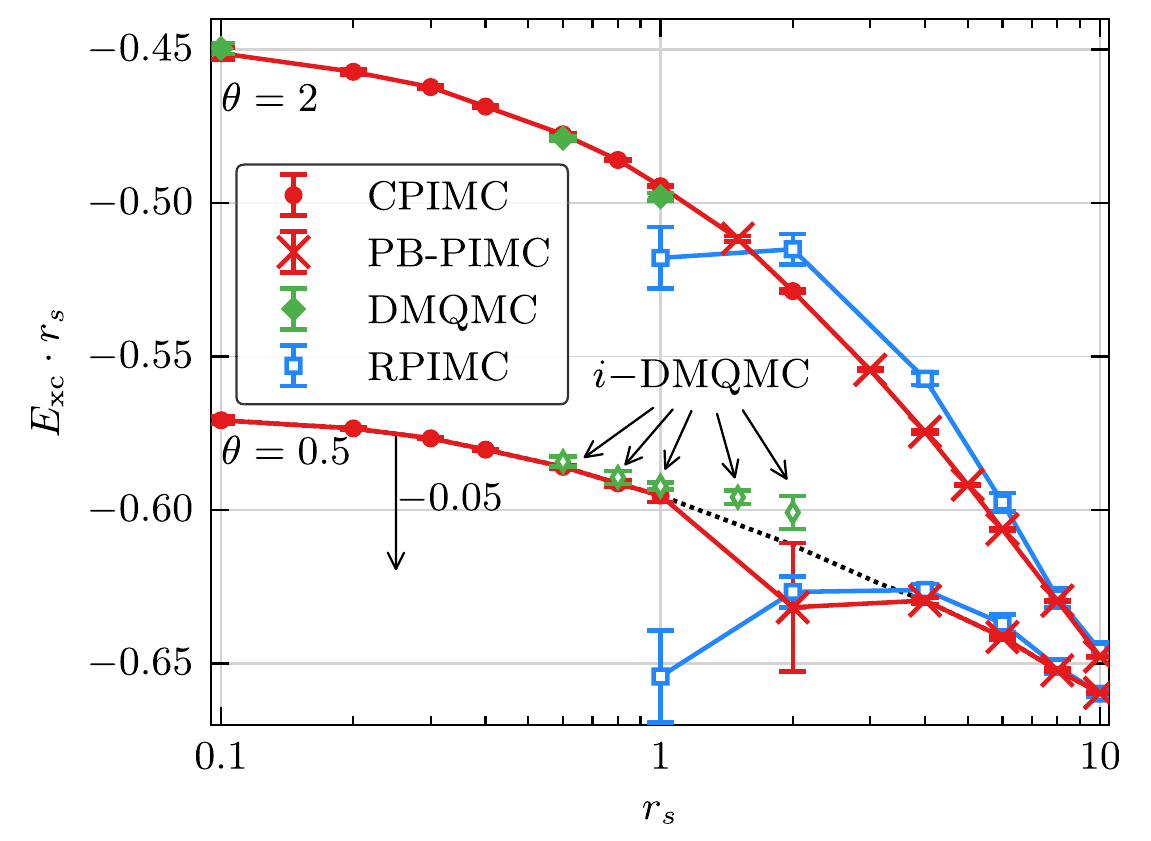}
\caption{\label{fig:pop}Comparison of all QMC methods for the spin-polarized electron gas at warm dense matter conditions.
Shown are results for the $r_s$-dependence of the exchange-correlation energy for $N=33$ electrons from CPIMC (red circles, data taken from Ref.~\cite{groth_abinitio_2016}), PB-PIMC (red crosses, data taken from Ref.~\cite{groth_abinitio_2016}), DMQMC (filled green diamonds) and initiator DMQMC (empty green diamonds, data taken from Ref.~\cite{malone_accurate_2016}) and RPIMC (blue squares, data taken from Ref.~\cite{brown_path-integral_2013}). For $\theta=0.5$, all data have been shifted by $0.05$ Hartree. Reproduced from Ref.~\cite{dornheim_abinitio_2017} with the permission of the authors.
}
\end{figure}

Shortly after the findings of the previous subsections had been reported, Malone and co-workers~\cite{malone_accurate_2016} achieved major breakthroughs regarding the application of the density matrix QMC method to the electron gas at WDM conditions. Their valuable set of additional, independent data has been included in Fig.~\ref{fig:pop} (green diamonds), where the $r_s$-dependence of $E_\textnormal{xc}$
is shown for all four QMC methods introduced above~\cite{dornheim_abinitio_2017}. Note that the $\theta=2$ data corresponds to the exact DMQMC algorithm whereas, for $\theta=0.5$, the initiator approximation was employed.
Evidently, the green points fully confirm our data up to $r_s=1$ within error bars, although, at larger values of $r_s$, the initiator approximation apparently cause $E_\text{xc}$ to be systematically to large.

We thus conclude that over the last two years there has emerged a consensus between different, independent QMC methods regarding the simulation of the UEG for a finite number of electrons.
Naturally, the next step that had to be accomplished was the extrapolation of these results to the thermodynamic limit without a significant loss of accuracy. This turned out to be a surprisingly challenging task, which will be discussed and explained in detail in the next section.

\begin{figure}\centering
\includegraphics[width=0.5\textwidth]{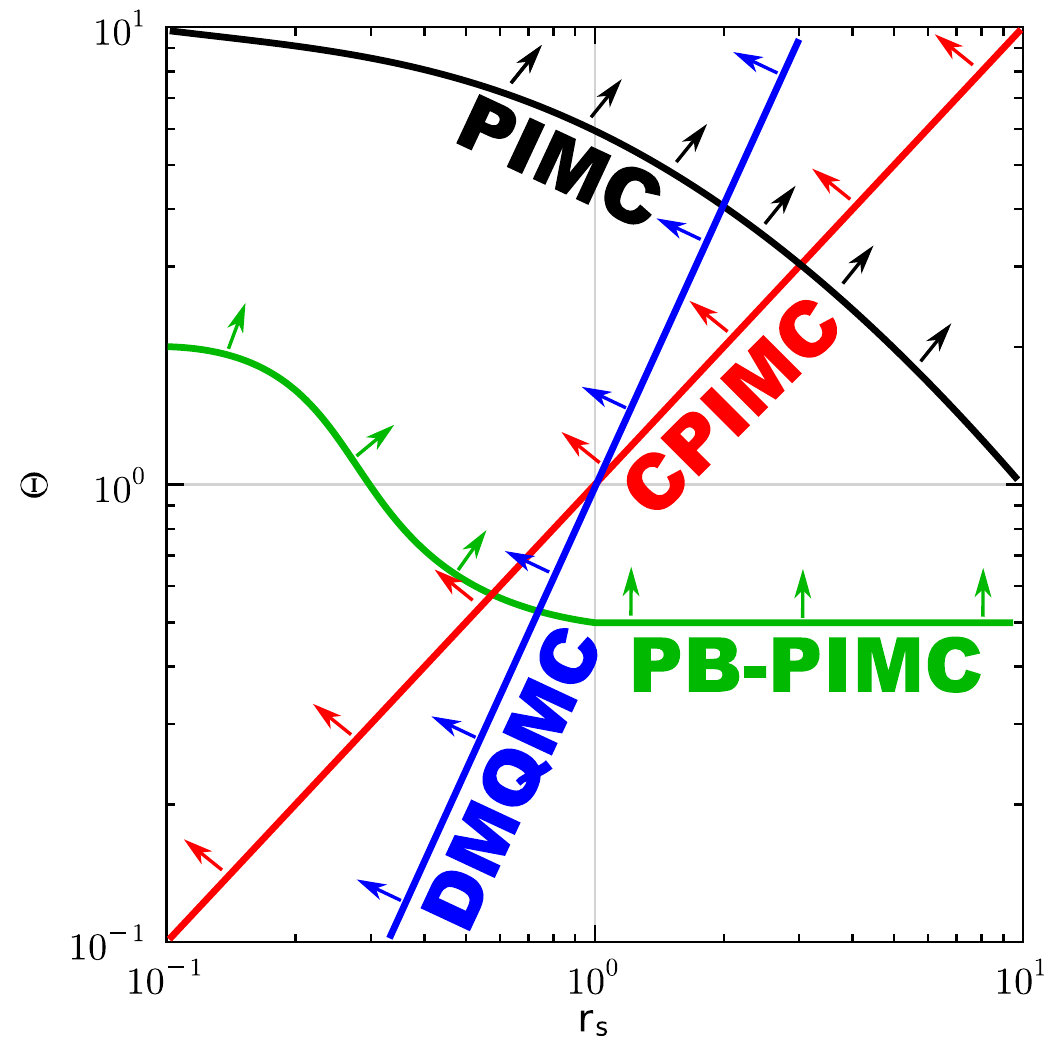}
\caption{\label{fig:pop2}Density-temperature plane around the warm dense matter regime. Shown are the parameter ranges where standard PIMC (black), DMQMC (blue), CPIMC (red) and PB-PIMC (green) are feasible. Reproduced from Ref.~\cite{dornheim_abinitio_2017} with the permission of the authors.
}
\end{figure}
Finally, in Fig.~\ref{fig:pop2}, we show the density-temperature combinations where the different QMC methods are feasible. Evidently, standard PIMC is only available at high temperature and strong coupling (due to the FSP). Our recent PB-PIMC method extends this regime significantly towards lower temperature and high density, i.e., towards strong degeneracy. In contrast, both the CPIMC and DMQMC methods, which are formulated in Fock space, excel at weak coupling but break down when correlation effects start to dominate. Observe that the apparent advantage of DMQMC over CPIMC at low temperature and intermediate $r_s$ is due to the utilized initiator approximation that can lead to a significant bias for quantities that do not commute with the Hamiltonian, see Sec.~\ref{sec:DMQMC} for details.



\section{Finite-size correction of QMC data\label{sec:FSC}}

\subsection{Introduction and problem statement\label{sec:quellbrunn}}
The big advantage of using the quantum Monte Carlo methods introduced in Sec.~\ref{sec:QMC} is that they -- in stark contrast to the dielectric approximations or quantum classical mappings -- allow to obtain an exact solution to the UEG Hamiltonian, Eq.~(\ref{eq:UEG_Ham}). However, this is only possible for a model system with a finite number of particles $N$ and a finite box length $L$. In practice, we are interested in the thermodynamic limit~\cite{lieb_thermodynamic_1975}, i.e., the limit where both $L$ and $N$ go to infinity while the density $n$ (and, therefore, the density parameter $r_s$) remain constant. To mimic as closely as possible the infinite electron gas in our QMC simulations, we employ periodic boundary conditions and incorporate the interaction of a single electron with an infinite array of periodic images via the Ewald interaction. Nevertheless, the interaction energy per particle, $V_N/N$, does not remain constant for different $N$ and is not equal to the thermodynamic limit, which is defined as
\begin{eqnarray}\label{eq:interact_fsc}
\nu = \lim_{N\to\infty} \left. \frac{V_N}{N} \right|_{r_s=\textnormal{const}}\ .
\end{eqnarray}
The difference between $\nu$ and $V/N$ is the so-called finite-size error
\begin{eqnarray}\label{eq:FSC}
\frac{\Delta V_N}{N} = \nu - \frac{V_N}{N}\ ,
\end{eqnarray}
which needs to be compensated for by adding a so-called finite-size correction to the QMC results, i.e., an estimation for $\Delta V_N/N$.
\begin{figure}
\includegraphics[width=0.45\textwidth]{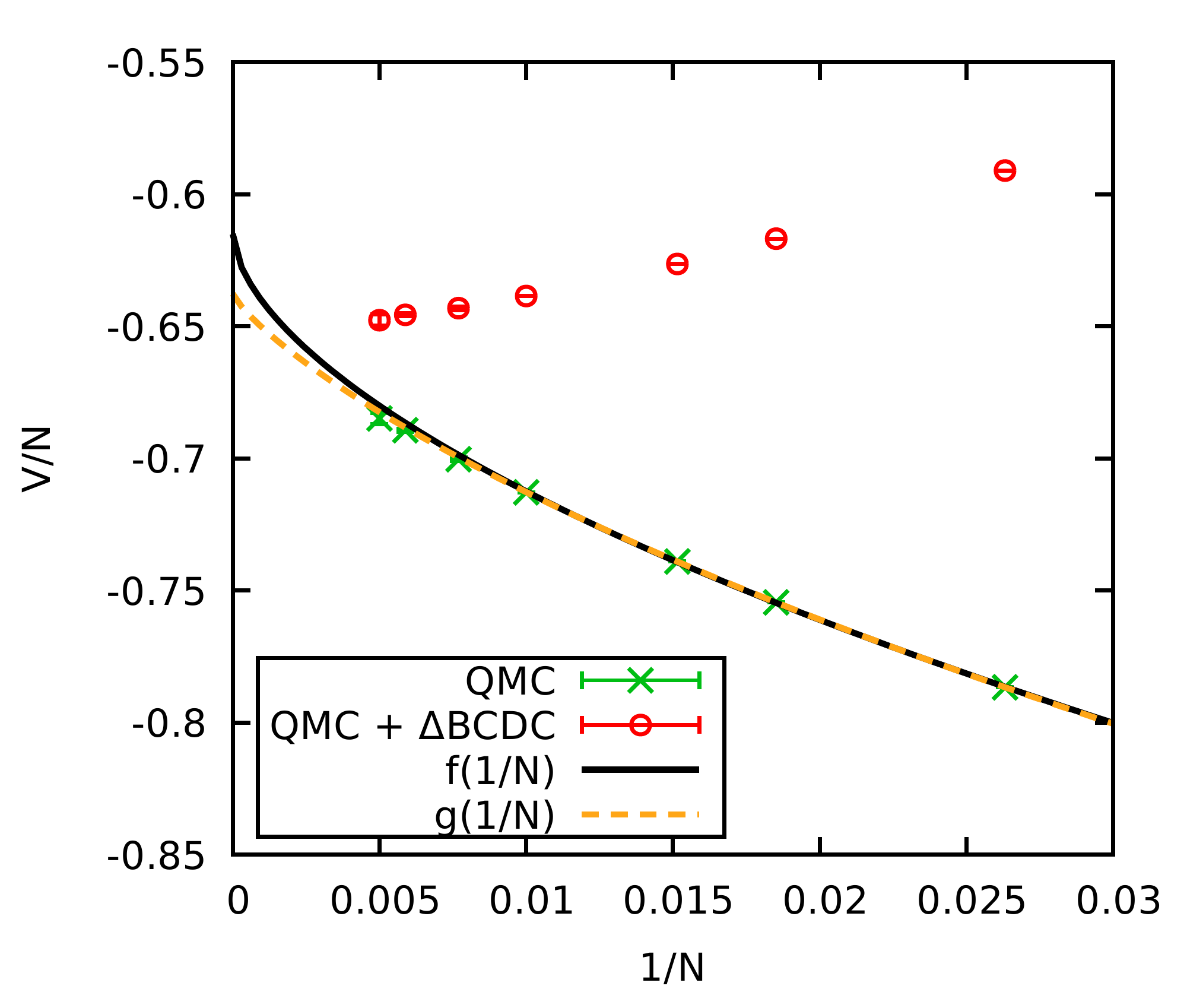}
\includegraphics[width=0.45\textwidth]{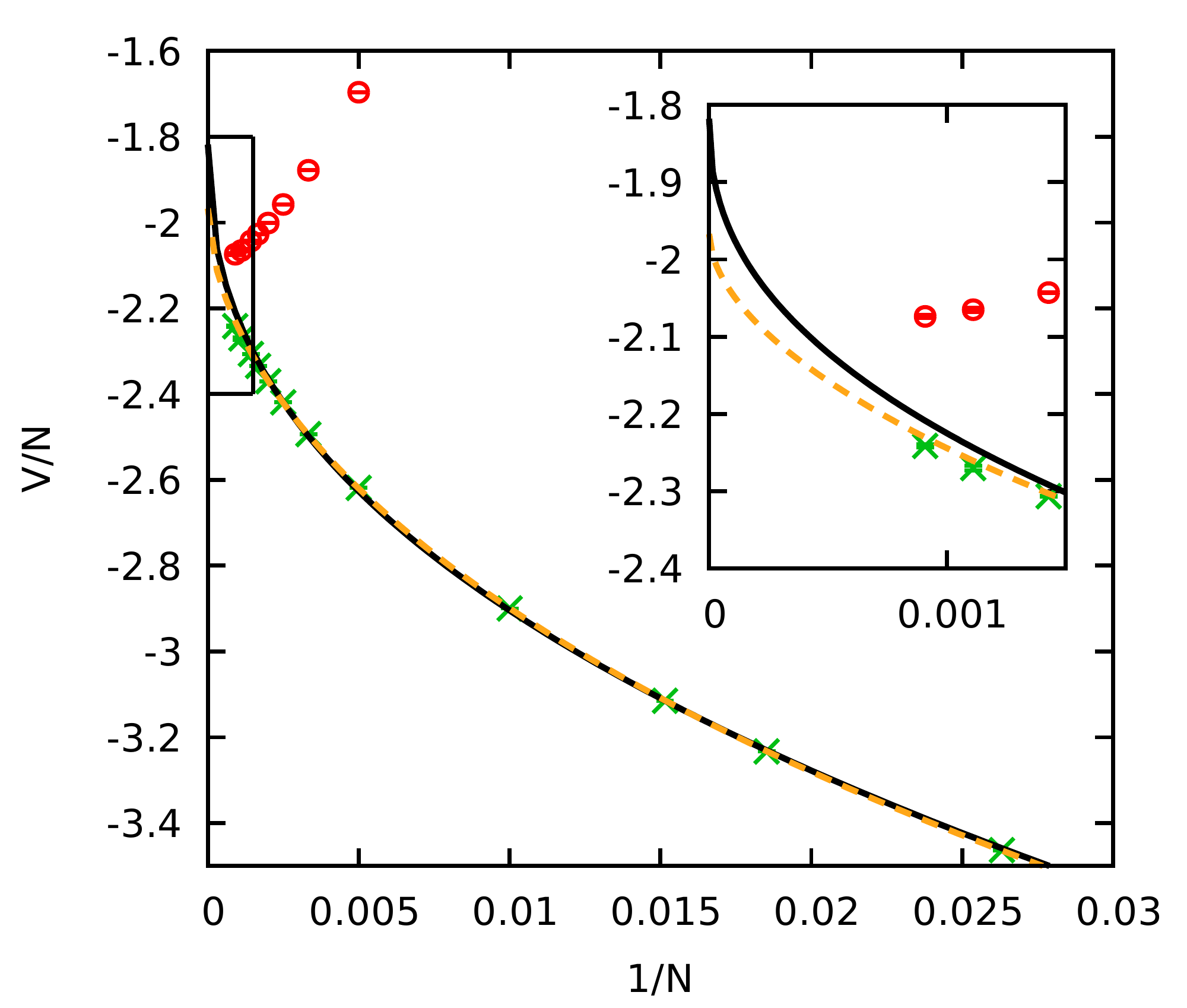}
\caption{\label{fig:FSC_intro}System size dependence of the potential energy per particle of the unpolarized UEG at $\theta=2$ and $r_s=0.5$ (left) and $r_s=0.1$ (right) -- Shown are bare QMC (CPIMC) results (green crosses) and the QMC results plus the finite-size correction proposed by Brown \textit{et al.}~\cite{brown_path-integral_2013} ($\Delta \textnormal{BCDC}$, red circles). The solid black and dashed yellow curves correspond to two equally reasonable fits to the QMC data of the form $f(x)=a+bx^c$ and $g(x)=a+bx+cx^d$, respectively.
The left panel has been adapted from Ref.~\cite{dornheim_abinitio_2016-1} with the permission of the authors.
}
\end{figure}
This is illustrated in Fig.~\ref{fig:FSC_intro}, where, in the left panel, we plot the interaction energy per particle of the unpolarized electron gas with $\theta=2$ and $r_s=0.5$ versus the inverse number of particles $1/N$. The green crosses correspond to the bare QMC results and, obviously, are not converged with respect to $N$. More precisely, for $N=38$ particles, there appears a finite-size error exceeding $10\%$. 
For a higher density, $r_s=0.1$ (see the right panel), things appear to be even more dire and, for $N=38$, $\Delta V_N/N$ is comparable in magnitude to $\nu$ and $V/N$ themselves. 
In this situation it might seem natural to perform a a direct extrapolation to the TDL by performing a fit to the QMC data. However, the problem is that the exact functional form of the finite-size error in dependence of $N$ is not known. The solid black and dashed yellow lines correspond to two fits with different functional forms, specifically
\begin{eqnarray}
f(N^{-1}) &=& a + \frac{b}{N^c}\ , \\
g(N^{-1}) &=& a + \frac{b}{N} + \frac{c}{N^d} \ ,
\end{eqnarray}
with $a, b, c$ and $d$ being the free parameters.
Evidently, for $r_s=0.5$ both fit functions are equally appropriate and reproduce the QMC data quite well. Still, the estimation of the value in the TDL differs by several per cent. This clearly demonstrates that a reliable extrapolation of the QMC data is not possible without knowing the exact $N$-dependence of the finite-size error, which is not the case.
Therefore, we need to derive a readily evaluable approximation to Eq.~(\ref{eq:FSC}).
In the ground state, finite-size effects are relatively well understood, see, e.g., Refs.~\cite{fraser_finite-size_1996,lin_twist-averaged_2001,chiesa_finite-size_2006,drummond_finite-size_2008,holzmann_theory_2016}.
In their pioneering work, Brown \textit{et al.}~\cite{brown_path-integral_2013} introduced a straightforward extension of the finite-size correction for the interaction energy by Chiesa \textit{et al.}~\cite{chiesa_finite-size_2006} to finite temperature [cf.~Eq.~(\ref{eq:BCDC})]. Adding this correction to the QMC results leads to the red circles in Fig.~\ref{fig:FSC_intro}. Obviously, the finite-size errors are overestimated and the remaining bias is of the same order as the original one. Even worse, for $r_s=0.1$ and $N<100$ the corrected data exhibit a larger $N$-dependence than the bare QMC results.
Hence, we conclude that in order to obtain accurate interaction energies in the thermodynamic limit we need to derive an improved finite-size correction. This requires us to analyze and understand the source of the finite-size error and find an accurate estimation for it.

\subsection{Theory of finite-size effects\label{sec:theory_of_finite_size_effects}}
\begin{figure}\centering
\includegraphics[width=0.55\textwidth]{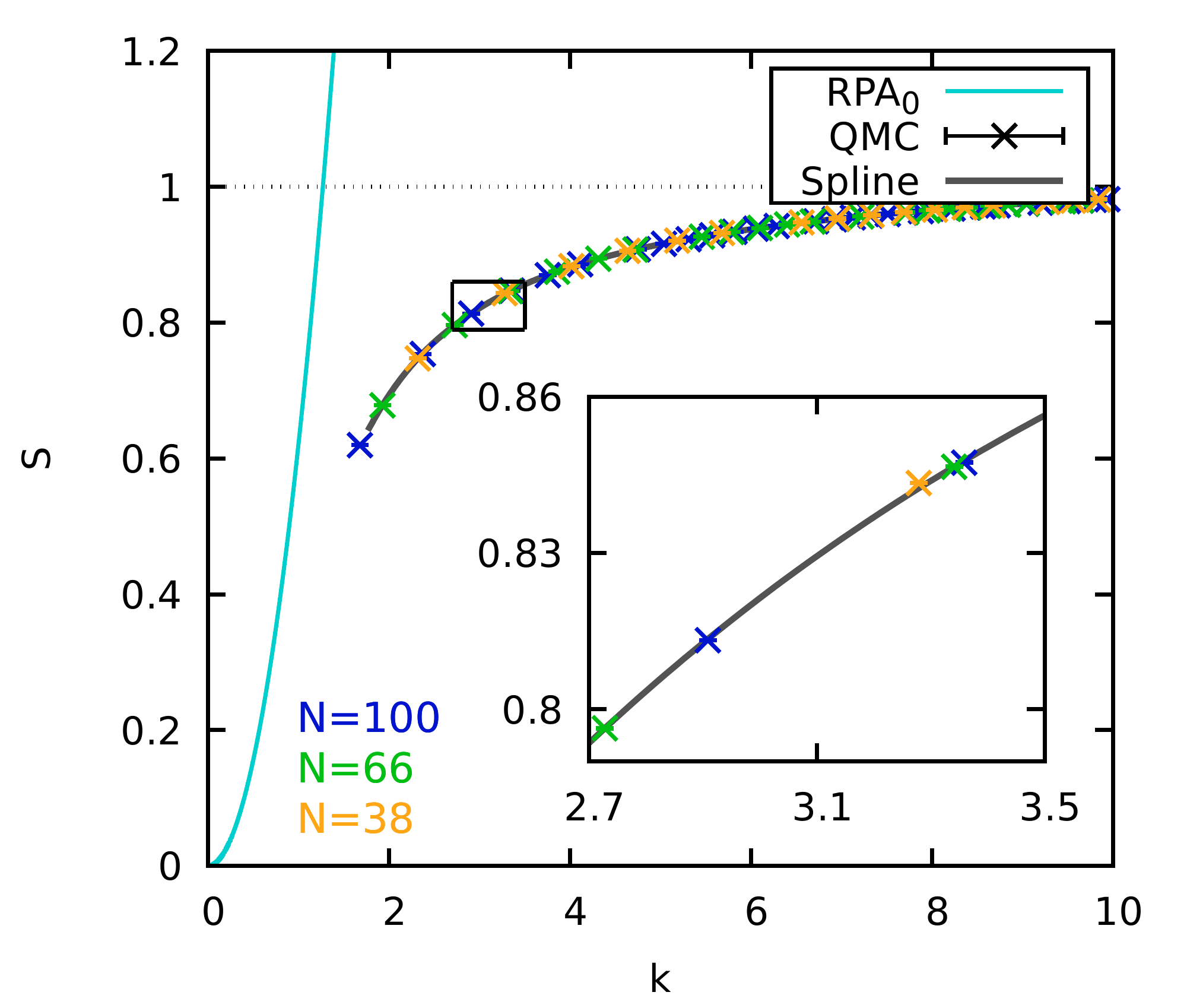}
\caption{\label{fig:FSC_S_illustration}Static structure factor of the unpolarized UEG at $\theta=2$ and $r_s=0.5$ -- Shown are QMC data for $N=100$ (blue), $N=66$ (green), and $N=38$ (yellow) particles and the parabolic RPA expansion around $k=0$ (light blue), cf.~Eq.~(\ref{eq:S0}). The solid grey line corresponds to a cubic spline fit to the $N=100$ data and the inset shows a magnified segment. Adapted from Ref.~\cite{dornheim_abinitio_2016-1} with the permission of the authors.
}
\end{figure}

To derive an expression for the finite-size error due to the final simulation box~\cite{chiesa_finite-size_2006,drummond_finite-size_2008,dornheim_abinitio_2016-1,dornheim_abinitio_2017}, it is convenient to express $V/N$ in terms of the static structure factor $S(\mathbf{k})$ 
\begin{eqnarray}\label{eq:VN}
\frac{V_N}{N} = \frac{1}{2L^3}\sum_{\mathbf{G}\neq \mathbf{0}} \left[ S_N(\mathbf{G}) - 1\right]\frac{4\pi}{G^2} + \frac{\xi_\textnormal{M}}{2}\ ,
\end{eqnarray}
where the subscripts '$N$' denote quantities computed for a finite number of particles, and the sum is to be carried out over the discrete reciprocal lattice vectors $\mathbf{G}$. In the thermodynamic limit, the Madelung constant vanishes, $\xi_\textnormal{M}\to 0$, and the potential energy per particle, Eq.~(\ref{eq:interact_fsc}), can be written as a continuous integral
\begin{eqnarray}\label{eq:v}
\nu = \frac{1}{2}\int_{k<\infty} \frac{ \textnormal{d}\mathbf{k} }{ (2\pi)^3 }\left[ S(k) - 1\right]\frac{4\pi}{k^2}\ ,
\end{eqnarray}
where we have made use of the fact that for a uniform system the static structure factor solely depends on the modulus of the wave vector, $S(\mathbf{k}) = S(k)$.
Obviously, the finite-size error is given by the difference of Eqs.~(\ref{eq:v}) and (\ref{eq:VN}),
\begin{eqnarray}\label{eq:V_difference}
\frac{ \Delta V_N }{N}\left[ S(k), S_N(k) \right] &=& \nu - \frac{V_N}{N} \\
&=& \underbrace{ \frac{1}{2}\int_{k<\infty}\frac{\textnormal{d}\mathbf{k}}{(2\pi)^3} \left[S(k)-1\right]\frac{4\pi}{k^2} }_{v} - \underbrace{ \left(
 \frac{1}{2L^3}\sum_{\mathbf{G}\ne\mathbf{0}}\left[S_N(\mathbf{G})-1\right]\frac{4\pi}{G^2}+\frac{\xi_\textnormal{M}}{2}\right) }_{V_N/N}\ , \nonumber
\end{eqnarray}
and, thus, is a functional of the SFs of the infinite and finite systems, respectively.
To derive a more easily workable expression for Eq.~(\ref{eq:V_difference}), we approximate the Madelung energy by~\cite{drummond_finite-size_2008}
\begin{eqnarray}\label{eq:Madelung_approximation}
\xi_\textnormal{M} \approx \frac{1}{L^3} \sum_{\mathbf{G}\neq \mathbf{0}} \frac{4\pi}{G^2} e^{-\epsilon G^2}
- \frac{1}{(2\pi)^3}\int_{k < \infty}\textnormal{d}\mathbf{k}\;\frac{4\pi}{k^2}e^{-\epsilon k^2}\ ,
\end{eqnarray}
which becomes exact for $\epsilon\to0$. 
Inserting Eq.~(\ref{eq:Madelung_approximation}) into (\ref{eq:V_difference}) gives
\begin{eqnarray}\label{eq:V_source}
\frac{ \Delta V_N }{N}\left[ S(k), S_N(k) \right] = \frac{1}{2}\int_{k<\infty}\frac{\textnormal{d}\mathbf{k}}{(2\pi)^3} S(k) \frac{4\pi}{k^2} - \frac{1}{2L^3}\sum_{\mathbf{G}\neq\mathbf{0}} S_N(\mathbf{G})\frac{4\pi}{G^2} \ .
\end{eqnarray}
Evidently, in Eq.~(\ref{eq:V_source}) there are two possible sources for the finite-size error of $V$: (i) the difference between the SFs of the finite and infinite system, i.e., a finite-size effect in the actual functional form of $S(k)$ itself, or (ii) the approximation of the continuous integral from Eq.~(\ref{eq:v}) by a discrete sum. Chiesa \textit{et al.}~\cite{chiesa_finite-size_2006} pointed out that, in the ground state, the SF converges remarkably fast with system size (this also holds at finite temperature, see Refs.~\cite{dornheim_abinitio_2016-1,dornheim_abinitio_2017} and the discussion of Fig.~\ref{fig:FSC_S_illustration}), leaving (ii) as the sole explanation. In fact, the same authors suggested that the main contribution to Eq.~(\ref{eq:V_source}) is the $\mathbf{G}=\mathbf{0}$ term, which is completely omitted from the sum.
To derive an analytic expression of this term, one makes use of the fact that the random phase approximation becomes exact in the long wave length limit, $k\to0$, which is valid at finite temperatures as well~\cite{kugler_bounds_1970}. In particular, an expansion of the RPA static structure factor around $k=0$ gives a parabolic expression,
\begin{eqnarray}\label{eq:S0}
S_0^\textnormal{RPA}(k) = \frac{k^2}{2\omega_p}\textnormal{coth}\left(\frac{\beta\omega_p}{2}\right)\ ,
\end{eqnarray}
with $\omega_p=\sqrt{3}/r_s^{3/2}$ being the plasma frequency. These considerations lead to the finite-$T$ extension of the FSC from Ref.~\cite{chiesa_finite-size_2006}, hereafter labelled as 'BCDC'~\cite{brown_path-integral_2013}
\begin{eqnarray}\label{eq:BCDC}
\Delta V_\textnormal{BCDC}(N) &=& \lim_{k\to 0} \frac{S_0^\textnormal{RPA}(k) }{2L^3}\ \frac{4\pi}{k^2} \\ \nonumber 
&=& \frac{ \omega_p }{4N} \textnormal{coth}\left(\frac{\beta\omega_p}{2}\right)\ .
\end{eqnarray}
Thus, the first order finite-size correction used by Brown and co-workers predicts a finite-size error with a simple $1/N$ behavior. However, this ansatz is not appropriate for the conditions encountered in Fig.~\ref{fig:FSC_intro}, as we shall now explain in detail.

In Fig.~\ref{fig:FSC_S_illustration}, we show the static structure factor for the unpolarized UEG at $\theta=2$ and $r_s=0.5$, i.e., the same conditions as in the left panel of Fig.~\ref{fig:FSC_intro} above. The blue, green, and yellow crosses correspond to QMC results for $N=100$, $N=66$, and $N=38$ electrons, respectively and the grey solid line to a cubic spline fit to the largest depicted particle number. Due to momentum quantization in a finite simulation cell, data for $S_N(k)$ are available on an $N$-dependent discrete $k$-grid, and restricted to $k\geq k_\textnormal{min}=2\pi/L$. Nevertheless, the functional form of $S_N(k)$ is remarkably well converged with system size for as few as $N=38$ electrons, see also the inset. This means that the finite-size errors in the interaction energy are indeed the consequence of a discretization error as explained above. 
The light blue curve in Fig.~\ref{fig:FSC_S_illustration} corresponds to the RPA expansion around $k=0$, i.e., Eq.~(\ref{eq:S0}).
Evidently, the parabola does not connect to the QMC data even for the largest particle number. Therefore, Eq.~(\ref{eq:BCDC}) is not sufficient to correct for the finite size error. In sum, the construction of a more accurate FSC requires accurate knowledge of $S(k)$ for $k<2\pi/L$, i.e., for those wave vectors that are not accessible within the QMC simulations.

\subsection{Improved finite-size correction of the interaction energy\label{sec:improved_fsc}}

\begin{figure}\centering
\includegraphics[width=0.55\textwidth]{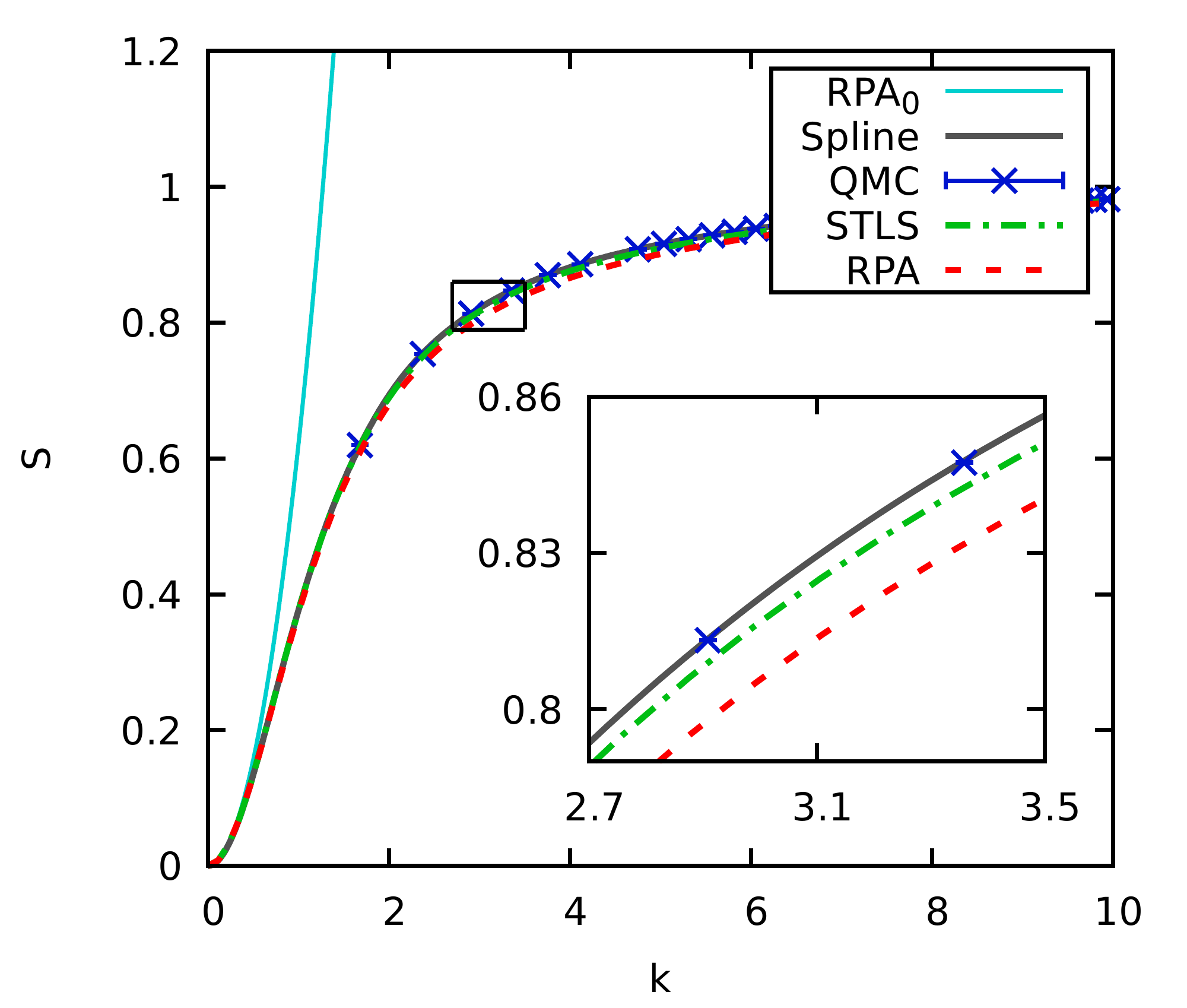}
\caption{\label{fig:FSC_S_full}Static structure factor of the unpolarized UEG at $\theta=2$ and $r_s=0.5$ -- Shown are QMC data for $N=100$ particles (blue crosses), the parabolic RPA expansion around $k=0$ (light blue), cf.~Eq.~(\ref{eq:S0}), full STLS and RPA data (green dash-dotted and red dashed lines, respectively), and a spline connecting STLS for small $k$ with QMC elsewhere (solid grey). 
The inset shows a magnified segment.
Adapted from Ref.~\cite{dornheim_abinitio_2016-1} with the permission of the authors.
}
\end{figure}

To obtain accurate data for the static structure factor for small $k$, we carry out full calculations within RPA and also with a static local field correction from the STLS formalism~\cite{tanaka_thermodynamics_1986,sjostrom_uniform_2013}, see Sec.~\ref{sec:LRT}. The results are shown in Fig.~\ref{fig:FSC_S_full}, where $S(k)$ is shown for the same conditions as in Fig.~\ref{fig:FSC_S_illustration}. The dashed red and dash-dotted green lines correspond to the full RPA and STLS data, respectively, and the blue crosses to the exact QMC results for $N=100$. In the limit $k\to0$, both the RPA and STLS curves are in perfect agreement with the parabolic form from Eq.~(\ref{eq:S0}), but strongly deviate for $k\gtrsim0.5$. Further, both dielectric approximations exhibit a fairly good agreement with the QMC point at $k_\textnormal{min}$ and the STLS result is within the statistical uncertainty.
Therefore, the combination of STLS at small $k$ with the exact QMC data elsewhere allows for exact, unbiased structure factor over the entire $k$-range. In practice, this is realized by a (cubic) spline, cf.~the solid grey line in Fig.~\ref{fig:FSC_S_full}.
Further, we note that the accuracy of both STLS and RPA decreases for larger $k$, see the inset, although the static local field correction from STLS constitutes a significant improvement.
This complementarity of QMC and the dielectric approximations allows for a rather vivid interpretation: Quantum Monte Carlo methods provide an exact treatment of all short-range exchange and correlation effects within the finite simulation box. However, due to the finite number of particles, the long-range limit cannot be resolved. In contrast, both RPA and STLS are formulated in the thermodynamic limit. Since the effect of correlations decreases for large distances, the small $k$-behavior is described accurately, whereas short-range XC effects are treated insufficiently.
For completeness, we note that an accurate knowledge of $S(k)$ would allow to obtain an unbiased result for the interaction energy per particle in the TDL by directly evaluating Eq.~(\ref{eq:v}). However, as we will see below, the detour over the finite-size corrections turns out to be advantageous for multiple reasons.

\begin{figure}\centering
\begin{minipage}{.45\textwidth}
\includegraphics[width=0.99\textwidth]{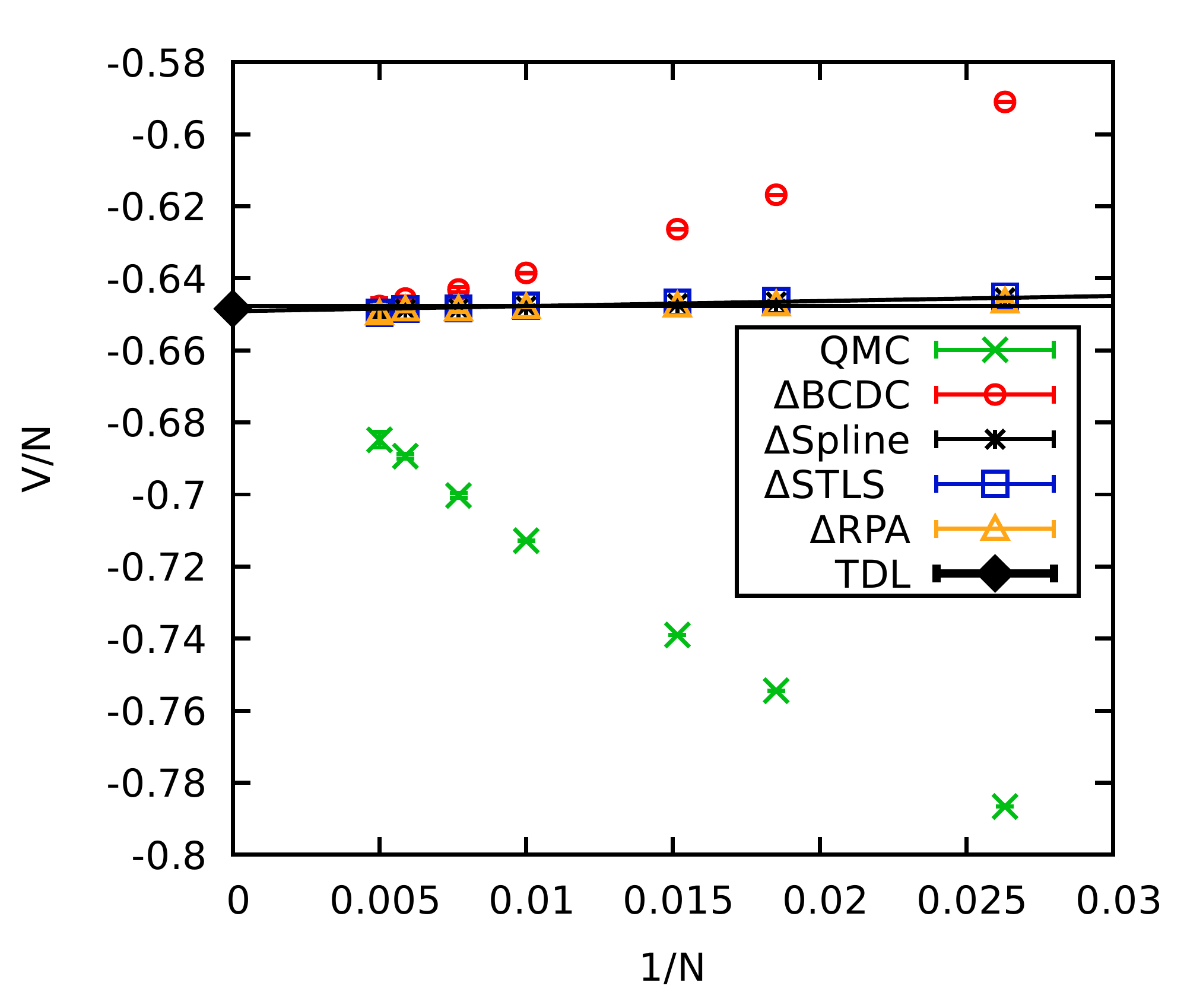}
\end{minipage}
\begin{minipage}{.47\textwidth}
\includegraphics[width=0.99\textwidth]{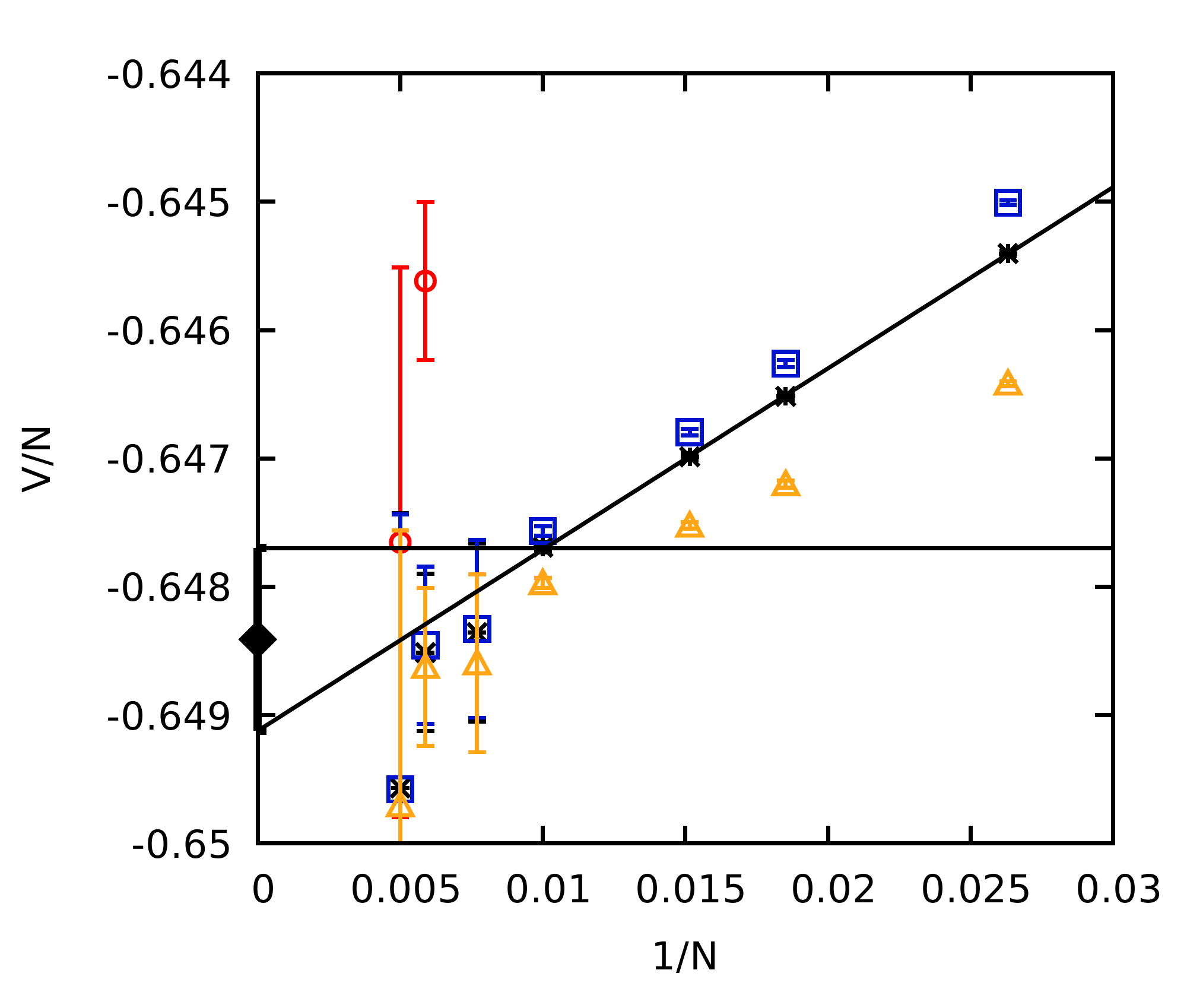}
\end{minipage}
\caption{\label{fig:FSC_RPA_STLS}Improved finite-size correction for the interaction energy per particle of the unpolarized electron gas at $\theta=2$ and $r_s=0.5$ -- Shown are the bare QMC data (green crosses) and the QMC data plus different finite-size corrections, namely $\Delta_\textnormal{BCDC}$ [red circles, see Eq.~(\ref{eq:BCDC})], and our new FSC from Eq.~(\ref{eq:our_FSC}) evaluated using the static structure factors from the spline (black stars), STLS (blue squares) and full RPA (yellow triangles). The solid black lines correspond to a linear and a constant fit to the black stars and the black diamond depicts our result for $V/N$ in the thermodynamic limit.
The right panel shows a magnified inset around the results obtained by adding our new FSCs and the subsequent extrapolation. Evidently, using the static SFs solely from full STLS or RPA is sufficient to accurately estimate the finite-size error.
Adapted from Ref.~\cite{dornheim_abinitio_2016-1} with the permission of the authors.
}
\end{figure}
The thusly obtained model function for the static structure factor [i.e., the spline, $S_\textnormal{Spline}(k)$] allows us to accurately estimate the finite-size error by straightforwardly evaluating Eq.~(\ref{eq:V_difference}) as
\begin{eqnarray}\label{eq:our_FSC}
\Delta V_N\bigg[S_\textnormal{model}(k)\bigg] = \frac{\Delta V_N}{N}\bigg[S_\textnormal{model}(k),S_\textnormal{model}(k)\bigg] \ ,
\end{eqnarray}
which we compute numerically. The resulting FSC is shown in Fig.~\ref{fig:FSC_RPA_STLS}, where we again show the $N$-dependence of the interaction energy per particle for the same conditions as above. Let us first consider the black stars, which have been obtained by adding to the bare QMC results $\Delta V_N[S_\textnormal{spline}(k)]$. Evidently, the dependence on system size has been decreased by two orders of magnitude. The right panel shows a magnified segment around the new corrected results and we detect a small remaining finite-size error with a linear behavior. The main source of this residual error is the small $N$-dependence of $S_N(k)$ itself. However, even for as few as $N=38$ particles, this bias is of the order of $\Delta V/V\sim 10^{-3}$. In practice, we always remove any residual errors by performing an additional extrapolation of the corrected data. In particular, we perform a linear fit over all $N$ and a constant fit to the last few points that are converged with $N$ within twice the error bars (the latter corresponds to the assumption that the small system size dependence in $S_N(k)$ vanishes for large $N$, which it might), see the solid black lines. Our final estimation of the interaction energy per particle in the thermodynamic limit is then obtained as the mean of both fits, and the difference between the two constitutes the remaining uncertainty interval. 
Let us now consider the blue squares and yellow triangles, which have been obtained by evaluating Eq.~(\ref{eq:our_FSC}) solely using the static structure factors from STLS and RPA, respectively, over the entire $k$-range. 
Surprisingly, both data sets are in good agreement with the black stars. This means that -- despite the rather significant bias for intermediate $k$ -- both the full RPA and STLS SFs are sufficient model functions to estimate the discretization error in the interaction energy per particle. Therefore, it is not necessary to perform a spline interpolation for each case, and, in the following, we will compute $\Delta V_N$ using STLS.
It is important to note that while the dielectric approximations allow to accurately estimate the discretization error in $V_N/N$, we still need a QMC result for $V_N/N$ itself, i.e.,
\begin{eqnarray}
\nu = \frac{V^\textnormal{QMC}_N}{N} + \Delta V_N\bigg[ S_\textnormal{STLS}(k) \bigg] \ .
\end{eqnarray}
Replacing $V_N/N$ by the STLS value, which is equivalent to evaluating Eq.~(\ref{eq:v}) using $S_\textnormal{STLS}(k)$, would neglect the short-range exchange-correlation effects and induce a systematic bias of the order of $\Delta V/V\sim 10^{-2}$, see Sec.~\ref{sec:comparison}.

\subsection{Examples of finite-size corrections of QMC data\label{sec:examples_fsc}}

\subsubsection{Coupling strength dependence of the finite-size correction of QMC data}

\begin{figure}\centering
\begin{minipage}{.45\textwidth}
\includegraphics[width=0.99\textwidth]{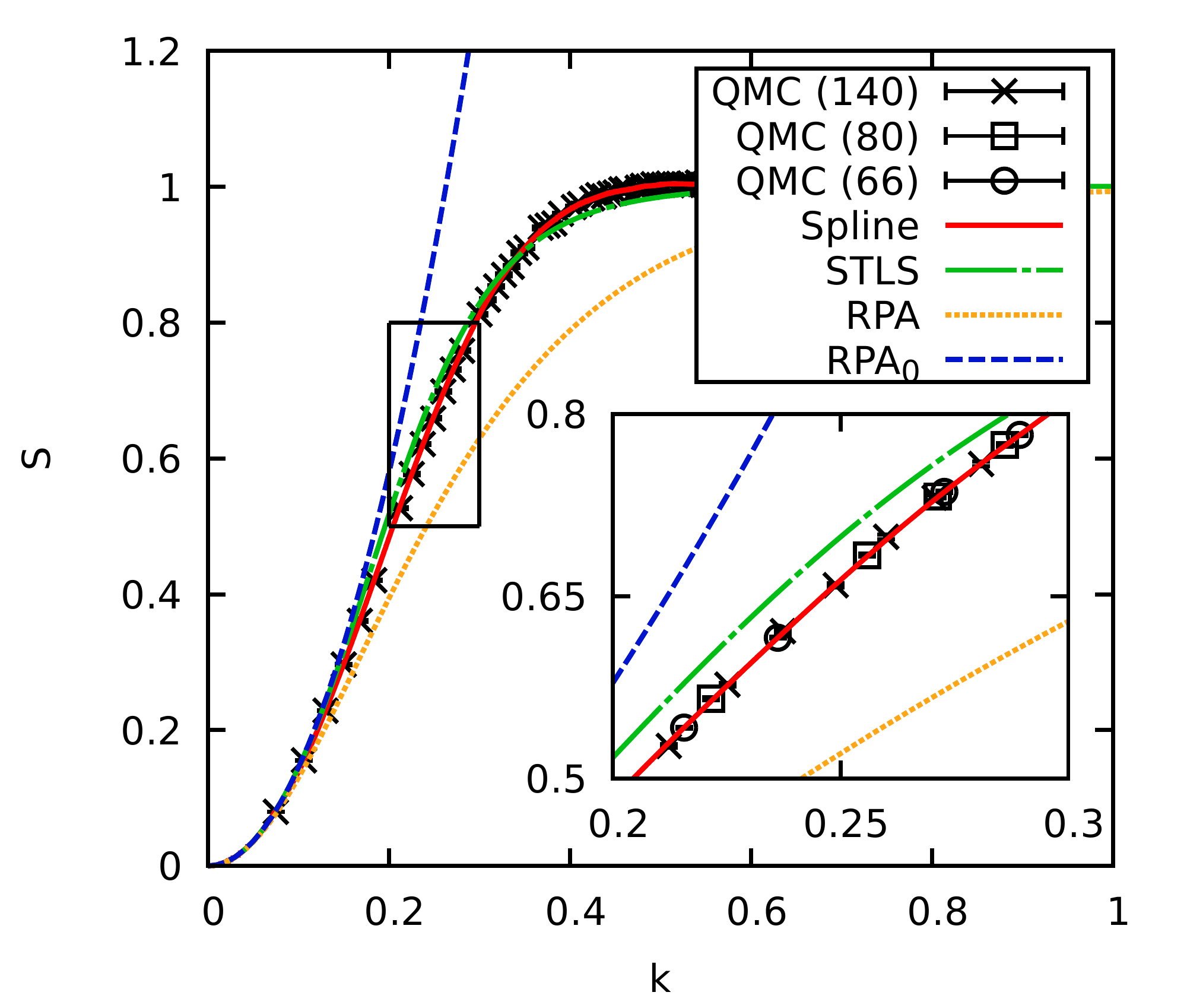}
\end{minipage}
\begin{minipage}{.49\textwidth}
\includegraphics[width=0.99\textwidth]{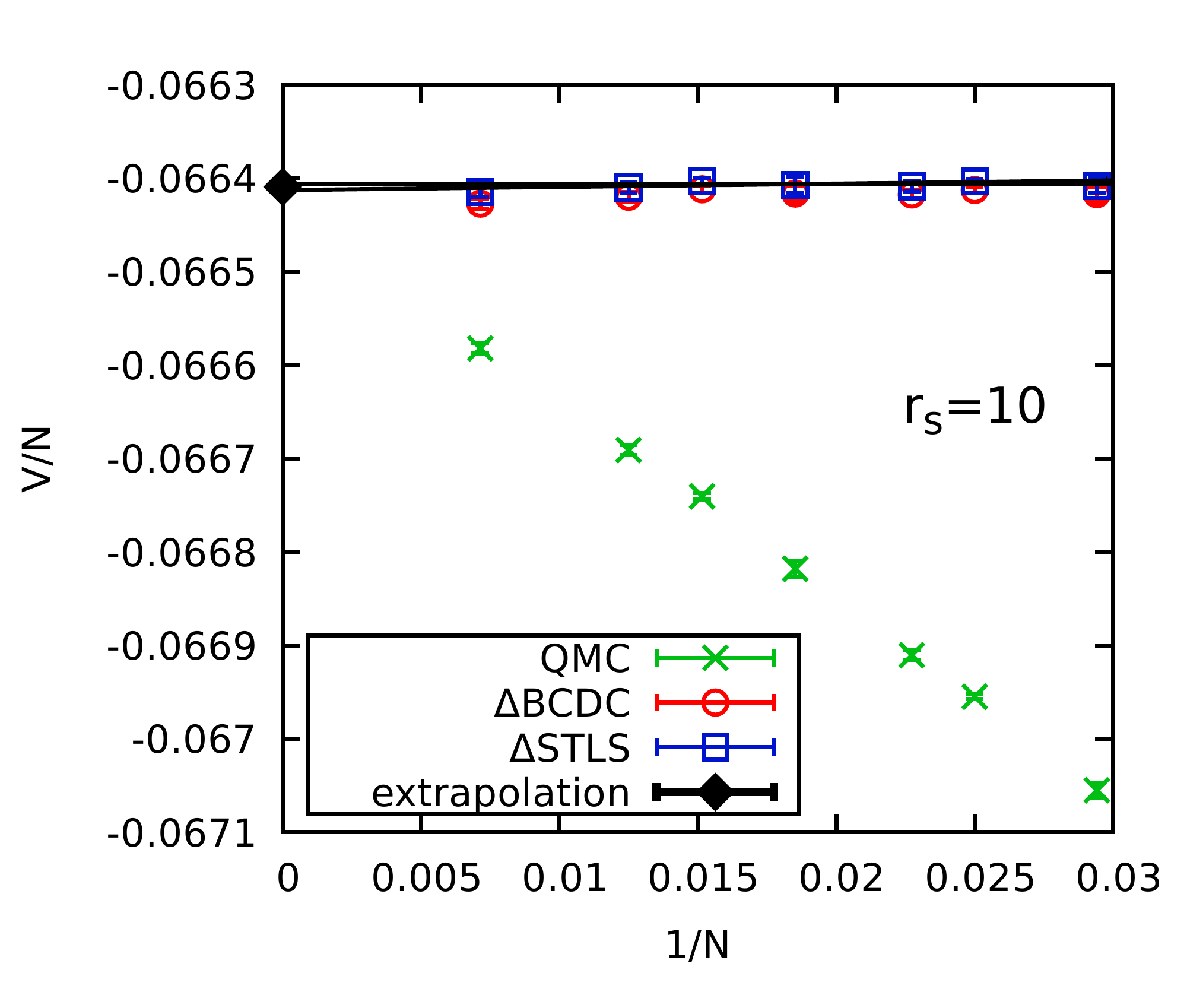}
\end{minipage}
\begin{minipage}{.45\textwidth}
\includegraphics[width=0.99\textwidth]{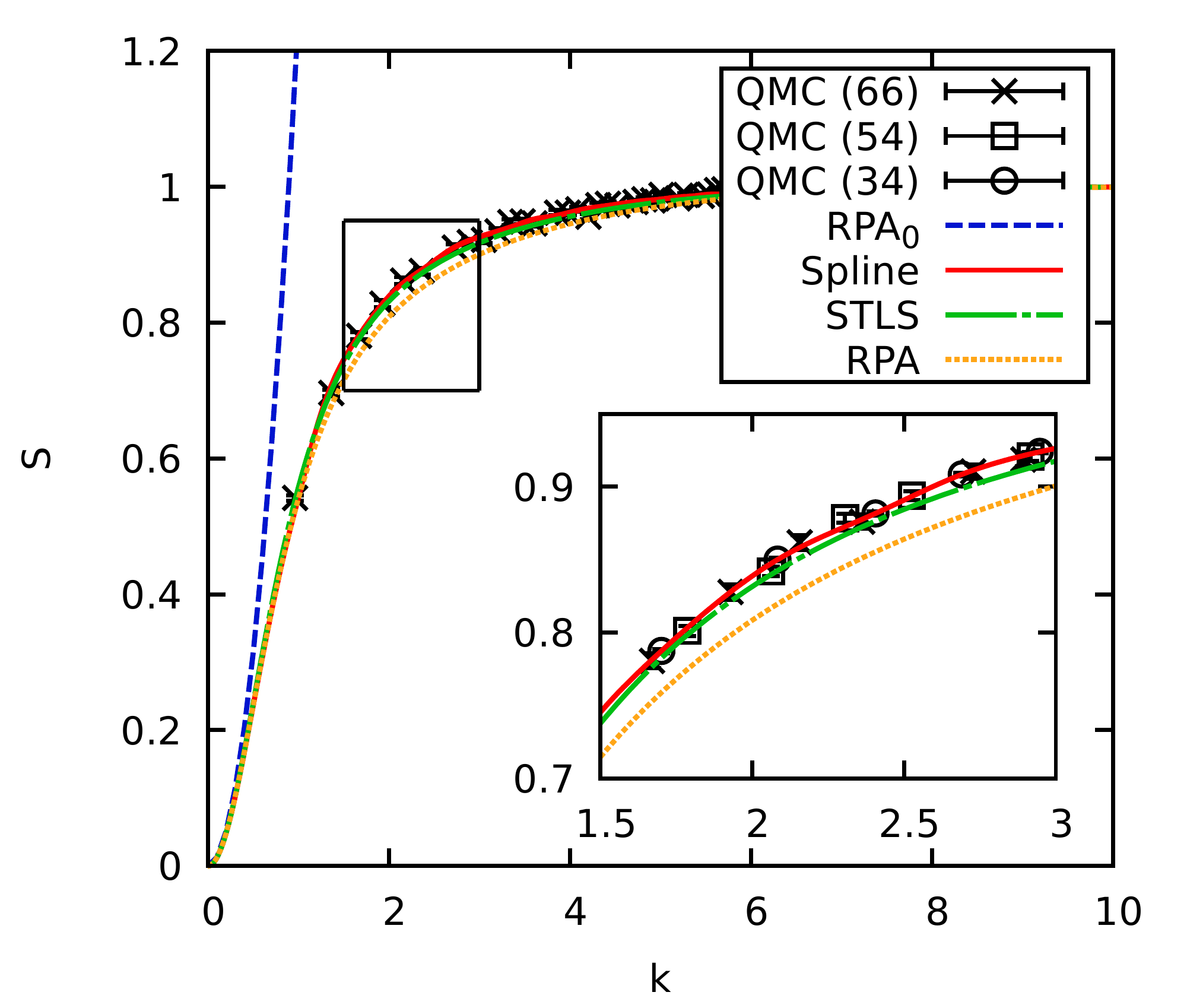}
\end{minipage}
\begin{minipage}{.02\textwidth}
$ $
\end{minipage}
\begin{minipage}{.47\textwidth}
\includegraphics[width=0.99\textwidth]{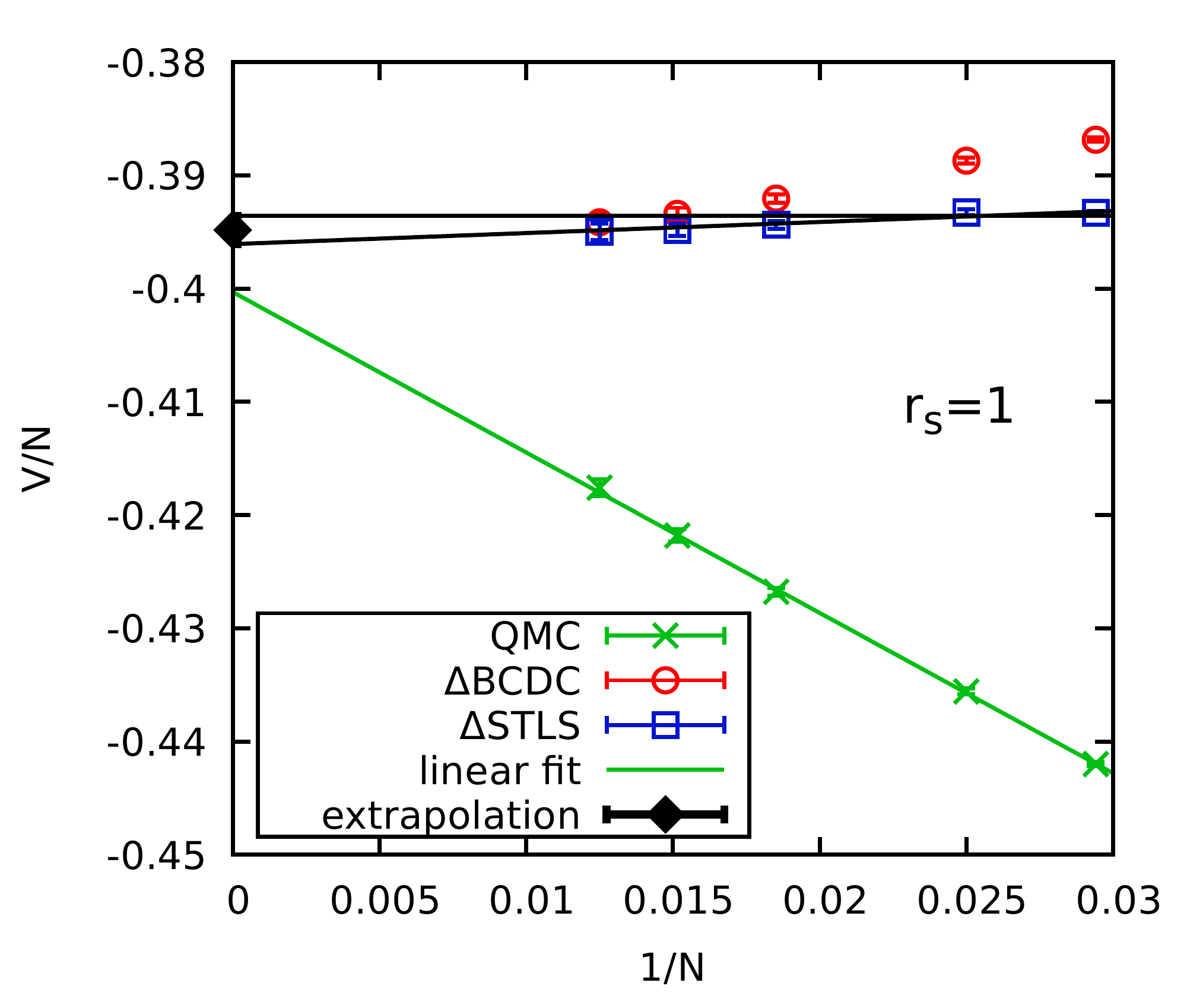}
\end{minipage}
\begin{minipage}{.45\textwidth}
\includegraphics[width=0.99\textwidth]{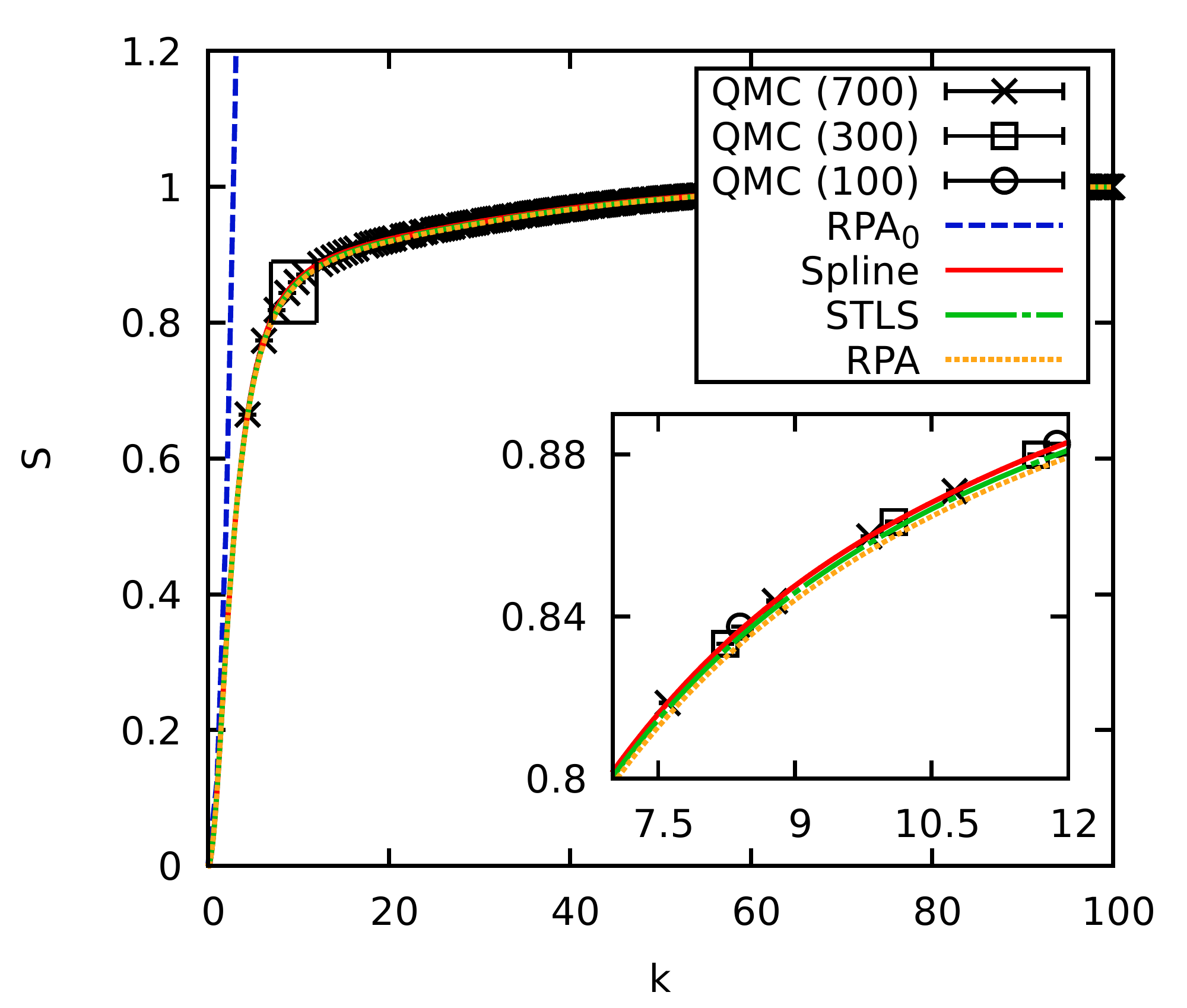}
\end{minipage}
\begin{minipage}{.03\textwidth}
$ $
\end{minipage}
\begin{minipage}{.45\textwidth}
\includegraphics[width=0.99\textwidth]{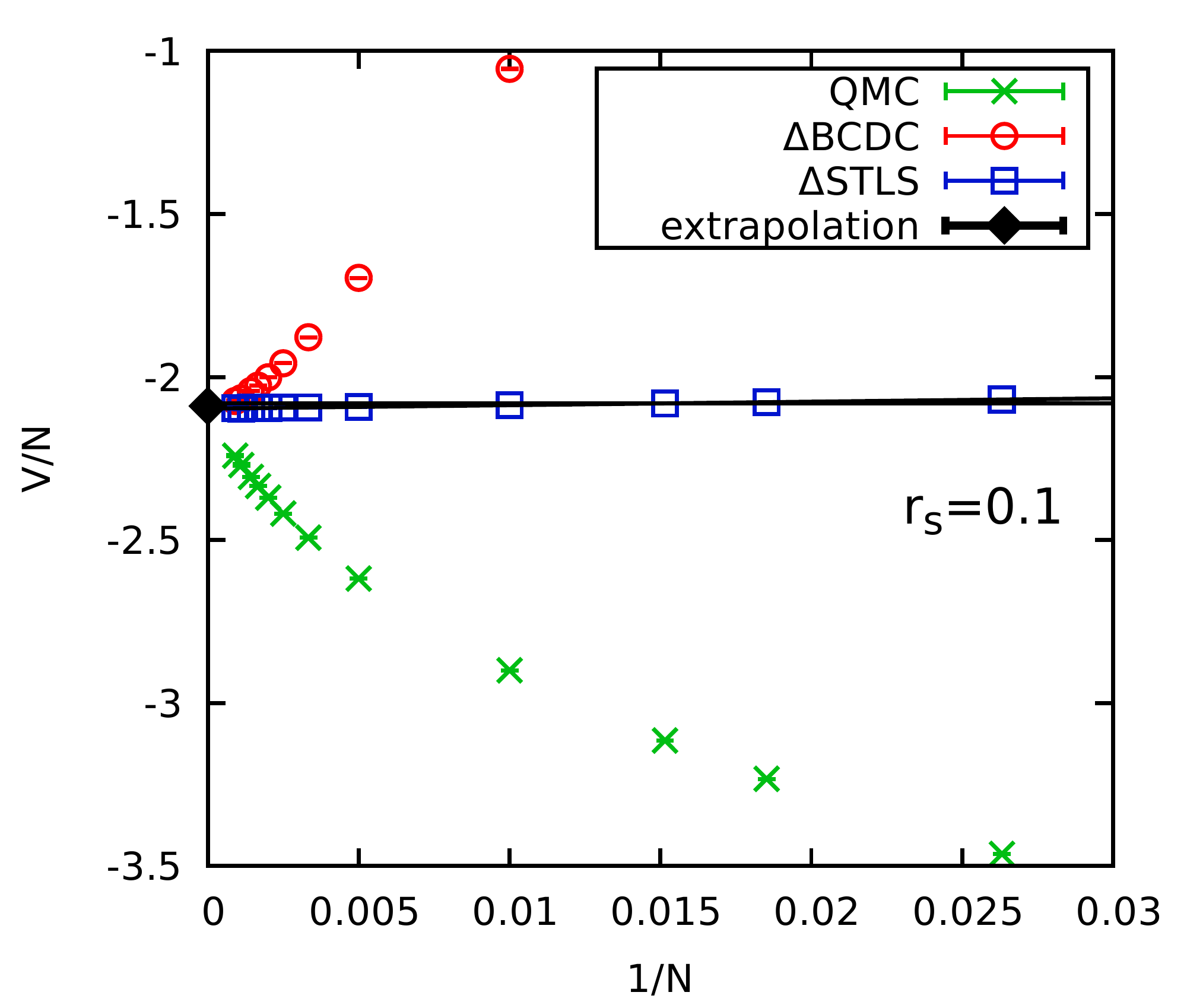}
\end{minipage}

\caption{\label{fig:FSC_panel_unpolarized}Coupling dependence of static structure factors (left) and interaction energies per particle (right) of the unpolarized electron gas at $\theta=2$ -- Top row: $r_s=10$, center row: $r_s=1$, bottom row: $r_s=0.1$. 
Shown are results for the static SF from QMC simulations with three different particle numbers (black symbols, the data for the two smallest $N$ appear in the inset only), the RPA expansion around $k=0$ (dashed blue), cf.~Eq.~(\ref{eq:S0}), and full RPA and STLS data (dotted yellow and dashed dotted green lines, respectively). The solid red line corresponds to a spline connecting STLS for small $k$ with QMC data elsewhere and the insets depict a magnified segment.
The interaction energies per particle correspond to the bare QMC results (green crosses), and finite-size corrected data using $\Delta_\textnormal{BCDC}$ (red circles) and the new improved FSC by Dornheim \textit{et al.}~\cite{dornheim_abinitio_2016-1} using $S_\textnormal{STLS}$ (blue squares). The solid black line corresponds to an extrapolation of the residual finite-size error and the black diamond depicts the extrapolated result for $V/N$ in the TDL.
}
\end{figure}

To demonstrate the universal applicability of the improved finite-size correction, in Fig.~\ref{fig:FSC_panel_unpolarized} we show results both for the static structure factor and the interaction energy per particle for the unpolarized UEG over three orders of magnitude of the coupling parameter $r_s$ at $\theta=2$.
In the top row, results are depicted for $r_s=10$, i.e., a relatively strongly coupled system. The left panel shows the static structure factor, where the QMC results for $N=140$ electrons are depicted by the black crosses. Furthermore, the dashed blue line corresponds to the parabolic RPA expansion around $k=0$ [see Eq.~(\ref{eq:S0})], the dash-dotted green and dotted yellow lines to the full STLS and RPA results, respectively, and the solid red line to the spline connecting STLS for small $k$ with QMC data elsewhere. For such parameters, QMC results for $S(k)$ range down to small $S$ and for $k_\textnormal{min}$ all depicted data sets -- even the RPA expansion -- are in excellent agreement. Therefore, the finite-size correction proposed by Brown \textit{et al.}~\cite{brown_path-integral_2013} is appropriate, cf.~the right panel.
Overall, we observe substantial errors in the RPA curve for intermediate $k$ starting around $k\gtrsim0.1$. The STLS curve is in much better agreement to the QMC data everywhere, although it is too large for $k\lesssim0.35$ and too small for larger $k$.
The inset shows a magnified segment where, in addition to the QMC data for $N=140$, we also show results for $N=80$ (squares) and $N=66$ (circles). Evidently, no system size dependence of $S_N(k)$ can be resolved within the given statistical uncertainty.
Let us now consider the interaction energy per particle, which is depicted as a function of $1/N$ in the right panel.
As usual, the green crosses correspond to the bare QMC results and, even for as few as $N=34$ electrons, the finite-size error does not exceed $\Delta V/V=1\%$. This can be explained by recalling the interpretation of finite-size effects as a discretization error in the integration of $S(k)$, which is densely sampled by the QMC points down to small values of $S$, cf.~the left panel.
Further, we note that the QMC points seem to exhibit a linear behavior as predicted by the BCDC-FSC, Eq.~(\ref{eq:BCDC}). Consequently, adding $\Delta_\textnormal{BCDC}$ to the QMC data (red circles) removes the finite-size error and no system size dependence can be resolved within the given statistical uncertainty.
Furthermore, we note that the improved FSC [Eq.~(\ref{eq:our_FSC})] using $S_\textnormal{STLS}$ as a model function leads to the same results.

In the center row, we show results for intermediate coupling, $r_s=1$. Here, in contrast to the previous case, the RPA expansion does clearly not connect to the QMC results, which are not available down to such small $S$-values as above. Furthermore, we note that both the full RPA and STLS curves exhibit much smaller deviation to the QMC data, as it is expected. In fact, the STLS curve is only seldom not within twice the statistical uncertainty of the QMC points. For completeness, we mention that again no difference between QMC data for different particle numbers can be resolved, see the inset.
The interaction energy per particle exhibits a rather peculiar behavior. First and foremost, we note that the finite-size error for $N=34$ is of the order of $10\%$ and, thus, larger than for the strong coupling case. Again this comes as no surprise when comparing the static structure factors and re-calling the discretization error. In addition, the bare QMC results seem to exhibit a linear dependence in $1/N$. This is further substantiated by a linear fit, cf.~the solid green line, which reproduces all points within error bars. Interestingly, however, the calculated slope is not equal to the BCDC prediction by Eq.~(\ref{eq:BCDC}). Consequently, the red circles exhibit a distinct system size dependence and are not in agreement with the linear extrapolation.
Finally, the improved FSC leads to significantly reduced finite-size errors, which we subsequently remove by an additional extrapolation as explained in the discussion of Fig.~\ref{fig:FSC_RPA_STLS}. The thusly obtained final result for the TDL significantly deviates from the linear extrapolation as well, which again demonstrates the problems with a direct extrapolation without knowing the exact functional form of the $N$-dependence.

Finally, in the bottom row we show results for $r_s=0.1$, which corresponds to weak coupling and high density. Even for as many as $N=700$ electrons, the QMC results are not available for the $k$-range where $S$ is small. Hence, the RPA expansion does come nowhere near the QMC point at $k_\textnormal{min}$ and the BCDC-FSC is not expected to work. Further, both the full RPA and STLS curves are in good agreement with the QMC data and each other over the entire $k$-range. Again, we note that $S_N(k)$ converges remarkably fast with system size, see the inset. 
The large value of $S_N(k)$ at $k_\textnormal{min}$ indicates that the wave vector range where $S$ varies most is not sampled sufficiently, or not accessed by QMC points at all. Consequently, the finite-size errors are substantially increased compared to $r_s=10$ and $r_s=1$ and, for $N=38$ particles, are comparable in magnitude to $V_N/N$ itself.
Furthermore, the BCDC-FSC is not useful and severly overestimates the discretization error. In particular, for $N\lesssim100$, the thusly 'corrected' data exhibit a larger system size dependence than the original bare QMC data.
The improved FSC computed from $S_\textnormal{STLS}$ again works remarkably well even for small $N$, and reduces the system-size dependence by two orders of magnitude.

\subsubsection{Temperature dependence  of the finite-size correction of QMC data}

\begin{figure}\centering
\begin{minipage}{.45\textwidth}
\includegraphics[width=0.99\textwidth]{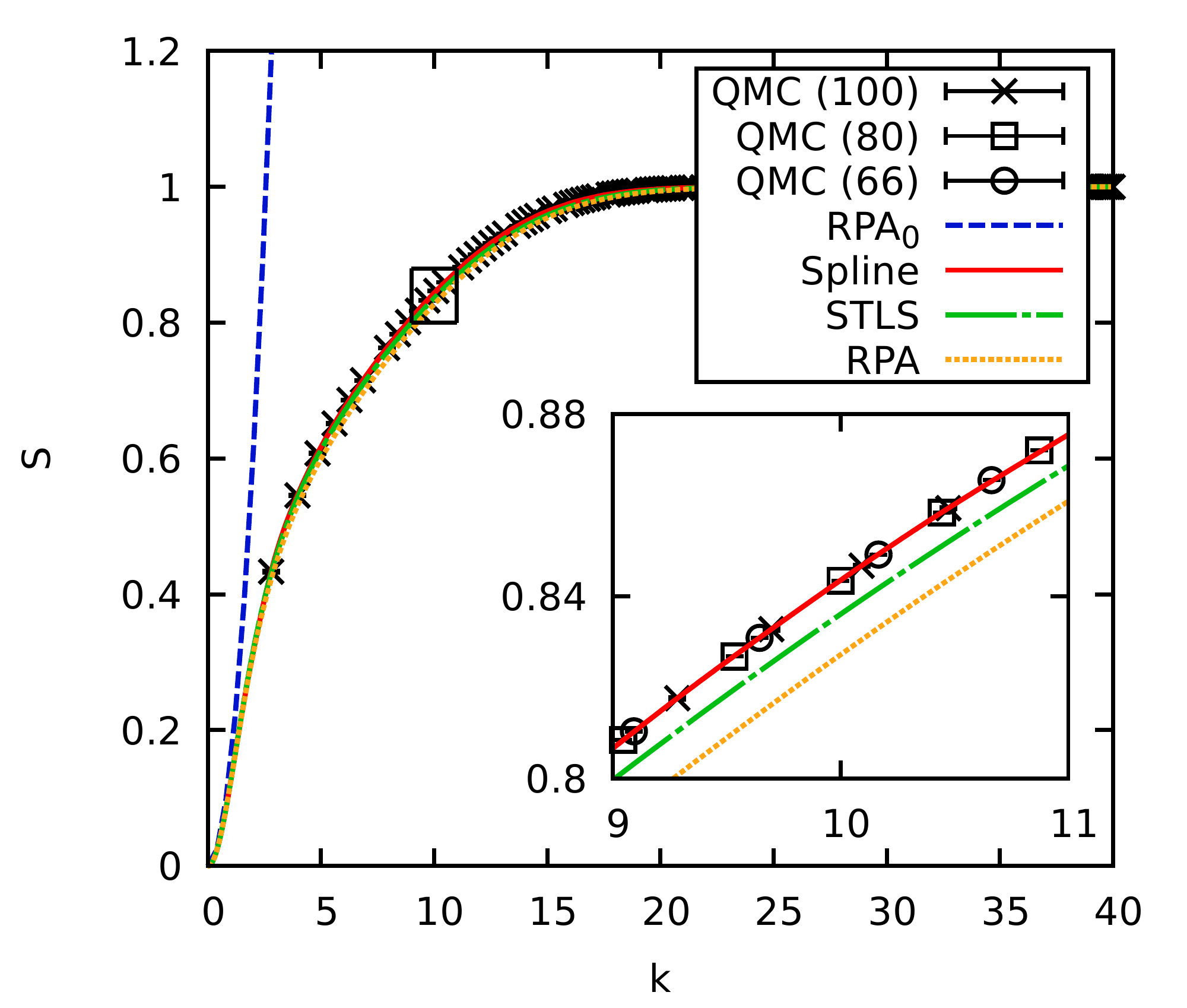}
\end{minipage}
\begin{minipage}{.02\textwidth}
$ $
\end{minipage}
\begin{minipage}{.47\textwidth}
\includegraphics[width=0.99\textwidth]{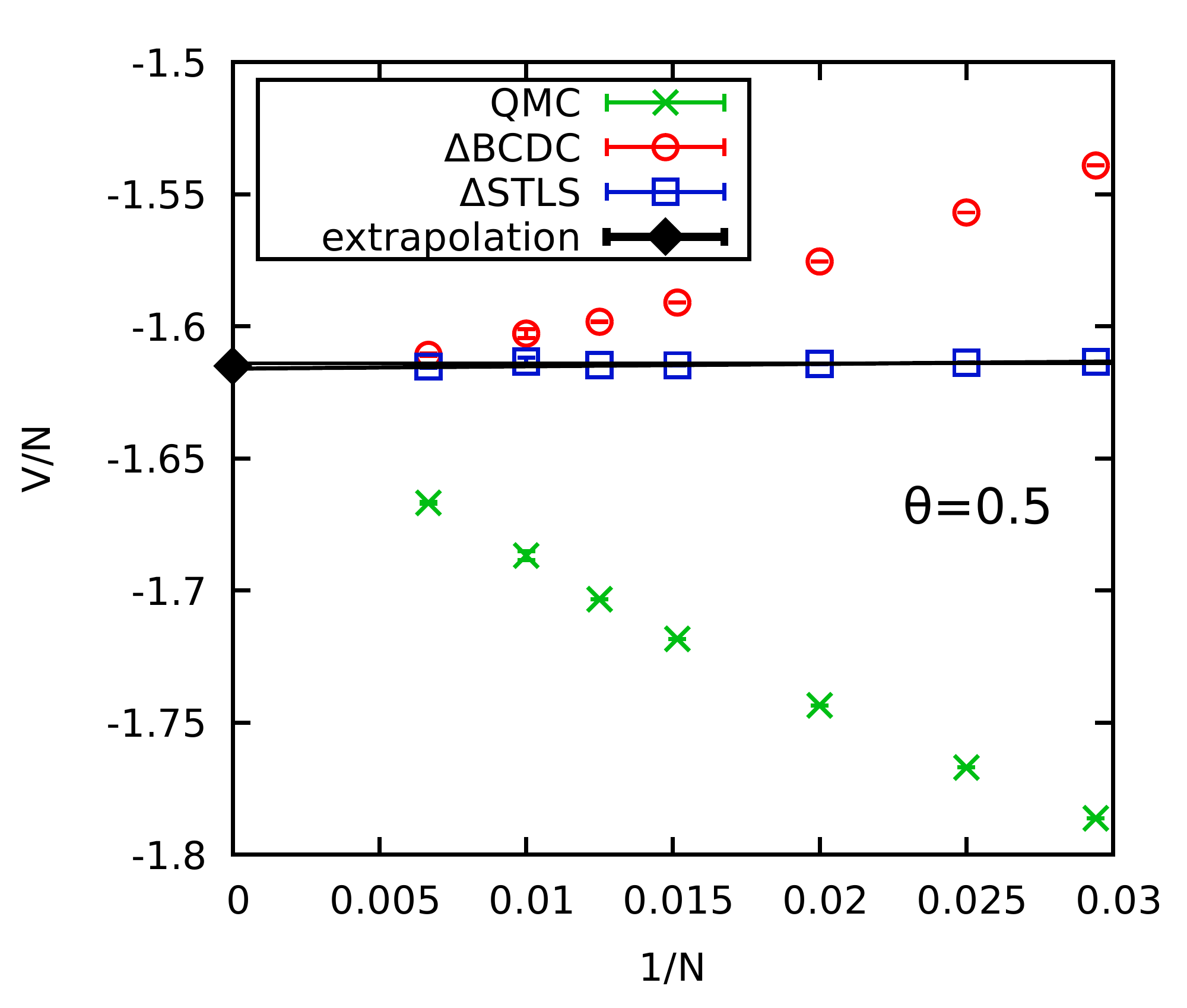}
\end{minipage}
\begin{minipage}{.45\textwidth}
\includegraphics[width=0.99\textwidth]{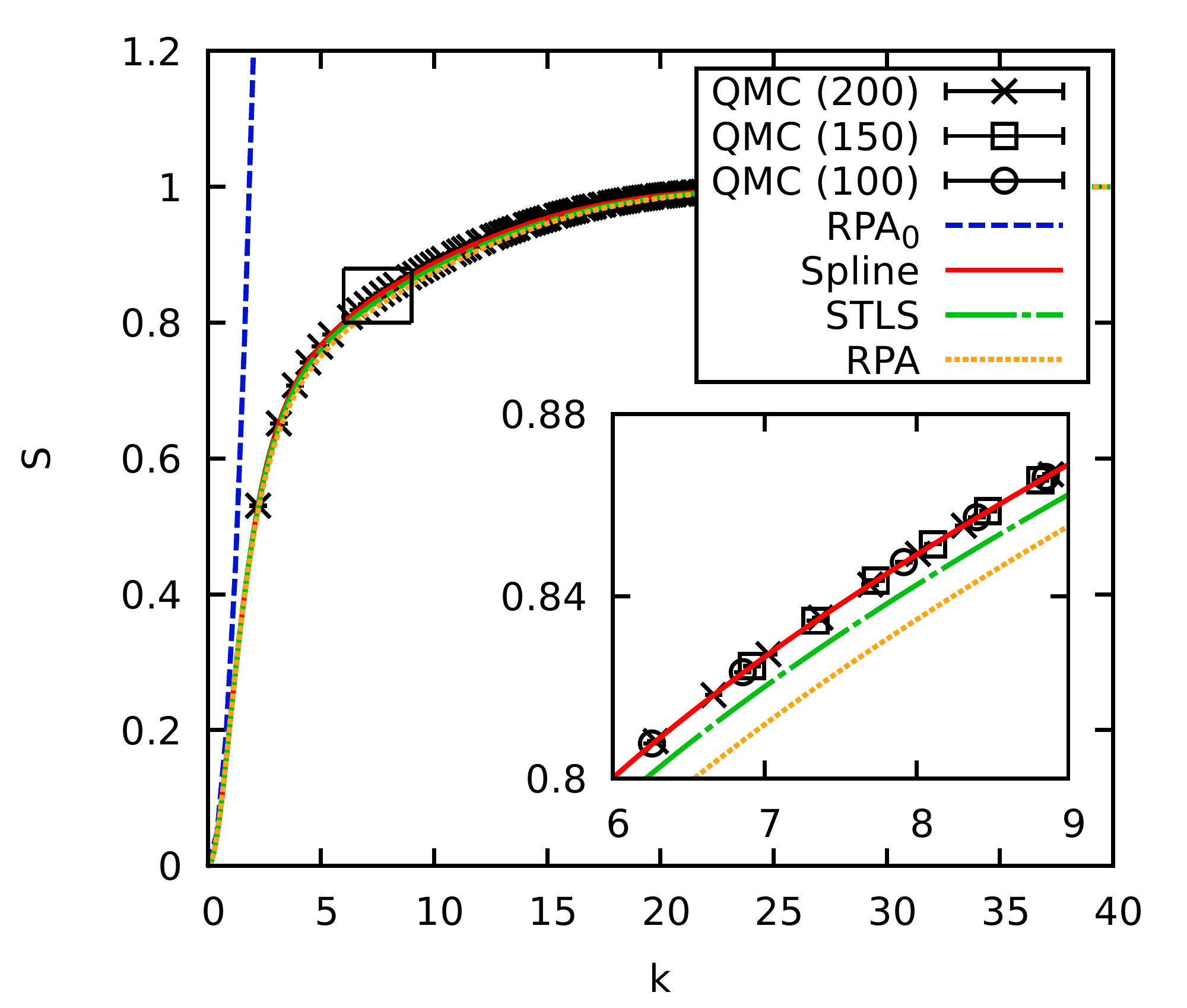}
\end{minipage}
\begin{minipage}{.02\textwidth}
$ $
\end{minipage}
\begin{minipage}{.47\textwidth}
\includegraphics[width=0.99\textwidth]{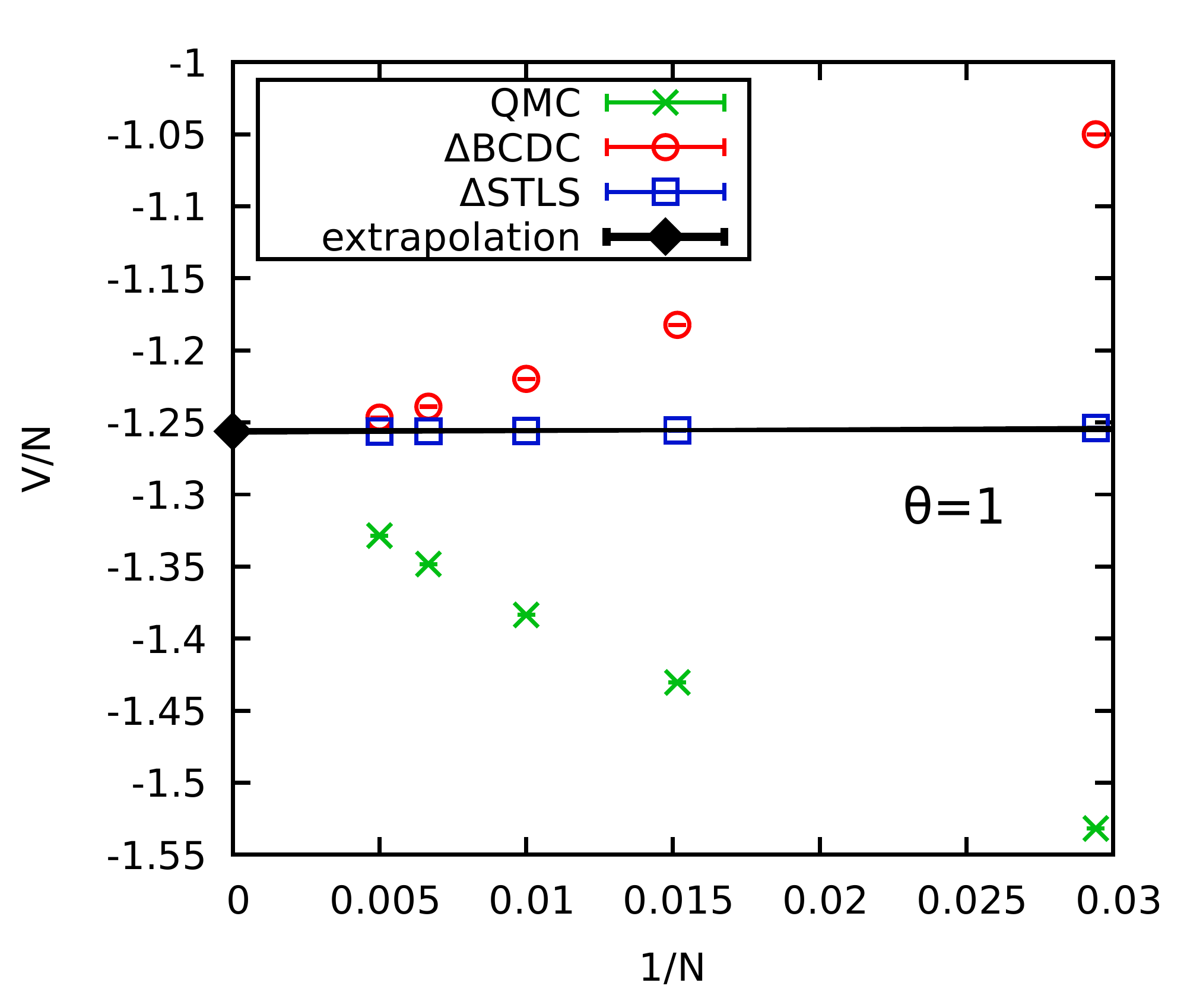}
\end{minipage}
\begin{minipage}{.45\textwidth}
\includegraphics[width=0.99\textwidth]{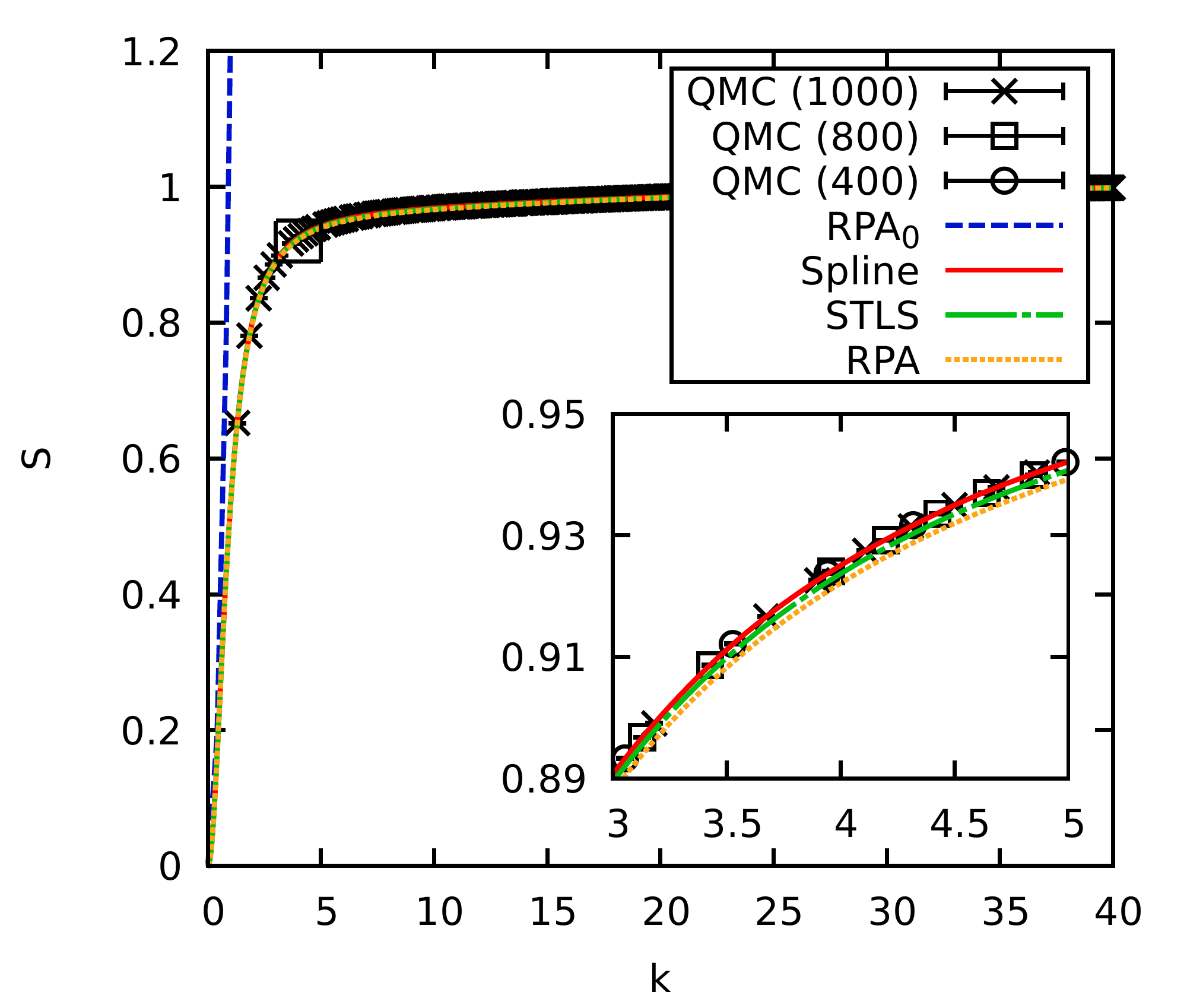}
\end{minipage}
\begin{minipage}{.03\textwidth}
$ $
\end{minipage}
\begin{minipage}{.46\textwidth}
\includegraphics[width=0.99\textwidth]{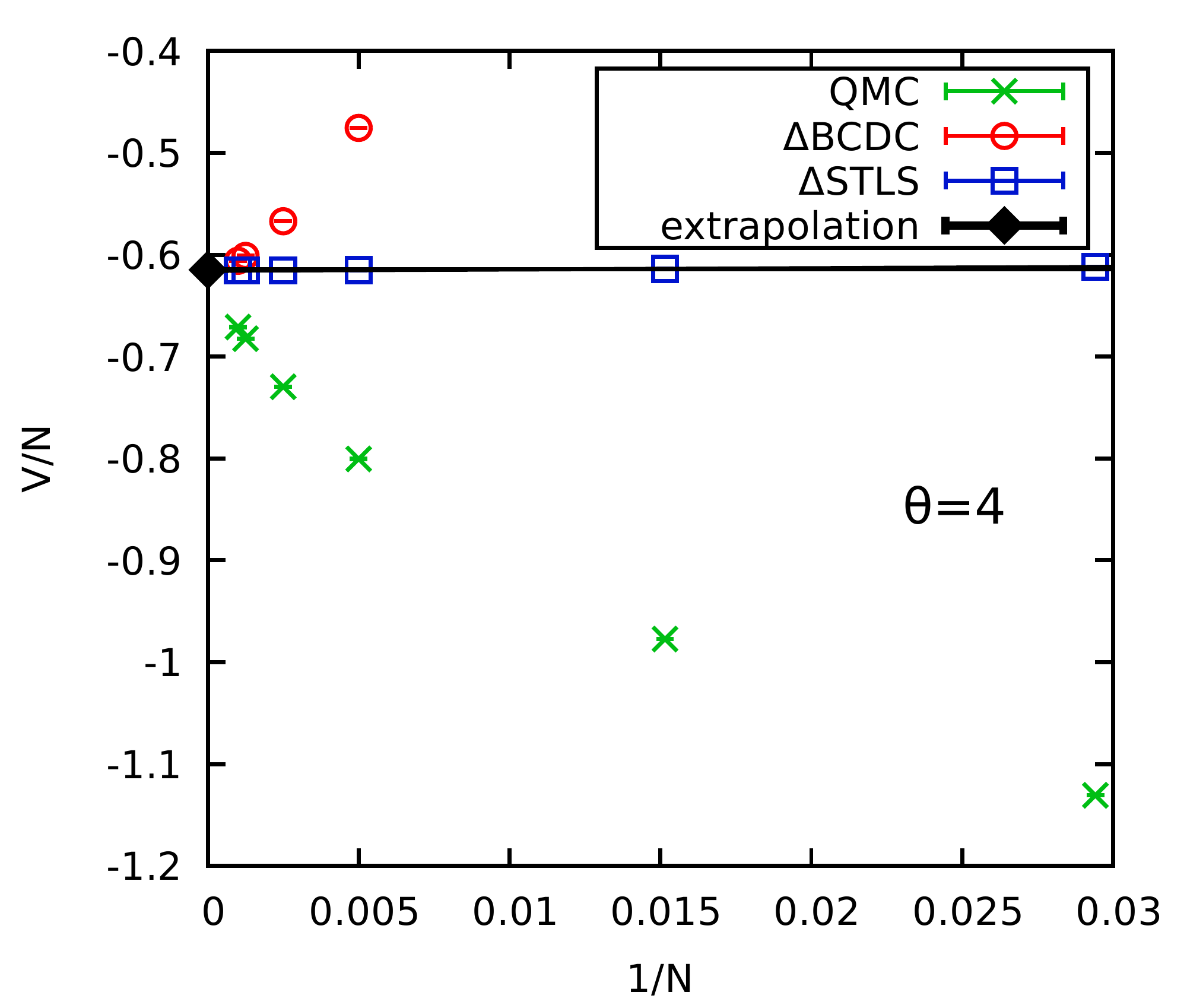}
\end{minipage}

\caption{\label{fig:FSC_panel_polarized}Temperature dependence of static structure factors (left) and interaction energies per particle (right) of the spin-polarized electron gas at $r_s=0.3$ -- Top row: $\theta=0.5$, center row: $\theta=1$, bottom row: $\theta=4$. 
Shown are results for the static SF from QMC simulations with three different particle numbers (black symbols, the data for the two smallest $N$ appear in the inset only), the RPA expansion around $k=0$ (dashed blue), cf.~Eq.~(\ref{eq:S0}), and full RPA and STLS data (dotted yellow and dashed dotted green lines, respectively). The solid red line corresponds to a spline connecting STLS for small $k$ with QMC data elsewhere and the insets depict a magnified segment.
The interaction energies per particle correspond to the bare QMC results (green crosses), and finite-size corrected data using $\Delta_\textnormal{BCDC}$ (red circles) and the new improved FSC from Ref.~\cite{dornheim_abinitio_2016-1} using $S_\textnormal{STLS}$ (blue squares). The solid black line corresponds to an extrapolation of the residual finite-size error and the black diamond depicts the extrapolated result for $V/N$ in the TDL.
}
\end{figure}

As a second demonstration of the versatility of the improved finite-size correction, in Fig.~\ref{fig:FSC_panel_polarized} we investigate the temperature dependence of the static structure factor and the interaction energy per particle of the spin-polarized UEG at $r_s=0.3$.
The top row shows results for $\theta=0.5$, which is the lowest temperature considered in the recent QMC simulations by Dornheim, Groth, and co-workers~\cite{dornheim_abinitio_2016-1,groth_ab_2017}. The QMC results for $S(k)$ range down to intermediate values of $S$, but do not connect to the RPA expansion. Further, we note that both the full RPA and STLS curves are in good agreement with each other and the QMC data over the entire $k$-range. As usual, the largest deviations occur for intermediate $k$ but are of the order of $0.1\%$. 
The bare QMC results for the interaction energy per particle seem to exhibit a linear behavior, but, similar to the observation in the center row of Fig.~\ref{fig:FSC_panel_unpolarized}, not with the slope predicted by Eq.~(\ref{eq:BCDC}). Consequently, adding the BCDC-FSC does not remove the system-size dependence, as expected from the discussion of the static structure factors. The improved FSC from Eq.~(\ref{eq:our_FSC}) using $S_\textnormal{STLS}$ as a model function to estimate the discretization error immediately improves the system size dependence by two orders of magnitude and no residual errors can be resolved with the naked eye.

The center and bottom rows show the same information for $\theta=1$ and $\theta=4$, respectively. First and foremost, we observe that the decline of $S(k)$ becomes steeper for increasing temperatures. This means that more QMC points are needed to accurately sample $S$, which, in turn, leads to increased discretization errors. In particular, for $\theta=4$ and $N=33$, the finite-size error is comparable in magnitude to $V_N/N$ itself, and, even for $N=1000$ electrons, no QMC results are available for $S\lesssim0.6$. Further, we note that both the full RPA and STLS results for the static structure factor become increasingly accurate for large $\theta$. This is, of course, expected as large temperatures render correlation effects less important. Finally, we mention that, while the BCDC-FSC becomes significantly less accurate, the improved FSC from Eq.~(\ref{eq:our_FSC}) works well for all temperatures (and densities).

\section{Benchmarks of other methods\label{sec:comparison}}

The improved finite-size correction introduced in this section has subsequently been used to obtain an exhaustive and very accurate data set for the interaction energy for different temperature-density combinations and four different spin-polarizations ($\xi=0$, $\xi=1/3$, $\xi=0.6$, and $\xi=1$), see Refs.~\cite{dornheim_abinitio_2016-1,groth_ab_2017}. This puts us, for the first time, in a position to gauge the accuracy of previously developed theories and approximations, most importantly that of the dielectric methods from Sec.~\ref{sec:LRT}.

\subsection{Benchmarks of the interaction energy}

\begin{figure}\centering
\begin{minipage}{.002\textwidth}
$ $
\end{minipage}
\begin{minipage}{.47\textwidth}
\includegraphics[width=0.99\textwidth]{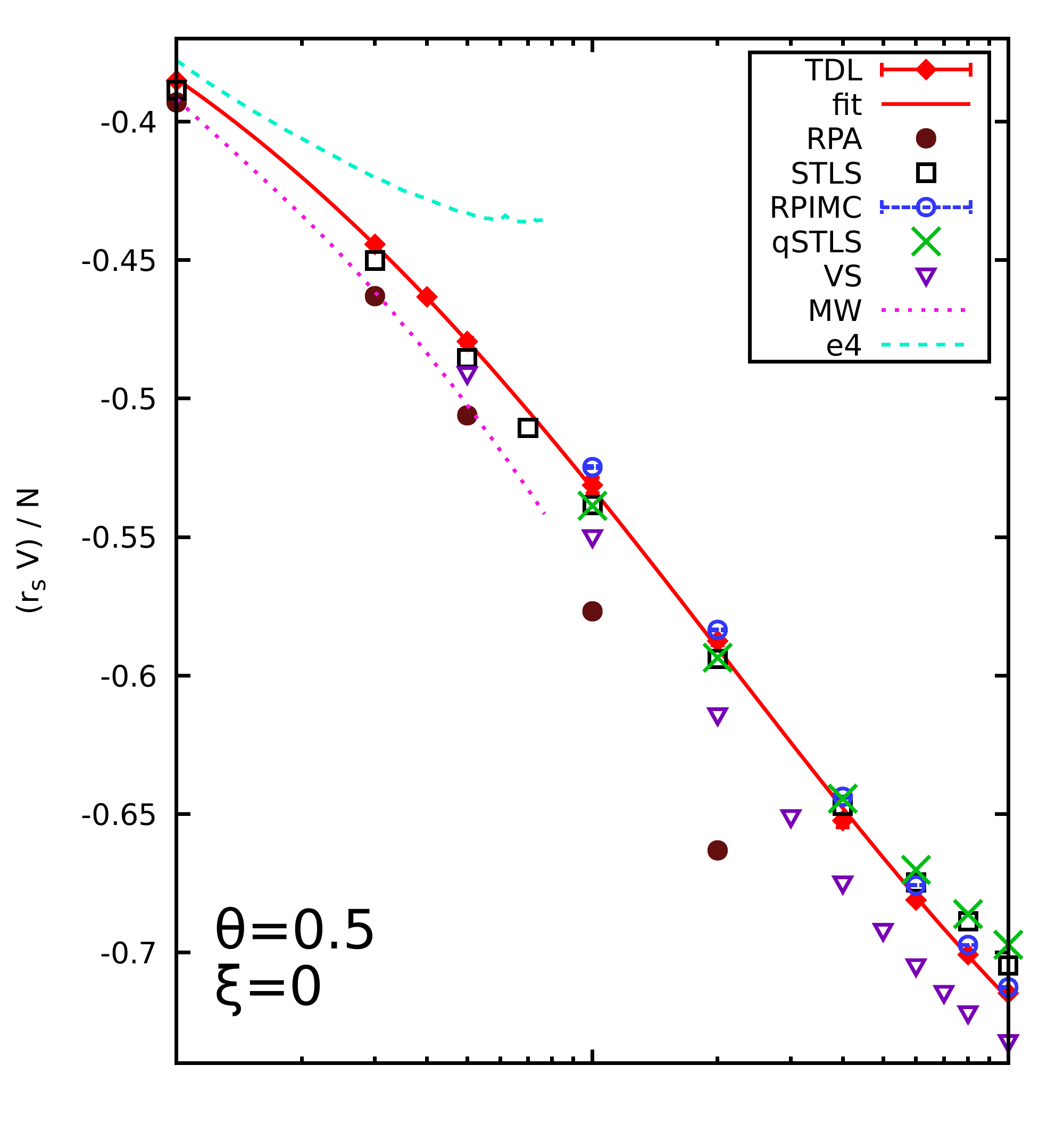}
\end{minipage}
\begin{minipage}{.02\textwidth}
$ $
\end{minipage}
\begin{minipage}{.47\textwidth}
\includegraphics[width=0.99\textwidth]{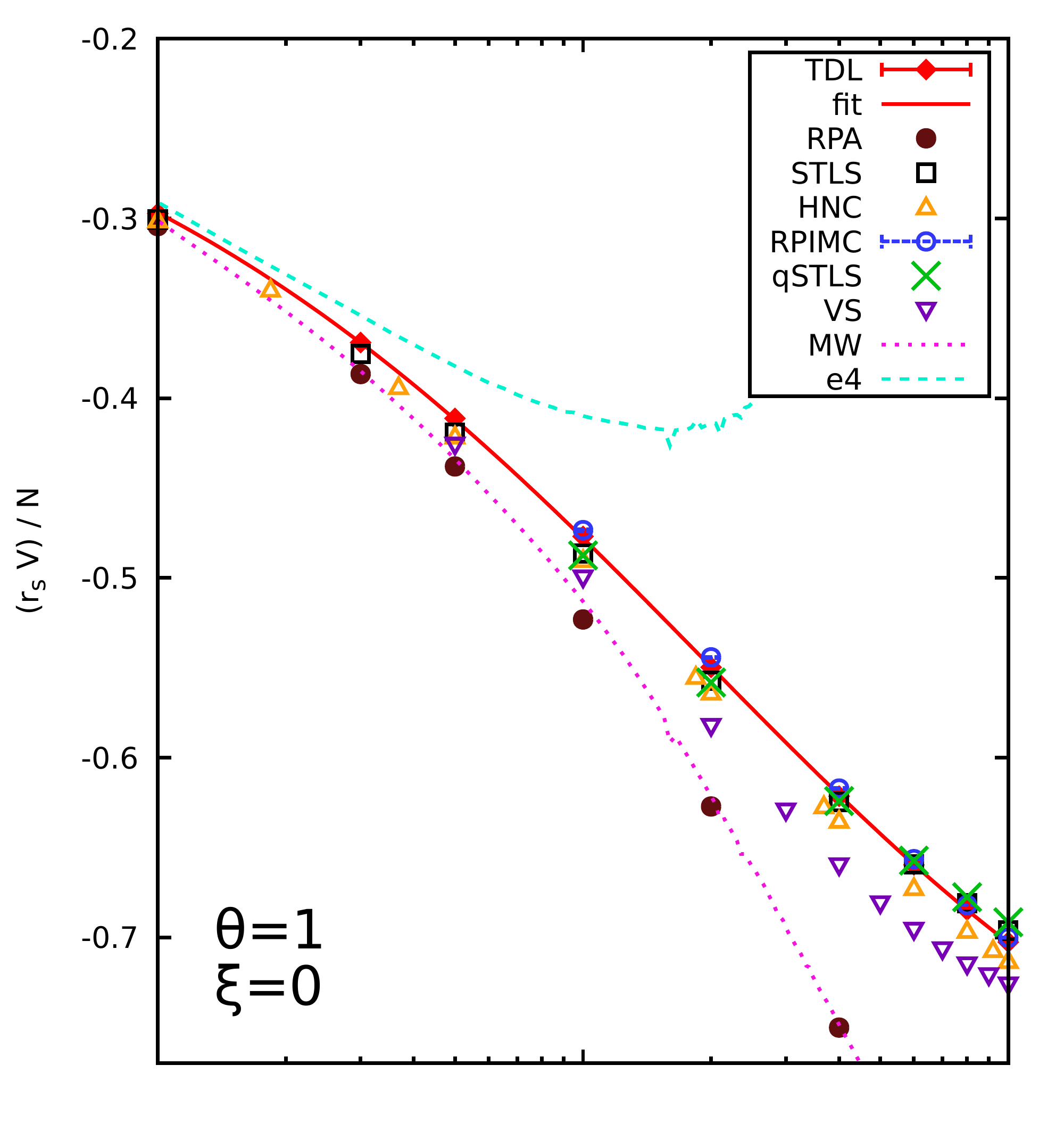}
\end{minipage}

\vspace*{-0.5cm}
\begin{minipage}{.00\textwidth}
$ $
\end{minipage}
\begin{minipage}{.47\textwidth}
\includegraphics[width=0.99\textwidth]{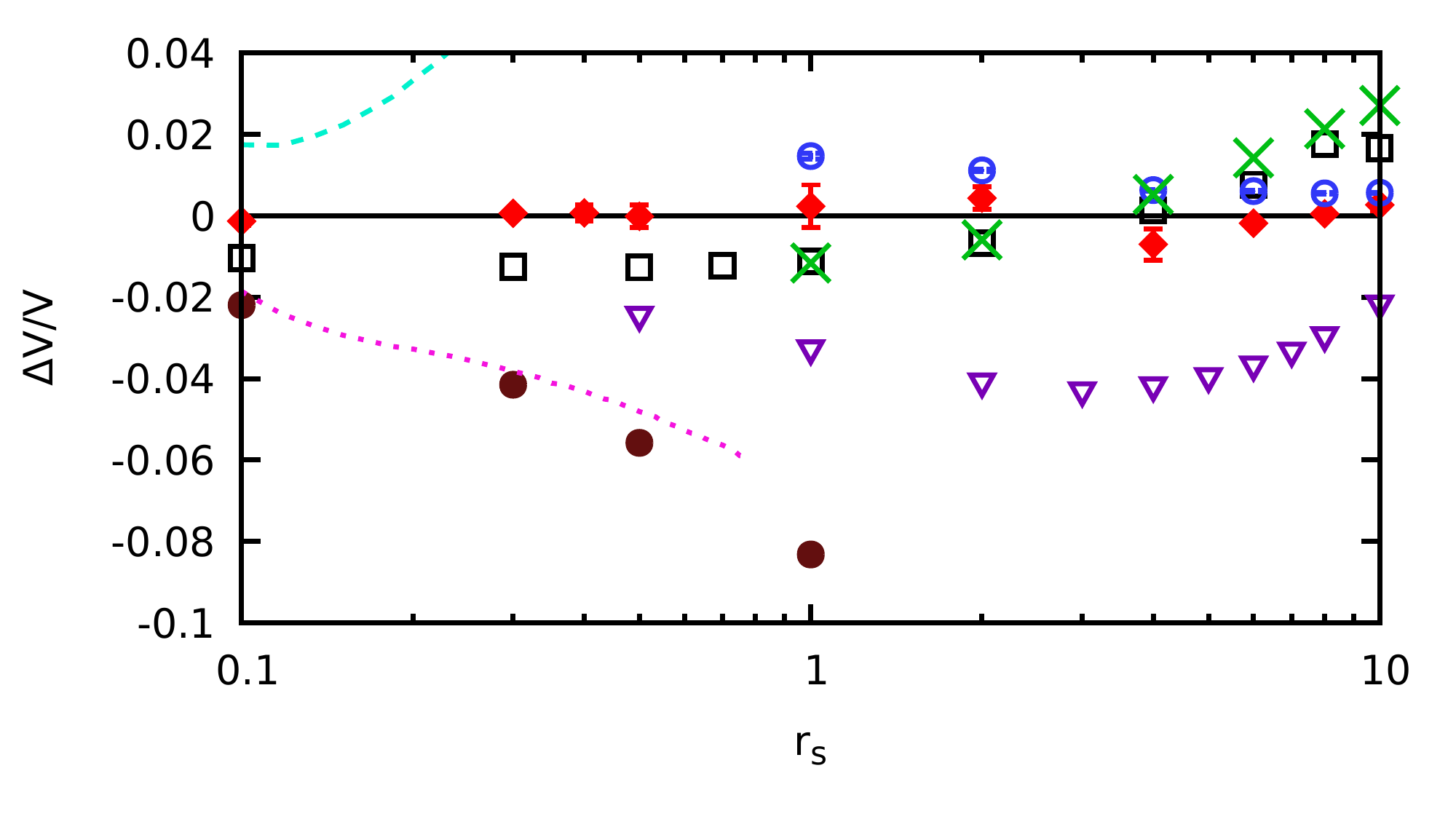}
\end{minipage}
\begin{minipage}{.01\textwidth}
$ $
\end{minipage}
\begin{minipage}{.48\textwidth}
\includegraphics[width=0.99\textwidth]{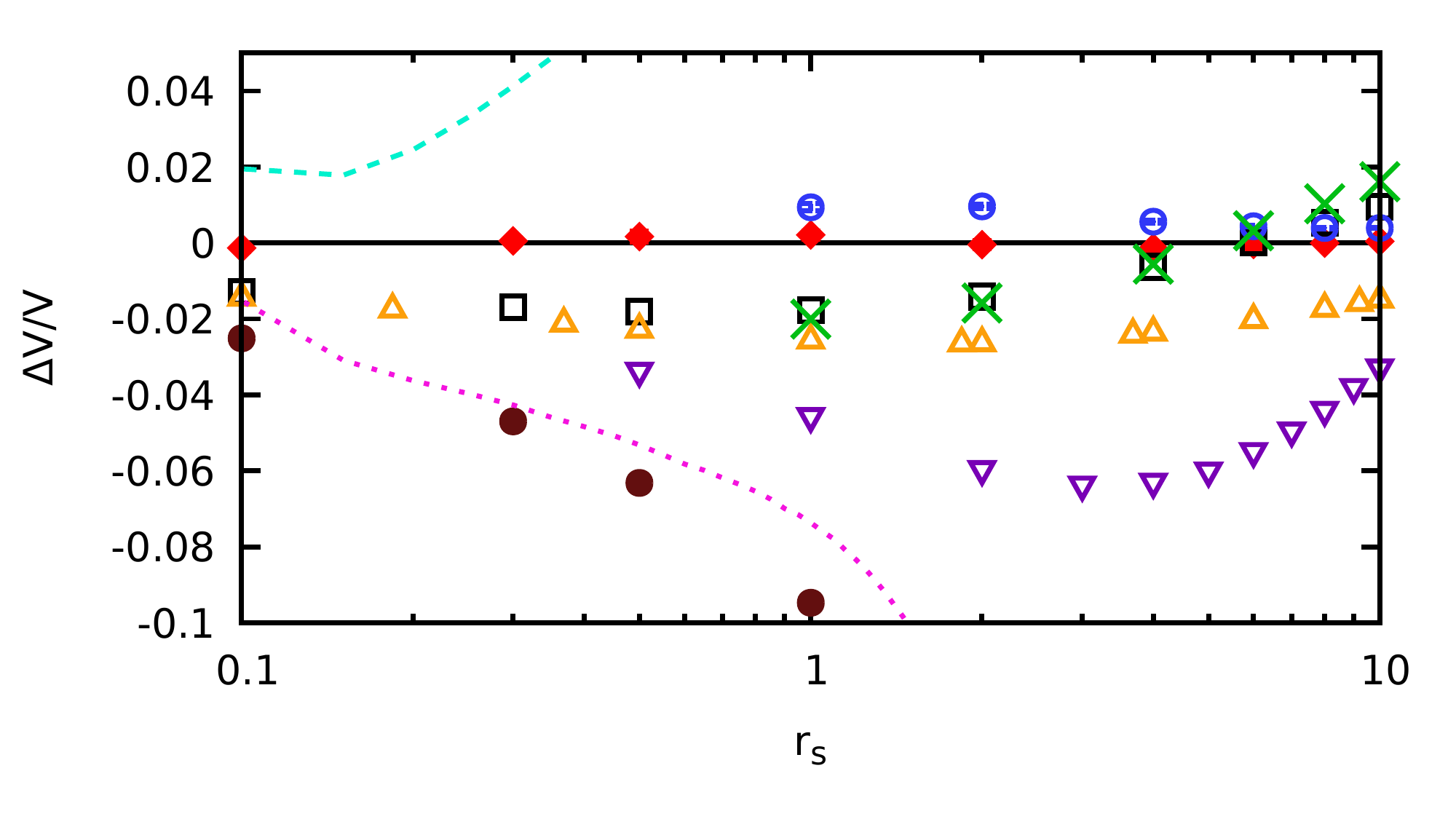}
\end{minipage}

\caption{\label{fig:comparison_interaction}Comparison of the interaction energies for the unpolarized electron gas at $\theta=0.5$ (left) and $\theta=1$ (right). The red diamonds correspond to the finite-size corrected QMC data by Dornheim, Groth and co-workers~\cite{dornheim_abinitio_2016-1} and the red lines depict fits to these data (see Ref.~\cite{dornheim_abinitio_2016-1}). Further shown are the RPIMC data by Brown \textit{et al.}~\cite{brown_path-integral_2013} (blue circles), finite-temperature Green function data computed in the Montroll-Ward (MW, dotted pink) and $e^4$-approximation (dashed light blue), cf.~Sec.~\ref{sec:other}, and various dielectric approximations, specifically RPA (brown dots), STLS (black squares), quantum STLS (green crosses, data obtained via integration of structure factors provided in Ref.~\cite{arora_spin-resolved_2017}), Vashista-Singwi (VS, purple downward triangles)~\cite{sjostrom_uniform_2013}, and the recent static local field correction based on the hypernetted chain (HNC) equation by Tanaka~\cite{tanaka_correlational_2016}. 
}
\end{figure}In Fig.~\ref{fig:comparison_interaction}, we show the $r_s$-dependence of the interaction energy per particle of the unpolarized electron gas at two relevant temperatures, $\theta=0.5$ (left) and $\theta=1$ (right).
The red diamonds correspond to our recent finite-size corrected QMC data and the solid red lines to simple fits at constant temperature $\theta$, see Ref.~\cite{dornheim_abinitio_2016-1} for details. 
Let us start our investigation by considering the most simple dielectric approach, i.e., the random phase approximation (brown dots). As expected, RPA only allows for a qualitative description at weak coupling, and even at extreme densities, $r_s=0.1$, there appear deviations exceeding $2\%$ in $v$. At moderate coupling, $r_s=1$, we find relative errors of $\Delta v/v\approx9\%$ for both depicted temperatures, indicating that RPA is of limited use for the description of electrons in the warm dense matter regime. The same applies for both depicted finite-temperature Green function data sets, where the Montroll-Ward approximation (MW, dotted pink line) closely follows RPA and the $e^4$-approximation exhibits a similar systematic error of the opposite sign (for more details on MW and $e^4$, see the Supplemental Material of Ref.~\cite{schoof_textitab_2015}).

Let us next consider the STLS approximation, both using the static (black squares) and dynamic (so-called quantum STLS or qSTLS, green crosses) versions of the local field correction. Obviously, this inclusion of correlation effects via $G(q)$ leads to a remarkable improvement in the interaction energy even up to relatively strong coupling, $r_s=10$. In particular, we find a maximum deviation of $\Delta v/v\approx 2\%$, which, for $\theta=1$, are most pronounced around $r_s=1$. This might seem surprising as the STLS closure relation for the LFC is expected to worsen towards increasing correlation effects. This is indeed the case both for the local field correction [and thus for the static density response function $\chi(q)$] as well as for the static structure factor. However, the interaction energy per particle is obtained from $S(k)$ via integration, cf.~Eq.~(\ref{eq:interaction_en_from_s}), and benefits from an error cancellation. For more details, see the investigation of the static structure factor in the next section.
Furthermore, we note that the inclusion of the frequency dependency of the STLS local field correction has only a minor effect on $v$ and even leads to slightly worse results compared to the static version introduced in Ref.~\cite{tanaka_thermodynamics_1986}.
At $\theta=1$, we can also investigate the performance of a recently introduced (static) local field correction that is based on the hypernetted chain equation~\cite{tanaka_correlational_2016}. The results for the interaction energy are shown as the yellow triangles in the right panel of Fig.~\ref{fig:comparison_interaction}. For weak coupling, $r_s<1$, the results are similar to both STLS versions, whereas for stronger coupling there appear differences between these dielectric methods of up to $\delta v/v=3\%$. However, while the SLTS results for $v$ intersect with the exact QMC results, the HNC data are always too low by up to $3\%$, making STLS the dielectric approximation of choice for the interaction energy. Again, this is in contrast to $S(k)$ and $G(k)$, where the new HNC-based formalism turns out to be superior, 
cf.~Figs.~\ref{fig:SSF_comparison_dielectric} and \ref{fig:LFC_compressibility}.
The purple downwards triangles correspond to the Vashista-Singwi formalism computed by Sjostrom and Dufty~\cite{sjostrom_uniform_2013}, which, for the present conditions, constitutes the least accurate dielectric approximation (excluding RPA) regarding $v$.
Finally, let us consider the restricted PIMC results by Brown \textit{et al.}~\cite{brown_path-integral_2013} (blue circles), which are available down to $r_s=1$. For the two depicted temperatures, these data are more accurate than the dielectric approximations with a maximum deviation of $\Delta v/v\approx1.5\%$ at $r_s=1$ and $\theta=0.5$.

\begin{figure}\centering
\begin{minipage}{.47\textwidth}
\includegraphics[width=0.99\textwidth]{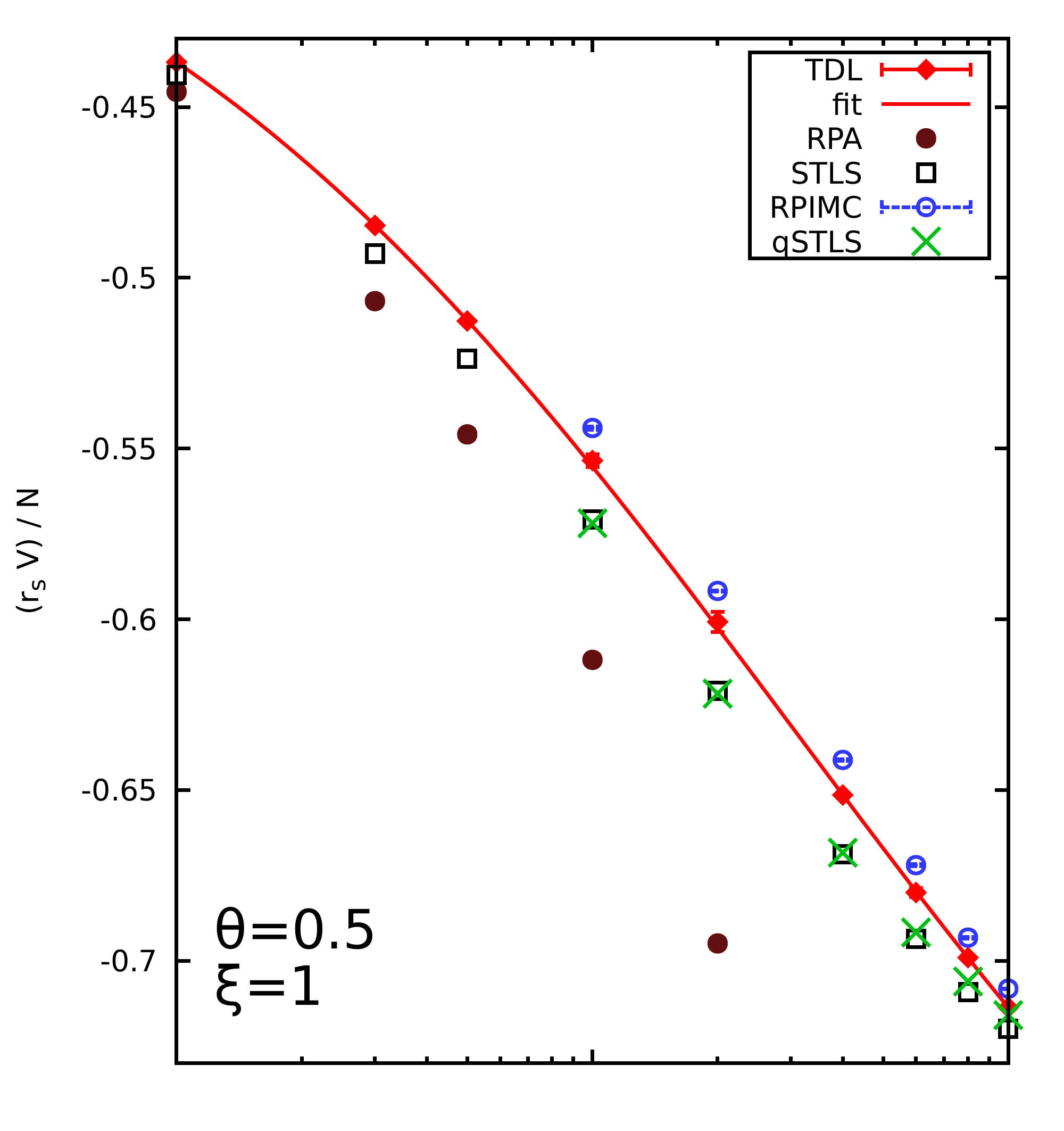}
\end{minipage}
\begin{minipage}{.02\textwidth}
$ $
\end{minipage}
\begin{minipage}{.47\textwidth}
\includegraphics[width=0.99\textwidth]{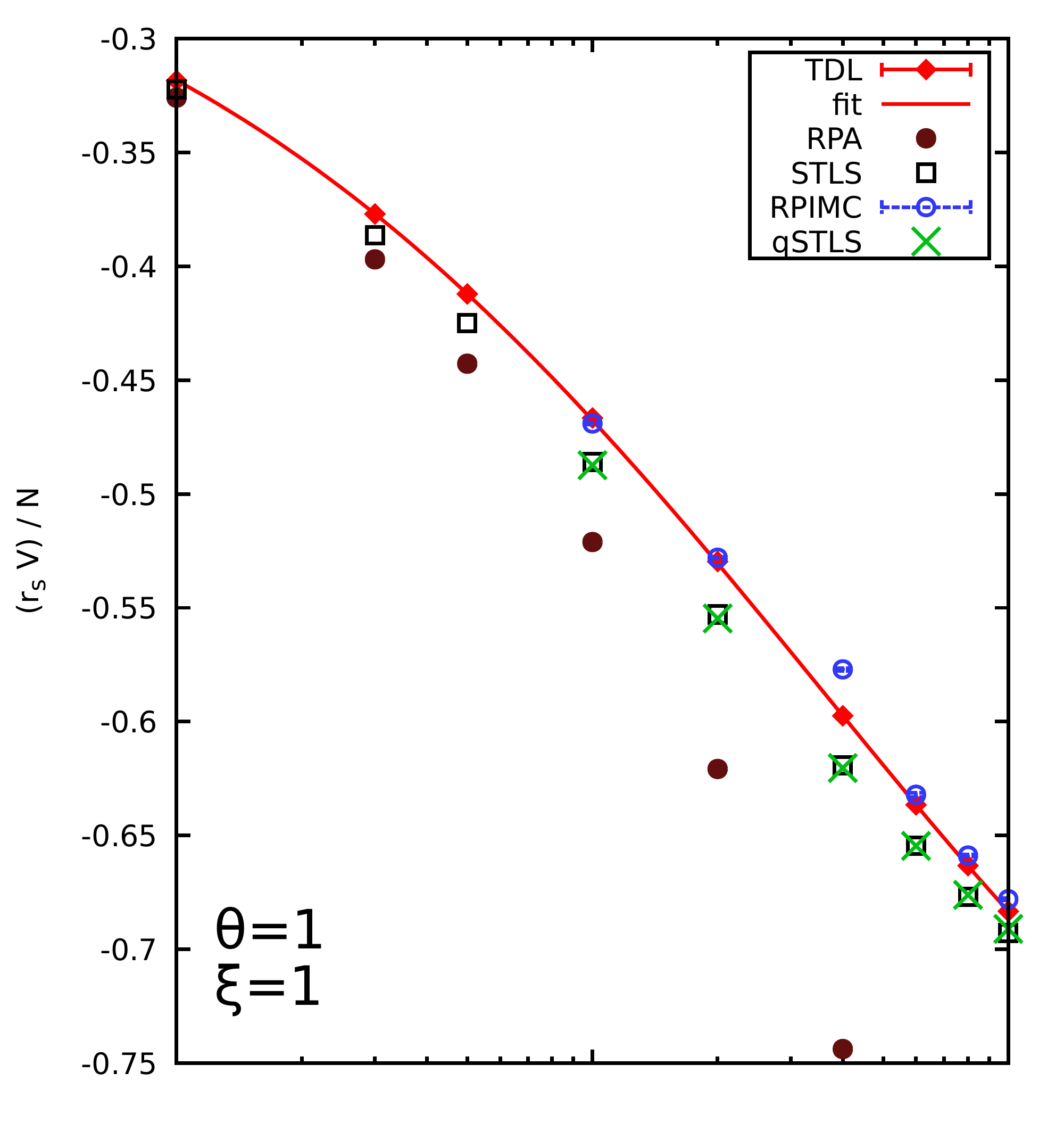}
\end{minipage}

\vspace*{-0.5cm}

\begin{minipage}{.47\textwidth}
\includegraphics[width=0.99\textwidth]{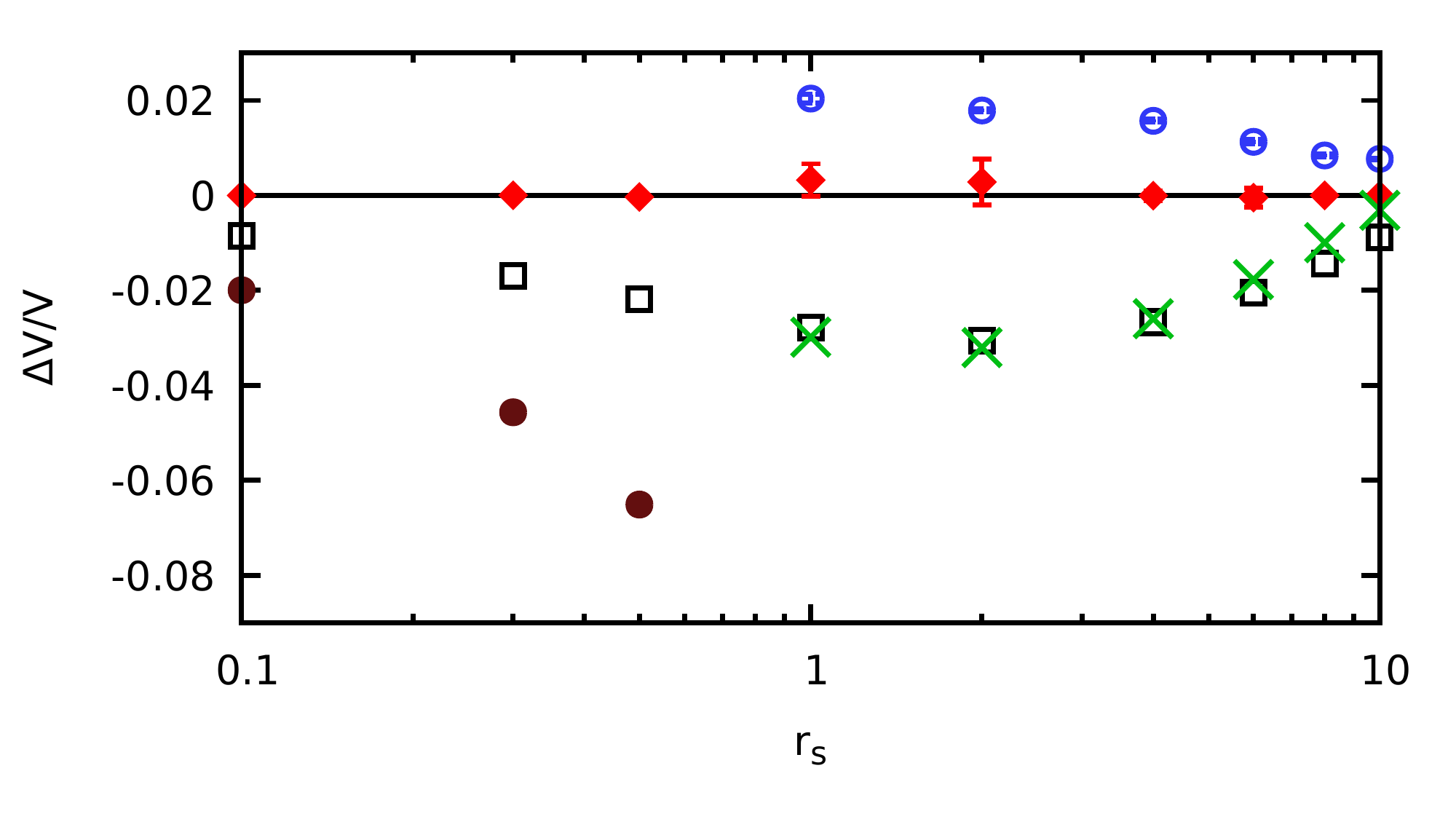}
\end{minipage}
\begin{minipage}{.02\textwidth}
$ $
\end{minipage}
\begin{minipage}{.47\textwidth}
\includegraphics[width=0.99\textwidth]{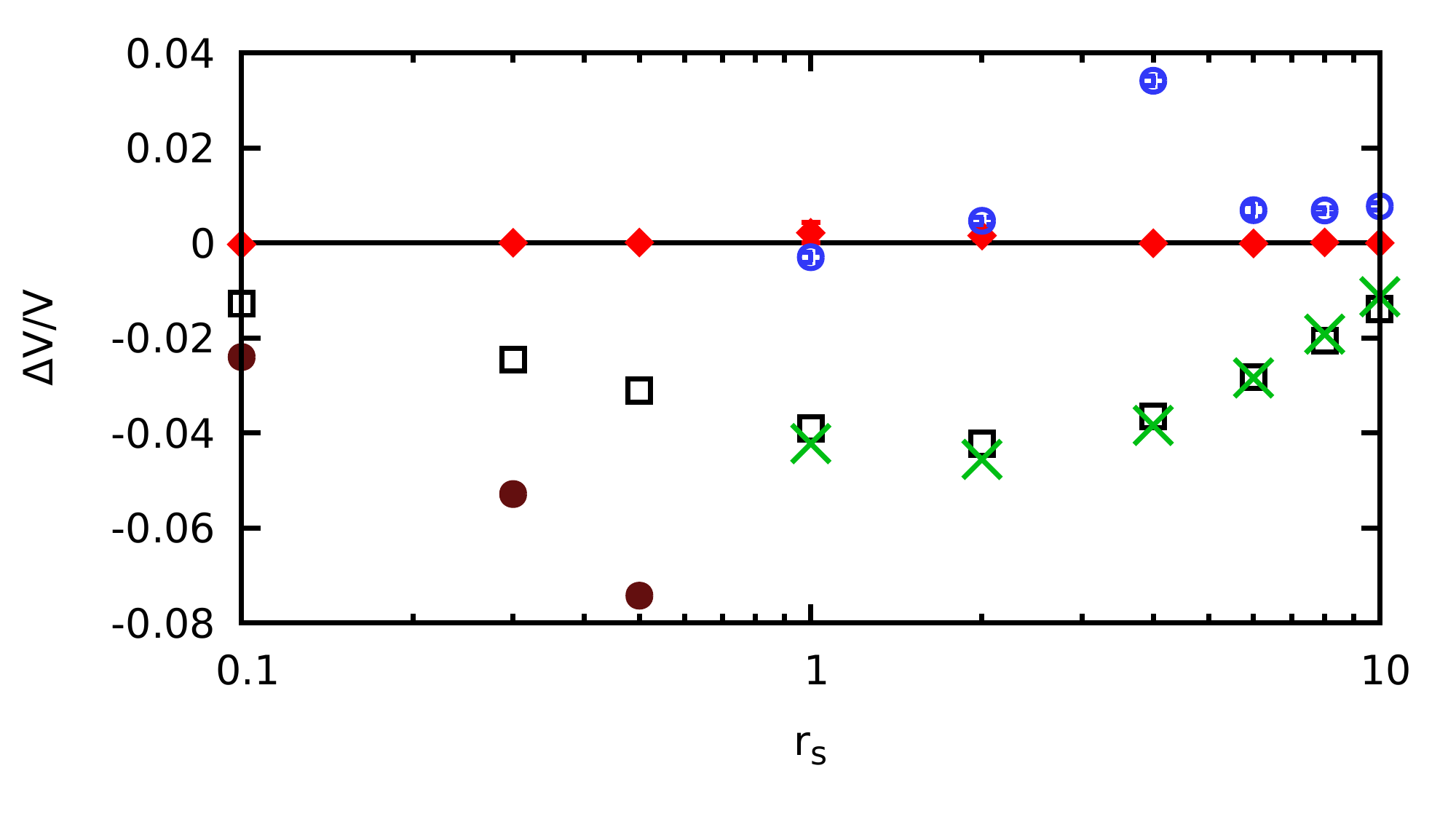}
\end{minipage}

\caption{\label{fig:pol_comparison_interaction}Gauging the accuracy of the interaction energy (per particle) of different approximations for the spin-polarized electron gas at $\theta=0.5$ (left) and $\theta=1$ (right). The red diamonds correspond to the finite-size corrected QMC data by Groth, Dornheim and co-workers~\cite{groth_ab_2017} and the red lines depict corresponding fits to these data (see the Supplemental Material of Ref.~\cite{dornheim_abinitio_2016-1} for more details). Further shown are the RPIMC data by Brown \textit{et al.}~\cite{brown_path-integral_2013} (blue circles) and various dielectric approximations, specifically RPA (brown dots), STLS (black squares), and quantum STLS (green crosses, data obtained via integration of structure factors provided in Ref.~\cite{arora_spin-resolved_2017}).
}
\end{figure}

Next, we consider the spin-polarized case, which is shown in Fig.~\ref{fig:pol_comparison_interaction}.
While RPA turns out to be similarly inaccurate as for the unpolarized case, we find a slightly worse performance of both STLS variants in this case. In particular, there appear maximum deviations of around $\Delta v/v=4\%$ at $r_s=2$, and the curves do not intersect with the exact results. Again, both STLS and qSTLS lead to almost indistinguishable results in the interaction energy, although at $\xi=1$ the qSTLS is slightly superior to STLS at large $r_s$.
The RPIMC data from Ref.~\cite{brown_path-integral_2013} are also slightly worse with a maximum deviation of $\Delta v/v\approx3.5\%$ at $r_s=4$ and $\theta=1$. In fact, this point constitutes an outlier, which has already been reported for the investigation of the finite model 
system~\cite{dornheim_permutation_2015-1}.

\begin{figure}\centering
\begin{minipage}{.002\textwidth}
$ $
\end{minipage}
\begin{minipage}{.47\textwidth}
\includegraphics[width=0.99\textwidth]{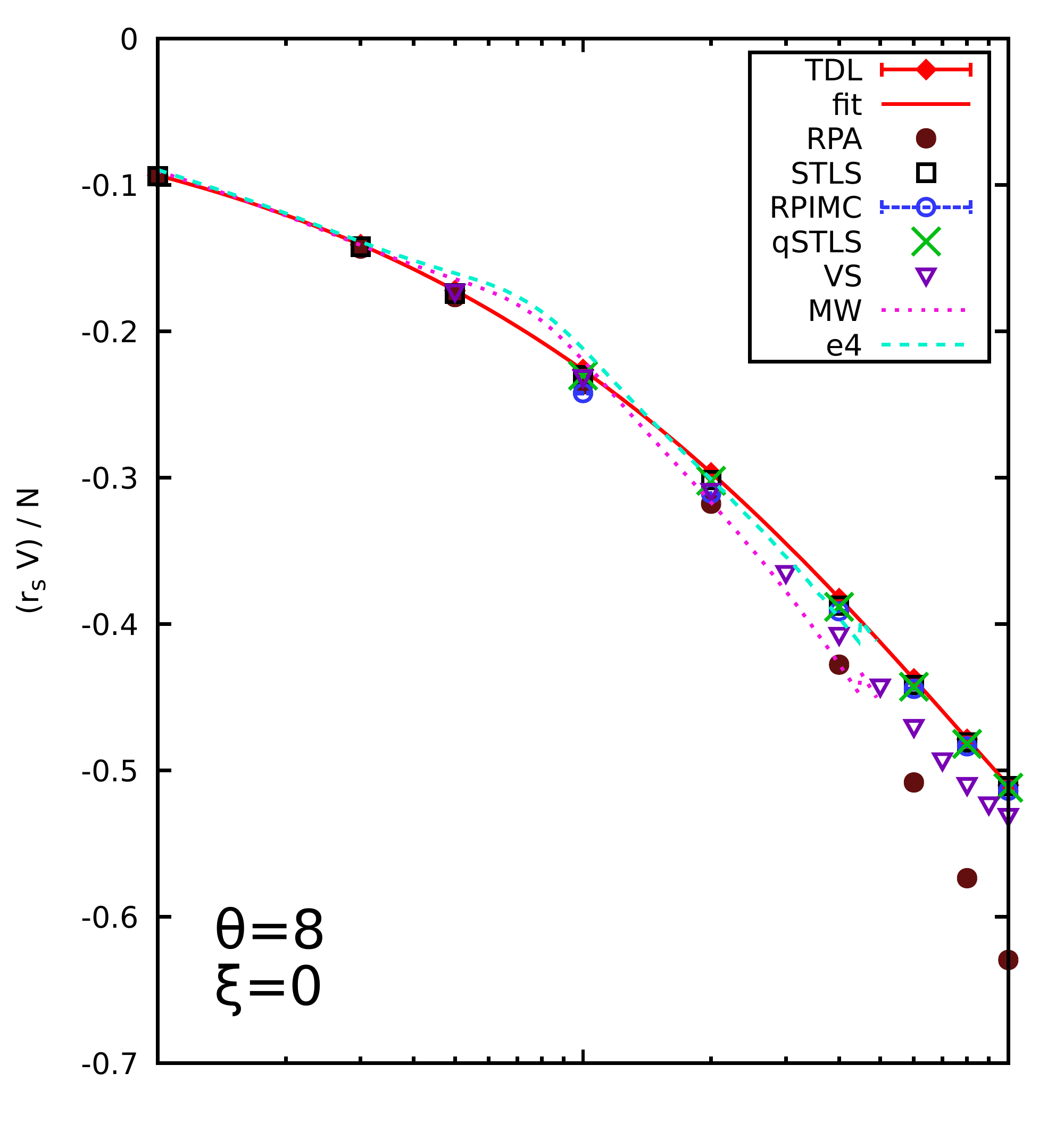}
\end{minipage}
\begin{minipage}{.02\textwidth}
$ $
\end{minipage}
\begin{minipage}{.47\textwidth}
\includegraphics[width=0.99\textwidth]{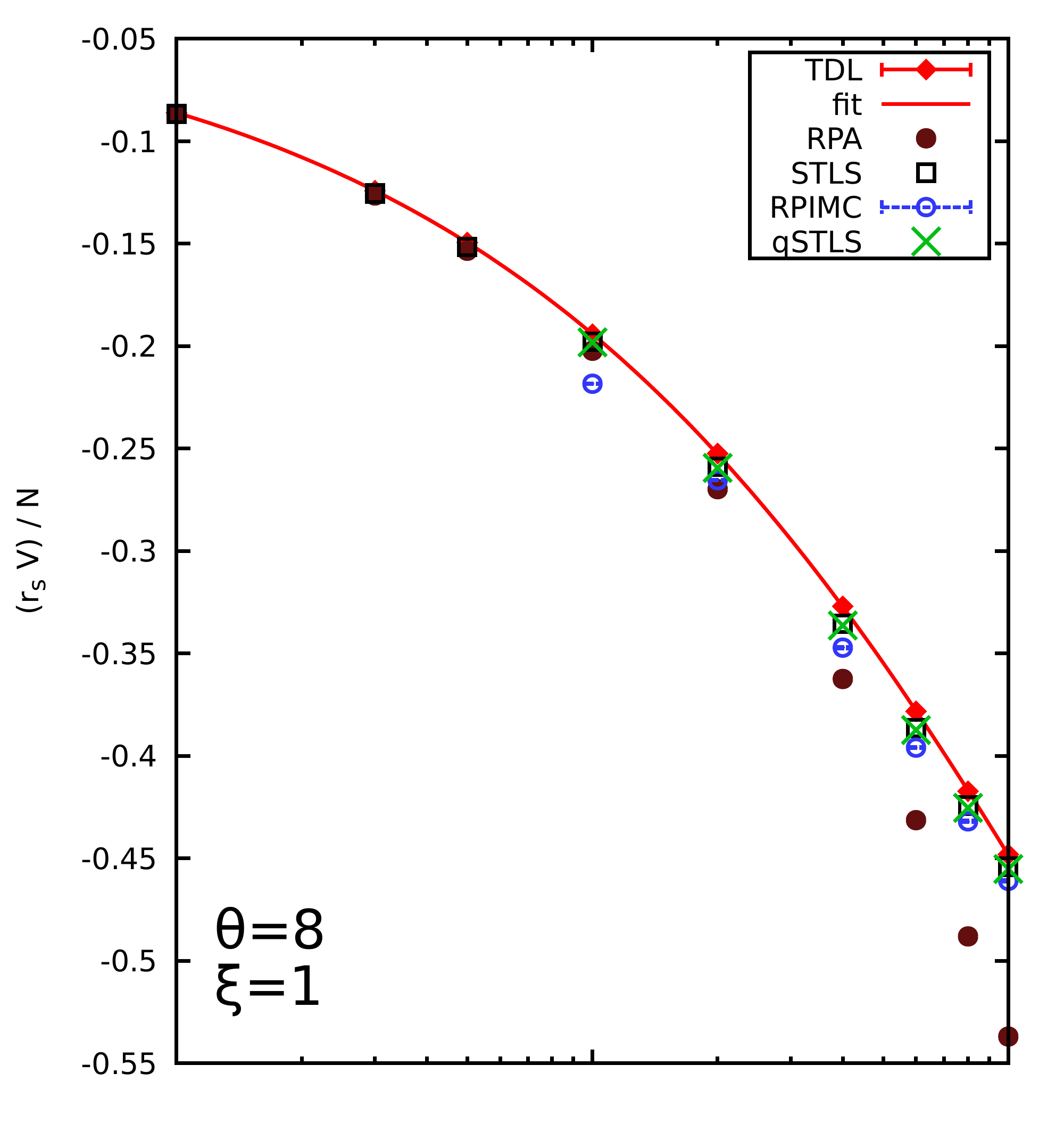}
\end{minipage}

\vspace*{-0.5cm}

\begin{minipage}{.48\textwidth}
\includegraphics[width=0.99\textwidth]{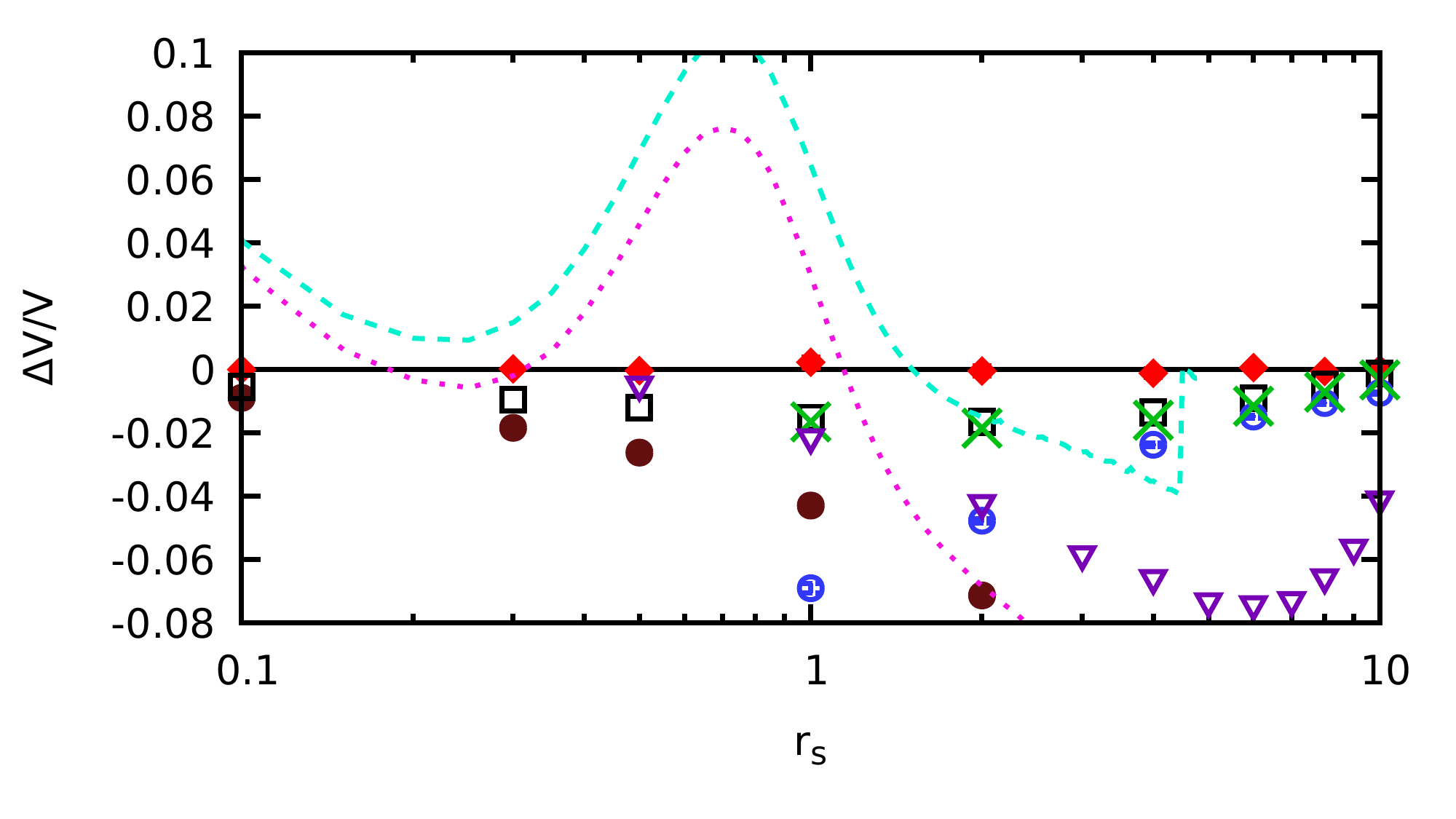}
\end{minipage}
\begin{minipage}{.02\textwidth}
$ $
\end{minipage}
\begin{minipage}{.47\textwidth}
\includegraphics[width=0.99\textwidth]{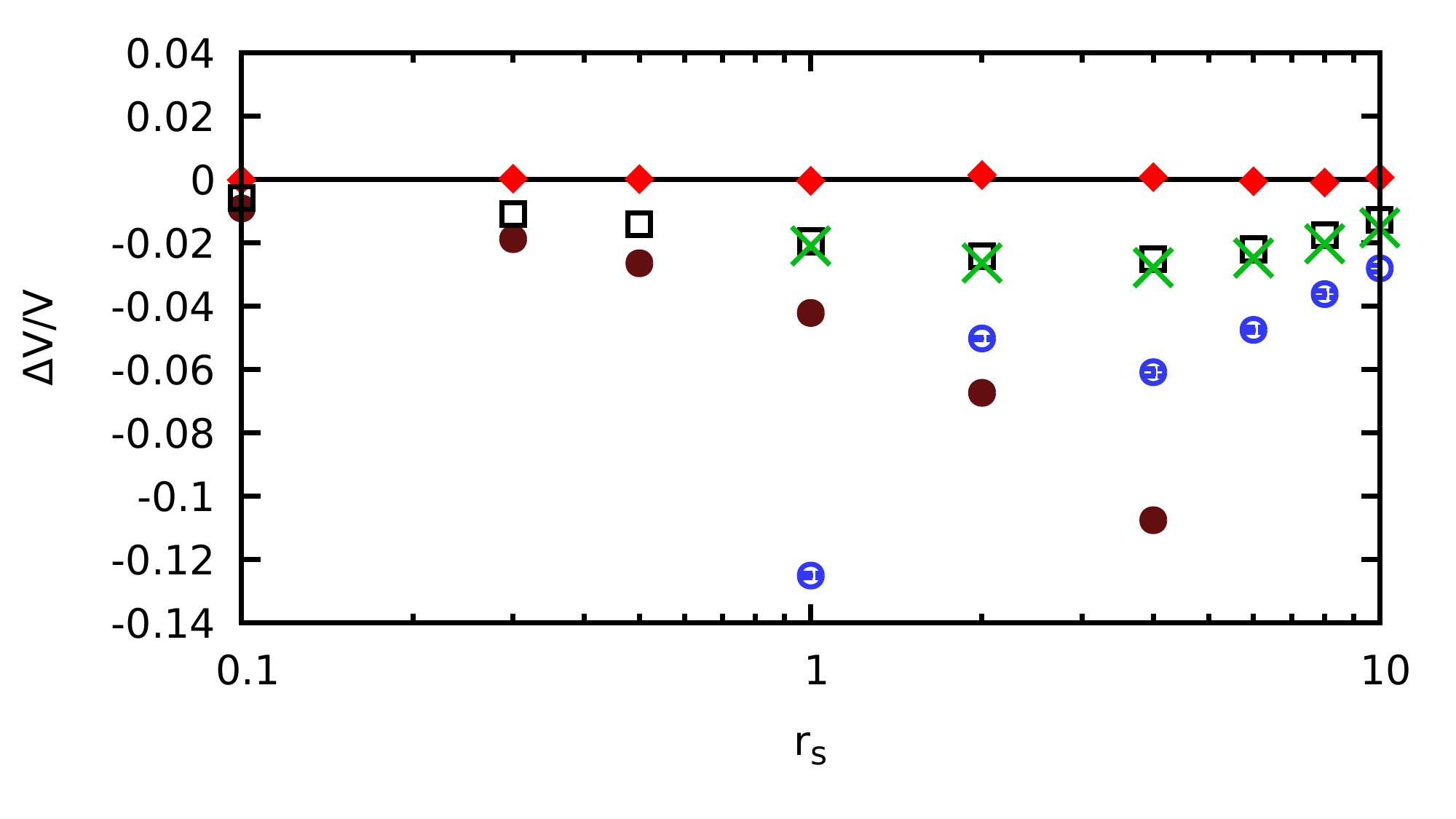}
\end{minipage}

\caption{\label{fig:high_comparison_interaction}Gauging the accuracy of the interaction energy (per particle) of different approximations for the unpolarized (left) and spin-polarized (right) electron gas at $\theta=8$. The red diamonds correspond to the finite-size corrected QMC data by Dornheim, Groth and co-workers~\cite{dornheim_abinitio_2016-1,groth_ab_2017} and the red lines depict corresponding fits to these data (see the Supplemental Material of Ref.~\cite{dornheim_abinitio_2016-1} for more details). Further shown are the RPIMC data by Brown \textit{et al.}~\cite{brown_path-integral_2013} (blue circles), finite-temperature Green function data computed in the Montroll-Ward (MW, dotted pink) and $e^4$-approximation (dashed light blue), cf.~Sec.~\ref{sec:other}, and various dielectric approximations, specifically RPA (brown dots), STLS (black squares), quantum STLS (green crosses, data obtained via integration of structure factors provided in Ref.~\cite{arora_spin-resolved_2017}), and, for $\xi=0$, recent Vashista-Singwi based data by Sjostrom and Dufty~\cite{sjostrom_uniform_2013} (purple downward triangles).
}
\end{figure}

Let us conclude this section with the investigation of the electron gas at high temperature, $\theta=8$, which is shown in Fig.~\ref{fig:high_comparison_interaction}. Both for the paramagnetic (left panel) and ferromagnetic (right panel) case, STLS and qSTLS lead to systematically too small results over the entire depicted density-range (the same is true for the VS data shown for $\xi=0$) with a maximum deviation slightly exceeding $2\%$ around $r_s=4$ for $\xi=1$. For completeness, we mention that coupling effects decrease with increasing $\theta$, leading to a large ratio of kinetic and interaction contribution to the total energy. However, this does not necessarily have to result in an improved relative accuracy in $v$ of the dielectric approximations, although, obviously, the total energy will be more accurate in this case.
The random phase approximation exhibits a significantly improved performance compared to the previous figures, although there still appear errors of $\Delta v/v\approx 4\%$ at $r_s=1$, which are rapidly increasing towards stronger coupling. 
In contrast to the lower temperature case, the finite-temperature Green function data, exhibits a pronounced unphysical bump in $v$ around $r_s=0.7$ with a maximum deviation of $7\%$ and $10\%$ for MW and $e^4$, respectively.
Finally, the RPIMC data are considerably less accurate at high temperature and exhibit an increasing systematic bias towards high density with a maxium error of $\Delta v/v\approx12\%$ at $r_s=1$ and $\xi=1$. This is mainly a consequence of the inappropriate finite-size correction, which becomes more severe both towards high density and temperature. The effect is more pronounced for the ferromagnetic case, since (i) $\theta=8$ constitutes a higher temperature than for $\xi=0$ due to the different Fermi energies, cf.~Eq.~(\ref{eq:theta_definition}), and (ii) only $N=33$ electrons were simulated in contrast to $N=66$ for the paramagnetic case.

\subsection{Static structure factor\label{sec:SSF_comparison}}

\begin{figure}\centering
\begin{minipage}{.47\textwidth}
\includegraphics[width=0.99\textwidth]{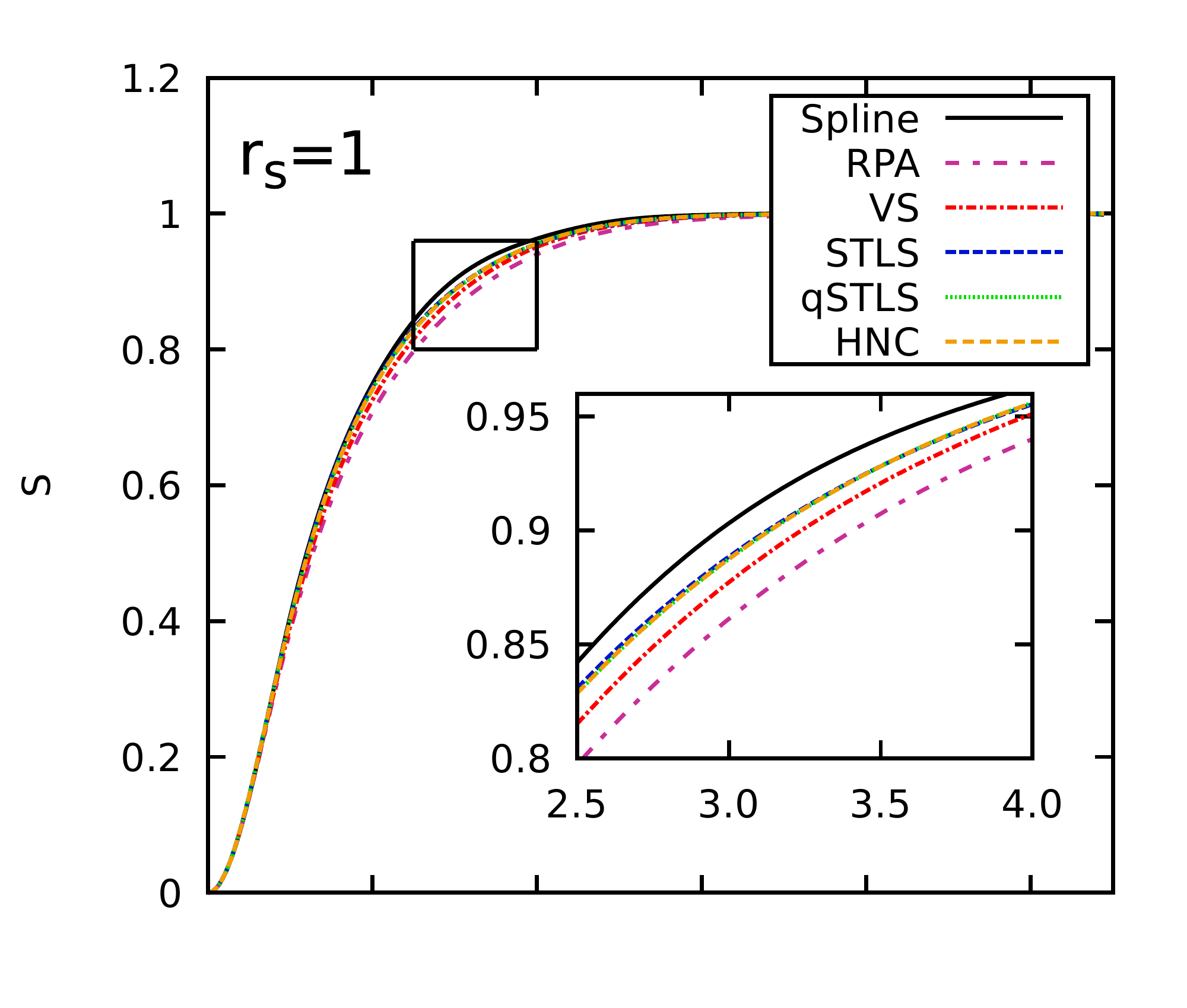}
\end{minipage}
\begin{minipage}{.02\textwidth}
$ $
\end{minipage}
\begin{minipage}{.47\textwidth}
\includegraphics[width=0.99\textwidth]{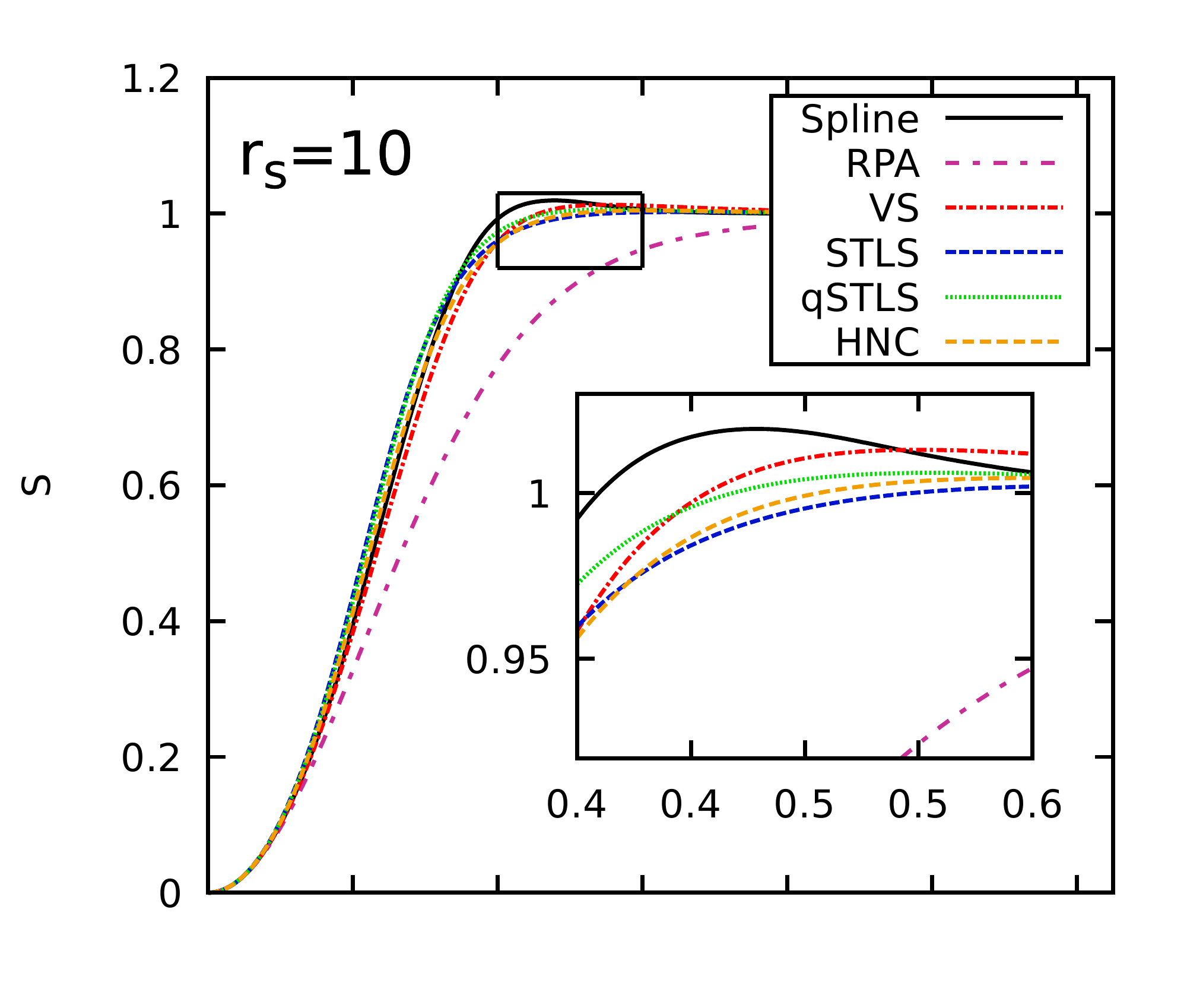}
\end{minipage}

\vspace*{-1.cm}

\begin{minipage}{.47\textwidth}
\hspace*{-0.5535cm}\includegraphics[width=1.09\textwidth]{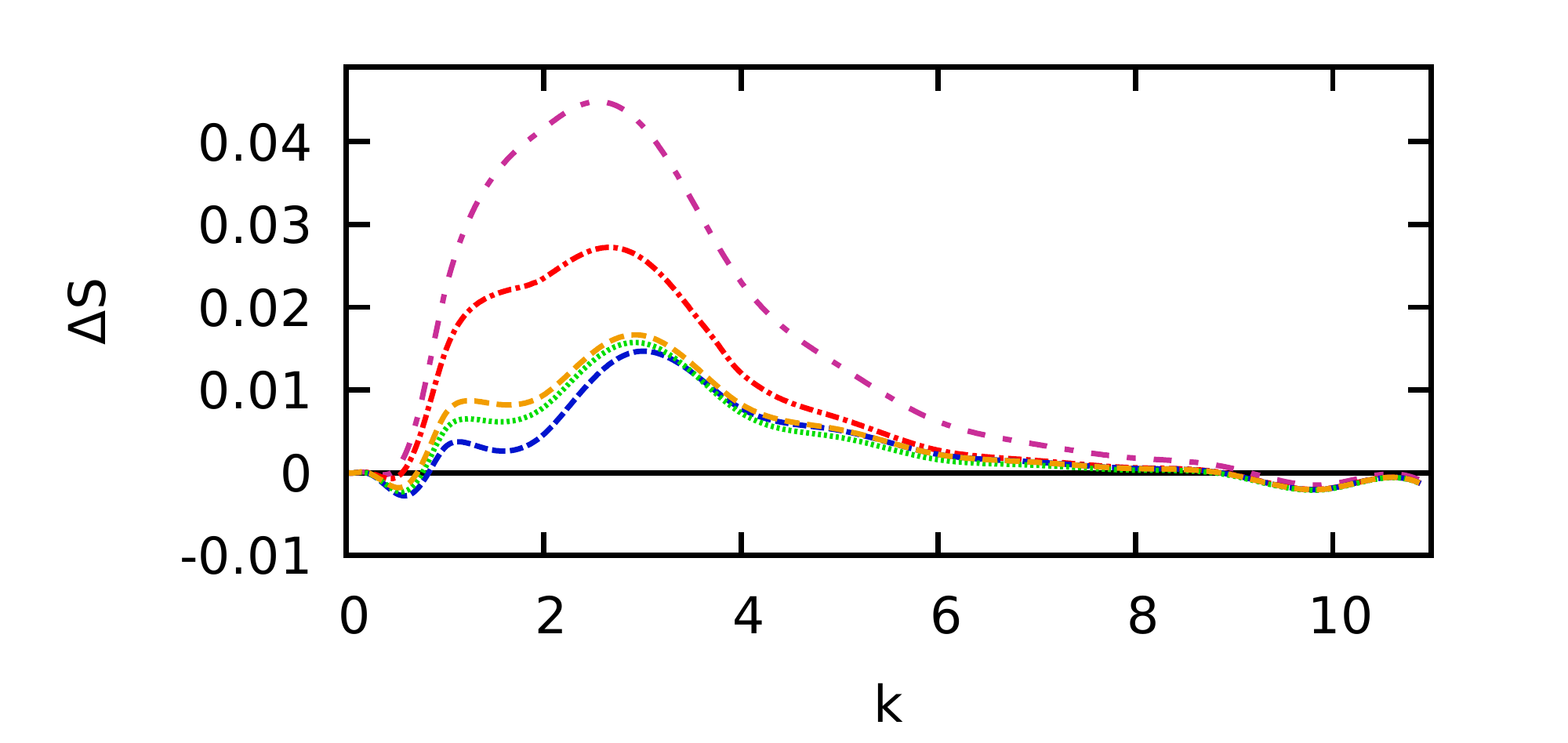}
\end{minipage}
\begin{minipage}{.02\textwidth}
$ $
\end{minipage}
\begin{minipage}{.47\textwidth}
\hspace*{-0.5535cm}\includegraphics[width=1.09\textwidth]{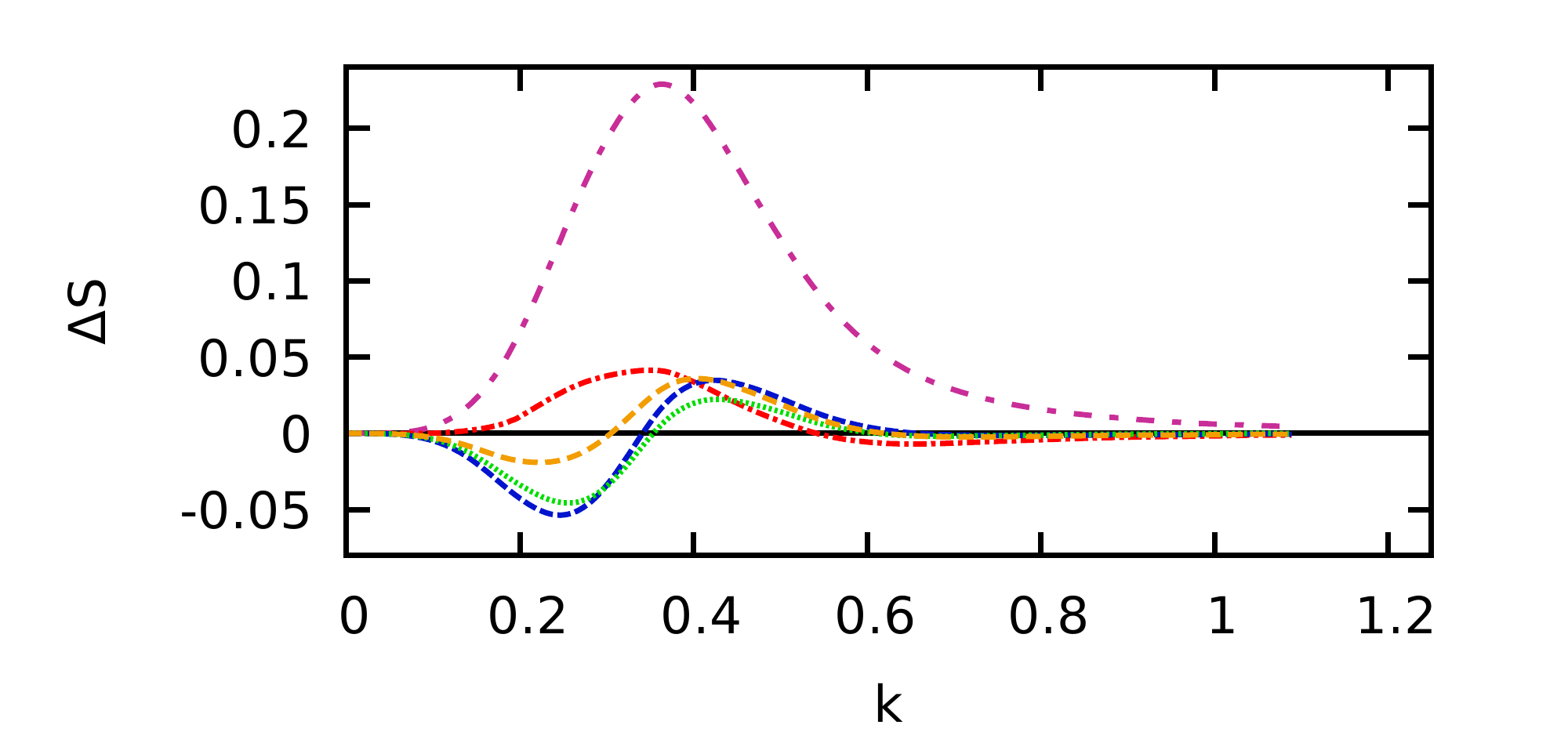}
\end{minipage}

\caption{\label{fig:SSF_comparison_dielectric}Gauging the accuracy of different approximations for the static structure factor of the unpolarized electron gas at $\theta=1$ and $r_s=1$ (left) and $r_s=10$ (right). The solid black line corresponds to cubic spline fits connecting STLS at small $k$ with our QMC data elsewhere~\cite{dornheim_textitab_2017}, the double-dashed purple line to RPA, the dash-dotted red line to Vashista-Singwi (VS)~\cite{sjostrom_uniform_2013}, the dashed blue line to STLS, the dotted green line to qSTLS~\cite{arora_spin-resolved_2017}, and the dashed orange line to the recent local field correction based on the hypernetted-chain (HNC) approximation by Tanaka~\cite{tanaka_correlational_2016}. The bottom panels depict the relative deviations to our spline.
}
\end{figure}

Finally, let us evaluate the accuracy of different theories regarding the static structure factor $S(k)$, which is of central importance for the dielectric approximations introduced in Sec.~\ref{sec:LRT}. In the left panel of Fig.~\ref{fig:SSF_comparison_dielectric}, we show $S(k)$ for the unpolarized electron gas at $\theta=1$ and $r_s=1$. The solid black line corresponds to a cubic basis spline connecting the STLS data for the limit of small $k$ with our QMC data elsewhere, see Ref.~\cite{dornheim_textitab_2017} and the explanation of finite-size effects in $v$ above. At these conditions, all dielectric approximations give the correct qualitative description of the SSF. The most pronounced systematic deviations occur for intermediate $k$, with a maximum deviation of $\Delta S/S\approx 10\%$ for RPA. On the other hand, STLS, qSTLS, and HNC exhibit a very similar behavior with maximum inaccuracies of $2\%$, and standard STLS being the most accurate approximation in this case. Further, the VS curve is significantly less accurate, albeit the overall behavior resembles the other LFC-based data.

In the right panel of Fig.~\ref{fig:SSF_comparison_dielectric}, the same information is shown for stronger coupling, $r_s=10$. In this case, our QMC-based spline exhibits a pronounced maximum around $k=0.45$, which is due to Coulomb correlation effects, and cannot be accurately resolved by any of the dielectric methods. The random phase approximation breaks down, with a systematically too small structure factor over the entire $k$-range and deviations exceeding $25\%$. Again, STLS and qSTLS are very similar and give too large results for $k\lesssim0.35$ and too small results elsewhere. The maximum deviations occur around $k=0.2$ with $\Delta S/S\approx10\%$, although qSTLS performs slightly better everywhere. The observed deviation $\Delta S$ (bottom panel) towards our spline is of high importance to understand the observed high performance of STLS in the interaction energy $v$. Since the latter is, for a uniform system, simply given by a one-dimensional integral over $S(k)-1$, the area under the $\Delta S$ curve is directly proportional to the error in $v$. Evidently, the negative area for small $k$ is of a similar magnitude as the positive one for larger $k$, which leads to a beneficial cancellation of errors and, thus, accurate results in $v$.
In contrast, the recent HNC results for $S(k)$ by Tanaka~\cite{tanaka_correlational_2016} are significantly better than STLS almost over the entire $k$-range. Nevertheless, the corresponding results for $v$ do not enjoy the error cancellation to the same degree.
Finally, let us consider the VS curve from Ref.~\cite{sjostrom_uniform_2013}, which exhibits a qualitatively different behavior from the other dielectric approximations. More specifically, the results for $S(k)$ are too low for small $k$ and slightly too large in the vicinity of large wave vectors. While the overall accuracy is again better than for STLS, there is almost no cancellation of errors when one is interested in $v$ or, via an additional coupling-constant integration, in $f_\text{xc}$.

\section{Parametrizations of the XC free energy\label{sec:fxc}}

\subsection{Introduction\label{sec:fsc_intro}}
In the ground state, the first parametrization of the exchange-correlation energy, $e_\text{xc}(r_s)$, of the unpolarized UEG on the basis of QMC data (by Ceperley and Alder~\cite{ceperley_ground_1978,ceperley_ground_1980}) has been obtained in 1981 by Perdew and Zunger~\cite{perdew_self-interaction_1981}. Shortly afterwards, Vosko, Wilk, and Nusair~\cite{vosko_accurate_1980} extended the parametrization to arbitrary spin-polarizations $\xi$, and provided a functional for $e_\text{xc}(r_s,\xi)$ in the entire parameter regime relevant to DFT calculations in the LSDA. 

At finite temperature, a parametrization of the exchange-correlation free energy, $f_\text{xc}(rs,\theta,\xi)$, in dependence of density, temperature and spin-polarization is required. In lieu of accurate finite temperature QMC data, in 1982, Ebeling \emph{et al.}~\cite{ebeling_thermodynamic_1982,richert_thermodynamic_1984,ebeling_plasma_1985,ebeling_nonideal_1989,ebeling_free_1990} carried out first attempts to obtain such a functional for the unpolarized case in terms of Pade approximations that interpolate between the known limits, i.e., the ground state limit,
$\lim_{\theta\to 0} f_\text{xc}(r_s,\theta)=e_\text{xc}(r_s,0)$, and the Debye-H\"uckel limit~\cite{dewitt_statistical_1966}, $\lim_{\theta\to\infty} f_\text{xc}(r_s,\theta)=-\frac{1}{\sqrt{3}}r_s^{-3/2}T^{-1/2}$. After that, various approximate functionals have been obatained on the basis of the results from different dielectric approaches (see Sec.~\ref{sec:LRT}).
Starting in the mid 1980s, Ichimaru, Tanaka and co-workers constructed a functional of $f_\text{xc}(r_s,\theta)$ by fitting a complex Pade approximation to the finite temperature STLS data~\cite{tanaka_parametrized_1985}, which has subsequently been improved (IIT) by incorporating the exact ground state limit via a suitable bridge function~\cite{ichimaru_statistical_1987}. Only very recently, this functional has been extended to arbitrary polarizations~\cite{tanaka_improved_2017}. In addition to the STLS approach,  the Vashishta-Singwi~\cite{sjostrom_uniform_2013}, hypernetted chain~\cite{tanaka_correlational_2016} (HNC), and the modified convolution approximation~\cite{tanaka_spin-dependent_1989} (MCA) have been successfully explored in the construction of parametrizations of the exchange correlation free energy. However, a suitable spin-interpolation function has only been deduced from the MCA results. This MCA spin-interpolation function is also utilized for the generalization of the IIT and HNC functionals to arbtitrary spin-polarizations.

Further, Dharama-wardana and Perrot presented~\cite{perrot_spin-polarized_2000,dharma-wardana_simple_2000} another widely used functional~\cite{dharma-wardana_spin-_2004,dharma-wardana_static_2006,dharma-wardana_pair-distribution_2008} based on data from their classical mapping approximation (see Sec.~\ref{sec:PDW}). Then, after the first finite temperature QMC data by Brown \emph{et al.}~\cite{brown_path-integral_2013} became available in 2013, several attempts have been made to obtain functionals from these~\cite{sjostrom_uniform_2013, brown_exchange-correlation_2013,karasiev_accurate_2014}. Among these, the most refined parametrization has been presented by Karasiev \emph{et al.}~\cite{karasiev_accurate_2014} (KSDT), who, following the IIT functional, incorporated all known limits: ground state, Debye-H\"uckel and the high-density Hartree-Fock limit~\cite{perrot_exchange_1984}. Yet, since Brown applied the RPIMC method solely to the fully polarized and unpolarized cases, the spin-interpolation of the KSDT functional has been constructed from the classical mapping data, for intermediate spin-polarizations. 
In addition, even for $\xi=0$ and $\xi=1$ the RPIMC data has turned out to be unreliable, as was shown in~Sec.~\ref{sec:comparison}.

Only recently, Groth, Dornheim \emph{et al.}~\cite{groth_ab_2017} presented a complete \emph{ab initio} parametrization of the exchange-correlation free energy, $f_\text{xc}(rs,\theta,\xi)$, that is based on highly accurate data obtained from two novel finite temperature QMC methods, CPIMC and PB-PIMC see Sec.~\ref{sec:CPIMC} and Sec.~\ref{sec:PB-PIMC} and references therein.

\subsection{Parametrizations\label{sec:questionable}}
In the following, we will provide the concrete functional form of all parametrizations, which are shown in the comparison plots in Sec.~\ref{sec:param_results}. Further, the precise way in which these were constructed as well as the included limits are discussed in detail. To be as concise as possible, we have restricted ourselves to the 5 most accurate functionals: IIT, HNC, PDW, KSDT, and GDB. For a discussion of the accuracy of the parametrization by Ebeling and co-workers, see Ref.~\cite{groth_free_2017}.

\subsubsection{IIT parametrization}
Since the dielectric methods are based on a self-consistency loop for the static structure factor and the local field correction, the natural thermodynamic quantity within this framework is given by the interaction energy computed from the static structure factor according to Eq.~(\ref{eq:interaction_en_from_s}). For fixed spin-polarization $\xi=(n_\uparrow-n_\downarrow)/n$ with the total electron density $n=(n_\uparrow+n_\downarrow)$, the interaction energy is linked to the exchange-correlation free energy via the well-known coupling constant integration formula
\begin{eqnarray}\label{eq:coupling_const_int_rs}
f^\xi_\text{xc}(r_s,\theta) &=& \frac{1}{r_s^2} \int_0^{r_s} \textnormal{d}\overline{r}_s\ \overline{r}_s\; v^\xi(\overline{r}_s,\theta) \ .
\end{eqnarray}
In the literature, the classical coupling parameter $\Gamma=1/(r_sa_\text{B}T)$ is often utilized, so that Eq.~(\ref{eq:coupling_const_int_rs}) reads
\begin{eqnarray}\label{eq:coupling_const_int_gamma}
f^\xi_\text{xc}(r_s,\theta)=\frac{1}{\Gamma^2}\int_0^\Gamma \mathrm{d}\overline{\Gamma}\ \overline{\Gamma}\ v^\xi(\overline{\Gamma},\theta)\ .
\end{eqnarray}
For the unpolarized ($\xi=0$) and polarized ($\xi=1$) case, Ichimaru, Tanaka and co-workers~\cite{ichimaru_statistical_1987,tanaka_improved_2017} proposed the following Pade fit function for the interaction energy:
\begin{align}
    v^\xi(\Gamma,\theta) = -\frac{1}{r_s} \frac{\omega_\xi a(\theta/\omega_\xi^2) + b^\xi(\theta)\sqrt{\theta}\sqrt{\Gamma}+c^\xi(\theta)\theta \Gamma }{ 1+d^\xi(\theta)\sqrt{\theta}\sqrt{\Gamma}+e^\xi(\theta)\theta\Gamma }\ , \label{eq:pade_ichi}
\end{align}
with the spin-factor $\omega_\xi=(1+\xi)^{1/3}$ and
\begin{align}\label{eq:HF_parametrization}
a(\theta) = &0.610887\tanh{(\theta^{-1})}\frac{0.75+3.04363\theta^2-0.09227\theta^3+1.7035\theta^4}{1 + 8.31051\theta^2+5.1105\theta^4}
\end{align}
ensures that the correct Hartree-Fock limit, i.e., $\lim_{r_s\to 0} v^\xi =-\frac{1}{r_s}\omega_\xi a(\theta/\omega_\xi^2)$, as parametrized in Ref.~\cite{perrot_exchange_1984} is fulfilled. The remaining functions $b,c,d,$ and $e$ are of the form
\begin{eqnarray}\label{eq:oxr_temperatuer_pade_coefficients}
b^\xi(\theta) &=& \text{tanh}\left(\frac{1}{\sqrt{\theta}}\right) \frac{ b^\xi_1 + b^\xi_2\theta^2 + b^\xi_3\theta^4 }{ 1+b^\xi_4\theta^2+b^\xi_5\theta^4 } 
\nonumber\\
c^\xi(\theta) &=& \left[ c^\xi_1 + c^\xi_2\cdot \text{exp}\left(-\theta^{-1}\right)\right] e^\xi(\theta) 
\nonumber\\
d^\xi(\theta) &=& \text{tanh}\left( \frac{1}{\sqrt{\theta}} \right) \frac{ d^\xi_1 + d^\xi_2\theta^2 + d^\xi_3\theta^4 }{ 1 + d^\xi_4\theta^2 + d^\xi_5\theta^4 } 
\nonumber\\
e^\xi(\theta) &=& \text{tanh}\left( \frac{1}{\theta} \right) \frac{ e^\xi_1 + e^\xi_2\theta^2 + e^\xi_3\theta^4 }{ 1 + e^\xi_4\theta^2 + e^\xi_5\theta^4 }\ ,
\nonumber
\end{eqnarray}
where the constants $b^\xi_1,\ldots,e^\xi_5$ are determined by a fit to the modified STLS data for the interaction energy. These modified results have been obtained by correcting the raw STLS interaction energy such that the exact ground state limit ($\theta=0$), that is known from the QMC simulations by Ceperly and Alder~\cite{ceperley_ground_1978,ceperley_ground_1980,vosko_accurate_1980}, and the classical limit ($\theta\to\infty$) are restored. This is achieved via a hypothetically assumed interpolation function that interpolates between these two limits~\cite{tanaka_improved_2017}, so that the accuracy for intermediate values of $\theta$ is naturally unclear. Once the fitting constants in Eq.~(\ref{eq:pade_ichi}) are known (for their concrete values see Ref.~\cite{tanaka_improved_2017}), the corresponding exchange-correlation free energy is immediately computed by analytically carrying out the coupling constant integration in Eq.~(\ref{eq:coupling_const_int_gamma}), yielding
\begin{eqnarray}\label{eq:ichi_fxc}
f^\xi_{xc}(r_s,\theta) = &-& \frac{1}{r_s}\frac{c(\theta)}{e(\theta)} \\ \nonumber
&-& \frac{ \theta}{2 e(\theta) r_s^2\lambda^2 } \left[ \left( \omega_\xi a(\theta/\omega_\xi^2)-\frac{c(\theta)}{e(\theta)}\right)
-\frac{d(\theta)}{e(\theta)}\left( b(\theta) - \frac{ c(\theta)d(\theta)}{e(\theta)}\right)\right] \\ \nonumber
&\times&\textnormal{log}\left|  \frac{ 2 e(\theta) \lambda^2 r_s }{ \theta} + \sqrt{2}d(\theta)\lambda r_s^{1/2} \theta^{-1/2} +1 \right| \\ \nonumber
&-& \frac{\sqrt{2}}{e(\theta)}\left( b(\theta) - \frac{ c(\theta)d(\theta) }{e(\theta)}\right) \frac{ \theta^{1/2} }{r_s^{1/2}\lambda}\\ \nonumber
&+& \frac{ \theta }{ r_s^2\lambda^2 e(\theta) \sqrt{4e(\theta)-d^2(\theta)}}\left[ d(\theta)\left(\omega_\xi a(\theta/\omega_\xi^2)-\frac{c(\theta)}{e(\theta)}\right)\right. \\ \nonumber &+& \left. \left(2-\frac{d^2(\theta)}{e(\theta)}\right)\left(b(\theta)-\frac{c(\theta)d(\theta)}{e(\theta)}\right)\right] \\ \nonumber
&\times&\left[ \textnormal{arctan}\left( \frac{ 2^{3/2} e(\theta) \lambda r_s^{1/2} \theta^{-1/2} + d(\theta) }{ \sqrt{4e(\theta)-d^2(\theta)} } \right) - \textnormal{arctan}\left( \frac{ d(\theta) }{ \sqrt{4e(\theta)-d^2(\theta)}}\right)\right] \ ,
\end{eqnarray}
where the relation $\Gamma\theta=2\lambda^2r_s$ with $\lambda=(4/(9\pi))^{1/3}$ may be used to recast this into a modified function $f_\text{xc}^\xi(\Gamma,\theta)$. We mention that although the IIT parametrization for the unpolarized case ($\xi=0$) has been provided long ago~\cite{ichimaru_statistical_1987}, the polarized case ($\xi=1$) became available only recently~\cite{tanaka_improved_2017}. Furthermore, we again stress that the IIT functional exactly fulfills all three know limits: classical, ground state and Hartree-Fock.

It is important to note that there are two different definitions of the degeneracy parameter for polarizations other than the fully unpolarized case. First, regardless of the polarization $\xi$, one may always use $\bar{\theta}=2k_\text{B}Tm_\text{e}/\hbar^2k_\text{F}^2$ with $k_\text{F}=(3\pi^2n)^{1/3}$ where $n=n_\uparrow+n_\downarrow$ is the total density of the system. This way, the parameter $\bar{\theta}$ is independent of the spin-polarization at constant values of $r_s$, but its physical meaning is somewhat unclear. The second possibility, which we employ, is to define the Fermi vector as $k_\text{F}^\uparrow=(6\pi^2 n_\uparrow)$, corresponding to the particle species with the higher density, $n_\uparrow\geq n_\downarrow$, cf.~Eq.~(\ref{eq:theta_definition}). Naturally, in the unpolarized case, where $n_\uparrow=n_\downarrow=n/2$, both definitions are equal, whereas at arbitrary polarizations the relation $\bar{\theta}=\theta(1+\xi)^{2/3}=\theta\omega_\xi^2$ holds. Since the authors of the IIT parametrizations chose the definition of $\bar{\theta}$ for the degeneracy parameter in the determination of the fitting constants\footnote{Note that the authors of Ref.~\cite{perrot_exchange_1984} also chose the definition of $\theta$ that is used here, which is the reason for the temperature scaling factor $\omega_\xi^{-2}$ in the Hartree-Fock parametrization $a$.}, we must evaluate Eq.~(\ref{eq:ichi_fxc}) at $\theta(1+\xi)^{2/3}$. For completeness, we mention that Sjostrom and Dufty~\cite{sjostrom_uniform_2013} used the same Pade ansatz for the interaction energy, Eq.~(\ref{eq:pade_ichi}), to obtain a functional of $f_\text{xc}$ both from the VS scheme (see Sec.~\ref{sec:LRT}) and the RPIMC data by Brown \textit{et al.}~\cite{brown_path-integral_2013}. 

\subsubsection{PDW parametrization}
Perrot and Dharma-wardana~\cite{perrot_spin-polarized_2000} came up with a different idea to parametrize $f_\text{xc}^0(r_s,\theta)$ that is more suitable for their classical mapping approach, see Sec.~\ref{sec:PDW}, which allows for the direct computation of the exchange-correlation free energy. These values have been directly fitted to the following parametrization
\begin{eqnarray}\label{eq:pdw}
f^0_\text{xc}(r_s,\theta) &=& \frac{ e_\text{xc}(r_s,0) - P_1(r_s,\theta) }{ P_2(r_s,\theta) }, \\ \nonumber
P_1(r_s,\theta) &=& \left(A_2(r_s)u_1(r_s) + A_3(r_s)u_2(r_s)\right) \theta^2 Q^2(r_s) + A_2(r_s)u_2(r_s)\theta^{5/2}Q^{5/2}(r_s), \\ \nonumber 
P_2(r_s,\theta) &=& 1 + A_1(r_s)\theta^2 Q^2(r_s) + A_3(r_s)\theta^{5/2}Q^{5/2}(r_s) + A_2(r_s)\theta^3Q^3(r_s), \\ \nonumber
Q(r_s) &=& \left( 2 r_s^2 \lambda^2 \right)^{-1}\ , \quad n(r_s) = \frac{3}{4\pi r_s^3}\ , \quad u_1(r_s) = \frac{\pi n(r_s)}{2}\ , \quad u_2(r_s) = \frac{2\sqrt{\pi n(r_s)}}{3}, \\ \nonumber
A_k(r_s) &=& \textnormal{exp}\left( \frac{ y_k(r_s) + \beta_k(r_s)z_k(r_s) }{ 1 + \beta_k(r_s) }
\right)\ , \quad \beta_k(r_s) = \textnormal{exp}\left( 5(r_s - r_k)
\right), \\ \nonumber
y_k(r_s) &=& \nu_k\ \textnormal{log}(r_s) + \frac{a_{1,k} + b_{1,k}r_s + c_{1,k}r_s^2  }{ 1 + r_s^2/5} \ , \quad
z_k(r_s) = r_s \frac{ a_{2,k} + b_{2,k}r_s }{ 1 + c_{2,k} r_s^2 } \quad ,
\end{eqnarray}
where the fitting constants are provided in Ref.~\cite{perrot_spin-polarized_2000}. This functional recovers the correct QMC ground state limit, $e_\text{xc}(r_s,0)$, as $\theta\to 0$ and the Debye-H\"uckel limit as $\theta\to \infty$. However, the Hartree-Fock limit at $r_s\to 0$ has not been included even though it were the very same authors who presented the Hartree-Fock parametrization~\cite{perrot_exchange_1984} 16 years earlier. For completeness, we mention that an ansatz of the form Eq.~(\ref{eq:pdw}) has also been utilized by Brown \emph{et al.}~\cite{brown_exchange-correlation_2013} to obtain the first parametrization from a fit to their RPIMC data~\cite{brown_path-integral_2013}, yet, the overall functional behavior of this parametrization has later been shown to be unsatisfactory~\cite{karasiev_accurate_2014}. 

\subsubsection{HNC parametrization}
In the recently proposed HNC functional~\cite{tanaka_correlational_2016}, Tanaka exploited the same Pade ansatz for the $r_s$-dependency of the HNC interaction energy as the IIT parametrization, cf.~Eq.~(\ref{eq:pade_ichi}):
\begin{align}
    v^\xi(r_s,\theta) = -\frac{1}{r_s} \frac{\bar{a}^\xi(\theta) + \bar{b}^\xi(\theta)\sqrt{r_s}+\bar{c}^\xi(\theta)r_s }{ 1+\bar{d}^\xi(\theta)\sqrt{r_s}+\bar{e}^\xi(\theta)r_s}\ , \label{eq:pade_HNC}
\end{align}
but slightly modified the $\theta$-dependence by using the general form
\begin{eqnarray}\label{eq:praise_the_lord}
g(\theta) = G(\theta) \frac{1 + x_2\theta^2 + x_3\theta^3 + x_4\theta^4}{1 + y_2\theta^2 + y_3\theta^3 + y_4\theta^4}\ ,
\end{eqnarray}
for all functions $\bar{b}^\xi, \bar{c}^\xi, \bar{d}^\xi, \bar{e}^\xi$. The major difference is that also terms with $\theta^3$ are included in the fit. The Hartree-Fock limit of the HNC parametrizations is incorporated in $\bar{a}^\xi(\theta)$. After fitting Eq.~(\ref{eq:pade_HNC}) to the interaction energy from the HNC scheme (values of the fitting constants can be found in Ref.~\cite{tanaka_correlational_2016}), the functional for the exchange-correlation free energy is again obtained by analytically carrying out the coupling constant integration, Eq.~(\ref{eq:coupling_const_int_rs}), which leads to a very similar expression as Eq.~(\ref{eq:ichi_fxc}). While the thus constructed HNC functional properly fulfills the classical Debye-H\"uckel limit, it does of course not include the exact QMC ground state limit. 

\subsubsection{KSDT parametrization\label{sec:KSDT_parametrization}}
The KSDT functional is based on the RPIMC data by Brown~\textit{et al.}~\cite{brown_exchange-correlation_2013}. These data have been obtained for the interaction, kinetic, and exchange-correlation energy covering the relevant warm dense matter regime of the UEG. Therefore, Karasiev~\emph{et al.} came up with a slightly different strategy to construct a parametrization by utilizing the IIT Pade anastz, Eq.~(\ref{eq:pade_ichi}), directly for the exchange-correlation free energy instead of the interaction energy, i.e.,
\begin{align}
f^\xi_\text{xc}(r_s,\theta) = -\frac{1}{r_s} \frac{\omega_\xi a(\theta) + b^\xi(\theta)\sqrt{r_s}+c^\xi(\theta)r_s }{ 1+d^\xi(\theta)\sqrt{r_s}+e^\xi(\theta)r_s }\ , \label{eq:pade_KSDT}
\end{align}
with the temperature Pade functions $b-e$ of Eq.~(\ref{eq:oxr_temperatuer_pade_coefficients}) and the Hartree-Fock parametrization, $a$, Eq.~(\ref{eq:HF_parametrization}). First, they fitted the ground state limit of Eq.~(\ref{eq:pade_KSDT})
\begin{eqnarray}
\lim_{\theta\to 0}f^\xi_\text{xc}(r_s,\theta)= e_\text{xc}^\xi(r_s,0) = -\frac{1}{r_s} \frac{\omega_\xi a_1 + b^\xi_1\sqrt{r_s}+c^\xi_1e^\xi_1r_s }{ 1+d^\xi_1\sqrt{r_s}+e^\xi_1r_s }\ ,
\end{eqnarray}
to the most recent QMC data by Spink \emph{et al.}~\cite{spink_quantum_2013}, separately for $\xi=0$ and $\xi=1$, which determines the four ground state coefficients $b_1^\xi,c_1^\xi,d_1^\xi,e_1^\xi$. The exchange-correlation free energy, $f_\text{xc}^\xi$, is linked to the interaction ($v^\xi$), kinetic, ($k^\xi$), and exchange-correlation energy $e_\text{xc}^\xi$  via the standard thermodynamic relations
\begin{align}\label{eq:fit}
v^\xi(r_s,\theta) &= \left. 2f^\xi_\text{xc}(r_s,\theta) + r_s \frac{ \partial f^\xi_\text{xc}(r_s,\theta) }{ \partial r_s }\right|_\theta\\
e^\xi_\text{xc}(r_s,\theta) &= f_\text{xc}^\xi(r_s,\theta) - \theta \left. \frac{ \partial f_\text{xc}^\xi(r_s,\theta) }{ \partial \theta}\right|_{r_s} \label{eq:exc_from_fxc}\\ 
k^\xi(r_s,\theta)&=k^\xi_s(r_s,\theta) - \theta \left. \frac{ \partial f_\text{xc}^\xi(r_s,\theta) }{ \partial \theta}\right|_{r_s}
\left. -f^\xi_\text{xc}(r_s,\theta) - r_s \frac{ \partial f^\xi_\text{xc}(r_s,\theta) }{ \partial r_s }\right|_\theta \quad ,\label{eq:k_from_fxc}
\end{align}
with $k^\xi_s(r_s,\theta)$ being the ideal kinetic energy. Therefore, the RPIMC data sets for each of these quantities can be used for a fit of the right hand sides to these data, thereby determining the remaining coefficients in Eq.~(\ref{eq:pade_KSDT}) that contain the temperature dependency. By carrying out all three of these fits both for $\xi=0$ and $\xi=1$, the authors of Ref.~\cite{karasiev_accurate_2014} found that using RPIMC data for $e_\text{xc}^\xi$ results in the smallest average and maximum deviation of the fit function to the data. Moreover, they performed the consistency checks of re-computing the two thermodynamic quantities from $f_\text{xc}^\xi$ that have not been used for the fit, and then compared the result to the corresponding RPIMC data. Again, the deviations were smallest when using $e^\xi_\text{xc}$ as input for the fit. In addition to the exact Hartree-Fock and ground state limit, the KSDT functional also fulfills the Debey-H\"uckel limit as $\theta\to\infty$ by simply fixing $b_5$ to $(3/2)^{1/2}\lambda^{-1}b_3$ for $\xi=0$ and to $(3/2)^{1/2}2^{1/3}\lambda^{-1}b_3$ with $\lambda=(4/(9\pi))^{1/3}$ for $\xi=1$. Finally, we mention that one of the temperature Pade functions, $c^\xi(\theta)$ [cf.~Eq.~(\ref{eq:oxr_temperatuer_pade_coefficients})], had to be modified in the KSDT functional to reproduce the RPIMC data sufficiently well. Naturally, this has been accomplished by adding an additional parameter $c_3$ in the exponent, i.e.,
\begin{eqnarray}\label{eq:additional_parameter}
c^\xi(\theta) &=& \left[ c^\xi_1 + c^\xi_2\cdot \text{exp}\left(-c_3\theta^{-1}\right)\right] e^\xi(\theta)\ .
\end{eqnarray}
The concrete values of all fitting constants of the KSDT functional are to be found in Ref.~\cite{karasiev_accurate_2014}.

\subsubsection{GDB Parametrization}
In the construction of the GDB parametrization~\cite{groth_ab_2017}, we followed the same strategy as the previously discussed KSDT functional (Sec.~\ref{sec:KSDT_parametrization}), but instead used our new finite size corrected QMC data for the interaction energy (see Sec.~\ref{sec:FSC}), which, due to the fermion sign problem, are available down to $\theta=0.5$. To close the remaining gap to the ground state, we computed a small temperature correction
\begin{eqnarray}
\Delta_\theta^\text{STLS}(r_s,\theta,\xi) = v^\text{STLS}(r_s,\theta,\xi) - v^\text{STLS}(r_s,0,\xi),
\label{eq:correction}
\end{eqnarray}
from the STLS method, (see Sec.~\ref{sec:LRT}), and added this onto the most accurate ground state QMC data by Spink \emph{et al.}~\cite{spink_quantum_2013} for temperatures $\theta\leq 0.25$. Thereby, we obtained a highly accurate data set for the interaction energy over the entire relevant warm dense matter regime, which we fitted to the right hand side of Eq.~(\ref{eq:fit}) with the Pade ansatz, Eq.~(\ref{eq:pade_KSDT}) for the exchange-correlation free energy. However, we found that the additional parameter $c_3$ in Eq.~(\ref{eq:additional_parameter}) is not necessary for a smooth fit through our data set. The values of the fitting constants in Eq.~(\ref{eq:pade_KSDT}) can be found in Ref.~\cite{groth_ab_2017}.

\subsection{Spin-interpolation\label{sec:spinnn}}
\subsubsection{Spin-interpolation of the KSDT and GDB functional}
To obtain an accurate parametrization of $f_\text{xc}$ at arbitrary spin polarization $0\leq \xi \leq 1$, the KSDT and GDB functional employ the ansatz~\cite{perrot_spin-polarized_2000}
\begin{eqnarray}
f_\text{xc}(r_s,\theta,\xi) = f^0_\text{xc}(r_s,\bar{\theta}) + \Big[f_\text{xc}^1(r_s,\bar{\theta}\cdot 2^{-2/3}) - f^0_\text{xc}(r_s,\bar{\theta})\Big]\Phi(r_s,\bar{\theta},\xi)\ , \label{eq:GDB_KSDT_full_fxc_xi}
\end{eqnarray}
with $\bar{\theta}= \theta (1+\xi)^{2/3}$ ensuring that the right hand side is evaluated at the same temperature $T$ for the given density parameter $r_s$. Knowing that the exact ground state spin-interpolation function in the ideal limit, $r_s\to0$, is given by
\begin{eqnarray}\label{eq:ground_state_ideal_xi}
\Phi(r_s=0,\theta=0,\xi) = \frac{ (1+\xi)^{4/3} + (1-\xi)^{4/3} -2 }{2^{4/3}-2} , 
\end{eqnarray}
Perrot and Dharama-wardana~\cite{perrot_spin-polarized_2000} proposed to extend this to the correlated system at finite temperature with the ansatz: 
\begin{align}
\Phi(r_s,\theta,\xi) &= \frac{ (1+\xi)^{\alpha(r_s,\theta)} + (1-\xi)^{\alpha(r_s,\theta)} -2 }{2^{\alpha(r_s,\theta)}-2} , \label{eq:phi}
\\
\alpha(r_s,\theta) &= 2 - h(r_s) e^{-\theta\lambda(r_s,\theta)}, 
\nonumber\\
h(r_s) &= \frac{2/3+h_1 r_s}{1+h_2 r_s}, \nonumber\\
\lambda(r_s,\theta) &= \lambda_1 + \lambda_2 \theta r_s^{1/2}\ ,
\nonumber
\end{align}
which fulfills the ground state limit of the ideal system, Eq.~(\ref{eq:ground_state_ideal_xi}). Both in the GDB and KSDT functional the parameters $h_1$ and $h_2$ are obtained by fitting $f_\text{xc}(r_s,0,\xi)$ to the ground state data of Ref.~\cite{spink_quantum_2013} for $\xi=0.34$ and $\xi=0.66$. Then, in the case of the KSDT functional, the remaining two parameters $\lambda_1$ and $\lambda_2$, which carry the temperature dependent information of the interpolation function, had to be determined by a subsequent fit to the approximate hypernetted chain data~\cite{perrot_spin-polarized_2000} of $f_\text{xc}$ at intermediate spin-polarization $\xi$ since Brown~\emph{et al.} did not provide these data. Whereas in case of the GDB functional~\cite{groth_ab_2017}, we performed vast additional QMC simulations to obtain \textit{ab initio} data for the interaction energy $v^\xi(r_s,\theta)$ at $\xi=1/3$ and $\xi=0.66$, which we utilized to determine the parameters $\lambda_1$ and $\lambda_2$ via Eq.~(\ref{eq:fit}). Interestingly, we find that the spin interpolation depends only very weakly on $\theta$, and in contrast to KSDT, $\lambda_2$ in fact vanishes within the accuracy of the fit and, thus, we set $\lambda_2=0$.

\subsubsection{Spin-interpolation of the IIT and HNC functional}
In 1989, Tanaka and Ichimaru~\cite{tanaka_spin-dependent_1989} introduced a different spin-interpolation for the warm dense electron gas on the basis of the modified convolution approximation (MCA) (see Sec.~\ref{sec:LRT}). Specifically, their ansatz for the interaction energy is given by
\begin{eqnarray}\label{eq:spin_interpolation_tanaka}
v(r_s,\theta,\xi) =  (1-\xi^6)v^0(r_s,\theta) + \xi^6 v^1(r_s,\theta) + \left( \frac{1}{2}\xi^2 + \frac{5}{108}\xi^4 - \frac{59}{108}\xi^6 \right) \frac{ s(r_s,\theta) }{ r_s }\quad ,
\end{eqnarray}
with the definition
\begin{eqnarray}\label{eq:magic_s}
s(r_s,\theta) = - \frac{ a_s(\theta) + b_s(\theta) r_s }{ 1 + c_s(\theta) r_s + d_s(\theta) r_s^2 }\quad .
\end{eqnarray}
Note that the temperature-dependent coefficients 
$a_s(\theta), b_s(\theta), c_s(\theta), d_s(\theta)$ are of the same form as Eq.~(\ref{eq:praise_the_lord}), see Ref.~\cite{tanaka_spin-dependent_1989} for the appropriate fitting constants.
This, in turn, leads to the spin-interpolation for the exchange-correlation free energy
\begin{eqnarray}\label{eq:spin_interpolation_tanaka_fxc}
f_\text{xc}(r_s,\theta,\xi) = (1-\xi^6)f_\text{xc}^0(r_s,\theta) + \xi^6 f_\text{xc}^1(r_s,\theta) + \left( \frac{1}{2}\xi^2 + \frac{5}{108}\xi^4 - \frac{59}{108}\xi^6 \right) \Sigma(r_s,\theta) \quad ,
\end{eqnarray}
and plugging Eq.~(\ref{eq:spin_interpolation_tanaka}) into (\ref{eq:coupling_const_int_rs}) immediately gives
\begin{eqnarray}
\Sigma(r_s,\theta) = \frac{1}{r_s^2}\int_0^{r_s} \textnormal{d}\overline{r}_s\ s(\overline{r}_s,\theta) \quad ,
\end{eqnarray}
which (up to moderate temperature, see the discussion of Fig.~\ref{fig:fxc_spin_dependence} below) can be evaluated analytically as
\begin{eqnarray}\label{eq:MCA_spin_alpha}
\Sigma(r_s,\theta) = 
 \begin{cases}
      \Sigma_{<}(r_s,\theta), & \text{if}\ c_s^2 < 4d_s \\
      \Sigma_{=}(r_s,\theta), & \text{if}\ c_s^2 = 4d_s \\
      \Sigma_{>}(r_s,\theta), & \text{otherwise}
    \end{cases}\quad ,
\end{eqnarray}
with
\begin{eqnarray}
\Sigma_{<}(r_s,\theta) = &-& \frac{1}{r_s^2}\left[ \frac{b_s}{2d_s}\text{log}\left| 1+c_sr_s+d_sr_s^2 \right| \right. \\ \nonumber & & + \frac{2a_s d_s - b_s c_s}{d_s \sqrt{4d_s-c_s^2}} \left[ \text{atan}\left( \frac{2d_sr_s+c_s}{\sqrt{4d_s-c_s^2}}\right) - \text{atan}\left( \frac{c_s}{\sqrt{4d_s-c_s^2}} \right) \right] \quad , \\
\Sigma_{=}(r_s,\theta) &=& - \frac{1}{r_s^2} \left[ \frac{b_s}{2d_s} \text{log}\left| 1+c_sr_s+d_sr_s^2\right| - \frac{2a_sd_s-b_sc_s}{d_s(2d_sr_s+c_s)} \right] \quad , \\
\Sigma_{>}(r_s,\theta) &=& - \frac{1}{r_s^2} \left[ 
\frac{b_s}{2d_s} \textnormal{log}| 1 + c_sr_s + d_sr_s^2|  \right. \\
& & + \left. \frac{ 2a_sd_s-b_sc_s }{ 2d_s\sqrt{ c_s^2 - 4d_s } }
\left( 
\textnormal{log}\left| 
\frac{ 2d_s r_s + c_s - \sqrt{c_s^2-4d_s} }{ 2d_sr_s+c_s+\sqrt{ c_s^2-4d_s} }
\right|
- \textnormal{log}\left|
\frac{ c_s - \sqrt{ c_s^2 - 4d_s} }{  c_s + \sqrt{ c_s^2 - 4d_s} }
\right|
\right)
\right] \quad . \nonumber 
\end{eqnarray}
As the spin-dependence of MCA is expected to be similar both to STLS and also the recent HNC-based LFC by Tanaka, Eqs.~(\ref{eq:spin_interpolation_tanaka}) and (\ref{eq:spin_interpolation_tanaka_fxc}) are used for both of these parametrization with the same fitting constants as in the original reference~\cite{tanaka_spin-dependent_1989}.

\subsection{Comparison of parametrizations\label{sec:param_results}}

\subsubsection{Interaction energy}

\begin{figure}
\includegraphics[width=0.4\textwidth]{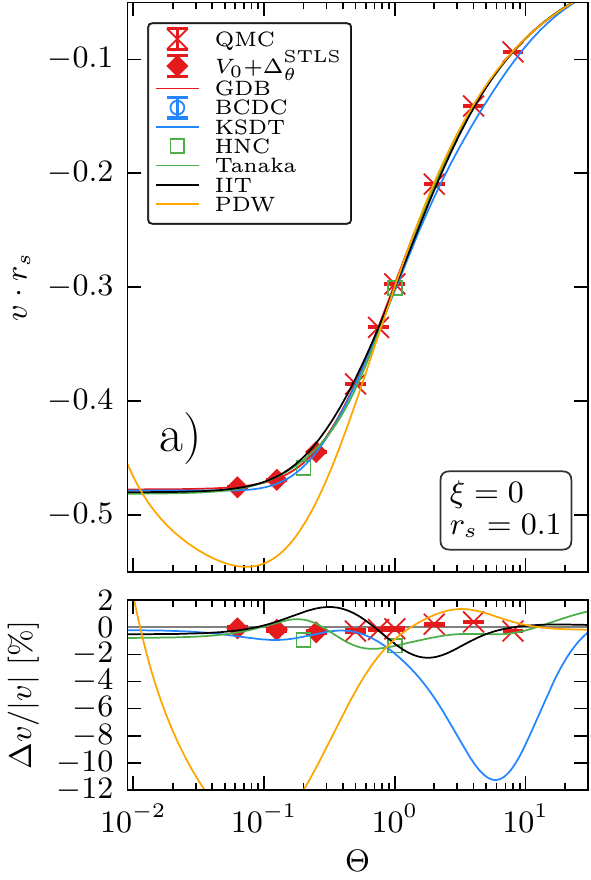}
\hspace*{1cm}\includegraphics[width=0.4\textwidth]{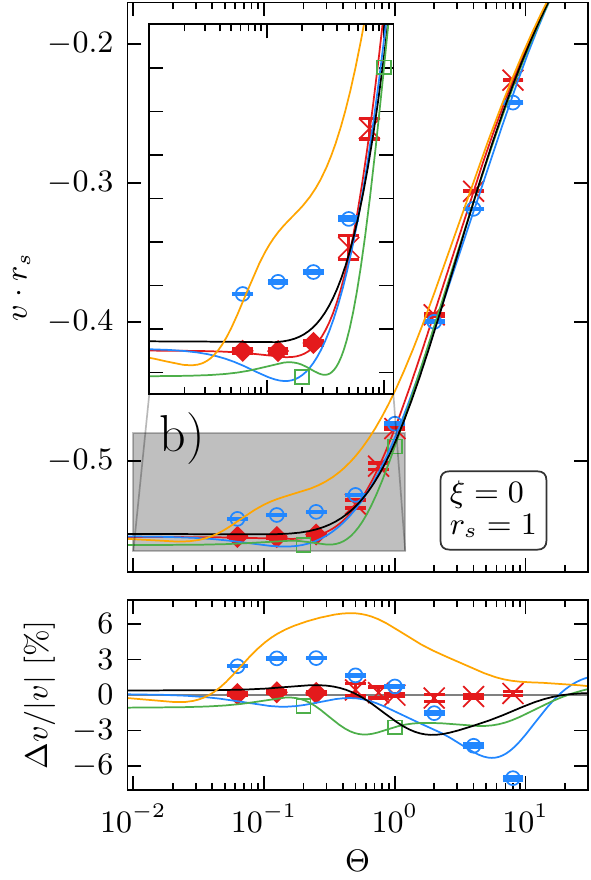}

\includegraphics[width=0.4\textwidth]{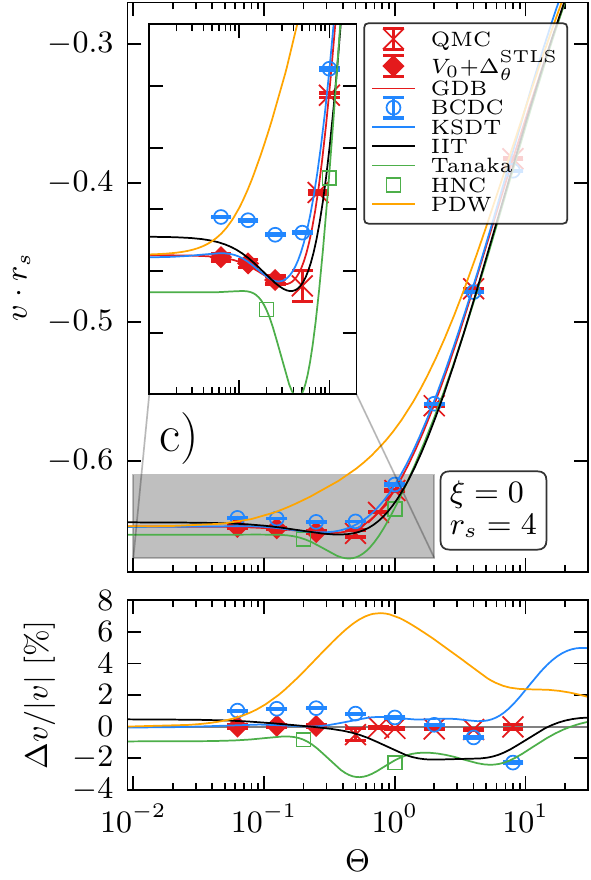}
\hspace*{1cm}\includegraphics[width=0.4\textwidth]{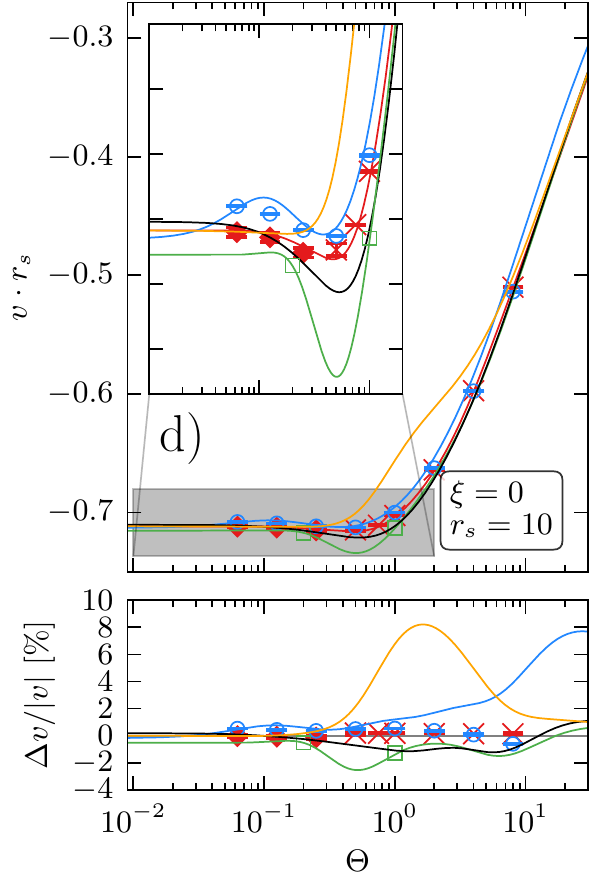}
\caption{\label{fig:panel_unpolarized_v}Temperature dependence of the interaction energy of the unpolarized electron gas for $r_s=0.1,1,4,10$ -- Shown are the recent QMC results from Refs.~\cite{dornheim_abinitio_2016-1,groth_ab_2017} (red crosses), STLS temperature-corrected ground state QMC data (see Eq.~(\ref{eq:correction}), red diamonds), the parametrization by Groth, Dornheim and co-workers (GDB, red line)~\cite{groth_ab_2017}, RPIMC data (blue circles, BCDC, Ref.~\cite{brown_path-integral_2013}) and the corresponding parametrization by Karasiev and co-workers (blue line, KSDT, Ref.~\cite{karasiev_accurate_2014}), data from an improved local field correction based on the hypernetted-chain approximation (green squares, HNC, Ref.~\cite{tanaka_correlational_2016}) and a corresponding parametrization (green line), the improved STLS parametriprzation by Ichimaru and co-workers (black line, IIT, Ref.~\cite{ichimaru_statistical_1987,tanaka_improved_2017}), and the parametrization by Perrot and Dharma-wardana (yellow line, PDW, Ref.~\cite{perrot_spin-polarized_2000}). The bottom panels depict the relative deviation towards the GDB curve and the insets correspond to magnified segments.
}
\end{figure}

In Fig.~\ref{fig:panel_unpolarized_v}, we compare various results for the temperature-dependence of the interaction energy per particle of the unpolarized electron gas for different densities.
The red crosses correspond to the finite-size corrected (using our new, improved finite-size correction, see Sec.~\ref{sec:FSC}) thermodynamic QMC results by Groth, Dornheim and co-workers~\cite{dornheim_abinitio_2016-1,groth_ab_2017} and the red diamonds to the ground state QMC data~\cite{spink_quantum_2013} with an STLS temperature correction obtained from Eq.~(\ref{eq:correction}). Observe the smooth connection between the two data sets over the entire density-range. Thus, in combination, these constitute the most accurate existing data for the interaction energy over the entire warm dense matter regime and have subsequently been used as input for our recent parametrization, i.e., the red line (GDB, Ref.~\cite{groth_ab_2017}). Evidently, the employed Pade ansatz is an appropriate fit function, as the input data are accurately reproduced with a mean and maximum deviation of $0.12\%$ and $0.63\%$, see also the corresponding bottom panels where we show the relative deviations of all data sets to the GDB curve.

Although the parametrization of the interaction energy is, for the most part, just a means to obtain the exchange-correlation free energy $f_\text{xc}$, cf.~Eq.~(\ref{eq:fit}), it is still worth to consider, at this point, $v$ itself to gauge the accuracy of various previous approximations and XC-functionals. The blue circles correspond to the RPIMC data from Ref.~\cite{brown_path-integral_2013} (BCDC) and the blue line to the corresponding parametrization by Karasiev \textit{et al.}~\cite{karasiev_accurate_2014} (KSDT). First and foremost, we note that the BCDC data are available for low to moderate densities, $r_s\geq1$, and exhibit the largest deviations for the smallest $r_s$-value. This is a combination of two different effects. At low temperature, the observed systematic bias is mostly a consequence of the employed fixed node approximation (and, possibly, related to ergodicity problems in the QMC algorithm, see Sec.~\ref{sec:RPIMC}), whereas at high temperature the effects of the inappropriate finite-size correction dominate (cf.~Sec.~\ref{sec:FSC}), leading to a maximum error of $\Delta v/v\approx7\%$ for the unpolarized case. In contrast, the BCDC data are substantially more accurate at stronger coupling, with maximum deviations of $2\%$ and $1\%$ for $r_s=4$ and $r_s=10$, respectively. 

The KSDT parametrization has been obtained from a fit to the BCDC data for $E_\text{xc}$, i.e., the sum of $v$ and the exchange-correlation part of the kinetic energy $k_\text{xc}$. However, the results for the interaction energy computed from $f_\text{xc}$ [cf.~Eq.~(\ref{eq:fit})] do not agree with the blue circles, which means that the parametrization and input data are not consistent as the exact thermodynamic relations, Eqs.~(\ref{eq:fit})-(\ref{eq:k_from_fxc}), are strongly violated. In particular, there appear pronounced deviations between the two at low temperature as the KSDT functional incorporates the correct ground state limit. The largest deviations ($\Delta V/V\approx11\%$) between the KSDT and GDB curves appear at high density, $r_s=0.1$. This is a consequence of the lack of input data for the former in this regime, which is bridged by an interpolation between the RPIMC data at $r_s\geq1$ and the correct Hartree Fock limit at $r_s=0$. Furthermore, we stress the surprisingly large errors at high temperature both for $r_s=4$ and $r_s=10$, and the unphysical bump at low temperature in the latter case.

The black line depicts the widely used improved STLS parametrization that is due to Ichimaru and co-workers (IIT, Ref.~\cite{ichimaru_statistical_1987,tanaka_improved_2017}). Given the incorporation of the exact behavior for $r_s\to0$, $\theta\to\infty$ and $\theta\to0$, and the remarkable accuracy of the STLS formalism inbetween (cf.~Sec.\ref{sec:FSC}), the overall good performance of this functional does not come as a surprise. In particular, the most severe systematic errors occur for intermediate density ($r_s=1$) and temperature, but do not exceed $\Delta v/v\approx 4\%$. 

Next, let us consider the green curve corresponding to a fit to the recent data based on the improved local field correction derived from the hypernetted-chain approximation (HNC, green squares) by Tanaka~\cite{tanaka_correlational_2016}. While this new LFC does constitute an improvement, both, for the static structure factor (see Sec.~\ref{sec:SSF_comparison}) and $G(q)$ itself (Sec.~\ref{sec:response}), the same does not apply for the interaction energy, as for this quantity STLS benefits from a fortunate error cancellation in the integration, in particular at large $r_s$, cf.~Fig.~\ref{fig:SSF_comparison_dielectric}.
Furthermore, the HNC parametrization exhibits a pronounced minimum around $\theta=0.5$, the origin of which is probably an artifact of the lack of HNC input data for these parameters, see the insets for $r_s=4$ and $r_s=10$. In addition, the ground state limit is obtained from the zero temperature HNC data and not from the more accurate QMC results, which leads to relative errors of around $1\%$ towards $\theta=0$. Hence, we conclude that the green curve does not improve the twenty years older IIT parametrization, although it exhibits an overall similar accuracy.

Finally, we include the interaction energy computed from the parametrization of classical-mapping data (cf.~Sec.~\ref{sec:PDW}) by Perrot and Dharma-wardana (yellow line, PDW, Ref.~\cite{perrot_spin-polarized_2000}). This curve was constructed from input data in the range $1\leq r_s\leq 10$, and, somewhat ironically, the Hartree-Fock limit that was parametrized by the same authors in 1984~\cite{perrot_exchange_1984}, was not incorporated. For this reason, the functional exhibits large deviations at high density and should not be used below $r_s=1$. While PDW did include the correct ground state limit, the lowest finite temperature values correspond to $\theta=0.25$,  which explains the unphysical behavior of the yellow curve at low temperature for $r_s=1$. Overall, we find that the PDW parametrization exhibits the largest systematic errors (with $\Delta v/v\gtrsim6\%$) at intermediate temperatures around $\theta=1$, which is not surprsing given the employed interpolation of the quantum temperature parameter in the classical mapping formalism, cf.~Eq.~(\ref{eq:PDW_T}).

\begin{figure}
\includegraphics[width=0.4\textwidth]{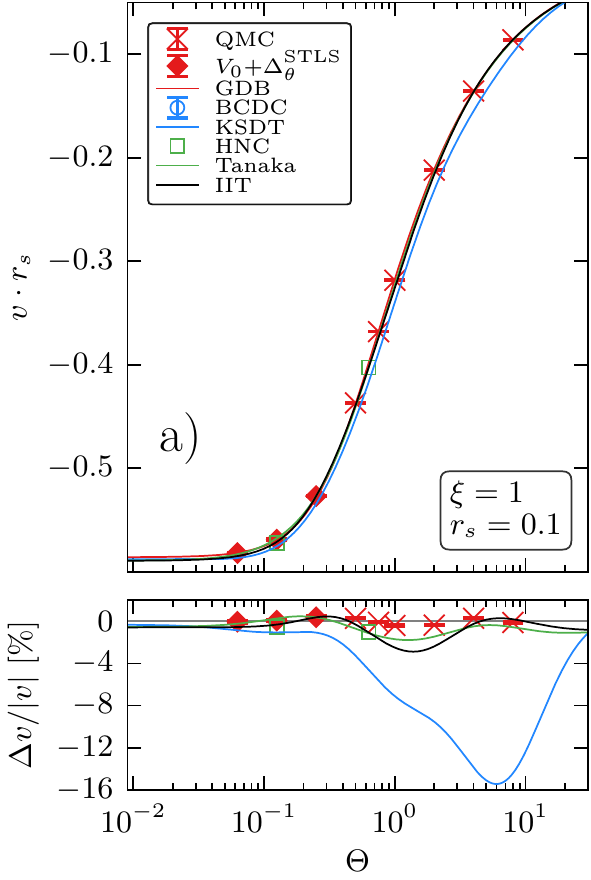}
\hspace*{1cm}\includegraphics[width=0.4\textwidth]{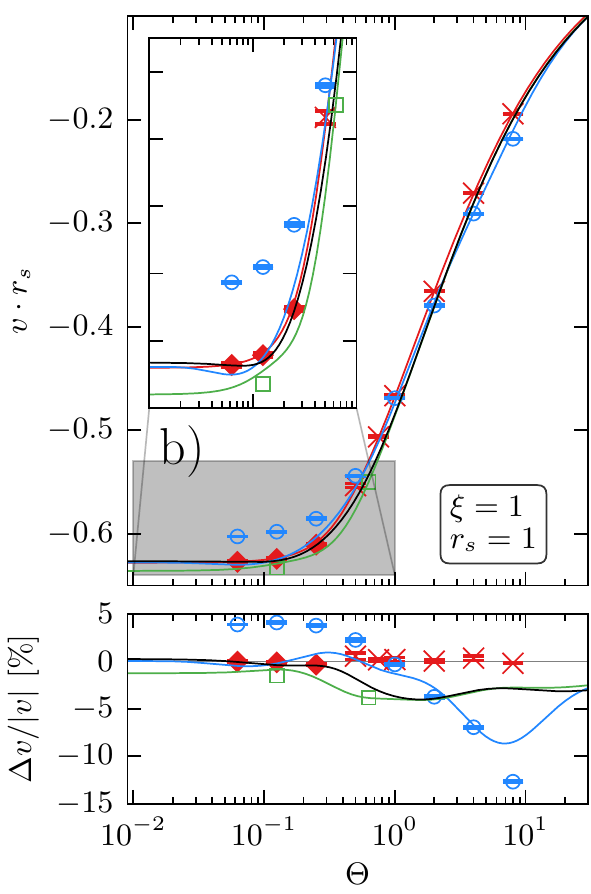}

\includegraphics[width=0.4\textwidth]{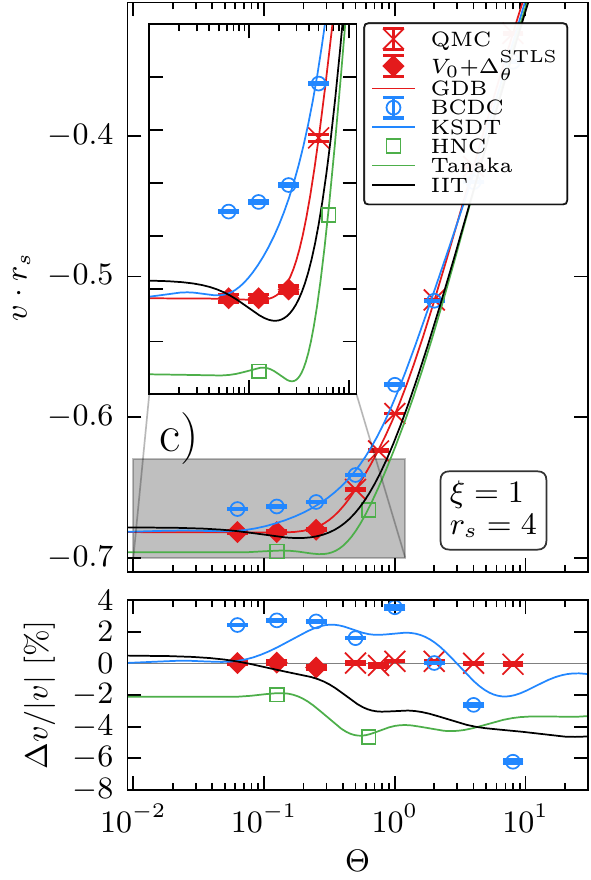}
\hspace*{1cm}\includegraphics[width=0.4\textwidth]{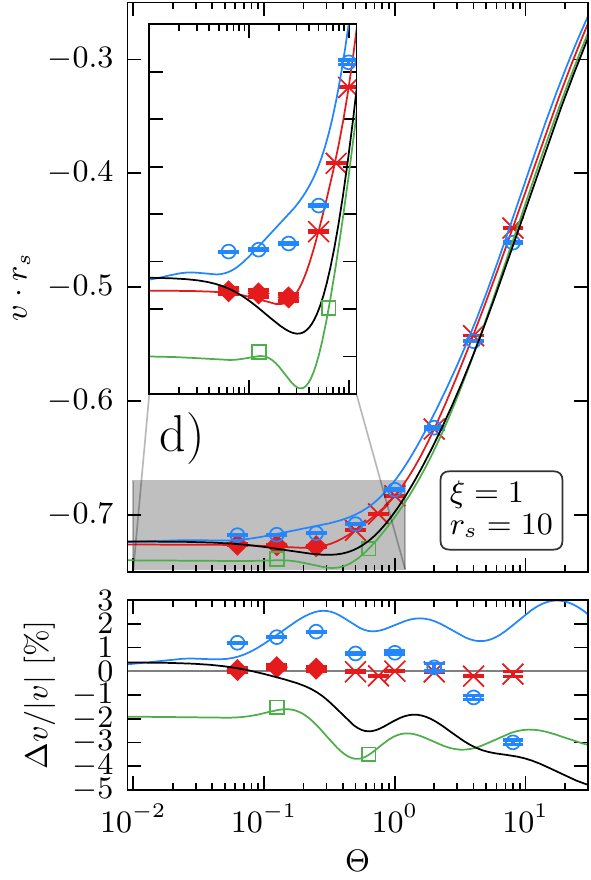}
\caption{\label{fig:panel_polarized_v}Temperature dependence of the interaction energy of the spin-polarized electron gas for $r_s=0.1,1,4,10$ -- Shown are the recent QMC results from Refs.~\cite{dornheim_abinitio_2016-1,groth_ab_2017} (red crosses), STLS temperature-corrected ground state QMC data (see Eq.~(\ref{eq:correction}), red diamonds), the parametrization by Groth, Dornheim and co-workers (GDB, red line)~\cite{groth_ab_2017}, RPIMC data (blue circles, BCDC, Ref.~\cite{brown_path-integral_2013}) and the corresponding parametrization by Karasiev and co-workers (blue line, KSDT, Ref.~\cite{karasiev_accurate_2014}), data from an improved local field correction based on the hypernetted-chain approximation (green squares, HNC, Ref.~\cite{tanaka_correlational_2016}) and a corresponding parametrization (green line), and the improved STLS parametriprzation by Ichimaru and co-workers (black line, IIT, Ref.~\cite{ichimaru_statistical_1987,tanaka_improved_2017}). The bottom panels depict the relative deviation towards the GDB curve and the insets correspond to magnified segments.
}
\end{figure}

In Fig.~\ref{fig:panel_polarized_v}, we show the same comparison but for the spin-polarized case, $\xi=1$. While we do find similar trends as in the previous figure, the relative biases of the different approximations are, overall, increased. In particular, the KSDT curve exhibits a maximum deviation exceeding $15\%$ at high density, and even at $r_s=1$ we find $\Delta v/v\approx8\%$ around $\theta=5$. Furthermore, this parametrization exhibits an unphysical plateau-like behavior in the low-temperature regime both at $r_s=4$ and $r_s=10$. In addition, the BCDC data are substantially more biased both at low and high temperature, with a maximum deviation of $\Delta v/v\approx 14\%$ at $r_s=1$ and $\theta=8$. The increased deviation for the latter case is a consequence of the definition of the reduced temperature, resulting in a larger temperature at equal $\theta$-values for the spin-polarized case. This, in turn, exacerbates the inaccuracy of the employed finite-size correction, cf.~Sec.~\ref{sec:FSC}. At low temperature, the fixed node approximation exhibits a worse performance even for a finite model system~\cite{dornheim_abinitio_2016}.
The HNC and IIT parametrizations are of a similar quality, but the latter appears to be superior due to the incorporation of the correct ground state limit. The main difference compared to the unpolarized case is the significantly larger deviation for large temperature at $r_s=10$. Interestingly, this is not a consequence of a worse performance of the STLS approximation itself, cf.~Fig.~\ref{fig:high_comparison_interaction} but, instead, of the \textit{a posteriori} modification of the STLS data to incorporate the exact high and low temperatuere limit.
Finally, we mention the excellent agreement between the GDB parametrization and its input data with a mean and maximum deviation of $0.17\%$ and $0.63\%$, respectively.

\subsubsection{Exchange-correlation free energy}

\begin{figure}
\includegraphics[width=0.4\textwidth]{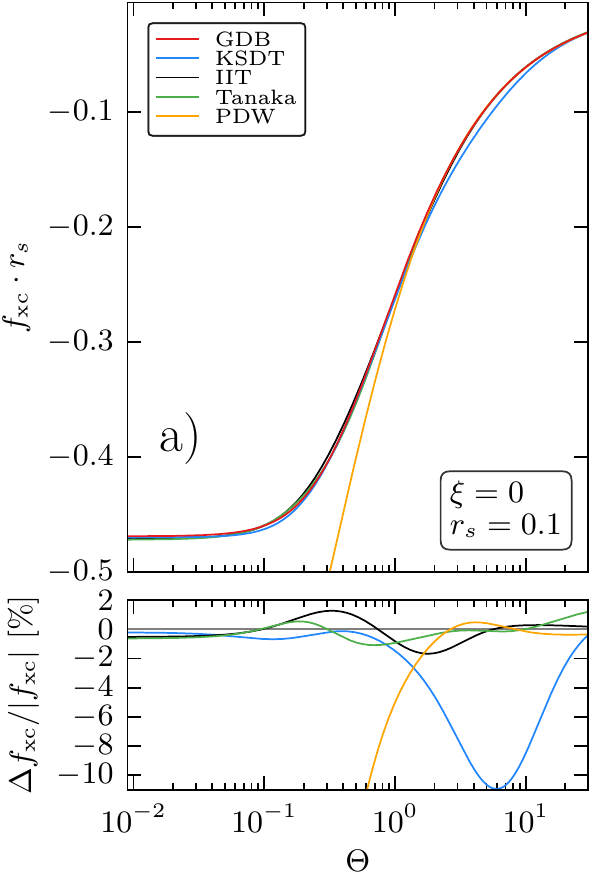}
\hspace*{1cm}\includegraphics[width=0.4\textwidth]{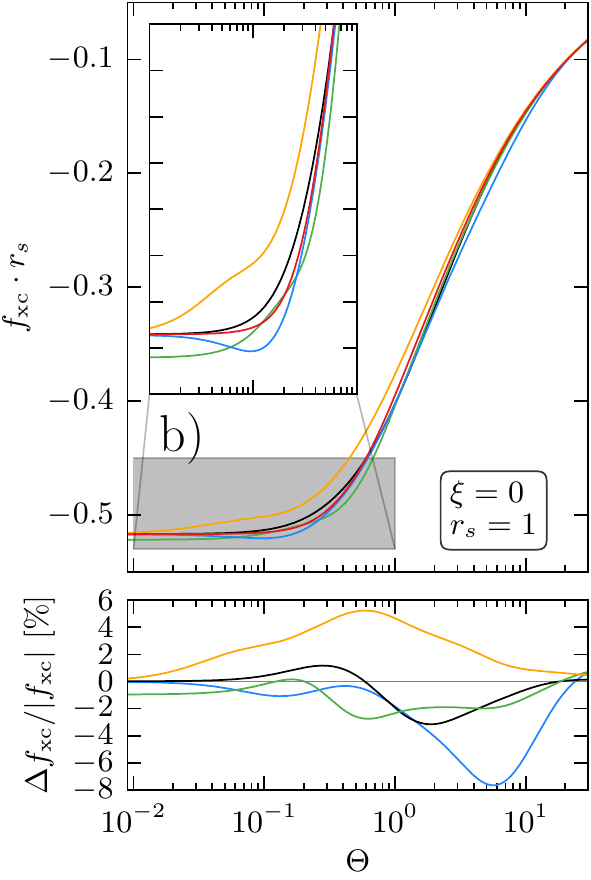}

\includegraphics[width=0.4\textwidth]{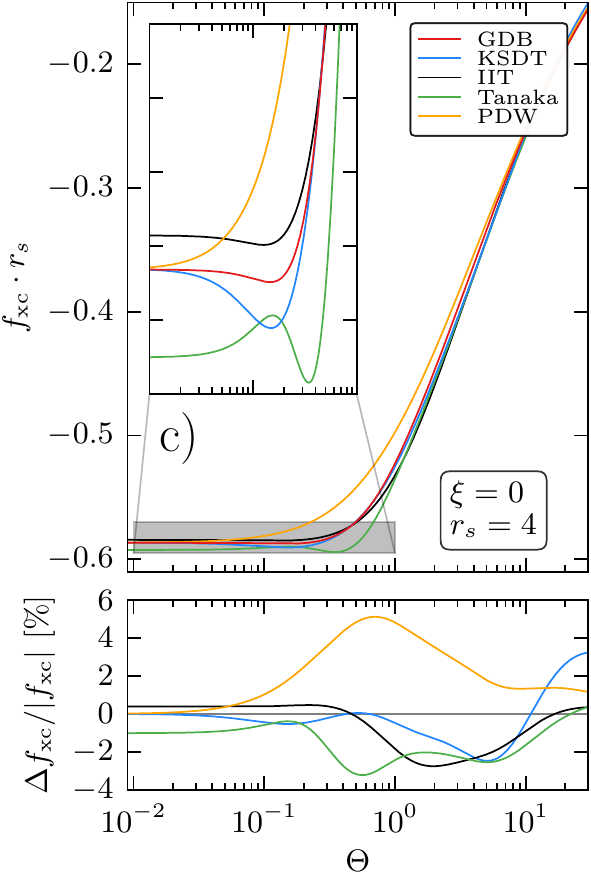}
\hspace*{1cm}\includegraphics[width=0.4\textwidth]{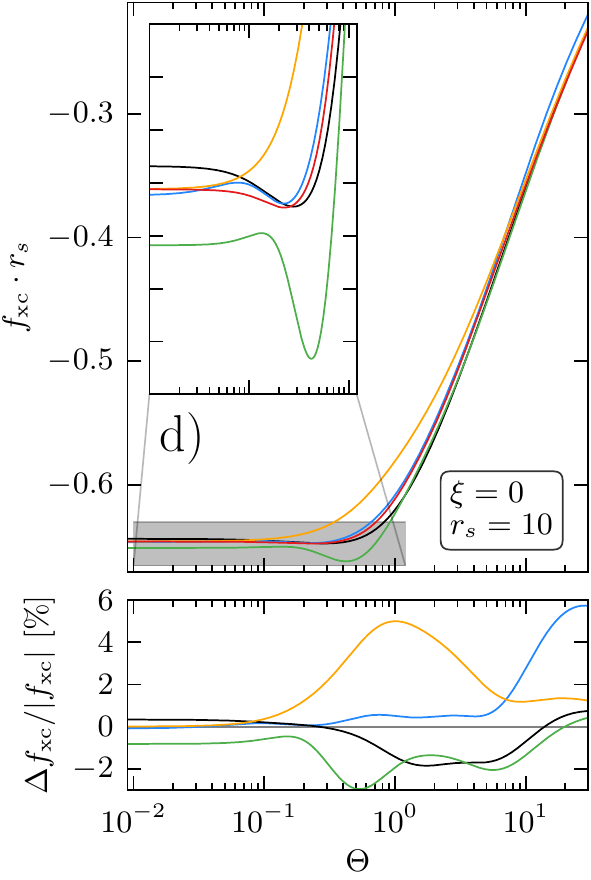}
\caption{\label{fig:panel_unpolarized_fxc}Temperature dependence of the exchange-correlation free energy of the unpolarized electron gas for $r_s=0.1,1,4,10$ -- Shown are the parametrizations by Groth, Dornheim \textit{et al.} (red line, GDB, Ref.~\cite{groth_ab_2017}),  Karasiev \textit{et al.} (blue line, KSDT, Ref.~\cite{karasiev_accurate_2014}), Tanaka (green line, HNC, Ref.~\cite{tanaka_correlational_2016}), Ichimaru \textit{et al.} (black line, IIT, Ref.~\cite{ichimaru_statistical_1987,tanaka_improved_2017}) and Perrot and Dharma-wardana (yellow line, PDW, Ref.~\cite{perrot_spin-polarized_2000}).
The bottom panels depict the relative deviation towards the GDB curve and the insets correspond to magnified segments.
}
\end{figure}

Let us now consider the main quantity of interest, i.e., the exchange-correlation free energy $f_\text{xc}$.
In Fig.~\ref{fig:panel_unpolarized_fxc}, we compare the temperature dependence of the five most accurate functionals for the unpolarized case and at the same densities as in the previous section. All curves exhibit a qualitatively similar behavior except PDW at $r_s=0.1$, which is again a consequence of the not incorporated Hartree-Fock limit and the density range of the input data ($1\leq r_s \leq 10$).
Overall, the KSDT parametrization is relatively accurate at low temperature ($\theta<1$) although there appears a bump in both $v$ and $f_\text{xc}$ at large $r_s$, which leads to an unphysical slightly negative entropy~\cite{burke_exact_2016}.
In contrast, at intermediate to high temperature we find substantial systematic deviations (exceeding $10\%$ at $r_s=0.1$), which are a direct consequence of the utilized RPIMC input data. Again, the IIT and HNC curves exhibit a very similar performance, with the former being superior due to the correct ground state limit. More specifically, for the unpolarized case we find maximum deviations of around $3\%$ at intermediate $r_s$-values and temperatures.
Finally, the classical-mapping based PDW parametrization by Perrot and Dharma-wardana~\cite{perrot_spin-polarized_2000} exhibits deviations of up to $\Delta f_\text{xc}/f_\text{xc}\approx5\%$ around the Fermi temperature.

\begin{figure}
\includegraphics[width=0.4\textwidth]{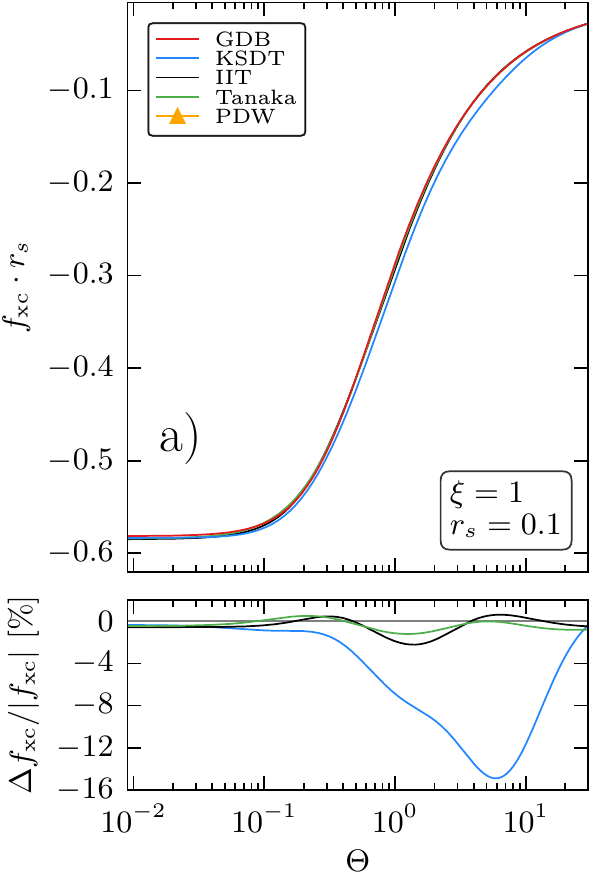}
\hspace*{1cm}\includegraphics[width=0.4\textwidth]{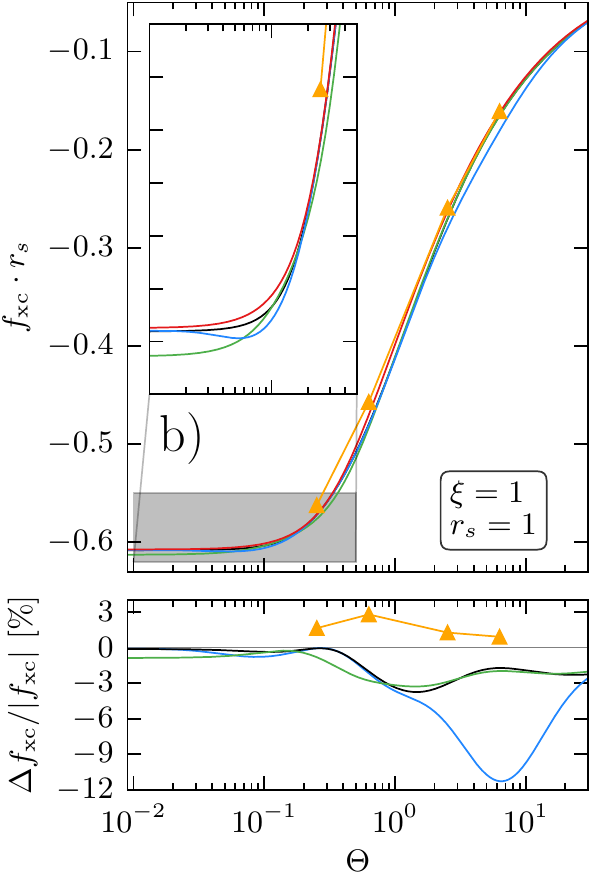}

\includegraphics[width=0.4\textwidth]{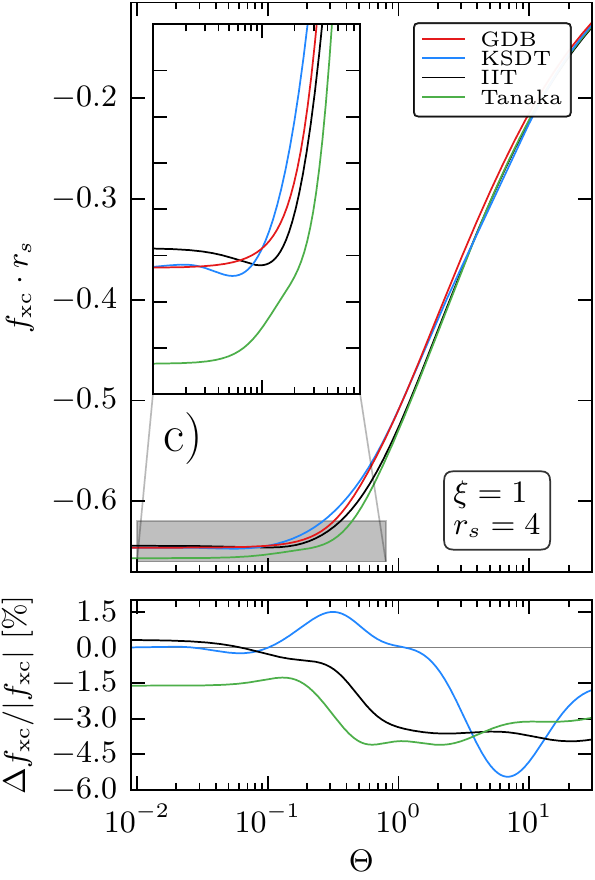}
\hspace*{1cm}\includegraphics[width=0.4\textwidth]{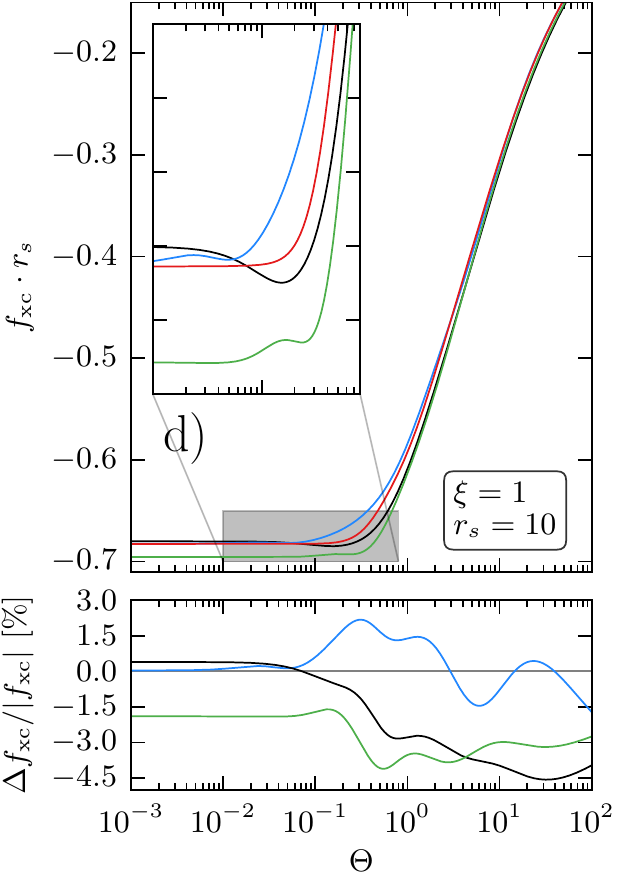}
\caption{\label{fig:panel_polarized_fxc}Temperature dependence of the exchange-correlation free energy of the spin-polarized electron gas for $r_s=0.1,1,4,10$ -- Shown are the parametrizations by Groth, Dornheim \textit{et al.} (red line, GDB, Ref.~\cite{groth_ab_2017}),  Karasiev \textit{et al.} (blue line, KSDT, Ref.~\cite{karasiev_accurate_2014}), Tanaka (green line, HNC, Ref.~\cite{tanaka_correlational_2016}), and Ichimaru \textit{et al.} (black line, IIT, Ref.~\cite{ichimaru_statistical_1987,tanaka_improved_2017}) and, for $r_s=1$, data points by Perrot and Dharma-wardana (yellow triangles, PDW, Ref.~\cite{perrot_spin-polarized_2000}).
The bottom panels depict the relative deviation towards the GDB curve and the insets correspond to magnified segments.
}
\end{figure}

For completeness, in Fig.~\ref{fig:panel_polarized_fxc} we show the same information for the spin-polarized electron gas.
Again, we find an overall qualitatively similar behavior as for $\xi=0$, but with increased systematic biases in the various approximations. The KSDT fit exhibits maximum deviations of up to $15\%$ and $12\%$ at the highest depicted densities, $r_s=0.1$ and $r_s=1$, respectively, around $\theta=6$. With increasing coupling strength, these errors decrease with a maximum of $\Delta f_\text{xc}/f_\text{xc}\approx2\%$ at $r_s=10$ around $\theta=0.4$. Moreover, there again appears an unphysical bump in the low temperature limit at low density.
The IIT and HNC parametrizations roughly follow the same behavior as the interaction energy for the ferromagnetic case, cf.~Fig.~\ref{fig:panel_polarized_v}. Interestingly, the maximum deviation of the IIT curve does not appear at intermediate temperature, as for the paramagnetic case, but towards $\theta>10$ at $r_s=10$. Further, we note that the green curve also exhibits some unphysical behavior towards low $\theta$ and large $r_s$, which is similar to the KSDT function.
Finally, let us consider the four PDW data points that are available at $r_s=1$. Somewhat surprisingly, at the present conditions the employed classical mapping constitutes the most accurate of all depicted approximations with a maximum error of $\Delta f_\text{xc}/f_\text{xc}\approx3\%$ around the Fermi temperature.

\subsubsection{Exchange-correlation energy}

\begin{figure}
\includegraphics[width=0.62\textwidth]{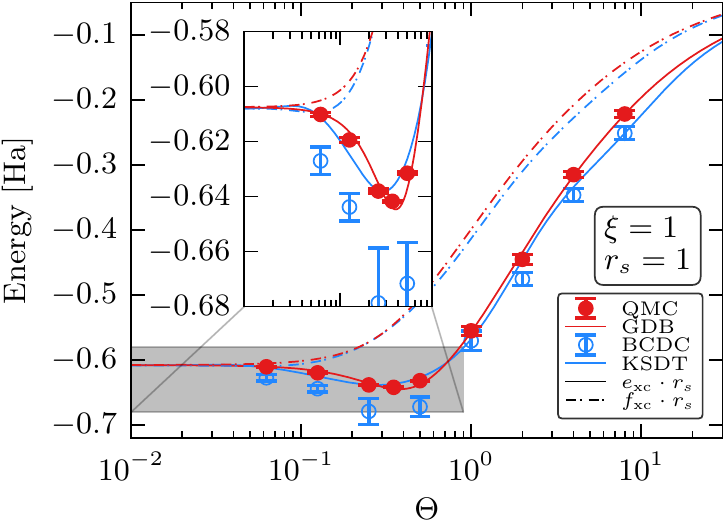}
\caption{\label{fig:exc_crosscheck_prl}
Cross-check of the GDB-parametrization via the exchange-correlation energy -- Shown are the temperature dependence of $e_\text{xc}$ (solid lines and points) and $f_\text{xc}$ (dash-dotted lines) for the spin-polarized electron gas at $r_s=1$.
The red and blue lines correspond to the parametrizations by Groth, Dornheim \textit{et al.}~\cite{groth_ab_2017} and Karasiev \textit{et al.}~\cite{karasiev_accurate_2014} and the red and blue points to our finite-size corrected QMC data (red) and the RPIMC data by Brown \textit{et al.}~\cite{brown_path-integral_2013}.
Reproduced from Ref.~\cite{groth_ab_2017} with the permission of the authors.
}
\end{figure}

Let us now consider another important thermodynamic quantity, i.e., the exchange-correlation energy $e_\text{xc}$, which is connected to $f_\text{xc}$ via Eq.~(\ref{eq:exc_from_fxc}).
Recall that the KSDT functional is actually based on the RPIMC data for $e_\text{xc}$, whereas our GDB parametrization was based on our QMC (and temperature corrected ground state QMC) data for the interaction energy alone.
The main reason for our choice was the, in general, higher statistical uncertainty and greater difficulty of the finite-size correction for the kinetic contribution to the total energy.
Nevertheless, for the ferromagnetic case we were able to obtain accurate QMC data for $e_\text{xc}$ (using CPIMC for $\theta\leq0.5$ and PB-PIMC elsewhere) over the entire temperature-range at $r_s=1$. For completeness, we mention that we applied a twist-averaging procedure~\cite{lin_twist-averaged_2001,drummond_finite-size_2008} for $\theta\leq0.5$ and added an additional finite-size correction onto the QMC data, see Ref.~\cite{groth_ab_2017} for details.
The results are depicted as the red points in Fig.~\ref{fig:exc_crosscheck_prl} and are compared to the exchange-correlation energy that has been computed from the GDB parametrization via Eq.~(\ref{eq:exc_from_fxc}) (solid red line). Evidently, those independent data are in striking agreement over the entire temperature-range. This is an important cross-check for our functional and, in particular, for the temperature-corrected ground state data used for $\theta\leq0.25$, see also the inset showing a magnified segments around the low-temperature regime. The blue circles correspond to the RPIMC data by Brown \textit{et al.}~\cite{brown_path-integral_2013} and are consistently too low over the entire depicted temperature range. The KSDT parametrization (blue solid line), which corresponds to a direct fit to these data, reproduces them for $\theta\geq1$, leading to an unphysical dent for $4\lesssim\theta\lesssim20$ until the correct Debye-H\"uckel limit is attained.
At low temperature, the KSDT curve does not reproduce the RPIMC input data, but performs significantly better, which is due to the incorporation of the exact ground state and high-density limits, which preclude this unphysically deep minimum at $r_s=1$.

\begin{figure}
\includegraphics[width=0.42\textwidth]{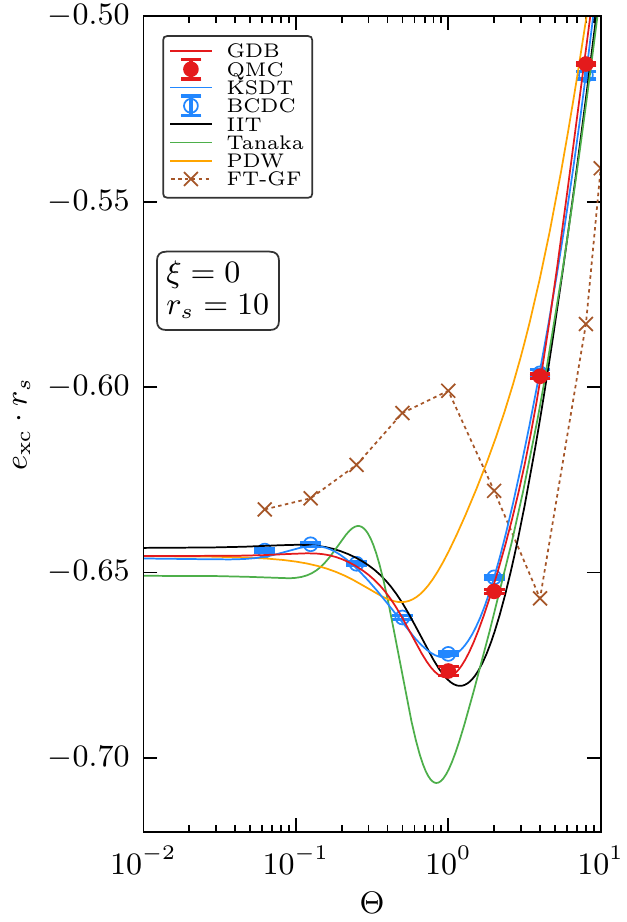}
\hspace*{1cm}\includegraphics[width=0.42\textwidth]{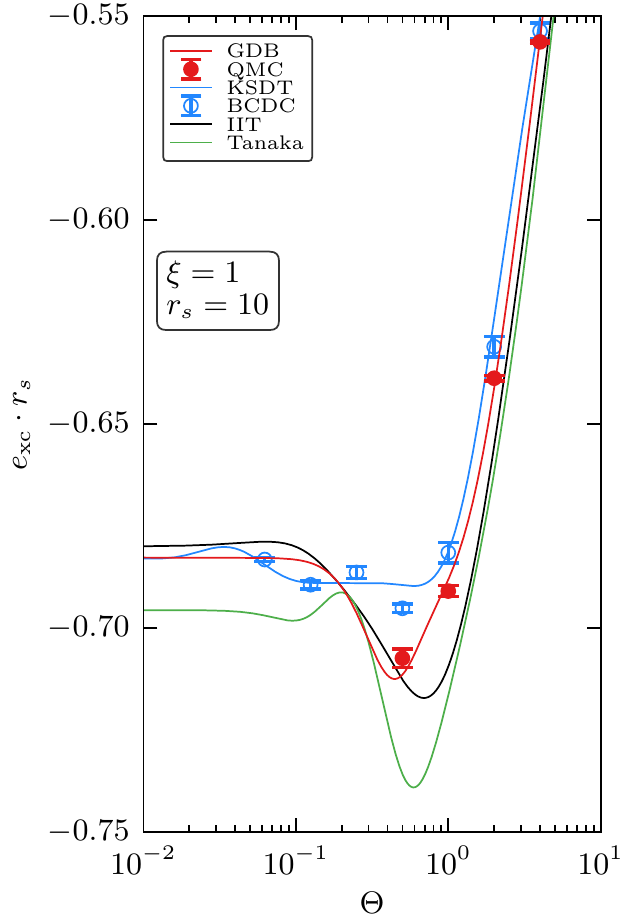}
\caption{\label{fig:exc_rs10}
Temperature dependence of the exchange-correlation energy of the unpolarized (left) and spin-polarized (right) electron gas at $r_s=10$ -- Shown are results computed from the parametrizations by Groth, Dornheim \textit{et al.} (red line, GDB, Ref.~\cite{groth_ab_2017}),  Karasiev \textit{et al.} (blue line, KSDT, Ref.~\cite{karasiev_accurate_2014}), Tanaka (green line, HNC, Ref.~\cite{tanaka_correlational_2016}), Ichimaru \textit{et al.} (black line, IIT, Ref.~\cite{ichimaru_statistical_1987,tanaka_improved_2017}), and Perrot and Dharma-wardana (yellow line, PDW, Ref.~\cite{perrot_spin-polarized_2000}).
In addition, we include the RPIMC data by Brown \textit{et al.}~\cite{brown_path-integral_2013} (BCDC, blue circles) and our recent finite-size corrected QMC results (red points, QMC). For completeness, we also compare with the very recent results of Kas and Rehr~\cite{kas_finite_2017}, which have been obtained from a refined finite temperature Green's function approach.
}
\end{figure}

Next, we investigate the performance and consistency of the various parametrizations with respect to $e_\text{xc}$ at $r_s=10$, starting with the unpolarized case (Fig.~\ref{fig:exc_rs10}, left panel). For these conditions, we were able to obtain independent finite-size corrected QMC data down to $\theta=0.5$ that has not been included in the construction of the functional. Again, the exchange-correlation energy computed from our GDB-parametrization via Eq.~(\ref{eq:exc_from_fxc}) is in perfect agreement with our QMC data for all temperatures. 
The RPIMC data (blue circles) and the KSDT fit to these data (blue line) are also in good agreement even at low temperature, which is in contrast to $r_s=1$. Overall, there occur only small deviations to our data, although there does appear a small bump towards low temperature, which is connected to an unphysical negative entropy~\cite{burke_exact_2016}.
The improved STLS parametrization by Ichimaru \textit{et al.}~\cite{ichimaru_statistical_1987,tanaka_improved_2017} (black line) is of a similar quality to the KSDT curve and gives systematically too low results for $\theta\gtrsim1$.
In constrast, the green line, which corresponds to the recent parametrization of the HNC-LFC data by Tanaka~\cite{tanaka_correlational_2016}, exhibits a substantially different behavior. While it is quite accurate for $\theta\gtrsim2$, it shows a significantly too deep minimum around $\theta=0.8$ followed by a pronounced unphysical bump at $\theta=0.25$. The classical-mapping based parametrization by Perrot and Dharma-wardana~\cite{perrot_spin-polarized_2000} (yellow curve) clearly gives the least accurate data for $\theta\geq1$. 

In addition, for the exchange-corrlation energy, we can also compare with the very recent results by Kas and Rehr~\cite{kas_finite_2017} (brown crosses), which have been computed via a refined finite temperature Green's function procedure (FT-GF). For the exchange-correlation energy of the unpolarized UEG, we can also perform the comparison for this quantity and thereby gauge the accuracy of this new approach. Surprisingly, at these parameters, the corresponding data exhibit a completely unphysical behavior with an additional local maximum in $e_\text{xc}$ at $\theta\sim 1$, where both our \textit{ab initio} functional and independent QMC data (red points) predict a minimum. Even at higher temperatures, the systematic bias of the FT-GF results is largest compared to all other depicted approaches.

Let us conclude this section with a brief discussion of the spin-polarized case, which is shown in the right panel of Fig.~\ref{fig:exc_rs10}. Again, we observe perfect agreement between our QMC data and the GDB-parametrization for all temperatures. While the KSDT curve is also in good agreement with the underlying RPIMC data, there appear significantly larger deviations towards our results. In particular, there abruptly appears a plateau between $\theta\approx 0.9$ and $\theta=0.1$, followed by an unphysical bump before the ground state limit is reached. In contrast, the IIT parametrization gives a qualitatively more similar behavior with respect to the red curve, although the overall accuracy is comparable to KSDT.
Finally, the HNC parametrization again exhibits a too deep minimum and, in addition, does not incorporate the correct ground state limit.

\subsubsection{Spin-dependency of the parametrizations}

\begin{figure}
\includegraphics[width=0.4\textwidth]{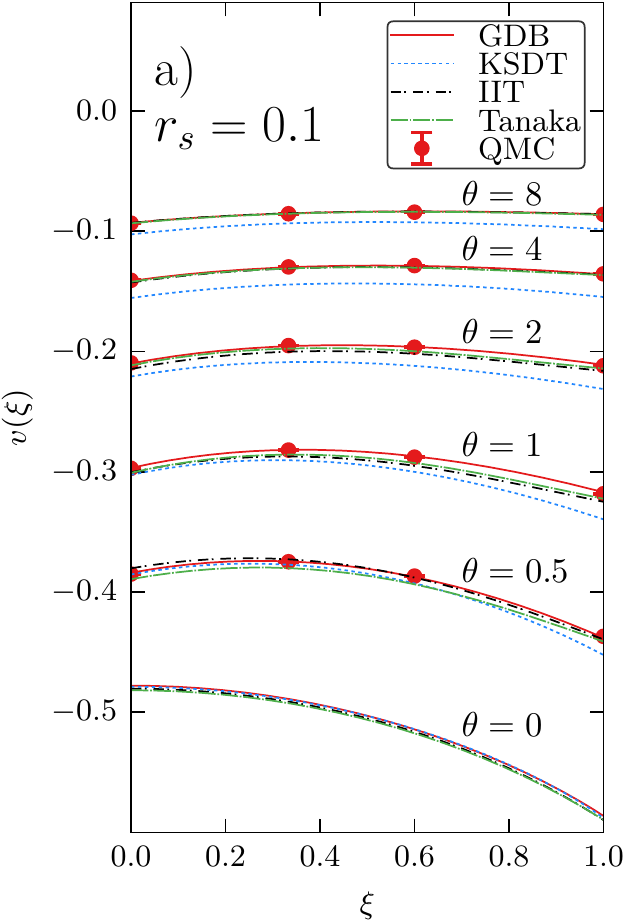}
\hspace*{1cm}\includegraphics[width=0.4\textwidth]{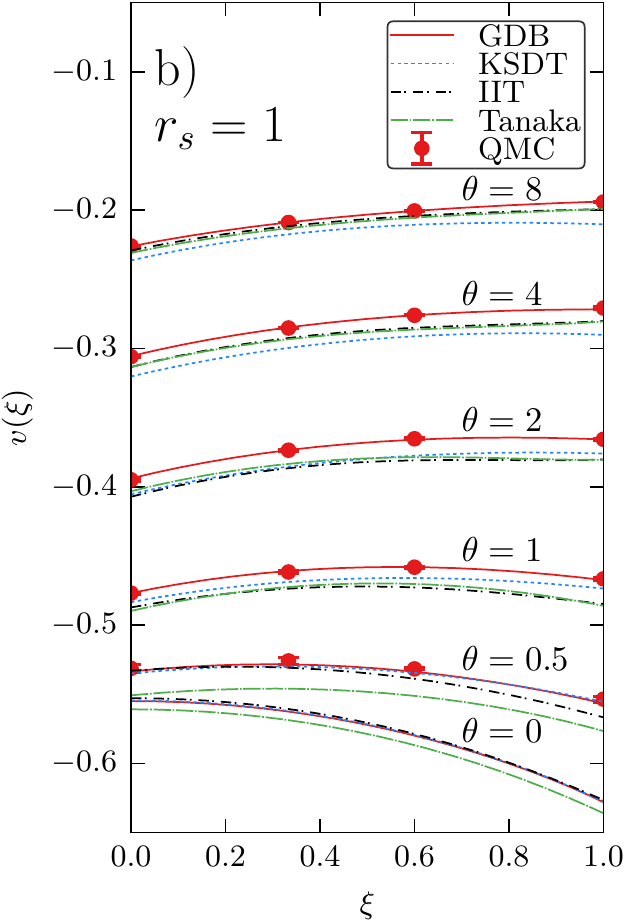}
\vspace*{1cm}

\includegraphics[width=0.4\textwidth]{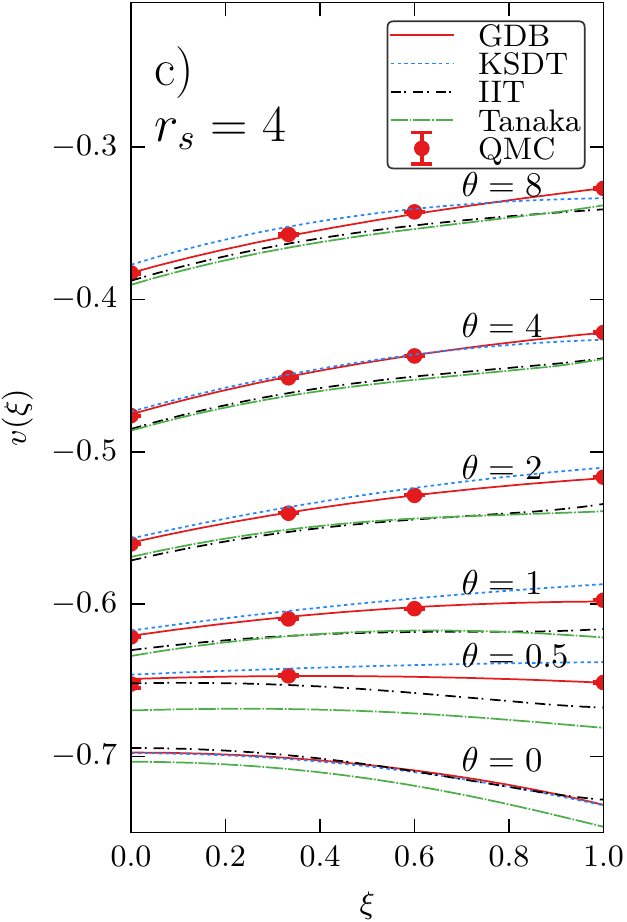}
\hspace*{1cm}\includegraphics[width=0.4\textwidth]{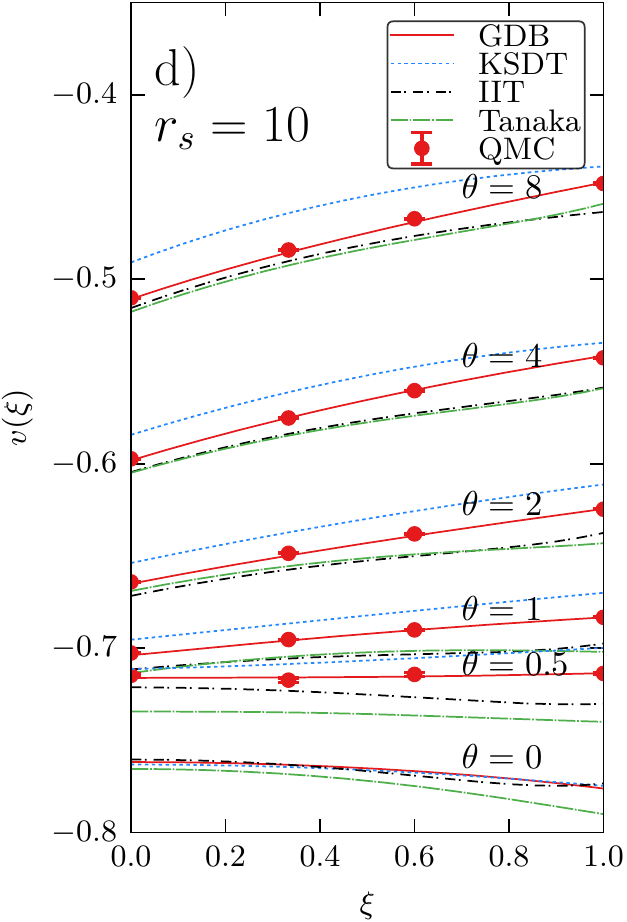}
\caption{\label{fig:v_spin_dependence}
Spin-dependency of the interaction energy of the uniform electron gas -- Shown are the parametrizations by Groth, Dornheim \textit{et al.}~(GDB, Ref.~\cite{groth_ab_2017}, red solid line), Karasiev \textit{et al.}~(KSDT, Ref.~\cite{karasiev_accurate_2014}, blue dotted line), Ichimaru, Tanaka~\textit{et al.}~(IIT, Refs.~\cite{ichimaru_statistical_1987,tanaka_spin-dependent_1989,tanaka_improved_2017}, black dash-dotted line), and the recent HNC-based function by Tanaka~(Ref.~\cite{tanaka_correlational_2016}, dashed green). The red points correspond to our finite-size corrected thermodynamic QMC data from Refs.~\cite{dornheim_abinitio_2016-1,groth_ab_2017}. Note that we define the Fermi energy in the reduced temperature with respect to the spin-up electrons for all polarizations, cf.~Eq.~(\ref{eq:theta_definition}), which is different from the definitions in parts of the literature~\cite{tanaka_spin-dependent_1989,karasiev_accurate_2014,perrot_spin-polarized_2000}.
At $r_s=4$ and $r_s=10$, the $\theta=0$ curves are shifted downward by $0.05$ Hartree for better visibility.
}
\end{figure}

In Fig.~\ref{fig:v_spin_dependence}, we show the spin-dependency of the interaction energy of the uniform electron gas for four different densities and six relevant temperatures. Note that we always define the Fermi energy entering the reduced temperature $\theta$ with respect to the spin-up electrons, cf.~Eq.~(\ref{eq:theta_definition}), which is different from definitions in parts of the relevant literature~\cite{tanaka_spin-dependent_1989,tanaka_correlational_2016,perrot_spin-polarized_2000,karasiev_accurate_2014}. The red points correspond to our recent finite-size corrected thermodynamic QMC data~\cite{dornheim_abinitio_2016-1,groth_ab_2017}, which is available at two intermediate spin-polarizations, $\xi=1/3$ and $\xi=0.6$. We stress that these data still constitute the only \textit{ab initio} investigation of the $\xi$-dependency of the warm dense electron gas. 
The solid red line depicts our GDB-parametrization~\cite{groth_ab_2017}, which utilizes the spin-interpolation between the para- and ferromagnetic limits from Eq.~(\ref{eq:phi}). Surprisingly, we find that a single free parameter [$\lambda_1$ in Eq.~(\ref{eq:phi})] is sufficient to accurately describe the temperature-dependence of the spin-interpolation, resulting in an average and maximum deviation between parametrization and QMC data of $0.15\%$ and $0.8\%$, respectively, at intermediate $\xi$. 
The dotted blue curve corresponds to the functional by Karasiev \textit{et al.}~\cite{karasiev_accurate_2014}, who used the same functional form as the GDB-parametrization. However, due to the lack of RPIMC data for $0<\xi<1$, they determined the $\theta$-dependent parameters in Eq.~(\ref{eq:phi}) from a fit to the sparse classical-mapping data from Perrot and Dharma-wardana~\cite{perrot_spin-polarized_2000} (12 values for $f_\text{xc}$ at $r_s=1, 3, 6$ and $\xi=0.6$).
At zero temperature, KSDT and GDB are in excellent agreement as both utilize the same ground state QMC data~\cite{spink_quantum_2013} to construct the ground state limit for all values of $\xi$. Towards higher temperatures, there occur increasing deviations that are most pronounced (in terms of the relative deviation) at $r_s=0.1$ and $\theta=4,8$. This is again a consequence of the lack of input data for the KSDT functional for $r_s<1$ at finite temperature, and the poor quality of the RPIMC data at $r_s=1$ for the $\xi=0,1$ limits.

\begin{figure}
\includegraphics[width=0.4\textwidth]{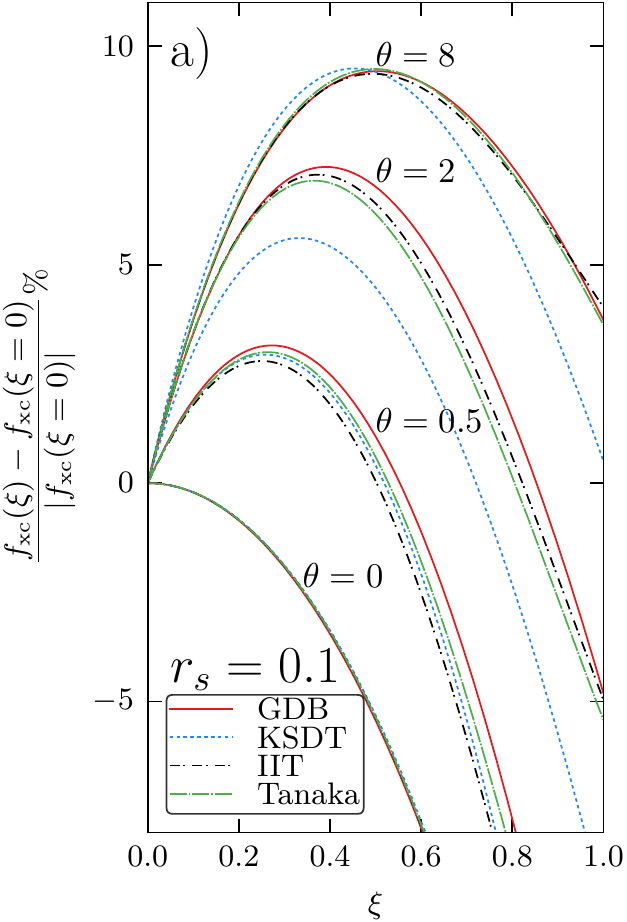}
\hspace*{1cm}\includegraphics[width=0.4\textwidth]{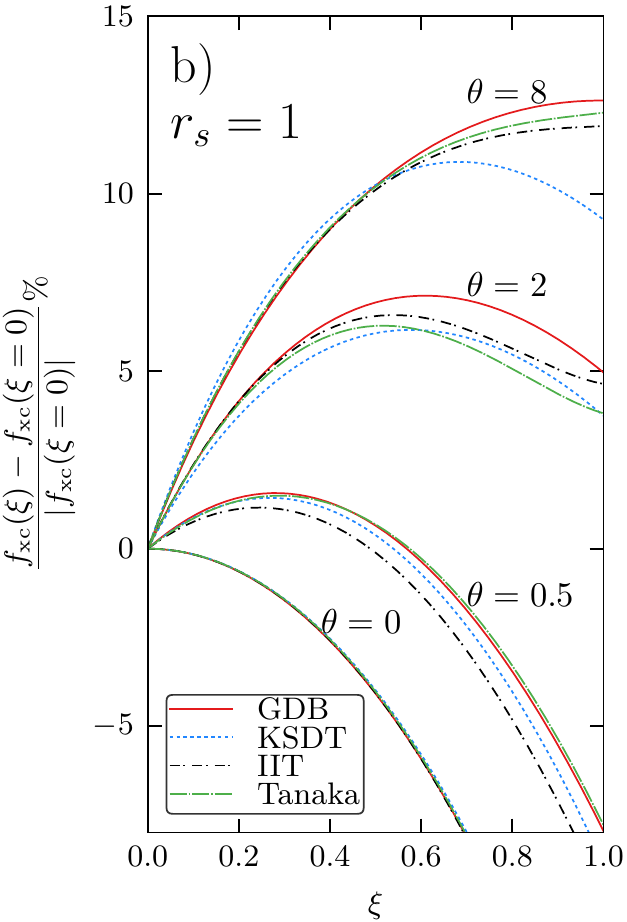}

\vspace*{1cm}
\includegraphics[width=0.4\textwidth]{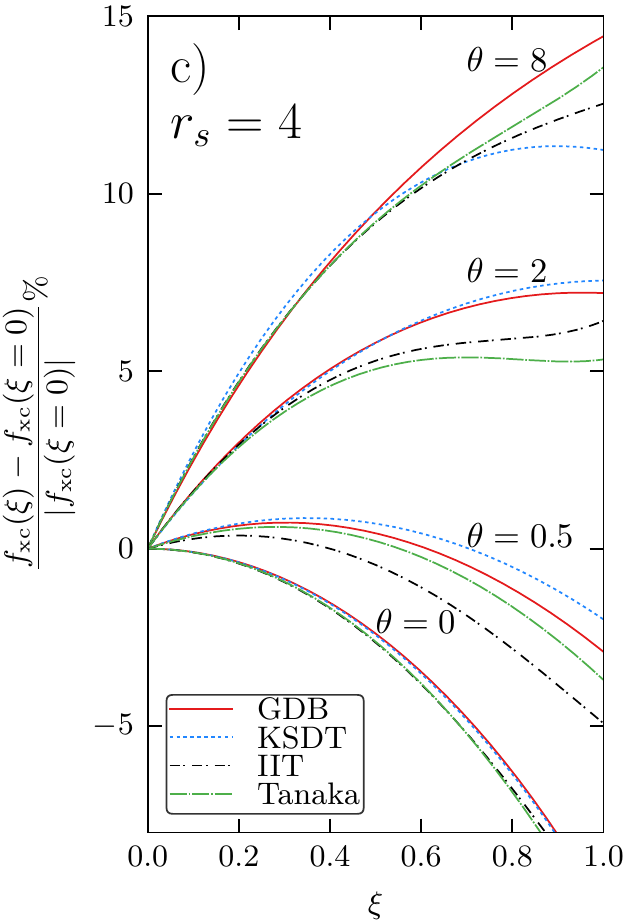}
\hspace*{1cm}\includegraphics[width=0.4\textwidth]{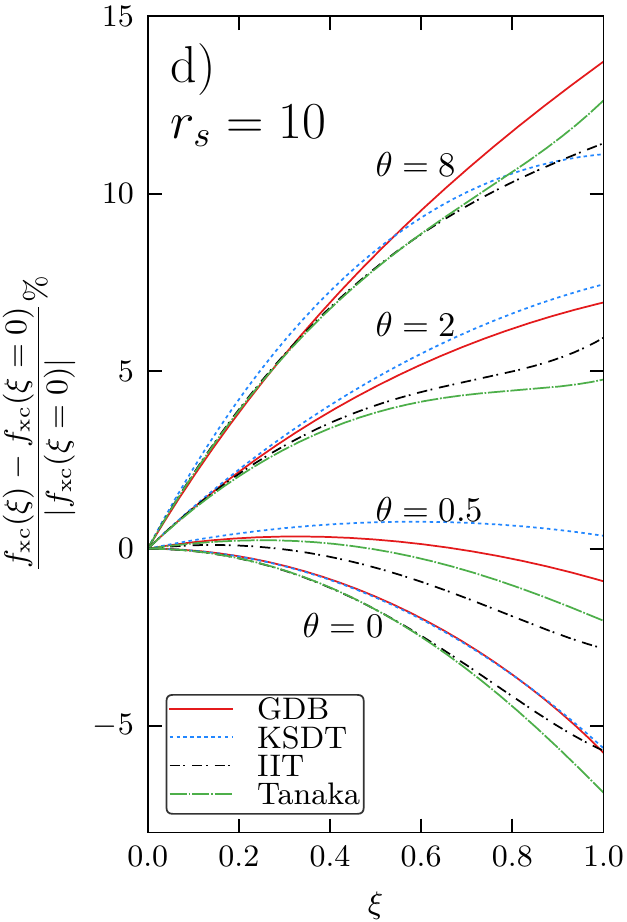}
\caption{\label{fig:fxc_spin_dependence}
Spin-dependency of the exchange-correlation free energy of the uniform electron gas -- Shown are the parametrizations by Groth, Dornheim \textit{et al.}~(GDB, Ref.~\cite{groth_ab_2017}, red solid line), Karasiev \textit{et al.}~(KSDT, Ref.~\cite{karasiev_accurate_2014}, blue dotted line), Ichimaru, Tanaka~\textit{et al.}~(IIT, Refs.~\cite{ichimaru_statistical_1987,tanaka_spin-dependent_1989,tanaka_improved_2017}, black dash-dotted line), and the recent HNC-based function by Tanaka~(Ref.~\cite{tanaka_correlational_2016}, dashed green). Note that we define the Fermi energy in the reduced temperature with respect to the spin-up electrons for all polarizations, cf.~Eq.~(\ref{eq:theta_definition}), which is different from the definitions in parts of the literature~\cite{tanaka_spin-dependent_1989,karasiev_accurate_2014,perrot_spin-polarized_2000}.
}
\end{figure}

The dashed-dotted black and dashed green lines correspond to the improved STLS parametrization by Ichimaru, Tanaka, \textit{et al.}~\cite{ichimaru_statistical_1987,tanaka_spin-dependent_1989,tanaka_improved_2017} and the recent HNC-based parametriztion by Tanaka~\cite{tanaka_correlational_2016}, respectively. Both use the finite-temperature spin-interpolation from Eq.~(\ref{eq:spin_interpolation_tanaka}) that has been constructed on the basis of the modified convolution approximation, see Ref.~\cite{tanaka_spin-dependent_1989}.
First and foremost, we note that the two curves do not agree, even in the ground state, since the $\xi=0$ and $\xi=1$ limits in IIT incorporate ground state QMC data, whereas the HNC limits have been constructed solely on the basis of the HNC data. Further, the IIT ground state limit for the $\xi$-dependence, at $r_s=10$, is slightly non-monotonic, with a shallow minimum around $\xi\approx0.8$.
Towards high temperature, the deviations between the IIT and Tanaka parametrizations vanish, since both  the STLS and HNC input data sets for the interaction energy eventually converge. At high density and temperature, we find an excellent agreement to our GDB curve, which is expected in this weak coupling regime. In contrast, towards lower density and temperature, there occur substantial deviations and, in addition, unphysical dents around $\xi\approx0.8$. In summary, we find that the KSDT, IIT and Tanaka curves exhibit, overall, a similar degree of accuracy.

Let us conclude the discussion of the different parametrizations with a comparison of the relative spin-dependency of the exchange-correlation free energy of the uniform electron gas at warm dense matter conditions which is presented in Fig.~\ref{fig:fxc_spin_dependence}.
In the ground state, all four depicted curves are close, although IIT and Tanaka substantially deviate from the other two at $r_s=10$. In this case, IIT attains the correct limit for $\xi=1$ due to the incorporation of ground state QMC data, which is lacking for Tanaka.
Furthermore,, similar to our findings for the interaction energy in Fig.~\ref{fig:v_spin_dependence}, there occur unphysical dents in $f_\text{xc}$ for IIT and Tanaka around $\xi=0.8$, even at $r_s=1$, which are caused by the MCA-based spin-interpolation for $f_\text{xc}$, cf.~Eq.~(\ref{eq:spin_interpolation_tanaka_fxc}). Finally, the KSDT results are best at $r_s=10$, whereas there occur substantial deviations, both, towards high temperature and high density.


\section{Inhomogeneous Electron Gas: QMC study of the static density response\label{sec:response}}

\subsection{Introduction\label{sec:response_intro}}
In Sec.~\ref{sec:LRT}, we gave a comprehensive introduction to the linear response theory of the uniform electron gas at warm dense matter conditions. In particular, we introduced several suitable approximations for the density response of the system to an external harmonic perturbation, which is fully characterized by the frequency-dependent response function $\chi(\mathbf{q},\omega)$, cf.~Eq.~(\ref{eq:chi_lfc}). The gist has been that the consideration of the perturbed system served as a means to an end, as complete knowledge of $\chi(\mathbf{q},\omega)$ allows to compute all static properties of the unperturbed electron gas, such as the static structure factor, $S(\mathbf{k})$, or the interaction energy, $v$.

In contrast, in the following we will consider the calculation of the density response function as an end in itself, as this information is of high importance for many applications~\cite{giuliani2005quantum}. First and foremost, the local field correction, $G(\mathbf{q},\omega)$, defined by Eq.~(\ref{eq:chi_lfc}) is directly related to the exchange-correlation kernel 
\begin{eqnarray} 
K_\text{xc}(\mathbf{q},\omega) = - \frac{4\pi}{q^2} G(\mathbf{q},\omega)\quad ,
\end{eqnarray}
which is the main input for density functional theory calculations in the adiabatic-connection fluctuation-dissipation formulation~\cite{lu_evaluation_2014,patrick_adiabatic-connection_2015,pribram-jones_thermal_2016}. Albeit computationally demanding, this formulation of a true \textit{non-local} XC-functional is a promising approach to go beyond widespread gradient approximations such as PBE~\cite{perdew_generalized_1996} or its recent extension to finite temperature by Karasiev \textit{et al.}~\cite{karasiev_nonempirical_2016}.
In addition, accurate data for the LFCs of the warm dense electron gas are needed for the calculation of the dynamic structure factor $S(\mathbf{q},\omega)$ of real systems (such as two-component plasmas), e.g.~Refs.~\cite{neumayer_plasmons_2010,plagemann_dynamic_2012,fortmann_influence_2010,graziani2014frontiers}. We stress that the cutting-edge theoretical description of $S(\mathbf{q},\omega)$ is among the most pressing goals of current warm dense matter research, as it is nowadays routinely obtained in experiments from x-ray Thomson scattering measurements for many systems, see Ref.~\cite{glenzer_x-ray_2009} for a review.
Further applications of $G(\mathbf{q},\omega)$ include the calculation of electrical and optical conductivities~\cite{reinholz_conductivity_2015,veysman_optical_2016}, energy transfer rates~\cite{vorberger_energy_2010,benedict_molecular_2017}, EOS models of ionized plasmas~\cite{kremp2006quantum,vorberger_equation_2013,chabrier_equation_1998}, and the construction of pseudo-potentials~\cite{starrett_simple_2014,souza_predictions_2014,senatore_local_1996,gravel_nonlinear_2007,moldabekov_ion_2017} that can be used, e.g., within simple molecular dynamics simulations.

In the following, we will explain how the static limit of the density response function,
\begin{eqnarray}\label{eq:static_response}
\chi(\mathbf{q}) \equiv \lim_{\omega\to0} \chi(\mathbf{q},\omega) \quad ,
\end{eqnarray}
can be obtained with high precision from \textit{ab initio} quantum Monte Carlo simulations at warm dense matter conditions.

\subsection{Theory}

At zero temperature, the static density response function was computed from ground state QMC simulations of the harmonically perturbed (and, thus, inhomogeneous) electron gas~\cite{sugiyama_static_1992,bowen_static_1994,moroni_static_1992,moroni_static_1995} back in the first half of the 1990s. Further, these accurate \textit{ab initio} data have subsequently been parametrized by Corradini \textit{et al.}~\cite{corradini_analytical_1998}. In contrast, at warm dense matter conditions, until very recently, there were no \textit{ab initio} data available, and one had to rely on interpolations between known limits, e.g.~Ref.~\cite{gregori_derivation_2007}.
In the following, we will demonstrate how this gap was  closed by extending the idea from Refs.~\cite{bowen_static_1994,sugiyama_static_1992} to finite temperature, in the recent work by Dornheim and co-workers~\cite{dornheim_permutation_2017}.

Consider a modified Hamiltonian of the form
\begin{eqnarray}
 \hat H = \hat H_0 + \hat H_\text{ext}(t) \quad ,
\end{eqnarray} 
where $\hat H_0$ corresponds to the standard Hamiltonian of the unperturbed uniform electron gas, cf.~Eq.~(\ref{eq:UEG_Ham}), and $\hat H_\text{ext}(t)$ denotes an, in general, time-dependent perturbation. In particular, we choose 
\begin{eqnarray}
 \hat H_\text{ext}(t) = 2 A\sum_{i=1}^N \text{cos}\left( \mathbf{r}_i\cdot\mathbf{q}-\Omega\ t\right)\quad,
\end{eqnarray}
i.e., a sinusoidal external charge density (of perturbation wave vector $\mathbf{q}$ and frequency $\Omega$) with the potential
\begin{eqnarray}
\phi_\text{ext}(\mathbf{r},t) = 2A\ \text{cos}\left(\mathbf{r}\cdot\mathbf{q}-\Omega\ t\right) \quad .
\end{eqnarray}
Let us recall the standard definition of the density response function as~\cite{giuliani2005quantum}
\begin{eqnarray}\label{eq:chi_standard_definition}
\tilde\chi(\mathbf{q},\tau) = \frac{-i}{\hbar} \braket{ \left[ \rho(\mathbf{q},\tau),\rho(-\mathbf{q},0) \right] }_0 \Theta(\tau)\ ,
\end{eqnarray}
with $\braket{\dots}_0$ indicating that the thermodynamic expectation value has to be carried out with respect to the unperturbed system. Naturally, Eq.~(\ref{eq:chi_standard_definition}) solely depends on the time difference, $\tau=t-t'$, and on the modulus of the wave vector, i.e. the wavenumber $q$. Further, it is often convenient to consider the Fourier transform of Eq.~(\ref{eq:chi_standard_definition}) with respect to the second argument,
\begin{eqnarray}\label{eq:chi_fft}
\chi(\mathbf{q},\omega) = \lim_{\eta\to0}\int_{-\infty}^\infty \text{d}\tau\ 
e^{(i\omega-\eta)\tau}\tilde\chi(\mathbf{q},\tau)\ .
\end{eqnarray}
However, at the time of this writing, time-dependent QMC simulations are still severely limited by an additional \textit{dynamical sign problem}, e.g.~Refs.~\cite{muhlbacher_real-time_2008,schiro_real-time_2009,schiro_real-time_2010}. Therefore, in the present work, we restrict ourselves to the static limit $\chi(\mathbf{q})$ [cf.~Eq.~(\ref{eq:static_response})], i.e., the density response to a constant (time-independent) perturbation,
\begin{eqnarray}
\phi_\text{ext}(\mathbf{r}) = 2A\ \textnormal{cos}(\mathbf{r}\cdot\mathbf{q})\ .
\end{eqnarray}
Still, we stress that the basic idea that is explained below can be applied within time-dependent simulations, such as the nonequilibrium Green functions technique~\cite{kwong_real-time_2000,stefanucci2013nonequilibrium,bonitz2015quantum}, in exactly the same way.
Note that all time-dependencies are, in the following, dropped for simplicity. 
In particular, $\chi(\mathbf{q})$ characterizes the linear relation between the induced and external charge densities,
\begin{eqnarray}\label{eq:chi}
\rho_\text{ind}(\mathbf{q}) = \frac{4\pi}{q^2}\,\chi(\mathbf{q})\,\rho_\text{ext}(\mathbf{q})\ .
\end{eqnarray}
The external density is straightforwardly obtained from the Poisson equation as
\begin{eqnarray}
\rho_\text{ext}(\mathbf{q}) = \frac{q^2 A}{4\pi}\left( \delta_{\mathbf{k},\mathbf{q}} + \delta_{\mathbf{k},\mathbf{-q}} \right)\ ,
\end{eqnarray}
and, by definition, the induced density is given by the difference between the densities of the perturbed and unperturbed systems,
\begin{eqnarray}\label{eq:rho_direct}
 \rho_\text{ind}(\mathbf{q}) &=& \braket{\hat\rho_\mathbf{q}}_A - \braket{\hat\rho_\mathbf{q}}_0 \\
&=& \nonumber \frac{1}{V}\left<\sum_{j=1}^Ne^{-i\mathbf{q}\cdot\mathbf{r}_j} \right>_A\ .
\end{eqnarray}
We note that the notation $\braket{\dots}_A$ indicates that the thermodynamic expectation value has to be computed in the perturbed system, and that, for the second equality in Eq.~(\ref{eq:rho_direct}) we made use of the fact that $\braket{\hat\rho_\mathbf{q}}_0=0$.
Finally, this gives a simple, direct expression for the static density response function,
\begin{eqnarray}\label{eq:chi_direct}
\chi(\mathbf{q}) = \frac{1}{A} \braket{\hat\rho_\mathbf{q}}_A \quad .
\end{eqnarray}

In practice, we carry out several \textit{ab initio} quantum Monte Carlo calculations of the harmonically perturbed electron gas for each perturbation wave vector $\mathbf{q}=2\pi L^{-1}(a,b,c)^T$ (with $a,b,c\in\mathbb{Z}$), for a variety of amplitudes $A$. This allows us to compute the exact induced density for arbitrarily strong perturbations. In the small $A$-regime, linear response theory is accurate and, thus, Eq.~(\ref{eq:chi_direct}) holds, implying that $\braket{\rho_\mathbf{q}}_A$ is linear in $A$, with $\chi(\mathbf{q})$ being the slope.

For completeness, we mention a second, closely related way to estimate $\chi(\mathbf{q})$ from a QMC simulation of the inhomogeneous system. In linear response theory, the perturbed density profile is given by
\begin{eqnarray}\label{eq:chi_density}
\braket{n(\mathbf{r})}_A = n_0 + 2A\ \text{cos}\left(\mathbf{q}\cdot\mathbf{r}\right)\chi(\mathbf{q})\ ,
\end{eqnarray}
with $n_0$ being the density of the unperturbed system. In particular, the LHS.~of Eq.~(\ref{eq:chi_density}) is easily obtained within a QMC simulation in coordinate space (such as PB-PIMC, but also standard PIMC), and a subsequent cosinusoidal fit gives another estimation of the desired static density response function.

\subsection{\textit{Ab initio} QMC results for the static density response}
In the following section, we will demonstrate the feasibility of computing \textit{ab initio} data for the static density response using QMC methods. In particular, we will focus on two exemplary test cases at low and high density and moderate temperatures to illustrate the range of validity of linear response theory. We will discuss the necessity of finite-size corrections at certain parameters and demonstrate how this can be accomplished, and compare our new data for $\chi(\mathbf{q})$ to the dielectric approximations introduced in Sec.~\ref{sec:LRT}.


\subsubsection{Strong coupling: PB-PIMC results \label{sec:response_pbpimc}}

\begin{figure}\centering
\includegraphics[width=0.44\textwidth]{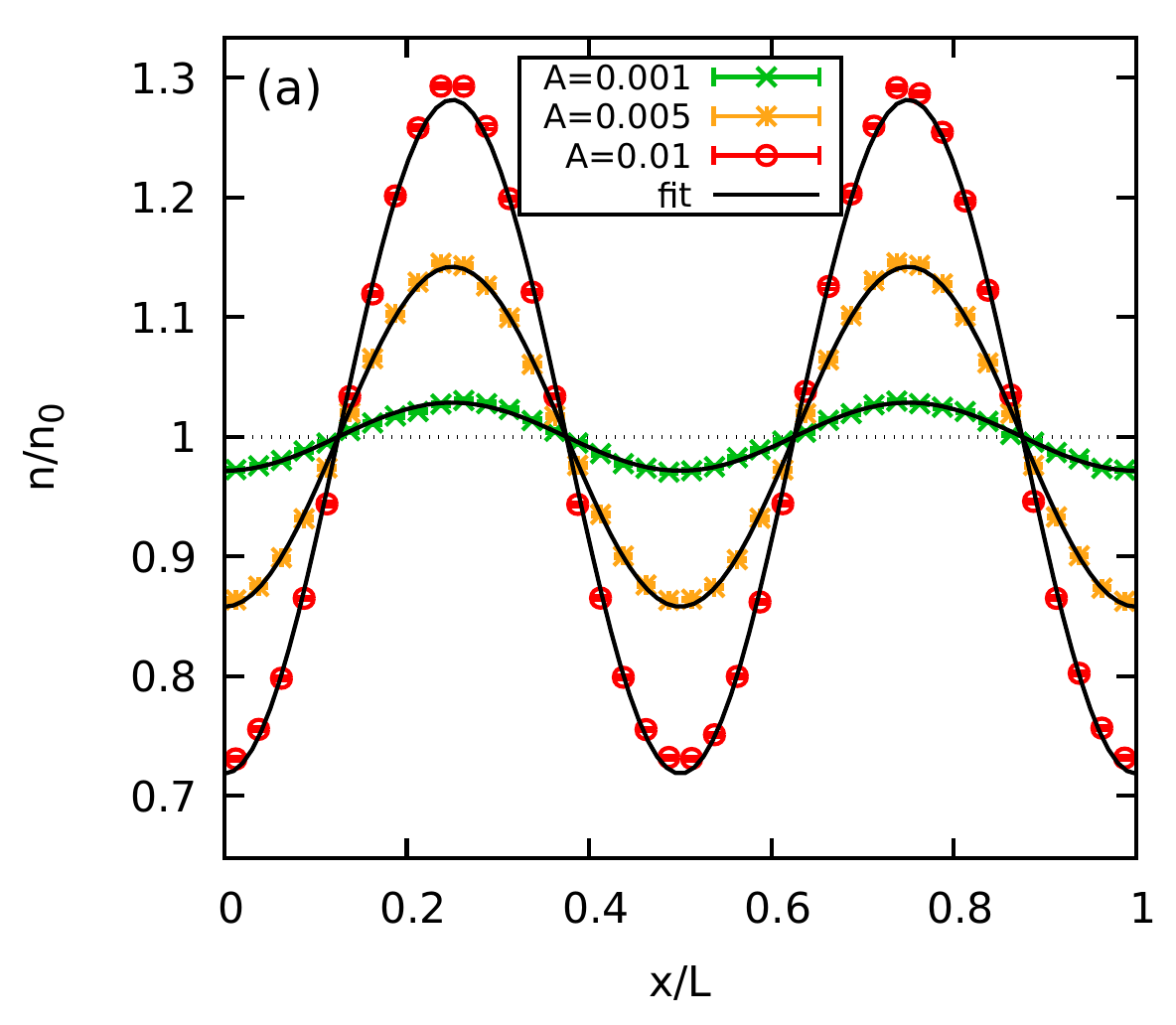}
\includegraphics[width=0.45\textwidth]{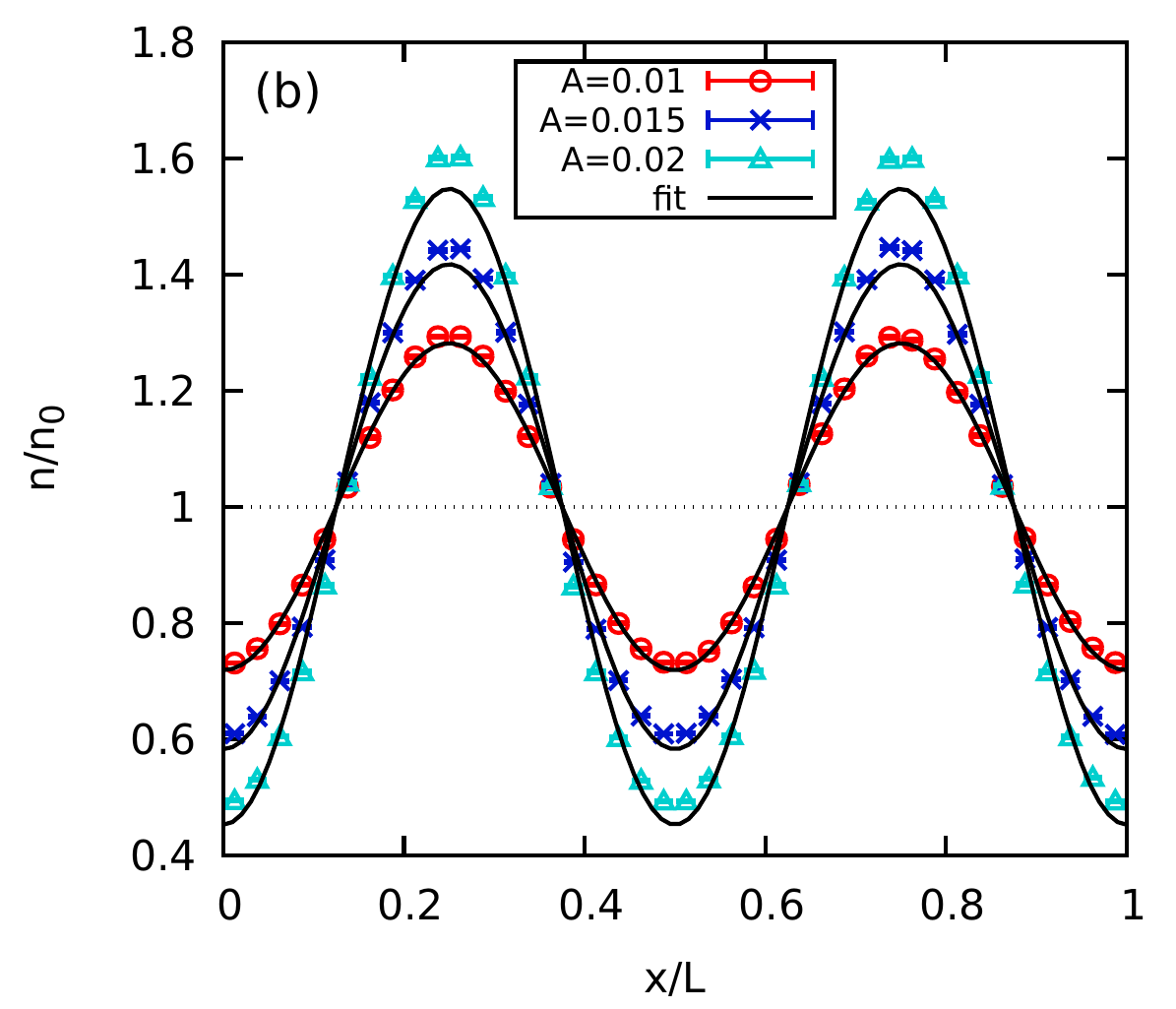}
\floatbox[{\capbeside\thisfloatsetup{capbesideposition={right,center},capbesidewidth=0.45\textwidth}}]{figure}[\FBwidth]
{\caption{Density profile $n(r)$ along the x-direction of a harmonically perturbed electron gas with $r_s=10$ and $\theta=1$ for $N=54$ unpolarized electrons. The results have been obtained using the PB-PIMC method for a wave vector of $\mathbf{q}=2\pi L^{-1}(2,0,0)^T$. The solid black lines depict fits according to Eq.~(\ref{eq:chi_density}) and panels (a), (b), and (c) correspond to  weak, medium, and strong perturbation amplitudes $A$, respectively. Reproduced from Ref.~\cite{dornheim_permutation_2017} with the permission of the authors.\label{fig:density_profile}}}
{\includegraphics[width=0.45\textwidth]{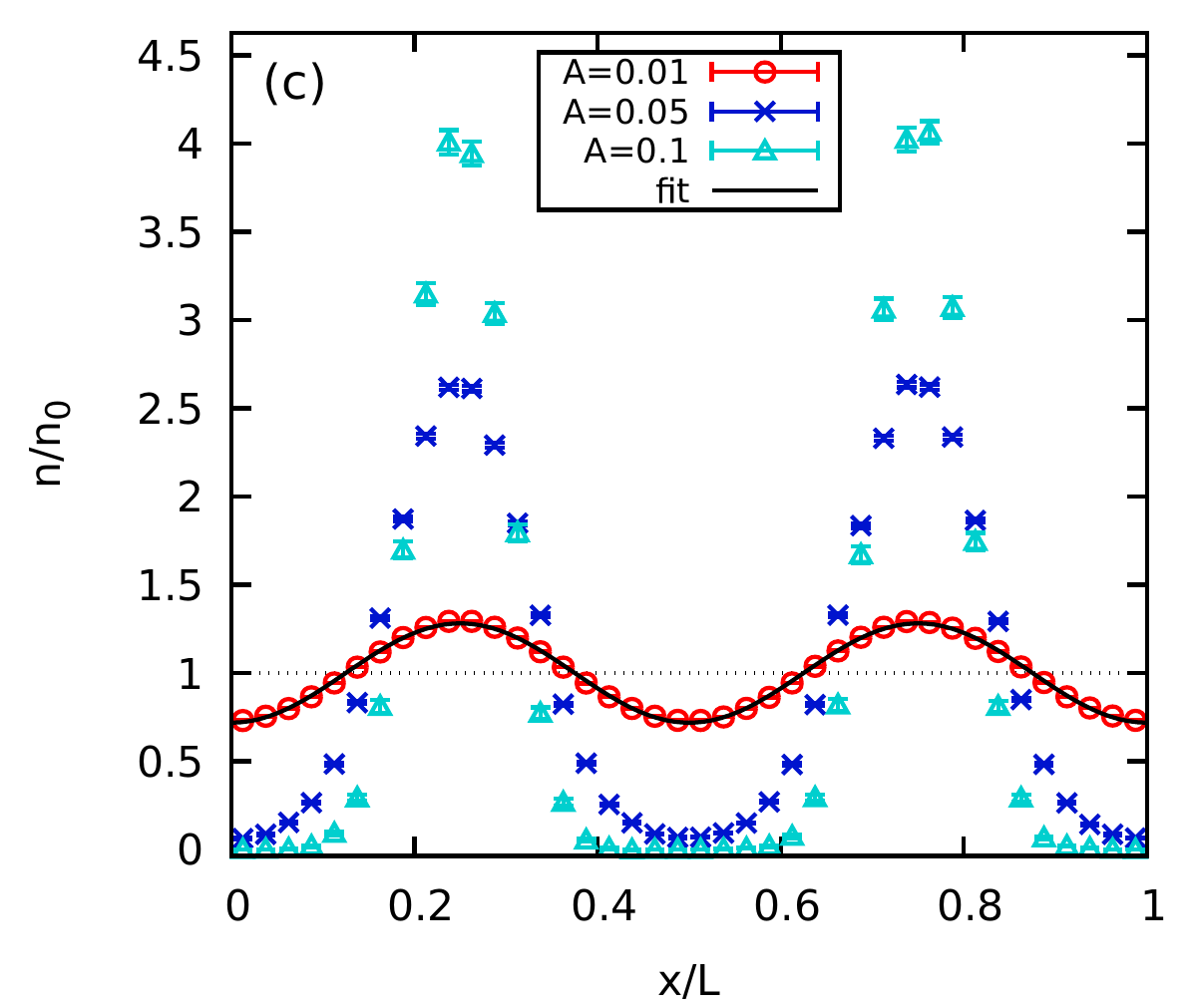}}
\end{figure}

In Fig.~\ref{fig:density_profile}, we show \textit{ab initio} PB-PIMC results~\cite{dornheim_permutation_2017} for the density profile along the $x$ direction (i.e., along the direction of the perturbation wave vector $\mathbf{q}=2\pi L^{-1}(2,0,0)^T$). The simulation was carried out for $N=54$ spin-unpolarized electrons at $r_s=10$ and $\theta=1$, which corresponds to moderate to strong coupling at a moderate temperature. The results for relatively weak perturbation amplitudes $A$ are depicted in panel a). The solid black lines correspond to the cosinusoidal fits according to Eq.~(\ref{eq:chi_density}). Evidently, for the two smallest perturbations ($A=0.001$, green crosses and $A=0.005$, yellow asterisks) no deviations from linear response theory can be resolved. This is a rather remarkable result, as the yellow points exhibit maximum deviations from the mean value, $n_0$, of more than $10\%$, i.e., the system is already moderately inhomogeneous. A doubling of the perturbation amplitude to $A=0.01$ (red circles) leads to density modulations of the order of $25\%$, and deviations from the cosine-fit are clearly visible around the maxima and minima. Still, these differences between data and fit are of the order of $1\%$. 
In panel b), we show density profiles for further increased perturbation amplitudes, $A=0.015$ (blue crosses) and $A=0.02$ (light blue triangles). Evidently, the observed shell structure further departs from the cosinusoidal prediction from LRT, as it is expected. Nevertheless, even at strong inhomogeneity, with density modulations exceeding $50\%$ of the mean value, LRT provides a good quantitative description as the maximum error around the maxima does still not exceed $10\%$.
Finally, in panel c) of Fig.~\ref{fig:density_profile}, we show results for strong perturbations, namely $A=0.05$ (blue crosses) and $A=0.1$ (light blue triangles). Eventually, the system is dominated by the external potential and, for the strongest depicted perturbation amplitude, two distinct shells with negligible overlap are formed. Obviously, Eq.~(\ref{eq:chi_density}) is no longer appropriate and LRT does not capture the dominant physical effects. 
For completeness, we mention that the relatively large statistical uncertainty in the light blue triangles, in particular around the maxima, is a consequence of the fact that the fermion sign problem becomes more severe towards increasing inhomogeneity, see Ref.~\cite{dornheim_permutation_2017} for a more detailed explanation.

\begin{figure}\centering
\includegraphics[width=0.45\textwidth]{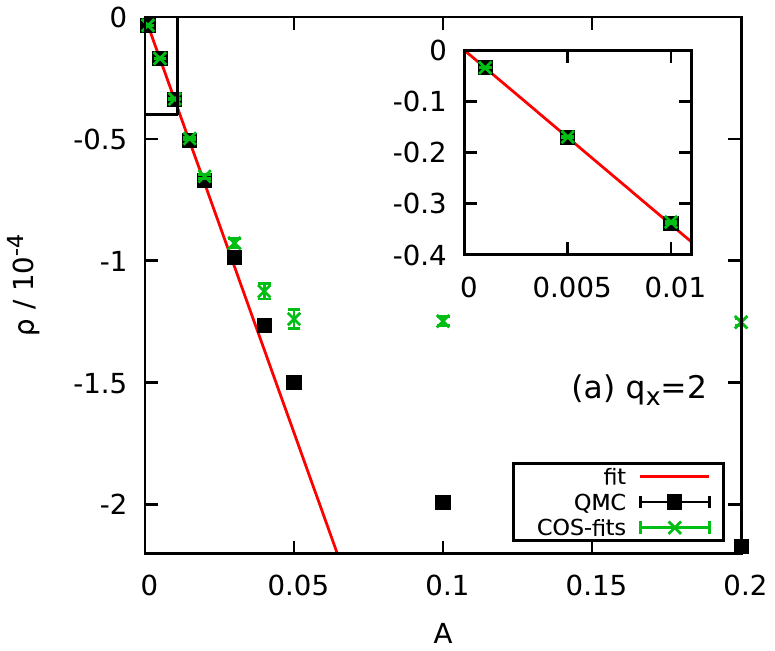}
\includegraphics[width=0.45\textwidth]{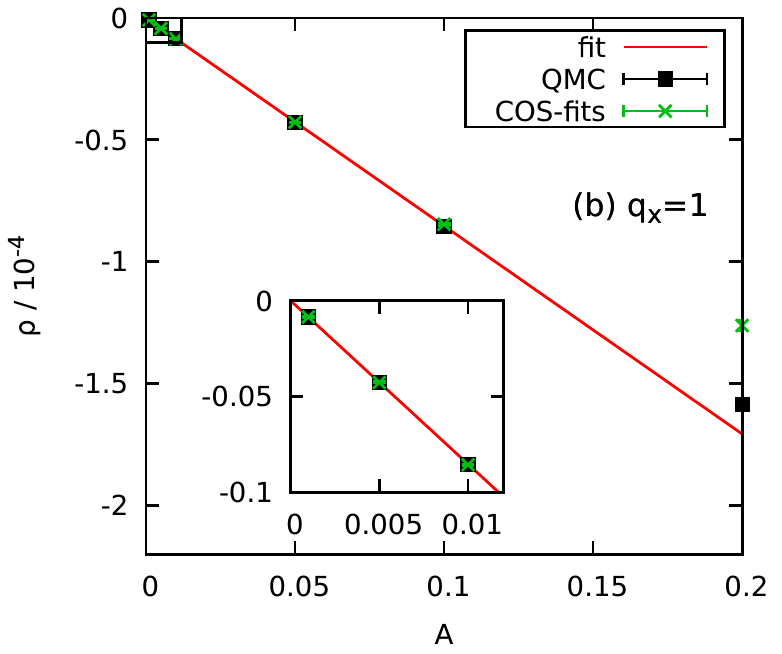}
\caption{\label{fig:A}PB-PIMC results for the perturbation strength dependence of the induced density modulation $\rho_q$ for $N=54$ unpolarized electrons at $r_s=10$ and $\theta=1$. The panels (a) and (b) correspond to the perturbation wave vectors $\mathbf{q}=2\pi L^{-1}(q_x,0,0)^T$ with $q_x=2$ and $q_x=1$, respectively. The black squares correspond to the direct evaluation of Eq.~(\ref{eq:chi_direct}), the green crosses have been obtained from the cosine-fits, cf.~Eq.~(\ref{eq:chi_density}), and the red lines depict linear fits to the QMC points. Reproduced from Ref.~\cite{dornheim_permutation_2017} with the permission of the authors.
} 
\end{figure}

In Fig.~\ref{fig:A}, we show a comparison of the QMC results for the static density response function $\chi(\mathbf{q})$ as obtained from cosinusoidal fits to the density profile (green crosses), cf.~Fig.~\ref{fig:density_profile}, or via the direct evaluation of Eq.~(\ref{eq:chi_direct}) (black squares). More specifically, we show the perturbation strength dependence of the induced density $\rho_\text{ind}(\mathbf{q})$ for two different wave vectors ($\mathbf{q}=2\pi L^{-1}(2,0,0)^T$, panel a) and $\mathbf{q}=2\pi L^{-1}(1,0,0)^T$, panel b).
Further, the solid red line corresponds to a linear fit of the black squares in the small $A$ regime ($A<0.01$). 
Let us start by considering the same $\mathbf{q}$-vector as in the previous figure, i.e., with panel a).
We note the perfect agreement between the cosine-fit and direct results for $\rho_\text{ind}$ for weak perturbations. Interestingly, this still holds for $A=0.01$, where we found visible deviations between density profile and fit, cf.~Fig.~\ref{fig:density_profile} a). With increasing $A$, both sets of data differ from the linear fit, although the deviations of the black squares are significantly smaller.
In panel b), the same information is shown for a smaller wave vector, $\mathbf{q}=2\pi L^{-1}(1,0,0)^T$. Overall, we observe the same trends as in panel a), although the density response is considerably smaller. This is a consequence of screening effects inherent to the uniform electron gas, e.g.~Ref.~\cite{kugler_bounds_1970}. As a consequence, the system is less inhomogeneous, and linear response theory holds up to larger $A$-values than in the former case.

\begin{figure}\centering
\includegraphics[width=0.45\textwidth]{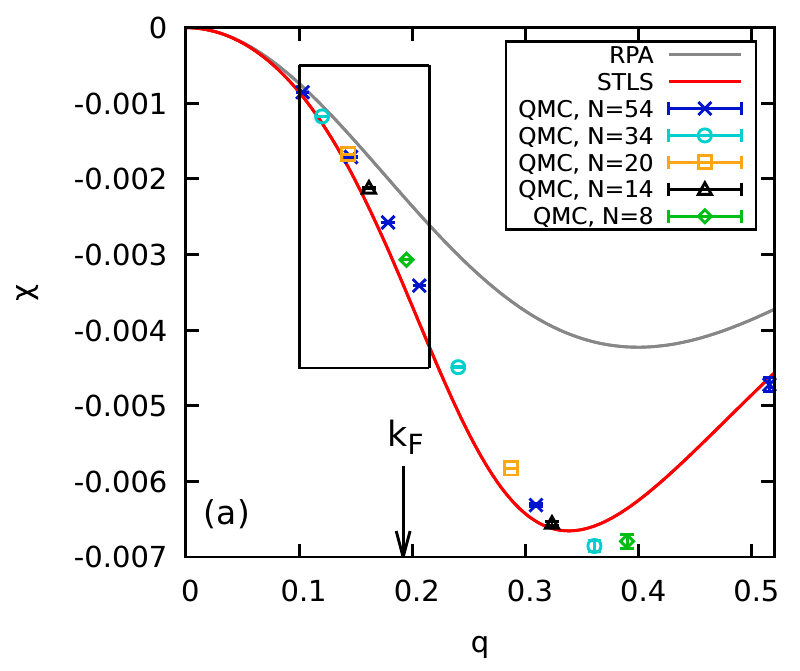}
\includegraphics[width=0.45\textwidth]{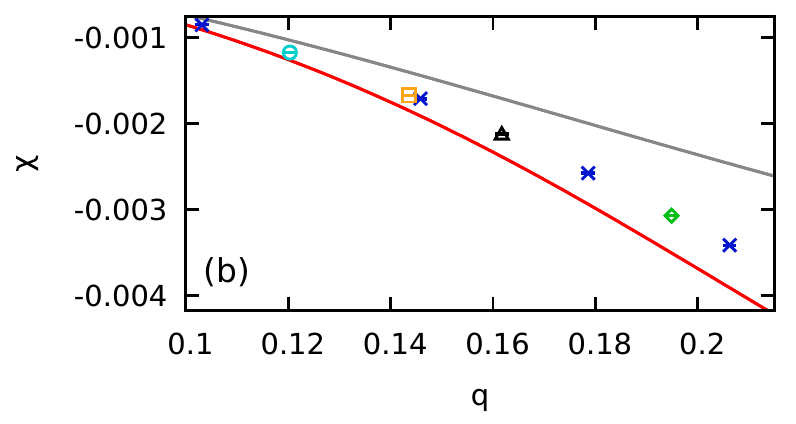}
\caption{\label{fig:curve}
PB-PIMC results for the wave vector dependence of the static density response function $\chi$ for the unpolarized electron gas at $r_s=10$ and $\theta=1$. Shown are QMC results [cf.~Eq.~(\ref{eq:chi_direct})] for different particle numbers $N$ (symbols), and results from dielectric approximations, namely RPA (grey line) and STLS (red line). Further, the black arrow indicated the Fermi wave vector, $k_\text{F}=(9\pi/4)^{1/3}/r_s$. Panel (b) shows a magnified segment.
Reproduced from Ref.~\cite{dornheim_permutation_2017} with the permission of the authors.
} 
\end{figure}

Let us now continue the discussion of the PB-PIMC results for the static density response function by considering the full wave vector dependence of $\chi(\mathbf{q})$, which is depicted in Fig.~\ref{fig:curve} for the same parameters as in the previous figures. The different symbols correspond to $N=54$ (blue crosses), $N=34$ (light blue circles), $N=20$ (yellow squares), $N=14$ (black triangles) and $N=8$ (green diamonds) unpolarized electrons. The main effect of the different system size is the unique $\mathbf{q}$-grid for each $N$, which is a direct consequence of the momentum quantization in the finite simulation cell, cf.~Sec.~\ref{sec:FSC}. The functional form of $\chi(\mathbf{q})$ itself, however, is, for the current set of parameters, remarkably well converged with system size. Even in the right panel, where a magnified segment around the smallest $\mathbf{q}$-values is shown, no finite-size effects in the density response function can be resolved (note that this changes for higher densities, see Sec.~\ref{sec:cpimc_response}). Furthermore, we note that the increased error bars towards large wave vectors are a consequence of the quickly oscillating nature of the external potential in this regime, see Ref.~\cite{dornheim_permutation_2017} for more details.
The solid grey and red lines correspond to dielectric approximations, namely RPA and STLS, respectively. In the small $q$-regime, both curves are in excellent agreement with each other and the parabolic asymptotic behavior known from the literature~\cite{kugler_bounds_1970}. With increasing $q$, however, they substantially deviate with a maximum difference of $\Delta \chi\sim 50\%$ around twice the Fermi vector $k_\text{F}$. In particular, we find that the inclusion of an appropriate local field correction is crucial to account for the pronounced coupling effects at these parameters. Consequently, the STLS approximation (see Sec.~\ref{sec:LRT}) gives significantly improved data for the static density response function, although the agreement with the QMC data is still only qualitative.

\begin{figure}\centering
\includegraphics[width=0.55\textwidth]{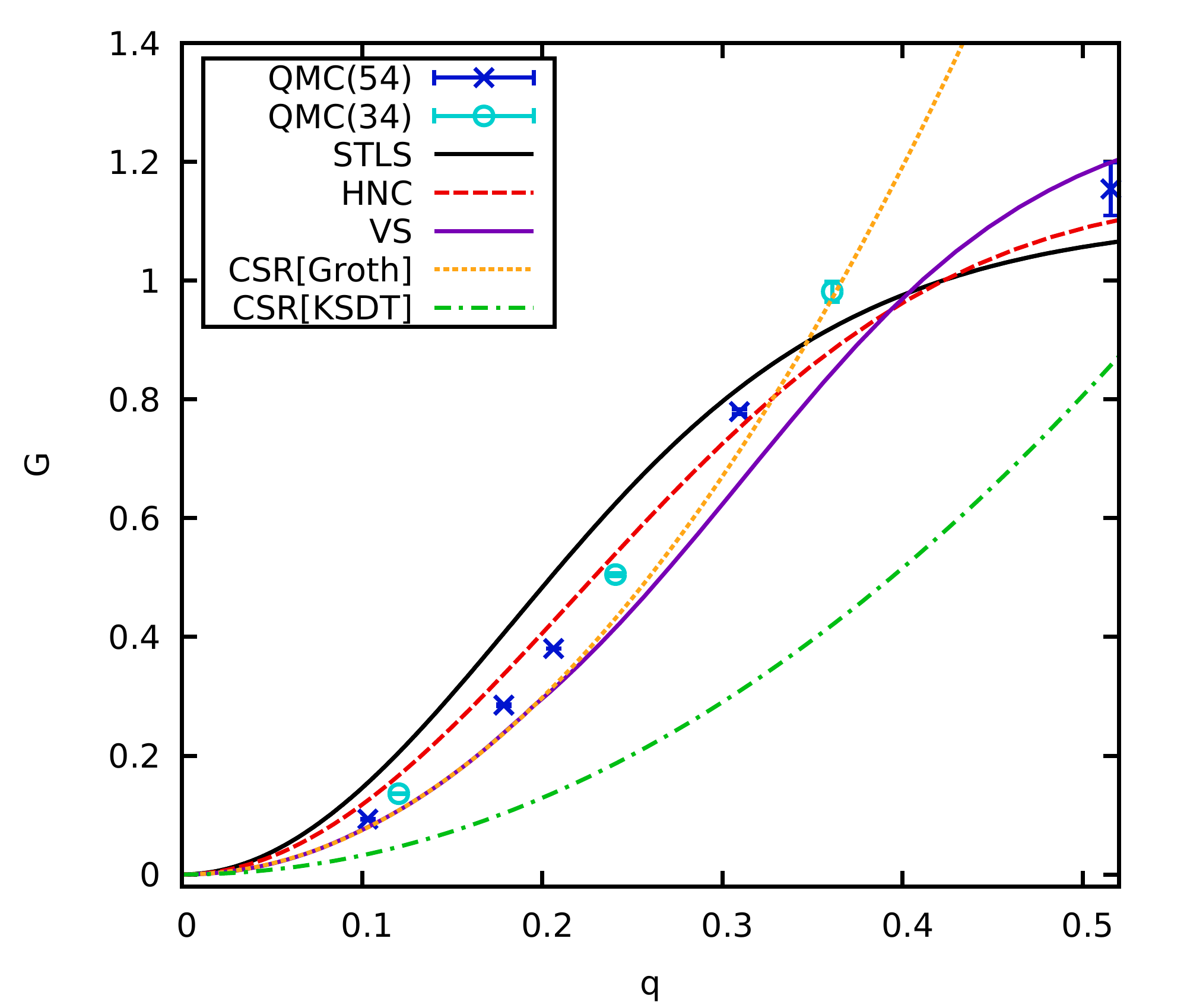}
\caption{\label{fig:LFC_compressibility}
PB-PIMC results for the wave vector dependence of the static local field correction $G$ for the unpolarized electron gas at $r_s=10$ and $\theta=1$. Shown are QMC results [cf.~Eq.~(\ref{eq:obtain_LFC})] for $N=54$ (blue crosses) and $N=34$ (light blue circles), data from STLS (solid black line), the recent LFC based on the HNC equation by Tanaka~\cite{tanaka_correlational_2016} (red dashed line) and Vashista-Singwi (solid purple line, Ref.~\cite{sjostrom_uniform_2013}), and asymptotic long-range predictions from the compressibility sum-rule [cf.~Eq.~(\ref{eq:CSR})] using the exchange-correlation functionals by Groth, Dornheim \textit{et al.}~\cite{groth_ab_2017} (yellow dotted) and Karasiev \textit{et al.}~\cite{karasiev_accurate_2014} (green dash-dotted, KSDT). 
} 
\end{figure}

We conclude this section with a discussion of the \textit{static local field correction}, $G(\mathbf{q})$, which is readily computed from $\chi(\mathbf{q})$, cf.~Eq.~(\ref{eq:obtain_LFC}) below. The results are shown in Fig.~\ref{fig:LFC_compressibility}, where we compare the QMC data for $N=34$ (light blue circles) and $N=54$ (blue crosses) to the static local field correction from STLS theory (solid black line). First and foremost, we note that no system-size dependence can be resolved within the given statistical uncertainty, as expected. Furthermore, the systematic bias in the STLS results is substantially larger than in the total density response function, since $G(\mathbf{q})$ is dominated by exchange-correlation effects. In addition, we note that the recent LFC based on the hypernetted chain equation by Tanaka~\cite{tanaka_correlational_2016} is significantly more accurate than STLS, which is in stark contrast to the corresponding results for the interaction energy $v$, cf.~Sec.~\ref{sec:FSC}. Moreover, the solid purple curve depicts the LFC obtained in the Vashista-Singwi scheme~\cite{sjostrom_uniform_2013} and, overall, exhibits a similar accuracy as the HNC curve.
The dotted yellow and dash-dotted green lines are predictions for the asymptotic behavior of the local field correction based on the compressibility sum-rule, cf.~Eq.~(\ref{eq:CSR}),
using as input the parametrization of $f_\text{xc}(r_s,\theta)$ by Groth, Dornheim \textit{et al.}~\cite{groth_ab_2017} or Karasiev \textit{et al.}~\cite{karasiev_accurate_2014}, respectively (for a review on sum rules in classical and quantum mechanical charged fluids, see Ref.~\cite{martin_sum_1988}).

For completeness, we mention that it is well known that the local field correction from STLS (and also from HNC) does not fulfill Eq.~(\ref{eq:CSR}) and, thus, does not give the correct long-range behavior [in contrast to the total static density response function $\chi(\mathbf{q})$]. In stark contrast, the VS curve is in perfect agreement to the asymptotic expansion, which is somewhat surprising given the systematic bias in the interaction energy itself.
Although both utilized parametrizations for $f_\text{xc}$ deviate by less than four percent, 
at the present conditions, the pre-factors of the parabolic behavior of $G$ differ by more than a factor of two. The reason for this striking deviation is that the compressibility sum-rule is not sensitive to $f_\text{xc}$ itself, but to its second derivative with respect to the density (or the density parameter $r_s$). Evidently, the yellow curve is consistent with the QMC results for the smallest wave vectors, whereas the KSDT prediction does not appear to be better than the STLS curve. Therefore, this investigation of the compressibility sum-rule convincingly demonstrates that a highly accurate parametrization of $f_\text{xc}$ is not only important as input to finite-temperature DFT calculations in the local density approximation. These data are equally important for observables that are related to derivatives of $f_\text{xc}$, e.g., Ref.~\cite{eich_effective_2017}.


\subsubsection{Moderate coupling: CPIMC results}\label{sec:cpimc_response}

\begin{figure}\centering
\includegraphics[width=0.45\textwidth]{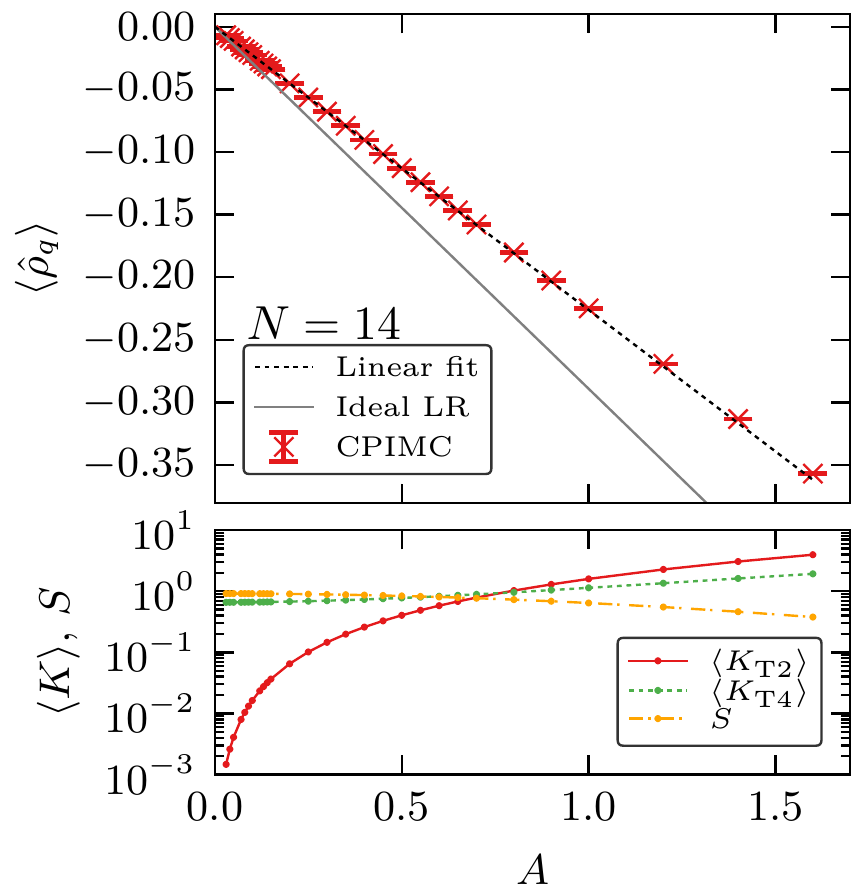}
\caption{\label{fig:A_CPIMC}In the top panel, we show CPIMC results for the perturbation strength dependence of the induced density modulation, $\rho_q$, for $N=14$ unpolarized electrons at $r_s=0.5$ and $\theta=0.5$ with the perturbation wave vector $\mathbf{q}=2\pi L^{-1}(1,0,0)^T$. The red crosses depict the CPIMC data and the dotted black line depicts a corresponding linear fit.
Also shown is the linear response of ideal fermions at the same parameters (solid grey line).
In the bottom panel, we show data for the average numbers of type-2 (red) and type-4 (green) kinks and the average sign (yellow) corresponding to the CPIMC simulations from the top panel.
Reproduced from Ref.~\cite{groth_configuration_2017} with the permission of the authors.
} 
\end{figure}
To obtain highly accurate data for the static density response function of the UEG at high densities, we also extended the CPIMC method to the simulation of the inhomogeneous electron gas, which leads to a significantly more complicated algorithm, compared to the unifrom electron gas. This is due to the substantially larger number of possible diagrams that have to be taken into account. Most importantly, in addition to the two-partical excitations (so-called type-4 kinks) in the CPIMC simulation paths, which are already present in the homogeneous case (see Sec.~\ref{sec:CPIMC}), there also occur one-partical excitations (type-2 kinks). For more details on the specific changes of the CPIMC algorithm we refer to Ref.~\cite{groth_configuration_2017}. 

In the top panel of Fig.~\ref{fig:A_CPIMC}, we show CPIMC results for the induced density, $\rho_\text{ind}(\mathbf{q})$, for $N=14$ unpolarized electrons at moderate coupling and temperature, $r_s=0.5$ and $\theta=0.5$ (red crosses), for the perturbation wave vector $\mathbf{q}=2\pi L^{-1}(1,0,0)^T$. The solid grey line corresponds to the linear response prediction for an ideal system and the dashed black line to a linear fit according to Eq.~(\ref{eq:chi_direct}). Clearly, linear response theory provides a remarkably accurate description of the static density response over the entire depicted $A$-range. 
The bottom panel of Fig.~\ref{fig:A_CPIMC} shows the corresponding simulation results for the average sign and the numbers of type-2 and type-4 kinks. First, we observe that the sign (yellow dash-dotted line) attains an almost constant value for $A<0.5$ and does not drop below $S=0.3$, even for the largest considered perturbation amplitude, explaining the small statistical uncertainty in the results for $\rho_\text{ind}$.
The average number of type-4 kinks (green dotted line) exhibits a qualitatively similar behavior, although with a slight increase towards increasing $A$. In stark contrast, the average number of type-2 kinks (red solid line) distinctly increases with the perturbation strength, as expected. Further, we note that, for weak inhomogeneity, the CPIMC simulation is dominated by Coulomb interaction effects, which manifests itself in the occurrence of type-4 kinks. With increasing $A$ (around $A\sim0.8$), there are on average more type-2 kinks present as the system becomes increasingly altered by the external potential.

\begin{figure}\centering
\includegraphics[width=0.8\textwidth]{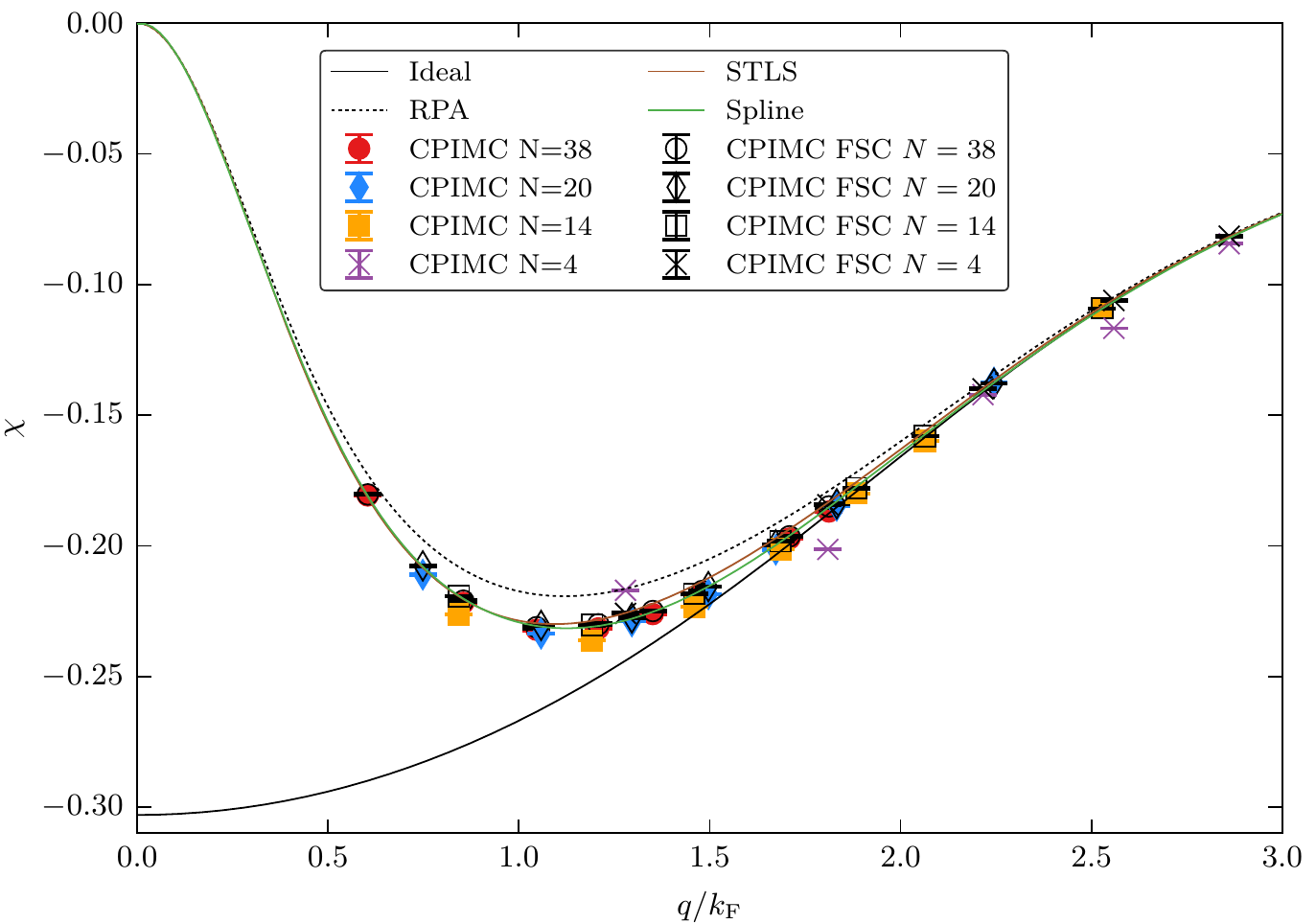}\\
\includegraphics[width=0.8\textwidth]{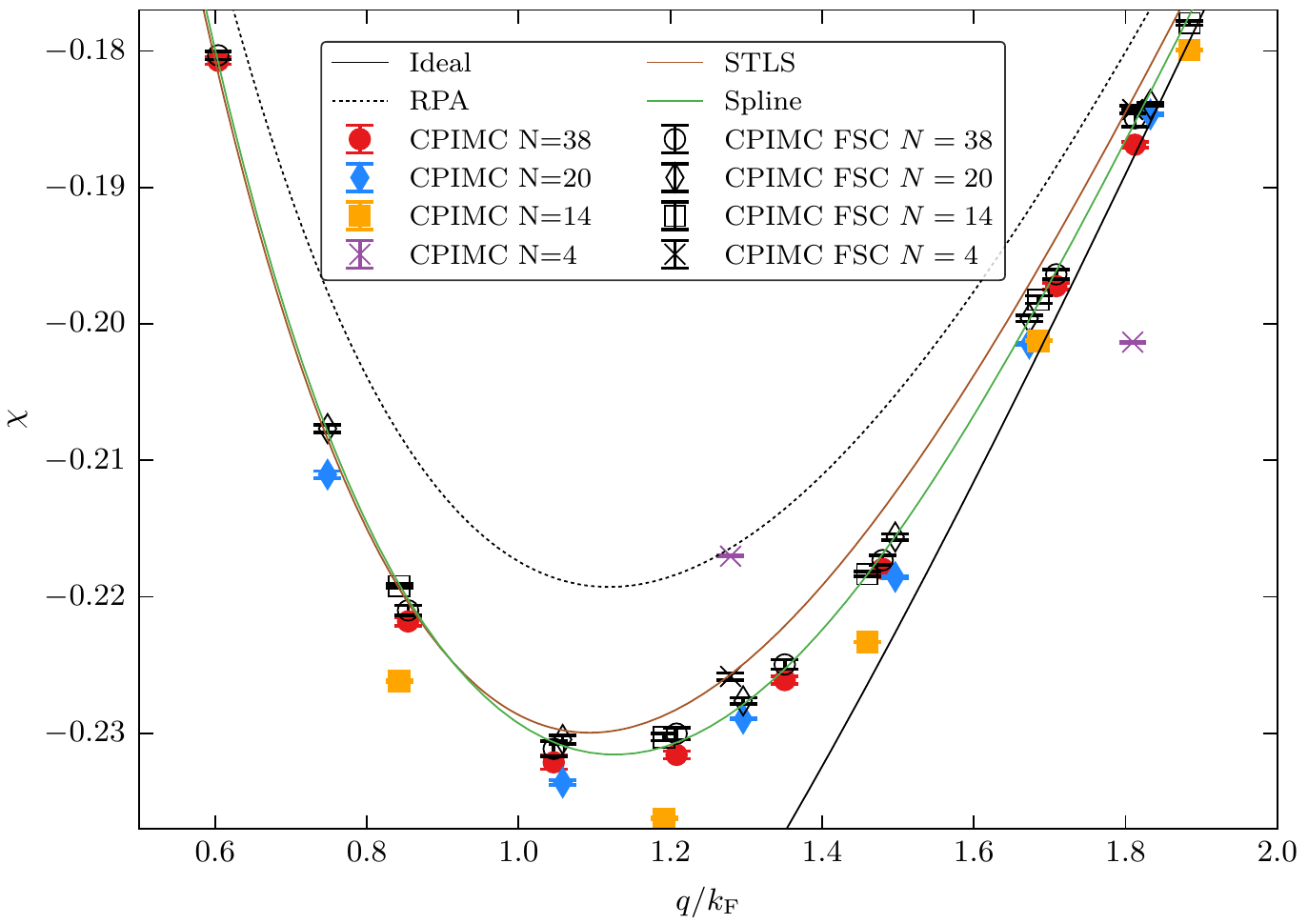}
\caption{\label{fig:wave_CPIMC}Wave vector dependence of the static density response function $\chi$ of the unpolarized electron gas at $r_s=0.5$ and $\theta=0.5$. The colored circles, diamonds, squared and crosses depict the bare CPIMC data for $N=38$, $N=20$, $N=14$, and $N=4$ electrons, respectively, and the corresponding black symbols have been obtained by applying the finite-size correction using the $N$-consistent data for the ideal density response function as explained in the text. The solid green line depicts a spline fit to the black points. Further shown are the ideal response function $\chi_0(q)$ (solid black), and dielectric approximations in RPA (dotted black) and STLS (solid brown). The bottom panel shows a magnified segment.
Reproduced from Ref.~\cite{groth_configuration_2017} with the permission of the authors.
} 
\end{figure}

As we have seen above (cf.~Fig.~\ref{fig:curve}), at $r_s=10$ no system size dependence has been resolved for as few as four electrons. However, it is well known that finite-size effects become more pronounced at higher densities. This is investigated in Fig.~\ref{fig:wave_CPIMC}, where we show the wave vector dependence of the static density response function for the same conditions as in Fig.~\ref{fig:A_CPIMC}. The red circles, blue diamonds, yellow squares, and purple crosses correspond to the raw CPIMC simulation results for $N=38$, $N=20$, $N=14$, and $N=4$ electrons, respectively.
Further, we show results from RPA (dashed black) and STLS (solid brown), as well as the static response function of the corresponding noninteracting system (solid black line). The dielectric approximations exhibit the same exact parabolic behavior for small $\mathbf{q}$ values~\cite{kugler_bounds_1970}, whereas the ideal function attains a maximum at $q=0$. This contrast is a consequence of the absence of Coulomb screening effects in the latter case. 
Further, we note that the inclusion of the static local field correction from STLS theory leads to differences in $\chi(\mathbf{q})$ of around $5\%$, which are most pronounced around the Fermi wave vector $k_\text{F}$. 
Let us now consider the uncorrected CPIMC simulation results. 
Evidently, these data are not converged with respect to system size (see in particular the bottom panel where we show a magnified segment) and, without further improvement, no systematic errors in the STLS curve can be resolved.

At the same time, it is well known from ground state QMC calculations of the static density response function~\cite{moroni_static_1992,moroni_static_1995} that the \textit{static local field correction}, $G$, which contains all information about short-range exchange-correlation effects, can be accurately obtained from simulations of few electrons in a small box, i.e., $G_N(\mathbf{q})\approx G(\mathbf{q})$. Therefore, the bulk of the system size dependence observed in Fig.~\ref{fig:wave_CPIMC} is due to finite-size effects in the ideal density response function, i.e., $\chi^N_0(\mathbf{q})\neq\chi_0(\mathbf{q})$.
In the following, we will exploit this fact to compute the density response function, $\chi^\text{TDL}(\mathbf{q})$, in the thermodynamic limit from the QMC result for a specific, finite number of electrons $N$, $\chi^N(\mathbf{q})$.
For this purpose, we rewrite Eq.~(\ref{eq:chi_lfc}) in terms of finite-size quantities,
\begin{eqnarray}\label{eq:finite_N_LFC}
\chi^N(\mathbf{q}) = \frac{ \chi^N_0(\mathbf{q}) }{ 1 - 4\pi/ q^2[1-G^N(\mathbf{q})] \chi^N_0(\mathbf{q}) } \quad ,
\end{eqnarray}
and solve Eq.~(\ref{eq:finite_N_LFC}) for the local field correction,
\begin{eqnarray}\label{eq:obtain_LFC}
G^N(\mathbf{q}) = 1 + \frac{q^2}{4\pi} \left( \frac{1}{\chi^N(\mathbf{q})} - \frac{1}{\chi^N_0(\mathbf{q}) } \right) \quad .
\end{eqnarray}
The finite-size corrected result for the density response function is then obtained by plugging the QMC result for the static local field correction, Eq.~(\ref{eq:obtain_LFC}), into Eq.~(\ref{eq:chi_lfc}),
\begin{eqnarray}\label{eq:chi_fsc}
\chi^\text{TDL}(\mathbf{q}) = \frac{ \chi_0(\mathbf{q}) }{ 1 + \frac{q^2}{4\pi} \left( \frac{1}{\chi^N(\mathbf{q})} - \frac{1}{\chi^N_0(\mathbf{q}) } \right) \chi_0(\mathbf{q}) } \quad .
\end{eqnarray}

\begin{figure}\centering
\includegraphics[width=0.8\textwidth]{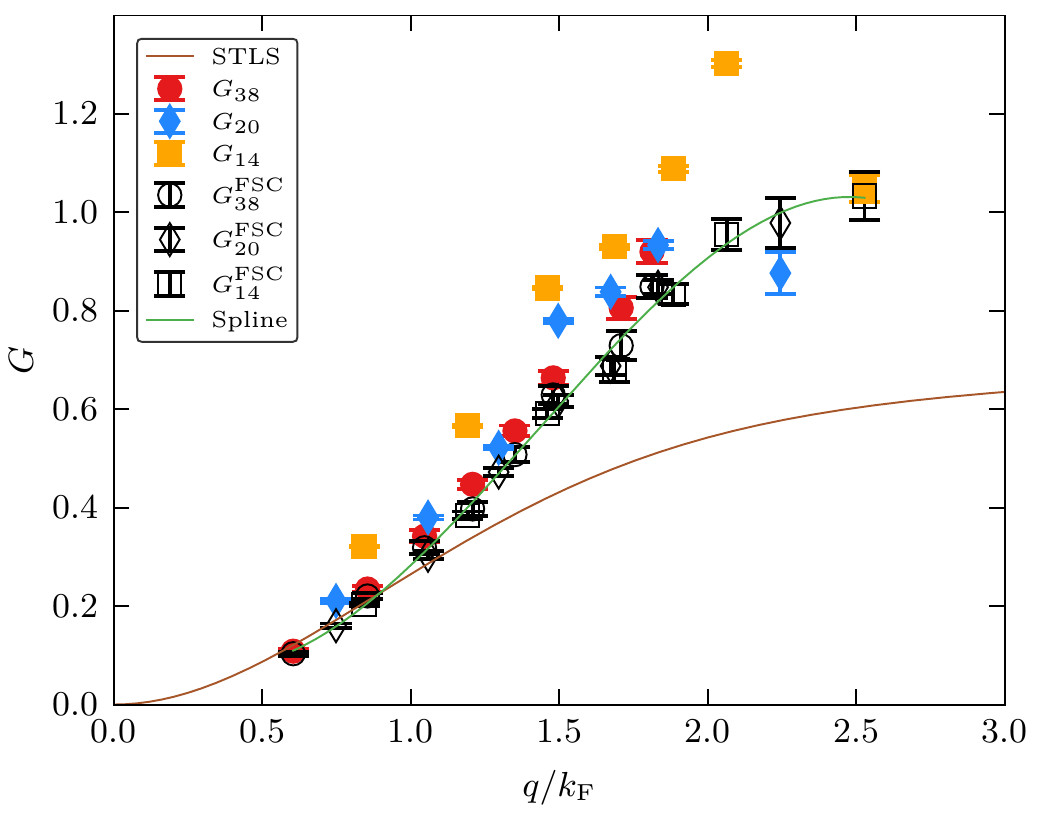}
\caption{\label{fig:G_CPIMC}Wave vector dependence of the static local field correction $G(q)$ for the unpolarized electron gas at $\theta=0.5$ and $r_s=0.5$. The circles, diamonds, and squares have been obtained from CPIMC calculations with $N=38$, $N=20$, and $M=14$ electrons, respectively. The colored symbols correspond to the results using the ideal response function in the thermodynamic limit [i.e., by replacing in Eq.~(\ref{eq:obtain_LFC}) $\chi^N_0$ by $\chi_0$] whereas the black symbols were computed directly from Eq.~(\ref{eq:obtain_LFC}) in a consistent manner by using the ideal response function with the same finite number of electrons as the CPIMC simulations.
Reproduced from Ref.~\cite{groth_configuration_2017} with the permission of the authors.
} 
\end{figure}

Let us now verify the underlying assumption of this finite-size correction procedure, i.e., that $G^N(\mathbf{q})$ does not depend on system size. In Fig.~\ref{fig:G_CPIMC}, we show the wave vector dependence of the local field correction computed from the QMC results for $\chi^N(\mathbf{q})$ depicted in Fig.~\ref{fig:wave_CPIMC}. The black symbols correspond to the direct evaluation of Eq.~(\ref{eq:chi_fsc}). Evidently, no finite-size effects can be resolved within the statistical uncertainty over the entire depicted $q$-range. We note that the increasing error bars towards large wave vectors are a consequence of the reduced impact of $G(\mathbf{q})$ on the total density response function, as it becomes the decreasing difference between two almost equal, large numbers, amplified by the factor of $q^2$. The green line corresponds to a spline fitted to the black symbols and the brown line depicts the local field correction from the STLS formalism. Again, note that the STLS theory does not give the correct asymptotic behavior for $G$ [in contrast to $\chi(\mathbf{q})$] as the compressibility sum-rule is violated, cf.~Sec.~\ref{sec:LRT}. In addition, we observe increasing deviations between the green and brown curves that start around the Fermi wave vector, $k_\text{F}$, and reach values of the order of $50\%$. Despite the good quality of STLS data for, e.g., interaction energies and static structure factors, this is not surprising since $G(\mathbf{q})$ constitutes one of the quantities in many-body theory that is most sensitive to exchange-correlation effects.
For completeness, we mention that the colored symbols in Fig.~\ref{fig:G_CPIMC} were obtained by replacing in Eq.~(\ref{eq:finite_N_LFC}) the size-consistent ideal density response function, $\chi^N_0(\mathbf{q})$, by the analogue in the thermodynamic limit, $\chi_0(\mathbf{q})$. This inconsistency results in significantly biased data for the local field correction, which highlights the necessity to use $\chi_0^N(\mathbf{q})$. We point out that the calculation of the latter is surprisingly involved at finite temperature as, to the best of our knowledge, no readily computable expression (such as the usual spectral representation in the ground state) exists. However, a detailed discussion of this issue is beyond the scope of the present work, for a comprehensive analysis, we refer to Ref.~\cite{groth_configuration_2017}.

Finally, let us examine the thus finite-size corrected data for the static density response function itself, i.e., the black symbols in Fig.~\ref{fig:wave_CPIMC}. Evidently, no system size dependence can be resolved for $N\geq14$, over the entire wave vector range. This allows us to construct a smooth spline fit of these data, which is depicted by the solid green line. In addition, we note that even the results obtained from a CPIMC simulation of as few as four electrons exhibit only minor deviations for intermediate $q$-values. 
We conclude this discussion with a brief comparison of our new accurate data for the static density response function to dielectric theories, namely the above mentioned RPA and STLS curves. Specifically, all curves (apart from the ideal result) exhibit the correct behavior for the limits $q\to 0$ and $q\to\infty$, as it is expected. Further, neglecting correlation effects causes substantial errors in the RPA results over a broad range of wave vectors, whereas the STLS data exhibit a maximum bias of around one percent between one and two Fermi wave vectors.

\section{Summary and Outlook\label{sec:outlook}}

\subsection{Summary and Discussion\label{sec:summary_and_discussion}}

The present work has been devoted to the thermodynamic description of the uniform electron gas at warm dense matter conditions -- a topic of high current interest in many fields including astrophysics, laser plasmas and material science. Accurate thermodynamic data for these systems are crucial for comparison with experiments and for the development of improved theoretical methods.
Of particular importance are such data
 as input for many-body simulations such as the ubiquitous  density functional theory. Our data are also highly valuable as input for other models such as quantum hydrodynamics, e.g. \cite{bonitz_qhd_pre_13, michta_quantum_2015},  in order to study screening effects and effective potentials, e.g. \cite{zhandos_pop_15} and transport and wave phenomena.
We have discussed a variety of theoretical approaches that are broadly used to compute the static properties of the electron gas, which include the dielectric formalism (Sec.~\ref{sec:LRT}), various quantum Monte Carlo methods (Sec.~\ref{sec:QMC}), quantum-classical mappings, and finite-temperature Green functions (Sec.~\ref{sec:other}). 
%
Among these approaches, the most accurate results are provided by path integral Monte Carlo (PIMC) calculations (Sec.~\ref{sec:QMC}), which, for the UEG, however, are severely limited by the fermion sign problem. For this reason, over the last years, much effort has been undertaken to develop improved fermionic QMC simulations at finite temperature that were reviewed in Sec.~\ref{sec:PIMC}. Particular progress was achieved by the present authors which we summarize in the following: 
\begin{enumerate}
    \item We introduced two novel QMC methods -- CPIMC (Configuration PIMC, Sec.~\ref{sec:CPIMC}) and PB-PIMC (Permutation blocking PIMC, Sec.~\ref{sec:PB-PIMC}) -- that are accurate and efficient in complementary parameter regions. 
    \item We have demonstrated in detail that the combination of CPIMC and PB-PIMC allows for a highly accurate description of electrons in the warm dense matter regime over the entire density range, down to half the Fermi temperature without the use of  uncontrolled approximations such as the fixed node approximation (RPIMC, see Sec.~\ref{sec:RPIMC}). 
    \item Our results have been fully confirmed by a third, independent new method---DMQMC (Density matrix QMC, Sec.~\ref{sec:DMQMC}), thereby leading to a consensus regarding the thermodynamic properties of the warm dense UEG for a finite number $N$ of electrons.
    \item The next natural step has been the extrapolation of the finite $N$-simulations to the thermodynamic limit (Sec.~\ref{sec:FSC}) -- a task that turned out to be surprisingly nontrivial. We have shown that the previously used finite-size correction is not appropriate over substantial parts of the WDM regime. Further, we demonstrated that the major finite-size error is due to the missing long-range contribution, which cannot be accessed directly within QMC simulations of a finite number of electrons in a finite simulation cell. 
    \item To compensate for this, we combined the exact treatment of short-range exchange-correlation effects from QMC with the dielectric formalism (specifically, with the STLS approximation), that is known to be exact precisely in the long wavelength limit, $q\to 0$. This combination of QMC and STLS data allows (i) for a highly accurate description of the static structure factor, $S(q)$, in the thermodynamic limit over the entire $q$-range, and (ii) for an improved finite-size correction that is efficient over the entire WDM regime. 
    \item Applying this scheme, we have performed extensive simulations for a broad parameter range and, thus, obtained the first \textit{ab initio} thermodynamic results for the warm dense UEG in the thermodynamic limit, with an unprecedented accuracy of $0.3\%$.
    \item Using these new data, for the first time, it became possible to benchmark previous approximations, including RPIMC and dielectric methods such as RPA, STLS, and the recent improved HNC-scheme by Tanaka (Sec.~\ref{sec:comparison}).
    \item For practical applications, we constructed -- based on an exhaustive  QMC data set -- a new parametrization (\textit{GDB parametrization}) of the exchange-correlation free energy of the warm dense UEG with respect to density, temperature, and spin-polarization, i.e., $f_\text{xc}(r_s,\theta,\xi)$, that
    bridges the gap to the well-known ground state limit, see Sec.~\ref{sec:fxc}. 
    \item Based on the new GDB parametrization we performed unambiguous tests of the accuracy and applicatbility limits of earlier parametrizations and fits.
    \item Finally, we have outlined strategies how to extend our \textit{ab initio} approach to the \textit{inhomogeneous} electron gas. This was achieved by performing, 
    both, PB-PIMC and the CPIMC simulations for harmonically perturbed systems (Sec.~\ref{sec:response}). 
    \item These simulations were utilized to compute the first \textit{ab inito} results for the static density response function, $\chi(q)$, and for the static local field correction, $G(q)$. 
\end{enumerate}
Even though the results for the inhomogeneous electron gas are still preliminary they demonstrate that the present approach is very promising. They also demonstrate that accurate QMC data are not only important for the exchange correlation free energy. Of possibly even greater importance is their use for quantities that are derivatives of the free energy that are much more sensitive to incaccuracies. This includes the compressibility and the local field corrections.


%



\subsection{Outlook}

A natural extension of our work is given by the thorough investigation of the static density response of the warm dense electron gas as outlined in Sec.~\ref{sec:response}. Similar to the parametrization of $f_\text{xc}$, the construction of a complete parametrization of the static local field correction with respect to density, temperature, and wave vector, $G(\mathbf{q},r_s,\theta)$, constitutes a highly desirable goal, since it allows, e.g., for the computation of a true nonlocal exchange-correlation functional within the adiabatic connection fluctuation dissipation formulation of density functional theory~\cite{pribram-jones_thermal_2016,lu_evaluation_2014,patrick_adiabatic-connection_2015}. Interesting open questions in this direction include the  large $q$-behavior of $G$ and the possible existence of charge- and spin-density waves~\cite{giuliani2005quantum,schweng_finite-temperature_1993}.

A further topic of high importance is the investigation of the \textit{dynamic properties} of warm dense electrons such as the single-particle spectrum~\cite{kutepov_one-electron_2017,kas_finite_2017}, $A(\mathbf{q},\omega)$, the single-particle dispersion, $\omega(\textbf{q})$, or the density of states. The spectral function $A(\mathbf{q},\omega)$ is a key quantity of many-body theories such as Matsubara and nonequilibrium Green functions theory, e.g. \cite{balzer-book, kremp2006quantum}, that are extensively applied to describe the properties of correlated macroscopic systems \cite{semkat_jmp_00}, atoms and molecules \cite{balzer_pra_10_gg}, Hubbard clusters \cite{schluenzen_cpp_16}, or ultracold atoms in traps \cite{schluenzen_prb_16}. Unbiased QMC results may play a crucial role to test and improve selfenergy approximations. Moreover, to probe the collective properties of correlated electrons, the dynamic structure factor, $S(\mathbf{q},\omega)$, plays a key role. It is of particular importance, e.g., for the description of collective excitations of realistic warm dense matter within the Chihara decomposition~\cite{chihara_difference_1987,glenzer_x-ray_2009}. Furthermore, the dynamic structure factor is directly linked to other dynamic and optical properties such as the dielectric function or the dynamic conductivity and reflectivity. Also, the dynamic structure factor yields the plasmon spectrum which is an important experimental diagnostic of warm dense matter. For correlated charged particles in traps, the plasmon spectrum transforms into discrete normal modes that contain important information on the state of the system. Of particular importance are the center of mass (dipole or Kohn) mode e.g. \cite{bonitz_prb_07}, and the breathing (monopole) mode \cite{bauch_prb_09, henning_prl_08}, and may serve as a diagnostic tool for electrons in quantum dots or ultracold atoms in traps, e.g. \cite{mcdonald_prl_13,schmelcher_prb_13} and references therein. Here, exact solutions of the Schr\"odinger equation are limited to a few particles, and QMC may provide the necessary \textit{ab initio results.}

In principle, dynamical properties and spectra of correlated electrons in equilibrium and nonequilibrium can be directly computed via \textit{time-propagation}, as demonstrated with nonequilibrium Green functions in Ref.~\cite{kwong_real-time_2000, balzer_epl_12}, calling for similar approaches using Monte Carlo methods.
Unfortunately, time-dependent QMC simulations are severely hampered by the so-called dynamical sign problem~\cite{schiro_real-time_2009,schiro_real-time_2010} that permits only very short simulations that are not suitable to generate spectra. An alternative strategy is given by the approximate method of moments~\cite{arkhipov_dielectric_2014}, where the possibility to include our \textit{ab initio} results for the static structure factor is currently being investigated.
In addition, it is straightforward to utilize our QMC methods to compute \textit{imaginary-time correlation functions}~\cite{berne_path_1986}. These can be used as the basis for the reconstruction of dynamic quantities~\cite{jarrell_bayesian_1996}, such as $S(\mathbf{q},\omega)$, which is a well established procedure for the investigation of bosons, e.g., Refs.~\cite{ferre_dynamic_2016,filinov_collective_2012,filinov_correlation_2016}. A particular advantage of this strategy is the exact treatment of correlation effects, which allows to benchmark other approaches including the above mentioned method of moments, (dynamic) RPA and STLS, or the interpolation between various limits proposed by Gregori \textit{et al.}~\cite{gregori_derivation_2007}. For completeness, we note that a similar strategy has recently been explored by Motta \textit{et al.}~\cite{motta_imaginary_2014,motta_imaginary_2015} for the $2D$ electron gas in the ground state, and the recent remarkable progress in the field of reconstruction, in general, Refs.~\cite{levy_implementation_2017,prokofev_spectral_2013,vitali_ab_2010,otsuki_sparse_2017,schott_comparison_2016}.

Another important quantity is the \textit{momentum distribution}, $n(k)$, of warm dense matter which is directly accessible experimentally via  photoionization of atoms and molecules \cite{hochstuhl_jcp_11, hochstuhl_epjst_14} or photoemission from solids and liquids, e.g. \cite{schattke-book}. The tail of $n(k)$ is crucial for impact excitation and ionization processes and directly reflects correlation and quantum effects in the system. Knowledge of the exact large-$k$ asymptotics of $n(k)$ is crucial for accurate predictions of impact excitation and ionization rates of chemical reactions and of nuclear fusion rates in a dense plasma environment, such as in the solar interior~\cite{starostin_quantum_2002}, in compact stars or in laser fusion experiments.
The momentum distribution of the UEG 
has been extensively investigated at zero temperature, e.g., Refs.~\cite{ortiz_correlation_1994,ortiz_erratum:_1997,holzmann_momentum_2011,kimball_short-range_1975,yasuhara_note_1976,starostin_quantum_2002,takada_momentum_1991,takada_kita_1991,ziesche_momentum_2012,ziesche_high-density_2010,ziesche_three-dimensional_2005,gori-giorgi_momentum_2002,maebashi_analysis_2011,takada_emergence_2016,PhysRevE.66.046405}. However, at warm dense matter conditions, to our knowledge, no similar studies exist. Due to its formulation in momentum space, the CPIMC method is perfectly suited to compute highly accurate results for the momentum distribution in dense quantum systems. 

We further note that, in many ultracompact astrophysical objects such as dwarf stars or neutron stars, densities are so high (small $r_s$ values), that  \textit{relativistic effects} become important~\cite{ichimaru2004statistical_1,ichimaru2004statistical_2}. For this task, one can extend our CPIMC method to the simulation of the relativistic Hamiltonian of the UEG (i.e., by using the appropriate modified dispersion relation).  

Finally, aside from its relevance as a model system in many-body physics and a benchmark tool for approximations and simulations, the warm dense electron gas constitutes the key contribution to real warm dense matter that contains, in addition, heavy particle species. The extension to realistic multi-component simulations can be done in various ways. One is to use the UEG data as an input to finite-temperature DFT simulations. Here the ab initio data for the exchange-correlation free energy of the warm dense electron gas and the analytical parametrization presented in this review are of direct importance. On the other hand, dense two-component plasmas have been successfully investigated by path integral Monte Carlo methods by Ceperley amd Militzer and co-workers (RPIMC), e.g.~\cite{pierleoni_computational_2005, militzer_path_2000, hu_first-principles_2011} and by Filinov and co-workers (direct fermionic PIMC), e.g. \cite{filinov_jetpl_00,filinov_thermodynamic_2004}. The problems analyzed include the thermodynamic functions, the pair distribution functions \cite{filinov_pla_00} and proton crystallization at high density \cite{bonitz_prl_05, filinov_proton_2012}. For two-component plasmas, of course, the fermion sign problem is even more severe than for the UEG. So the accuracy of the commonly used fixed node approximation remains to be verified against unbiased methods. A powerful tool for these simulations is the use of effective quantum pair potentials, that incorporate many-body and quantum effects and have been derived by Kelbg~\cite{kelbg_63_1,kelbg_63_2,kelbg_64}, Ebeling and co-workers and many others, see e.g. refs.~\cite{filinov_jpa_03, filinov_pre_04} and references therein. Another promising strategy is to extend the coupled electron-ion Monte Carlo method \cite{pierleoni_coupled_2004} to finite temperatures. Yet the high complexity and the vast parameter space of warm dense matter requires the parallel development of independent theoretical and computational methods that can be used to benchmark one against the other. The present \textit{ab initio} data is expected to be a valuable reference for these developments.

\subsection{Open resources}\label{ss:git}
Finally, we mention the paramount value of the UEG as a test bed for the development of simulation techniques, as it requires an accurate treatment of (i) fermionic exchange, (ii) Coulomb correlation, and (iii) thermal excitations at the same time. For this reason, our extensive QMC data set (for various energies and the static structure factor) and the GDB parametrization of the free energy are openly available~\cite{our_git}.

\section*{Acknowledgements}

We are grateful to Tim Schoof and W.M.C.~Foulkes for many stimulating 
discussions. Moreover, we acknowledge Jan Vorberger for providing the 
Montroll-Ward and $e^4$ data for the interaction energy shown in 
Figs.~\ref{fig:comparison_interaction} and 
\ref{fig:high_comparison_interaction}, Travis Sjostrom for the Vashista-Singwi 
data for the interaction energy shown in Figs.~\ref{fig:comparison_interaction} 
and \ref{fig:high_comparison_interaction}, the static structure factor, 
Fig.~\ref{fig:SSF_comparison_dielectric}, and the local field correction 
depicted in Fig.~\ref{fig:LFC_compressibility}, and Shigenori Tanaka for the 
results from his HNC-based dielectric method for the static structure factor, 
Fig.~\ref{fig:SSF_comparison_dielectric} and the static local field correction, 
Fig.~\ref{fig:LFC_compressibility}. We also thank Shigenori Tanaka and Jan 
Vorberger for helpful comments on the manuscript. 
This work was supported by the Deutsche Forschungsgemeinschaft via project BO1366-10, as well as grant shp00015 for CPU time at the Norddeutscher Verbund f\"ur Hoch- und H\"ochstleistungsrechnen (HLRN).



\section{References}


\begin{thebibliography}{100}
\expandafter\ifx\csname url\endcsname\relax
  \def\url#1{\texttt{#1}}\fi
\expandafter\ifx\csname urlprefix\endcsname\relax\def\urlprefix{URL }\fi
\expandafter\ifx\csname href\endcsname\relax
  \def\href#1#2{#2} \def\path#1{#1}\fi

\bibitem{giuliani2005quantum}
G.~Giuliani, G.~Vignale,
  \href{https://books.google.de/books?id=kFkIKRfgUpsC}{Quantum Theory of the
  Electron Liquid}, Masters Series in Physics and Astronomy, Cambridge
  University Press, 2005.
\newline\urlprefix\url{https://books.google.de/books?id=kFkIKRfgUpsC}

\bibitem{ott_2018}
T.~Ott, H.~Thomsen, J.~Abraham, T.~Dornheim, M.~Bonitz, {Recent progress in the
  theory and simulation of strongly correlated plasmas: phase transitions,
  transport, quantum, and magnetic field effects}, Eur.~Phys.~J.~D (in print).

\bibitem{loos_uniform_2016}
P.-F. Loos, P.~M.~W. Gill,
  \href{http://onlinelibrary.wiley.com/doi/10.1002/wcms.1257/abstract}{The
  uniform electron gas}, Comp.~Mol.~Sci. 6~(4) (2016) 410--429.
\newblock \href {http://dx.doi.org/10.1002/wcms.1257}
  {\path{doi:10.1002/wcms.1257}}.
\newline\urlprefix\url{http://onlinelibrary.wiley.com/doi/10.1002/wcms.1257/abstract}

\bibitem{mahan1990many}
G.~Mahan, \href{https://books.google.de/books?id=v8du6cp0vUAC}{Many-Particle
  Physics}, Physics of Solids and Liquids, Springer US, 1990.
\newline\urlprefix\url{https://books.google.de/books?id=v8du6cp0vUAC}

\bibitem{bardeen_theory_1957}
J.~Bardeen, L.~N. Cooper, J.~R. Schrieffer,
  \href{http://link.aps.org/doi/10.1103/PhysRev.108.1175}{Theory of
  {Superconductivity}}, Phys.~Rev. 108~(5) (1957) 1175--1204.
\newblock \href {http://dx.doi.org/10.1103/PhysRev.108.1175}
  {\path{doi:10.1103/PhysRev.108.1175}}.
\newline\urlprefix\url{http://link.aps.org/doi/10.1103/PhysRev.108.1175}

\bibitem{baym2008landau}
G.~Baym, C.~Pethick,
  \href{https://books.google.de/books?id=xmiV4YSEjE4C}{Landau Fermi-Liquid
  Theory: Concepts and Applications}, Wiley, 2008.
\newline\urlprefix\url{https://books.google.de/books?id=xmiV4YSEjE4C}

\bibitem{pines_collective_1952}
D.~Pines, D.~Bohm, \href{http://link.aps.org/doi/10.1103/PhysRev.85.338}{A
  {Collective} {Description} of {Electron} {Interactions}: {II}. {Collective}
  $\mathrm{vs}$ {Individual} {Particle} {Aspects} of the {Interactions}},
  Phys.~Rev. 85~(2) (1952) 338--353.
\newblock \href {http://dx.doi.org/10.1103/PhysRev.85.338}
  {\path{doi:10.1103/PhysRev.85.338}}.
\newline\urlprefix\url{http://link.aps.org/doi/10.1103/PhysRev.85.338}

\bibitem{bohm_collective_1953}
D.~Bohm, D.~Pines, \href{http://link.aps.org/doi/10.1103/PhysRev.92.609}{A
  {Collective} {Description} of {Electron} {Interactions}: {III}. {Coulomb}
  {Interactions} in a {Degenerate} {Electron} {Gas}}, Phys.~Rev. 92~(3) (1953)
  609--625.
\newblock \href {http://dx.doi.org/10.1103/PhysRev.92.609}
  {\path{doi:10.1103/PhysRev.92.609}}.
\newline\urlprefix\url{http://link.aps.org/doi/10.1103/PhysRev.92.609}

\bibitem{singwi_electron_1968}
K.~S. Singwi, M.~P. Tosi, R.~H. Land, A.~Sj\"olander,
  \href{http://link.aps.org/doi/10.1103/PhysRev.176.589}{Electron
  {Correlations} at {Metallic} {Densities}}, Phys.~Rev. 176~(2) (1968)
  589--599.
\newblock \href {http://dx.doi.org/10.1103/PhysRev.176.589}
  {\path{doi:10.1103/PhysRev.176.589}}.
\newline\urlprefix\url{http://link.aps.org/doi/10.1103/PhysRev.176.589}

\bibitem{vashishta_electron_1972}
P.~Vashishta, K.~S. Singwi,
  \href{http://link.aps.org/doi/10.1103/PhysRevB.6.875}{Electron {Correlations}
  at {Metallic} {Densities}. {V}}, Phys.~Rev.~B 6~(3) (1972) 875--887.
\newblock \href {http://dx.doi.org/10.1103/PhysRevB.6.875}
  {\path{doi:10.1103/PhysRevB.6.875}}.
\newline\urlprefix\url{http://link.aps.org/doi/10.1103/PhysRevB.6.875}

\bibitem{kugler_theory_1975}
A.~A. Kugler, \href{http://link.springer.com/article/10.1007/BF01024183}{Theory
  of the local field correction in an electron gas}, J.~Stat.~Phys. 12~(1)
  (1975) 35--87.
\newblock \href {http://dx.doi.org/10.1007/BF01024183}
  {\path{doi:10.1007/BF01024183}}.
\newline\urlprefix\url{http://link.springer.com/article/10.1007/BF01024183}

\bibitem{kugler_collective_????}
A.~A. Kugler,
  \href{http://link.springer.com/article/10.1007/BF01008535}{Collective modes,
  damping, and the scattering function in classical liquids}, J.~Stat.~Phys.
  8~(2) (1973) 107--153.
\newblock \href {http://dx.doi.org/10.1007/BF01008535}
  {\path{doi:10.1007/BF01008535}}.
\newline\urlprefix\url{http://link.springer.com/article/10.1007/BF01008535}

\bibitem{ichimaru_strongly_1982}
S.~Ichimaru, \href{http://link.aps.org/doi/10.1103/RevModPhys.54.1017}{Strongly
  coupled plasmas: high-density classical plasmas and degenerate electron
  liquids}, Rev.~Mod.~Phys. 54~(4) (1982) 1017--1059.
\newblock \href {http://dx.doi.org/10.1103/RevModPhys.54.1017}
  {\path{doi:10.1103/RevModPhys.54.1017}}.
\newline\urlprefix\url{http://link.aps.org/doi/10.1103/RevModPhys.54.1017}

\bibitem{nozieres_theory_1999}
P.~Nozieres, D.~Pines,
  \href{https://books.google.de/books?id=q3wCwaV-gmUC}{Theory Of Quantum
  Liquids}, Advanced Books Classics, Avalon Publishing, 1999.
\newline\urlprefix\url{https://books.google.de/books?id=q3wCwaV-gmUC}

\bibitem{ceperley_ground_1978}
D.~Ceperley, \href{http://link.aps.org/doi/10.1103/PhysRevB.18.3126}{Ground
  state of the fermion one-component plasma: {A} {Monte} {Carlo} study in two
  and three dimensions}, Phys.~Rev.~B 18~(7) (1978) 3126--3138.
\newblock \href {http://dx.doi.org/10.1103/PhysRevB.18.3126}
  {\path{doi:10.1103/PhysRevB.18.3126}}.
\newline\urlprefix\url{http://link.aps.org/doi/10.1103/PhysRevB.18.3126}

\bibitem{ceperley_ground_1980}
D.~M. Ceperley, B.~J. Alder,
  \href{http://link.aps.org/doi/10.1103/PhysRevLett.45.566}{Ground {State} of
  the {Electron} {Gas} by a {Stochastic} {Method}}, Phys.~Rev.~Lett. 45~(7)
  (1980) 566.
\newblock \href {http://dx.doi.org/10.1103/PhysRevLett.45.566}
  {\path{doi:10.1103/PhysRevLett.45.566}}.
\newline\urlprefix\url{http://link.aps.org/doi/10.1103/PhysRevLett.45.566}

\bibitem{foulkes_quantum_2001}
W.~M.~C. Foulkes, L.~Mitas, R.~J. Needs, G.~Rajagopal,
  \href{http://link.aps.org/doi/10.1103/RevModPhys.73.33}{Quantum {Monte}
  {Carlo} simulations of solids}, Rev.~Mod.~Phys. 73~(1) (2001) 33--83.
\newblock \href {http://dx.doi.org/10.1103/RevModPhys.73.33}
  {\path{doi:10.1103/RevModPhys.73.33}}.
\newline\urlprefix\url{http://link.aps.org/doi/10.1103/RevModPhys.73.33}

\bibitem{shepherd_convergence_2012}
J.~J. Shepherd, A.~Gr\"uneis, G.~H. Booth, G.~Kresse, A.~Alavi,
  \href{http://link.aps.org/doi/10.1103/PhysRevB.86.035111}{Convergence of
  many-body wave-function expansions using a plane-wave basis: {From}
  homogeneous electron gas to solid state systems}, Phys.~Rev.~B 86~(3) (2012)
  035111.
\newblock \href {http://dx.doi.org/10.1103/PhysRevB.86.035111}
  {\path{doi:10.1103/PhysRevB.86.035111}}.
\newline\urlprefix\url{http://link.aps.org/doi/10.1103/PhysRevB.86.035111}

\bibitem{shepherd_full_2012}
J.~J. Shepherd, G.~Booth, A.~Gr\"uneis, A.~Alavi,
  \href{http://link.aps.org/doi/10.1103/PhysRevB.85.081103}{Full configuration
  interaction perspective on the homogeneous electron gas}, Phys.~Rev.~B 85~(8)
  (2012) 081103.
\newblock \href {http://dx.doi.org/10.1103/PhysRevB.85.081103}
  {\path{doi:10.1103/PhysRevB.85.081103}}.
\newline\urlprefix\url{http://link.aps.org/doi/10.1103/PhysRevB.85.081103}

\bibitem{shepherd_investigation_2012}
J.~J. Shepherd, G.~H. Booth, A.~Alavi,
  \href{http://scitation.aip.org/content/aip/journal/jcp/136/24/10.1063/1.4720076}{Investigation
  of the full configuration interaction quantum {Monte} {Carlo} method using
  homogeneous electron gas models}, J.~Chem.~Phys. 136~(24) (2012) 244101.
\newblock \href {http://dx.doi.org/10.1063/1.4720076}
  {\path{doi:10.1063/1.4720076}}.
\newline\urlprefix\url{http://scitation.aip.org/content/aip/journal/jcp/136/24/10.1063/1.4720076}

\bibitem{lopez_rios_inhomogeneous_2006}
P.~L\'opez~R\'ios, A.~Ma, N.~D. Drummond, M.~D. Towler, R.~J. Needs,
  \href{http://link.aps.org/doi/10.1103/PhysRevE.74.066701}{Inhomogeneous
  backflow transformations in quantum {Monte} {Carlo} calculations},
  Phys.~Rev.~E 74~(6) (2006) 066701.
\newblock \href {http://dx.doi.org/10.1103/PhysRevE.74.066701}
  {\path{doi:10.1103/PhysRevE.74.066701}}.
\newline\urlprefix\url{http://link.aps.org/doi/10.1103/PhysRevE.74.066701}

\bibitem{holzmann_backflow_2003}
M.~Holzmann, D.~M. Ceperley, C.~Pierleoni, K.~Esler,
  \href{http://link.aps.org/doi/10.1103/PhysRevE.68.046707}{Backflow
  correlations for the electron gas and metallic hydrogen}, Phys.~Rev.~E 68~(4)
  (2003) 046707.
\newblock \href {http://dx.doi.org/10.1103/PhysRevE.68.046707}
  {\path{doi:10.1103/PhysRevE.68.046707}}.
\newline\urlprefix\url{http://link.aps.org/doi/10.1103/PhysRevE.68.046707}

\bibitem{kohn_self-consistent_1965}
W.~Kohn, L.~J. Sham,
  \href{http://link.aps.org/doi/10.1103/PhysRev.140.A1133}{Self-{Consistent}
  {Equations} {Including} {Exchange} and {Correlation} {Effects}}, Phys.~Rev.
  140~(4A) (1965) A1133--A1138.
\newblock \href {http://dx.doi.org/10.1103/PhysRev.140.A1133}
  {\path{doi:10.1103/PhysRev.140.A1133}}.
\newline\urlprefix\url{http://link.aps.org/doi/10.1103/PhysRev.140.A1133}

\bibitem{hohenberg_inhomogeneous_1964}
P.~Hohenberg, W.~Kohn,
  \href{http://link.aps.org/doi/10.1103/PhysRev.136.B864}{Inhomogeneous
  {Electron} {Gas}}, Phys.~Rev. 136~(3B) (1964) B864--B871.
\newblock \href {http://dx.doi.org/10.1103/PhysRev.136.B864}
  {\path{doi:10.1103/PhysRev.136.B864}}.
\newline\urlprefix\url{http://link.aps.org/doi/10.1103/PhysRev.136.B864}

\bibitem{jones_density_2015}
R.~Jones, \href{http://link.aps.org/doi/10.1103/RevModPhys.87.897}{Density
  functional theory: {Its} origins, rise to prominence, and future},
  Rev.~Mod.~Phys. 87~(3) (2015) 897--923.
\newblock \href {http://dx.doi.org/10.1103/RevModPhys.87.897}
  {\path{doi:10.1103/RevModPhys.87.897}}.
\newline\urlprefix\url{http://link.aps.org/doi/10.1103/RevModPhys.87.897}

\bibitem{burke_perspective_2012}
K.~Burke,
  \href{http://scitation.aip.org/content/aip/journal/jcp/136/15/10.1063/1.4704546}{Perspective
  on density functional theory}, J.~Chem.~Phys. 136~(15) (2012) 150901.
\newblock \href {http://dx.doi.org/10.1063/1.4704546}
  {\path{doi:10.1063/1.4704546}}.
\newline\urlprefix\url{http://scitation.aip.org/content/aip/journal/jcp/136/15/10.1063/1.4704546}

\bibitem{jones_density_1989}
R.~O. Jones, O.~Gunnarsson,
  \href{http://link.aps.org/doi/10.1103/RevModPhys.61.689}{The density
  functional formalism, its applications and prospects}, Rev.~Mod.~Phys. 61~(3)
  (1989) 689--746.
\newblock \href {http://dx.doi.org/10.1103/RevModPhys.61.689}
  {\path{doi:10.1103/RevModPhys.61.689}}.
\newline\urlprefix\url{http://link.aps.org/doi/10.1103/RevModPhys.61.689}

\bibitem{vosko_accurate_1980}
S.~H. Vosko, L.~Wilk, M.~Nusair,
  \href{http://www.nrcresearchpress.com/doi/abs/10.1139/p80-159}{Accurate
  spin-dependent electron liquid correlation energies for local spin density
  calculations: a critical analysis}, Can.~J.~Phys. 58~(8) (1980) 1200--1211.
\newblock \href {http://dx.doi.org/10.1139/p80-159}
  {\path{doi:10.1139/p80-159}}.
\newline\urlprefix\url{http://www.nrcresearchpress.com/doi/abs/10.1139/p80-159}

\bibitem{perdew_self-interaction_1981}
J.~P. Perdew, A.~Zunger,
  \href{http://link.aps.org/doi/10.1103/PhysRevB.23.5048}{Self-interaction
  correction to density-functional approximations for many-electron systems},
  Phys.~Rev.~B 23~(10) (1981) 5048--5079.
\newblock \href {http://dx.doi.org/10.1103/PhysRevB.23.5048}
  {\path{doi:10.1103/PhysRevB.23.5048}}.
\newline\urlprefix\url{http://link.aps.org/doi/10.1103/PhysRevB.23.5048}

\bibitem{chachiyo_communication:_2016}
T.~Chachiyo,
  \href{http://aip.scitation.org/doi/10.1063/1.4958669}{Communication: {Simple}
  and accurate uniform electron gas correlation energy for the full range of
  densities}, J.~Chem.~Phys. 145~(2) (2016) 021101.
\newblock \href {http://dx.doi.org/10.1063/1.4958669}
  {\path{doi:10.1063/1.4958669}}.
\newline\urlprefix\url{http://aip.scitation.org/doi/10.1063/1.4958669}

\bibitem{perdew_generalizedd_1996}
J.~P. Perdew, K.~Burke, Y.~Wang,
  \href{https://link.aps.org/doi/10.1103/PhysRevB.54.16533}{Generalized
  gradient approximation for the exchange-correlation hole of a many-electron
  system}, Phys.~Rev.~B 54~(23) (1996) 16533--16539.
\newblock \href {http://dx.doi.org/10.1103/PhysRevB.54.16533}
  {\path{doi:10.1103/PhysRevB.54.16533}}.
\newline\urlprefix\url{https://link.aps.org/doi/10.1103/PhysRevB.54.16533}

\bibitem{perdew_generalized_1996}
J.~P. Perdew, K.~Burke, M.~Ernzerhof,
  \href{http://link.aps.org/doi/10.1103/PhysRevLett.77.3865}{Generalized
  {Gradient} {Approximation} {Made} {Simple}}, Phys.~Rev.~Lett. 77~(18) (1996)
  3865--3868.
\newblock \href {http://dx.doi.org/10.1103/PhysRevLett.77.3865}
  {\path{doi:10.1103/PhysRevLett.77.3865}}.
\newline\urlprefix\url{http://link.aps.org/doi/10.1103/PhysRevLett.77.3865}

\bibitem{ortiz_correlation_1993}
G.~Ortiz, P.~Ballone, \href{http://stacks.iop.org/0295-5075/23/i=1/a=002}{The
  {Correlation} {Energy} of the {Spin}-{Polarized} {Uniform} {Electron} {Gas}},
  Europhys.~Lett. 23~(1) (1993) 7.
\newblock \href {http://dx.doi.org/10.1209/0295-5075/23/1/002}
  {\path{doi:10.1209/0295-5075/23/1/002}}.
\newline\urlprefix\url{http://stacks.iop.org/0295-5075/23/i=1/a=002}

\bibitem{ortiz_correlation_1994}
G.~Ortiz, P.~Ballone,
  \href{http://link.aps.org/doi/10.1103/PhysRevB.50.1391}{Correlation energy,
  structure factor, radial distribution function, and momentum distribution of
  the spin-polarized uniform electron gas}, Phys.~Rev.~B 50~(3) (1994)
  1391--1405.
\newblock \href {http://dx.doi.org/10.1103/PhysRevB.50.1391}
  {\path{doi:10.1103/PhysRevB.50.1391}}.
\newline\urlprefix\url{http://link.aps.org/doi/10.1103/PhysRevB.50.1391}

\bibitem{ortiz_erratum:_1997}
G.~Ortiz, P.~Ballone,
  \href{https://link.aps.org/doi/10.1103/PhysRevB.56.9970}{Erratum:
  {Correlation} energy, structure factor, radial distribution function, and
  momentum distribution of the spin-polarized uniform electron gas [{Phys}.
  {Rev}. {B} 50, 1391 (1994)]}, Phys.~Rev.~B 56~(15) (1997) 9970--9970.
\newblock \href {http://dx.doi.org/10.1103/PhysRevB.56.9970}
  {\path{doi:10.1103/PhysRevB.56.9970}}.
\newline\urlprefix\url{https://link.aps.org/doi/10.1103/PhysRevB.56.9970}

\bibitem{ortiz_zero_1999}
G.~Ortiz, M.~Harris, P.~Ballone,
  \href{http://link.aps.org/doi/10.1103/PhysRevLett.82.5317}{Zero {Temperature}
  {Phases} of the {Electron} {Gas}}, Phys.~Rev.~Lett. 82~(26) (1999)
  5317--5320.
\newblock \href {http://dx.doi.org/10.1103/PhysRevLett.82.5317}
  {\path{doi:10.1103/PhysRevLett.82.5317}}.
\newline\urlprefix\url{http://link.aps.org/doi/10.1103/PhysRevLett.82.5317}

\bibitem{drummond_diffusion_2004}
N.~D. Drummond, Z.~Radnai, J.~R. Trail, M.~D. Towler, R.~J. Needs,
  \href{http://link.aps.org/doi/10.1103/PhysRevB.69.085116}{Diffusion quantum
  {Monte} {Carlo} study of three-dimensional {Wigner} crystals}, Phys.~Rev.~B
  69~(8) (2004) 085116.
\newblock \href {http://dx.doi.org/10.1103/PhysRevB.69.085116}
  {\path{doi:10.1103/PhysRevB.69.085116}}.
\newline\urlprefix\url{http://link.aps.org/doi/10.1103/PhysRevB.69.085116}

\bibitem{spink_quantum_2013}
G.~G. Spink, R.~J. Needs, N.~D. Drummond,
  \href{http://link.aps.org/doi/10.1103/PhysRevB.88.085121}{Quantum {Monte}
  {Carlo} study of the three-dimensional spin-polarized homogeneous electron
  gas}, Phys.~Rev.~B 88~(8) (2013) 085121.
\newblock \href {http://dx.doi.org/10.1103/PhysRevB.88.085121}
  {\path{doi:10.1103/PhysRevB.88.085121}}.
\newline\urlprefix\url{http://link.aps.org/doi/10.1103/PhysRevB.88.085121}

\bibitem{overhauser_1995}
A.~W. Overhauser, \href{https://doi.org/10.1139/p95-101}{Pair-correlation
  function of an electron gas}, Can.~J.~Phys. 73~(11-12) (1995) 683--686.
\newblock \href {http://arxiv.org/abs/https://doi.org/10.1139/p95-101}
  {\path{arXiv:https://doi.org/10.1139/p95-101}}, \href
  {http://dx.doi.org/10.1139/p95-101} {\path{doi:10.1139/p95-101}}.
\newline\urlprefix\url{https://doi.org/10.1139/p95-101}

\bibitem{perdew_pair-distribution_1992}
J.~P. Perdew, Y.~Wang,
  \href{http://link.aps.org/doi/10.1103/PhysRevB.46.12947}{Pair-distribution
  function and its coupling-constant average for the spin-polarized electron
  gas}, Phys.~Rev.~B 46~(20) (1992) 12947--12954.
\newblock \href {http://dx.doi.org/10.1103/PhysRevB.46.12947}
  {\path{doi:10.1103/PhysRevB.46.12947}}.
\newline\urlprefix\url{http://link.aps.org/doi/10.1103/PhysRevB.46.12947}

\bibitem{gori-giorgi_analytic_2000}
P.~Gori-Giorgi, F.~Sacchetti, G.~B. Bachelet,
  \href{http://link.aps.org/doi/10.1103/PhysRevB.61.7353}{Analytic static
  structure factors and pair-correlation functions for the unpolarized
  homogeneous electron gas}, Phys.~Rev.~B 61~(11) (2000) 7353--7363.
\newblock \href {http://dx.doi.org/10.1103/PhysRevB.61.7353}
  {\path{doi:10.1103/PhysRevB.61.7353}}.
\newline\urlprefix\url{http://link.aps.org/doi/10.1103/PhysRevB.61.7353}

\bibitem{gori-giorgi_pair_2002}
P.~Gori-Giorgi, J.~P. Perdew,
  \href{http://link.aps.org/doi/10.1103/PhysRevB.66.165118}{Pair distribution
  function of the spin-polarized electron gas: {A} first-principles analytic
  model for all uniform densities}, Phys.~Rev.~B 66~(16) (2002) 165118.
\newblock \href {http://dx.doi.org/10.1103/PhysRevB.66.165118}
  {\path{doi:10.1103/PhysRevB.66.165118}}.
\newline\urlprefix\url{http://link.aps.org/doi/10.1103/PhysRevB.66.165118}

\bibitem{holzmann_momentum_2011}
M.~Holzmann, B.~Bernu, C.~Pierleoni, J.~McMinis, D.~M. Ceperley, V.~Olevano,
  L.~Delle~Site,
  \href{http://link.aps.org/doi/10.1103/PhysRevLett.107.110402}{Momentum
  {Distribution} of the {Homogeneous} {Electron} {Gas}}, Phys.~Rev.~Lett.
  107~(11) (2011) 110402.
\newblock \href {http://dx.doi.org/10.1103/PhysRevLett.107.110402}
  {\path{doi:10.1103/PhysRevLett.107.110402}}.
\newline\urlprefix\url{http://link.aps.org/doi/10.1103/PhysRevLett.107.110402}

\bibitem{kimball_short-range_1975}
J.~C. Kimball, \href{http://stacks.iop.org/0305-4470/8/i=9/a=021}{Short-range
  correlations and the structure factor and momentum distribution of
  electrons}, J.~Phys.~A 8~(9) (1975) 1513.
\newblock \href {http://dx.doi.org/10.1088/0305-4470/8/9/021}
  {\path{doi:10.1088/0305-4470/8/9/021}}.
\newline\urlprefix\url{http://stacks.iop.org/0305-4470/8/i=9/a=021}

\bibitem{yasuhara_note_1976}
H.~Yasuhara, Y.~Kawazoe,
  \href{http://www.sciencedirect.com/science/article/pii/0378437176900601}{A
  note on the momentum distribution function for an electron gas}, Phys.~A
  85~(2) (1976) 416--424.
\newblock \href {http://dx.doi.org/10.1016/0378-4371(76)90060-1}
  {\path{doi:10.1016/0378-4371(76)90060-1}}.
\newline\urlprefix\url{http://www.sciencedirect.com/science/article/pii/0378437176900601}

\bibitem{starostin_quantum_2002}
A.~N. Starostin, A.~B. Mironov, N.~L. Aleksandrov, N.~J. Fisch, R.~M. Kulsrud,
  \href{http://www.sciencedirect.com/science/article/pii/S037843710100677X}{Quantum
  corrections to the distribution function of particles over momentum in dense
  media}, Phys.~A 305~(1–2) (2002) 287--296.
\newblock \href {http://dx.doi.org/10.1016/S0378-4371(01)00677-X}
  {\path{doi:10.1016/S0378-4371(01)00677-X}}.
\newline\urlprefix\url{http://www.sciencedirect.com/science/article/pii/S037843710100677X}

\bibitem{takada_momentum_1991}
Y.~Takada, H.~Yasuhara,
  \href{https://link.aps.org/doi/10.1103/PhysRevB.44.7879}{Momentum
  distribution function of the electron gas at metallic densities},
  Phys.~Rev.~B 44~(15) (1991) 7879--7887.
\newblock \href {http://dx.doi.org/10.1103/PhysRevB.44.7879}
  {\path{doi:10.1103/PhysRevB.44.7879}}.
\newline\urlprefix\url{https://link.aps.org/doi/10.1103/PhysRevB.44.7879}

\bibitem{takada_kita_1991}
Y.~Takada, T.~Kita, New self-consistency relation between the correlation
  energy and the momentum distribution function with application to the
  one-dimensional hubbard model, J.~Phys.~Soc.~Jpn. 60 (1991) 25.

\bibitem{ziesche_momentum_2012}
P.~Ziesche, \href{http://link.aps.org/doi/10.1103/PhysRevA.86.012508}{Momentum
  distribution and structure factors of a high-density homogeneous electron gas
  from its cumulant two-body reduced density matrix}, Phys.~Rev.~A 86~(1)
  (2012) 012508.
\newblock \href {http://dx.doi.org/10.1103/PhysRevA.86.012508}
  {\path{doi:10.1103/PhysRevA.86.012508}}.
\newline\urlprefix\url{http://link.aps.org/doi/10.1103/PhysRevA.86.012508}

\bibitem{ziesche_high-density_2010}
P.~Ziesche,
  \href{http://onlinelibrary.wiley.com/doi/10.1002/andp.201000022/abstract}{The
  high-density electron gas: {How} momentum distribution $n(k)$ and static
  structure factor $s(q)$ are mutually related through the off-shell
  self-energy $\sum(k, \omega)$}, Ann.~Phys. 522~(10) (2010) 739--765.
\newblock \href {http://dx.doi.org/10.1002/andp.201000022}
  {\path{doi:10.1002/andp.201000022}}.
\newline\urlprefix\url{http://onlinelibrary.wiley.com/doi/10.1002/andp.201000022/abstract}

\bibitem{ziesche_three-dimensional_2005}
P.~Ziesche, J.~Cioslowski,
  \href{http://www.sciencedirect.com/science/article/pii/S0378437105003407}{The
  three-dimensional electron gas at the weak-correlation limit: how
  peculiarities of the momentum distribution and the static structure factor
  give rise to logarithmic non-analyticities in the kinetic and potential
  correlation energies}, Phys.~A 356~(2–4) (2005) 598--608.
\newblock \href {http://dx.doi.org/10.1016/j.physa.2005.04.006}
  {\path{doi:10.1016/j.physa.2005.04.006}}.
\newline\urlprefix\url{http://www.sciencedirect.com/science/article/pii/S0378437105003407}

\bibitem{gori-giorgi_momentum_2002}
P.~Gori-Giorgi, P.~Ziesche,
  \href{https://link.aps.org/doi/10.1103/PhysRevB.66.235116}{Momentum
  distribution of the uniform electron gas: {Improved} parametrization and
  exact limits of the cumulant expansion}, Phys.~Rev.~B 66~(23) (2002) 235116.
\newblock \href {http://dx.doi.org/10.1103/PhysRevB.66.235116}
  {\path{doi:10.1103/PhysRevB.66.235116}}.
\newline\urlprefix\url{https://link.aps.org/doi/10.1103/PhysRevB.66.235116}

\bibitem{maebashi_analysis_2011}
H.~Maebashi, Y.~Takada,
  \href{https://link.aps.org/doi/10.1103/PhysRevB.84.245134}{{Analysis of exact
  vertex function for improving on the {GW}$\Gamma$ scheme for first-principles
  calculation of electron self-energy}}, Phys.~Rev.~B 84~(24) (2011) 245134.
\newblock \href {http://dx.doi.org/10.1103/PhysRevB.84.245134}
  {\path{doi:10.1103/PhysRevB.84.245134}}.
\newline\urlprefix\url{https://link.aps.org/doi/10.1103/PhysRevB.84.245134}

\bibitem{takada_emergence_2016}
Y.~Takada, \href{http://link.aps.org/doi/10.1103/PhysRevB.94.245106}{Emergence
  of an excitonic collective mode in the dilute electron gas}, Phys.~Rev.~B
  94~(24) (2016) 245106.
\newblock \href {http://dx.doi.org/10.1103/PhysRevB.94.245106}
  {\path{doi:10.1103/PhysRevB.94.245106}}.
\newline\urlprefix\url{http://link.aps.org/doi/10.1103/PhysRevB.94.245106}

\bibitem{moroni_static_1992}
S.~Moroni, D.~M. Ceperley, G.~Senatore,
  \href{http://link.aps.org/doi/10.1103/PhysRevLett.69.1837}{Static response
  from quantum {Monte} {Carlo} calculations}, Phys.~Rev.~Lett. 69~(13) (1992)
  1837--1840.
\newblock \href {http://dx.doi.org/10.1103/PhysRevLett.69.1837}
  {\path{doi:10.1103/PhysRevLett.69.1837}}.
\newline\urlprefix\url{http://link.aps.org/doi/10.1103/PhysRevLett.69.1837}

\bibitem{moroni_static_1995}
S.~Moroni, D.~M. Ceperley, G.~Senatore,
  \href{http://link.aps.org/doi/10.1103/PhysRevLett.75.689}{Static {Response}
  and {Local} {Field} {Factor} of the {Electron} {Gas}}, Phys.~Rev.~Lett.
  75~(4) (1995) 689--692.
\newblock \href {http://dx.doi.org/10.1103/PhysRevLett.75.689}
  {\path{doi:10.1103/PhysRevLett.75.689}}.
\newline\urlprefix\url{http://link.aps.org/doi/10.1103/PhysRevLett.75.689}

\bibitem{sugiyama_static_1992}
G.~Sugiyama, C.~Bowen, B.~J. Alder,
  \href{http://link.aps.org/doi/10.1103/PhysRevB.46.13042}{Static dielectric
  response of charged bosons}, Phys.~Rev.~B 46~(20) (1992) 13042--13050.
\newblock \href {http://dx.doi.org/10.1103/PhysRevB.46.13042}
  {\path{doi:10.1103/PhysRevB.46.13042}}.
\newline\urlprefix\url{http://link.aps.org/doi/10.1103/PhysRevB.46.13042}

\bibitem{bowen_static_1994}
C.~Bowen, G.~Sugiyama, B.~J. Alder,
  \href{https://link.aps.org/doi/10.1103/PhysRevB.50.14838}{Static dielectric
  response of the electron gas}, Phys.~Rev.~B 50~(20) (1994) 14838--14848.
\newblock \href {http://dx.doi.org/10.1103/PhysRevB.50.14838}
  {\path{doi:10.1103/PhysRevB.50.14838}}.
\newline\urlprefix\url{https://link.aps.org/doi/10.1103/PhysRevB.50.14838}

\bibitem{corradini_analytical_1998}
M.~Corradini, R.~Del~Sole, G.~Onida, M.~Palummo,
  \href{http://link.aps.org/doi/10.1103/PhysRevB.57.14569}{{Analytical
  expressions for the local-field factor $G(q)$ and the exchange-correlation
  kernel $K_\text{xc}(r)$ of the homogeneous electron gas}}, Phys.~Rev.~B
  57~(23) (1998) 14569--14571.
\newblock \href {http://dx.doi.org/10.1103/PhysRevB.57.14569}
  {\path{doi:10.1103/PhysRevB.57.14569}}.
\newline\urlprefix\url{http://link.aps.org/doi/10.1103/PhysRevB.57.14569}

\bibitem{fortov_extreme_2009}
V.~E. Fortov,
  \href{http://iopscience.iop.org/article/10.3367/UFNe.0179.200906h.0653/meta}{Extreme
  states of matter on {Earth} and in space}, Phys.-Usp. 52~(6) (2009) 615.
\newblock \href {http://dx.doi.org/10.3367/UFNe.0179.200906h.0653}
  {\path{doi:10.3367/UFNe.0179.200906h.0653}}.
\newline\urlprefix\url{http://iopscience.iop.org/article/10.3367/UFNe.0179.200906h.0653/meta}

\bibitem{gov}
U.~S. {D}epartment~of {E}nergy, Basic research needs for high energy density
  laboratory physics,
  https://nnsa.energy.gov/sites/default/files/nnsa/01-13-inlinefiles/Basic
  office of Science and National Nuclear Security Administration (2009).

\bibitem{wigner_interaction_1934}
E.~Wigner, \href{https://link.aps.org/doi/10.1103/PhysRev.46.1002}{On the
  {Interaction} of {Electrons} in {Metals}}, Phys.~Rev. 46~(11) (1934)
  1002--1011.
\newblock \href {http://dx.doi.org/10.1103/PhysRev.46.1002}
  {\path{doi:10.1103/PhysRev.46.1002}}.
\newline\urlprefix\url{https://link.aps.org/doi/10.1103/PhysRev.46.1002}

\bibitem{filinov_wigner_2001}
A.~Filinov, M.~Bonitz, Y.~Lozovik,
  \href{http://link.aps.org/doi/10.1103/PhysRevLett.86.3851}{Wigner
  {Crystallization} in {Mesoscopic} 2d {Electron} {Systems}}, Phys.~Rev.~Lett.
  86~(17) (2001) 3851--3854.
\newblock \href {http://dx.doi.org/10.1103/PhysRevLett.86.3851}
  {\path{doi:10.1103/PhysRevLett.86.3851}}.
\newline\urlprefix\url{http://link.aps.org/doi/10.1103/PhysRevLett.86.3851}

\bibitem{filinov_pss_2000}
A.~Filinov, Y.~Lozovik, M.~Bonitz, Path integral simulations of crystallization
  of quantum confined electrons, phys. stat. sol. (b) 221 (2000) 231.

\bibitem{bonitz_ropp_10}
M.~Bonitz, C.~Henning, D.~Block,
  \href{http://stacks.iop.org/0034-4885/73/i=6/a=066501}{Complex plasmas: a
  laboratory for strong correlations}, Reports on Progress in Physics 73~(6)
  (2010) 066501.
\newline\urlprefix\url{http://stacks.iop.org/0034-4885/73/i=6/a=066501}

\bibitem{graziani2014frontiers}
F.~Graziani, M.~Desjarlais, R.~Redmer, S.~Trickey,
  \href{https://books.google.de/books?id=Hdm4BAAAQBAJ}{Frontiers and Challenges
  in Warm Dense Matter}, Lecture Notes in Computational Science and
  Engineering, Springer International Publishing, 2014.
\newline\urlprefix\url{https://books.google.de/books?id=Hdm4BAAAQBAJ}

\bibitem{pustow_h/he_2016}
R.~P\"ustow, N.~Nettelmann, W.~Lorenzen, R.~Redmer,
  \href{http://www.sciencedirect.com/science/article/pii/S0019103515005606}{H/{He}
  demixing and the cooling behavior of {Saturn}}, Icarus 267~(Supplement C)
  (2016) 323--333.
\newblock \href {http://dx.doi.org/10.1016/j.icarus.2015.12.009}
  {\path{doi:10.1016/j.icarus.2015.12.009}}.
\newline\urlprefix\url{http://www.sciencedirect.com/science/article/pii/S0019103515005606}

\bibitem{nettelmann_saturn_2013}
N.~Nettelmann, R.~P\"ustow, R.~Redmer,
  \href{http://www.sciencedirect.com/science/article/pii/S0019103513001784}{Saturn
  layered structure and homogeneous evolution models with different {EOSs}},
  Icarus 225~(1) (2013) 548--557.
\newblock \href {http://dx.doi.org/10.1016/j.icarus.2013.04.018}
  {\path{doi:10.1016/j.icarus.2013.04.018}}.
\newline\urlprefix\url{http://www.sciencedirect.com/science/article/pii/S0019103513001784}

\bibitem{nettelmann_uranus_2016}
N.~Nettelmann, K.~Wang, J.~J. Fortney, S.~Hamel, S.~Yellamilli,
  M.~Bethkenhagen, R.~Redmer,
  \href{http://www.sciencedirect.com/science/article/pii/S0019103516300537}{Uranus
  evolution models with simple thermal boundary layers}, Icarus 275~(Supplement
  C) (2016) 107--116.
\newblock \href {http://dx.doi.org/10.1016/j.icarus.2016.04.008}
  {\path{doi:10.1016/j.icarus.2016.04.008}}.
\newline\urlprefix\url{http://www.sciencedirect.com/science/article/pii/S0019103516300537}

\bibitem{militzer_massive_2008}
B.~Militzer, W.~B. Hubbard, J.~Vorberger, I.~Tamblyn, S.~A. Bonev,
  \href{http://iopscience.iop.org/1538-4357/688/1/L45}{A {Massive} {Core} in
  {Jupiter} {Predicted} from {First}-{Principles} {Simulations}},
  Astrophys.~J.~Lett. 688~(1) (2008) L45.
\newblock \href {http://dx.doi.org/10.1086/594364} {\path{doi:10.1086/594364}}.
\newline\urlprefix\url{http://iopscience.iop.org/1538-4357/688/1/L45}

\bibitem{wilson_sequestration_2010}
H.~F. Wilson, B.~Militzer,
  \href{https://link.aps.org/doi/10.1103/PhysRevLett.104.121101}{Sequestration
  of {Noble} {Gases} in {Giant} {Planet} {Interiors}}, Phys.~Rev.~Lett.
  104~(12) (2010) 121101.
\newblock \href {http://dx.doi.org/10.1103/PhysRevLett.104.121101}
  {\path{doi:10.1103/PhysRevLett.104.121101}}.
\newline\urlprefix\url{https://link.aps.org/doi/10.1103/PhysRevLett.104.121101}

\bibitem{soubiran_properties_2017}
F.~Soubiran, B.~Militzer, K.~P. Driver, S.~Zhang,
  \href{http://aip.scitation.org/doi/abs/10.1063/1.4978618}{Properties of
  hydrogen, helium, and silicon dioxide mixtures in giant planet interiors},
  Phys.~Plasmas 24~(4) (2017) 041401.
\newblock \href {http://dx.doi.org/10.1063/1.4978618}
  {\path{doi:10.1063/1.4978618}}.
\newline\urlprefix\url{http://aip.scitation.org/doi/abs/10.1063/1.4978618}

\bibitem{vorberger_hydrogen-helium_2007}
J.~Vorberger, I.~Tamblyn, B.~Militzer, S.~A. Bonev,
  \href{https://link.aps.org/doi/10.1103/PhysRevB.75.024206}{Hydrogen-helium
  mixtures in the interiors of giant planets}, Phys.~Rev.~B 75~(2) (2007)
  024206.
\newblock \href {http://dx.doi.org/10.1103/PhysRevB.75.024206}
  {\path{doi:10.1103/PhysRevB.75.024206}}.
\newline\urlprefix\url{https://link.aps.org/doi/10.1103/PhysRevB.75.024206}

\bibitem{vorberger_properties_2007}
J.~Vorberger, I.~Tamblyn, S.~A. Bonev, B.~Militzer,
  \href{http://onlinelibrary.wiley.com/doi/10.1002/ctpp.200710050/abstract}{Properties
  of {Dense} {Fluid} {Hydrogen} and {Helium} in {Giant} {Gas} {Planets}},
  Contrib.~Plasma Phys. 47~(4-5) (2007) 375--380.
\newblock \href {http://dx.doi.org/10.1002/ctpp.200710050}
  {\path{doi:10.1002/ctpp.200710050}}.
\newline\urlprefix\url{http://onlinelibrary.wiley.com/doi/10.1002/ctpp.200710050/abstract}

\bibitem{nettelmann_ab_2008}
N.~Nettelmann, B.~Holst, A.~Kietzmann, M.~French, R.~Redmer, D.~Blaschke,
  \href{http://stacks.iop.org/0004-637X/683/i=2/a=1217}{Ab {Initio} {Equation}
  of {State} {Data} for {Hydrogen}, {Helium}, and {Water} and the {Internal}
  {Structure} of {Jupiter}}, Astrophys.~J. 683~(2) (2008) 1217.
\newblock \href {http://dx.doi.org/10.1086/589806} {\path{doi:10.1086/589806}}.
\newline\urlprefix\url{http://stacks.iop.org/0004-637X/683/i=2/a=1217}

\bibitem{french_equation_2009}
M.~French, T.~R. Mattsson, N.~Nettelmann, R.~Redmer,
  \href{https://link.aps.org/doi/10.1103/PhysRevB.79.054107}{Equation of state
  and phase diagram of water at ultrahigh pressures as in planetary interiors},
  Phys.~Rev.~B 79~(5) (2009) 054107.
\newblock \href {http://dx.doi.org/10.1103/PhysRevB.79.054107}
  {\path{doi:10.1103/PhysRevB.79.054107}}.
\newline\urlprefix\url{https://link.aps.org/doi/10.1103/PhysRevB.79.054107}

\bibitem{knudson_probing_2012}
M.~D. Knudson, M.~P. Desjarlais, R.~W. Lemke, T.~R. Mattsson, M.~French,
  N.~Nettelmann, R.~Redmer,
  \href{http://link.aps.org/doi/10.1103/PhysRevLett.108.091102}{{Probing the
  {Interiors} of the {Ice} {Giants}: {Shock} {Compression} of {Water} to 700
  GPa and $3.8~\text{g}/\text{cm}^3$}}, Phys.~Rev.~Lett. 108~(9) (2012) 091102.
\newblock \href {http://dx.doi.org/10.1103/PhysRevLett.108.091102}
  {\path{doi:10.1103/PhysRevLett.108.091102}}.
\newline\urlprefix\url{http://link.aps.org/doi/10.1103/PhysRevLett.108.091102}

\bibitem{saumon_the_role_1992}
D.~{Saumon}, W.~B. {Hubbard}, G.~{Chabrier}, H.~M. {van Horn}, {The role of the
  molecular-metallic transition of hydrogen in the evolution of Jupiter,
  Saturn, and brown dwarfs}, Astrophys.~J. 391 (1992) 827--831.
\newblock \href {http://dx.doi.org/10.1086/171391} {\path{doi:10.1086/171391}}.

\bibitem{hubbard_liquid_1997}
W.~B. Hubbard, T.~Guillot, J.~I. Lunine, A.~Burrows, D.~Saumon, M.~S. Marley,
  R.~S. Freedman,
  \href{http://aip.scitation.org/doi/abs/10.1063/1.872570}{Liquid metallic
  hydrogen and the structure of brown dwarfs and giant planets}, Phys.~Plasmas
  4~(5) (1997) 2011--2015.
\newblock \href {http://dx.doi.org/10.1063/1.872570}
  {\path{doi:10.1063/1.872570}}.
\newline\urlprefix\url{http://aip.scitation.org/doi/abs/10.1063/1.872570}

\bibitem{collins_measurements_1998}
G.~W. Collins, L.~B.~D. Silva, P.~Celliers, D.~M. Gold, M.~E. Foord, R.~J.
  Wallace, A.~Ng, S.~V. Weber, K.~S. Budil, R.~Cauble,
  \href{http://science.sciencemag.org/content/281/5380/1178}{Measurements of
  the {Equation} of {State} of {Deuterium} at the {Fluid} {Insulator}-{Metal}
  {Transition}}, Science 281~(5380) (1998) 1178--1181.
\newblock \href {http://dx.doi.org/10.1126/science.281.5380.1178}
  {\path{doi:10.1126/science.281.5380.1178}}.
\newline\urlprefix\url{http://science.sciencemag.org/content/281/5380/1178}

\bibitem{glenzer_matter_2016}
S.~H. Glenzer, L.~B. Fletcher, E.~Galtier, B.~Nagler, R.~Alonso-Mori,
  B.~Barbrel, S.~B. Brown, D.~A. Chapman, Z.~Chen, C.~B. Curry, F.~Fiuza,
  E.~Gamboa, M.~Gauthier, D.~O. Gericke, A.~Gleason, S.~Goede, E.~Granados,
  P.~Heimann, J.~Kim, D.~Kraus, M.~J. MacDonald, A.~J. Mackinnon, R.~Mishra,
  A.~Ravasio, C.~Roedel, P.~Sperling, W.~Schumaker, Y.~Y. Tsui, J.~Vorberger,
  {U Zastrau}, A.~Fry, W.~E. White, J.~B. Hasting, H.~J. Lee,
  \href{http://stacks.iop.org/0953-4075/49/i=9/a=092001}{Matter under extreme
  conditions experiments at the {Linac} {Coherent} {Light} {Source}},
  J.~Phys.~B 49~(9) (2016) 092001.
\newblock \href {http://dx.doi.org/10.1088/0953-4075/49/9/092001}
  {\path{doi:10.1088/0953-4075/49/9/092001}}.
\newline\urlprefix\url{http://stacks.iop.org/0953-4075/49/i=9/a=092001}

\bibitem{chabrier_cooling_2000}
G.~Chabrier, P.~Brassard, G.~Fontaine, D.~Saumon,
  \href{http://iopscience.iop.org/article/10.1086/317092/meta}{Cooling
  {Sequences} and {Color}-{Magnitude} {Diagrams} for {Cool} {White} {Dwarfs}
  with {Hydrogen} {Atmospheres}}, Astrophys.~J. 543~(1) (2000) 216.
\newblock \href {http://dx.doi.org/10.1086/317092} {\path{doi:10.1086/317092}}.
\newline\urlprefix\url{http://iopscience.iop.org/article/10.1086/317092/meta}

\bibitem{daligault_electronion_2009}
J.~Daligault, S.~Gupta,
  \href{http://stacks.iop.org/0004-637X/703/i=1/a=994}{Electron-{Ion}
  {Scattering} in {Dense} {Multi}-{Component} {Plasmas}: {Application} to the
  {Outer} {Crust} of an {Accreting} {Neutron} {Star}}, Astrophys.~J. 703~(1)
  (2009) 994.
\newblock \href {http://dx.doi.org/10.1088/0004-637X/703/1/994}
  {\path{doi:10.1088/0004-637X/703/1/994}}.
\newline\urlprefix\url{http://stacks.iop.org/0004-637X/703/i=1/a=994}

\bibitem{shukla_colloquium_2011}
P.~K. Shukla, B.~Eliasson,
  \href{https://link.aps.org/doi/10.1103/RevModPhys.83.885}{Colloquium},
  Rev.~Mod.~Phys. 83~(3) (2011) 885--906.
\newblock \href {http://dx.doi.org/10.1103/RevModPhys.83.885}
  {\path{doi:10.1103/RevModPhys.83.885}}.
\newline\urlprefix\url{https://link.aps.org/doi/10.1103/RevModPhys.83.885}

\bibitem{brumfiel_nuclear_2010}
G.~Brumfiel,
  \href{http://www.nature.com/news/2010/100310/full/464156a.html}{Nuclear
  weapons physics: {Welcome} to the {Atomic} {Weapons} {Establishment}}, Nature
  News 464~(7286) (2010) 156--157.
\newblock \href {http://dx.doi.org/10.1038/464156a}
  {\path{doi:10.1038/464156a}}.
\newline\urlprefix\url{http://www.nature.com/news/2010/100310/full/464156a.html}

\bibitem{hu_first-principles_2011}
S.~X. Hu, B.~Militzer, V.~N. Goncharov, S.~Skupsky,
  \href{http://link.aps.org/doi/10.1103/PhysRevB.84.224109}{First-principles
  equation-of-state table of deuterium for inertial confinement fusion
  applications}, Phys.~Rev.~B 84~(22) (2011) 224109.
\newblock \href {http://dx.doi.org/10.1103/PhysRevB.84.224109}
  {\path{doi:10.1103/PhysRevB.84.224109}}.
\newline\urlprefix\url{http://link.aps.org/doi/10.1103/PhysRevB.84.224109}

\bibitem{kritcher_-flight_2011}
A.~L. Kritcher, T.~D\"oppner, C.~Fortmann, T.~Ma, O.~L. Landen, R.~Wallace,
  S.~H. Glenzer,
  \href{http://link.aps.org/doi/10.1103/PhysRevLett.107.015002}{In-{Flight}
  {Measurements} of {Capsule} {Shell} {Adiabats} in {Laser}-{Driven}
  {Implosions}}, Phys.~Rev.~Lett. 107~(1) (2011) 015002.
\newblock \href {http://dx.doi.org/10.1103/PhysRevLett.107.015002}
  {\path{doi:10.1103/PhysRevLett.107.015002}}.
\newline\urlprefix\url{http://link.aps.org/doi/10.1103/PhysRevLett.107.015002}

\bibitem{gomez_experimental_2014}
M.~Gomez, S.~Slutz, A.~Sefkow, D.~Sinars, K.~Hahn, S.~Hansen, E.~Harding,
  P.~Knapp, P.~Schmit, C.~Jennings, T.~Awe, M.~Geissel, D.~Rovang, G.~Chandler,
  G.~Cooper, M.~Cuneo, A.~Harvey-Thompson, M.~Herrmann, M.~Hess, O.~Johns,
  D.~Lamppa, M.~Martin, R.~McBride, K.~Peterson, J.~Porter, G.~Robertson,
  G.~Rochau, C.~Ruiz, M.~Savage, I.~Smith, W.~Stygar, R.~Vesey,
  \href{http://link.aps.org/doi/10.1103/PhysRevLett.113.155003}{Experimental
  {Demonstration} of {Fusion}-{Relevant} {Conditions} in {Magnetized} {Liner}
  {Inertial} {Fusion}}, Phys.~Rev.~Lett. 113~(15) (2014) 155003.
\newblock \href {http://dx.doi.org/10.1103/PhysRevLett.113.155003}
  {\path{doi:10.1103/PhysRevLett.113.155003}}.
\newline\urlprefix\url{http://link.aps.org/doi/10.1103/PhysRevLett.113.155003}

\bibitem{schmit_understanding_2014}
P.~Schmit, P.~Knapp, S.~Hansen, M.~Gomez, K.~Hahn, D.~Sinars, K.~Peterson,
  S.~Slutz, A.~Sefkow, T.~Awe, E.~Harding, C.~Jennings, G.~Chandler, G.~Cooper,
  M.~Cuneo, M.~Geissel, A.~Harvey-Thompson, M.~Herrmann, M.~Hess, O.~Johns,
  D.~Lamppa, M.~Martin, R.~McBride, J.~Porter, G.~Robertson, G.~Rochau,
  D.~Rovang, C.~Ruiz, M.~Savage, I.~Smith, W.~Stygar, R.~Vesey,
  \href{http://link.aps.org/doi/10.1103/PhysRevLett.113.155004}{Understanding
  {Fuel} {Magnetization} and {Mix} {Using} {Secondary} {Nuclear} {Reactions} in
  {Magneto}-{Inertial} {Fusion}}, Phys.~Rev.~Lett. 113~(15) (2014) 155004.
\newblock \href {http://dx.doi.org/10.1103/PhysRevLett.113.155004}
  {\path{doi:10.1103/PhysRevLett.113.155004}}.
\newline\urlprefix\url{http://link.aps.org/doi/10.1103/PhysRevLett.113.155004}

\bibitem{nora_gigabar_2015}
R.~Nora, W.~Theobald, R.~Betti, F.~Marshall, D.~Michel, W.~Seka, B.~Yaakobi,
  M.~Lafon, C.~Stoeckl, J.~Delettrez, A.~Solodov, A.~Casner, C.~Reverdin,
  X.~Ribeyre, A.~Vallet, J.~Peebles, F.~Beg, M.~Wei,
  \href{http://link.aps.org/doi/10.1103/PhysRevLett.114.045001}{Gigabar
  {Spherical} {Shock} {Generation} on the {OMEGA} {Laser}}, Phys.~Rev.~Lett.
  114~(4) (2015) 045001.
\newblock \href {http://dx.doi.org/10.1103/PhysRevLett.114.045001}
  {\path{doi:10.1103/PhysRevLett.114.045001}}.
\newline\urlprefix\url{http://link.aps.org/doi/10.1103/PhysRevLett.114.045001}

\bibitem{hurricane_inertially_2016}
O.~A. Hurricane, D.~A. Callahan, D.~T. Casey, E.~L. Dewald, T.~R. Dittrich,
  T.~D\"oppner, S.~Haan, D.~E. Hinkel, L.~F. Berzak~Hopkins, O.~Jones, A.~L.
  Kritcher, S.~Le~Pape, T.~Ma, A.~G. MacPhee, J.~L. Milovich, J.~Moody, A.~Pak,
  H.-S. Park, P.~K. Patel, J.~E. Ralph, H.~F. Robey, J.~S. Ross, J.~D.
  Salmonson, B.~K. Spears, P.~T. Springer, R.~Tommasini, F.~Albert, L.~R.
  Benedetti, R.~Bionta, E.~Bond, D.~K. Bradley, J.~Caggiano, P.~M. Celliers,
  C.~Cerjan, J.~A. Church, R.~Dylla-Spears, D.~Edgell, M.~J. Edwards,
  D.~Fittinghoff, M.~A. Barrios~Garcia, A.~Hamza, R.~Hatarik, H.~Herrmann,
  M.~Hohenberger, D.~Hoover, J.~L. Kline, G.~Kyrala, B.~Kozioziemski, G.~Grim,
  J.~E. Field, J.~Frenje, N.~Izumi, M.~Gatu~Johnson, S.~F. Khan, J.~Knauer,
  T.~Kohut, O.~Landen, F.~Merrill, P.~Michel, A.~Moore, S.~R. Nagel, A.~Nikroo,
  T.~Parham, R.~R. Rygg, D.~Sayre, M.~Schneider, D.~Shaughnessy, D.~Strozzi,
  R.~P.~J. Town, D.~Turnbull, P.~Volegov, A.~Wan, K.~Widmann, C.~Wilde,
  C.~Yeamans,
  \href{http://www.nature.com/nphys/journal/vaop/ncurrent/full/nphys3720.html}{Inertially
  confined fusion plasmas dominated by alpha-particle self-heating}, Nat.~Phys.
  advance online publication.
\newblock \href {http://dx.doi.org/10.1038/nphys3720}
  {\path{doi:10.1038/nphys3720}}.
\newline\urlprefix\url{http://www.nature.com/nphys/journal/vaop/ncurrent/full/nphys3720.html}

\bibitem{moses_national_2009}
E.~I. Moses, R.~N. Boyd, B.~A. Remington, C.~J. Keane, R.~Al-Ayat,
  \href{http://aip.scitation.org/doi/abs/10.1063/1.3116505}{The {National}
  {Ignition} {Facility}: {Ushering} in a new age for high energy density
  science}, Phys.~Plasmas 16~(4) (2009) 041006.
\newblock \href {http://dx.doi.org/10.1063/1.3116505}
  {\path{doi:10.1063/1.3116505}}.
\newline\urlprefix\url{http://aip.scitation.org/doi/abs/10.1063/1.3116505}

\bibitem{hammel_high-mode_2010}
B.~A. Hammel, S.~W. Haan, D.~S. Clark, M.~J. Edwards, S.~H. Langer, M.~M.
  Marinak, M.~V. Patel, J.~D. Salmonson, H.~A. Scott,
  \href{http://www.sciencedirect.com/science/article/pii/S1574181809001323}{High-mode
  {Rayleigh}-{Taylor} growth in {NIF} ignition capsules}, High Energy Density
  Phys. 6~(2) (2010) 171--178.
\newblock \href {http://dx.doi.org/10.1016/j.hedp.2009.12.005}
  {\path{doi:10.1016/j.hedp.2009.12.005}}.
\newline\urlprefix\url{http://www.sciencedirect.com/science/article/pii/S1574181809001323}

\bibitem{root_shock_2010}
S.~Root, R.~J. Magyar, J.~H. Carpenter, D.~L. Hanson, T.~R. Mattsson,
  \href{https://link.aps.org/doi/10.1103/PhysRevLett.105.085501}{Shock
  {Compression} of a {Fifth} {Period} {Element}: {Liquid} {Xenon} to 840
  {GPa}}, Phys.~Rev.~Lett. 105~(8) (2010) 085501.
\newblock \href {http://dx.doi.org/10.1103/PhysRevLett.105.085501}
  {\path{doi:10.1103/PhysRevLett.105.085501}}.
\newline\urlprefix\url{https://link.aps.org/doi/10.1103/PhysRevLett.105.085501}

\bibitem{knudson_use_2003}
M.~D. Knudson, D.~L. Hanson, J.~E. Bailey, C.~A. Hall, J.~R. Asay,
  \href{https://link.aps.org/doi/10.1103/PhysRevLett.90.035505}{Use of a {Wave}
  {Reverberation} {Technique} to {Infer} the {Density} {Compression} of
  {Shocked} {Liquid} {Deuterium} to 75 {GPa}}, Phys.~Rev.~Lett. 90~(3) (2003)
  035505.
\newblock \href {http://dx.doi.org/10.1103/PhysRevLett.90.035505}
  {\path{doi:10.1103/PhysRevLett.90.035505}}.
\newline\urlprefix\url{https://link.aps.org/doi/10.1103/PhysRevLett.90.035505}

\bibitem{knudson_direct_2015}
M.~D. Knudson, M.~P. Desjarlais, A.~Becker, R.~W. Lemke, K.~R. Cochrane, M.~E.
  Savage, D.~E. Bliss, T.~R. Mattsson, R.~Redmer,
  \href{http://science.sciencemag.org/content/348/6242/1455}{Direct observation
  of an abrupt insulator-to-metal transition in dense liquid deuterium},
  Science 348~(6242) (2015) 1455--1460.
\newblock \href {http://dx.doi.org/10.1126/science.aaa7471}
  {\path{doi:10.1126/science.aaa7471}}.
\newline\urlprefix\url{http://science.sciencemag.org/content/348/6242/1455}

\bibitem{matzen_pulsed-power-driven_2005}
M.~K. Matzen, M.~A. Sweeney, R.~G. Adams, J.~R. Asay, J.~E. Bailey, G.~R.
  Bennett, D.~E. Bliss, D.~D. Bloomquist, T.~A. Brunner, R.~B. Campbell, G.~A.
  Chandler, C.~A. Coverdale, M.~E. Cuneo, J.-P. Davis, C.~Deeney, M.~P.
  Desjarlais, G.~L. Donovan, C.~J. Garasi, T.~A. Haill, C.~A. Hall, D.~L.
  Hanson, M.~J. Hurst, B.~Jones, M.~D. Knudson, R.~J. Leeper, R.~W. Lemke,
  M.~G. Mazarakis, D.~H. McDaniel, T.~A. Mehlhorn, T.~J. Nash, C.~L. Olson,
  J.~L. Porter, P.~K. Rambo, S.~E. Rosenthal, G.~A. Rochau, L.~E. Ruggles,
  C.~L. Ruiz, T.~W.~L. Sanford, J.~F. Seamen, D.~B. Sinars, S.~A. Slutz, I.~C.
  Smith, K.~W. Struve, W.~A. Stygar, R.~A. Vesey, E.~A. Weinbrecht, D.~F.
  Wenger, E.~P. Yu,
  \href{http://aip.scitation.org/doi/10.1063/1.1891746}{Pulsed-power-driven
  high energy density phys. and inertial confinement fusion research},
  Phys.~Plasmas 12~(5) (2005) 055503.
\newblock \href {http://dx.doi.org/10.1063/1.1891746}
  {\path{doi:10.1063/1.1891746}}.
\newline\urlprefix\url{http://aip.scitation.org/doi/10.1063/1.1891746}

\bibitem{magyar_equations_2012}
R.~Magyar, \href{http://aip.scitation.org/doi/abs/10.1063/1.3686494}{Equations
  of state of mixtures: {Density} functional theory ({DFT}) simulations and
  experiments on {Sandia}'s z machine}, AIP Conf.~Proc. 1426~(1) (2012)
  1195--1198.
\newblock \href {http://dx.doi.org/10.1063/1.3686494}
  {\path{doi:10.1063/1.3686494}}.
\newline\urlprefix\url{http://aip.scitation.org/doi/abs/10.1063/1.3686494}

\bibitem{ding_measurements_2009}
Y.~Ding, A.~Brachmann, F.-J. Decker, D.~Dowell, P.~Emma, J.~Frisch,
  S.~Gilevich, G.~Hays, P.~Hering, Z.~Huang, R.~Iverson, H.~Loos, A.~Miahnahri,
  H.-D. Nuhn, D.~Ratner, J.~Turner, J.~Welch, W.~White, J.~Wu,
  \href{https://link.aps.org/doi/10.1103/PhysRevLett.102.254801}{Measurements
  and {Simulations} of {Ultralow} {Emittance} and {Ultrashort} {Electron}
  {Beams} in the {Linac} {Coherent} {Light} {Source}}, Phys.~Rev.~Lett.
  102~(25) (2009) 254801.
\newblock \href {http://dx.doi.org/10.1103/PhysRevLett.102.254801}
  {\path{doi:10.1103/PhysRevLett.102.254801}}.
\newline\urlprefix\url{https://link.aps.org/doi/10.1103/PhysRevLett.102.254801}

\bibitem{fletcher_ultrabright_2015}
L.~B. Fletcher, H.~J. Lee, T.~D\"oppner, E.~Galtier, B.~Nagler, P.~Heimann,
  C.~Fortmann, S.~LePape, T.~Ma, M.~Millot, A.~Pak, D.~Turnbull, D.~A. Chapman,
  D.~O. Gericke, J.~Vorberger, T.~White, G.~Gregori, M.~Wei, B.~Barbrel, R.~W.
  Falcone, C.-C. Kao, H.~Nuhn, J.~Welch, U.~Zastrau, P.~Neumayer, J.~B.
  Hastings, S.~H. Glenzer,
  \href{https://www.nature.com/nphoton/journal/v9/n4/abs/nphoton.2015.41.html}{Ultrabright
  {X}-ray laser scattering for dynamic warm dense matter physics},
  Nat.~Photonics 9~(4) (2015) 274--279.
\newblock \href {http://dx.doi.org/10.1038/nphoton.2015.41}
  {\path{doi:10.1038/nphoton.2015.41}}.
\newline\urlprefix\url{https://www.nature.com/nphoton/journal/v9/n4/abs/nphoton.2015.41.html}

\bibitem{sperling_free-electron_2015}
P.~Sperling, E.~Gamboa, H.~Lee, H.~Chung, E.~Galtier, Y.~Omarbakiyeva,
  H.~Reinholz, G.~R\"opke, U.~Zastrau, J.~Hastings, L.~Fletcher, S.~Glenzer,
  \href{http://link.aps.org/doi/10.1103/PhysRevLett.115.115001}{Free-{Electron}
  {X}-{Ray} {Laser} {Measurements} of {Collisional}-{Damped} {Plasmons} in
  {Isochorically} {Heated} {Warm} {Dense} {Matter}}, Phys.~Rev.~Lett. 115~(11)
  (2015) 115001.
\newblock \href {http://dx.doi.org/10.1103/PhysRevLett.115.115001}
  {\path{doi:10.1103/PhysRevLett.115.115001}}.
\newline\urlprefix\url{http://link.aps.org/doi/10.1103/PhysRevLett.115.115001}

\bibitem{zastrau_resolving_2014}
U.~Zastrau, P.~Sperling, M.~Harmand, A.~Becker, T.~Bornath, R.~Bredow,
  S.~Dziarzhytski, T.~Fennel, L.~Fletcher, E.~F\"orster, S.~G\"ode, G.~Gregori,
  V.~Hilbert, D.~Hochhaus, B.~Holst, T.~Laarmann, H.~Lee, T.~Ma, J.~Mithen,
  R.~Mitzner, C.~Murphy, M.~Nakatsutsumi, P.~Neumayer, A.~Przystawik,
  S.~Roling, M.~Schulz, B.~Siemer, S.~Skruszewicz, J.~Tiggesb\"aumker,
  S.~Toleikis, T.~Tschentscher, T.~White, M.~W\"ostmann, H.~Zacharias,
  T.~D\"oppner, S.~Glenzer, R.~Redmer,
  \href{https://link.aps.org/doi/10.1103/PhysRevLett.112.105002}{Resolving
  {Ultrafast} {Heating} of {Dense} {Cryogenic} {Hydrogen}}, Phys.~Rev.~Lett.
  112~(10) (2014) 105002.
\newblock \href {http://dx.doi.org/10.1103/PhysRevLett.112.105002}
  {\path{doi:10.1103/PhysRevLett.112.105002}}.
\newline\urlprefix\url{https://link.aps.org/doi/10.1103/PhysRevLett.112.105002}

\bibitem{tschentscher_photon_2017}
T.~Tschentscher, C.~Bressler, J.~Gr\"unert, A.~Madsen, A.~P. Mancuso, M.~Meyer,
  A.~Scherz, H.~Sinn, U.~Zastrau,
  \href{http://www.mdpi.com/2076-3417/7/6/592}{Photon {Beam} {Transport} and
  {Scientific} {Instruments} at the {European} {XFEL}}, Appl.~Sci. 7~(6) (2017)
  592.
\newblock \href {http://dx.doi.org/10.3390/app7060592}
  {\path{doi:10.3390/app7060592}}.
\newline\urlprefix\url{http://www.mdpi.com/2076-3417/7/6/592}

\bibitem{fortov_phase_2007}
V.~E. Fortov, R.~I. Ilkaev, V.~A. Arinin, V.~V. Burtzev, V.~A. Golubev, I.~L.
  Iosilevskiy, V.~V. Khrustalev, A.~L. Mikhailov, M.~A. Mochalov, V.~Y.
  Ternovoi, M.~V. Zhernokletov,
  \href{https://link.aps.org/doi/10.1103/PhysRevLett.99.185001}{Phase
  {Transition} in a {Strongly} {Nonideal} {Deuterium} {Plasma} {Generated} by
  {Quasi}-{Isentropical} {Compression} at {Megabar} {Pressures}},
  Phys.~Rev.~Lett. 99~(18) (2007) 185001.
\newblock \href {http://dx.doi.org/10.1103/PhysRevLett.99.185001}
  {\path{doi:10.1103/PhysRevLett.99.185001}}.
\newline\urlprefix\url{https://link.aps.org/doi/10.1103/PhysRevLett.99.185001}

\bibitem{fortov_shock_2010}
V.~E. Fortov, I.~V. Lomonosov,
  \href{https://link.springer.com/article/10.1007/s00193-009-0224-8}{Shock
  waves and equations of state of matter}, Shock Waves 20~(1) (2010) 53--71.
\newblock \href {http://dx.doi.org/10.1007/s00193-009-0224-8}
  {\path{doi:10.1007/s00193-009-0224-8}}.
\newline\urlprefix\url{https://link.springer.com/article/10.1007/s00193-009-0224-8}

\bibitem{glenzer_observations_2007}
S.~H. Glenzer, O.~L. Landen, P.~Neumayer, R.~W. Lee, K.~Widmann, S.~W.
  Pollaine, R.~J. Wallace, G.~Gregori, A.~H\"oll, T.~Bornath, R.~Thiele,
  V.~Schwarz, W.-D. Kraeft, R.~Redmer,
  \href{https://link.aps.org/doi/10.1103/PhysRevLett.98.065002}{Observations of
  {Plasmons} in {Warm} {Dense} {Matter}}, Phys.~Rev.~Lett. 98~(6) (2007)
  065002.
\newblock \href {http://dx.doi.org/10.1103/PhysRevLett.98.065002}
  {\path{doi:10.1103/PhysRevLett.98.065002}}.
\newline\urlprefix\url{https://link.aps.org/doi/10.1103/PhysRevLett.98.065002}

\bibitem{fortmann_theory_2012}
C.~Fortmann, C.~Niemann, S.~H. Glenzer,
  \href{https://link.aps.org/doi/10.1103/PhysRevB.86.174116}{Theory of x-ray
  scattering in high-pressure electrides}, Phys.~Rev.~B 86~(17) (2012) 174116.
\newblock \href {http://dx.doi.org/10.1103/PhysRevB.86.174116}
  {\path{doi:10.1103/PhysRevB.86.174116}}.
\newline\urlprefix\url{https://link.aps.org/doi/10.1103/PhysRevB.86.174116}

\bibitem{clerouin_evidence_2015}
J.~Clerouin, G.~Robert, P.~Arnault, C.~Ticknor, J.~D. Kress, L.~A. Collins,
  \href{https://link.aps.org/doi/10.1103/PhysRevE.91.011101}{Evidence for
  out-of-equilibrium states in warm dense matter probed by x-ray {Thomson}
  scattering}, Phys.~Rev.~E 91~(1) (2015) 011101.
\newblock \href {http://dx.doi.org/10.1103/PhysRevE.91.011101}
  {\path{doi:10.1103/PhysRevE.91.011101}}.
\newline\urlprefix\url{https://link.aps.org/doi/10.1103/PhysRevE.91.011101}

\bibitem{kritcher_ultrafast_2008}
A.~L. Kritcher, P.~Neumayer, J.~Castor, T.~D\"oppner, R.~W. Falcone, O.~L.
  Landen, H.~J. Lee, R.~W. Lee, E.~C. Morse, A.~Ng, S.~Pollaine, D.~Price,
  S.~H. Glenzer,
  \href{http://science.sciencemag.org/content/322/5898/69}{Ultrafast {X}-ray
  {Thomson} {Scattering} of {Shock}-{Compressed} {Matter}}, Science 322~(5898)
  (2008) 69--71.
\newblock \href {http://dx.doi.org/10.1126/science.1161466}
  {\path{doi:10.1126/science.1161466}}.
\newline\urlprefix\url{http://science.sciencemag.org/content/322/5898/69}

\bibitem{kraus_nanosecond_2016}
D.~Kraus, A.~Ravasio, M.~Gauthier, D.~O. Gericke, J.~Vorberger, S.~Frydrych,
  J.~Helfrich, L.~B. Fletcher, G.~Schaumann, B.~Nagler, B.~Barbrel,
  B.~Bachmann, E.~J. Gamboa, S.~G\"ode, E.~Granados, G.~Gregori, H.~J. Lee,
  P.~Neumayer, W.~Schumaker, T.~D\"oppner, R.~W. Falcone, S.~H. Glenzer,
  M.~Roth,
  \href{https://www.ncbi.nlm.nih.gov/pmc/articles/PMC4793081/}{Nanosecond
  formation of diamond and lonsdaleite by shock compression of graphite},
  Nat.~Commun. 7.
\newblock \href {http://dx.doi.org/10.1038/ncomms10970}
  {\path{doi:10.1038/ncomms10970}}.
\newline\urlprefix\url{https://www.ncbi.nlm.nih.gov/pmc/articles/PMC4793081/}

\bibitem{davis_x-ray_2016}
P.~Davis, T.~D\"oppner, J.~R. Rygg, C.~Fortmann, L.~Divol, A.~Pak, L.~Fletcher,
  A.~Becker, B.~Holst, P.~Sperling, R.~Redmer, M.~P. Desjarlais, P.~Celliers,
  G.~W. Collins, O.~L. Landen, R.~W. Falcone, S.~H. Glenzer,
  \href{https://www.ncbi.nlm.nih.gov/pmc/articles/PMC4835540/}{X-ray scattering
  measurements of dissociation-induced metallization of dynamically compressed
  deuterium}, Nat.~Commun. 7.
\newblock \href {http://dx.doi.org/10.1038/ncomms11189}
  {\path{doi:10.1038/ncomms11189}}.
\newline\urlprefix\url{https://www.ncbi.nlm.nih.gov/pmc/articles/PMC4835540/}

\bibitem{glenzer_x-ray_2009}
S.~H. Glenzer, R.~Redmer,
  \href{http://link.aps.org/doi/10.1103/RevModPhys.81.1625}{X-ray {Thomson}
  scattering in high energy density plasmas}, Rev.~Mod.~Phys. 81~(4) (2009)
  1625--1663.
\newblock \href {http://dx.doi.org/10.1103/RevModPhys.81.1625}
  {\path{doi:10.1103/RevModPhys.81.1625}}.
\newline\urlprefix\url{http://link.aps.org/doi/10.1103/RevModPhys.81.1625}

\bibitem{ng_outstanding_2012}
A.~Ng,
  \href{http://onlinelibrary.wiley.com/doi/10.1002/qua.23197/abstract}{Outstanding
  questions in electron--ion energy relaxation, lattice stability, and
  dielectric function of warm dense matter}, Int.~J.~Quantum Chem. 112~(1)
  (2012) 150--160.
\newblock \href {http://dx.doi.org/10.1002/qua.23197}
  {\path{doi:10.1002/qua.23197}}.
\newline\urlprefix\url{http://onlinelibrary.wiley.com/doi/10.1002/qua.23197/abstract}

\bibitem{ping_broadband_2006}
Y.~Ping, D.~Hanson, I.~Koslow, T.~Ogitsu, D.~Prendergast, E.~Schwegler,
  G.~Collins, A.~Ng,
  \href{https://link.aps.org/doi/10.1103/PhysRevLett.96.255003}{Broadband
  {Dielectric} {Function} of {Nonequilibrium} {Warm} {Dense} {Gold}},
  Phys.~Rev.~Lett. 96~(25) (2006) 255003.
\newblock \href {http://dx.doi.org/10.1103/PhysRevLett.96.255003}
  {\path{doi:10.1103/PhysRevLett.96.255003}}.
\newline\urlprefix\url{https://link.aps.org/doi/10.1103/PhysRevLett.96.255003}

\bibitem{ng_dc_2016}
A.~Ng, P.~Sterne, S.~Hansen, V.~Recoules, Z.~Chen, Y.~Y. Tsui, B.~Wilson,
  \href{https://link.aps.org/doi/10.1103/PhysRevE.94.033213}{dc conductivity of
  two-temperature warm dense gold}, Phys.~Rev.~E 94~(3) (2016) 033213.
\newblock \href {http://dx.doi.org/10.1103/PhysRevE.94.033213}
  {\path{doi:10.1103/PhysRevE.94.033213}}.
\newline\urlprefix\url{https://link.aps.org/doi/10.1103/PhysRevE.94.033213}

\bibitem{ping_differential_2015}
Y.~Ping, A.~Fernandez-Panella, H.~Sio, A.~Correa, R.~Shepherd, O.~Landen, R.~A.
  London, P.~A. Sterne, H.~D. Whitley, D.~Fratanduono, T.~R. Boehly, G.~W.
  Collins, \href{http://aip.scitation.org/doi/10.1063/1.4929797}{Differential
  heating: {A} versatile method for thermal conductivity measurements in
  high-energy-density matter}, Phys.~Plasmas 22~(9) (2015) 092701.
\newblock \href {http://dx.doi.org/10.1063/1.4929797}
  {\path{doi:10.1063/1.4929797}}.
\newline\urlprefix\url{http://aip.scitation.org/doi/10.1063/1.4929797}

\bibitem{chen_evolution_2013}
Z.~Chen, B.~Holst, S.~E. Kirkwood, V.~Sametoglu, M.~Reid, Y.~Y. Tsui,
  V.~Recoules, A.~Ng,
  \href{https://link.aps.org/doi/10.1103/PhysRevLett.110.135001}{Evolution of
  ac {Conductivity} in {Nonequilibrium} {Warm} {Dense} {Gold}},
  Phys.~Rev.~Lett. 110~(13) (2013) 135001.
\newblock \href {http://dx.doi.org/10.1103/PhysRevLett.110.135001}
  {\path{doi:10.1103/PhysRevLett.110.135001}}.
\newline\urlprefix\url{https://link.aps.org/doi/10.1103/PhysRevLett.110.135001}

\bibitem{chen_single-shot_2016}
Z.~Chen, P.~Hering, S.~B. Brown, C.~Curry, Y.~Y. Tsui, S.~H. Glenzer,
  \href{http://aip.scitation.org/doi/10.1063/1.4962057}{A single-shot spatial
  chirp method for measuring initial {AC} conductivity evolution of femtosecond
  laser pulse excited warm dense matter}, Rev.~Sci.~Instrum. 87~(11) (2016)
  11E548.
\newblock \href {http://dx.doi.org/10.1063/1.4962057}
  {\path{doi:10.1063/1.4962057}}.
\newline\urlprefix\url{http://aip.scitation.org/doi/10.1063/1.4962057}

\bibitem{hartley_electron-ion_2015}
N.~J. Hartley, P.~Belancourt, D.~A. Chapman, T.~D\"oppner, R.~P. Drake, D.~O.
  Gericke, S.~H. Glenzer, D.~Khaghani, S.~LePape, T.~Ma, P.~Neumayer, A.~Pak,
  L.~Peters, S.~Richardson, J.~Vorberger, T.~G. White, G.~Gregori,
  \href{http://www.sciencedirect.com/science/article/pii/S1574181814000639}{Electron-ion
  temperature equilibration in warm dense tantalum}, High Energy Density Phys.
  14~(Supplement C) (2015) 1--5.
\newblock \href {http://dx.doi.org/10.1016/j.hedp.2014.10.003}
  {\path{doi:10.1016/j.hedp.2014.10.003}}.
\newline\urlprefix\url{http://www.sciencedirect.com/science/article/pii/S1574181814000639}

\bibitem{ernstorfer_formation_2009}
R.~Ernstorfer, M.~Harb, C.~T. Hebeisen, G.~Sciaini, T.~Dartigalongue, R.~J.~D.
  Miller, \href{http://www.sciencemag.org/content/323/5917/1033}{The
  {Formation} of {Warm} {Dense} {Matter}: {Experimental} {Evidence} for
  {Electronic} {Bond} {Hardening} in {Gold}}, Science 323~(5917) (2009)
  1033--1037.
\newblock \href {http://dx.doi.org/10.1126/science.1162697}
  {\path{doi:10.1126/science.1162697}}.
\newline\urlprefix\url{http://www.sciencemag.org/content/323/5917/1033}

\bibitem{ebeling_fortov_filinov_17}
W.~Ebeling, V.~Fortov, V.~Filinov, Quantum Statistics of Dense Gases and
  Nonideal Plasmas, Springer Series in Plasma Science and Technology, Springer,
  2017.

\bibitem{desilva_electrical_1998}
A.~W. DeSilva, J.~D. Katsouros,
  \href{https://link.aps.org/doi/10.1103/PhysRevE.57.5945}{Electrical
  conductivity of dense copper and aluminum plasmas}, Phys.~Rev.~E 57~(5)
  (1998) 5945--5951.
\newblock \href {http://dx.doi.org/10.1103/PhysRevE.57.5945}
  {\path{doi:10.1103/PhysRevE.57.5945}}.
\newline\urlprefix\url{https://link.aps.org/doi/10.1103/PhysRevE.57.5945}

\bibitem{mostovych_reflective_1997}
A.~N. Mostovych, Y.~Chan,
  \href{https://link.aps.org/doi/10.1103/PhysRevLett.79.5094}{Reflective
  {Probing} of the {Electrical} {Conductivity} of {Hot} {Aluminum} in the
  {Solid}, {Liquid}, and {Plasma} {Phases}}, Phys.~Rev.~Lett. 79~(25) (1997)
  5094--5097.
\newblock \href {http://dx.doi.org/10.1103/PhysRevLett.79.5094}
  {\path{doi:10.1103/PhysRevLett.79.5094}}.
\newline\urlprefix\url{https://link.aps.org/doi/10.1103/PhysRevLett.79.5094}

\bibitem{filinov_construction_1986}
V.~S. Filinov,
  \href{http://www.sciencedirect.com/science/article/pii/004155538690176X}{Construction
  of a {Monte}-{Carlo} method for calculating {Feynman} integrals}, USSR
  Comput.~Math.~Math.~Phys. 26~(1) (1986) 21--29.
\newblock \href {http://dx.doi.org/10.1016/0041-5553(86)90176-X}
  {\path{doi:10.1016/0041-5553(86)90176-X}}.
\newline\urlprefix\url{http://www.sciencedirect.com/science/article/pii/004155538690176X}

\bibitem{filinov_thermodynamics_2001}
V.~S. Filinov, M.~Bonitz, W.~Ebeling, V.~E. Fortov,
  \href{http://stacks.iop.org/0741-3335/43/i=6/a=301}{Thermodynamics of hot
  dense {H}-plasmas: path integral {Monte} {Carlo} simulations and analytical
  approximations}, Plasma Phys.~Contr.~Fusion 43~(6) (2001) 743.
\newblock \href {http://dx.doi.org/10.1088/0741-3335/43/6/301}
  {\path{doi:10.1088/0741-3335/43/6/301}}.
\newline\urlprefix\url{http://stacks.iop.org/0741-3335/43/i=6/a=301}

\bibitem{filinov_phase_2001}
V.~S. Filinov, V.~E. Fortov, M.~Bonitz, P.~R. Levashov,
  \href{http://link.springer.com/article/10.1134/1.1427127}{Phase transition in
  strongly degenerate hydrogen plasma}, JETP Lett. 74~(7) (2001) 384--387.
\newblock \href {http://dx.doi.org/10.1134/1.1427127}
  {\path{doi:10.1134/1.1427127}}.
\newline\urlprefix\url{http://link.springer.com/article/10.1134/1.1427127}

\bibitem{filinov_thermodynamic_2004}
V.~S. Filinov, M.~Bonitz, V.~E. Fortov, W.~Ebeling, P.~Levashov, M.~Schlanges,
  \href{http://onlinelibrary.wiley.com/doi/10.1002/ctpp.200410057/abstract}{Thermodynamic
  {Properties} and {Plasma} {Phase} {Transition} in dense {Hydrogen}},
  Contrib.~Plasma Phys. 44~(5-6) (2004) 388--394.
\newblock \href {http://dx.doi.org/10.1002/ctpp.200410057}
  {\path{doi:10.1002/ctpp.200410057}}.
\newline\urlprefix\url{http://onlinelibrary.wiley.com/doi/10.1002/ctpp.200410057/abstract}

\bibitem{filinov_correlation_2007}
V.~S. Filinov, H.~Fehske, M.~Bonitz, V.~E. Fortov, P.~Levashov,
  \href{http://link.aps.org/doi/10.1103/PhysRevE.75.036401}{Correlation effects
  in partially ionized mass asymmetric electron-hole plasmas}, Phys.~Rev.~E
  75~(3) (2007) 036401.
\newblock \href {http://dx.doi.org/10.1103/PhysRevE.75.036401}
  {\path{doi:10.1103/PhysRevE.75.036401}}.
\newline\urlprefix\url{http://link.aps.org/doi/10.1103/PhysRevE.75.036401}

\bibitem{filinov_proton_2012}
V.~S. Filinov, M.~Bonitz, H.~Fehske, V.~E. Fortov, P.~R. Levashov,
  \href{http://onlinelibrary.wiley.com/doi/10.1002/ctpp.201100085/abstract}{Proton
  {Crystallization} in a {Dense} {Hydrogen} {Plasma}}, Contrib.~Plasma Phys.
  52~(3) (2012) 224--228.
\newblock \href {http://dx.doi.org/10.1002/ctpp.201100085}
  {\path{doi:10.1002/ctpp.201100085}}.
\newline\urlprefix\url{http://onlinelibrary.wiley.com/doi/10.1002/ctpp.201100085/abstract}

\bibitem{filinov_fermionic_2015}
V.~S. Filinov, V.~E. Fortov, M.~Bonitz, Z.~Moldabekov,
  \href{http://link.aps.org/doi/10.1103/PhysRevE.91.033108}{Fermionic
  path-integral {Monte} {Carlo} results for the uniform electron gas at finite
  temperature}, Phys.~Rev.~E 91~(3) (2015) 033108.
\newblock \href {http://dx.doi.org/10.1103/PhysRevE.91.033108}
  {\path{doi:10.1103/PhysRevE.91.033108}}.
\newline\urlprefix\url{http://link.aps.org/doi/10.1103/PhysRevE.91.033108}

\bibitem{filinov_thermodynamics_2015}
V.~Filinov, M.~Bonitz, Y.~Ivanov, E.-M. Ilgenfritz, V.~Fortov,
  \href{http://onlinelibrary.wiley.com/doi/10.1002/ctpp.201400056/abstract}{Thermodynamics
  of the {Quark}-{Gluon} {Plasma} at {Finite} {Chemical} {Potential}: {Color}
  {Path} {Integral} {Monte} {Carlo} {Results}}, Contrib.~Plasma Phys. 55~(2-3)
  (2015) 203--208.
\newblock \href {http://dx.doi.org/10.1002/ctpp.201400056}
  {\path{doi:10.1002/ctpp.201400056}}.
\newline\urlprefix\url{http://onlinelibrary.wiley.com/doi/10.1002/ctpp.201400056/abstract}

\bibitem{filinov_total_2015}
V.~S. Filinov, V.~E. Fortov, M.~Bonitz, Z.~Moldabekov,
  \href{http://stacks.iop.org/1742-6596/653/i=1/a=012113}{Total and correlation
  energy of the uniform polarized electron gas at finite temperature: {Direct}
  path integral simulations}, J.~Phys.~Conf.~Ser. 653~(1) (2015) 012113.
\newblock \href {http://dx.doi.org/10.1088/1742-6596/653/1/012113}
  {\path{doi:10.1088/1742-6596/653/1/012113}}.
\newline\urlprefix\url{http://stacks.iop.org/1742-6596/653/i=1/a=012113}

\bibitem{ceperley_fermion_1991}
D.~M. Ceperley, \href{http://link.springer.com/10.1007/BF01030009}{Fermion
  nodes}, J.~Stat.~Phys. 63~(5-6) (1991) 1237--1267.
\newblock \href {http://dx.doi.org/10.1007/BF01030009}
  {\path{doi:10.1007/BF01030009}}.
\newline\urlprefix\url{http://link.springer.com/10.1007/BF01030009}

\bibitem{ceperley_path-integral_1992}
D.~M. Ceperley,
  \href{http://link.aps.org/doi/10.1103/PhysRevLett.69.331}{{Path-integral
  calculations of normal liquid $^3$He}}, Phys.~Rev.~Lett. 69~(2) (1992)
  331--334.
\newblock \href {http://dx.doi.org/10.1103/PhysRevLett.69.331}
  {\path{doi:10.1103/PhysRevLett.69.331}}.
\newline\urlprefix\url{http://link.aps.org/doi/10.1103/PhysRevLett.69.331}

\bibitem{militzer_path_2000}
B.~Militzer, D.~M. Ceperley,
  \href{http://link.aps.org/doi/10.1103/PhysRevLett.85.1890}{Path {Integral}
  {Monte} {Carlo} {Calculation} of the {Deuterium} {Hugoniot}},
  Phys.~Rev.~Lett. 85~(9) (2000) 1890--1893.
\newblock \href {http://dx.doi.org/10.1103/PhysRevLett.85.1890}
  {\path{doi:10.1103/PhysRevLett.85.1890}}.
\newline\urlprefix\url{http://link.aps.org/doi/10.1103/PhysRevLett.85.1890}

\bibitem{mermin_thermal_1965}
N.~D. Mermin, \href{http://link.aps.org/doi/10.1103/PhysRev.137.A1441}{Thermal
  {Properties} of the {Inhomogeneous} {Electron} {Gas}}, Phys.~Rev. 137~(5A)
  (1965) A1441--A1443.
\newblock \href {http://dx.doi.org/10.1103/PhysRev.137.A1441}
  {\path{doi:10.1103/PhysRev.137.A1441}}.
\newline\urlprefix\url{http://link.aps.org/doi/10.1103/PhysRev.137.A1441}

\bibitem{gupta_density_1982}
U.~Gupta, A.~K. Rajagopal,
  \href{http://www.sciencedirect.com/science/article/pii/0370157382900771}{Density
  functional formalism at finite temperatures with some applications},
  Phys.~Rep. 87~(6) (1982) 259--311.
\newblock \href {http://dx.doi.org/10.1016/0370-1573(82)90077-1}
  {\path{doi:10.1016/0370-1573(82)90077-1}}.
\newline\urlprefix\url{http://www.sciencedirect.com/science/article/pii/0370157382900771}

\bibitem{Pribram-Jones2014}
A.~Pribram-Jones, S.~Pittalis, E.~K.~U. Gross, K.~Burke,
  \href{https://doi.org/10.1007/978-3-319-04912-0_2}{Thermal Density Functional
  Theory in Context}, Springer International Publishing, Cham, 2014, pp.
  25--60.
\newblock \href {http://dx.doi.org/10.1007/978-3-319-04912-0_2}
  {\path{doi:10.1007/978-3-319-04912-0_2}}.
\newline\urlprefix\url{https://doi.org/10.1007/978-3-319-04912-0_2}

\bibitem{balbuena1999molecular}
P.~Balbuena, J.~Seminario,
  \href{https://books.google.de/books?id=dpgXPzTLSpYC}{Molecular Dynamics: From
  Classical to Quantum Methods}, Theoretical and Computational Chemistry,
  Elsevier Science, 1999.
\newline\urlprefix\url{https://books.google.de/books?id=dpgXPzTLSpYC}

\bibitem{desjarlais_density-functional_2003}
M.~P. Desjarlais,
  \href{https://link.aps.org/doi/10.1103/PhysRevB.68.064204}{Density-functional
  calculations of the liquid deuterium {Hugoniot}, reshock, and reverberation
  timing}, Phys.~Rev.~B 68~(6) (2003) 064204.
\newblock \href {http://dx.doi.org/10.1103/PhysRevB.68.064204}
  {\path{doi:10.1103/PhysRevB.68.064204}}.
\newline\urlprefix\url{https://link.aps.org/doi/10.1103/PhysRevB.68.064204}

\bibitem{holst_thermophysical_2008}
B.~Holst, R.~Redmer, M.~P. Desjarlais,
  \href{http://link.aps.org/doi/10.1103/PhysRevB.77.184201}{Thermophysical
  properties of warm dense hydrogen using quantum molecular dynamics
  simulations}, Phys.~Rev.~B 77~(18) (2008) 184201.
\newblock \href {http://dx.doi.org/10.1103/PhysRevB.77.184201}
  {\path{doi:10.1103/PhysRevB.77.184201}}.
\newline\urlprefix\url{http://link.aps.org/doi/10.1103/PhysRevB.77.184201}

\bibitem{holst_electronic_2011}
B.~Holst, M.~French, R.~Redmer,
  \href{http://link.aps.org/doi/10.1103/PhysRevB.83.235120}{Electronic
  transport coefficients from ab initio simulations and application to dense
  liquid hydrogen}, Phys.~Rev.~B 83~(23) (2011) 235120.
\newblock \href {http://dx.doi.org/10.1103/PhysRevB.83.235120}
  {\path{doi:10.1103/PhysRevB.83.235120}}.
\newline\urlprefix\url{http://link.aps.org/doi/10.1103/PhysRevB.83.235120}

\bibitem{witte_warm_2017}
B.~Witte, L.~Fletcher, E.~Galtier, E.~Gamboa, H.~Lee, U.~Zastrau, R.~Redmer,
  S.~Glenzer, P.~Sperling,
  \href{https://link.aps.org/doi/10.1103/PhysRevLett.118.225001}{Warm {Dense}
  {Matter} {Demonstrating} {Non}-{Drude} {Conductivity} from {Observations} of
  {Nonlinear} {Plasmon} {Damping}}, Phys.~Rev.~Lett. 118~(22) (2017) 225001.
\newblock \href {http://dx.doi.org/10.1103/PhysRevLett.118.225001}
  {\path{doi:10.1103/PhysRevLett.118.225001}}.
\newline\urlprefix\url{https://link.aps.org/doi/10.1103/PhysRevLett.118.225001}

\bibitem{clay_benchmarking_2014}
R.~C. Clay, J.~Mcminis, J.~M. McMahon, C.~Pierleoni, D.~M. Ceperley, M.~A.
  Morales,
  \href{http://link.aps.org/doi/10.1103/PhysRevB.89.184106}{Benchmarking
  exchange-correlation functionals for hydrogen at high pressures using quantum
  {Monte} {Carlo}}, Phys.~Rev.~B 89~(18) (2014) 184106.
\newblock \href {http://dx.doi.org/10.1103/PhysRevB.89.184106}
  {\path{doi:10.1103/PhysRevB.89.184106}}.
\newline\urlprefix\url{http://link.aps.org/doi/10.1103/PhysRevB.89.184106}

\bibitem{clay_benchmarking_2016}
R.~C. Clay, M.~Holzmann, D.~M. Ceperley, M.~A. Morales,
  \href{http://link.aps.org/doi/10.1103/PhysRevB.93.035121}{Benchmarking
  density functionals for hydrogen-helium mixtures with quantum {Monte}
  {Carlo}: {Energetics}, pressures, and forces}, Phys.~Rev.~B 93~(3) (2016)
  035121.
\newblock \href {http://dx.doi.org/10.1103/PhysRevB.93.035121}
  {\path{doi:10.1103/PhysRevB.93.035121}}.
\newline\urlprefix\url{http://link.aps.org/doi/10.1103/PhysRevB.93.035121}

\bibitem{karasiev_importance_2016}
V.~V. Karasiev, L.~Calder\'in, S.~B. Trickey,
  \href{http://link.aps.org/doi/10.1103/PhysRevE.93.063207}{Importance of
  finite-temperature exchange correlation for warm dense matter calculations},
  Phys.~Rev.~E 93~(6) (2016) 063207.
\newblock \href {http://dx.doi.org/10.1103/PhysRevE.93.063207}
  {\path{doi:10.1103/PhysRevE.93.063207}}.
\newline\urlprefix\url{http://link.aps.org/doi/10.1103/PhysRevE.93.063207}

\bibitem{dharma-wardana_current_2016}
M.~W.~C. Dharma-wardana, \href{http://www.mdpi.com/2079-3197/4/2/16}{Current
  {Issues} in {Finite}-{T} {Density}-{Functional} {Theory} and
  {Warm}-{Correlated} {Matter}}, Computation 4~(2) (2016) 16.
\newblock \href {http://dx.doi.org/10.3390/computation4020016}
  {\path{doi:10.3390/computation4020016}}.
\newline\urlprefix\url{http://www.mdpi.com/2079-3197/4/2/16}

\bibitem{pribram-jones_dft:_2015}
A.~Pribram-Jones, D.~A. Gross, K.~Burke,
  \href{https://doi.org/10.1146/annurev-physchem-040214-121420}{{DFT}: {A}
  {Theory} {Full} of {Holes}?}, Annu.~Rev.~Phys.~Chem. 66~(1) (2015) 283--304.
\newblock \href {http://dx.doi.org/10.1146/annurev-physchem-040214-121420}
  {\path{doi:10.1146/annurev-physchem-040214-121420}}.
\newline\urlprefix\url{https://doi.org/10.1146/annurev-physchem-040214-121420}

\bibitem{driver_all-electron_2012}
K.~P. Driver, B.~Militzer,
  \href{http://link.aps.org/doi/10.1103/PhysRevLett.108.115502}{All-{Electron}
  {Path} {Integral} {Monte} {Carlo} {Simulations} of {Warm} {Dense} {Matter}:
  {Application} to {Water} and {Carbon} {Plasmas}}, Phys.~Rev.~Lett. 108~(11)
  (2012) 115502.
\newblock \href {http://dx.doi.org/10.1103/PhysRevLett.108.115502}
  {\path{doi:10.1103/PhysRevLett.108.115502}}.
\newline\urlprefix\url{http://link.aps.org/doi/10.1103/PhysRevLett.108.115502}

\bibitem{militzer_development_2015}
B.~Militzer, K.~P. Driver,
  \href{http://link.aps.org/doi/10.1103/PhysRevLett.115.176403}{Development of
  {Path} {Integral} {Monte} {Carlo} {Simulations} with {Localized} {Nodal}
  {Surfaces} for {Second}-{Row} {Elements}}, Phys.~Rev.~Lett. 115~(17) (2015)
  176403.
\newblock \href {http://dx.doi.org/10.1103/PhysRevLett.115.176403}
  {\path{doi:10.1103/PhysRevLett.115.176403}}.
\newline\urlprefix\url{http://link.aps.org/doi/10.1103/PhysRevLett.115.176403}

\bibitem{driver_first-principles_2016}
K.~P. Driver, B.~Militzer,
  \href{https://link.aps.org/doi/10.1103/PhysRevB.93.064101}{First-principles
  equation of state calculations of warm dense nitrogen}, Phys.~Rev.~B 93~(6)
  (2016) 064101.
\newblock \href {http://dx.doi.org/10.1103/PhysRevB.93.064101}
  {\path{doi:10.1103/PhysRevB.93.064101}}.
\newline\urlprefix\url{https://link.aps.org/doi/10.1103/PhysRevB.93.064101}

\bibitem{zhang_first-principles_2017}
S.~Zhang, K.~P. Driver, F.~Soubiran, B.~Militzer,
  \href{https://link.aps.org/doi/10.1103/PhysRevE.96.013204}{First-principles
  equation of state and shock compression predictions of warm dense
  hydrocarbons}, Phys.~Rev.~E 96~(1) (2017) 013204.
\newblock \href {http://dx.doi.org/10.1103/PhysRevE.96.013204}
  {\path{doi:10.1103/PhysRevE.96.013204}}.
\newline\urlprefix\url{https://link.aps.org/doi/10.1103/PhysRevE.96.013204}

\bibitem{driver_comparison_2017}
K.~P. Driver, F.~Soubiran, S.~Zhang, B.~Militzer,
  \href{https://www.sciencedirect.com/science/article/pii/S1574181817300228}{Comparison
  of path integral {Monte} {Carlo} simulations of helium, carbon, nitrogen,
  oxygen, water, neon, and silicon plasmas}, High Energy Density Phys. 23
  (2017) 81--89.
\newblock \href {http://dx.doi.org/10.1016/j.hedp.2017.03.003}
  {\path{doi:10.1016/j.hedp.2017.03.003}}.
\newline\urlprefix\url{https://www.sciencedirect.com/science/article/pii/S1574181817300228}

\bibitem{driver_first-principles_2017}
K.~P. Driver, B.~Militzer,
  \href{https://link.aps.org/doi/10.1103/PhysRevE.95.043205}{First-principles
  simulations of warm dense lithium fluoride}, Phys.~Rev.~E 95~(4) (2017)
  043205.
\newblock \href {http://dx.doi.org/10.1103/PhysRevE.95.043205}
  {\path{doi:10.1103/PhysRevE.95.043205}}.
\newline\urlprefix\url{https://link.aps.org/doi/10.1103/PhysRevE.95.043205}

\bibitem{lambert_structural_2006}
F.~Lambert, J.~Cl\'erouin, S.~Mazevet,
  \href{http://iopscience.iop.org/article/10.1209/epl/i2006-10184-7/meta}{Structural
  and dynamical properties of hot dense matter by a {Thomas}-{Fermi}-{Dirac}
  molecular dynamics}, Europhys.~Lett. 75~(5) (2006) 681.
\newblock \href {http://dx.doi.org/10.1209/epl/i2006-10184-7}
  {\path{doi:10.1209/epl/i2006-10184-7}}.
\newline\urlprefix\url{http://iopscience.iop.org/article/10.1209/epl/i2006-10184-7/meta}

\bibitem{lambert_properties_2007}
F.~Lambert, J.~Cl\'erouin, S.~Mazevet, D.~Gilles,
  \href{http://onlinelibrary.wiley.com/doi/10.1002/ctpp.200710037/abstract}{Properties
  of {Hot} {Dense} {Plasmas} by {Orbital}-{Free} {Molecular} {Dynamics}},
  Contrib.~Plasma Phys. 47~(4-5) (2007) 272--280.
\newblock \href {http://dx.doi.org/10.1002/ctpp.200710037}
  {\path{doi:10.1002/ctpp.200710037}}.
\newline\urlprefix\url{http://onlinelibrary.wiley.com/doi/10.1002/ctpp.200710037/abstract}

\bibitem{karasiev_generalized-gradient-approximation_2012}
V.~V. Karasiev, T.~Sjostrom, S.~B. Trickey,
  \href{http://link.aps.org/doi/10.1103/PhysRevB.86.115101}{Generalized-gradient-approximation
  noninteracting free-energy functionals for orbital-free density functional
  calculations}, Phys.~Rev.~B 86~(11) (2012) 115101.
\newblock \href {http://dx.doi.org/10.1103/PhysRevB.86.115101}
  {\path{doi:10.1103/PhysRevB.86.115101}}.
\newline\urlprefix\url{http://link.aps.org/doi/10.1103/PhysRevB.86.115101}

\bibitem{sjostrom_fast_2014}
T.~Sjostrom, J.~Daligault,
  \href{http://link.aps.org/doi/10.1103/PhysRevLett.113.155006}{Fast and
  {Accurate} {Quantum} {Molecular} {Dynamics} of {Dense} {Plasmas} {Across}
  {Temperature} {Regimes}}, Phys.~Rev.~Lett. 113~(15) (2014) 155006.
\newblock \href {http://dx.doi.org/10.1103/PhysRevLett.113.155006}
  {\path{doi:10.1103/PhysRevLett.113.155006}}.
\newline\urlprefix\url{http://link.aps.org/doi/10.1103/PhysRevLett.113.155006}

\bibitem{karasiev_finite-temperature_2014}
V.~V. Karasiev, T.~Sjostrom, S.~B. Trickey,
  \href{http://www.sciencedirect.com/science/article/pii/S001046551400304X}{Finite-temperature
  orbital-free {DFT} molecular dynamics: {Coupling} {Profess} and {Quantum}
  {Espresso}}, Comp.~Phys.~Comm. 185~(12) (2014) 3240--3249.
\newblock \href {http://dx.doi.org/10.1016/j.cpc.2014.08.023}
  {\path{doi:10.1016/j.cpc.2014.08.023}}.
\newline\urlprefix\url{http://www.sciencedirect.com/science/article/pii/S001046551400304X}

\bibitem{gao_validity_2016}
C.~Gao, S.~Zhang, W.~Kang, C.~Wang, P.~Zhang, X.~T. He,
  \href{http://link.aps.org/doi/10.1103/PhysRevB.94.205115}{Validity boundary
  of orbital-free molecular dynamics method corresponding to thermal ionization
  of shell structure}, Phys.~Rev.~B 94~(20) (2016) 205115.
\newblock \href {http://dx.doi.org/10.1103/PhysRevB.94.205115}
  {\path{doi:10.1103/PhysRevB.94.205115}}.
\newline\urlprefix\url{http://link.aps.org/doi/10.1103/PhysRevB.94.205115}

\bibitem{dufty_scaling_2011}
J.~W. Dufty, S.~B. Trickey,
  \href{https://link.aps.org/doi/10.1103/PhysRevB.84.125118}{Scaling, bounds,
  and inequalities for the noninteracting density functionals at finite
  temperature}, Phys.~Rev.~B 84~(12) (2011) 125118.
\newblock \href {http://dx.doi.org/10.1103/PhysRevB.84.125118}
  {\path{doi:10.1103/PhysRevB.84.125118}}.
\newline\urlprefix\url{https://link.aps.org/doi/10.1103/PhysRevB.84.125118}

\bibitem{zhang_link_2016}
S.~Zhang, S.~Zhao, W.~Kang, P.~Zhang, X.-T. He,
  \href{http://link.aps.org/doi/10.1103/PhysRevB.93.115114}{Link between $k$
  absorption edges and thermodynamic properties of warm dense plasmas
  established by an improved first-principles method}, Phys.~Rev.~B 93~(11)
  (2016) 115114.
\newblock \href {http://dx.doi.org/10.1103/PhysRevB.93.115114}
  {\path{doi:10.1103/PhysRevB.93.115114}}.
\newline\urlprefix\url{http://link.aps.org/doi/10.1103/PhysRevB.93.115114}

\bibitem{zhang_extended_2016}
S.~Zhang, H.~Wang, W.~Kang, P.~Zhang, X.~T. He,
  \href{http://aip.scitation.org/doi/abs/10.1063/1.4947212}{Extended
  application of {Kohn}-{Sham} first-principles molecular dynamics method with
  plane wave approximation at high energy -- {From} cold materials to hot dense
  plasmas}, Phys.~Plasmas 23~(4) (2016) 042707.
\newblock \href {http://dx.doi.org/10.1063/1.4947212}
  {\path{doi:10.1063/1.4947212}}.
\newline\urlprefix\url{http://aip.scitation.org/doi/abs/10.1063/1.4947212}

\bibitem{pierleoni_computational_2005}
C.~Pierleoni, D.~M. Ceperley,
  \href{http://onlinelibrary.wiley.com/doi/10.1002/cphc.200400587/abstract}{Computational
  {Methods} in {Coupled} {Electron}-{Ion} {Monte} {Carlo} {Simulations}},
  Chem.~Phys.~Chem. 6~(9) (2005) 1872--1878.
\newblock \href {http://dx.doi.org/10.1002/cphc.200400587}
  {\path{doi:10.1002/cphc.200400587}}.
\newline\urlprefix\url{http://onlinelibrary.wiley.com/doi/10.1002/cphc.200400587/abstract}

\bibitem{ceperley_coupled_2002}
D.~Ceperley, M.~Dewing, C.~Pierleoni,
  \href{http://arxiv.org/abs/physics/0207006}{The {Coupled}
  {Electronic}-{Ionic} {Monte} {Carlo} {Simulation} {Method}},
  arXiv:physics/0207006ArXiv: physics/0207006.
\newline\urlprefix\url{http://arxiv.org/abs/physics/0207006}

\bibitem{pierleoni_coupled_2004}
C.~Pierleoni, D.~M. Ceperley, M.~Holzmann,
  \href{http://link.aps.org/doi/10.1103/PhysRevLett.93.146402}{Coupled
  {Electron}-{Ion} {Monte} {Carlo} {Calculations} of {Dense} {Metallic}
  {Hydrogen}}, Phys.~Rev.~Lett. 93~(14) (2004) 146402.
\newblock \href {http://dx.doi.org/10.1103/PhysRevLett.93.146402}
  {\path{doi:10.1103/PhysRevLett.93.146402}}.
\newline\urlprefix\url{http://link.aps.org/doi/10.1103/PhysRevLett.93.146402}

\bibitem{tubman_molecular-atomic_2015}
N.~M. Tubman, E.~Liberatore, C.~Pierleoni, M.~Holzmann, D.~M. Ceperley,
  \href{http://link.aps.org/doi/10.1103/PhysRevLett.115.045301}{Molecular-{Atomic}
  {Transition} along the {Deuterium} {Hugoniot} {Curve} with {Coupled}
  {Electron}-{Ion} {Monte} {Carlo} {Simulations}}, Phys.~Rev.~Lett. 115~(4)
  (2015) 045301.
\newblock \href {http://dx.doi.org/10.1103/PhysRevLett.115.045301}
  {\path{doi:10.1103/PhysRevLett.115.045301}}.
\newline\urlprefix\url{http://link.aps.org/doi/10.1103/PhysRevLett.115.045301}

\bibitem{dias_observation_2017}
R.~P. Dias, I.~F. Silvera,
  \href{http://science.sciencemag.org/content/early/2017/01/25/science.aal1579}{Observation
  of the {Wigner}-{Huntington} transition to metallic hydrogen}, Science (2017)
  eaal1579\href {http://dx.doi.org/10.1126/science.aal1579}
  {\path{doi:10.1126/science.aal1579}}.
\newline\urlprefix\url{http://science.sciencemag.org/content/early/2017/01/25/science.aal1579}

\bibitem{morales_equation_2010}
M.~A. Morales, C.~Pierleoni, D.~M. Ceperley,
  \href{http://link.aps.org/doi/10.1103/PhysRevE.81.021202}{Equation of state
  of metallic hydrogen from coupled electron-ion {Monte} {Carlo} simulations},
  Phys.~Rev.~E 81~(2) (2010) 021202.
\newblock \href {http://dx.doi.org/10.1103/PhysRevE.81.021202}
  {\path{doi:10.1103/PhysRevE.81.021202}}.
\newline\urlprefix\url{http://link.aps.org/doi/10.1103/PhysRevE.81.021202}

\bibitem{pierleoni_liquidliquid_2016}
C.~Pierleoni, M.~A. Morales, G.~Rillo, M.~Holzmann, D.~M. Ceperley,
  Liquid-liquid phase transition in hydrogen by coupled electron–ion {Monte}
  {Carlo} simulations, Proc.~Natl.~Acad.~Sci.~U.S.A 113~(18) (2016) 4953--4957.
\newblock \href {http://dx.doi.org/10.1073/pnas.1603853113}
  {\path{doi:10.1073/pnas.1603853113}}.

\bibitem{luo_ab_2015}
Y.~Luo, S.~Sorella,
  \href{http://journal.frontiersin.org/article/10.3389/fmats.2015.00029/full}{Ab
  initio molecular dynamics with quantum {Monte} {Carlo}}, Mech.~Mat. 2 (2015)
  29.
\newblock \href {http://dx.doi.org/10.3389/fmats.2015.00029}
  {\path{doi:10.3389/fmats.2015.00029}}.
\newline\urlprefix\url{http://journal.frontiersin.org/article/10.3389/fmats.2015.00029/full}

\bibitem{attaccalite_stable_2008}
C.~Attaccalite, S.~Sorella,
  \href{http://link.aps.org/doi/10.1103/PhysRevLett.100.114501}{Stable {Liquid}
  {Hydrogen} at {High} {Pressure} by a {Novel} \textit{Ab Initio}
  {Molecular}-{Dynamics} {Calculation}}, Phys.~Rev.~Lett. 100~(11) (2008)
  114501.
\newblock \href {http://dx.doi.org/10.1103/PhysRevLett.100.114501}
  {\path{doi:10.1103/PhysRevLett.100.114501}}.
\newline\urlprefix\url{http://link.aps.org/doi/10.1103/PhysRevLett.100.114501}

\bibitem{mazzola_distinct_2015}
G.~Mazzola, S.~Sorella,
  \href{http://link.aps.org/doi/10.1103/PhysRevLett.114.105701}{Distinct
  {Metallization} and {Atomization} {Transitions} in {Dense} {Liquid}
  {Hydrogen}}, Phys.~Rev.~Lett. 114~(10) (2015) 105701.
\newblock \href {http://dx.doi.org/10.1103/PhysRevLett.114.105701}
  {\path{doi:10.1103/PhysRevLett.114.105701}}.
\newline\urlprefix\url{http://link.aps.org/doi/10.1103/PhysRevLett.114.105701}

\bibitem{zen_ab_2015}
A.~Zen, Y.~Luo, G.~Mazzola, L.~Guidoni, S.~Sorella,
  \href{http://scitation.aip.org/content/aip/journal/jcp/142/14/10.1063/1.4917171}{Ab
  initio molecular dynamics simulation of liquid water by quantum {Monte}
  {Carlo}}, J.~Chem.~Phys. 142~(14) (2015) 144111.
\newblock \href {http://dx.doi.org/10.1063/1.4917171}
  {\path{doi:10.1063/1.4917171}}.
\newline\urlprefix\url{http://scitation.aip.org/content/aip/journal/jcp/142/14/10.1063/1.4917171}

\bibitem{mazzola_unexpectedly_2014}
G.~Mazzola, S.~Yunoki, S.~Sorella,
  \href{http://www.nature.com/ncomms/2014/140319/ncomms4487/full/ncomms4487.html}{Unexpectedly
  high pressure for molecular dissociation in liquid hydrogen by electronic
  simulation}, Nat.~Commun. 5 (2014) 3487.
\newblock \href {http://dx.doi.org/10.1038/ncomms4487}
  {\path{doi:10.1038/ncomms4487}}.
\newline\urlprefix\url{http://www.nature.com/ncomms/2014/140319/ncomms4487/full/ncomms4487.html}

\bibitem{mazzola_finite-temperature_2012}
G.~Mazzola, A.~Zen, S.~Sorella,
  \href{http://scitation.aip.org/content/aip/journal/jcp/137/13/10.1063/1.4755992}{Finite-temperature
  electronic simulations without the {Born}-{Oppenheimer} constraint},
  J.~Chem.~Phys. 137~(13) (2012) 134112.
\newblock \href {http://dx.doi.org/10.1063/1.4755992}
  {\path{doi:10.1063/1.4755992}}.
\newline\urlprefix\url{http://scitation.aip.org/content/aip/journal/jcp/137/13/10.1063/1.4755992}

\bibitem{ullrich2012time}
C.~Ullrich, \href{https://books.google.de/books?id=hCNNsC4sEtkC}{Time-Dependent
  Density-Functional Theory: Concepts and Applications}, Oxford Graduate Texts,
  OUP Oxford, 2012.
\newline\urlprefix\url{https://books.google.de/books?id=hCNNsC4sEtkC}

\bibitem{Ullrich2014}
C.~A. Ullrich,
  \href{https://doi.org/10.1007/978-3-319-04912-0_1}{Time-Dependent
  Density-Functional Theory: Features and Challenges, with a Special View on
  Matter Under Extreme Conditions}, Springer International Publishing, Cham,
  2014, pp. 1--23.
\newblock \href {http://dx.doi.org/10.1007/978-3-319-04912-0_1}
  {\path{doi:10.1007/978-3-319-04912-0_1}}.
\newline\urlprefix\url{https://doi.org/10.1007/978-3-319-04912-0_1}

\bibitem{baczewski_x-ray_2016}
A.~Baczewski, L.~Shulenburger, M.~Desjarlais, S.~Hansen, R.~Magyar,
  \href{http://link.aps.org/doi/10.1103/PhysRevLett.116.115004}{X-ray {Thomson}
  {Scattering} in {Warm} {Dense} {Matter} without the {Chihara}
  {Decomposition}}, Phys.~Rev.~Lett. 116~(11) (2016) 115004.
\newblock \href {http://dx.doi.org/10.1103/PhysRevLett.116.115004}
  {\path{doi:10.1103/PhysRevLett.116.115004}}.
\newline\urlprefix\url{http://link.aps.org/doi/10.1103/PhysRevLett.116.115004}

\bibitem{magyar_stopping_2016}
R.~J. Magyar, L.~Shulenburger, A.~D. Baczewski,
  \href{http://onlinelibrary.wiley.com/doi/10.1002/ctpp.201500143/abstract}{Stopping
  of {Deuterium} in {Warm} {Dense} {Deuterium} from {Ehrenfest}
  {Time}-{Dependent} {Density} {Functional} {Theory}}, Contrib.~Plasma Phys.
  56~(5) (2016) 459--466.
\newblock \href {http://dx.doi.org/10.1002/ctpp.201500143}
  {\path{doi:10.1002/ctpp.201500143}}.
\newline\urlprefix\url{http://onlinelibrary.wiley.com/doi/10.1002/ctpp.201500143/abstract}

\bibitem{dharma-wardana_spin-_2004}
M.~W.~C. Dharma-wardana, F.~Perrot,
  \href{https://link.aps.org/doi/10.1103/PhysRevB.70.035308}{Spin- and
  valley-dependent analysis of the two-dimensional low-density electron system
  in $\mathrm{Si}$ {MOSFETs}}, Phys.~Rev.~B 70~(3) (2004) 035308.
\newblock \href {http://dx.doi.org/10.1103/PhysRevB.70.035308}
  {\path{doi:10.1103/PhysRevB.70.035308}}.
\newline\urlprefix\url{https://link.aps.org/doi/10.1103/PhysRevB.70.035308}

\bibitem{dharma-wardana_static_2006}
M.~W.~C. Dharma-wardana,
  \href{https://link.aps.org/doi/10.1103/PhysRevE.73.036401}{Static and dynamic
  conductivity of warm dense matter within a density-functional approach:
  {Application} to aluminum and gold}, Phys.~Rev.~E 73~(3) (2006) 036401.
\newblock \href {http://dx.doi.org/10.1103/PhysRevE.73.036401}
  {\path{doi:10.1103/PhysRevE.73.036401}}.
\newline\urlprefix\url{https://link.aps.org/doi/10.1103/PhysRevE.73.036401}

\bibitem{dharma-wardana_pair-distribution_2008}
M.~W.~C. Dharma-wardana, M.~S. Murillo,
  \href{https://link.aps.org/doi/10.1103/PhysRevE.77.026401}{Pair-distribution
  functions of two-temperature two-mass systems: {Comparison} of molecular
  dynamics, classical-map hypernetted chain, quantum {Monte} {Carlo}, and
  {Kohn}-{Sham} calculations for dense hydrogen}, Phys.~Rev.~E 77~(2) (2008)
  026401.
\newblock \href {http://dx.doi.org/10.1103/PhysRevE.77.026401}
  {\path{doi:10.1103/PhysRevE.77.026401}}.
\newline\urlprefix\url{https://link.aps.org/doi/10.1103/PhysRevE.77.026401}

\bibitem{dharma-wardana_classical-map_2012}
M.~W.~C. Dharma-Wardana,
  \href{http://onlinelibrary.wiley.com/doi/10.1002/qua.23170/abstract}{The
  classical-map hyper-netted-chain ({CHNC}) method and associated novel
  density-functional techniques for warm dense matter}, Int.~J.~Quantum Chem.
  112~(1) (2012) 53--64.
\newblock \href {http://dx.doi.org/10.1002/qua.23170}
  {\path{doi:10.1002/qua.23170}}.
\newline\urlprefix\url{http://onlinelibrary.wiley.com/doi/10.1002/qua.23170/abstract}

\bibitem{karasiev_nonempirical_2016}
V.~V. Karasiev, J.~W. Dufty, S.~B. Trickey,
  \href{http://arxiv.org/abs/1612.06266}{Nonempirical semi-local free-energy
  density functional for matter under extreme conditions}, arXiv:1612.06266
  [cond-mat]ArXiv: 1612.06266.
\newline\urlprefix\url{http://arxiv.org/abs/1612.06266}

\bibitem{saumon_fluid_1991}
D.~Saumon, G.~Chabrier,
  \href{https://link.aps.org/doi/10.1103/PhysRevA.44.5122}{Fluid hydrogen at
  high density: {Pressure} dissociation}, Phys.~Rev.~A 44~(8) (1991)
  5122--5141.
\newblock \href {http://dx.doi.org/10.1103/PhysRevA.44.5122}
  {\path{doi:10.1103/PhysRevA.44.5122}}.
\newline\urlprefix\url{https://link.aps.org/doi/10.1103/PhysRevA.44.5122}

\bibitem{saumon_fluid_1992}
D.~Saumon, G.~Chabrier,
  \href{http://link.aps.org/doi/10.1103/PhysRevA.46.2084}{Fluid hydrogen at
  high density: {Pressure} ionization}, Phys.~Rev.~A 46~(4) (1992) 2084--2100.
\newblock \href {http://dx.doi.org/10.1103/PhysRevA.46.2084}
  {\path{doi:10.1103/PhysRevA.46.2084}}.
\newline\urlprefix\url{http://link.aps.org/doi/10.1103/PhysRevA.46.2084}

\bibitem{chabrier_quantum_1993}
G.~Chabrier, \href{http://adsabs.harvard.edu/doi/10.1086/173115}{Quantum
  effects in dense {Coulumbic} matter - {Application} to the cooling of white
  dwarfs}, Astrophys.~J. 414 (1993) 695.
\newblock \href {http://dx.doi.org/10.1086/173115} {\path{doi:10.1086/173115}}.
\newline\urlprefix\url{http://adsabs.harvard.edu/doi/10.1086/173115}

\bibitem{chabrier_equation_1998}
G.~Chabrier, A.~Y. Potekhin,
  \href{https://link.aps.org/doi/10.1103/PhysRevE.58.4941}{Equation of state of
  fully ionized electron-ion plasmas}, Phys.~Rev.~E 58~(4) (1998) 4941--4949.
\newblock \href {http://dx.doi.org/10.1103/PhysRevE.58.4941}
  {\path{doi:10.1103/PhysRevE.58.4941}}.
\newline\urlprefix\url{https://link.aps.org/doi/10.1103/PhysRevE.58.4941}

\bibitem{potekhin_thermodynamic_2010}
A.~Y. Potekhin, G.~Chabrier,
  \href{http://onlinelibrary.wiley.com/doi/10.1002/ctpp.201010017/abstract}{Thermodynamic
  {Functions} of {Dense} {Plasmas}: {Analytic} {Approximations} for
  {Astrophysical} {Applications}}, Contrib.~Plasma Phys. 50~(1) (2010) 82--87.
\newblock \href {http://dx.doi.org/10.1002/ctpp.201010017}
  {\path{doi:10.1002/ctpp.201010017}}.
\newline\urlprefix\url{http://onlinelibrary.wiley.com/doi/10.1002/ctpp.201010017/abstract}

\bibitem{potekhin_equation_2013}
A.~Y. Potekhin, G.~Chabrier,
  \href{http://dx.doi.org/10.1051/0004-6361/201220082}{Equation of state for
  magnetized {Coulomb} plasmas}, Astron.~Astrophys. 550 (2013) A43.
\newblock \href {http://dx.doi.org/10.1051/0004-6361/201220082}
  {\path{doi:10.1051/0004-6361/201220082}}.
\newline\urlprefix\url{http://dx.doi.org/10.1051/0004-6361/201220082}

\bibitem{crouseilles_quantum_2008}
N.~Crouseilles, P.-A. Hervieux, G.~Manfredi,
  \href{http://link.aps.org/doi/10.1103/PhysRevB.78.155412}{Quantum
  hydrodynamic model for the nonlinear electron dynamics in thin metal films},
  Phys.~Rev.~B 78~(15) (2008) 155412.
\newblock \href {http://dx.doi.org/10.1103/PhysRevB.78.155412}
  {\path{doi:10.1103/PhysRevB.78.155412}}.
\newline\urlprefix\url{http://link.aps.org/doi/10.1103/PhysRevB.78.155412}

\bibitem{michta_quantum_2015}
D.~Michta, F.~Graziani, M.~Bonitz,
  \href{http://onlinelibrary.wiley.com/doi/10.1002/ctpp.201500024/abstract}{Quantum
  {Hydrodynamics} for {Plasmas} – a {Thomas}-{Fermi} {Theory} {Perspective}},
  Contrib.~Plasma Phys. 55~(6) (2015) 437--443.
\newblock \href {http://dx.doi.org/10.1002/ctpp.201500024}
  {\path{doi:10.1002/ctpp.201500024}}.
\newline\urlprefix\url{http://onlinelibrary.wiley.com/doi/10.1002/ctpp.201500024/abstract}

\bibitem{diaw_viscous_2017}
A.~Diaw, M.~S. Murillo,
  \href{https://www.nature.com/articles/s41598-017-14414-9}{A viscous quantum
  hydrodynamics model based on dynamic density functional theory}, Sci.~Rep.
  7~(1) (2017) 15352.
\newblock \href {http://dx.doi.org/10.1038/s41598-017-14414-9}
  {\path{doi:10.1038/s41598-017-14414-9}}.
\newline\urlprefix\url{https://www.nature.com/articles/s41598-017-14414-9}

\bibitem{kremp2006quantum}
D.~Kremp, M.~Schlanges, W.~Kraeft,
  \href{https://books.google.de/books?id=savMPLWUFfkC}{Quantum Statistics of
  Nonideal Plasmas}, Springer Series on Atomic, Optical, and Plasma Physics,
  Springer Berlin Heidelberg, 2006.
\newline\urlprefix\url{https://books.google.de/books?id=savMPLWUFfkC}

\bibitem{vorberger_equation_2004}
J.~Vorberger, M.~Schlanges, W.~D. Kraeft,
  \href{http://link.aps.org/doi/10.1103/PhysRevE.69.046407}{Equation of state
  for weakly coupled quantum plasmas}, Phys.~Rev.~E 69~(4) (2004) 046407.
\newblock \href {http://dx.doi.org/10.1103/PhysRevE.69.046407}
  {\path{doi:10.1103/PhysRevE.69.046407}}.
\newline\urlprefix\url{http://link.aps.org/doi/10.1103/PhysRevE.69.046407}

\bibitem{kas_finite_2017}
J.~Kas, J.~Rehr,
  \href{https://link.aps.org/doi/10.1103/PhysRevLett.119.176403}{Finite
  {Temperature} {Green}'s {Function} {Approach} for {Excited} {State} and
  {Thermodynamic} {Properties} of {Cool} to {Warm} {Dense} {Matter}},
  Phys.~Rev.~Lett. 119~(17) (2017) 176403.
\newblock \href {http://dx.doi.org/10.1103/PhysRevLett.119.176403}
  {\path{doi:10.1103/PhysRevLett.119.176403}}.
\newline\urlprefix\url{https://link.aps.org/doi/10.1103/PhysRevLett.119.176403}

\bibitem{ebeling_thermodynamic_1982}
W.~Ebeling, W.~Richert,
  \href{http://onlinelibrary.wiley.com/doi/10.1002/andp.19824940508/abstract}{Thermodynamic
  {Functions} of {Nonideal} {Hydrogen} {Plasmas}}, Ann.~Phys. 494~(5) (1982)
  362--370.
\newblock \href {http://dx.doi.org/10.1002/andp.19824940508}
  {\path{doi:10.1002/andp.19824940508}}.
\newline\urlprefix\url{http://onlinelibrary.wiley.com/doi/10.1002/andp.19824940508/abstract}

\bibitem{richert_thermodynamic_1984}
W.~Richert, W.~Ebeling,
  \href{http://onlinelibrary.wiley.com/doi/10.1002/pssb.2221210222/abstract}{Thermodynamic
  {Functions} of the {Electron} {Fluid} for a {Wide} {Density}-{Temperature}
  {Range}}, physica status solidi (b) 121~(2) (1984) 633--639.
\newblock \href {http://dx.doi.org/10.1002/pssb.2221210222}
  {\path{doi:10.1002/pssb.2221210222}}.
\newline\urlprefix\url{http://onlinelibrary.wiley.com/doi/10.1002/pssb.2221210222/abstract}

\bibitem{ebeling_plasma_1985}
W.~Ebeling, W.~Richert,
  \href{http://www.sciencedirect.com/science/article/pii/0375960185905213}{Plasma
  phase transition in hydrogen}, Phys.~Lett.~A 108~(2) (1985) 80--82.
\newblock \href {http://dx.doi.org/10.1016/0375-9601(85)90521-3}
  {\path{doi:10.1016/0375-9601(85)90521-3}}.
\newline\urlprefix\url{http://www.sciencedirect.com/science/article/pii/0375960185905213}

\bibitem{ebeling_nonideal_1989}
W.~Ebeling, H.~Lehmann,
  \href{http://onlinelibrary.wiley.com/doi/10.1002/ctpp.2150290406/abstract}{Nonideal
  {Ionization} in {Plasmas} with {Higher} {Charges}}, Contrib.~Plasma Phys.
  29~(4-5) (1989) 365--371.
\newblock \href {http://dx.doi.org/10.1002/ctpp.2150290406}
  {\path{doi:10.1002/ctpp.2150290406}}.
\newline\urlprefix\url{http://onlinelibrary.wiley.com/doi/10.1002/ctpp.2150290406/abstract}

\bibitem{ebeling_free_1990}
W.~Ebeling,
  \href{http://onlinelibrary.wiley.com/doi/10.1002/ctpp.2150300502/abstract}{Free
  {Energy} and {Ionization} in {Dense} {Plasmas} of the {Light} {Elements}},
  Contrib.~Plasma Phys. 30~(5) (1990) 553--561.
\newblock \href {http://dx.doi.org/10.1002/ctpp.2150300502}
  {\path{doi:10.1002/ctpp.2150300502}}.
\newline\urlprefix\url{http://onlinelibrary.wiley.com/doi/10.1002/ctpp.2150300502/abstract}

\bibitem{tanaka_parametrized_1985}
S.~Tanaka, S.~Mitake, S.~Ichimaru,
  \href{http://link.aps.org/doi/10.1103/PhysRevA.32.1896}{Parametrized equation
  of state for electron liquids in the {Singwi}-{Tosi}-{Land}-{Sj\"olander
  approximation}}, Phys. Rev. A 32~(3) (1985) 1896--1899.
\newblock \href {http://dx.doi.org/10.1103/PhysRevA.32.1896}
  {\path{doi:10.1103/PhysRevA.32.1896}}.
\newline\urlprefix\url{http://link.aps.org/doi/10.1103/PhysRevA.32.1896}

\bibitem{tanaka_thermodynamics_1986}
S.~Tanaka, S.~Ichimaru,
  \href{http://journals.jps.jp/doi/abs/10.1143/JPSJ.55.2278}{Thermodynamics and
  {Correlational} {Properties} of {Finite}-{Temperature} {Electron} {Liquids}
  in the {Singwi}-{Tosi}-{Land}-{Sj\"olander} {Approximation}}, J. Phys. Soc.
  Jpn. 55~(7) (1986) 2278--2289.
\newblock \href {http://dx.doi.org/10.1143/JPSJ.55.2278}
  {\path{doi:10.1143/JPSJ.55.2278}}.
\newline\urlprefix\url{http://journals.jps.jp/doi/abs/10.1143/JPSJ.55.2278}

\bibitem{tanaka_spin-dependent_1989}
S.~Tanaka, S.~Ichimaru,
  \href{http://link.aps.org/doi/10.1103/PhysRevB.39.1036}{Spin-dependent
  correlations and thermodynamic functions for electron liquids at arbitrary
  degeneracy and spin polarization}, Phys. Rev. B 39~(2) (1989) 1036--1051.
\newblock \href {http://dx.doi.org/10.1103/PhysRevB.39.1036}
  {\path{doi:10.1103/PhysRevB.39.1036}}.
\newline\urlprefix\url{http://link.aps.org/doi/10.1103/PhysRevB.39.1036}

\bibitem{ichimaru_statistical_1987}
S.~Ichimaru, H.~Iyetomi, S.~Tanaka,
  \href{http://www.sciencedirect.com/science/article/pii/0370157387901256}{Statistical
  physics of dense plasmas: {Thermodynamics}, transport coefficients and
  dynamic correlations}, Phys.~Rep. 149~(2) (1987) 91--205.
\newblock \href {http://dx.doi.org/10.1016/0370-1573(87)90125-6}
  {\path{doi:10.1016/0370-1573(87)90125-6}}.
\newline\urlprefix\url{http://www.sciencedirect.com/science/article/pii/0370157387901256}

\bibitem{ichimaru2004statistical_1}
S.~Ichimaru, \href{https://books.google.de/books?id=CWssAAAAYAAJ}{Statistical
  Plasma Physics: Basic principles}, Frontiers in physics, Westview Press,
  2004.
\newline\urlprefix\url{https://books.google.de/books?id=CWssAAAAYAAJ}

\bibitem{ichimaru2004statistical_2}
S.~Ichimaru, \href{https://books.google.de/books?id=T2ssAAAAYAAJ}{Statistical
  Plasma Physics: Condensed plasmas}, Frontiers in physics, Westview Press,
  2004.
\newline\urlprefix\url{https://books.google.de/books?id=T2ssAAAAYAAJ}

\bibitem{dharma-wardana_simple_2000}
M.~W.~C. Dharma-wardana, F.~Perrot,
  \href{http://link.aps.org/doi/10.1103/PhysRevLett.84.959}{Simple {Classical}
  {Mapping} of the {Spin}-{Polarized} {Quantum} {Electron} {Gas}:
  {Distribution} {Functions} and {Local}-{Field} {Corrections}},
  Phys.~Rev.~Lett. 84~(5) (2000) 959--962.
\newblock \href {http://dx.doi.org/10.1103/PhysRevLett.84.959}
  {\path{doi:10.1103/PhysRevLett.84.959}}.
\newline\urlprefix\url{http://link.aps.org/doi/10.1103/PhysRevLett.84.959}

\bibitem{perrot_spin-polarized_2000}
F.~Perrot, M.~W.~C. Dharma-wardana,
  \href{http://link.aps.org/doi/10.1103/PhysRevB.62.16536}{Spin-polarized
  electron liquid at arbitrary temperatures: {Exchange}-correlation energies,
  electron-distribution functions, and the static response functions},
  Phys.~Rev.~B 62~(24) (2000) 16536--16548.
\newblock \href {http://dx.doi.org/10.1103/PhysRevB.62.16536}
  {\path{doi:10.1103/PhysRevB.62.16536}}.
\newline\urlprefix\url{http://link.aps.org/doi/10.1103/PhysRevB.62.16536}

\bibitem{brown_path-integral_2013}
E.~W. Brown, B.~K. Clark, J.~L. DuBois, D.~M. Ceperley,
  \href{http://link.aps.org/doi/10.1103/PhysRevLett.110.146405}{Path-{Integral}
  {Monte}-{Carlo} {Simulation} of the {Warm} {Dense} {Homogeneous} {Electron}
  {Gas}}, Phys.~Rev.~Lett. 110~(14) (2013) 146405.
\newblock \href {http://dx.doi.org/10.1103/PhysRevLett.110.146405}
  {\path{doi:10.1103/PhysRevLett.110.146405}}.
\newline\urlprefix\url{http://link.aps.org/doi/10.1103/PhysRevLett.110.146405}

\bibitem{karasiev_accurate_2014}
V.~V. Karasiev, T.~Sjostrom, J.~Dufty, S.~Trickey,
  \href{http://link.aps.org/doi/10.1103/PhysRevLett.112.076403}{Accurate
  {Homogeneous} {Electron} {Gas} {Exchange}-{Correlation} {Free} {Energy} for
  {Local} {Spin}-{Density} {Calculations}}, Phys.~Rev.~Lett. 112~(7) (2014)
  076403.
\newblock \href {http://dx.doi.org/10.1103/PhysRevLett.112.076403}
  {\path{doi:10.1103/PhysRevLett.112.076403}}.
\newline\urlprefix\url{http://link.aps.org/doi/10.1103/PhysRevLett.112.076403}

\bibitem{sjostrom_uniform_2013}
T.~Sjostrom, J.~Dufty,
  \href{http://link.aps.org/doi/10.1103/PhysRevB.88.115123}{Uniform electron
  gas at finite temperatures}, Phys.~Rev.~B 88~(11) (2013) 115123.
\newblock \href {http://dx.doi.org/10.1103/PhysRevB.88.115123}
  {\path{doi:10.1103/PhysRevB.88.115123}}.
\newline\urlprefix\url{http://link.aps.org/doi/10.1103/PhysRevB.88.115123}

\bibitem{brown_exchange-correlation_2013}
E.~W. Brown, J.~L. DuBois, M.~Holzmann, D.~M. Ceperley,
  \href{http://link.aps.org/doi/10.1103/PhysRevB.88.081102}{Exchange-correlation
  energy for the three-dimensional homogeneous electron gas at arbitrary
  temperature}, Phys.~Rev.~B 88~(8) (2013) 081102.
\newblock \href {http://dx.doi.org/10.1103/PhysRevB.88.081102}
  {\path{doi:10.1103/PhysRevB.88.081102}}.
\newline\urlprefix\url{http://link.aps.org/doi/10.1103/PhysRevB.88.081102}

\bibitem{schoof_configuration_2011}
T.~Schoof, M.~Bonitz, A.~Filinov, D.~Hochstuhl, J.~Dufty,
  \href{http://onlinelibrary.wiley.com/doi/10.1002/ctpp.201100012/abstract}{Configuration
  {Path} {Integral} {Monte} {Carlo}}, Contrib.~Plasma Phys. 51~(8) (2011)
  687--697.
\newblock \href {http://dx.doi.org/10.1002/ctpp.201100012}
  {\path{doi:10.1002/ctpp.201100012}}.
\newline\urlprefix\url{http://onlinelibrary.wiley.com/doi/10.1002/ctpp.201100012/abstract}

\bibitem{schoof_towards_2015}
T.~Schoof, S.~Groth, M.~Bonitz,
  \href{http://onlinelibrary.wiley.com/doi/10.1002/ctpp.201400072/abstract}{Towards
  ab {Initio} {Thermodynamics} of the {Electron} {Gas} at {Strong}
  {Degeneracy}}, Contrib.~Plasma Phys. 55~(2-3) (2015) 136--143.
\newblock \href {http://dx.doi.org/10.1002/ctpp.201400072}
  {\path{doi:10.1002/ctpp.201400072}}.
\newline\urlprefix\url{http://onlinelibrary.wiley.com/doi/10.1002/ctpp.201400072/abstract}

\bibitem{groth_abinitio_2016}
S.~Groth, T.~Schoof, T.~Dornheim, M.~Bonitz,
  \href{http://link.aps.org/doi/10.1103/PhysRevB.93.085102}{\textit{{Ab}
  initio} quantum {Monte} {Carlo} simulations of the uniform electron gas
  without fixed nodes}, Phys.~Rev.~B 93~(8) (2016) 085102.
\newblock \href {http://dx.doi.org/10.1103/PhysRevB.93.085102}
  {\path{doi:10.1103/PhysRevB.93.085102}}.
\newline\urlprefix\url{http://link.aps.org/doi/10.1103/PhysRevB.93.085102}

\bibitem{schoof_textitab_2015}
T.~Schoof, S.~Groth, J.~Vorberger, M.~Bonitz,
  \href{http://link.aps.org/doi/10.1103/PhysRevLett.115.130402}{\textit{{Ab}
  {Initio}} {Thermodynamic} {Results} for the {Degenerate} {Electron} {Gas} at
  {Finite} {Temperature}}, Phys.~Rev.~Lett. 115~(13) (2015) 130402.
\newblock \href {http://dx.doi.org/10.1103/PhysRevLett.115.130402}
  {\path{doi:10.1103/PhysRevLett.115.130402}}.
\newline\urlprefix\url{http://link.aps.org/doi/10.1103/PhysRevLett.115.130402}

\bibitem{dornheim_permutation_2015}
T.~Dornheim, S.~Groth, A.~Filinov, M.~Bonitz,
  \href{http://iopscience.iop.org/1367-2630/17/7/073017}{Permutation blocking
  path integral {Monte} {Carlo}: a highly efficient approach to the simulation
  of strongly degenerate non-ideal fermions}, New J.~Phys. 17~(7) (2015)
  073017.
\newblock \href {http://dx.doi.org/10.1088/1367-2630/17/7/073017}
  {\path{doi:10.1088/1367-2630/17/7/073017}}.
\newline\urlprefix\url{http://iopscience.iop.org/1367-2630/17/7/073017}

\bibitem{dornheim_permutation_2015-1}
T.~Dornheim, T.~Schoof, S.~Groth, A.~Filinov, M.~Bonitz,
  \href{http://scitation.aip.org/content/aip/journal/jcp/143/20/10.1063/1.4936145}{Permutation
  blocking path integral {Monte} {Carlo} approach to the uniform electron gas
  at finite temperature}, J.~Chem.~Phys. 143~(20) (2015) 204101.
\newblock \href {http://dx.doi.org/10.1063/1.4936145}
  {\path{doi:10.1063/1.4936145}}.
\newline\urlprefix\url{http://scitation.aip.org/content/aip/journal/jcp/143/20/10.1063/1.4936145}

\bibitem{dornheim_abinitio_2016}
T.~Dornheim, S.~Groth, T.~Schoof, C.~Hann, M.~Bonitz,
  \href{http://link.aps.org/doi/10.1103/PhysRevB.93.205134}{\textit{{Ab}
  initio} quantum {Monte} {Carlo} simulations of the uniform electron gas
  without fixed nodes: {The} unpolarized case}, Phys.~Rev.~B 93~(20) (2016)
  205134.
\newblock \href {http://dx.doi.org/10.1103/PhysRevB.93.205134}
  {\path{doi:10.1103/PhysRevB.93.205134}}.
\newline\urlprefix\url{http://link.aps.org/doi/10.1103/PhysRevB.93.205134}

\bibitem{malone_interaction_2015}
F.~D. Malone, N.~S. Blunt, J.~J. Shepherd, D.~K.~K. Lee, J.~S. Spencer,
  W.~M.~C. Foulkes,
  \href{http://scitation.aip.org/content/aip/journal/jcp/143/4/10.1063/1.4927434}{Interaction
  picture density matrix quantum {Monte} {Carlo}}, J.~Chem.~Phys. 143~(4)
  (2015) 044116.
\newblock \href {http://dx.doi.org/10.1063/1.4927434}
  {\path{doi:10.1063/1.4927434}}.
\newline\urlprefix\url{http://scitation.aip.org/content/aip/journal/jcp/143/4/10.1063/1.4927434}

\bibitem{malone_accurate_2016}
F.~D. Malone, N.~Blunt, E.~W. Brown, D.~Lee, J.~Spencer, W.~Foulkes, J.~J.
  Shepherd,
  \href{http://link.aps.org/doi/10.1103/PhysRevLett.117.115701}{Accurate
  {Exchange}-{Correlation} {Energies} for the {Warm} {Dense} {Electron} {Gas}},
  Phys.~Rev.~Lett. 117~(11) (2016) 115701.
\newblock \href {http://dx.doi.org/10.1103/PhysRevLett.117.115701}
  {\path{doi:10.1103/PhysRevLett.117.115701}}.
\newline\urlprefix\url{http://link.aps.org/doi/10.1103/PhysRevLett.117.115701}

\bibitem{blunt_density-matrix_2014}
N.~S. Blunt, T.~W. Rogers, J.~S. Spencer, W.~M.~C. Foulkes,
  \href{http://link.aps.org/doi/10.1103/PhysRevB.89.245124}{Density-matrix
  quantum {Monte} {Carlo} method}, Phys.~Rev.~B 89~(24) (2014) 245124.
\newblock \href {http://dx.doi.org/10.1103/PhysRevB.89.245124}
  {\path{doi:10.1103/PhysRevB.89.245124}}.
\newline\urlprefix\url{http://link.aps.org/doi/10.1103/PhysRevB.89.245124}

\bibitem{dornheim_abinitio_2017}
T.~Dornheim, S.~Groth, F.~D. Malone, T.~Schoof, T.~Sjostrom, W.~M.~C. Foulkes,
  M.~Bonitz, \href{http://aip.scitation.org/doi/full/10.1063/1.4977920}{Ab
  initio quantum {Monte} {Carlo} simulation of the warm dense electron gas},
  Phys.~Plasmas 24~(5) (2017) 056303.
\newblock \href {http://dx.doi.org/10.1063/1.4977920}
  {\path{doi:10.1063/1.4977920}}.
\newline\urlprefix\url{http://aip.scitation.org/doi/full/10.1063/1.4977920}

\bibitem{dornheim_abinitio_2016-1}
T.~Dornheim, S.~Groth, T.~Sjostrom, F.~D. Malone, W.~Foulkes, M.~Bonitz,
  \href{http://link.aps.org/doi/10.1103/PhysRevLett.117.156403}{\textit{{Ab}
  {Initio}} {Quantum} {Monte} {Carlo} {Simulation} of the {Warm} {Dense}
  {Electron} {Gas} in the {Thermodynamic} {Limit}}, Phys.~Rev.~Lett. 117~(15)
  (2016) 156403.
\newblock \href {http://dx.doi.org/10.1103/PhysRevLett.117.156403}
  {\path{doi:10.1103/PhysRevLett.117.156403}}.
\newline\urlprefix\url{http://link.aps.org/doi/10.1103/PhysRevLett.117.156403}

\bibitem{groth_ab_2017}
S.~Groth, T.~Dornheim, T.~Sjostrom, F.~D. Malone, W.~Foulkes, M.~Bonitz,
  \href{https://link.aps.org/doi/10.1103/PhysRevLett.119.135001}{\textit{Ab
  initio} exchange-correlation free energy of the uniform electron gas at warm
  dense matter conditions}, Phys. Rev. Lett. 119~(13) (2017) 135001.
\newblock \href {http://dx.doi.org/10.1103/PhysRevLett.119.135001}
  {\path{doi:10.1103/PhysRevLett.119.135001}}.
\newline\urlprefix\url{https://link.aps.org/doi/10.1103/PhysRevLett.119.135001}

\bibitem{leeuw_simulation_1980}
S.~W.~d. Leeuw, J.~W. Perram, E.~R. Smith,
  \href{http://rspa.royalsocietypublishing.org/content/373/1752/27}{Simulation
  of electrostatic systems in periodic boundary conditions. {I}. {Lattice} sums
  and dielectric constants}, Proc. R. Soc. Lond. A 373~(1752) (1980) 27--56.
\newblock \href {http://dx.doi.org/10.1098/rspa.1980.0135}
  {\path{doi:10.1098/rspa.1980.0135}}.
\newline\urlprefix\url{http://rspa.royalsocietypublishing.org/content/373/1752/27}

\bibitem{ballenegger_communication:_2014}
V.~Ballenegger,
  \href{http://scitation.aip.org/content/aip/journal/jcp/140/16/10.1063/1.4872019}{Communication:
  {On} the origin of the surface term in the {Ewald} formula}, J.~Chem.~Phys.
  140~(16) (2014) 161102.
\newblock \href {http://dx.doi.org/10.1063/1.4872019}
  {\path{doi:10.1063/1.4872019}}.
\newline\urlprefix\url{http://scitation.aip.org/content/aip/journal/jcp/140/16/10.1063/1.4872019}

\bibitem{toukmaji_ewald_1996}
A.~Y. Toukmaji, J.~A. Board,
  \href{http://www.sciencedirect.com/science/article/pii/0010465596000161}{Ewald
  summation techniques in perspective: a survey}, Comp.~Phys.~Comm. 95~(2)
  (1996) 73--92.
\newblock \href {http://dx.doi.org/10.1016/0010-4655(96)00016-1}
  {\path{doi:10.1016/0010-4655(96)00016-1}}.
\newline\urlprefix\url{http://www.sciencedirect.com/science/article/pii/0010465596000161}

\bibitem{fraser_finite-size_1996}
L.~M. Fraser, W.~M.~C. Foulkes, G.~Rajagopal, R.~J. Needs, S.~D. Kenny, A.~J.
  Williamson,
  \href{http://link.aps.org/doi/10.1103/PhysRevB.53.1814}{Finite-size effects
  and {Coulomb} interactions in quantum {Monte} {Carlo} calculations for
  homogeneous systems with periodic boundary conditions}, Phys.~Rev.~B 53~(4)
  (1996) 1814--1832.
\newblock \href {http://dx.doi.org/10.1103/PhysRevB.53.1814}
  {\path{doi:10.1103/PhysRevB.53.1814}}.
\newline\urlprefix\url{http://link.aps.org/doi/10.1103/PhysRevB.53.1814}

\bibitem{duan_ewald_2000}
Z.-H. Duan, R.~Krasny,
  \href{http://aip.scitation.org/doi/abs/10.1063/1.1289918}{An {Ewald}
  summation based multipole method}, J.~Chem.~Phys. 113~(9) (2000) 3492--3495.
\newblock \href {http://dx.doi.org/10.1063/1.1289918}
  {\path{doi:10.1063/1.1289918}}.
\newline\urlprefix\url{http://aip.scitation.org/doi/abs/10.1063/1.1289918}

\bibitem{natoli_optimized_1995}
V.~Natoli, D.~M. Ceperley,
  \href{http://www.sciencedirect.com/science/article/pii/S0021999185710546}{An
  {Optimized} {Method} for {Treating} {Long}-{Range} {Potentials}},
  J.~Comp.~Phys. 117~(1) (1995) 171--178.
\newblock \href {http://dx.doi.org/10.1006/jcph.1995.1054}
  {\path{doi:10.1006/jcph.1995.1054}}.
\newline\urlprefix\url{http://www.sciencedirect.com/science/article/pii/S0021999185710546}

\bibitem{fennell_is_2006}
C.~J. Fennell, J.~D. Gezelter,
  \href{http://aip.scitation.org/doi/abs/10.1063/1.2206581}{Is the {Ewald}
  summation still necessary? {Pairwise} alternatives to the accepted standard
  for long-range electrostatics}, J.~Chem.~Phys. 124~(23) (2006) 234104.
\newblock \href {http://dx.doi.org/10.1063/1.2206581}
  {\path{doi:10.1063/1.2206581}}.
\newline\urlprefix\url{http://aip.scitation.org/doi/abs/10.1063/1.2206581}

\bibitem{yakub_efficient_2003}
E.~Yakub, C.~Ronchi,
  \href{http://scitation.aip.org/content/aip/journal/jcp/119/22/10.1063/1.1624364}{An
  efficient method for computation of long-ranged {Coulomb} forces in computer
  simulation of ionic fluids}, J.~Chem.~Phys. 119~(22) (2003) 11556--11560.
\newblock \href {http://dx.doi.org/10.1063/1.1624364}
  {\path{doi:10.1063/1.1624364}}.
\newline\urlprefix\url{http://scitation.aip.org/content/aip/journal/jcp/119/22/10.1063/1.1624364}

\bibitem{yakub_new_2005}
E.~Yakub, C.~Ronchi,
  \href{http://link.springer.com/article/10.1007/s10909-005-5451-5}{A {New}
  {Method} for {Computation} of {Long} {Ranged} {Coulomb} {Forces} in
  {Computer} {Simulation} of {Disordered} {Systems}}, J.~Low Temp.~Phys.
  139~(5-6) (2005) 633--643.
\newblock \href {http://dx.doi.org/10.1007/s10909-005-5451-5}
  {\path{doi:10.1007/s10909-005-5451-5}}.
\newline\urlprefix\url{http://link.springer.com/article/10.1007/s10909-005-5451-5}

\bibitem{yakub_effective_2006}
E.~S. Yakub, \href{http://stacks.iop.org/0305-4470/39/i=17/a=S51}{Effective
  computer simulation of strongly coupled {Coulomb} fluids}, J.~Phys.~A 39~(17)
  (2006) 4643.
\newblock \href {http://dx.doi.org/10.1088/0305-4470/39/17/S51}
  {\path{doi:10.1088/0305-4470/39/17/S51}}.
\newline\urlprefix\url{http://stacks.iop.org/0305-4470/39/i=17/a=S51}

\bibitem{vernizzi_coulomb_2011}
G.~Vernizzi, G.~I. Guerrero-García, M.~Olvera de~la Cruz,
  \href{http://link.aps.org/doi/10.1103/PhysRevE.84.016707}{Coulomb
  interactions in charged fluids}, Phys.~Rev.~E 84~(1) (2011) 016707.
\newblock \href {http://dx.doi.org/10.1103/PhysRevE.84.016707}
  {\path{doi:10.1103/PhysRevE.84.016707}}.
\newline\urlprefix\url{http://link.aps.org/doi/10.1103/PhysRevE.84.016707}

\bibitem{bonitz2015quantum}
M.~Bonitz, \href{https://books.google.de/books?id=wW7\_CgAAQBAJ}{Quantum
  Kinetic Theory}, Springer International Publishing, 2016.
\newline\urlprefix\url{https://books.google.de/books?id=wW7\_CgAAQBAJ}

\bibitem{klimontowich_1952}
Y.~Klimontovich, V.~Silin, J. Exptl. Theor. Phys. (U.S.S.R.) 23 (1952) 151.

\bibitem{gell-mann_correlation_1957}
M.~Gell-Mann, K.~A. Brueckner,
  \href{http://link.aps.org/doi/10.1103/PhysRev.106.364}{Correlation {Energy}
  of an {Electron} {Gas} at {High} {Density}}, Phys.~Rev. 106~(2) (1957)
  364--368.
\newblock \href {http://dx.doi.org/10.1103/PhysRev.106.364}
  {\path{doi:10.1103/PhysRev.106.364}}.
\newline\urlprefix\url{http://link.aps.org/doi/10.1103/PhysRev.106.364}

\bibitem{tanaka_correlational_2016}
S.~Tanaka,
  \href{http://scitation.aip.org/content/aip/journal/jcp/145/21/10.1063/1.4969071}{Correlational
  and thermodynamic properties of finite-temperature electron liquids in the
  hypernetted-chain approximation}, J.~Chem.~Phys. 145~(21) (2016) 214104.
\newblock \href {http://dx.doi.org/10.1063/1.4969071}
  {\path{doi:10.1063/1.4969071}}.
\newline\urlprefix\url{http://scitation.aip.org/content/aip/journal/jcp/145/21/10.1063/1.4969071}

\bibitem{hayashi_electron_1980}
H.~Hayashi, M.~Shimizu,
  \href{http://journals.jps.jp/doi/abs/10.1143/JPSJ.48.16}{Electron
  {Correlations} at {Metallic} {Densities}. {IV}}, J.~Phys.~Soc.~Jpn. 48~(1)
  (1980) 16--23.
\newblock \href {http://dx.doi.org/10.1143/JPSJ.48.16}
  {\path{doi:10.1143/JPSJ.48.16}}.
\newline\urlprefix\url{http://journals.jps.jp/doi/abs/10.1143/JPSJ.48.16}

\bibitem{hasegawa_electron_1975}
T.~Hasegawa, M.~Shimizu,
  \href{http://journals.jps.jp/doi/abs/10.1143/JPSJ.38.965}{Electron
  {Correlations} at {Metallic} {Densities}, {II}. {Quantum} {Mechanical}
  {Expression} of {Dielectric} {Function} with {Wigner} {Distribution}
  {Function}}, J.~Phys.~Soc.~Jpn. 38~(4) (1975) 965--973.
\newblock \href {http://dx.doi.org/10.1143/JPSJ.38.965}
  {\path{doi:10.1143/JPSJ.38.965}}.
\newline\urlprefix\url{http://journals.jps.jp/doi/abs/10.1143/JPSJ.38.965}

\bibitem{holas_dynamic_1987}
A.~Holas, S.~Rahman,
  \href{http://link.aps.org/doi/10.1103/PhysRevB.35.2720}{Dynamic local-field
  factor of an electron liquid in the quantum versions of the
  {Singwi}-{Tosi}-{Land}-{Sj\"olander} and {Vashishta}-{Singwi} theories},
  Phys.~Rev.~B 35~(6) (1987) 2720--2731.
\newblock \href {http://dx.doi.org/10.1103/PhysRevB.35.2720}
  {\path{doi:10.1103/PhysRevB.35.2720}}.
\newline\urlprefix\url{http://link.aps.org/doi/10.1103/PhysRevB.35.2720}

\bibitem{gupta_inhomogeneous_1980}
U.~Gupta, A.~K. Rajagopal,
  \href{https://link.aps.org/doi/10.1103/PhysRevA.21.2064}{Inhomogeneous
  electron gas at nonzero temperatures: {Exchange} effects}, Phys. Rev. A
  21~(6) (1980) 2064--2068.
\newblock \href {http://dx.doi.org/10.1103/PhysRevA.21.2064}
  {\path{doi:10.1103/PhysRevA.21.2064}}.
\newline\urlprefix\url{https://link.aps.org/doi/10.1103/PhysRevA.21.2064}

\bibitem{gupta_exchange-correlation_1980}
U.~Gupta, A.~K. Rajagopal,
  \href{http://link.aps.org/doi/10.1103/PhysRevA.22.2792}{Exchange-correlation
  potential for inhomogeneous electron systems at finite temperatures}, Phys.
  Rev. A 22~(6) (1980) 2792--2797.
\newblock \href {http://dx.doi.org/10.1103/PhysRevA.22.2792}
  {\path{doi:10.1103/PhysRevA.22.2792}}.
\newline\urlprefix\url{http://link.aps.org/doi/10.1103/PhysRevA.22.2792}

\bibitem{perrot_exchange_1984}
F.~Perrot, M.~W.~C. Dharma-wardana,
  \href{http://link.aps.org/doi/10.1103/PhysRevA.30.2619}{Exchange and
  correlation potentials for electron-ion systems at finite temperatures},
  Phys.~Rev.~A 30~(5) (1984) 2619--2626.
\newblock \href {http://dx.doi.org/10.1103/PhysRevA.30.2619}
  {\path{doi:10.1103/PhysRevA.30.2619}}.
\newline\urlprefix\url{http://link.aps.org/doi/10.1103/PhysRevA.30.2619}

\bibitem{tanaka_improved_2017}
S.~Tanaka,
  \href{http://onlinelibrary.wiley.com/doi/10.1002/ctpp.201600096/abstract}{Improved
  equation of state for finite-temperature spin-polarized electron liquids on
  the basis of {Singwi}-{Tosi}-{Land}-{Sj\"olander} approximation},
  Contrib.~Plasma Phys. (2017) n/a--n/a\href
  {http://dx.doi.org/10.1002/ctpp.201600096}
  {\path{doi:10.1002/ctpp.201600096}}.
\newline\urlprefix\url{http://onlinelibrary.wiley.com/doi/10.1002/ctpp.201600096/abstract}

\bibitem{stolzmann_static_2001}
W.~Stolzmann, M.~R\"osler,
  \href{http://onlinelibrary.wiley.com/doi/10.1002/1521-3986(200103)41:2/3<203::AID-CTPP203>3.0.CO;2-S/abstract}{Static
  {Local}-{Field} {Corrected} {Dielectric} and {Thermodynamic} {Functions}},
  Contrib. Plasma Phys. 41~(2-3) (2001) 203--206.
\newblock \href
  {http://dx.doi.org/10.1002/1521-3986(200103)41:2/3<203::AID-CTPP203>3.0.CO;2-S}
  {\path{doi:10.1002/1521-3986(200103)41:2/3<203::AID-CTPP203>3.0.CO;2-S}}.
\newline\urlprefix\url{http://onlinelibrary.wiley.com/doi/10.1002/1521-3986(200103)41:2/3<203::AID-CTPP203>3.0.CO;2-S/abstract}

\bibitem{schweng_finite-temperature_1993}
H.~K. Schweng, H.~M. B\"ohm,
  \href{http://link.aps.org/doi/10.1103/PhysRevB.48.2037}{Finite-temperature
  electron correlations in the framework of a dynamic local-field correction},
  Phys. Rev. B 48~(4) (1993) 2037--2045.
\newblock \href {http://dx.doi.org/10.1103/PhysRevB.48.2037}
  {\path{doi:10.1103/PhysRevB.48.2037}}.
\newline\urlprefix\url{http://link.aps.org/doi/10.1103/PhysRevB.48.2037}

\bibitem{arora_spin-resolved_2017}
P.~Arora, K.~Kumar, R.~K. Moudgil,
  \href{https://link.springer.com/article/10.1140/epjb/e2017-70532-y}{Spin-resolved
  correlations in the warm-dense homogeneous electron gas}, Eur. Phys. J. B
  90~(4) (2017) 76.
\newblock \href {http://dx.doi.org/10.1140/epjb/e2017-70532-y}
  {\path{doi:10.1140/epjb/e2017-70532-y}}.
\newline\urlprefix\url{https://link.springer.com/article/10.1140/epjb/e2017-70532-y}

\bibitem{kahlert_dynamics_2014}
H.~K\"ahlert, G.~J. Kalman, M.~Bonitz,
  \href{https://link.aps.org/doi/10.1103/PhysRevE.90.011101}{Dynamics of
  strongly correlated and strongly inhomogeneous plasmas}, Phys.~Rev.~E 90~(1)
  (2014) 011101.
\newblock \href {http://dx.doi.org/10.1103/PhysRevE.90.011101}
  {\path{doi:10.1103/PhysRevE.90.011101}}.
\newline\urlprefix\url{https://link.aps.org/doi/10.1103/PhysRevE.90.011101}

\bibitem{kahlert_linear_2015}
H.~K\"ahlert, G.~J. Kalman, M.~Bonitz,
  \href{http://onlinelibrary.wiley.com/doi/10.1002/ctpp.201400085/abstract}{Linear
  {Fluid} {Theory} for {Weakly} {Inhomogeneous} {Plasmas} with {Strong}
  {Correlations}}, Contrib.~Plasma Phys. 55~(5) (2015) 352--359.
\newblock \href {http://dx.doi.org/10.1002/ctpp.201400085}
  {\path{doi:10.1002/ctpp.201400085}}.
\newline\urlprefix\url{http://onlinelibrary.wiley.com/doi/10.1002/ctpp.201400085/abstract}

\bibitem{springer_integral_1973}
J.~F. Springer, M.~A. Pokrant, F.~A. Stevens,
  \href{http://aip.scitation.org/doi/10.1063/1.1679070}{Integral equation
  solutions for the classical electron gas}, J.~Chem.~Phys. 58~(11) (1973)
  4863--4867.
\newblock \href {http://dx.doi.org/10.1063/1.1679070}
  {\path{doi:10.1063/1.1679070}}.
\newline\urlprefix\url{http://aip.scitation.org/doi/10.1063/1.1679070}

\bibitem{ng_hypernetted_1974}
K.-C. Ng,
  \href{http://scitation.aip.org/content/aip/journal/jcp/61/7/10.1063/1.1682399}{Hypernetted
  chain solutions for the classical one-component plasma up to $\gamma=7000$},
  J.~Chem.~Phys. 61~(7) (1974) 2680--2689.
\newblock \href {http://dx.doi.org/10.1063/1.1682399}
  {\path{doi:10.1063/1.1682399}}.
\newline\urlprefix\url{http://scitation.aip.org/content/aip/journal/jcp/61/7/10.1063/1.1682399}

\bibitem{mermin_lindhard_1970}
N.~D. Mermin, \href{https://link.aps.org/doi/10.1103/PhysRevB.1.2362}{Lindhard
  {Dielectric} {Function} in the {Relaxation}-{Time} {Approximation}},
  Phys.~Rev.~B 1~(5) (1970) 2362--2363.
\newblock \href {http://dx.doi.org/10.1103/PhysRevB.1.2362}
  {\path{doi:10.1103/PhysRevB.1.2362}}.
\newline\urlprefix\url{https://link.aps.org/doi/10.1103/PhysRevB.1.2362}

\bibitem{selchow_extended_2002}
A.~Selchow, G.~R\"opke, A.~Wierling,
  \href{http://onlinelibrary.wiley.com/doi/10.1002/1521-3986(200201)42:1<43::AID-CTPP43>3.0.CO;2-3/abstract}{Extended
  {Mermin}-like {Dielectric} {Function} for a {Two}-{Component} {Plasma}},
  Contrib.~Plasma Phys. 42~(1) (2002) 43--54.
\newblock \href
  {http://dx.doi.org/10.1002/1521-3986(200201)42:1<43::AID-CTPP43>3.0.CO;2-3}
  {\path{doi:10.1002/1521-3986(200201)42:1<43::AID-CTPP43>3.0.CO;2-3}}.
\newline\urlprefix\url{http://onlinelibrary.wiley.com/doi/10.1002/1521-3986(200201)42:1<43::AID-CTPP43>3.0.CO;2-3/abstract}

\bibitem{stefanucci2013nonequilibrium}
G.~Stefanucci, R.~van Leeuwen,
  \href{https://books.google.de/books?id=6GsrjPFXLDYC}{Nonequilibrium Many-Body
  Theory of Quantum Systems: A Modern Introduction}, Cambridge University
  Press, 2013.
\newline\urlprefix\url{https://books.google.de/books?id=6GsrjPFXLDYC}

\bibitem{montroll_quantum_1958}
E.~W. Montroll, J.~C. Ward,
  \href{http://scitation.aip.org/content/aip/journal/pof1/1/1/10.1063/1.1724337}{Quantum
  {Statistics} of {Interacting} {Particles}; {General} {Theory} and {Some}
  {Remarks} on {Properties} of an {Electron} {Gas}}, Phys.~Fluids 1~(1) (1958)
  55--72.
\newblock \href {http://dx.doi.org/10.1063/1.1724337}
  {\path{doi:10.1063/1.1724337}}.
\newline\urlprefix\url{http://scitation.aip.org/content/aip/journal/pof1/1/1/10.1063/1.1724337}

\bibitem{dutta_uniform_2013}
S.~Dutta, J.~Dufty,
  \href{http://iopscience.iop.org/0295-5075/102/6/67005}{Uniform electron gas
  at warm, dense matter conditions}, Europhys.~Lett. 102~(6) (2013) 67005.
\newblock \href {http://dx.doi.org/10.1209/0295-5075/102/67005}
  {\path{doi:10.1209/0295-5075/102/67005}}.
\newline\urlprefix\url{http://iopscience.iop.org/0295-5075/102/6/67005}

\bibitem{dutta_classical_2013}
S.~Dutta, J.~Dufty,
  \href{http://link.aps.org/doi/10.1103/PhysRevE.87.032102}{Classical
  representation of a quantum system at equilibrium: {Applications}},
  Phys.~Rev.~E 87~(3) (2013) 032102.
\newblock \href {http://dx.doi.org/10.1103/PhysRevE.87.032102}
  {\path{doi:10.1103/PhysRevE.87.032102}}.
\newline\urlprefix\url{http://link.aps.org/doi/10.1103/PhysRevE.87.032102}

\bibitem{dufty_classical_2013}
J.~Dufty, S.~Dutta,
  \href{http://link.aps.org/doi/10.1103/PhysRevE.87.032101}{Classical
  representation of a quantum system at equilibrium: {Theory}}, Phys.~Rev.~E
  87~(3) (2013) 032101.
\newblock \href {http://dx.doi.org/10.1103/PhysRevE.87.032101}
  {\path{doi:10.1103/PhysRevE.87.032101}}.
\newline\urlprefix\url{http://link.aps.org/doi/10.1103/PhysRevE.87.032101}

\bibitem{binder1995monte}
K.~Binder, \href{https://books.google.de/books?id=NLJe2ypFM7IC}{Monte Carlo and
  Molecular Dynamics Simulations in Polymer Science}, Oxford University Press,
  1995.
\newline\urlprefix\url{https://books.google.de/books?id=NLJe2ypFM7IC}

\bibitem{liu_bridge-functional-based_2014}
Y.~Liu, J.~Wu,
  \href{http://scitation.aip.org/content/aip/journal/jcp/140/8/10.1063/1.4865935}{A
  bridge-functional-based classical mapping method for predicting the
  correlation functions of uniform electron gases at finite temperature},
  J.~Chem.~Phys. 140~(8) (2014) 084103.
\newblock \href {http://dx.doi.org/10.1063/1.4865935}
  {\path{doi:10.1063/1.4865935}}.
\newline\urlprefix\url{http://scitation.aip.org/content/aip/journal/jcp/140/8/10.1063/1.4865935}

\bibitem{liu_improved_2014}
Y.~Liu, J.~Wu,
  \href{http://scitation.aip.org/content/aip/journal/jcp/141/6/10.1063/1.4892587}{An
  improved classical mapping method for homogeneous electron gases at finite
  temperature}, J.~Chem.~Phys. 141~(6) (2014) 064115.
\newblock \href {http://dx.doi.org/10.1063/1.4892587}
  {\path{doi:10.1063/1.4892587}}.
\newline\urlprefix\url{http://scitation.aip.org/content/aip/journal/jcp/141/6/10.1063/1.4892587}

\bibitem{dufty_classical_2012}
J.~W. Dufty, S.~Dutta,
  \href{http://onlinelibrary.wiley.com/doi/10.1002/ctpp.201100066/abstract}{Classical
  {Representation} of a {Quantum} {System} at {Equilibrium}}, Contrib.~Plasma
  Phys. 52~(1) (2012) 100--103.
\newblock \href {http://dx.doi.org/10.1002/ctpp.201100066}
  {\path{doi:10.1002/ctpp.201100066}}.
\newline\urlprefix\url{http://onlinelibrary.wiley.com/doi/10.1002/ctpp.201100066/abstract}

\bibitem{wrighton_finite-temperature_2016}
J.~Wrighton, J.~Dufty, S.~Dutta,
  \href{https://link.aps.org/doi/10.1103/PhysRevE.94.053208}{Finite-temperature
  quantum effects on confined charges}, Phys.~Rev.~E 94~(5) (2016) 053208.
\newblock \href {http://dx.doi.org/10.1103/PhysRevE.94.053208}
  {\path{doi:10.1103/PhysRevE.94.053208}}.
\newline\urlprefix\url{https://link.aps.org/doi/10.1103/PhysRevE.94.053208}

\bibitem{metropolis_equation_1953}
N.~Metropolis, A.~W. Rosenbluth, M.~N. Rosenbluth, A.~H. Teller, E.~Teller,
  \href{http://scitation.aip.org/content/aip/journal/jcp/21/6/10.1063/1.1699114}{Equation
  of {State} {Calculations} by {Fast} {Computing} {Machines}}, J.~Chem.~Phys.
  21~(6) (1953) 1087--1092.
\newblock \href {http://dx.doi.org/10.1063/1.1699114}
  {\path{doi:10.1063/1.1699114}}.
\newline\urlprefix\url{http://scitation.aip.org/content/aip/journal/jcp/21/6/10.1063/1.1699114}

\bibitem{mak_multilevel_1998}
C.~H. Mak, R.~Egger, H.~Weber-Gottschick,
  \href{http://link.aps.org/doi/10.1103/PhysRevLett.81.4533}{Multilevel
  {Blocking} {Approach} to the {Fermion} {Sign} {Problem} in {Path}-{Integral}
  {Monte} {Carlo} {Simulations}}, Phys.~Rev.~Lett. 81~(21) (1998) 4533--4536.
\newblock \href {http://dx.doi.org/10.1103/PhysRevLett.81.4533}
  {\path{doi:10.1103/PhysRevLett.81.4533}}.
\newline\urlprefix\url{http://link.aps.org/doi/10.1103/PhysRevLett.81.4533}

\bibitem{egger_path-integral_2000}
R.~Egger, L.~M\"uhlbacher, C.~H. Mak,
  \href{http://link.aps.org/doi/10.1103/PhysRevE.61.5961}{Path-integral {Monte}
  {Carlo} simulations without the sign problem: {Multilevel} blocking approach
  for effective actions}, Phys.~Rev.~E 61~(5) (2000) 5961--5966.
\newblock \href {http://dx.doi.org/10.1103/PhysRevE.61.5961}
  {\path{doi:10.1103/PhysRevE.61.5961}}.
\newline\urlprefix\url{http://link.aps.org/doi/10.1103/PhysRevE.61.5961}

\bibitem{egger_multilevel_2001}
R.~Egger, C.~H. Mak,
  \href{http://www.worldscientific.com/doi/abs/10.1142/S021797920100591X}{Multilevel
  blocking monte carlo simulations for quantum dots}, Int.~J.~Mod.~Phys.~B
  15~(10n11) (2001) 1416--1425.
\newblock \href {http://dx.doi.org/10.1142/S021797920100591X}
  {\path{doi:10.1142/S021797920100591X}}.
\newline\urlprefix\url{http://www.worldscientific.com/doi/abs/10.1142/S021797920100591X}

\bibitem{dikovsky_analysis_2001}
M.~V. Dikovsky, C.~H. Mak,
  \href{http://link.aps.org/doi/10.1103/PhysRevB.63.235105}{Analysis of the
  multilevel blocking approach to the fermion sign problem: {Accuracy}, errors,
  and practice}, Phys.~Rev.~B 63~(23) (2001) 235105.
\newblock \href {http://dx.doi.org/10.1103/PhysRevB.63.235105}
  {\path{doi:10.1103/PhysRevB.63.235105}}.
\newline\urlprefix\url{http://link.aps.org/doi/10.1103/PhysRevB.63.235105}

\bibitem{muhlbacher_crossover_2003}
L.~M\"uhlbacher, R.~Egger,
  \href{http://scitation.aip.org/content/aip/journal/jcp/118/1/10.1063/1.1523014}{Crossover
  from nonadiabatic to adiabatic electron transfer reactions: {Multilevel}
  blocking {Monte} {Carlo} simulations}, J.~Chem.~Phys. 118~(1) (2003)
  179--191.
\newblock \href {http://dx.doi.org/10.1063/1.1523014}
  {\path{doi:10.1063/1.1523014}}.
\newline\urlprefix\url{http://scitation.aip.org/content/aip/journal/jcp/118/1/10.1063/1.1523014}

\bibitem{vorontsov-velyaminov_path_2006}
P.~N. Vorontsov-Velyaminov, M.~A. Voznesenski, D.~V. Malakhov, A.~P.
  Lyubartsev, A.~V. Broukhno,
  \href{http://stacks.iop.org/0305-4470/39/i=17/a=S62}{Path integral method in
  quantum statistics problems: generalized ensemble {Monte} {Carlo} and density
  functional approach}, J.~Phys.~A 39~(17) (2006) 4711.
\newblock \href {http://dx.doi.org/10.1088/0305-4470/39/17/S62}
  {\path{doi:10.1088/0305-4470/39/17/S62}}.
\newline\urlprefix\url{http://stacks.iop.org/0305-4470/39/i=17/a=S62}

\bibitem{voznesenskiy_path-integralchar21expanded-ensemble_2009}
M.~A. Voznesenskiy, P.~N. Vorontsov-Velyaminov, A.~P. Lyubartsev,
  \href{http://link.aps.org/doi/10.1103/PhysRevE.80.066702}{Path-integral-expanded-ensemble
  {Monte} {Carlo} method in treatment of the sign problem for fermions},
  Phys.~Rev.~E 80~(6) (2009) 066702.
\newblock \href {http://dx.doi.org/10.1103/PhysRevE.80.066702}
  {\path{doi:10.1103/PhysRevE.80.066702}}.
\newline\urlprefix\url{http://link.aps.org/doi/10.1103/PhysRevE.80.066702}

\bibitem{herman_path_1982}
M.~F. Herman, E.~J. Bruskin, B.~J. Berne,
  \href{http://aip.scitation.org/doi/abs/10.1063/1.442815}{On path integral
  {Monte} {Carlo} simulations}, J.~Chem.~Phys. 76~(10) (1982) 5150--5155.
\newblock \href {http://dx.doi.org/10.1063/1.442815}
  {\path{doi:10.1063/1.442815}}.
\newline\urlprefix\url{http://aip.scitation.org/doi/abs/10.1063/1.442815}

\bibitem{ceperley_path_1995}
D.~M. Ceperley, \href{http://link.aps.org/doi/10.1103/RevModPhys.67.279}{Path
  integrals in the theory of condensed helium}, Rev.~Mod.~Phys. 67~(2) (1995)
  279--355.
\newblock \href {http://dx.doi.org/10.1103/RevModPhys.67.279}
  {\path{doi:10.1103/RevModPhys.67.279}}.
\newline\urlprefix\url{http://link.aps.org/doi/10.1103/RevModPhys.67.279}

\bibitem{chandler_exploiting_1981}
D.~Chandler, P.~Wolynes, \href{http://dx.doi.org/10.1063/1.441588}{Exploiting
  the isomorphism between quantum theory and classical statistical mechanics of
  polyatomic fluids}, J.~Chem.~Phys. 74 (1981) 4078.
\newblock \href {http://dx.doi.org/10.1063/1.441588}
  {\path{doi:10.1063/1.441588}}.
\newline\urlprefix\url{http://dx.doi.org/10.1063/1.441588}

\bibitem{pollock_simulation_1984}
E.~Pollock, D.~Ceperley,
  \href{http://link.aps.org/doi/10.1103/PhysRevB.30.2555}{Simulation of quantum
  many-body systems by path-integral methods}, Phys.~Rev.~B 30~(5) (1984)
  2555--2568.
\newblock \href {http://dx.doi.org/10.1103/PhysRevB.30.2555}
  {\path{doi:10.1103/PhysRevB.30.2555}}.
\newline\urlprefix\url{http://link.aps.org/doi/10.1103/PhysRevB.30.2555}

\bibitem{pollock_path-integral_1987}
E.~L. Pollock, D.~M. Ceperley,
  \href{http://link.aps.org/doi/10.1103/PhysRevB.36.8343}{Path-integral
  computation of superfluid densities}, Phys.~Rev.~B 36~(16) (1987) 8343--8352.
\newblock \href {http://dx.doi.org/10.1103/PhysRevB.36.8343}
  {\path{doi:10.1103/PhysRevB.36.8343}}.
\newline\urlprefix\url{http://link.aps.org/doi/10.1103/PhysRevB.36.8343}

\bibitem{sindzingre_path-integral_1989}
P.~Sindzingre, M.~L. Klein, D.~M. Ceperley,
  \href{http://link.aps.org/doi/10.1103/PhysRevLett.63.1601}{{Path-integral
  {Monte} {Carlo} study of low-temperature $^4$He clusters}}, Phys.~Rev.~Lett.
  63~(15) (1989) 1601--1604.
\newblock \href {http://dx.doi.org/10.1103/PhysRevLett.63.1601}
  {\path{doi:10.1103/PhysRevLett.63.1601}}.
\newline\urlprefix\url{http://link.aps.org/doi/10.1103/PhysRevLett.63.1601}

\bibitem{kwon_local_2006}
Y.~Kwon, F.~Paesani, K.~Whaley,
  \href{http://link.aps.org/doi/10.1103/PhysRevB.74.174522}{Local superfluidity
  in inhomogeneous quantum fluids}, Phys.~Rev.~B 74~(17).
\newblock \href {http://dx.doi.org/10.1103/PhysRevB.74.174522}
  {\path{doi:10.1103/PhysRevB.74.174522}}.
\newline\urlprefix\url{http://link.aps.org/doi/10.1103/PhysRevB.74.174522}

\bibitem{dornheim_superfluidity_2015}
T.~Dornheim, A.~Filinov, M.~Bonitz,
  \href{http://link.aps.org/doi/10.1103/PhysRevB.91.054503}{Superfluidity of
  strongly correlated bosons in two- and three-dimensional traps}, Phys.~Rev.~B
  91~(5) (2015) 054503.
\newblock \href {http://dx.doi.org/10.1103/PhysRevB.91.054503}
  {\path{doi:10.1103/PhysRevB.91.054503}}.
\newline\urlprefix\url{http://link.aps.org/doi/10.1103/PhysRevB.91.054503}

\bibitem{gruter_critical_1997}
P.~Gr\"uter, D.~Ceperley, F.~Laloë,
  \href{https://link.aps.org/doi/10.1103/PhysRevLett.79.3549}{Critical
  {Temperature} of {Bose}-{Einstein} {Condensation} of {Hard}-{Sphere}
  {Gases}}, Phys.~Rev.~Lett. 79~(19) (1997) 3549--3552.
\newblock \href {http://dx.doi.org/10.1103/PhysRevLett.79.3549}
  {\path{doi:10.1103/PhysRevLett.79.3549}}.
\newline\urlprefix\url{https://link.aps.org/doi/10.1103/PhysRevLett.79.3549}

\bibitem{pilati_dilute_2010}
S.~Pilati, S.~Giorgini, M.~Modugno, N.~Prokof'ev,
  \href{http://stacks.iop.org/1367-2630/12/i=7/a=073003}{Dilute {Bose} gas with
  correlated disorder: a path integral {Monte} {Carlo} study}, New J.~Phys.
  12~(7) (2010) 073003.
\newblock \href {http://dx.doi.org/10.1088/1367-2630/12/7/073003}
  {\path{doi:10.1088/1367-2630/12/7/073003}}.
\newline\urlprefix\url{http://stacks.iop.org/1367-2630/12/i=7/a=073003}

\bibitem{noauthor_path-integral_2016}
H.~Saito,
  \href{http://journals.jps.jp/doi/full/10.7566/JPSJ.85.053001}{Path-{Integral}
  {Monte} {Carlo} {Study} on a {Droplet} of a {Dipolar} {Bose}--{Einstein}
  {Condensate} {Stabilized} by {Quantum} {Fluctuation}}, J.~Phys.~Soc.~Jpn.
  85~(5) (2016) 053001.
\newblock \href {http://dx.doi.org/10.7566/JPSJ.85.053001}
  {\path{doi:10.7566/JPSJ.85.053001}}.
\newline\urlprefix\url{http://journals.jps.jp/doi/full/10.7566/JPSJ.85.053001}

\bibitem{filinov_collective_2012}
A.~Filinov, M.~Bonitz,
  \href{http://link.aps.org/doi/10.1103/PhysRevA.86.043628}{Collective and
  single-particle excitations in two-dimensional dipolar {Bose} gases},
  Phys.~Rev.~A 86~(4).
\newblock \href {http://dx.doi.org/10.1103/PhysRevA.86.043628}
  {\path{doi:10.1103/PhysRevA.86.043628}}.
\newline\urlprefix\url{http://link.aps.org/doi/10.1103/PhysRevA.86.043628}

\bibitem{filinov_correlation_2016}
A.~Filinov,
  \href{http://link.aps.org/doi/10.1103/PhysRevA.94.013603}{Correlation effects
  and collective excitations in bosonic bilayers: {Role} of quantum statistics,
  superfluidity, and the dimerization transition}, Phys.~Rev.~A 94~(1) (2016)
  013603.
\newblock \href {http://dx.doi.org/10.1103/PhysRevA.94.013603}
  {\path{doi:10.1103/PhysRevA.94.013603}}.
\newline\urlprefix\url{http://link.aps.org/doi/10.1103/PhysRevA.94.013603}

\bibitem{loh_sign_1990}
E.~Loh, J.~Gubernatis, R.~Scalettar, S.~White, D.~Scalapino, R.~Sugar,
  \href{http://link.aps.org/doi/10.1103/PhysRevB.41.9301}{Sign problem in the
  numerical simulation of many-electron systems}, Phys.~Rev.~B 41~(13) (1990)
  9301--9307.
\newblock \href {http://dx.doi.org/10.1103/PhysRevB.41.9301}
  {\path{doi:10.1103/PhysRevB.41.9301}}.
\newline\urlprefix\url{http://link.aps.org/doi/10.1103/PhysRevB.41.9301}

\bibitem{troyer_computational_2005}
M.~Troyer, U.-J. Wiese,
  \href{http://link.aps.org/doi/10.1103/PhysRevLett.94.170201}{Computational
  {Complexity} and {Fundamental} {Limitations} to {Fermionic} {Quantum} {Monte}
  {Carlo} {Simulations}}, Phys.~Rev.~Lett. 94~(17).
\newblock \href {http://dx.doi.org/10.1103/PhysRevLett.94.170201}
  {\path{doi:10.1103/PhysRevLett.94.170201}}.
\newline\urlprefix\url{http://link.aps.org/doi/10.1103/PhysRevLett.94.170201}

\bibitem{kleinert_path_2009}
H.~Kleinert, Path Integrals in Quantum Mechanics, Statistics, Polymer Physics,
  and Financial Markets, World Scientific, 2009.

\bibitem{trotter_product_1959}
H.~F. Trotter, \href{http://www.jstor.org/stable/2033649}{On the {Product} of
  {Semi}-{Groups} of {Operators}}, Proc.~Am.~Math.~Soc. 10~(4) (1959) 545--551.
\newblock \href {http://dx.doi.org/10.2307/2033649}
  {\path{doi:10.2307/2033649}}.
\newline\urlprefix\url{http://www.jstor.org/stable/2033649}

\bibitem{de_raedt_applications_1983}
H.~De~Raedt, B.~De~Raedt,
  \href{https://link.aps.org/doi/10.1103/PhysRevA.28.3575}{Applications of the
  generalized {Trotter} formula}, Phys.~Rev.~A 28~(6) (1983) 3575--3580.
\newblock \href {http://dx.doi.org/10.1103/PhysRevA.28.3575}
  {\path{doi:10.1103/PhysRevA.28.3575}}.
\newline\urlprefix\url{https://link.aps.org/doi/10.1103/PhysRevA.28.3575}

\bibitem{sign_cite}
D.~Ceperley, Path integral monte carlo methods for fermions, in: K.~Binder,
  G.~Ciccotti (Eds.), Monte Carlo and Molecular Dynamics of Condensed Matter
  Systems, Italian Physical Society, Bologna, 1996.

\bibitem{boninsegni_worm_2006}
M.~Boninsegni, N.~Prokof’ev, B.~Svistunov,
  \href{http://link.aps.org/doi/10.1103/PhysRevLett.96.070601}{Worm {Algorithm}
  for {Continuous}-{Space} {Path} {Integral} {Monte}-{Carlo} {Simulations}},
  Phys.~Rev.~Lett. 96~(7) (2006) 070601.
\newblock \href {http://dx.doi.org/10.1103/PhysRevLett.96.070601}
  {\path{doi:10.1103/PhysRevLett.96.070601}}.
\newline\urlprefix\url{http://link.aps.org/doi/10.1103/PhysRevLett.96.070601}

\bibitem{boninsegni_worm_2006-1}
M.~Boninsegni, N.~V. Prokof’ev, B.~V. Svistunov,
  \href{http://link.aps.org/doi/10.1103/PhysRevE.74.036701}{Worm algorithm and
  diagrammatic {Monte} {Carlo}: {A} new approach to continuous-space path
  integral {Monte} {Carlo} simulations}, Phys.~Rev.~E 74~(3) (2006) 036701.
\newblock \href {http://dx.doi.org/10.1103/PhysRevE.74.036701}
  {\path{doi:10.1103/PhysRevE.74.036701}}.
\newline\urlprefix\url{http://link.aps.org/doi/10.1103/PhysRevE.74.036701}

\bibitem{kraeft_quantum_1986}
W.-D. Kraeft, D.~Kremp, W.~Ebeling, G.~R\"opke, Quantum Statistics of Charged
  Particle Systems, Akademie-Verlag Berlin, 1986.

\bibitem{dornheim_analyzing_2016}
T.~Dornheim, H.~Thomsen, P.~Ludwig, A.~Filinov, M.~Bonitz,
  \href{http://onlinelibrary.wiley.com/doi/10.1002/ctpp.201500120/abstract}{Analyzing
  {Quantum} {Correlations} {Made} {Simple}}, Contrib.~Plasma Phys. 56~(5)
  (2016) 371--379.
\newblock \href {http://dx.doi.org/10.1002/ctpp.201500120}
  {\path{doi:10.1002/ctpp.201500120}}.
\newline\urlprefix\url{http://onlinelibrary.wiley.com/doi/10.1002/ctpp.201500120/abstract}

\bibitem{filinov_book_1977}
V.~M. Zamalin, G.~E. Norman, V.~S. Filinov, The Monte Carlo Method in
  Statistical Thermodynamics, Nauka, Moscow, 1977.

\bibitem{takahashi_monte_1984}
M.~Takahashi, M.~Imada,
  \href{http://journals.jps.jp/doi/abs/10.1143/JPSJ.53.963}{Monte {Carlo}
  {Calculation} of {Quantum} {Systems}}, J.~Phys.~Soc.~Jpn. 53~(3) (1984)
  963--974.
\newblock \href {http://dx.doi.org/10.1143/JPSJ.53.963}
  {\path{doi:10.1143/JPSJ.53.963}}.
\newline\urlprefix\url{http://journals.jps.jp/doi/abs/10.1143/JPSJ.53.963}

\bibitem{lyubartsev_simulation_2005}
A.~P. Lyubartsev,
  \href{http://iopscience.iop.org/0305-4470/38/30/003}{Simulation of excited
  states and the sign problem in the path integral {Monte} {Carlo} method},
  J.~Phys.~A 38~(30) (2005) 6659.
\newblock \href {http://dx.doi.org/10.1088/0305-4470/38/30/003}
  {\path{doi:10.1088/0305-4470/38/30/003}}.
\newline\urlprefix\url{http://iopscience.iop.org/0305-4470/38/30/003}

\bibitem{chin_high-order_2015}
S.~A. Chin,
  \href{http://link.aps.org/doi/10.1103/PhysRevE.91.031301}{High-order
  path-integral {Monte} {Carlo} methods for solving quantum dot problems},
  Phys.~Rev.~E 91~(3) (2015) 031301.
\newblock \href {http://dx.doi.org/10.1103/PhysRevE.91.031301}
  {\path{doi:10.1103/PhysRevE.91.031301}}.
\newline\urlprefix\url{http://link.aps.org/doi/10.1103/PhysRevE.91.031301}

\bibitem{takahashi_monte_1984-1}
M.~Takahashi, M.~Imada,
  \href{http://journals.jps.jp/doi/abs/10.1143/JPSJ.53.3765}{Monte {Carlo}
  {Calculation} of {Quantum} {Systems}. {II}. {Higher} {Order} {Correction}},
  J.~Phys.~Soc.~Jpn. 53~(11) (1984) 3765--3769.
\newblock \href {http://dx.doi.org/10.1143/JPSJ.53.3765}
  {\path{doi:10.1143/JPSJ.53.3765}}.
\newline\urlprefix\url{http://journals.jps.jp/doi/abs/10.1143/JPSJ.53.3765}

\bibitem{chin_gradient_2002}
S.~A. Chin, C.~R. Chen,
  \href{http://scitation.aip.org/content/aip/journal/jcp/117/4/10.1063/1.1485725}{Gradient
  symplectic algorithms for solving the {Schr\"odinger} equation with
  time-dependent potentials}, J.~Chem.~Phys. 117~(4) (2002) 1409--1415.
\newblock \href {http://dx.doi.org/10.1063/1.1485725}
  {\path{doi:10.1063/1.1485725}}.
\newline\urlprefix\url{http://scitation.aip.org/content/aip/journal/jcp/117/4/10.1063/1.1485725}

\bibitem{brualla_higher_2004}
L.~Brualla, K.~Sakkos, J.~Boronat, J.~Casulleras,
  \href{http://scitation.aip.org/content/aip/journal/jcp/121/2/10.1063/1.1760512}{Higher
  order and infinite {Trotter}-number extrapolations in path integral {Monte}
  {Carlo}}, J.~Chem.~Phys. 121~(2) (2004) 636--643.
\newblock \href {http://dx.doi.org/10.1063/1.1760512}
  {\path{doi:10.1063/1.1760512}}.
\newline\urlprefix\url{http://scitation.aip.org/content/aip/journal/jcp/121/2/10.1063/1.1760512}

\bibitem{sakkos_high_2009}
K.~Sakkos, J.~Casulleras, J.~Boronat,
  \href{http://scitation.aip.org/content/aip/journal/jcp/130/20/10.1063/1.3143522}{High
  order {Chin} actions in path integral {Monte} {Carlo}}, J.~Chem.~Phys.
  130~(20) (2009) 204109.
\newblock \href {http://dx.doi.org/10.1063/1.3143522}
  {\path{doi:10.1063/1.3143522}}.
\newline\urlprefix\url{http://scitation.aip.org/content/aip/journal/jcp/130/20/10.1063/1.3143522}

\bibitem{zillich_extrapolated_2010}
R.~E. Zillich, J.~M. Mayrhofer, S.~A. Chin,
  \href{http://scitation.aip.org/content/aip/journal/jcp/132/4/10.1063/1.3297888}{Extrapolated
  high-order propagators for path integral {Monte} {Carlo} simulations},
  J.~Chem.~Phys. 132~(4) (2010) 044103.
\newblock \href {http://dx.doi.org/10.1063/1.3297888}
  {\path{doi:10.1063/1.3297888}}.
\newline\urlprefix\url{http://scitation.aip.org/content/aip/journal/jcp/132/4/10.1063/1.3297888}

\bibitem{prokofev_exact_1996}
N.~V. Prokofâ€™ev, B.~V. Svistunov, I.~S. Tupitsyn,
  \href{http://www.springerlink.com/content/613449m532p78470/}{Exact quantum
  {Monte} {Carlo} process for the statistics of discrete systems}, JETP Lett.
  64~(12) (1996) 911--916.
\newblock \href {http://dx.doi.org/10.1134/1.567243}
  {\path{doi:10.1134/1.567243}}.
\newline\urlprefix\url{http://www.springerlink.com/content/613449m532p78470/}

\bibitem{beard_simulations_1996}
B.~B. Beard, U.-J. Wiese,
  \href{https://link.aps.org/doi/10.1103/PhysRevLett.77.5130}{Simulations of
  discrete quantum systems in continuous euclidean time}, Phys. Rev. Lett. 77
  (1996) 5130--5133.
\newblock \href {http://dx.doi.org/10.1103/PhysRevLett.77.5130}
  {\path{doi:10.1103/PhysRevLett.77.5130}}.
\newline\urlprefix\url{https://link.aps.org/doi/10.1103/PhysRevLett.77.5130}

\bibitem{gull_continuous-time_2011}
E.~Gull, A.~J. Millis, A.~I. Lichtenstein, A.~N. Rubtsov, M.~Troyer, P.~Werner,
  \href{http://link.aps.org/doi/10.1103/RevModPhys.83.349}{Continuous-time
  {Monte} {Carlo} methods for quantum impurity models}, Rev.~Mod.~Phys. 83~(2)
  (2011) 349--404.
\newblock \href {http://dx.doi.org/10.1103/RevModPhys.83.349}
  {\path{doi:10.1103/RevModPhys.83.349}}.
\newline\urlprefix\url{http://link.aps.org/doi/10.1103/RevModPhys.83.349}

\bibitem{sandvik_quantum_1991}
A.~W. Sandvik, J.~Kurkij\"arvi,
  \href{https://link.aps.org/doi/10.1103/PhysRevB.43.5950}{Quantum monte carlo
  simulation method for spin systems}, Phys. Rev. B 43 (1991) 5950--5961.
\newblock \href {http://dx.doi.org/10.1103/PhysRevB.43.5950}
  {\path{doi:10.1103/PhysRevB.43.5950}}.
\newline\urlprefix\url{https://link.aps.org/doi/10.1103/PhysRevB.43.5950}

\bibitem{sandvik_finite_1997}
A.~W. Sandvik,
  \href{https://link.aps.org/doi/10.1103/PhysRevB.56.11678}{Finite-size scaling
  of the ground-state parameters of the two-dimensional heisenberg model},
  Phys. Rev. B 56 (1997) 11678--11690.
\newblock \href {http://dx.doi.org/10.1103/PhysRevB.56.11678}
  {\path{doi:10.1103/PhysRevB.56.11678}}.
\newline\urlprefix\url{https://link.aps.org/doi/10.1103/PhysRevB.56.11678}

\bibitem{sandvik_stochastic_1999}
A.~W. Sandvik,
  \href{https://link.aps.org/doi/10.1103/PhysRevB.59.R14157}{Stochastic series
  expansion method with operator-loop update}, Phys. Rev. B 59 (1999)
  R14157--R14160.
\newblock \href {http://dx.doi.org/10.1103/PhysRevB.59.R14157}
  {\path{doi:10.1103/PhysRevB.59.R14157}}.
\newline\urlprefix\url{https://link.aps.org/doi/10.1103/PhysRevB.59.R14157}

\bibitem{sandvik_mulichain_1999}
A.~W. Sandvik,
  \href{https://link.aps.org/doi/10.1103/PhysRevLett.83.3069}{Multichain
  mean-field theory of quasi-one-dimensional quantum spin systems}, Phys. Rev.
  Lett. 83 (1999) 3069--3072.
\newblock \href {http://dx.doi.org/10.1103/PhysRevLett.83.3069}
  {\path{doi:10.1103/PhysRevLett.83.3069}}.
\newline\urlprefix\url{https://link.aps.org/doi/10.1103/PhysRevLett.83.3069}

\bibitem{shevchenko_double_2000}
P.~V. Shevchenko, A.~W. Sandvik, O.~P. Sushkov,
  \href{https://link.aps.org/doi/10.1103/PhysRevB.61.3475}{Double-layer
  {H}eisenberg antiferromagnet at finite temperature: Brueckner theory and
  quantum {M}onte {C}arlo simulations}, Phys. Rev. B 61 (2000) 3475--3487.
\newblock \href {http://dx.doi.org/10.1103/PhysRevB.61.3475}
  {\path{doi:10.1103/PhysRevB.61.3475}}.
\newline\urlprefix\url{https://link.aps.org/doi/10.1103/PhysRevB.61.3475}

\bibitem{booth_fermion_2009}
G.~H. Booth, A.~J.~W. Thom, A.~Alavi,
  \href{http://scitation.aip.org/content/aip/journal/jcp/131/5/10.1063/1.3193710}{Fermion
  {Monte} {Carlo} without fixed nodes: {A} game of life, death, and
  annihilation in {Slater} determinant space}, J.~Chem.~Phys. 131~(5) (2009)
  054106.
\newblock \href {http://dx.doi.org/10.1063/1.3193710}
  {\path{doi:10.1063/1.3193710}}.
\newline\urlprefix\url{http://scitation.aip.org/content/aip/journal/jcp/131/5/10.1063/1.3193710}

\bibitem{booth_towards_2013}
G.~H. Booth, A.~Gr\"uneis, G.~Kresse, A.~Alavi,
  \href{http://www.nature.com/nature/journal/v493/n7432/abs/nature11770.html}{Towards
  an exact description of electronic wavefunctions in real solids}, Nature
  493~(7432) (2013) 365--370.
\newblock \href {http://dx.doi.org/10.1038/nature11770}
  {\path{doi:10.1038/nature11770}}.
\newline\urlprefix\url{http://www.nature.com/nature/journal/v493/n7432/abs/nature11770.html}

\bibitem{umrigar_diffusion_1993}
C.~J. Umrigar, M.~P. Nightingale, K.~J. Runge,
  \href{http://aip.scitation.org/doi/abs/10.1063/1.465195}{A diffusion {Monte}
  {Carlo} algorithm with very small time‐step errors}, J.~Chem.~Phys. 99~(4)
  (1993) 2865--2890.
\newblock \href {http://dx.doi.org/10.1063/1.465195}
  {\path{doi:10.1063/1.465195}}.
\newline\urlprefix\url{http://aip.scitation.org/doi/abs/10.1063/1.465195}

\bibitem{dubois_overcoming_2014}
J.~L. DuBois, E.~W. Brown, B.~J. Alder,
  \href{http://arxiv.org/abs/1409.3262}{Overcoming the fermion sign problem in
  homogeneous systems}, arXiv:1409.3262 [cond-mat]ArXiv: 1409.3262.
\newline\urlprefix\url{http://arxiv.org/abs/1409.3262}

\bibitem{sjostrom_gradient_2014}
T.~Sjostrom, J.~Daligault,
  \href{http://link.aps.org/doi/10.1103/PhysRevB.90.155109}{Gradient
  corrections to the exchange-correlation free energy}, Phys.~Rev.~B 90~(15)
  (2014) 155109.
\newblock \href {http://dx.doi.org/10.1103/PhysRevB.90.155109}
  {\path{doi:10.1103/PhysRevB.90.155109}}.
\newline\urlprefix\url{http://link.aps.org/doi/10.1103/PhysRevB.90.155109}

\bibitem{filinov_cluster_2001}
V.~Filinov, \href{http://iopscience.iop.org/0305-4470/34/8/312}{Cluster
  expansion for ideal {Fermi} systems in the `fixed-node approximation'},
  J.~Phys.~A 34~(8) (2001) 1665.
\newblock \href {http://dx.doi.org/10.1088/0305-4470/34/8/312}
  {\path{doi:10.1088/0305-4470/34/8/312}}.
\newline\urlprefix\url{http://iopscience.iop.org/0305-4470/34/8/312}

\bibitem{filinov_analytical_2014}
V.~S. Filinov,
  \href{http://link.springer.com/article/10.1134/S0018151X14040105}{Analytical
  contradictions of the fixed-node density matrix}, High Temp. 52~(5) (2014)
  615--620.
\newblock \href {http://dx.doi.org/10.1134/S0018151X14040105}
  {\path{doi:10.1134/S0018151X14040105}}.
\newline\urlprefix\url{http://link.springer.com/article/10.1134/S0018151X14040105}

\bibitem{clark_hexatic_2009}
B.~Clark, M.~Casula, D.~Ceperley,
  \href{http://link.aps.org/doi/10.1103/PhysRevLett.103.055701}{Hexatic and
  {Mesoscopic} {Phases} in a 2d {Quantum} {Coulomb} {System}}, Phys.~Rev.~Lett.
  103~(5).
\newblock \href {http://dx.doi.org/10.1103/PhysRevLett.103.055701}
  {\path{doi:10.1103/PhysRevLett.103.055701}}.
\newline\urlprefix\url{http://link.aps.org/doi/10.1103/PhysRevLett.103.055701}

\bibitem{gaudoin_hellman-feynman_2007}
R.~Gaudoin, J.~M. Pitarke,
  \href{https://link.aps.org/doi/10.1103/PhysRevLett.99.126406}{Hellman-{Feynman}
  {Operator} {Sampling} in {Diffusion} {Monte} {Carlo} {Calculations}},
  Phys.~Rev.~Lett. 99~(12) (2007) 126406.
\newblock \href {http://dx.doi.org/10.1103/PhysRevLett.99.126406}
  {\path{doi:10.1103/PhysRevLett.99.126406}}.
\newline\urlprefix\url{https://link.aps.org/doi/10.1103/PhysRevLett.99.126406}

\bibitem{gurtubay_benchmark_2010}
I.~G. Gurtubay, R.~Gaudoin, J.~M. Pitarke,
  \href{http://stacks.iop.org/0953-8984/22/i=6/a=065501}{Benchmark quantum
  {Monte} {Carlo} calculations of the ground-state kinetic, interaction and
  total energy of the three-dimensional electron gas}, J.~Phys.~Condens.~Matter
  22~(6) (2010) 065501.
\newblock \href {http://dx.doi.org/10.1088/0953-8984/22/6/065501}
  {\path{doi:10.1088/0953-8984/22/6/065501}}.
\newline\urlprefix\url{http://stacks.iop.org/0953-8984/22/i=6/a=065501}

\bibitem{lieb_thermodynamic_1975}
E.~H. Lieb, H.~Narnhofer,
  \href{https://link.springer.com/article/10.1007/BF01012066}{The thermodynamic
  limit for jellium}, J.~Stat.~Phys. 12~(4) (1975) 291--310.
\newblock \href {http://dx.doi.org/10.1007/BF01012066}
  {\path{doi:10.1007/BF01012066}}.
\newline\urlprefix\url{https://link.springer.com/article/10.1007/BF01012066}

\bibitem{lin_twist-averaged_2001}
C.~Lin, F.~H. Zong, D.~M. Ceperley,
  \href{http://link.aps.org/doi/10.1103/PhysRevE.64.016702}{Twist-averaged
  boundary conditions in continuum quantum {Monte} {Carlo} algorithms},
  Phys.~Rev.~E 64~(1) (2001) 016702.
\newblock \href {http://dx.doi.org/10.1103/PhysRevE.64.016702}
  {\path{doi:10.1103/PhysRevE.64.016702}}.
\newline\urlprefix\url{http://link.aps.org/doi/10.1103/PhysRevE.64.016702}

\bibitem{chiesa_finite-size_2006}
S.~Chiesa, D.~M. Ceperley, R.~M. Martin, M.~Holzmann,
  \href{http://link.aps.org/doi/10.1103/PhysRevLett.97.076404}{Finite-{Size}
  {Error} in {Many}-{Body} {Simulations} with {Long}-{Range} {Interactions}},
  Phys.~Rev.~Lett. 97~(7) (2006) 076404.
\newblock \href {http://dx.doi.org/10.1103/PhysRevLett.97.076404}
  {\path{doi:10.1103/PhysRevLett.97.076404}}.
\newline\urlprefix\url{http://link.aps.org/doi/10.1103/PhysRevLett.97.076404}

\bibitem{drummond_finite-size_2008}
N.~D. Drummond, R.~J. Needs, A.~Sorouri, W.~M.~C. Foulkes,
  \href{http://link.aps.org/doi/10.1103/PhysRevB.78.125106}{Finite-size errors
  in continuum quantum {Monte} {Carlo} calculations}, Phys.~Rev.~B 78~(12)
  (2008) 125106.
\newblock \href {http://dx.doi.org/10.1103/PhysRevB.78.125106}
  {\path{doi:10.1103/PhysRevB.78.125106}}.
\newline\urlprefix\url{http://link.aps.org/doi/10.1103/PhysRevB.78.125106}

\bibitem{holzmann_theory_2016}
M.~Holzmann, R.~C. Clay, M.~A. Morales, N.~M. Tubman, D.~M. Ceperley,
  C.~Pierleoni,
  \href{http://link.aps.org/doi/10.1103/PhysRevB.94.035126}{Theory of finite
  size effects for electronic quantum {Monte} {Carlo} calculations of liquids
  and solids}, Phys.~Rev.~B 94~(3) (2016) 035126.
\newblock \href {http://dx.doi.org/10.1103/PhysRevB.94.035126}
  {\path{doi:10.1103/PhysRevB.94.035126}}.
\newline\urlprefix\url{http://link.aps.org/doi/10.1103/PhysRevB.94.035126}

\bibitem{kugler_bounds_1970}
A.~A. Kugler, \href{http://link.aps.org/doi/10.1103/PhysRevA.1.1688}{Bounds for
  {Some} {Equilibrium} {Properties} of an {Electron} {Gas}}, Phys.~Rev.~A 1~(6)
  (1970) 1688--1696.
\newblock \href {http://dx.doi.org/10.1103/PhysRevA.1.1688}
  {\path{doi:10.1103/PhysRevA.1.1688}}.
\newline\urlprefix\url{http://link.aps.org/doi/10.1103/PhysRevA.1.1688}

\bibitem{dornheim_textitab_2017}
T.~Dornheim, S.~Groth, M.~Bonitz,
  \href{http://dx.doi.org/10.1002/ctpp.201700096}{Ab initio results for the
  static structure factor of the warm dense electron gas}, Contrib.~Plasma
  Phys. 57~(10) (2017) 468--478.
\newblock \href {http://dx.doi.org/10.1002/ctpp.201700096}
  {\path{doi:10.1002/ctpp.201700096}}.
\newline\urlprefix\url{http://dx.doi.org/10.1002/ctpp.201700096}

\bibitem{dewitt_statistical_1966}
H.~E. DeWitt,
  \href{http://scitation.aip.org/content/aip/journal/jmp/7/4/10.1063/1.1704974;jsessionid=8AjO6DUAODnHhlQ1CPoUsjbY.x-aip-live-03}{Statistical
  {Mechanics} of {High}‐{Temperature} {Quantum} {Plasmas} {Beyond} the {Ring}
  {Approximation}}, J.~Math.~Phys. 7~(4) (1966) 616--626.
\newblock \href {http://dx.doi.org/10.1063/1.1704974}
  {\path{doi:10.1063/1.1704974}}.
\newline\urlprefix\url{http://scitation.aip.org/content/aip/journal/jmp/7/4/10.1063/1.1704974;jsessionid=8AjO6DUAODnHhlQ1CPoUsjbY.x-aip-live-03}

\bibitem{groth_free_2017}
S.~Groth, T.~Dornheim, M.~Bonitz,
  \href{http://onlinelibrary.wiley.com/doi/10.1002/ctpp.201600082/abstract}{Free
  energy of the uniform electron gas: {Testing} analytical models against
  first-principles results}, Contrib.~Plasma Phys. 57~(3) (2017) 137--146.
\newblock \href {http://dx.doi.org/10.1002/ctpp.201600082}
  {\path{doi:10.1002/ctpp.201600082}}.
\newline\urlprefix\url{http://onlinelibrary.wiley.com/doi/10.1002/ctpp.201600082/abstract}

\bibitem{burke_exact_2016}
K.~Burke, J.~C. Smith, P.~E. Grabowski, A.~Pribram-Jones,
  \href{http://link.aps.org/doi/10.1103/PhysRevB.93.195132}{Exact conditions on
  the temperature dependence of density functionals}, Phys.~Rev.~B 93~(19)
  (2016) 195132.
\newblock \href {http://dx.doi.org/10.1103/PhysRevB.93.195132}
  {\path{doi:10.1103/PhysRevB.93.195132}}.
\newline\urlprefix\url{http://link.aps.org/doi/10.1103/PhysRevB.93.195132}

\bibitem{lu_evaluation_2014}
D.~Lu,
  \href{http://scitation.aip.org/content/aip/journal/jcp/140/18/10.1063/1.4867538}{Evaluation
  of model exchange-correlation kernels in the adiabatic connection
  fluctuation-dissipation theorem for inhomogeneous systems}, J.~Chem.~Phys.
  140~(18) (2014) 18A520.
\newblock \href {http://dx.doi.org/10.1063/1.4867538}
  {\path{doi:10.1063/1.4867538}}.
\newline\urlprefix\url{http://scitation.aip.org/content/aip/journal/jcp/140/18/10.1063/1.4867538}

\bibitem{patrick_adiabatic-connection_2015}
C.~E. Patrick, K.~S. Thygesen,
  \href{http://scitation.aip.org/content/aip/journal/jcp/143/10/10.1063/1.4919236}{Adiabatic-connection
  fluctuation-dissipation {DFT} for the structural properties of solids --
  {The} renormalized {ALDA} and electron gas kernels}, J.~Chem.~Phys. 143~(10)
  (2015) 102802.
\newblock \href {http://dx.doi.org/10.1063/1.4919236}
  {\path{doi:10.1063/1.4919236}}.
\newline\urlprefix\url{http://scitation.aip.org/content/aip/journal/jcp/143/10/10.1063/1.4919236}

\bibitem{pribram-jones_thermal_2016}
A.~Pribram-Jones, P.~E. Grabowski, K.~Burke,
  \href{http://link.aps.org/doi/10.1103/PhysRevLett.116.233001}{Thermal
  {Density} {Functional} {Theory}: {Time}-{Dependent} {Linear} {Response} and
  {Approximate} {Functionals} from the {Fluctuation}-{Dissipation} {Theorem}},
  Phys.~Rev.~Lett. 116~(23) (2016) 233001.
\newblock \href {http://dx.doi.org/10.1103/PhysRevLett.116.233001}
  {\path{doi:10.1103/PhysRevLett.116.233001}}.
\newline\urlprefix\url{http://link.aps.org/doi/10.1103/PhysRevLett.116.233001}

\bibitem{neumayer_plasmons_2010}
P.~Neumayer, C.~Fortmann, T.~D\"oppner, P.~Davis, R.~W. Falcone, A.~L.
  Kritcher, O.~L. Landen, H.~J. Lee, R.~W. Lee, C.~Niemann, S.~Le~Pape, S.~H.
  Glenzer,
  \href{http://link.aps.org/doi/10.1103/PhysRevLett.105.075003}{Plasmons in
  {Strongly} {Coupled} {Shock}-{Compressed} {Matter}}, Phys.~Rev.~Lett. 105~(7)
  (2010) 075003.
\newblock \href {http://dx.doi.org/10.1103/PhysRevLett.105.075003}
  {\path{doi:10.1103/PhysRevLett.105.075003}}.
\newline\urlprefix\url{http://link.aps.org/doi/10.1103/PhysRevLett.105.075003}

\bibitem{plagemann_dynamic_2012}
K.-U. Plagemann, P.~Sperling, R.~Thiele, M.~P. Desjarlais, C.~Fortmann,
  T.~D\"oppner, H.~J. Lee, S.~H. Glenzer, R.~Redmer,
  \href{http://stacks.iop.org/1367-2630/14/i=5/a=055020}{Dynamic structure
  factor in warm dense beryllium}, New J.~Phys. 14~(5) (2012) 055020.
\newblock \href {http://dx.doi.org/10.1088/1367-2630/14/5/055020}
  {\path{doi:10.1088/1367-2630/14/5/055020}}.
\newline\urlprefix\url{http://stacks.iop.org/1367-2630/14/i=5/a=055020}

\bibitem{fortmann_influence_2010}
C.~Fortmann, A.~Wierling, G.~R\"opke,
  \href{http://link.aps.org/doi/10.1103/PhysRevE.81.026405}{Influence of
  local-field corrections on {Thomson} scattering in collision-dominated
  two-component plasmas}, Phys.~Rev.~E 81~(2) (2010) 026405.
\newblock \href {http://dx.doi.org/10.1103/PhysRevE.81.026405}
  {\path{doi:10.1103/PhysRevE.81.026405}}.
\newline\urlprefix\url{http://link.aps.org/doi/10.1103/PhysRevE.81.026405}

\bibitem{reinholz_conductivity_2015}
H.~Reinholz, G.~R\"opke, S.~Rosmej, R.~Redmer,
  \href{https://link.aps.org/doi/10.1103/PhysRevE.91.043105}{Conductivity of
  warm dense matter including electron-electron collisions}, Phys.~Rev.~E
  91~(4) (2015) 043105.
\newblock \href {http://dx.doi.org/10.1103/PhysRevE.91.043105}
  {\path{doi:10.1103/PhysRevE.91.043105}}.
\newline\urlprefix\url{https://link.aps.org/doi/10.1103/PhysRevE.91.043105}

\bibitem{veysman_optical_2016}
M.~Veysman, G.~R\"opke, M.~Winkel, H.~Reinholz,
  \href{https://link.aps.org/doi/10.1103/PhysRevE.94.013203}{Optical
  conductivity of warm dense matter within a wide frequency range using quantum
  statistical and kinetic approaches}, Phys.~Rev.~E 94~(1) (2016) 013203.
\newblock \href {http://dx.doi.org/10.1103/PhysRevE.94.013203}
  {\path{doi:10.1103/PhysRevE.94.013203}}.
\newline\urlprefix\url{https://link.aps.org/doi/10.1103/PhysRevE.94.013203}

\bibitem{vorberger_energy_2010}
J.~Vorberger, D.~O. Gericke, T.~Bornath, M.~Schlanges,
  \href{https://link.aps.org/doi/10.1103/PhysRevE.81.046404}{Energy relaxation
  in dense, strongly coupled two-temperature plasmas}, Phys.~Rev.~E 81~(4)
  (2010) 046404.
\newblock \href {http://dx.doi.org/10.1103/PhysRevE.81.046404}
  {\path{doi:10.1103/PhysRevE.81.046404}}.
\newline\urlprefix\url{https://link.aps.org/doi/10.1103/PhysRevE.81.046404}

\bibitem{benedict_molecular_2017}
L.~X. Benedict, M.~P. Surh, L.~G. Stanton, C.~R. Scullard, A.~A. Correa, J.~I.
  Castor, F.~R. Graziani, L.~A. Collins, O.~Cert\'ik, J.~D. Kress, M.~S.
  Murillo, \href{https://link.aps.org/doi/10.1103/PhysRevE.95.043202}{Molecular
  dynamics studies of electron-ion temperature equilibration in hydrogen
  plasmas within the coupled-mode regime}, Phys.~Rev.~E 95~(4) (2017) 043202.
\newblock \href {http://dx.doi.org/10.1103/PhysRevE.95.043202}
  {\path{doi:10.1103/PhysRevE.95.043202}}.
\newline\urlprefix\url{https://link.aps.org/doi/10.1103/PhysRevE.95.043202}

\bibitem{vorberger_equation_2013}
J.~Vorberger, D.~O. Gericke, W.~D. Kraeft,
  \href{http://www.sciencedirect.com/science/article/pii/S1574181813000517}{The
  equation of state for hydrogen at high densities}, High Energy Density Phys.
  9~(3) (2013) 448--456.
\newblock \href {http://dx.doi.org/10.1016/j.hedp.2013.04.011}
  {\path{doi:10.1016/j.hedp.2013.04.011}}.
\newline\urlprefix\url{http://www.sciencedirect.com/science/article/pii/S1574181813000517}

\bibitem{starrett_simple_2014}
C.~E. Starrett, D.~Saumon,
  \href{http://www.sciencedirect.com/science/article/pii/S1574181813001900}{A
  simple method for determining the ionic structure of warm dense matter}, High
  Energy Density Phys. 10 (2014) 35--42.
\newblock \href {http://dx.doi.org/10.1016/j.hedp.2013.12.001}
  {\path{doi:10.1016/j.hedp.2013.12.001}}.
\newline\urlprefix\url{http://www.sciencedirect.com/science/article/pii/S1574181813001900}

\bibitem{souza_predictions_2014}
A.~N. Souza, D.~J. Perkins, C.~E. Starrett, D.~Saumon, S.~B. Hansen,
  \href{https://link.aps.org/doi/10.1103/PhysRevE.89.023108}{Predictions of
  x-ray scattering spectra for warm dense matter}, Phys.~Rev.~E 89~(2) (2014)
  023108.
\newblock \href {http://dx.doi.org/10.1103/PhysRevE.89.023108}
  {\path{doi:10.1103/PhysRevE.89.023108}}.
\newline\urlprefix\url{https://link.aps.org/doi/10.1103/PhysRevE.89.023108}

\bibitem{senatore_local_1996}
G.~Senatore, S.~Moroni, D.~M. Ceperley,
  \href{http://www.sciencedirect.com/science/article/pii/S002230939600316X}{Local
  field factor and effective potentials in liquid metals}, J.~Non-Cryst.~Solids
  205 (1996) 851--854.
\newblock \href {http://dx.doi.org/10.1016/S0022-3093(96)00316-X}
  {\path{doi:10.1016/S0022-3093(96)00316-X}}.
\newline\urlprefix\url{http://www.sciencedirect.com/science/article/pii/S002230939600316X}

\bibitem{gravel_nonlinear_2007}
S.~Gravel, N.~W. Ashcroft,
  \href{https://link.aps.org/doi/10.1103/PhysRevB.76.144103}{Nonlinear response
  theories and effective pair potentials}, Phys.~Rev.~B 76~(14) (2007) 144103.
\newblock \href {http://dx.doi.org/10.1103/PhysRevB.76.144103}
  {\path{doi:10.1103/PhysRevB.76.144103}}.
\newline\urlprefix\url{https://link.aps.org/doi/10.1103/PhysRevB.76.144103}

\bibitem{moldabekov_ion_2017}
Z.~Moldabekov, S.~Groth, T.~Dornheim, M.~Bonitz, T.~Ramazanov,
  \href{http://dx.doi.org/10.1002/ctpp.201700109}{Ion potential in non-ideal
  dense quantum plasmas}, Contrib.~Plasma Phys. 57~(10) (2017) 532--538.
\newblock \href {http://dx.doi.org/10.1002/ctpp.201700109}
  {\path{doi:10.1002/ctpp.201700109}}.
\newline\urlprefix\url{http://dx.doi.org/10.1002/ctpp.201700109}

\bibitem{gregori_derivation_2007}
G.~Gregori, A.~Ravasio, A.~H\"oll, S.~H. Glenzer, S.~J. Rose,
  \href{http://www.sciencedirect.com/science/article/pii/S157418180700016X}{Derivation
  of the static structure factor in strongly coupled non-equilibrium plasmas
  for {X}-ray scattering studies}, High Energy Density Phys. 3~(1–2) (2007)
  99--108.
\newblock \href {http://dx.doi.org/10.1016/j.hedp.2007.02.006}
  {\path{doi:10.1016/j.hedp.2007.02.006}}.
\newline\urlprefix\url{http://www.sciencedirect.com/science/article/pii/S157418180700016X}

\bibitem{dornheim_permutation_2017}
T.~Dornheim, S.~Groth, J.~Vorberger, M.~Bonitz,
  \href{https://link.aps.org/doi/10.1103/PhysRevE.96.023203}{Permutation-blocking
  path-integral {Monte} {Carlo} approach to the static density response of the
  warm dense electron gas}, Phys.~Rev. E 96~(2) (2017) 023203.
\newblock \href {http://dx.doi.org/10.1103/PhysRevE.96.023203}
  {\path{doi:10.1103/PhysRevE.96.023203}}.
\newline\urlprefix\url{https://link.aps.org/doi/10.1103/PhysRevE.96.023203}

\bibitem{muhlbacher_real-time_2008}
L.~M\"uhlbacher, E.~Rabani,
  \href{https://link.aps.org/doi/10.1103/PhysRevLett.100.176403}{Real-{Time}
  {Path} {Integral} {Approach} to {Nonequilibrium} {Many}-{Body} {Quantum}
  {Systems}}, Phys.~Rev.~Lett. 100~(17) (2008) 176403.
\newblock \href {http://dx.doi.org/10.1103/PhysRevLett.100.176403}
  {\path{doi:10.1103/PhysRevLett.100.176403}}.
\newline\urlprefix\url{https://link.aps.org/doi/10.1103/PhysRevLett.100.176403}

\bibitem{schiro_real-time_2009}
M.~Schir\'o, M.~Fabrizio,
  \href{https://link.aps.org/doi/10.1103/PhysRevB.79.153302}{Real-time
  diagrammatic {Monte} {Carlo} for nonequilibrium quantum transport},
  Phys.~Rev.~B 79~(15) (2009) 153302.
\newblock \href {http://dx.doi.org/10.1103/PhysRevB.79.153302}
  {\path{doi:10.1103/PhysRevB.79.153302}}.
\newline\urlprefix\url{https://link.aps.org/doi/10.1103/PhysRevB.79.153302}

\bibitem{schiro_real-time_2010}
M.~Schir\'o,
  \href{https://link.aps.org/doi/10.1103/PhysRevB.81.085126}{Real-time dynamics
  in quantum impurity models with diagrammatic {Monte} {Carlo}}, Phys.~Rev.~B
  81~(8) (2010) 085126.
\newblock \href {http://dx.doi.org/10.1103/PhysRevB.81.085126}
  {\path{doi:10.1103/PhysRevB.81.085126}}.
\newline\urlprefix\url{https://link.aps.org/doi/10.1103/PhysRevB.81.085126}

\bibitem{kwong_real-time_2000}
N.-H. Kwong, M.~Bonitz,
  \href{https://link.aps.org/doi/10.1103/PhysRevLett.84.1768}{Real-{Time}
  {Kadanoff}-{Baym} {Approach} to {Plasma} {Oscillations} in a {Correlated}
  {Electron} {Gas}}, Phys.~Rev.~Lett. 84~(8) (2000) 1768--1771.
\newblock \href {http://dx.doi.org/10.1103/PhysRevLett.84.1768}
  {\path{doi:10.1103/PhysRevLett.84.1768}}.
\newline\urlprefix\url{https://link.aps.org/doi/10.1103/PhysRevLett.84.1768}

\bibitem{martin_sum_1988}
P.~A. Martin, \href{https://link.aps.org/doi/10.1103/RevModPhys.60.1075}{Sum
  rules in charged fluids}, Rev.~Mod.~Phys. 60~(4) (1988) 1075--1127.
\newblock \href {http://dx.doi.org/10.1103/RevModPhys.60.1075}
  {\path{doi:10.1103/RevModPhys.60.1075}}.
\newline\urlprefix\url{https://link.aps.org/doi/10.1103/RevModPhys.60.1075}

\bibitem{eich_effective_2017}
F.~G. Eich, M.~Holzmann, G.~Vignale,
  \href{https://link.aps.org/doi/10.1103/PhysRevB.96.035132}{Effective mass of
  quasiparticles from thermodynamics}, Phys.~Rev.~B 96~(3) (2017) 035132.
\newblock \href {http://dx.doi.org/10.1103/PhysRevB.96.035132}
  {\path{doi:10.1103/PhysRevB.96.035132}}.
\newline\urlprefix\url{https://link.aps.org/doi/10.1103/PhysRevB.96.035132}

\bibitem{groth_configuration_2017}
S.~Groth, T.~Dornheim, M.~Bonitz,
  \href{http://aip.scitation.org/doi/10.1063/1.4999907}{Configuration path
  integral {Monte} {Carlo} approach to the static density response of the warm
  dense electron gas}, J.~Chem.~Phys. 147~(16) (2017) 164108.
\newblock \href {http://dx.doi.org/10.1063/1.4999907}
  {\path{doi:10.1063/1.4999907}}.
\newline\urlprefix\url{http://aip.scitation.org/doi/10.1063/1.4999907}

\bibitem{bonitz_qhd_pre_13}
M.~Bonitz, E.~Pehlke, T.~Schoof,
  \href{https://link.aps.org/doi/10.1103/PhysRevE.87.033105}{Attractive forces
  between ions in quantum plasmas: Failure of linearized quantum
  hydrodynamics}, Phys. Rev. E 87 (2013) 033105.
\newblock \href {http://dx.doi.org/10.1103/PhysRevE.87.033105}
  {\path{doi:10.1103/PhysRevE.87.033105}}.
\newline\urlprefix\url{https://link.aps.org/doi/10.1103/PhysRevE.87.033105}

\bibitem{zhandos_pop_15}
Z.~Moldabekov, T.~Schoof, P.~Ludwig, M.~Bonitz, T.~Ramazanov,
  \href{https://doi.org/10.1063/1.4932051}{Statically screened ion potential
  and bohm potential in a quantum plasma}, Phys.~Plasmas 22~(10) (2015) 102104.
\newblock \href {http://arxiv.org/abs/https://doi.org/10.1063/1.4932051}
  {\path{arXiv:https://doi.org/10.1063/1.4932051}}, \href
  {http://dx.doi.org/10.1063/1.4932051} {\path{doi:10.1063/1.4932051}}.
\newline\urlprefix\url{https://doi.org/10.1063/1.4932051}

\bibitem{kutepov_one-electron_2017}
A.~L. Kutepov, G.~Kotliar,
  \href{https://link.aps.org/doi/10.1103/PhysRevB.96.035108}{One-electron
  spectra and susceptibilities of the three-dimensional electron gas from
  self-consistent solutions of {Hedin}'s equations}, Phys.~Rev.~B 96~(3) (2017)
  035108.
\newblock \href {http://dx.doi.org/10.1103/PhysRevB.96.035108}
  {\path{doi:10.1103/PhysRevB.96.035108}}.
\newline\urlprefix\url{https://link.aps.org/doi/10.1103/PhysRevB.96.035108}

\bibitem{balzer-book}
K.~Balzer, M.~Bonitz, Nonequilibrium Green's Functions Approach to
  Inhomogeneous Systems, Springer, 2013.

\bibitem{semkat_jmp_00}
D.~Semkat, D.~Kremp, M.~Bonitz,
  \href{https://doi.org/10.1063/1.1286204}{{Kadanoff-Baym equations and
  non-Markovian Boltzmann equation in generalized T-matrix approximation}},
  Journal of Mathematical Physics 41~(11) (2000) 7458--7467.
\newblock \href {http://arxiv.org/abs/https://doi.org/10.1063/1.1286204}
  {\path{arXiv:https://doi.org/10.1063/1.1286204}}, \href
  {http://dx.doi.org/10.1063/1.1286204} {\path{doi:10.1063/1.1286204}}.
\newline\urlprefix\url{https://doi.org/10.1063/1.1286204}

\bibitem{balzer_pra_10_gg}
K.~Balzer, S.~Bauch, M.~Bonitz,
  \href{https://link.aps.org/doi/10.1103/PhysRevA.81.022510}{Efficient
  grid-based method in nonequilibrium {G}reen's function calculations:
  Application to model atoms and molecules}, Phys. Rev. A 81 (2010) 022510.
\newblock \href {http://dx.doi.org/10.1103/PhysRevA.81.022510}
  {\path{doi:10.1103/PhysRevA.81.022510}}.
\newline\urlprefix\url{https://link.aps.org/doi/10.1103/PhysRevA.81.022510}

\bibitem{schluenzen_cpp_16}
N.~Schl\"unzen, M.~Bonitz, Nonequilibrium {G}reen functions approach to
  strongly correlated fermions in lattice systems, Contrib. Plasma Phys. B 56
  (2016) 5--91.
\newblock \href {http://dx.doi.org/10.1002/ctpp.201610003}
  {\path{doi:10.1002/ctpp.201610003}}.

\bibitem{schluenzen_prb_16}
N.~Schl\"unzen, S.~Hermanns, M.~Bonitz, C.~Verdozzi,
  \href{https://link.aps.org/doi/10.1103/PhysRevB.93.035107}{Dynamics of
  strongly correlated fermions: \textit{Ab initio} results for two and three
  dimensions}, Phys. Rev. B 93 (2016) 035107.
\newblock \href {http://dx.doi.org/10.1103/PhysRevB.93.035107}
  {\path{doi:10.1103/PhysRevB.93.035107}}.
\newline\urlprefix\url{https://link.aps.org/doi/10.1103/PhysRevB.93.035107}

\bibitem{chihara_difference_1987}
J.~Chihara, \href{http://stacks.iop.org/0305-4608/17/i=2/a=002}{Difference in
  {X}-ray scattering between metallic and non-metallic liquids due to
  conduction electrons}, J.~Phys.~F 17~(2) (1987) 295.
\newblock \href {http://dx.doi.org/10.1088/0305-4608/17/2/002}
  {\path{doi:10.1088/0305-4608/17/2/002}}.
\newline\urlprefix\url{http://stacks.iop.org/0305-4608/17/i=2/a=002}

\bibitem{bonitz_prb_07}
M.~Bonitz, K.~Balzer, R.~van Leeuwen,
  \href{https://link.aps.org/doi/10.1103/PhysRevB.76.045341}{Invariance of the
  {K}ohn center-of-mass mode in a conserving theory}, Phys. Rev. B 76 (2007)
  045341.
\newblock \href {http://dx.doi.org/10.1103/PhysRevB.76.045341}
  {\path{doi:10.1103/PhysRevB.76.045341}}.
\newline\urlprefix\url{https://link.aps.org/doi/10.1103/PhysRevB.76.045341}

\bibitem{bauch_prb_09}
S.~Bauch, K.~Balzer, C.~Henning, M.~Bonitz,
  \href{https://link.aps.org/doi/10.1103/PhysRevB.80.054515}{Quantum breathing
  mode of trapped bosons and fermions at arbitrary coupling}, Phys. Rev. B 80
  (2009) 054515.
\newblock \href {http://dx.doi.org/10.1103/PhysRevB.80.054515}
  {\path{doi:10.1103/PhysRevB.80.054515}}.
\newline\urlprefix\url{https://link.aps.org/doi/10.1103/PhysRevB.80.054515}

\bibitem{henning_prl_08}
C.~Henning, K.~Fujioka, P.~Ludwig, A.~Piel, A.~Melzer, M.~Bonitz,
  \href{https://link.aps.org/doi/10.1103/PhysRevLett.101.045002}{Existence and
  vanishing of the breathing mode in strongly correlated finite systems}, Phys.
  Rev. Lett. 101 (2008) 045002.
\newblock \href {http://dx.doi.org/10.1103/PhysRevLett.101.045002}
  {\path{doi:10.1103/PhysRevLett.101.045002}}.
\newline\urlprefix\url{https://link.aps.org/doi/10.1103/PhysRevLett.101.045002}

\bibitem{mcdonald_prl_13}
C.~R. McDonald, G.~Orlando, J.~W. Abraham, D.~Hochstuhl, M.~Bonitz, T.~Brabec,
  \href{https://link.aps.org/doi/10.1103/PhysRevLett.111.256801}{Theory of the
  quantum breathing mode in harmonic traps and its use as a diagnostic tool},
  Phys. Rev. Lett. 111 (2013) 256801.
\newblock \href {http://dx.doi.org/10.1103/PhysRevLett.111.256801}
  {\path{doi:10.1103/PhysRevLett.111.256801}}.
\newline\urlprefix\url{https://link.aps.org/doi/10.1103/PhysRevLett.111.256801}

\bibitem{schmelcher_prb_13}
R.~Schmitz, S.~Kr\"onke, L.~Cao, P.~Schmelcher,
  \href{https://link.aps.org/doi/10.1103/PhysRevA.88.043601}{Quantum breathing
  dynamics of ultracold bosons in one-dimensional harmonic traps: Unraveling
  the pathway from few- to many-body systems}, Phys. Rev. A 88 (2013) 043601.
\newblock \href {http://dx.doi.org/10.1103/PhysRevA.88.043601}
  {\path{doi:10.1103/PhysRevA.88.043601}}.
\newline\urlprefix\url{https://link.aps.org/doi/10.1103/PhysRevA.88.043601}

\bibitem{balzer_epl_12}
K.~Balzer, S.~Hermanns, M.~Bonitz,
  \href{http://stacks.iop.org/0295-5075/98/i=6/a=67002}{Electronic double
  excitations in quantum wells: Solving the two-time {Kadanoff-Baym}
  equations}, EPL (Europhysics Letters) 98~(6) (2012) 67002.
\newline\urlprefix\url{http://stacks.iop.org/0295-5075/98/i=6/a=67002}

\bibitem{arkhipov_dielectric_2014}
Y.~V. Arkhipov, A.~B. Ashikbayeva, A.~Askaruly, A.~E. Davletov, I.~M.
  Tkachenko,
  \href{http://link.aps.org/doi/10.1103/PhysRevE.90.053102}{Dielectric function
  of dense plasmas, their stopping power, and sum rules}, Phys.~Rev.~E 90~(5)
  (2014) 053102.
\newblock \href {http://dx.doi.org/10.1103/PhysRevE.90.053102}
  {\path{doi:10.1103/PhysRevE.90.053102}}.
\newline\urlprefix\url{http://link.aps.org/doi/10.1103/PhysRevE.90.053102}

\bibitem{berne_path_1986}
B.~J. Berne, \href{https://link.springer.com/article/10.1007/BF02628319}{Path
  integral {Monte} {Carlo} methods: {Static}- and time-correlation functions},
  J.~Stat.~Phys. 43~(5-6) (1986) 911--929.
\newblock \href {http://dx.doi.org/10.1007/BF02628319}
  {\path{doi:10.1007/BF02628319}}.
\newline\urlprefix\url{https://link.springer.com/article/10.1007/BF02628319}

\bibitem{jarrell_bayesian_1996}
M.~Jarrell, J.~E. Gubernatis,
  \href{http://www.sciencedirect.com/science/article/pii/0370157395000747}{Bayesian
  inference and the analytic continuation of imaginary-time quantum {Monte}
  {Carlo} data}, Phys.~Rep. 269~(3) (1996) 133--195.
\newblock \href {http://dx.doi.org/10.1016/0370-1573(95)00074-7}
  {\path{doi:10.1016/0370-1573(95)00074-7}}.
\newline\urlprefix\url{http://www.sciencedirect.com/science/article/pii/0370157395000747}

\bibitem{ferre_dynamic_2016}
G.~Ferr\'e, J.~Boronat,
  \href{https://link.aps.org/doi/10.1103/PhysRevB.93.104510}{{Dynamic structure
  factor of liquid $^4$He across the normal-superfluid transition}},
  Phys.~Rev.~B 93~(10) (2016) 104510.
\newblock \href {http://dx.doi.org/10.1103/PhysRevB.93.104510}
  {\path{doi:10.1103/PhysRevB.93.104510}}.
\newline\urlprefix\url{https://link.aps.org/doi/10.1103/PhysRevB.93.104510}

\bibitem{motta_imaginary_2014}
M.~Motta, D.~E. Galli, S.~Moroni, E.~Vitali,
  \href{http://scitation.aip.org/content/aip/journal/jcp/140/2/10.1063/1.4861227}{Imaginary
  time correlations and the phaseless auxiliary field quantum {Monte} {Carlo}},
  J.~Chem.~Phys. 140~(2) (2014) 024107.
\newblock \href {http://dx.doi.org/10.1063/1.4861227}
  {\path{doi:10.1063/1.4861227}}.
\newline\urlprefix\url{http://scitation.aip.org/content/aip/journal/jcp/140/2/10.1063/1.4861227}

\bibitem{motta_imaginary_2015}
M.~Motta, D.~E. Galli, S.~Moroni, E.~Vitali,
  \href{http://scitation.aip.org/content/aip/journal/jcp/143/16/10.1063/1.4934666}{Imaginary
  time density-density correlations for two-dimensional electron gases at high
  density}, J.~Chem.~Phys. 143~(16) (2015) 164108.
\newblock \href {http://dx.doi.org/10.1063/1.4934666}
  {\path{doi:10.1063/1.4934666}}.
\newline\urlprefix\url{http://scitation.aip.org/content/aip/journal/jcp/143/16/10.1063/1.4934666}

\bibitem{levy_implementation_2017}
R.~Levy, J.~P.~F. LeBlanc, E.~Gull,
  \href{http://www.sciencedirect.com/science/article/pii/S0010465517300309}{Implementation
  of the maximum entropy method for analytic continuation}, Comp.~Phys.~Comm.
  215~(Supplement C) (2017) 149--155.
\newblock \href {http://dx.doi.org/10.1016/j.cpc.2017.01.018}
  {\path{doi:10.1016/j.cpc.2017.01.018}}.
\newline\urlprefix\url{http://www.sciencedirect.com/science/article/pii/S0010465517300309}

\bibitem{prokofev_spectral_2013}
N.~V. Prokof’ev, B.~V. Svistunov,
  \href{https://link.springer.com/article/10.1134/S002136401311009X}{Spectral
  analysis by the method of consistent constraints}, JETP Letters 97~(11)
  (2013) 649--653.
\newblock \href {http://dx.doi.org/10.1134/S002136401311009X}
  {\path{doi:10.1134/S002136401311009X}}.
\newline\urlprefix\url{https://link.springer.com/article/10.1134/S002136401311009X}

\bibitem{vitali_ab_2010}
E.~Vitali, M.~Rossi, L.~Reatto, D.~E. Galli,
  \href{http://link.aps.org/doi/10.1103/PhysRevB.82.174510}{{Ab initio
  low-energy dynamics of superfluid and solid $^4$He}}, Phys.~Rev.~B 82~(17).
\newblock \href {http://dx.doi.org/10.1103/PhysRevB.82.174510}
  {\path{doi:10.1103/PhysRevB.82.174510}}.
\newline\urlprefix\url{http://link.aps.org/doi/10.1103/PhysRevB.82.174510}

\bibitem{otsuki_sparse_2017}
J.~Otsuki, M.~Ohzeki, H.~Shinaoka, K.~Yoshimi,
  \href{https://link.aps.org/doi/10.1103/PhysRevE.95.061302}{Sparse modeling
  approach to analytical continuation of imaginary-time quantum {Monte} {Carlo}
  data}, Phys.~Rev.~E 95~(6) (2017) 061302.
\newblock \href {http://dx.doi.org/10.1103/PhysRevE.95.061302}
  {\path{doi:10.1103/PhysRevE.95.061302}}.
\newline\urlprefix\url{https://link.aps.org/doi/10.1103/PhysRevE.95.061302}

\bibitem{schott_comparison_2016}
J.~Sch\"ott, E.~G. C.~P. van Loon, I.~L.~M. Locht, M.~I. Katsnelson,
  I.~Di~Marco,
  \href{https://link.aps.org/doi/10.1103/PhysRevB.94.245140}{Comparison between
  methods of analytical continuation for bosonic functions}, Phys.~Rev.~B
  94~(24) (2016) 245140.
\newblock \href {http://dx.doi.org/10.1103/PhysRevB.94.245140}
  {\path{doi:10.1103/PhysRevB.94.245140}}.
\newline\urlprefix\url{https://link.aps.org/doi/10.1103/PhysRevB.94.245140}

\bibitem{hochstuhl_jcp_11}
D.~Hochstuhl, M.~Bonitz, \href{https://doi.org/10.1063/1.3553176}{Two-photon
  ionization of helium studied with the multiconfigurational time-dependent
  {Hartree-Fock} method}, The Journal of Chemical Physics 134~(8) (2011)
  084106.
\newblock \href {http://arxiv.org/abs/https://doi.org/10.1063/1.3553176}
  {\path{arXiv:https://doi.org/10.1063/1.3553176}}, \href
  {http://dx.doi.org/10.1063/1.3553176} {\path{doi:10.1063/1.3553176}}.
\newline\urlprefix\url{https://doi.org/10.1063/1.3553176}

\bibitem{hochstuhl_epjst_14}
D.~Hochstuhl, C.~Hinz, M.~Bonitz,
  \href{https://doi.org/10.1140/epjst/e2014-02092-3}{Time-dependent
  multiconfiguration methods for the numerical simulation of photoionization
  processes of many-electron atoms}, The European Physical Journal Special
  Topics 223~(2) (2014) 177--336.
\newblock \href {http://dx.doi.org/10.1140/epjst/e2014-02092-3}
  {\path{doi:10.1140/epjst/e2014-02092-3}}.
\newline\urlprefix\url{https://doi.org/10.1140/epjst/e2014-02092-3}

\bibitem{schattke-book}
W.~Schattke, M.~van Howe (Eds.), Solid-State Photoemission and Related Methods:
  Theory and Experiment, Wiley-VCH, 2003.

\bibitem{PhysRevE.66.046405}
W.~D. Kraeft, M.~Schlanges, J.~Vorberger, H.~E. DeWitt,
  \href{https://link.aps.org/doi/10.1103/PhysRevE.66.046405}{Kinetic and
  correlation energies and distribution functions of dense plasmas}, Phys. Rev.
  E 66 (2002) 046405.
\newblock \href {http://dx.doi.org/10.1103/PhysRevE.66.046405}
  {\path{doi:10.1103/PhysRevE.66.046405}}.
\newline\urlprefix\url{https://link.aps.org/doi/10.1103/PhysRevE.66.046405}

\bibitem{filinov_jetpl_00}
V.~S. Filinov, M.~Bonitz, V.~E. Fortov,
  \href{http://link.springer.com/article/10.1134/1.1427127}{High-density
  pheomena in hydrogen plasma}, JETP Lett. 72~(5) (2000) 361--365.
\newline\urlprefix\url{http://link.springer.com/article/10.1134/1.1427127}

\bibitem{filinov_pla_00}
V.~S. Filinov, V.~E. Fortov, M.~Bonitz, D.~Kremp,
  \href{http://link.springer.com/article/10.1134/1.1427127}{Pair distribution
  functions of dense partially ionized hydrogen plasma}, Phys. Lett. A 72~(5)
  (2000) 228--235.
\newline\urlprefix\url{http://link.springer.com/article/10.1134/1.1427127}

\bibitem{bonitz_prl_05}
M.~Bonitz, V.~S. Filinov, V.~E. Fortov, P.~R. Levashov, H.~Fehske,
  \href{https://link.aps.org/doi/10.1103/PhysRevLett.95.235006}{Crystallization
  in two-component coulomb systems}, Phys. Rev. Lett. 95 (2005) 235006.
\newblock \href {http://dx.doi.org/10.1103/PhysRevLett.95.235006}
  {\path{doi:10.1103/PhysRevLett.95.235006}}.
\newline\urlprefix\url{https://link.aps.org/doi/10.1103/PhysRevLett.95.235006}

\bibitem{kelbg_63_1}
G.~Kelbg, \href{http://dx.doi.org/10.1002/andp.19634670308}{{Theorie des
  Quanten-Plasmas}}, Ann. Phys., Lpz. 467~(3-4) (1963) 219--224.
\newblock \href {http://dx.doi.org/10.1002/andp.19634670308}
  {\path{doi:10.1002/andp.19634670308}}.
\newline\urlprefix\url{http://dx.doi.org/10.1002/andp.19634670308}

\bibitem{kelbg_63_2}
G.~Kelbg, \href{http://dx.doi.org/10.1002/andp.19634670703}{{Quantenstatistik
  der Gase mit Coulomb-Wechselwirkung}}, Ann. Phys., Lpz. 467~(7-8) (1963)
  354--360.
\newblock \href {http://dx.doi.org/10.1002/andp.19634670703}
  {\path{doi:10.1002/andp.19634670703}}.
\newline\urlprefix\url{http://dx.doi.org/10.1002/andp.19634670703}

\bibitem{kelbg_64}
G.~Kelbg, \href{http://dx.doi.org/10.1002/andp.19644690705}{{Klassische
  statistische Mechanik der Teilchen-Mischungen mit sortenabhängigen
  weitreichenden zwischenmolekularen Wechselwirkungen}}, Ann. Phys., Lpz.
  469~(7-8) (1964) 394--403.
\newblock \href {http://dx.doi.org/10.1002/andp.19644690705}
  {\path{doi:10.1002/andp.19644690705}}.
\newline\urlprefix\url{http://dx.doi.org/10.1002/andp.19644690705}

\bibitem{filinov_jpa_03}
A.~V. Filinov, M.~Bonitz, W.~Ebeling, Improved {K}elbg potential for correlated
  {C}oulomb systems, J. Phys. A: Math. General 36 (2003) 5957--5962.

\bibitem{filinov_pre_04}
A.~V. Filinov, V.~O. Golubnychiy, M.~Bonitz, W.~Ebeling, J.~W. Dufty,
  \href{https://link.aps.org/doi/10.1103/PhysRevE.70.046411}{Temperature-dependent
  quantum pair potentials and their application to dense partially ionized
  hydrogen plasmas}, Phys. Rev. E 70 (2004) 046411.
\newblock \href {http://dx.doi.org/10.1103/PhysRevE.70.046411}
  {\path{doi:10.1103/PhysRevE.70.046411}}.
\newline\urlprefix\url{https://link.aps.org/doi/10.1103/PhysRevE.70.046411}

\bibitem{our_git}
\href{https://github.com/agbonitz/xc_functional}{Git-repository: Ab-initio
  thermodynamic description of the warm dense electron gas} (2017).
\newline\urlprefix\url{https://github.com/agbonitz/xc_functional}

\end{thebibliography}

\end{document}